\documentclass[12pt,a4paper]{article}
\pdfoutput=1
\usepackage{adjustbox}
\usepackage{color}
\usepackage{amssymb,amsmath,bm,bbm}
\usepackage{epsf}
\usepackage{epsfig}
\usepackage{afterpage}
\usepackage{longtable}
\usepackage[dvipsnames]{xcolor}
\usepackage[linktoc=page,bookmarks=false,colorlinks=false,
linkbordercolor=RoyalBlue,citebordercolor=ForestGreen,urlbordercolor=CornflowerBlue]{hyperref}
\usepackage{latexsym,mathrsfs,dsfont}
\usepackage[normalem]{ulem} 
\usepackage[compress]{cite}
\usepackage{graphicx}
\usepackage{url}
\usepackage{paralist}
\usepackage{bbold}

\newcommand{\as}{\alpha_s}
\newcommand{\aem}{\alpha}

\newcommand{\GSM}{{\mathrm{G_{SM}}}}
\newcommand{\GSMUpr}{{\mathrm{G^\prime_{SM}}}}
\definecolor{green1}{rgb}{0.06,0.66,0.06}
\definecolor{red}{cmyk}{0,1,1,0.4}
\newcommand{\green}{{\color{green1}$\bigstar$}}
\newcommand{\red}{{\color{red}\small \protect\raisebox{-0.05em}{$\blacksquare$}}}
\newcommand{\sgreen}{{\color{green1}$\star$}}
\newcommand{\sred}{{\color{red}\tiny \protect\raisebox{-0.05em}{$\blacksquare$}}}
\definecolor{darkgreen}{rgb}{0.0,0.6,0.0}
\definecolor{darkblue}{RGB}{12,13,115}
\definecolor{darkred}{RGB}{204,6,0}

\newcommand{\rSt}{{\color{darkblue}$\bigstar$}}
\newcommand{\bSt}{{\color{darkred}$\bigstar$}}
\newcommand{\rst}{{\color{darkblue}$\star$}}
\newcommand{\bst}{{\color{darkred}$\star$}}

\newcommand{\mvs}{\vbox{\vskip 8mm}}

\newcommand{\Lms}{\Lambda_{\overline{\rm MS}}}

\newcommand{\nnb}{\nonumber}
\newcommand{\ord}{\mathcal{O}}
\usepackage{slashed}

\newcommand{\IM}{{\rm Im}}
\newcommand{\RE}{{\rm Re}}

\newcommand{\tev}{\, {\rm TeV}}
\newcommand{\gev}{\, {\rm GeV}}
\newcommand{\mev}{\, {\rm MeV}}

\newcommand{\vcb}{|V_{cb}|}
\newcommand{\vtd}{|V_{td}|}
\newcommand{\vub}{|V_{ub}|}
\newcommand{\vts}{|V_{ts}|}
\newcommand{\vus}{|V_{us}|}

\newcommand{\bsi}{B_6^{(1/2)}}
\newcommand{\bei}{B_8^{(3/2)}}

\def\epe{\varepsilon'/\varepsilon}
\newcommand{\beq}{\begin{equation}}
\newcommand{\eeq}{\end{equation}}
\newcommand{\be}{\begin{equation}}
\newcommand{\ee}{\end{equation}}
\newcommand{\bi}{\begin{itemize}}
\newcommand{\ei}{\end{itemize}}
\newcommand{\ba}{\begin{array}}
\newcommand{\ea}{\end{array}}
\newcommand{\beqa}{\begin{eqnarray}}
\newcommand{\eeqa}{\end{eqnarray}}
\newcommand{\bea}{\begin{eqnarray}}
\newcommand{\eea}{\end{eqnarray}}
\newcommand{\beqn}{\begin{eqnarray}}
\newcommand{\eeqn}{\end{eqnarray}}

\newcommand{\D}{\Delta}

\newcommand{\eps}{\epsilon}
\newcommand{\nn}{\nonumber}
\newcommand{\ov}{\overline}
\newcommand{\mc}{\mathcal}

\newcommand{\im}{{\rm Im}}

\newcommand{\epsK}{\varepsilon_K}

\definecolor{red}{cmyk}{0,1,1,0.4}

\def\kpn{K^+\rightarrow\pi^+\nu\bar\nu}
\def\klpn{K_{L}\rightarrow\pi^0\nu\bar\nu}
\def\klpll{K_L\to\pi^0\ell\bar\ell}

\def\ksm{K_S\to\mu\bar\mu}
\def\klm{K_L\to\mu\bar\mu}

\newcommand{\keps}{\kappa_{\varepsilon}}
\newcommand{\muLow}{{\mu_\text{had}}}
\newcommand{\muEW}{{\mu_\text{ew}}}

\begin{document}

\begin{flushleft}
{\em Version of \today}
\end{flushleft}

\vspace{-14mm}
\begin{flushright}
        {AJB-26-1}
\end{flushright}

\vspace{8mm}

\begin{center}
{\LARGE\bf
\boldmath{Addicted to Flavour: 1976-2026}
\\[8mm]
{\large\bf Andrzej~J.~Buras \\[0.3cm]}}
{\small 
      TUM Institute for Advanced Study, Lichtenbergstr.~2a, D-85748 Garching, Germany\\
      Physik Department, TUM School of Natural Sciences, TU M\"unchen,\\ James-Franck-Stra{\ss}e, D-85748 Garching, Germany\\
      E-mail: aburas@ph.tum.de }\\
\vspace{2mm}
{\bf Dedicated to My Flavour Collaborators}
\end{center}

\vspace{4mm}

\begin{abstract}
  \noindent
I describe my activities in Flavour Physics from 1976 to 2026. However,  this 50th anniversary is not the only motivation for this writing. The second reason
  is the 350th anniversary of the discovery of the first animalcula by van Leeuvanhoek in 1676.
Flavour physics makes it possible to search for new animalcula at distance scales far shorter than those resolved by van Leeuwenhoek in 1676 and even shorter than those directly accessible at the Large Hadron Collider. Achieving this goal requires not only precise measurements of a wide variety of processes, but also equally precise theoretical calculations, both within the Standard Model (SM) and beyond it. In this respect, next-to-leading-order (NLO) and next-to-next-to-leading-order (NNLO) QCD calculations of various Wilson coefficients  in the SM and beyond it,  in which I was involved for two decades, as well as reliable treatments of non-perturbative QCD effects, are indispensable.
Equally important is the proper choice of observables that are best suited to revealing these new animalcula of particle physics. Moreover, in my view it is crucial to develop strategies for the search for New Physics (NP) that go beyond the global fits that are very popular today. While effective field theories such as WET and SMEFT are formulated in terms of Wilson coefficients of the relevant operators, with correlations characteristic of the SM and of specific NP scenarios, the most direct tests of the SM and its extensions are, in my opinion, correlations among different observables that are characteristic of particular new animalcula at work. Numerous colourful plots in this article illustrate this point.
I hope that these ideas are clearly conveyed in my Flavour Autobiography, which also includes my memories of many conferences, workshops, and schools, as well as related anecdotes that are not always directly connected to physics.
\end{abstract}

\vspace{4mm}
\setcounter{page}{0}
\thispagestyle{empty}
\newpage

\tableofcontents
\vspace{4mm}
       {\bf \large Modifications relative to V1 of 7 January  2026}

       \vspace{2mm}
       In addition to correcting obvious misprints and new references the following modifications have been
       made:
       \begin{itemize}
       \item
         A few additions to the main text, indicated in boldface.
         They can be found on pages 30, 106, 150, 153, 233, 162, 245, 280, 281.
         New references can be found there.
       \item
         Update of citations in Table~\ref{tab:ranking}.
       \item
         New Appendix~\ref{SEF}: TUM Senior Excellence Faculty.
       \item
         New Appendix~\ref{Oppenheimer}: Meeting Robert Oppenheimer.
         \end{itemize}

\newpage

\part{Overture and Outline}
\section{Overture}
The year 2026 is rather special for me. It is the 50th anniversary of my activities in flavour physics. Strictly speaking after two flavour papers in 1976, 
between 1976 and 1981 I was working on deep-inelastic scattering (DIS),  but these five years had also an impact on my strategies in flavour physics. The point is
that being of the old generation of physicists my computer skills were always
rather limited, although in the 1980s I was using Fortran for numerical evaluations and in the 1990s Wolfram's Mathematica for the evaluation of traces of the chains of many Dirac-$\gamma$ matrices present
in two-loop calculations and for the addition of analytic results for numerous
two-loop diagrams that I always obtained by using my right hand with some help from my brains.

Similarly, until today my skills in  making  plots remained rather limited. Yet, in the projects in which  plots were required I had always
excellent younger collaborators, my diploma students, my PhD students, postdocs
and other collaborators with higher computer skills than mine. On three occasions   my older son Robert did such plots for me, in particular for my book 
\cite{Buras:2020xsm}.

Looking back on my research in the last 50 years all these deficiencies were
fortunate for me and this motivated me to strees it here and to describe my research to demonstrate
to younger generations that also in the phenomenology one can obtain useful results without knowing how to use computer codes. Simply, because then one is forced
to take other routes and make shortcuts which often produce the results that one would not get if one used computer codes only.

Returning to my activities in  deep-inelastic scattering (1976-1981) there were two projects
that had an impact on my strategies in flavour physics:
\begin{itemize}
  \item
    As I was unable to perform a Mellin transformation by computer to obtain
    from the moments of parton distributions with QCD effects their numerical
    dependence on Bjorken-x, I modified the known Feynman-Field parametrization
    of valence quark distributions without these effects \cite{Field:1976ve}. I simply introduced
    logarithmic $Q^2$ dependence in two exponents present in their
    formula with two parameters that I determined  by calculating their moments by hand
    \cite{Buras:1977ra}. More details can be found in Appendix~\ref{PD77}. This was in fact the first analytic expression
    for valence quark  distribution presented in the literature that included QCD effects.  The ones for gluon and sea distributions
    have been obtained by means of this method in collaboration with my friend Karel Gaemers \cite{Buras:1977yj,Buras:1977hj}.
    For many years  these distributions were called Buras-Gaemers distributions. They were significantly improved numerically by other researchers since then but still earned 747 citations until now.
  \item
    Already in the 1970s it was possible to use computers for loop calculations,
    in particular by means of Veltman's Schoonship. But in my main contribution
    to deep-inelastic scattering with Bardeen, Duke and Muta \cite{Bardeen:1978yd}, that led to the $\overline{\text{MS}}$ scheme, we did all two-loop calculations by hand.
    This was nothing special  because most of two-loop calculations at that time have been done by hand as Wolfram's Mathematica did not exist yet. But this
    was a very good training for future NLO  QCD calculations in flavour physics. In particular the realization that it is a great fun in doing such calculations by hand and not giving this pleasure to a computer. While for deep-inelastic
    scattering this fun was only moderate, it increased a decade later for flavour physics because of the presence of $\gamma_5$ and related evanescent operators.
    \end{itemize}

\begin{figure}[t]
\centering%
\includegraphics[width=0.8\textwidth]{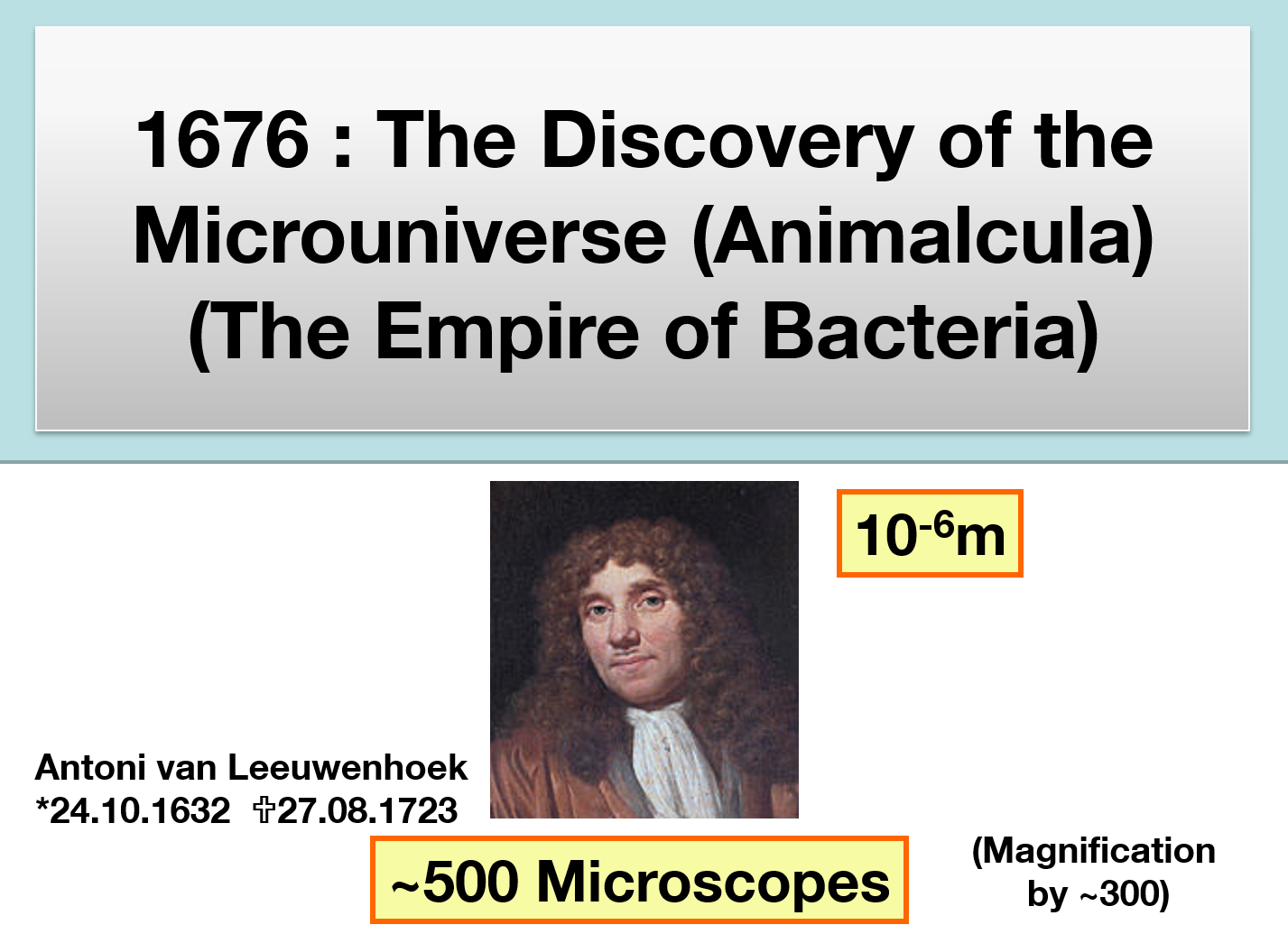}%
\caption{\it Antoni van Leeuwenhoek.\label{Antoni}}
\end{figure}

\begin{figure}[t]
\centering%
\includegraphics[width=0.8\textwidth]{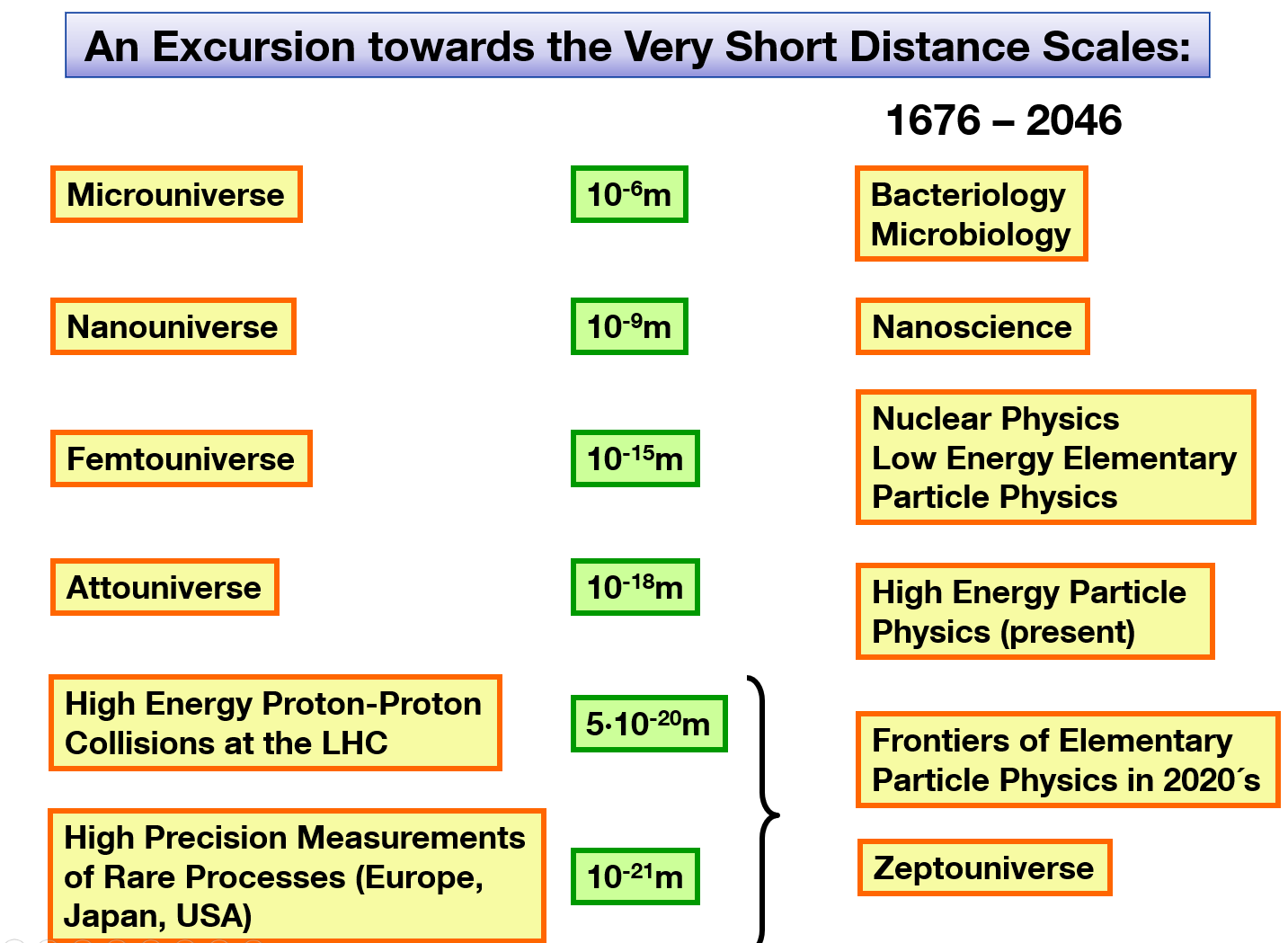}%
\caption{\it Progress in Resolving Short Distance Scales 1676-2026.\label{Progress}}
\end{figure}

As there are very many topics in particle physics one could ask why I have
concentrated for such a long time entirely on flavour physics and in particular
on processes that in the Standard Model (SM) are governed by loops being
forbidden by GIM mechanism \cite{Glashow:1970gm} to appear already at tree-level. There were in particular two reasons:

\begin{itemize}
\item
  In 1980 I read a very interesting science book by Paul de Kruif: {\em Microbe Hunters}\footnote{The book was published in 1926, another anniversary. I read its polish translation of 1948 that I found among the books of my father.}.
  It described first of all the discovery of the empire of bacteria made in 1676
by Antoni van Leeuvenhoek (1632-1723) (See Fig.~\ref{Antoni}).  
He called these small creatures {\it animalcula} (small animals). This 
discovery was a mile stone in our civilization  for several reasons.
 He discovered invisible to us creatures which over thousands of years 
 were systematcally killing the humans, often responsible for millions 
 of death in one year. While Antoni van Leeuvanhoek did not know that 
 bacteria could be dangerous for humans, his followers like L. Pasteur (1822-1895),
 Robert Koch (1843-1910) and other {\it microbe hunters} not only realized 
 the danger coming from this tiny creatures but also developed weapons against
 this empire. Moreover, he was the first human who looked at short distance scales invisible to  us, discovering thereby
 a new {\it underground world}. 

 While van Leeuvanhoek could reach the resolution down to roughly 
$10^{-6}$m, over the last  350 years this resolution could be improved 
 by many orders of magnitude reaching already in 1980 the resolution
 of $10^{-15}$m  in the context of low energy elementary particle physics. As seen in Fig.~\ref{Progress} it has further
 been improved since then and I will elaborate on it as we proceed. But
 this book motivated me to look at loop induced processes in the SM which
 could allow to sense the presence of the top quark well before its discovery.
 In fact the top quark was discovered 15 years later: 1995. Since 1995
 my main motivation for exploring flavour physics was the search for new particles beyond the
 SM ones and this will play an important role in my story. 
\item
  The second reason was the multitude of flavour violating processes for
  which one could derive elegant and often simple expressions by calculating
  first tree and then one-loop and two-loop diagrams with internal virtual SM
  particles or new particles. Having them one can easily play with them
  eliminating poorly known parameters and finding correlations between various
  observables which would provide good test of the theory involved.
  \end{itemize}

As already stated above, in 2026 there is still another anniversary: the 350th
anniversary of the discovery of the first animalcula by van Leeuwenhoek in 1676. This is still
another reason why I wanted to describe my strategies for hunting new animalcula
beyond the ones of the SM, in particular those strategies that differ from
those of other particle theorists and which do not require sophisticated
computer codes from the start.

Yet, I should not give the wrong impression that I am against computer codes.
In the intermediated  stages of the analyses and in particular in the final ones
efficient codes are very important and in the case of the Standard Model Effective Field Theory (SMEFT) \cite{Buchmuller:1985jz,Grzadkowski:2010es} they are often indispensable. This is evident
from numerous papers written since 2010 with the most recent reviews
in \cite{Brivio:2017vri,Isidori:2023pyp,Aebischer:2025qhh}.

Similarly, I do not wish to give the impression that physics was the only thing I did during the last fifty years. A substantial part of my life was devoted to my family: my wife Gurli; our children Robert, Karin, and Allan; our daughters-in-law Franzi and Ute; and our six grandchildren, Freya, Falk, Elisabeth, Inga, Janosch, and Arvid. We have had—and continue to have—a wonderful time together, especially as they all live in Munich or nearby. Over the years, we have undertaken many trips to different countries and spent numerous vacations together.
Since my family preferred that this aspect of my life not be described in the present autobiography, I have recorded these experiences in a separate autobiographical account intended exclusively for family members.

\section{Outline}
After this long introduction let me make the outline of 
my flavour story. It it given by the table of contents. I had a dilemma whether
I should follow precisely the chronological order or divide the presentation
in topics. As I often worked parallel on different topics I decided to
divide the presentations in different topics that are represented by the
parts of my story. However, in each part I tried to follow chronological order
as far as possible. The express description of these parts is as follows.

{\bf Part II} I will describe my  first steps in flavour physics, first at CERN in 1976
and then in the last years at Fermilab and first two years at MPI for Physics in
Munich. This is the period 1980-1984.

{\bf Part III} This part describes possibly my best achievement at MPI,
that is the development in collaboration with Bill Bardeen and Jean-Marc G\'erard of a non-perturbative approach to Kaon physics: Dual QCD (DQCD). This allowed to identify the dynamics behind the $\Delta I=1/2$ rule 26 years before this was possible
with Lattice QCD (LQCD). Also our pioneering calculation of the parameter $\hat B_K$ in QCD, relevant for CP violation in $K^0-\bar K^0$ mixing was an important result.

{\bf Part IV} It deals with heavy top quark effects in FCNC processes prior to
top quark discovery. In this context the ratio $\epe$ played an important role
due to first measurements of this ratio. The calculation of isospin breaking
corrections to QCD penguin contributions to $\epe$ and the enhancement of the suppression of this ratio by the electroweak penguin contributions for large $m_t$
were the most important results of this part. Also the formulation of
the gauge-invariant Penguin-Box Expansion turned out to be very useful
for the phenomenology of flavour changing neutral current processes (FCNC) both in the SM and beyond it.

It also contains  a detailed update of strategies for CKM and FCNC processes
after the top quark discovery in 1995. 

{\bf Part V} This part describes my most important contributions not only
to Flavour Physics but to particle physics as a whole even more important
than the development of the $\overline{\text{MS}}$ scheme for QCD calculations.
It deals with the calculations of NLO and NNLO QCD calculations to most important rare $K$ and $B$ decays and particle-antiparticle mixing.

{\bf Part VI} In this part various insights in the CKM matrix \cite{Cabibbo:1963yz,Kobayashi:1973fv}  and the Unitarity Triangle (UT) obtained in my group will be presented.
 This includes in particular the improved Wolfenstein
parametrization and the introduction of the parameters $(\bar\varrho,\bar\eta)$
that describy the apex of the UT. In this part I will also describe various
strategies for the determination of the CKM matrix and their tests which took
place not only in 1994 but also in this decade.

{\bf Part VII} This part decribes numerous strategies for identifying NP
in FCNC processes that have been developed in my group during the first two decades of this millennium which go beyond the analyses of specific models
that will be discussed in {\bf Part X.}

{\bf Part VIII} This part deals predominantly with the parameter $\varepsilon_K$. In particular it deals with the calculation of long distance contributions to this important quantity.  Moreover the importance of  $\varepsilon_K$
in the tests of the Minimal Flavour Violation (MFV) will be emphasized.

{\bf Part IX} This part deals with my activities in non-leptonic two-body B decays
that were concentrated in the second half of the 1990s and in the first years
of this millennium. It describes the papers with Robert Fleischer and few other
collaborators.

{\bf Part X} This part describes possibly my main contributions to FCNC
processes except for Part V that I rate higher. It deals with detailed
analyses of specific NP models. It contains 15 sections. My PhD students and
Postdocs contributed in a crucial manner to these analyses. There is no
group worldwide that analyzed so many NP models in details.
This means deriving Feynman Rules, calculating flavour observables and
studying phenomenological implications of these models. In particular through
correlations between different observables.

{\bf Part XI} It describes the Renaissance of DQCD which took place from 2014 to 2018.
 Starting in 2014 we have improved
and generalized in six papers our studies of 1985-1987 (Part III).

{\bf Part XII} This part deals exclusively with the master formulae and
anatomies of $\epe$ beyond the SM that have been developed from 2018 til 2020. It also summarizes the present status
of $\epe$ in the SM.

{\bf Part XIII} This part deals with my  activities in the context
of the Weak Effective Theory (WET) and the Standard Model Effective Theory (SMEFT). Except for Part XIV it describes my main activities
in the 2020s until now.

{\bf Part XIV} Here I describe strategies on the elimination of the $\vcb$ dependence in the SM  phenomenology of FCNC processes. While this idea goes back to
my solo 2003 paper, the latest papers with Elena Venturini generalized it to
all K and B decays. In my view this is the most efficient strategy to find
out whether there is any NP at work in FCNC processes without doing a global analysis and getting involved in controversial strategies related to the exclusive and inclusive determinations of $\vcb$.

{\bf Part XV} The final part includes the summary of my flavour expedition,
my shopping list and my general view on the present status of flavour physics.

I decided to avoid technical details as much as possible. For those readers
less familiar with flavour physics I collected some information
in several {\bf Appendices}. In App~\ref{BFramework} the basic formulae based on operator product expansion (OPE),  in App.~ \ref{HeffSM} the
operators in the SM and in App.~ \ref{HeffBSM} most important operators
beyond the SM.
In App.~\ref{UsefulEX}
I list a number of useful expressions for the most interesting
branching ratios within the SM that being theoretically clean should play an important role in the indirect search for NP in the rest of this decade.

 During my flavour expedition I  have been invited to several conferences to report on
  our work  and to review the field of Rare Decays and CP violation as a whole, in particular at the ICHEP 96 conference in Warsaw and the
  EPS-HEP 2009 in Cracow. Moreover
  I have been invited to Winter and Summer Schools. The list of them will be given in App.~\ref{Lectures}. In App.~\ref{Book} I will describe the realization of my book    and in App.~\ref{Conferences} the most important
  review talks at various conferences and workshops are listed.
In App.~\ref{Memo} I collected some additional
memories from conferences, schools and non-flavour collaborations in addition to the ones presented in   the main text. Among them are some personal
experiences that are not known even to my friends. This is also the case of the Appendices I-K.

 \section{My Theory Collaborators}
\begin{figure}[t]
\centering%
\includegraphics[width=0.8\textwidth]{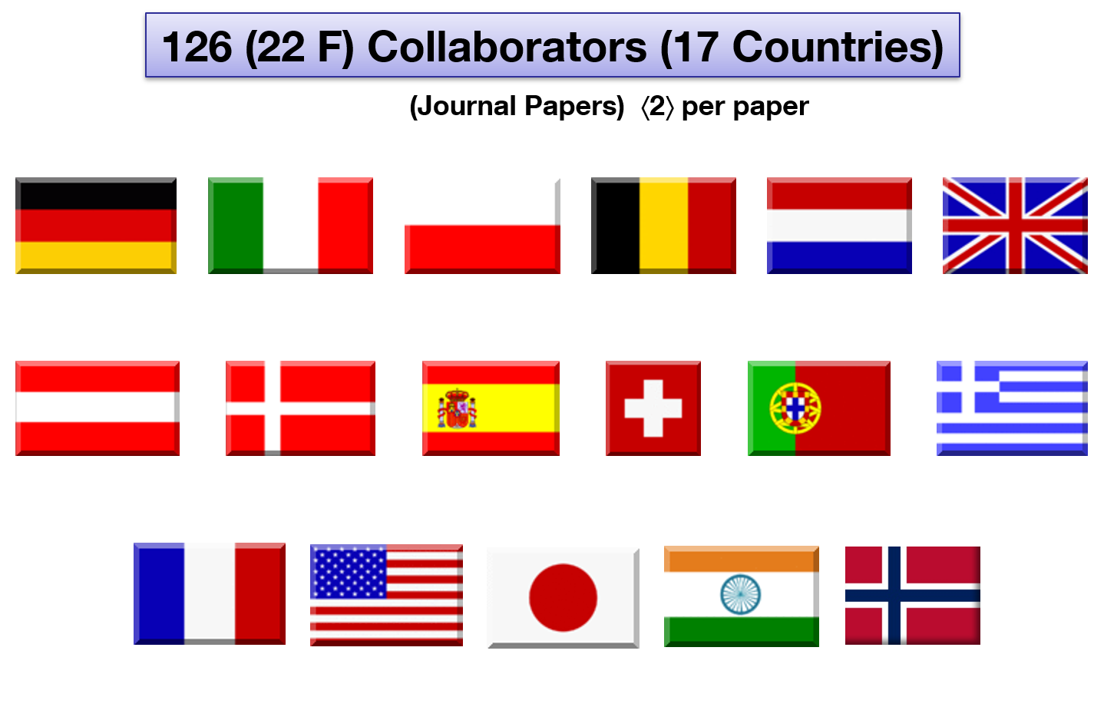}%
\caption{\it The Countries of My Collaborators.\label{FIG1}}
\end{figure}

  As stated on the front page, my flavour  story is dedicated to my flavour collaborators. As summarized in Fig.~\ref{FIG1}, I had until now 126 collaborators from 17 countries, where only papers with less than 10 authors are taken into account. On the
  average I had only two collaborators per paper. 22 among these collaborators were females,   which is an important achievement in theoretical particle physics  as their number is unfortunately low. While
  the largest number of my collaborators were German, my PhD students (35 in total),  I had many Italian   collaborators, 20 in total, among them 5 females. I did not write a single paper with two of my excellent PhD students: Stefan Herrlich and Alex Lenz.
  
  Most of my collaborators played an important role in my research, although I
  have written of about 30 papers alone not counting conference proceedings.
  But very many results obtained in my research would not be possible
  without their participation  as will be evident from my flavour story.
  Their names related to particular projects will be mentioned  as we proceed but for fun I already now present  in Table~\ref{tab:ranking} a ranking of my collaborators as far as the number of papers with me is concerned and a second ranking   of joined citations.
Only papers with less than 10
authors were taken into account and as a source INSPIRE has been used on
the day shown on the front page.

  I did not include in the first column  authors with whom I have written less than four   papers on flavour physics because then the table would not fit
  on one page. Among  them are my three excellent PhD students Ulrich Haisch,
  Ulrich Nierste and Emmanuel Stamou. They contributed to  flavour physics in an important manner anyway.  However, it should be noted that Manfred M\"unz,
  Michael Wick,  Paolo Gambino, Herbert Steger and Dario Buttazzo  are in the right column despite having less than four papers with me.
  Similarly, I did not include in this list authors with whom I have written a single
  paper in different fields that where cited more than 1000 times.
  These are John Ellis, Mary K. Gaillard and Dimitri Nanopoulos \cite{Buras:1977yy}   and Dennis Duke and Taizo Muta \cite{Bardeen:1978yd}. Also my great
  collaborator from my CERN times, Karel Gaemers, already mentioned before, is
  not on this list. To compensate for this I list the names of all my
  collaborators in Table~\ref{tab:collaborators}.

 As this year I will turn 80 and several of my colleagues of similar age are not among
 us anymore, it is not evident that I will reach 100 as is my plan for the time
 being. In addition it is not clear that in 10 or 20 years from now I will remember my research in any details. Therefore I thought  that it is the right time to thank my  collaborators already now for all the time we spent together.
 This is  despite the hope  that I will be updating this flavour  story at least until 2046. In this context those of the readers who want to witness
 discoveries of new particles at the FCC should read the book\footnote{Given to me by our daughter Karin.}
 by Thomas Schulz:
 {\bf Project LIFE Extension: How 100 healthy years of life are already becoming possible thanks to cutting-edge research and high-tech medicine — and what we ourselves must do to achieve this.} Only in German until now.

 Except for Figs.~\ref{Koba} and \ref{F2005} I do not show any  photos of my collaborators.
 They can be found in my talks on the webcites of various conferences and
 on {\em YouTube}, in particular in my talk related to the Max Planck Medal 2020
 that I gave in Dresden at the 86th Annual Meeting of the German Physical Society  in March 2023. The list of my talks that appear in YouTube can be found in
 Section~\ref{YouTube}.

\newpage

{\small
  \begin{table}[thb]
\centering  
\renewcommand{\arraystretch}{1.3}
\begin{adjustbox}{width=0.8\textwidth}
\begin{tabular}{|ll||ll|}
\hline
Place  &  Number of Joint Papers    & Place &  Number of Joint Citations    \\
\hline
1. & Jean-Marc G\'erard (21) & 1.   &  Markus Lautenbacher (6036)   \\
2. & Monika Blanke (20) & 2.   &  Gerhard Buchalla (5907)   \\
3. & Jennifer Girrbach-Noe (19) & 3.   &  William Bardeen (3659)   \\
4. & Robert Fleischer (17) & 4.   &  Matthias Jamin (2962)   \\
5. & Stefan Recksiegel (17) & 5.   &  Jean-Marc G\'erard (2876)   \\
6. & Jason Aebischer (15) & 6.   &  Jennifer Girrbach-Noe (2774)   \\
7. & Christoph Bobeth (14)& 7.   &    Monika Blanke (2503)                    \\
8. & Markus Lautenbacher (13) &      8.   &  Robert Fleischer (2285)       \\
9. & Gerhard Buchalla  (12) & 9.   &  Peter Weisz (2152)   \\
11. & Fulvia De Fazio  (11) & 10.   &  Stefan Recksiegel (2065)   \\
11. & Matthias Jamin  (10) & 11.   &  Luca Silvestrini (1949)   \\
11. & Luca Silvestrini  (10) & 12.   &   David Straub (1821)  \\
11. & Cecila Tarantino  (10) & 13.   &   Sebastian J\"ager (1645) \\  
11. & Selma Uhlig (10)       &   14. & Martin Gorbahn (1608)\\       
11. & Anton Poschenrieder (10) & 15.   & Wolfgang Altmannshofer (1507) \\
12. & William Bardeen (9)  &  16. &         Andreas Weiler (1503)  \\
12. & Andreas Weiler (9)   &  17. & Bj\"orn Duling (1476)\\
12. & Bj\"orn Duling (9)    &  18.   & Mikolaj Misiak (1397)  \\
12. & Jacky Kumar (9)      &  19. & Cecilia Tarantino (1348) \\
13. & Wolfgang Altmannshofer (8) & 20.   & Gino Isidori (1336) \\
13. & Gino Isidori (8)     &   21.  & Manfred M\"unz (1294)  \\
13. & Diego Guadagnoli (8) &   22.  & Diego Guadagnoli (1283) \\
13.  & Felix Schwab (8)     &  23.  & Felix Schwab  (1173) \\
14. & David Straub (7) & 24. & Stefania Gori (1167) \\
14. & Sebastian J\"ager (7) & 25.  & Anton Poschenrieder (1160) \\
14. & Mikolaj Misiak (7) & 26. &    Selma Uhlig (1158)         \\
15. & Martin Gorbahn (6)  & 27.  & Fulvia De Fazio (1086)  \\
15.  & Paride Paradisi (6) &   28.  &  J\"org Urban  (1048) \\
16. & Peter Weisz (5)   &  29.     & Michael Wick (1009)  \\
16. & Stefania Gori (5) &  30.     &  Robert Knegjens (951)  \\
16. &    J\"org Urban (5)  &   31.     & Christoph Bobeth (940)  \\  
16. &  Robert Knegjens (5)     &  32. & Paolo Gambino (932) \\         
17. &      Wojtek Slominski (4)    &  33. & Wojtek Slominski (827) \\
17. & Janusz Rosiek (4)    &  34. & Janusz Rosiek (803) \\
17. & Katrin Gemmler (4)  & 35. & Herbert Steger (760) \\
17. & Michaela Harlander (4)  &  36. & Dario Buttazzo (728) \\
\hline
\end{tabular}
\end{adjustbox}
\caption{\small Ranking of my collaborators in the number of joint papers and joint citations.}
\label{tab:ranking}
\end{table}
}

\newpage

{
  \begin{table}[thb]
\centering  
\renewcommand{\arraystretch}{1.3}
\begin{adjustbox}{width=0.9\textwidth}
\begin{tabular}{|l|l|l|l|}
\hline
Jason Aebischer & Michaela Albrecht  & W. Alles & {\em Wolfgang Altmannshofer} \\
Patricia Ball   &  
William Bardeen & A.  Bassetto & Ulrich Baur\\
Aoife Bharucha & Andrzej Bialas & Ikaros Bigi &  James Bjorken \\
 {\em Monika Blanke} &
 {\em Christoph Bobeth} &  Marcella Bona & Stefan Bosch\\
 C. Boyer &Gustavo Branco & Douglas Bryman &{\em Gerhard Buchalla}\\
 Robert Buras &  Dario Buttazzo & Lorenzo Calibbi& {\em Maria Carlucci}\\
 Alejandro Celis &Maria Cerd\'a-Sevilla & Marco Ciuchini & Piotr Chankowski\\
 Gilberto Colangelo & Pietro Colangelo & Andreas Crivellin & Andrzej Czarnecki\\ Sally Dawson  &
 Fulvia De Fazio & J.M. Dethlefsen & Jorge Dias de Deus\\
 {\em Bj\"orn Duling} & Dennis Duke & John Ellis & {\em Thorsten Ewerth}\\
  Thorsten Feldmann &
  {\em Robert Fleischer} & Emmanuel Floratos & Enrico Franco \\
    Mary K. Gaillard & Karel Gaemers&
    Paolo Gambino & {\em Katrin Gemmler}\\
    Jean-Marc G\'erard & Jennifer Girrbach-Noe & {\em Martin Gorbahn} & {\em Stefania Gori} \\ Christophe Grojean &
    Diego Guadagnoli & {\em Michaela Harlander} & Julia Harz\\
    {\em Ulrich Haisch} & {\em Tillmann Heidsieck} & P.Q. Hung & Gino Isidori\\
     {\em Sebastian J\"ager}&
     Matthias Jamin & Martin Jung & Fiona Kirk\\
      Robert Knegjens&
      Ziro Koba & Pawel Krawczyk & Frank Kr\"uger \\
      Jacky Kumar & Axel Kwiatkowski&
      {\em Markus Lautenbacher} & {\em Manfred Lindner}\\
      Laurence Littenberg &Francesco Loparco &
      Thomas Mannel & Claudio Manzari\\
       Guido Martinelli & Luca Merlo &
Mikolaj Misiak & Martin Mojahed\\  Rasmus Moller & Marc Montuli&
{\em Manfred M\"unz} & Taizo Muta \\ Minoru Nagai & Dimitri Nanopoulos &Christoph Niehoff&
{\em Ulrich Nierste}\\ {\em Gaby Ostermaier} & Paride Paradisi & Fabrizio Parodi &
Maurizio Pierini \\ Stefan Pokorski & {\em Anton Poschenrieder}& {\em Nicolas Pott} &
{\em Christoph Promberger}\\
Stefan Recksiegel & Andrea Romanino &Janusz Rosiek&
Douglas Ross\\  Reinhold R\"uckl&Chris Sachrajda &{\em Claudia Salazar}&
A.N. Schellekens\\  {\em Felix Schwab} & Joao Silva & Luca Silvestrini&
Lucja Slawianowska \\ Wojtek Slominski& {\em Michael Spranger}& Peter Stangl&
{\em Emmanuel Stamou}\\  {\em Herbert Steger} & Achille Stocchi & {\em David Straub} &
{\em Selma Uhlig}\\  Cecilia Tarantino & Ngoc Tran & J\"org Urban &
Elena Venturini\\ {\em Andreas Weiler}  &Peter Weisz & {\em Michael Wick}&
{\em Elmar Wyszomirski}\\ T Yanagida & Robert Ziegler & & \\
\hline
\end{tabular}
\end{adjustbox}
\caption{\small My Collaborators. In particular my {\em PhD Students}.}
\label{tab:collaborators}
\end{table}
}

Before we start it is useful to be equipped with the meaning of many abbreviations. These are:
{\bf BSM} Beyond the SM,
{\bf CKM} Cabibbo-Kobayashi-Maskawa, 
{\bf CP} Charge Parity,
{\bf EFT} Effective Field Theory,
{\bf EW} Electroweak,
{\bf FCNC} Flavour-Changing Neutral Current, 
{\bf GIM} Glashow-Iliopoulos-Maiani,
{\bf LHC} Large Hadron Collider,
{\bf LO} Leading Order,
{\bf MS} Minimal Subtraction,
{\bf NLO} Next to Leading Order,   {\bf NNLO} Next-Next to Leading Order,
{\bf NP} New Physics,  
{\bf OPE} Operator Product Expansion,
{\bf RGE} Renormalization Group Equations,
{\bf SM} Standard Model,
{\bf SMEFT} Standard Model Effective Field Theory,
{\bf VLL} Vector-Like Lepton,  
{\bf VLQ} Vector-Like Quark,  
{\bf WC} Wilson Coefficient,
{\bf WET} Weak Effective Theory.   

\section{My Experimental Collaborators and Colleagues}

In the search for NP close cooperation between theorists and experimentalists is essential. In the context of flavour physics, it is therefore appropriate to mention some of my contacts with experimentalists. I will focus here on those related to the processes I have studied.

First of all, Konrad Kleinknecht should be mentioned. I have been in contact with him since the mid-1980s, not only in connection with the ratio 
$\epe$
measured by the NA31 and NA48 experiments at CERN, but also through service on a number of committees in Germany. Our interactions became even more frequent after he moved from Mainz to Munich.

Still in connection with 
$\epe$, I would like to mention Bruce Winstein from the University of Chicago, whom I first met at Fermilab. I still remember when he entered my office at Fermilab in 1981 and asked for my opinion on the work of Gilman and Wise \cite{Gilman:1978wm}. They had correlated the explanation of the
$\Delta I=1/2$ rule by means of QCD penguins proposed by Shifman et al \cite{Shifman:1975tn} with a value of $\epe$ as large as $10^{-2}$.

I told Bruce that this is in principle possible, as both Fred and Mark are
first-class scientists. Bruce replied that if $\epe$
was indeed so large, he would be able to measure it.
Eventually, it was shown in 1986 by Bardeen, G{\'e}rard and myself \cite{Bardeen:1986vz} 
that the $\Delta I=1/2$ rule is dominated by current-current operators rather than 
QCD penguins and such  strong correlation of with $\epe$ does not  occur
in the SM. It then took
 20 years to measure this ratio, which turned out to be  an order of magnitude smaller than predicted by Fred and Mark. Bruce led the  E731 and KTEV
experiments at Fermilab which similarly  to the NA31 and NA48 experiments at CERN
measured this ratio. I will report on the present status of $\epe$ in
Section~\ref{Statusepe}.

In connection with $K\to\pi\nu\bar\nu$ decays I had long-standing  contacts
with Douglas Bryman and Laurence Littenberg which eventually led in 2005
to a joint paper with them and Gino Isidori \cite{Bryman:2005xp} that I will briefly describe later on.

Another important experimentalist with whom I had greatly appreciated
contacts for many years both in connection with $\epe$ and $K\to\pi\nu\bar\nu$ decays, is Augusto Ceccucci from CERN. The same applies to 
Taku Yamanaka whom I met already at Fermilab
and who later moved to Osaka as well as 
to Takeshi Komatsubara from J-PARC KOTO experiment. 

In recent years, in connection with the NA64 experiment at CERN, I
exchanged many Emails with Evgueni Goudzovski and Cristina Lazzeroni,
in particular in relation to the articles \cite{Goudzovski:2022scl,Aebischer:2022vky}.

Next, I would like to mention
Fabrizio Parodi and Achille Stocchi with whom  I
wrote a paper in 2001 presenting a new perspective on the Unitarity Triangle and the CKM matrix \cite{Buras:2002yj}. This is in fact, my most important paper written jointly with
experimentalists and I will report on it in Section~\ref{BPS}.

B-physics experimentalists such as  Marina Artuso and Sheldon Stone as well as  high energy collider 
physicists like Uta and Max Klein,  Robin Erbacher and John Conway should also be mentioned here. Sadly, Sheldon and Max are no  longer among us.

No doubt I have forgotten someone and will try to improve on this in the
future. In particular, I met many experimentalists not only at conferences
but also at various advisory committee meetings at CERN, DESY, Fermilab and KEK.

\part{First Steps}

\boldmath
\section{CERN: $1976$}
\unboldmath

My flavour story starts in the
CERN Theory Division where I was CERN fellow from October 1975 till
September 1977.
In early 1976 I made the first steps in flavour physics by investigating
 the decays of charmed baryons even before the discovery of $D$ mesons
 \cite{Buras:1976dg,Buras:1976ub}. I arrived at CERN in October 1975 having
 a very poor knowledge of the SM and  definitely having desire  to learn more about it.

 After studying the book of Okun which
 taught me  simple calculations of Feynman diagrams, 
 I simply asked John Ellis what I could calculate in the SM.
He suggested this project which I completed by myself 
and published the paper in Nuclear Physics B without any problems.
Subsequently a short paper with John followed. 

Truth be told I am not particularly proud of my first flavour papers
but these were my first steps in the SM and I am grateful to John for
 this suggestion and the subsequent one to calculate the process $e^+e^-\to W^+W^-$  both with two on-shell $W^\pm$ and one with one off-shell $W^\pm$
 \cite{Alles:1976qv}. The second one could be measured in principle
 at LEP I, the first one at LEP II. As the second process
 turned out to have a very small cross-section, John used to say that
 because of our paper LEP II has been built. A nice statement
 but not really true. More on this paper in App.~\ref{Alles}.

 But this paper and the subsequent two on deep-inelastic
 scattering 
 \cite{Buras:1977ra,Buras:1977yj}, mentioned already before, were much more successful than my two first  flavour papers. Moreover these two DIS papers were
 my idea and not John's. Despite this success I have not reported on this work at CERN for various reasons. When I suggested it to John he told me that I would be destroyed by Bruno
Zumino who apparently claimed that asymptotic freedom was mathematically
inconsistent and by Giuliano Preparata, who claimed until his death in 2000
that there is mounting evidence against QCD.

But they were not the only scientists at CERN who were against QCD. I still
remember meeting by chance in front of my office John Ellis just walking with
the director general Leon van Hove. Seeing me John introduced me to van Hove
telling him that here is a fellow doing detailed calculations of QCD.
It was clear from van Hove's reaction that he was unhappy that CERN fellows
are wasting time and CERN resources for such a research. But he wanted to know
more about my work and next day he invited me, John Ellis and Mary Kay Gaillard
to his office on the top of the main building.

I was the only one talking, trying to explain what I was doing but I doubt
that I convinced van Hove that this was an important work. This doubt was undermined
by the fact that after my presentation was finished he invited Mary Kay and
John for lunch but not me. I went for lunch with other fellows worrying that
I will get some difficulties from the director general and after lunch I went
to Mary Kay to ask her how the lunch was. She didn't seem  to be amused.
She just said they had to pay for their lunch.

I am also grateful to John for inviting me to the last paper I wrote at CERN.
The 
 paper with him, Mary K. Gaillard and Dimitri Nanopolous \cite{Buras:1977yy}, on the $\text{SU(5)}$
 Grand Unification, was of much higher caliber than my first paper with him and technically much more involved.

The main goal of this project was a detailed analysis of the phenomenological implications of the Georgi-Glashow $\text{SU(5)}$ Grand Unification model \cite{Georgi:1974sy}. The idea
was to calculate the rate for proton decay, investigate the rate of fine-tuning
of parameters to keep the Higgs mass around the electroweak symmetry scale
in the presence of unification scale of strong and electroweak interactions $\ord(10^{15})\gev$ as required in 1977, perform a more detailed analysis of
$\sin^2\theta_W$ than done before by Howard Georgi, Helen Quinn and
Steven Weinberg \cite{Georgi:1974yf} and a more detailed prediction for the $b$-quark mass from
the $\tau$-lepton mass improving on the work of Chanowitz, Ellis and Gaillard \cite{Chanowitz:1977ye}.

 In particular in the case of $\sin^2\theta_W$ and the b-quark mass
 I could contribute with my knowledge of the Renormalization Group (RG)  technology
 gained already in the context of the Reggeon Calculus at Niels Bohr Institute
 before coming to CERN (see App.~\ref{Deus}) and through my studies of the DIS mentioned
 above. 

The idea was to perform this project within six weeks and therefore
the assignments of different
parts to different members of the team had to be made. John was supposed to analyze
proton decay, Mary Kay the issue of fine-tuning pointed out already by Gildner
\cite{Gildener:1976ai}
and I was assigned, as expected, to do a detailed renormalization group
study of the  $\sin^2\theta_W$ and the $b$-quark mass that both were
predicted at the unification scale: $3/8$ and $m_b=m_\tau$, respectively.
Dimitri Nanopoulos was also involved in this project but I do not know
which assignment he got and in fact I never discussed with him this
project.

I will not discuss the results because they are well known \cite{Buras:1977yy}. In particular
it is well known that this model has been ruled out later on because
it predicted the proton to decay faster than the existing lower bound
on its life-time and the value of  $\sin^2\theta_W$ turned out to be
significantly smaller than its experimental value when the accuracy of
its measurement increased a few years later.

However, there are two additional results in my view that were equally important.
The fine-tuning of parameters to keep the Higgs mass around the electroweak
symmetry breaking scale in the presence of large GUT scale  was huge,
fine-tuning to eleven decimal places. Even if this hierarchy problem was already pointed out by Gildner one year earlier \cite{Gildener:1976ai},  our much more detailed analysis of this problem  had surely a significant impact on the community, although our paper became
only well known in 1979 when Steven Weinberg made apparently a big advertising of it. This motivated  many theorists like Savas Dimopoulos, Steward Raby and Frank Wilczek \cite{Dimopoulos:1981yj} to suggest
supersymmetry as a possible solution or Lenny Susskind \cite{Susskind:1978ms} and Steven Weinberg \cite{Weinberg:1975gm} to invent a new strong force,  {\em technicolour.}

In fact Larry Hall told me a couple of years later that every PhD student
in particle physics at Harvard was obliged to have our paper on her/his desk. Probably
it was an exaggeration. Yet when a few years later I met at the 1980 Grand Unification Conference several former
Harvard PhD students, it was evident that they all knew my name because of this paper. At this conference I met Steven Weinberg for the first time and when
he saw my conference badge, he said: ``Oh, you are Buras et al''. 

The second result was the prediction for the b-quark mass from the $\tau$-lepton
mass that was proposed earlier by Chanowitz, Ellis and Gaillard \cite{Chanowitz:1977ye} but not analyzed in details by them.
My renormalization group  analysis of the the b-quark mass has shown that for four generations of
fermions it was not possible to obtain correct b-quark mass. In view
of other problems of this model one cannot claim that our paper excluded
fourth generation but still our analysis has shown that in principle
one could exclude it on the basis of the renormalization group effects.

One year later Dimitri Nanopoulos and Douglas Ross strengthened my
findings by extending my calculation to the NLO level \cite{Nanopoulos:1978hh}.
In any case fourth generation of the standard quarks is ruled out
because of the Higgs width.
But this is still another story and I move to my activities after CERN.

   \boldmath
\section{Next Period: $1980-1984$}
\unboldmath
\subsection{Preface}
Let us go back to the fall of 1976 when I had to apply for a new postdoc position.
My CERN fellowship was for two years and I had to find a new position after September 1977. Many young postdocs send applications typically to twenty or more institutions. But I applied only to three. Initially I wanted to apply only to two.
First one was Bern University, where Peter Minkowski,  whom I met at CERN, was working  and Heinrich Leutwyler who was sitting next to me in Schladming in 1972 as described in one of the appendices. Truth to tell one of the reasons to apply there were surroundings. My wife Gurli and me enjoyed hiking in the Alps in the summer and skiing in the winter. The second Institute was Rutherford Laboratory which at that time was
a rather strong place in particle physics. The reason was that I was quite impressed by the atmosphere in London during the ICHEP conference in London in 1974 (see App.~\ref{London74}) and I thought spending a couple of years in England would be a good idea.

In addition
Mary Kay Gaillard suggested that I apply to Fermilab. She spent there two years (1973-1975) working with Benjamin Lee, one of the top gauge theorists worldwide,
writing two important papers in 1974 with him. One was the first attempt to explain within QCD the $\Delta I=1/2$ rule of Gell-Mann and Pais \cite{Gell-Mann:1955ipe} which already at that time was a puzzle for twenty years. This attempt was unsuccessful \cite{Gaillard:1974nj}  which was also an independent finding of Altarelli and Maiani \cite{Altarelli:1974exa}. In the second paper, which was one of the first extensive papers on rare Kaon decays, in addition to presenting estimates for
branching ratios of several Kaon decays in gauge theories, including GIM mechanism, Mary Kay and Ben
predicted correctly the mass of the charm quark by calculating $K_L-K_S$ mass difference \cite{Gaillard:1974hs}\footnote{It is interesting that a similar estimate has also been made by Arkady Vainshtein and I. Khriplovich \cite{Vainshtein:1973md}, that being published in a Russian journal is practically unknown to
the community. It has been cited 30 times to be compared with 1315 citations
of \cite{Gaillard:1974hs}. However, it should be noted that the latter paper
discussed many other topics beyond the determination of the charm quark mass.}.

I was a bit surprised that she suggested it because in the fall of 1976 I had only two known papers, one on geometric scaling in proton-proton collisions written  with Dias de Deus shortly after completing my PhD at the Niels Bohr Institute in Copenhagen and the second on $e^+e^-\to W^+W^-$ mentioned above. But I was in contact with her in connection with my charmed baryon calculations and apparently her opinion about my skills was good. I applied then to Fermilab after Gurli agreed to go there for at most two years.

I went in November to Bern for a seminar, where I presented our calculation of
$e^+e^-\to W^+W^-$. It went well and two days after returning to CERN I received
an offer of a postdoc position for two years. A few weeks later came the offer from Rutherford
laboratory, this time this was a three year position. But there was such a big
difference in salary between Bern's position and the later one that talking
to Gurli I rejected Rutherford position and decided to accept Bern position
because I thought there was no chance to get a position at Fermilab.

Two days after I accepted position in Bern, Mary Kay entered my office and has
shown me a telegram from Ben Lee with the following content:

``Convince Buras to come for a year as a visitor to Fermilab to work with me''

I was amazed by this news, contacted Gurli and she agreed to go for a year to Fermilab but then return to Denmark. I called Bern and told them about the Fermilab offer. Fortunately they understood that for me at this stage of career it was important to go to USA, which in the 1970s was much stronger in theoretical particle physics than Europe. I accepted Fermilab offer, continued with my study
of asymptotic freedom but simultaneously I was very excited about my future
collaboration with Ben Lee.

But then on June 17th 1977 I entered Mary Kay's office.
She was crying. She just got the message from Fermilab that one day before Ben Lee died in a car accident. This was a shock for his family, for Fermilab
and particle physicists around the world but also for me who lost an important collaborator
before our collaboration started. In fact I told John Ellis that as far
as my collaborators are concerned I was until now not very lucky: Ziro Koba with
whom I collaborated in Copenhagen
died six months after our collaboration, my young collaborator on
 $e^+e^-\to W^+W^-$, Boyer, ended in the hospital after
he collaborated with me and Ben Lee died even before our collaboration started.

When I told John  about it, he told me that I must be a dangerous person and he has
to think over whether he will continue collaboration with me after the paper
on $\text{SU(5)}$, described above, is finished. I thought it was a joke but in fact we never
collaborated again. Since then I had over hundred collaborators and all survived working with me.

At the end of September 1977 we (my wife Gurli, our 2 year old son Robert and myself) moved
to Fermilab.
After the death of Ben Lee, it was not clear to me with whom I would collaborate
at Fermilab. Fortunately, I found a great collaborator there, Bill Bardeen. After NLO QCD calculations for DIS with him Dennis Duke and Taizo Muta \cite{Bardeen:1978yd} in which we introduced the $\overline{\text{MS}}$ scheme (see App.\ref{MSbar}) and
  NLO QCD calculation for Photon-Photon scattering with Bill alone  \cite{Bardeen:1978hg} I have written a long review
on Asymptotic Freedom in Deep Inelastic Scattering  \cite{Buras:1979yt}. It was my first long review published  in Reviews of Modern Physics. Subsequently I started to look for other projects.

 Rather accidentally in 1980 I read the book  by P. de Kruif {\em Microbe Hunters}
 which I liked very much and realized that particle physicists are also
 microbe hunters but at much shorter distance scales,
 at the level of a femtometer $10^{-15}$m reached in 1980.

 As I knew that in 1974 Gaillard and Lee predicted the mass of the charm quark prior to its discovery \cite{Gaillard:1974hs} by studying  the $K_L-K_S$
 mass difference $\Delta M_K$ caused by the $K^0-\bar K^0$ mixing,  I asked  myself how the top quark mass could be predicted in a similar manner prior
 to its discovery. In 1980 the lower limit for its mass was of about $20\gev$.

 As I did not read this paper until then, I decided to study it in details. This paper considers not only $K^0-\bar K^0$ mixing but also 
 rare Kaon decays, like  $K_L\to\mu^+\mu^-$, $\kpn$, $\klpn$
 and several other decays. Yet, I realized quickly that their calculations could not
 be used for the top quark because they were done for $m_c^2/M_W^2$, much smaller
 than unity, and in the case of the top quark one knew already in 1980
 that this approximation  could not be trusted. Here begins my first encounter with flavour changing neutral current (FCNC) processes.

\boldmath
\subsection{Top Quark Mass Dependence in FCNC Processes}\label{Fermilab}
\unboldmath
In the fall of 1980 I noticed 
some papers calculating the top quark contribution to $K^0-\bar K^0$ mixing
for arbitrary top quark mass, in particular by Shrock and Treiman \cite{Shrock:1978dm}.  These
papers motivated me 
to calculate such contributions for several rare Kaon decays, like the ones
considered by Gaillard and Lee.
This was a nice exercise in one-loop calculations, involving in addition
to box diagrams relevant for $K^0-\bar K^0$ mixing also $Z^0$-penguin and photon-penguin diagrams dominating rare $K$ decays. While doing these calculations 
I was excited because 
I thought I was the first one to perform such a complete study of FCNC processes
for an arbitrary top quark mass.  But a couple of weeks before I completed writing up my results,
Chris Hill entered my office at Fermilab and handed me a  paper by Inami  and Lim \cite{Inami:1980fz}
where my calculations have already been done. This was a similar expirience to the $e^+e^-\to W^+W^-$ calculation performed at CERN when a Russian group  was ahead of us (see App.~\ref{Alles}). But that days papers from Russia and Japan  were arriving in Europe and USA with a significant delay, not through internet but by ordinary
mail.

It was a real disappontment even if fortunately my results agreed with the
ones presented by Inami and Lim. Still I decided to  publish my results and  to use them
 for a simple phenomenology. As the values of the elements of the CKM
matrix, except for $V_{us}$ and $V_{ud}$, were practically unknown at that time, I eliminated them by combining
various observables, in particular the $K_L-K_S$ mass difference $\Delta M_K$ and the
branching ratio for $K_L\to\mu^+\mu^-$. These were the two best measured
rare transitions at that time.
Eliminating free parameters by taking the ratios of observables is a good idea,
that I used for the following 45 years from time to time,
but can be risky as we will see in a moment.

The point is that the long distance
contributions to  $K_L-K_S$ mass difference $\Delta M_K$ and to the
branching ratio for $K_L\to\mu^+\mu^-$ were practically unknown at that time
so that I did not take them into account. Also the parameter $B_K$ that is crucial for $\Delta M_K$ and for the CP violation parameter $\varepsilon_K$ was only
known in the MIT-Bag model, $B_K=1/3$, \cite{Donoghue:1982cq} that is by a factor of 2.3 lower than what we know today. I did not think too much about all these omissions because
in their prediction of the charm quark mass Gaillard and Lee also omitted
long distance contributions and they were only lucky because they have
chosen $B_K=1$, closer to the best value of this parameter known today.
I will report on it later. Moreover, while in the case of charm mass prediction from  $\Delta M_K$ only $\vus$ was needed, in my case $\vts$ and $\vtd$ played
an important role.

Playing numerically by means of Fortran with my
expressions for $\Delta M_K$, $K_L\to\mu^+\mu^-$ and $\varepsilon_K$
I noticed an upper bound on $m_t$ of $27\gev$. I still remember that there was
no output for higher values of $m_t$ and got very excited.
Indeed the experimental {\em lower bound} on $m_t$ from DESY and SLAC in 1981 was about  $20\gev$ so finding an {\em upper bound} on $m_t$  of $27\gev$ was
an interesting result. I wrote a letter and submitted it to Phys.Rev.Lett.
After the change of the title and a few modifications requested by a referee, my first FCNC paper was published \cite{Buras:1981cr}, moreover in PRL!

At the Grand Unification Conference in 1981  several physicists were interested in my
paper including Steven Weinberg and Tiny Veltman. Subsequently I was invited
to several universities in USA to report on this result. Moreover my paper collected
over 100 citations. As I was told by a Japanese experimentalist many years
later, the motivation for building Tristan collider increased due
to my paper by much.

However not everybody was positive about my result. In particular Gordy Kane
claimed that although I eliminated the CKM parameters, a closer look at my expressions indicated
that for the analysis to be self-consistent the values of the CKM elements
$\vtd$, $\vts$ and $\vcb$  had to be only by a small amount smaller than $\vus$.
This was indeed true but at that time it was not known what they are. Two years
later it was found that $\vcb$ is by a factor of five lower than $\vus$
and $\vtd$ lower even by a factor of about twenty. Also the neglect of long distance contributions to $\Delta M_K$ and $K_L\to\mu^+\mu^-$
was totally unjustified. But this also became clearer only a few years later. Similarly the value of $B_K$ was very poorly known at that time.

In summary, it was real fun to do this analysis but the phenomenological analysis in my first paper on rare transitions in the Kaon system turned out to be simply wrong. But what was really wrong was the neglect of long distance contributions to $\Delta M_K$ and $K_L\to\mu^+\mu^-$
and the wrong value of $B_K$. The idea of eliminating CKM parameters was a good one. In addition the independent confirmation of Inami-Lim results for the one-loop
$m_t$-dependent functions, used since then in all papers on FCNCs, was in my
view an important result even if full credit for these functions is given in the literature to
Inami and Lim.

Fortunately, already
two years later I have written papers on flavour physics that had significant
impact on phenomenology of FCNC processes in the SM.
I will describe them subsequently. In fact the strategy of eliminating  CKM parameters by taking ratios of various branching ratios turned out to be useful for the tests of the Standard Model but only years later  when various non-perturbative parameters like $B_K$, long distance QCD contributions and QCD short distance corrections to various decays not only
of Kaons but also of  $B$ mesons and in $B^0_{s,d}-\bar B^0_{s,d}$ mixings have been put under control. As I was involved in these calculations they will enter the scene in several parts of this writing. But for the time being let us describe my next 
activities in FCNC processes that took place at the Max-Planck Institute for Physics in Munich. We (my wife Gurli, our son Robert 7, our daughter Karin 4 and myself)  moved there in September 1982. In October our second son Allan was born.

\boldmath
\subsection{$\varepsilon_K$, $\epe$ and $B_d^0-\bar B_d$ mixing }
\unboldmath
In the summer of 1983 I visited Fermilab and subsequently attended the Photon-Lepton conference in Cornell. In that summer new results from
SLAC, Cornell and DESY on $B$ meson life-times $\tau_B$ implied that the CKM element $\vcb$ was significantly smaller than the $\vus$ one, typically by a factor of five
smaller. The element $\vub$ being still by roughly one-order of magnitude  smaller than $\vcb$ could not be measured at that time.

This rather unexpected result, which disproved my upper bound on the top-quark mass,
assured me that flavour physics could be an interesting direction and I
should invest more time on it. In fact from 1983 until today I worked only on flavour physics, dominantly in the first twenty years in the quark sector but later also in the lepton sector. Moreover I studied also flavour conserving decays like electric dipole moments and anomalous magnetic moments. 

These new experimental results motivated me to perform two analyses soon after
I returned to Munich. I had two collaborators. One of them was  
Wojtek Slominski from Cracow, 
a post doc at MPI from 1982 to 1984. The second one was  my first PhD student, Herbert Steger.

Basically, the main strategy of these papers, that distinguished them from most
of other papers dealing with the implications of these new results,  was the
calculation of various observables in $K$ and $B$ systems as functions
of
\be
\tau_B, \qquad R=\frac{\Gamma(b\to u)}{\Gamma(b\to c)},\qquad m_t, \qquad \hat B_i
  \ee
  with $B_d$-lifetime $\tau_B$ and $R$ constrained by
    recent experiments, $m_t$ being the top quark mass and $\hat B_i$ being
    non-perturbative factors in the relevant hadronic matrix elements with
    $i=K,s,d$ for $K^0-\bar K^0$ and $B^0_{s,d}-\bar B^0_{s,d}$ mixings, respectively.

    Bounding first CKM parameters by these observables and by $\varepsilon_K$ we have calculated:
    \begin{itemize}
    \item
      The ratio $\epe$\footnote{$\varepsilon^\prime$ gives the size of direct CP violation in $K\to\pi\pi$ decays while $\varepsilon$, also denoted by $\varepsilon_K$, the size of indirect CP violation in these decays caused by $K^0-\bar K^0$ mixing.} as a function of $\tau_B$ for various values of $R$, $m_t$ and $\hat B_K$ \cite{Buras:1983ap},
    \item
      Several observables in the $B_d$ system:
      \be
      \frac{\Delta M_d}{\Gamma_d},\qquad \frac{\Delta\Gamma_d}{\Gamma_d},\qquad S_{\psi K_S}\,,
      \ee
      again as functions of of $\tau_B$ for various values of $R$, $m_t$ and $\hat B_d$ parameter relevant for the $B_d^0-\bar B_d$ mixing \cite{Buras:1984pq}.
    \end{itemize}
    I will not describe the numerical results of these papers because they are totally out of date. Their detailed summary can be found in \cite{Buras:1984ur}. This summary
    shows that dramatic progress has been made over last 40 years in this field.
    On the front  page of my contribution one can also find my photo of 1984
    which also  demonstrates  dramatic changes in my appearance.

    In spite of being out of date as far as numerics is concerned,  both papers became very well known, in particular the one on 
    $B_d^0-\bar B_d^0$ mixing \cite{Buras:1984pq} that was written well before
    the discovery of this mixing. It  has been cited more than 450 times by now.

    The reasons for this success are possibly the following ones.
    \begin{itemize}
    \item
      First collection of numerous formulae for processes considered  that have been derived partly
      by other authors as well,  but checked by us and in few instances corrected.
    \item
      Introduction of the renormalization group invariant parameters representing hadronic matrix elements of $\Delta F=2$ processes:
      $\hat B_K$ for $K^0-\bar K^0$ mixing and  $\hat B_{s,d}$ for $B_{s,d}^0-\bar B_{s,d}$ mixings.
    \item
      Introduction of the one-loop functions
      \be
      S(x_c),\qquad S(x_c,x_t), \qquad S(x_t)
      \ee
      for the particle-antiparticle mixing processes with $x_i=m^2_i/M_W^2$. These functions are often called
      Inami-Lim functions but in fact they have been introduced in \cite{Buras:1983ap}. The letter ``$S$'' has been chosen by my collaborators for obvious reasons.
    \end{itemize}

    Our analytical results have been adopted in the literature since then.
On the other hand 
    the numerical results in these papers are out of date in particular for the following reasons.
      \begin{itemize}
    \item
      The top quark mass was unknown in 1984 and we have varied it below $100\gev$, significantly below $m_t=173\gev$ known today.
    \item
      CKM parameters were poorly known.
    \item
      $\hat B_K$ was basically unknown and we made calculations for $\hat B_K=1/3,2/3,1$.
    \item
      Most importantly, the hadronic matrix element of the QCD penguin operator
      contributing to $\epe$ was based on the Gilman-Wise paper \cite{Gilman:1978wm}  who following
      Shifman et al \cite{Shifman:1975tn} assumed  that
      this operator is responsible for $\Delta I=1/2$ rule in $K\to\pi\pi$ decays.
      This led to $\epe$ as high as $10^{-2}$.
      Not only this assumption has been ruled out in 1986 by Bardeen, G\'erard and myself as described soon. Also the contribution of electroweak penguins to $\epe$,  not included by
      us, was found only in 1989 to suppress significantly 
      $\epe$ for a heavy top quark. I have
described the $\epe$ story until 2021 in \cite{Buras:2020wyv,Buras:2022cyc} and subsequently discussed $\epe$  in several papers. I will return to these topics as we proceed. 
\end{itemize}

\part{Dual QCD: 1985-1987}

\boldmath
\section{Weak Decays in the Strict Large N Limit}\label{LargeN}
\unboldmath
My interest in large N (colours) ideas arose by reading the papers by Berthold Stech, 
Wirbel and Bauer\cite{Wirbel:1985ji,Bauer:1986bm}, who performed an extensive analysis of non-leptonic two-body charm meson decays. While in 1985 the Wilson coefficients relevant for these decays were
known in the LO of renormalization group improved perturbation theory there was no idea
how to calculate the hadronic matrix elements of the corresponding four-quark
operators in QCD. Therefore the standard method for the calculation of such matrix elements was the so-called vacuum insertion approximation (VIA) in which
hadronic matrix elements of four-quark operators are factorized into the
products of matrix elements of quark currents that for vector-currents can be parametrized in terms of weak decay constants and masses of mesons in the initial and final states. But some of the operators involved have non-singlet colour structure
and to be able to factorize them into a product of colour singlet current matrix
elements one has to perform suitable Fierz transformations. This indeed brings
the hadronic matrix element into the product of matrix elements of quark currents but
in addition brings in a second product with different current matrix elements multiplied
by a factor $1/3$.

This VIA method implied for a certain class of two-body charm decays branching ratios in clear disagreement with experiment. This was for decays in which both terms were of a similar magnitude but of opposite sign, cancelling partly each other
and implying  branching ratios that were ways below their experimental values.

What Bauer, Stech and Wirbel noticed is that the removal of these additional Fierz terms improved the agreement of the theory with experiment significantly. But they
did not give any justification for this procedure which by some was called
{\it the method of discarding Fierz terms.}

{\bf Stech reported on his work at the 1985 Moriond Workshop on Heavy Quarks, Flavour Mixing and CP-Violation which was attended by a young Belgian post-coc at the MPI, Jean-Marc G{\'e}rard. Jean-Marc after returning to Munich and talking to
Reinhold Rückl realized that this 1/3 terms were simply $1/N$ terms. Therefore, 
 in fact what Stech and Wirbel were calculating were hadronic matrix elements in the
large $N$ limit} and one could give strong theoretical arguments for discarding
Fierz terms. The latter do not represent the full $1/N$ contributions
and keeping them as done in all papers until then was simply inconsistent.

{\bf Jean-Marc, Rheinhold and myself  decided then} to give these arguments in print and to perform a new analysis of
two-body charm decays. Our extensive analysis was soon very well accepted by the community
and collected of about 500 citations \cite{Buras:1985xv} by now.
 Parallel to this work Jean-Marc and me applied this method to Kaon decays
 finding that also there discarding $1/N$ Fierz terms brought the theory closer
 to experiment \cite{Buras:1985yx}. This paper has also been successful.

 However, it was clear we could not stop here for a simple reason. The matrix elements of current-current operators calculated in this manner were renormalization scale ($\mu$) independent in contradiction to what QCD was telling us. In fact the $\mu$-dependence of the hadronic matrix elements cancels the one of the  WCs. Without this cancellation
 the decay amplitudes exhibit non-physical  $\mu$ dependence present in WCs
 which is simply wrong.

 It was clear that we had to calculate $1/N$ corrections to the hadronic matrix elements in question. 
 I started then reading papers by Gerald 't Hooft
 \cite{'tHooft:1973jz,'tHooft:1974hx}  who was
 the first to apply large $N$ ideas to QCD. But in his formulation one had
 to calculate all non-planar diagrams which nobody succeeded until then.
 Moreover, even the summation of planar diagrams corresponding to leading order would be very difficult. Yet, we succeeded to find this result by just factorizing  the matrix elements without any summation of planar diagrams.

 I understood better our results reading the  articles of Edward Witten \cite{Witten:1979kh,Treiman:1986ep} who
 following 't Hooft ideas conjectured that in the large $N$ limit
 QCD as a theory of quarks and gluons becomes dual to the theory of weakly
 interacting mesons with the coupling given by $1/f^2_\pi$ with $f_\pi$ being
 the pion decay constant. As $f^2_\pi=\ord(N)$, this coupling vanishes in the large $N$ limit and in this limit QCD is dual to a free theory of mesons.
 This automatically implies the factorization of operator matrix elements and
 the leading order corresponds to the tree-diagrams in this dual theory
 which is fully formulated in terms of mesons. Colour is confined and $N$
 is hidden in $f^2_\pi=\ord(N)$.

\boldmath
\section{Weak Decays in the Dual QCD Beyond Leading Order}\label{DQCD}
\unboldmath
In 1986, Bill Bardeen visited the MPI for several months and became interested in our work. Together with Jean-Marc, he and I began considering how to extend this approach beyond the large $N$ limit. It was Bill who first realized that, in this framework, doing so would require calculating meson loops.

 Such meson loops have been calculated already in the
 framework of Chiral Perturbation Theory (ChPT) as proposed by
 Weinberg in 1979 \cite{Weinberg:1978kz} and
 executed in a series of papers by Gasser and Leutwyler \cite{Gasser:1983yg,Gasser:1984gg}. They used dimensional
 regularization for meson loop calculations and introduced a number of
 counter terms in order to remove the divergences. These counter terms
 were represented in their final formulae by low-energy constants which
 were supposed to be determined from experiment and possibly calculated
 one day by Lattice QCD (LQCD).

 We know by now that ChPT turned out to be successful for
 leptonic and semi-leptonic decays of Kaons but not for non-leptonic ones
 like $K\to\pi\pi$. Simply because in this case it is practically
 impossible to determine the relevant low-energy constants from experiment.
 Also LQCD was unable to calculate them until today with a satisfactory
 precision.

 Fortunately, Bill noticed  that to avoid this problem  we should calculate  the meson-loops with a physical cut-off
 $\Lambda$ and presented to us an explicit calculation of two
 hadronic matrix elements. This turned out to be a real breakthrough.
 As far as I remember these were the only matrix elements he calculated but this was sufficient for Jean-Marc and me to calculate
 the remaining matrix elements. I still remember that I was so excited about
 these calculations that I was performing them also in the U-Bahn when going
 to the Max Planck Institute. Few times people were asking me what it was I was doing 
 but the only thing I told them was that it was a very important work.
 
 In 1986 we included in  the loops for $\Lambda\le 0.7\gev$ only pseudoscalar
 mesons \cite{Bardeen:1986vp,Bardeen:1986uz,Bardeen:1986vz} and for $0.7\gev\le\Lambda\le 1 \gev$ 28 years later vector mesons \cite{Buras:2014maa}. In addition to $K\to\pi\pi$ decays we analyzed
 also $K^0-\bar K^0$ mixing and calculated for the first time $\hat B_K$ in QCD
 finding  $\hat B_K=0.67\pm 0.06$ \cite{Bardeen:1987vg}, close to one of the values considered by Slominski, Steger and myself in 1984.
 As in  1986-1987 we
 included only pseudoscalars, Jean-Marc started including vector mesons
 one year later \cite{Gerard:1988it}  but the full analysis with vector mesons has been done only  in 2013 with the paper published early 2014 \cite{Buras:2014maa}. I will describe this paper in Part XI.

 This approach
allowed also for the first time a meaningful calculation of the hadronic
matrix elements of QCD penguin operators
entering $\epe$ \cite{Buras:1987qa}. Interestingly, the notation $\bsi$ and
$\bei$ used these days for dominant QCD penguin and electroweak penguin operators has not been used in this paper and in several  papers on $\epe$ written by us in the following
years. It has been introduced only in 1993 in our first NLO anatomy of
$\epe$ \cite{Buras:1993dy}. See Part~\ref{NLOQCD}.

 As fas as non-leptonic transitions are concerned, the important advantage of our approach over ChPT was that our results
 were only functions of masses of pseudoscalar and vector mesons and had
 only one free parameter, the physical cut-off $\Lambda$ whose value
 we new approximately as stated above.

 Probably the most important result we got in 1986 \cite{Bardeen:1986vz} was the identification
 of the dominant dynamics behind the so-called $\Delta I=1/2$ rule, a huge
 enhancement of the $A_0$ amplitude over $A_2$ one in $K\to\pi\pi$ decays:
 $22.4$ compared to $\sqrt{2}$ which one finds in the large $N$ limit
 in which QCD interactions are absent. This rule, proposed by Gell-Mann and Pais
 in 1955 was a big puzzle \cite{Gell-Mann:1955ipe}. The attempts by Altarelli and Maiani \cite{Altarelli:1974exa} and Gaillard and Lee \cite{Gaillard:1974nj}, in 1974, one year after the discovery of Asymptotic Freedom, to explain
 this rule by QCD effects above $1\gev$, failed. They could obtain this ratio
 in the ballpark of $3-4$. The history of this rule has been summarized in chapter 7 of my book  \cite{Buras:2020xsm}.

 But calculating the meson loops that implied {\em meson evolution} of operator matrix elements from very low momentum scales to 1 $\gev$, we found that the QCD dynamics below $1\gev$
 could enhance in our approach this ratio to roughly $16\pm 2$ \cite{Bardeen:1986vz,Buras:2014maa}. Our result ruled out already in 1986 also the claim of Shifman et al, \cite{Shifman:1975tn} 
 that QCD penguins are responsible for the $\Delta I=1/2$ rule. These are
 the current-current operators with QCD penguins enhancing this ratio by
 at most $10\%$.
These findings have been confirmed by lattice QCD
only in 2020.

 In my view this 
 was definitely an important result, in particular that it took
 us only three months to get it in 1986 and the inclusion of vector mesons in
 2013 took another four months. As far as LQCD is concerned the first result has
 been presented by RBC-UKQCD collaboration in 2012 with the value of $12\pm10$
 \cite{RBC:2012ynq},
the second in 2015 with the value $33\pm11$ \cite{RBC:2015gro} and the last one in 2020 with $20\pm4$ \cite{Abbott:2020hxn}.

It is not only that our result was presented roughly 30 years before the LQCD
one but moreover we could find out what was missing in  Altarelli and Maiani and Gaillard and
Lee calculations of 1974. They had to stop the QCD evolution in terms of quarks and gluons from the electroweak scale down to low energy scales at $1\gev$ because otherwise the perturbation theory would break down. Switching around $1\gev$ to the meson theory, the QCD evolution
in DQCD, the  meson evolution,  would allow them to go
down to very low momenta at which the large $N$ limit allows to factorize
the hadronic matrix elements and calculate them.
 They would then find the enhancement of $A_0/A_2$
ratio we found.

In the case of LQCD, the ratio in question is a result of a very complicated
and time consuming numerical calculation\footnote{The simplicity of our numerical DQCD calculations is underlined by the fact that I led these evaluations despite having rather limited computer skills.}.
One can only state that dominantly QCD is responsible for this rule but the physical picture behind
it, as identified by us, is missing.

I was very satisfied with this result despite the fact that we could not fully
explain the $\Delta I=1/2$ rule (more about it in Part~\ref{RDQCD} ).
But both ChPT experts and also LQCD
experts were very critical of our approach. At the ICHEP conference in
1986 and also at EPS High-Energy Conference in 1987, at which
I presented our results, I was strongly criticized  by Eduardo de Rafael
representing ChPT and Guido Martinelli representing LQCD. Also at
a Ringberg Workshop in February 1988 at which Bill Bardeen presented our results, 
Martinelli, Maiani, de Rafael, Ecker and Donoghue were very critical about
our approach. Similar at a Summer School in Jaca (Spain) in 1988, where I was giving
lectures and also when I gave a seminar in Marseille the same year, being invited by de Rafael. The summary of our work on large N in this period can be found
in \cite{Buras:1988kp,Buras:1988ky}.

The main criticism was the $\Lambda^2$ dependence of our matrix elements
as opposed to the logarithmic scale dependence of WCs. What I think was not
realized by the opponents that the qualitative behaviour of this dependence
(the signs) was as expected from QCD. Otherwise we would not be able
to explain the dominant part of the $\Delta I=1/2$ rule. We gave also the
arguments that including vector mesons and higher meson states would turn
 this dependence into a logarithmic one but we demonstrated it
 only in 2014.  

In fact the DQCD approach has been refined many years later in
\cite{Buras:2014maa,Buras:2015xba,Buras:2016fys,Buras:2018evv,Buras:2018lgu,Aebischer:2018rrz,Aebischer:2018quc} and several new results have been presented
in these papers. I will describe them in Part~\ref{RDQCD}.

Yet, there were some physicists who were truly impressed by our results.
In April 1987 I was invited to a conference in Bad Honnef. It was organized
by nuclear physicists but dealt with strangeness and I was asked to talk
about my work on DQCD. My talk was very well received and one of
the world leading nuclear physicist, Gerry Brown, told me several times
that our work was as beautiful as Venus. This was very encouraging because 
he was an excellent scientist.

This event had some impact on my future soon after this conference. In July 1987  I was supposed to give a talk 
in connection with my application for a full professorship at the Technical University Munich (TUM).
There was no theoretical particle physics at TUM but there was a strong
nuclear physics group. After the success in Bad Honnef I decided to repeat
my Bad Honnef talk with a few additions on $\epe$. When I arrived at TUM and
met the chairman of the committee, Klaus Dietrich, he told me that 
he heard about my excellent talk in Bad Honnef. It was Gerry Brown who
told him this and I was now becoming confident that I could get this position.
I had to give this talk in German, the one in Bad Honnef was in English,
but the talk went very well. Three months later, at the beginning of October,
two weeks before my birthday, I was informed that I won the competition.
It took one year before I was satisfied with the offer and on
November 1st 1988 I began to work at the Physics Department of the Technical University Munich, situated in Garching, roughly 13 km north of Munich down town.

\boldmath
\part{Analyses of FCNC Processes within the SM (1987-1991)}
\unboldmath
\boldmath
\section{Isospin Breaking Corrections to $\epe$}\label{NIB}
\unboldmath
One of my last papers at MPI was the calculation of isospin breaking
effects to the contributions of QCD penguins to $\epe$ done with Jean-Marc \cite{Buras:1987wc}. In this paper we reanalyzed the $\pi^0-\eta-\eta^\prime$ mixing
and the QED penguin operator contributions to $\epe$ using the DQCD approach
just described. This paper was motivated by an analysis of Donoghue and
collaborators \cite{Donoghue:1985jkn}  who found a suppression of QCD penguin contribution to $\epe$
by this mixing by $40\%$. Our analysis implied $27\%$ suppression and moreover
only small contribution  of QED penguin operators to $\epe$ in contrast
to what these authors found. Eventually they found an error in their
calculations and subsequently confirmed our results within the uncertainties.

Having  this result we have performed an improved analysis of $\epe$ with $55\gev\le m_t\le 130\gev $ \cite{Buras:1987qa}. I do not describe the outcome
of this analysis because after 38 years it is out of date.
Moreover, it should be emphasized that the role of electroweak penguins, that is the
contribution of $Z^0$ penguin to $\epe$, has not been taken into account
in our analysis. At these values of $m_t$ it is rather small but
as we will see soon it increases by much for the true top quark mass.

Our analysis of isospin breaking corrections to $\epe$ 
has been updated in 2020 \cite{Buras:2020pjp} due to change of various
input parameters during 33 years. We basically confirmed our 1987 results
with the suppression of QCD penguin contribution by $(29\pm7)\%$ compared 
 to $(17\pm9)\%$ obtained using Chiral Perturbation theory \cite{Cirigliano:2019cpi},
which however does not take the presence of $\eta^\prime$ into account.
Removing the $\eta^\prime$ contribution from our calculation we indeed
found similar result.

As already stressed by me in several reviews, in particular in \cite{Buras:2021ane}, it is very important
that LQCD also includes isospin breaking effects on the QCD penguin contribution to $\epe$, in particular
the effect from $\eta^\prime$. Without this their 2020 result, that agrees
within the theoretical errors with the experimental one, cannot be trusted. Unfortunately, it looks like it could still
take some time until this  happens. See Section~\ref{Statusepe} for the summary
of the present status of $\epe$ in the SM.

\boldmath
\section{First Anatomy of $\epe$ for Arbitrary Top Quark Mass}\label{Anatomy1}
\unboldmath
In December 1988 I attended the first Kaon conference in Vancouver. Fred Gilman during
a lunch told me about his paper with Claudio Dib and Isard Dunietz \cite{Dib:1988md}, in  which they calculated electroweak contributions to $K_L\to\pi^0e^+ e^-$ for an arbitrary top quark mass finding for $m_t> 150\gev$
a large contribution from the $Z^0$-penguin that increased with $m_t$ roughly
 as $m_t^2$. A similar result has been obtained by Jonathan Flynn and Lisa Randall \cite{Flynn:1988ve}.  This gave me the idea to calculate  $Z^0$-penguin contribution  to $\epe$ for large $m_t$. The QCDP and photon penguin
contributions were known already at that time to have a very weak $m_t$ dependence.

I moved to TUM in November 1988, but already in January 1989 I got a number
of very good diploma students. Two of them, Gerhard Buchalla and Michala
Harlander were supposed to perform the calculation of $\epe$ together with me
for an arbitrary top quark mass. But this was their first calculation of that
type and I was simultaneously busy preparing first university lectures in my life
 as well as starting the project on NLO QCD corrections to all flavour
violating processes in collaboration with Peter Weisz \cite{Buras:1989xd}. 
Consequently my project with Gerhard and Michaela took longer than
I initially expected and in April 1989\footnote{The paper appeared already in March but in 1989 the arXiv did not exist and we learned about it several weeks later.} we were surprised by a paper by
Flynn and Randall  \cite{Flynn:1989iu} in which $\epe$ including 
 QCDP,  EWP ($\gamma$ and $Z^0$ penguins)
and the relevant box diagrams were calculated for an arbitrary
top quark mass. Significant suppression of
$\epe$ for $m_t > 150\gev$ by EWPs has been found by them. While definitely Flynn and Randall
should be given the credit for pointing out the importance of $Z^0$ penguins
in $\epe$ in print, fortunately for us their calculation of the QCDP contribution for large $m_t$
was incorrect so that in fact the first correct calculation of $\epe$ 
in the SM for an arbitrary $m_t$ including  LO QCD corrections has
been presented by us in \cite{Buchalla:1989we}. In fact the suppression of
$\epe$ for large $m_t$ turned out to be significantly stronger
than reported by Jonathan and Lisa in \cite{Flynn:1989iu}.
The strong cancellation between QCDP
and EWP for $m_t > 150~\gev$ found by us was soon confirmed in the erratum
to \cite{Flynn:1989iu} and subsequently by other authors \cite{Paschos:1991as,Lusignoli:1991bm}. In fact at that 
time, due to the aforementioned cancellation between QCDP and EWP contributions, the vanishing of $\epe$ in the SM and negative values for it could not be excluded.

As the upper bound on $m_t$ from electroweak precision tests was still
unknown in 1989, one can find  in our paper plots of $\epe$ as a function
of $m_t$ with $\epe$ vanishing for $m_t\approx 200\gev$. I was visiting
CERN in November 1989, one month after our paper appeared. In the CERN
cafeteria  I met Jack Steinberger and asked him the standard question
how he was doing. He told me that NA31 collaboration was shocked by
the theory. I asked him which theory. Your paper with Buchalla and
Harlander he replied. In fact at that time only the first result from NA31
collaboration was known \cite{NA31:1988eyf} implying $\epe=(33\pm 11)\times 10^{-4}$. The second NA31 analysis in 1992 gave  $\epe=(23\pm 7)\times 10^{-4}$
\cite{NA31:1993tha} and
the one from the E731 experiment at Fermilab  $\epe=(7.4\pm 5.9)\times 10^{-4}$ \cite{E731:1993niw}. Therefore, the experimental situation of $\epe$ was still unclear in the first half
of the 1990s.  More on the history of $\epe$ until 2021 can be found in
\cite{Buras:2020wyv}.

As a side remark I want to mention my 1993 analysis \cite{Buras:1993wr} on the lower bound on $m_t$
from the CP violation parameter $\varepsilon_K$ which was important for $\epe$.  Using Wolfenstein parametrization of the CKM matrix I derived an analytic expression for this bound as a function of $\vcb$, $\vub/\vcb$ and the non-perturbative parameter $\hat B_K$.
For $\hat B_K\le 0.80$, $\vcb\le 0.040$ and $\vub/\vcb \le 0.10$, I found
$m_t\ge 130\gev$. However, for
$\hat B_K\le 0.70$, $\vcb\le 0.040$ and $\vub/\vcb \le 0.08$
the bound is raised to $m_t\ge 205\gev$. 

These papers demonstrated very clearly that the top-quark and in particular
the value of its mass is crucial for FCNC processes in the SM.

\boldmath
\section{Penguin-Box Expansion for FCNC Processes within the SM}
\unboldmath
The first calculations of all top quark mass dependent one-loop functions within the SM that enter
FCNC processes in the SM have been presented by Inami and Lim \cite{Inami:1980fz} and soon after by myself \cite{Buras:1981cr}. I have reported on it in
Section~\ref{Fermilab}.

In 1990 I decided to look at these functions again with the goal to
calculate both rare Kaon and B meson decays for arbitrary top quark mass.
Again Gerhard Buchalla and Michaela Harlander joined me in these efforts.
The one-loop functions to be considered in this case are:
\begin{itemize}
\item
  $C_0(x_t)$ resulting from $Z$-penguin diagrams,
\item
  $D_0(x_t)$   resulting from Photon-penguin diagrams,
\item
  $B_0(x_t)$, resulting from box diagrams with two external quarks and two external leptons
  as opposed to $S_0(x_t)$ that results from box diagrams with four external quarks.
\end{itemize}

It was Gerhard who noticed that these three functions, introduced by Inami and Lim, 
are gauge dependent and only their gauge independent combinations enter physical observables. 
The full
literature of the 1980s used these functions, given in the Feynman gauge, in phenomenology.

We have calculated them in an arbitrary covariant gauge and demonstrated
that their combinations appearing in the literature were indeed gauge independent. This led us to the formulation of FCNC
processes in terms of manifestly gauge independent functions 
\cite{Buchalla:1990qz}
\be\label{PB1}
X_0(x_t)=C_0(x_t)-4 B_0(x_t), \quad Y_0(x_t)=C_0(x_t)-B_0(x_t),\quad Z_0(x_t)=C_0(x_t)+\frac{1}{4} D_0(x_t),
\ee
where $x_t={m_t^2}/{M_W^2}$ and the subscript indicates that no QCD corrections to $Z$-penguin $C_0(x_t)$,  photon-penguin $D_0(x_t)$ and box-diagrams $B_0(x_t)$ have been included.

The functions $X_0(x_t)$, $Y_0(x_t)$ and $Z_0(x_t)$, that dominate since then
the literature on FCNC processes in the SM,  are called Inami-Lim functions,
but in fact they have been first proposed in \cite{Buchalla:1990qz}. In the latter
paper we have also expressed the amplitudes for all FCNC processes in terms of these
functions with the QCD factors and hadronic matrix elements multiplying them
evaluated at the electroweak scale. To this end the QCD renormalization group
evolution from hadronic scale at which these matrix elements are evaluated
to the electroweak scale have to be performed. Such expressions for FCNC
amplitudes were called in \cite{Buchalla:1990qz} {\em Penguin-Box Expansions} (PBE).
They are used in the full FCNC literature these days but
this name is used only seldom. QCD corrections to these functions at the NLO
and NNLO level have been calculated dominantly by the Munich NLO-Club (MNLC) and
will be reported later on. The usefulness of the PBE in the context of new physics (NP)
will be stressed in Part~\ref{NPModels}.

In 1992 I have written together with Michaela Harlander a  review on the heavy top quark effects in FCNC processes \cite{Buras:1992uf}. Some of my colleagues
liked it. Unfortunately, it was not put in the arXiv and can only be
found in the Volume {\em Heavy Flavours} edited with Manfred Lindner \cite{Buras:1992qm}. As the review was
written before the top quark discovery and before I have written several
phenomenological papers in the following years, it became in 1995  out of date.
Yet, it contains many plots and tables as functions of $m_t$ and CKM
parameters and comparing it with my review with Robert Fleischer \cite{Buras:1997fb}, published
in {\em Heavy Flavours II} in 1998 \cite{Buras:1998vd} , one can see not only dramatic impact of top quark
discovery on flavour physics but also theory progress in this field which
will be described in the following sections of this write-up.

\boldmath
\part{Weak Decays Beyond Leading Order}\label{NLOQCD}
\unboldmath
\section{Preface}\label{MPI}
In April 1988 I gave a series of lectures on Dual QCD at a Spanish school
 in Jaca.  As already mentioned in Section~\ref{DQCD}
 I was again criticized by the same people, Martinelli, De Rafael, Gavela,.. which was rather unpleasant. Yet
 there were some interesting aspects beyond my lectures which remained in my memory. 

 First there were basically only Spanish students who made no effort to speak
 in English at breakfasts, lunches or dinners even in the presence of foreign
 lecturers who did not know Spanish. As a cosequence  not knowing
 Spanish I spent most of my time with Peter Hasenfratz, an excellent
 Lattice Gauge Theorist and Tiny Veltmann, one of the heroes of the 1960s and
 1970s who in 1999 received Nobel Prize for his work with 't Hooft on the
 renormalization of non-abelian spontaneously broken gauge theories. But
 in 1988 he did not get this prize yet and it was clear that he was unhappy
 about it. Basically at each lunch and dinner he critizised the work of
 other scientists. This was clearly interesting but eventually rather
 tiring. In certain cases I disagreed with him and then he became angry.
 I knew Tiny still since my Fermilab times and actually I liked him but
 his way of attacking and criticizing  people was sometimes unacceptable.

 Peter was a different person, also born in 1946 like me. I knew him from
 Erice school in 1972. He was a polite person and we could very well
 talk to each other. After the school in Jaca we went together by bus to Barcelona,
 where we spent two days together before he went back to Bern and I returned to Munich.
 
 During the long drive by bus from Jaca to Barcelona we discussed a lot of
 physics including also Dual QCD.  He advised me not to work using
 this method for  a while. He told me that he cannot judge how good this
 approach was, but there was just waste of time to discuss it with
 Martinelli, Gavela and other lattice gauge theorists because according
 to him as of 1988 they had no idea how to calculate hadronic matrix elements
 relevant for non-leptonic Kaon decays. He expected that it will take
 at least a decade before LQCD calculations for these decays will be
 of any value and then it will be interesting to see whether LQCD will
 confirm the approximate results of Bardeen, G{\'e}rard and myself.

 I followed his advice and basically stopped working on DQCD after
 returning to Munich. Peter was only partly right. It took 25 years
 before one could meaningfully compare LQCD results with ours. Basically
 in all cases they confirmed our results within the uncertainties of both approaches but did not give us proper credit
 for them.

 As LQCD started obtaining respectable results 28 years after our work in 1986, I
 returned to DQCD in 2013 and from 2013 to 2018 performed, mainly with
 Jean-Marc, new calculations
 improving our old results through the inclusion of vector meson contributions
 and extending them with Jean-Marc and a young Swiss theorist, Jason Aebischer,
 to the hadronic matrix elements of new operators absent
 in the SM but present in various extensions of this model. Presently LQCD
 is far behind us in the latter case. The related papers can be found in
 \cite{Buras:2014maa,Buras:2015xba,Buras:2016fys,Buras:2018evv,Buras:2018lgu,Aebischer:2018rrz} and I will describe them
 in Part~\ref{RDQCD}.

 Changing the subject, something amusing happened during our visit in Barcelona.
 One afternoon we noticed many people standing in a queue and entering
 a building which was more of a palace than an ordinary building. We decided
 to check what it was and eventually entered the building and followed
 others in the queue upstairs. It was a bit strange that everybody was rather elegantly dressed and every body was
 serious. We realized finally that an important person must have died and
 these people came to say goodbye to this person. It turned out to be the major
 of Barcelona who died a few days before and was now lying in an open
 sarcophagus in the middle of a big room. The people were approaching  the  sarcophagus one by one, stopping for 5 seconds, making a bowing and leaving the
 room. We had no idea who this major was but there was no way out, we had
 to do the same.

 I did not have much contact with Peter afterward which I regret because he
 was a very nice person and his advice to pause in the exploration of Dual QCD
 was very useful for me. Unfortunately Peter passed away already in 2016, much too early.

 In my last two  years at MPI I thought one should bring more life into the group. Therefore,
  I arranged twice meetings for the full theory group in which most of the
 members gave short 5 min talks in which they presented their research. I even
 succeeded in convince the three directors, Hans Peter Dürr, Wolfhart Zimmermann
 and Leo Stodolsky to present their work. It was fun but I am not sure it helped
 in bringing people to collaborate with each other except for one case as I
 described at the beginning of my review \cite{Buras:2011we}: a very important
 collaboration with Peter Weisz on higher order QCD corrections to weak decays was initiated in this manner. Eventually we have written four papers together
 which by now earned more than 2100 citations with one, together with Matthias Jamin, cited roughly  800 times. It is now time to review my main activities at TUM
 in the 1990s which were dominated by NLO QCD calculations for rare K and B decays and the phenomenological analyses of the latter in the Standard Model and in
 a few cases beyond it. As described in \cite{Buras:2011we} the motivation
 for going in this direction came from one discussion with Guido Martinelli
 at the 1988 Ringberg Workshop and from my interest in climbing high summits 
 as described in Appendix~\ref{Climb}.

\section{Contributions of the Munich NLO Club (MNLC)}
The review of the NLO and NNLO QCD calculations performed in Munich
and other places in the world appeared first in 2011 \cite{Buras:2011we}.
But expecting further progress, in particular in NNLO QCD calculations and
the ones for operators contributing beyond the SM, I published its updated
version in Physics Reports only in 2023. I am very satisfied that most of the
pioneering NLO and NNLO QCD calculations have been carried out in my group in
Munich that consisted of postdocs, PhD students and diploma students.
Moreover, several of them contributed to NNLO QCD calculations after leaving
my group.

As my book \cite{Buras:2020xsm}, the review with Gerhard Buchalla and Markus Lautenbacher \cite{Buchalla:1995vs}, my Les Houches lectures \cite{Buras:1998raa}
     and the review in \cite{Buras:2011we} describe  our calculations in some
detail, including several anecdotes related to them, I will  only briefly describe the calculations  in which I took part and will only list other
calculations from our club, with short comments, in which I did not participate. My 2023 review in
\cite{Buras:2011we} lists in Tables 1-5 all NLO and NNLO calculations
that appeared until the Spring of 2023. Several new calculation since then
have been performed and will be listed below. I plan to update
Tables 1-5 in \cite{Buras:2011we} in the fall of this year.

\subsection{Non-leptonic Current-Current Operators (1989)}\label{NLO1}
  The first paper of my grand project to calculate NLO QCD corrections to all weak decays
  was written in collaboration with Peter Weisz \cite{Buras:1989xd}. It dealt with the
  NLO QCD corrections to the WCs of the four-quark current-current operators. The main
  achievements in this paper are the calculation of these corrections in the 't Hooft-Veltman (HV) scheme for $\gamma_5$ and the introduction of the so-called
  Naive Dimensional Renormalization (NDR) scheme which allows to perform calculations with an anti-commuting $\gamma_5$ provided the evanescent operators are properly taken   into account. The latter calculations are much simpler than in  the HV scheme and also in the DRED scheme used several years before by an Italian group \cite{Altarelli:1980fi}. The two-loop anomalous dimensions depend on the
  renormalization scheme (RS) and are different in these three schemes. We have demonstrated how this dependence
  is canceled by the one present in the hadronic matrix elements of current-current operators so that physical observables are RS-independent. This is similar
  to my previous analysis in DIS in 1978 but the new complication is the treatment of $\gamma_5$ in $D\not=4$ dimensions. The NDR scheme was used predominantly
  in our subsequent NLO and NNLO papers although in some of them HV scheme has been used to make some checks of the final results. The results of this paper
  have been used in the subsequent papers by other authors.
  \boldmath
  \subsection{$\Delta F=2$ Non-Leptonic Operators (1990)}\label{NLO2}
  \unboldmath
  Our NLO team increased in the fall of 1989 when Matthias Jamin, an excellent postdoctoral fellow, joined my group at TUM. He was already well trained in two-loop
  calculations in a different context and could right away contribute in an important manner to our second project: calculations of NLO QCD corrections to
  heavy top quark contributions to the parameter $\varepsilon_K$ and the $\Delta M_{d,s}$ in 
  $B_{d,s}^0-\bar B^0_{d,s}$ mixings \cite{Buras:1990fn}. To my knowledge these were the first NLO QCD calculations
  of  FCNC processes. In this paper we have considered the top-quark to be
  heavier than $W$ and $Z$ bosons which made in this sector the RG evolution to low energies 
  much simpler both at the LO and NLO that would be the case if top quark
  was much lighter. It could be integrated out right away in contrast to the much earlier LO calculations of Gilman and Wise \cite{Gilman:1982ap} which implied more
  complicated formulae.

  For $\Delta M_{d,s}$ this was the complete NLO QCD calculation because charm contributions could be neglected. For  $\varepsilon_K$ this was by far the dominant
  contribution, in particular for the true top quark mass which was unknown
  to us in 1990. But  in this case pure charm contributions and mixed top-charm
  contributions cannot be neglected. These much harder calculations have been performed by my PhD students Stefan Herrlich  and Ulrich Nierste \cite{Herrlich:1993yv,Herrlich:1995hh,Herrlich:1996vf} in 1993-1996. In this context Herrlich and Nierste
  analyzed the issue of the evanescent operators more generally than I had done with Peter Weisz \cite{Herrlich:1994kh}. They have shown that one can make different choices
  of them but that the one made by Peter and me was the most convenient one.
  See also the paper by Dugan and Grinstein \cite{Dugan:1990df}.

  The NLO QCD calculations in  \cite{Buras:1990fn} and \cite{Herrlich:1993yv,Herrlich:1995hh,Herrlich:1996vf} have been extended years later to the NNLO level
  by other members of my physics family \cite{Brod:2011ty,Brod:2019rzc} with
  Martin Gorbahn and Emanuel Stamou being my physics sons and Joachim Brod my
  physics grandson. Most recent summary of the status of quark mixing can
  be found in the article of Ulrich Nierste \cite{Nierste:2025jxe} where many
  references can be found.

  All these papers played an essential role in the phenomenology since they were published  and are highly
  cited. In particular my paper with Jamin and Weisz \cite{Buras:1990fn}
  collected roughly 800 citations and among my papers on flavour physics
  only my review with Gerhard Buchalla and Markus Lautenbacher on NLO QCD
  corrections \cite{Buchalla:1995vs} and my Les Houches lectures \cite{Buras:1998raa} have more citations.
    \subsection{Non-Leptonic QCD-Penguin and Electroweak Penguin\\ Operators (1991-1992)}
  In \cite{Buras:1989xd} only current-current four-quark operators have been considered. In order to study non-leptonic decays, in particular the ratio $\epe$ in $K_L\to\pi\pi$ decays and decays like $B\to\pi K$, $B\to\pi\pi$ and $B\to KK$, NLO QCD corrections to WCs of non-leptonic QCD-Penguin and Electroweak Penguin operators have to be calculated.
  For this project our NLO team has been increased. An excellent new PhD student, Markus Lautenbacher, joined our efforts which resulted in several papers.
  In the first paper we set up the general structure of $\Delta S=1$ and $\Delta B=1$ effective Hamiltonians beyond the leading logarithmic approximation and
  calculated two loop $6\times 6$ anomalous dimension matrix of current-current
  and QCD penguin operators \cite{Buras:1991jm}. Working by hand I performed
  this calculation only in the NDR scheme while my collaborators using Mathematica did this calculation both in the NDR and the HV scheme. The two-loop calculations in the HV scheme where too involved to be done in this case by hand but
  I was able to perform a one-loop calculation that related these two schemes demonstrating that our two-loop calculations in NDR and HV schemes where consistent with each other.

  In   \cite{Buras:1992tc} and 
  \cite{Buras:1992zv} we extended these calculations to Electroweak Penguin operators by calculating $10\times 10$ ADM $\ord(\alpha_s^2)$ and
  $\ord(\alpha_s\alpha)$, respectively. In the case of the $\ord(\alpha_s^2)$ ADM we could confirm the consistency of our NDR and HV calculations. However,
  there was a problem with it in the case of $\ord(\alpha_s\alpha)$. Possibly, if
  Peter Weisz would still take part in this calculation  this problem in the HV scheme would not exist. Unfortunately spending few months looking for the reason of   this inconsistency Markus and Matthias were unable to find the error.

  However, as described in more details in \cite{Buras:2011we}, we learned
  in the fall of 1992 that Guido Martinelli and his PhD students Marco Ciuchini, Enrico Franco and Laura Reina  were performing the same calculation in the HV
  scheme. In order to be the first team that published these ADMs not only in the NDR scheme but also
  in the HV scheme, I suggested that we should publish the ADM in the HV
  scheme that was obtained from the NDR scheme  by adding a one-loop shift.
  I was rather confident that our NDR calculation, that I was also involved in, was correct and   as the one-loop shift is easy to calculate I was sure that the
  resulting  $\ord(\alpha_s\alpha)$ ADM in the HV scheme obtained by us was also correct.

  In fact few months later the Italian competitors published their paper
  with the $10\times 10$ ADMs at $\ord(\alpha_s^2)$ and
  $\ord(\alpha_s\alpha)$ in the HV scheme that agreed with ours
\cite{Ciuchini:1992tj,Ciuchini:1993vr}. Studying
  their papers we could identify the error Markus and Matthias made in their HV
  calculation of $\ord(\alpha_s\alpha)$ ADM 
  (nobody is perfect). Thus while a credit for the first direct
  NDR calculation should be given to us, the one for the direct calculation of
$\ord(\alpha_s\alpha)$ ADM
  in the HV scheme should be given to the Rome group.

  The agreement on these two ADMs between the Munich and Rome groups
  was an important result with a significant impact on all phenomenological
  analyses of non-leptonic meson decays since then.

\subsection{The Anatomy of $\epe$ Beyond Leading Logarithms with\\ Improved Hadronic Matrix Elements (1993)}
  Our first phenomenological analysis using the above NLO  results
  appeared in March 1993 \cite{Buras:1993dy}. This was another anatomy
  of $\epe$ with not only improved ADMs but also improved hadronic matrix elements stressing in particular the scale and renormalization scheme dependences
  of these matrix elements.

  We stressed that the scheme
dependence of the Wilson coefficients $C_i(\mu)$ can only be cancelled
by the one present in the hadronic matrix elements $\langle Q_i(\mu)
\rangle$.  This requires also the calculation of $\ord(\aem)$
corrections to $\langle Q_i(\mu) \rangle$. We proposed a new
semi-phenomenological approach to hadronic matrix elements which
incorporates the data for $CP$-conserving $K\to\pi\pi$ amplitudes and
allows to determine the matrix elements of all $(V-A)\otimes (V-A)$
operators in any renormalization scheme. In particular
the $\Delta I=1/2$ rule was automatically taken into account.   

Calculating $\langle Q_6 \rangle$ and $\langle Q_8 \rangle$ in the
$1/N$ approach we found in the NDR scheme
$\epe = (6.7 \pm 2.6)\times 10^{-4}$.
  Clearly after more than 30 years the details of the numerical analysis of this anatomy
  is out of date\footnote{Despite the fact that the final value for $\epe$ agrees very well with my claim of 2026.} but several expressions presented there were basic for future
  anatomies that we will meet on our route, in particular in Part~\ref{Anatomies}.

  In fact the parameters $\bsi$ and $\bei$ and analogous parameters
  for other operators have been introduced in this paper. Most importantly
  the expression
  \begin{equation}
\epe = \; 10^{-4}\,\left[\frac{\IM\lambda_{ t}}
{1.7\cdot 10^{-4}}\right]\, \left[\,P^{(1/2)}-P^{(3/2)}\,\right] \, ,
\label{eq:1.4}
\end{equation}
where $\lambda_{t}=V_{ts}^* V_{td}$ and
\begin{eqnarray}
P^{(1/2)} & = & a_0^{(1/2)} + a_2^{(1/2)}\,B_2^{(1/2)} +
                  a_6^{(1/2)}\,B_6^{(1/2)} \, ,
\label{eq:1.5} \\
\mvs
P^{(3/2)} & = & a_0^{(3/2)} + a_8^{(3/2)}\,B_8^{(3/2)} \, ,
\label{eq:1.6}
\end{eqnarray}
was used in several subsequent papers. With $a_6^{(1/2)}$ and $a_8^{(3/2)}$
being positive and $\bsi=\bei=1$ in the large N limit it expresses the
positive contribution of QCD penguin operators to $\epe$ that is suppressed
by electroweak penguin operators as found in 1989 in \cite{Flynn:1989iu}
and \cite{Buchalla:1989we}.
We calculated the coefficients $a_i$ as functions of $\Lms$ and $m_t$
for the leading order and the two renormalization schemes (NDR, HV)
considered. The $1/N$ correction to $\bsi$ and $\bei$ have 
been calculated by Jean-Marc and myself only in 2015 in \cite{Buras:2015xba}
as described in Section~\ref{B6B8}. We found $\bsi \le \bei<1 $.

  \subsection{Electroweak Penguin Contributions to Non-Leptonic\\ $\Delta F=1$  Decays at NNLO (1999)}
  In order to increase the precision one has to go beyond the NLO and calculate NNLO
  contributions. In 1999 the first NNLO calculation in weak decays has been
  performed by Paolo Gambino (a well known postdoc in my group), Ulrich Haisch (my diploma student at that time) and myself \cite{Buras:1999st}. It was the calculation of Electroweak Penguin contribution to non-leptonic $\Delta F=1$ decays at the NNLO. It is particularly relevant for the ratio $\epe$, increasing the role of EW penguins in the suppression of this ratio. More about it in Section~\ref{epe}.

 In more details we calculated the $\ord(\alpha_s)$ corrections to the $Z^0$-penguin and electroweak box diagrams relevant 
  for non-leptonic $\Delta S=1$ and $\Delta B=1$  decays. This calculation provided the complete 
$\ord (\alpha_W\alpha_s)$ 
and $\ord(\alpha_W\alpha_s \sin^2\theta_W m_t^2)$ corrections  
($\alpha_W= \alpha/\sin^2\theta_W$)
to the Wilson coefficients of the electroweak penguin four quark operators relevant for non-leptonic K- and B-decays. We argued that this is the dominant part of the NNLO contributions to these coefficients. Our results allowed to reduce considerably the uncertainty due to the definition of the top quark mass present in the existing NLO calculations of non-leptonic decays. The NNLO corrections to the coefficient of the color
singlet $(V-A)\otimes ((V-A)$ electroweak penguin operator $Q_9$ relevant for B-decays are generally moderate, amount to a few percent for the choice
$m_t(\mu_t=m_t)$  and depend only
weakly on the renormalization scheme. Larger NNLO corrections with substantial scheme dependence were found for the coefficients of the remaining electroweak penguin operators $Q_7, Q_8$ and $Q_{10}$. In particular, the strong scheme dependence of the NNLO corrections to
$C_8$ allows to reduce considerably the scheme dependence of $C_8\langle Q_8\rangle_2$ relevant for the ratio $\epe$. I should stress that although I
derived many expressions in this paper and calculated several contributions,
Paolo and Uli were leading this project. In fact, as far as I remember, Uli Haisch was the only diploma student I had, who during his diploma calculated
two-loop diagrams.

\subsection{Rare $K$ and $B$ Decays (1992-2005)}
Parallel to the calculations of NLO QCD corrections to non-leptonic decays I
started to calculate such corrections to rare K and B decays. My excellent PhD
student Gerhard Buchalla joined me in this effort and several papers resulted
from this collaboration.

Already in December 1992 we have completed the calculation of QCD corrections to the $\bar s dZ$ vertex for arbitrary top quark mass \cite{Buchalla:1992zm}.
This vertex is crucial for these decays as it enters the $Z$-penguin diagram and
the function $C(x_t)$ in (\ref{PB1}).

In a subsequent paper we have calculated QCD corrections to the box function
$B_0(x_t)$ in (\ref{PB1} \cite{Buchalla:1993bv}, which allowed us to obtain QCD corrections to the gauge independent $X(x_t)$ and $Y(x_t)$ functions so that now
we could write
\be\label{PB2}
X(x_t)=C(x_t)-4 B(x_t), \qquad Y(x_t)=C(x_t)-B(x_t)
\ee
without ``0'' index, meaning that QCD corrections have now been included.
Having this we analyzed the size of these corrections in $K\to\pi\nu\bar\nu$,
$B\to X_s\nu\bar\nu$ and $B\to\ell^+\ell^-$ decays.

In the third paper of this period, in which only decays governed by functions
$X(x_t)$ and $Y(x_t)$ have been considered, a detailed NLO analysis of $\kpn$ and
$K_L\to \mu^+\mu^-$ has been performed in \cite{Buchalla:1993wq}. Most
importantly charm contributions to both decays have been calculated at the NLO level
and the reduction of various scale uncertainties through the inclusion
of NLO QCD corrections has been demonstrated.

There were still several phenomenological papers written with Gerhard to be listed in Part~\ref{Rare}. But here I just want to mention the paper on two-loop
large $m_t$ electroweak corrections to $\kpn$ for arbitrary Higgs boson mass which was unknown at that time \cite{Buchalla:1997kz}.

Unfortunately, as pointed out by Misiak and Urban in 1998 \cite{Misiak:1999yg}, in calculating QCD corrections to box diagrams we have missed the presence of an evanescent operator which
modified our results by $1\%$. A tiny effect but still an error. Misiak and
Urban considered only $\kpn$ decay. We have then presented an update of
all our calculations in which this small error appeared \cite{Buchalla:1998ba}.

My efforts to calculate $\kpn$ in the SM as precisely as possible culminated in 2005 in the calculation of charm contribution to this decay at the NNLO level in QCD.
This rather advanced calculation was done in collaboration with three of my PhD students: Martin Gorbahn, Ulrich Haisch and Ulrich Nierste \cite{Buras:2005gr,Buras:2006gb}.

This new contribution reduced the theoretical uncertainty in the relevant
parameter $P_c$  from $\pm 10.1\% $ down to $\pm 2.4\%$. After this calculation
the error in $P_c = 0.37 \pm 0.04$ was fully dominated by the uncertainty in the charm quark mass $m_c$. Including contributions from light-quark loops \cite{Isidori:2005xm}
and electroweak corrections \cite{Brod:2008ss}, one finds $P_c = 0.405 \pm 0.024$.

The rare decays described so far involved the $Z$ penguin but not the photon
penguin which is the case of decays like $K_L\to\pi^0e^+e^-$, $B\to X_s e^+e^-$ and alike. In early 1994 Markus Lautenbacher, Mikolaj Misiak, Manfred Münz
and myself calculated NLO QCD corrections to $K_L\to\pi^0e^+e^-$ \cite{Buras:1994qa}. Manfred was
another excellent PhD student of mine and Mikolaj one of the leading experts in NLO and NNLO QCD calculations to weak decays, in particular $B\to X_s\gamma$, as will be evident soon. Having this result Gerhard Buchalla and myself could
investigate parity violating longitudinal polarization in $K^+\to\pi^+\mu^+\mu^-$ at the NLO level \cite{Buchalla:1994ix}. Hopefully it will be measured one day.

Somewhat surprisingly, in \cite{Buras:1994qa} we did not calculate such corrections to $B\to X_s e^+e^-$ decay. 
 Possibly because Markus was busy finishing his PhD and Mikolaj went
back to Warsaw. But in the fall of 1994 walking one day after lunch from MENSA to the Physics Department Manfred and I realized that one could easily extend our NLO calculation
for $K_L\to\pi^0e^+e^-$ to the $B\to X_s e^+e^-$ decay that of course could be used with few modifications for $B\to X_s \mu^+\mu^-$ as well. We performed this NLO
calculation both in the NDR and HV scheme verifying the consistency of these
calculations \cite{Buras:1994dj}. It is really amazing that this paper, with over 600 citations, is the most cited of my papers written with a single collaborator. While I think it is a good paper, the main reason for this was the popularity of the $b\to s\mu^+\mu^-$ transitions in the last 30 years.

   \boldmath
 \subsection{$B\to X_s\gamma$ (1993-2002)}\label{bsgamma}
 \unboldmath
 As described in details in \cite{Buras:2011we} and \cite{Buras:1998raa},
 at the end of the 1980s
 there were several attempts
 by different groups to calculate  QCD corrections to $B\to X_s\gamma$
 decay with different results depending on the renormalization scheme used.
 This was surprising as one-loop ADMs should be RS independent. While this was
 one motivation of Peter Weisz and myself to enter QCD calculations in 1988,
 being busy with the NLO QCD calculations for other decays, we did not have time
 to look at this issue. This problem has been solved by the Rome group
 \cite{Ciuchini:1993ks,Ciuchini:1993fk}  who
 noticed that this scheme dependence cancels in physical observables already at
 the level of the WCs. This motivated me to look at this decay. This time
 with Mikolaj Misiak, Manfred Münz and Stefan Pokorski. The resulting
 paper \cite{Buras:1993xp} became a bestseller for two reasons:
 \begin{itemize}
 \item
   We have introduced two effective WCs for the magnetic penguin operators $Q_7$
   and $Q_8$ 
   that in the LO were RS independent. $C_7^{eff}$ was a sum  of
   the usual WC $C_7$ and the linear combination of $C_i$ with $=1,..6$.
   Similar for $C_8$. This effective WCs have been used in the $B\to X_s\gamma$
   literature since then.
 \item
   We have outlined the procedure for the calculation of NLO QCD corrections to this decay and
   demonstrated how the strong scale dependence at LO pointed out earlier
   by Ali, Greub and Mannel \cite{Ali:1993cj} is reduced after the inclusion of NLO corrections. 
 \end{itemize}

 Few years later, in 1997, I returned to this issue with Axel Kwiatkowski (a postdoc in my group) and Nicolas Pott (my PhD student).
 In this context we have calculated
 QCD corrections to matching conditions at the electroweak scale and demonstrated now explicitly
 how the uncertainty in the choice of the matching scale is reduced through
 NLO QCD corrections  \cite{Buras:1997bk,Buras:1997xf}. At this stage it is
 appropriate to mention numerous contributions of  Christoph Greub to this decay
 that are summarized in \cite{Aebischer:2025wzl}.

 Parallel to our calculations of QCD corrections to the matching, Misiak and M\"unz \cite{Misiak:1994zw} calculated two-loop anomalous dimensions of magnetic penguin operators. See also their important 1996 paper with Chetyrkin \cite{Chetyrkin:1996vx}.
 However, in order to complete the calculation of NLO QCD corrections to this decay one had to calculate still 
 two-loop matrix elements of current-current and QCD penguin operators relevant
 for this decay. This has been done in 
 \cite{Buras:2001mq} and \cite{Buras:2002tp}  respectively\footnote{The story of similar calculations by other groups can be found in Section 6.3.4 of \cite{Buras:2011we}.}. Mikolaj Misiak
 was again in our team and two of my new collaborators: Andrzej Czarnecki
 from Alberta University, who was the leader in \cite{Buras:2001mq} due to some
 technique developed by him which was very useful for these calculations, and
 J\"org Urban, an excellent postdoc in my group. In 2002 Mikolaj and
 I wrote a paper summarizing the completion of NLO QCD corrections to
 $B\to X_s\gamma$ \cite{Buras:2002er}. We dedicated it to Stefan Pokorski
 on the occasion of his 60th birthday.

\boldmath
\subsection{Two-Loop Non-Leptonic $\Delta F=1$ and $\Delta F=2$ Anomalous Dimensions Beyond the SM (2000)}\label{BMU}
\unboldmath
In 1998 the Rome Group involving Ciuchini, Franco, Lubicz, Martinelli, Scimemi and Silvestrini \cite{Ciuchini:1997bw},
calculated two-Loop ADMs of non-leptonic $\Delta F=1$ and $\Delta F=2$ operators present beyond the SM. They calculated them in a particular Lattice scheme and the NDR
scheme. The calculation in the lattice scheme was in $D=4$ while the one in the NDR in $D\not=4$. Mikolaj Misiak noticed rather strange pattern of the obtained ADMs in the NDR scheme: some special entries in the ADMs that on general grounds should be vanishing in the NDR scheme were non-zero. From my studies with Peter Weisz in 1989 it was rather clear what has happened. The Italian masters have
not included evanescent operators. Mikolaj, J\"org Urban and myself  decided to check it and calculated these
ADMs in the NDR scheme properly taking evanescent operators into account. We obtained
rather different ADMs then the Italian Group with vanishing elements in question. Subsequently we calculated the ADMs in the Lattice scheme confirming the result of the Italian group. We have also demonstrated that these two results are consistent with each other provided also the one-loop shift is properly calculated.
The details can be found in  \cite{Buras:2000if}. In fact not all BSM $\Delta F=1$ operators have been considered by the Italian group so that our results were more complete. This was fortunate 25 years later in the context of the WET and the SMEFT
as I will discuss in Part~\ref{SMEFT}.

Subsequently in \cite{Buras:2001ra} Jörg, Sebastian J\"ager (my excelent PhD student) and myself using the NDR results in \cite{Buras:2000if} presented
analytical formulae for the NLO QCD RG factors relevant for $\Delta F=2$
processes. Such factors are very useful for phenomenology. The increased
number of $\Delta F=1$ operators relative to the $\Delta F=2$ case precludes
useful analytical expressions for the corresponding factors and numerical
computer calculations are preferable.
\subsection{NLO QCD Corrections  
for Tree Level Non-Leptonic\\
$\Delta F = 2$ and $\Delta F=1$ Transitions Beyond the 
SM (2012)                       }\label{BG12}
\unboldmath
In 2011 my granddaughter in physics Jennifer Girrbach-Noe, PhD student of
Ulrich Nierste joined my group. Somehow Ulrich, despite being an NLO-QCD
expert, did not teach Jennifer how to calculate NLO QCD corrections.
But her physics grandfather did it.

Approaching 66 by then I did not want to perform again two-loop calculations 
but fortunately I noticed that for tree-level FCNC processes like the ones
mediated by heavy $Z^\prime$ or Scalars, what remained to be done where calculations of one-loop QCD corrections to these processes. These were simply
one-loop QCD corrections to the matching of the full theory to the effective
theory. The relevant two-loop anomalous dimensions where known already by that
time as described previously.

First in \cite{Buras:2012fs} tree-level $\Delta F=2$ transitions mediated
by $Z^\prime$ or Neutral Scalars were considered. In these models  new local operators absent in 
the SM are generated having Wilson coefficients that are generally much 
stronger affected by RG QCD effects than it is the case 
of the SM operators.

We performed the calculation of $\mathcal{O}(\alpha_s)$
corrections to matching conditions for the Wilson coefficients of the contributing operators in the NDR-$\overline{\text{MS}}$ scheme. 
This
allowed to reduce certain unphysical scale and renormalization scheme
dependences in the existing  NLO calculations that did not take these contributions into account. We showed explicitly how our results can be combined
with the analytic formulae for the so-called $P_i^a$ QCD factors (see \cite{Buras:2001ra})  that include both hadronic matrix
elements of contributing operators and RG  evolution from high energy to low
energy scales. For the masses of heavy gauge bosons and scalars $\ord(1)\tev$ 
the remaining unphysical scale dependences for the mixing 
amplitudes $M_{12}$ 
are reduced typically from $10-25\%$, depending on the operator  considered, 
down to $1-2\%$.

The corresponding calculations for $\Delta F=1$ proceses have been performed in
\cite{Buras:2012gm}.
Prominent examples in this case are the right-handed $W^\prime$ bosons in left-right symmetric models and charged Higgs ($H^\pm$) particles in models with extended scalar sector like  two Higgs doublet models and supersymmetric models. 
Even in the case of decays with four different quark flavours involved, 
to which penguin operators cannot contribute, twenty linearly independent 
operators, instead of two in the SM, have to be considered.
Anticipating the important role of such decays at the LHCb and Belle  and having in mind future improved lattice computations, we completed
the existing NLO QCD  formulae for these processes by  calculating
$\mathcal{O}(\alpha_s)$
corrections to matching conditions for the Wilson coefficients of all 
contributing operators in the NDR-$\overline{\text{MS}}$ scheme.
This again 
allowed to reduce unphysical scale and renormalization scheme
dependences in the existing NLO calculations. Our results can also be 
applied to models with tree-level heavy neutral gauge boson and scalar 
exchanges in $\Delta F = 1$ transitions and constitute an important part of NLO analyses of those non-leptonic
 decays to which 
also penguin operators contribute. 

\section{NLO and NNLO Flavour Calculations of my Physics Family without my Participation}\label{NLOfamily}
As in mountain climbing and sports, there comes a stage in physics when younger generations take the lead.
My two papers with Jennifer, just described, were my last
 NLO calculations in the SM or specific NP models but fortunately my physics sons and one grandson (Joachim Brod) took
over such calculations and dominantly concentrated on NNLO QCD calculations and
 NLO electroweak ones within the SM. I just want to list these
calculations and summarize the most important results of each paper,  leaving
their detailed description to the authors of these papers. However, as I
will describe in Part~\ref{SMEFT}, in the last years I took part in NLO QCD
calculations for WET and SMEFT as we  could use some of our older results
without being forced to do explicit two-loop calculations.

The list of the NLO and NNLO calculation in flavour physics performed in my physics family without
my participation is given below. I list only those which I did not describe
already. We have then the following impressive list.

\begin{itemize}
\item
  M. Beneke, G.  Buchalla,  C. Greub,  A. Lenz, and U. Nierste:
  Next-to-leading order QCD Corrections to the Lifetime Difference of $B_s$ Mesons \cite{Beneke:1998sy}.

The authors computed the QCD corrections to the decay rate difference
in the $B_s-\bar B_s$ system, $\Delta\Gamma_{B_s}$, in the 
NLO logarithmic approximation using the heavy quark 
expansion approach. Going beyond leading order in QCD is essential
to obtain a proper matching of the Wilson coefficients to the
matrix elements of local operators from lattice gauge theory. The lifetime 
difference is reduced considerably at the NLO order.
As a further application of their analysis they also derived the NLO result for the mixing-induced CP asymmetry in inclusive
$b\to u\bar ud$ decays, which measures $\sin 2\alpha$.
\item
  M. Beneke, G.  Buchalla,  C. Greub,  A. Lenz, and U. Nierste:
  The $B^+ -B^0_d$ Lifetime Difference Beyond Leading Logarithms \cite{Beneke:2002rj}.

The authors computed perturbative QCD corrections to the lifetime splitting
between the charged and neutral $B$ meson in the framework of
the heavy quark expansion. These NLO logarithmic
corrections are necessary for a meaningful use of hadronic matrix
elements of local operators from lattice gauge theory. They implied
significant reduction of the 
uncertainties associated with the choices of renormalization scale and
scheme compared to the LO result. 
They included the full dependence on the charm-quark mass $m_c$ without any
approximations.
\item
P. Gambino, M. Gorbahn, and U. Haisch: Anomalous Dimension Matrix for
Radiative and Rare Semileptonic B Decays Up to Three Loops,
\cite{Gambino:2003zm}.

The authors computed the complete $\ord (\as^2)$ anomalous dimension \ matrix relevant for the
$b\to s\gamma$, $b\to s g$ and $ b\to s \ell^+\ell^-$ transitions in the
SM and some of its extensions. For radiative decays they
confirmed the results of Misiak and M\"unz \cite{Misiak:1994zw} and of 
Chetyrkin, Misiak and M\"unz \cite{Chetyrkin:1997gb}. The $\ord (\as^2)$ mixing of four-quark
into  semileptonic operators was instead a new result and represented
one of the last missing ingredients of the
NNLO analysis of rare semileptonic $B$ meson
decays.
\item
  C. Bobeth, P. Gambino, M. Gorbahn, and U. Haisch: Complete NNLO QCD
Analysis of $\bar B\to X_s\ell^+\ell^-$ and Higher Order Electroweak Effects,
\cite{Bobeth:2003at}.

The authors completed the NNLO QCD calculation of the 
branching ratio for $\bar B\to X_s\ell^+\ell^-$ including recent results for
the three-loop anomalous dimension matrix and two-loop matrix
elements. These new contributions  modified the branching ratio in the
low-$q^2$ region by about $+1 \%$ and $-4 \%$,
respectively. They also discussed the appropriate normalization of
the electromagnetic coupling $\aem$ and calculated the dominant higher
order electroweak effects, showing that, due to accidental
cancellations, they change the branching ratio by only $-1.5 \%$ if $\aem (\mu)$
is normalized at $\mu =\ord(m_b)$, while they shift it by about $-8.5 \%$ if
one uses a high scale normalization $\mu = \ord(M_W)$. The position
of the zero of the forward-backward asymmetry, $q_0^2$,  changed by
around $+2 \%$. After introducing a few additional improvements in
order to reduce the theoretical error, they performed a comprehensive
study of the uncertainties that likely have been changed in the last 20 years.
\item
M. Gorbahn and U. Haisch: Effective Hamiltonian for non-leptonic $|\Delta F=1|$
decays at NNLO in QCD, 
\cite{Gorbahn:2004my}.

Martin and Uli computed the effective hamiltonian for non-leptonic
$| \Delta F | = 1$ decays in the SM including NNLO
QCD corrections. In particular, they presented the complete
three-loop anomalous dimension matrix describing the mixing of current-current and
QCD penguin operators. The calculation was performed in an operator
basis which allowed to consistently use fully anticommuting $\gamma_5$
in dimensional regularization at an arbitrary number of loops. The
renormalization scheme dependences and their cancellation in physical
quantities was discussed in detail. Furthermore other technicalities
were discussed in detail. Finally, they solved
the renormalization group equation and gave the analytic expressions
for the low-energy Wilson coefficients relevant for non-leptonic $B$
meson decays beyond NNLO in two renormalization
schemes.    
\item
M. Gorbahn, U. Haisch, and M. Misiak: Three-loop mixing of dipole operators, 
\cite{Gorbahn:2005sa}.

This trio  calculated the complete three-loop $\ord (\as^3)$ anomalous
dimension matrix for the dimension-five dipole operators that arise in
the SM after integrating out the top quark and the heavy
electroweak bosons. Their computation completed the three-loop anomalous
dimension matrix of operators that govern low-energy $| \Delta F | =
1$ flavor-changing processes, and represents an important ingredient
of the NNLO QCD analysis of the $B\to X_s\gamma$
decay.   
\item
M. Gorbahn and U. Haisch: Charm Quark Contribution to $K_L\to\mu^+\mu^-$ at
NNLO, 
\cite{Gorbahn:2006bm}.

This time Martin and Uli  calculated the charm-quark contribution to the decay
$K_L\to\mu^+\mu^-$
in
NNLO of QCD. This new contribution reduced
the theoretical uncertainty in the relevant parameter $P_C$ from $\pm
22 \%$ down to $\pm 7 \%$, corresponding to scale uncertainties of
$\pm 3 \%$ and $\pm 6 \%$ in the short-distance part of the branching
ratio and the determination of the Wolfenstein parameter $\bar{\rho}$
from $K_L\to\mu^+\mu^-$, respectively.  The error in $P_c = 0.115 \pm 0.018$ originates
from the remaining scale uncertainties and the 2006
uncertainty in the charm-quark mass. They found $\mathcal{B}(K_L\to\mu^+\mu^-)_{\rm SD}= \left ( 0.79 \pm 0.12 \right ) \times 10^{-9}$. The dominant
uncertainty originates in $V_{td}$ or equivalently $V_{cb}$. This is still the case in 2026.
\item
J. Brod and M. Gorbahn: Electroweak Corrections to the Charm Quark
Contribution to $\kpn$,
\cite{Brod:2008ss}.

In this paper the leading-log QED, the next-to-leading-log QED-QCD, and
the electroweak corrections to the charm quark contribution to $\kpn$
have been calculated. The corresponding
  parameter $P_c(X)$  increased by up to $2\%$ with respect to the
  pure QCD estimate to $P_c(X)=0.372\pm 0.015$.
\item
J. Brod and M. Gorbahn, $\varepsilon_K$  at NNLO: The
Charm-Top-Quark Contribution, 
\cite{Brod:2010mj}.

In this paper NNLO QCD analysis of
  the charm-top-quark contribution $\eta_{ct}$ to the effective
  $|\Delta S|=2$ Hamiltonian in the SM has been performed. This required
  the computation of various three-loop anomalous dimensions and of corresponding two-loop matching conditions at the 
 electroweak, the bottom-quark, and the charm-quark scale. This implied
  $\eta_{ct} = 0.496 \pm 0.047$,
  which is roughly $7 \%$ larger than the NLO
  value $\eta_{ct}^{\text{NLO}} = 0.457 \pm 0.073$. The reduction of the error
  should be noticed.
\item
J. Brod and M. Gorbahn: NNLO Charm-Quark
Contribution to the CP Violation Parameter $\varepsilon_K$  and $\Delta M_K$, 
\cite{Brod:2011ty}.

In this paper NNLO QCD corrections to the charm-quark
  contribution $\eta_{cc}$ to the effective $|\Delta S|=2$ Hamiltonian
  in the SM have been calculated. A large positive shift of $36\%$,   leading to $\eta_{cc} = 1.87(76)$ has been found. This result might cast doubt on   the validity of the perturbative expansion and possible
  solutions have been discussed.
  \item
  C. Bobeth, M. Gorbahn, and E. Stamou: Electroweak Corrections to
  $B_{s,d}\to \ell^+\ell^-$, \cite{Bobeth:2013tba}.

In the present paper 
 the full two-loop electroweak matching corrections to the
  operator governing the decays $B_{s,d}\to \ell^+\ell^-$ in the SM have been calculated. Their inclusion removed an electroweak scheme and scale uncertainty of
  about $\pm 7\%$ of the branching ratio. The  residual perturbative
  electroweak and QED uncertainties were estimated to be less than $\pm 1\%$ at the level of   the branching ratio. 
\item
  T. Hermann, M. Misiak and M. Steinhauser: Three-loop QCD corrections to $B_s \to \mu^+ \mu^-$, \cite{Hermann:2013kca}.

  In this paper  QCD
  corrections of order $\alpha_s^2$ to the Wilson coefficient $C_A$
  governing the decays $B_{s(d)} \to \ell^+ \ell^-$ have been computed.
  It required performing three-loop matching between the full and effective
  theories. Including these corrections reduced the unphysical  dependence 
 on the matching scale $\mu_0$ at which the top-quark mass and
      $\alpha_s$ are renormalized. The corresponding uncertainty in
      $|C_A|^2$ got reduced from around $1.8\%$ to less than $0.2\%$. 
  \item
C. Bobeth, M. Gorbahn, T. Hermann, M. Misiak, E. Stamou, and M. Steinhauser:
$B_{s,d}\to \ell^+\ell^-$ in the Standard Model with Reduced Theoretical Uncertainty, 
\cite{Bobeth:2013uxa}.

In this  paper, 
new calculations of the NLO~EW~\cite{Bobeth:2013tba} and NNLO~QCD~\cite{Hermann:2013kca} corrections to $B_{s,d}\to \ell^+\ell^-$, mentioned above, have been combined
resulting in the significant reduction of theoretical perturbative
uncertainties and implying $\overline{\mathcal B}(B_s \to \mu^+ \mu^-) = (3.65 \pm 0.23) \times 10^{-9}$.
\item
J. Brod and E. Stamou: Electric dipole moment constraints on CP-violating
heavy-quark Yukawas at NLO,
\cite{Brod:2018pli}.

EDMs are sensitive probes of new phases in the
Higgs Yukawa couplings. In this paper complete two-loop QCD
anomalous dimension matrix for the mixing of CP-odd scalar and tensor
operators has been calculated.  Applying these results for a phenomenological study of CP
violation in the bottom and charm Yukawa couplings one finds large
shifts of the induced Wilson coefficients at
NLO order. This allowed to update the
constraints on CP-violating phases in the bottom and charm quark
Yukawas by using the experimental bound on the
electric dipole moments of the neutron and mercury. 
\item
J. Brod, M. Gorbahn, and E. Stamou: Standard-Model Prediction of $\varepsilon_K$  with
Manifest Quark-Mixing Unitarity,  
\cite{Brod:2019rzc}.

This is a very interesting paper that allowed by just reschuffling various
contributions in the effective $\Delta S=2$ Hamiltonian to reduce
significantly perturbative uncertainties in $\varepsilon_K$ in the
SM. I describe it in some detail in Part~\ref{eKunitarity}. This paper shows
that sometimes a good idea is more powerful than multi-loop calculations.
\item
F. Bishara, J. Brod, M. Gorbahn, and U. Moldanazarova: Generic one-loop
matching conditions for rare meson decays,
\cite{Bishara:2021buy}.

In this paper explicit results for
the Wilson coefficients of the weak effective Lagrangian for
leptonic and semileptonic meson decays
in any perturbative model, in which these processes
proceed via one-loop contributions, have been presented. A Mathematica code
that implements these results in an easily usable form was provided.
\item
  M. Gerlach, U. Nierste, V. Shtabovenko and M. Steinhauser:
  Two-loop QCD penguin contribution to the width difference in $B_s-\bar B_s$,
  \cite{Gerlach:2021xtb}.

In this paper two-loop QCD corrections to the element $\Gamma_{12}^q$ of the
  decay matrix in  $B_q-\bar{B}_q$ mixing, $q=d,s$, in the leading power
  of the Heavy Quark Expansion have been computed. The calculated
  contributions  involve one current-current and one penguin operator
  and constitute the next step towards a theory prediction for 
  the width difference $\Delta\Gamma_s$ matching the precise
  experimental data.  Useful  compact analytic results for all
  matching coefficients were presented.
  The new corrections are comparable in size to the current
  experimental error and slightly increase  $\Delta\Gamma_s$.
  \item
J. Brod, S. Kvedarait\'e, and Z. Polonsky: Two-loop electroweak corrections to the
Top-Quark Contribution to $\varepsilon_K$,
\cite{Brod:2021qvc}.

 In this paper the calculation of the full
two-loop electroweak corrections to the top-quark contribution to
$\varepsilon_K$, including the resummation of QED-QCD logarithms has been presented. Different renormalization prescriptions for the electroweak
input parameters were discussed. In the traditional normalization of the weak
Hamiltonian with two powers of the Fermi constant $G_F$, the top-quark
contribution is shifted by $-1\%$.
\item
M. Gerlach, U. Nierste, V. Shtabovenko and M. Steinhauser:
The width difference in $B - \bar B$ mixing at order $\alpha_s$ and beyond   \cite{Gerlach:2022wgb}.

In this paper the calculation of the element $\Gamma_{12}^q$ of the
  decay matrix in $B_q-\bar{B}_q$ mixing, $q=d,s$, to order $\alpha_s$
  in the leading power of the Heavy Quark Expansion has been completed.
  To this end 
  one- and two-loop contributions involving two four-quark
  penguin operators have been computed. Furthermore, the authors presented two-loop QCD corrections
  involving a chromomagnetic operator and either a current-current or
  four-quark penguin operator. Such contributions are of order
  $\alpha_s^2$, that is NNLO contributions. They  also
  presented one-loop and two-loop results involving two chromomagnetic
  operators which are formally of NNLO and NNNLO, respectively.
\item
J. Brod, J. M. Cornell, D. Skodras, and E. Stamou: Global constraints on Yukawa
operators in the SMEFT,
\cite{Brod:2022bww}.

CP-violating contributions to Higgs--fermion couplings are
absent in the SM, but are
motivated by models of electroweak baryogenesis. Working within
the SMEFT to parameterise
deviations from SM Yukawa couplings, the authors  presented the leading
contributions of the relevant operators to the fermionic EDMs.
They  obtained constraints on the SMEFT Wilson
coefficients from the combination of LHC data and experimental bounds
on the electron, neutron, and mercury EDMs, and for the first time, 
performed a combined fit to LHC and EDM data allowing the presence of
CP-violating contributions from several fermion species
simultaneously. Among other results, they found non-trivial correlations
between EDM and LHC constraints even in the multi-parameter scans, for
instance, when floating the CP-even and CP-odd couplings to all
third-generation fermions.
    \item
M. Gerlach, U. Nierste,  V. Shtabovenko and M. Steinhauser:
Width Difference in the $B-\bar B$ System at NNLO of QCD \cite{Gerlach:2022hoj}.

The authors  extended the theoretical prediction for the width difference
  $\Delta \Gamma_q$ in the mixing of neutral $B$ mesons in the SM
to NNLO in $\alpha_s$. To this end they calculated various three-loop and
two-loop diagrams.
  Their theoretical prediction reads
  $\Delta\Gamma_s/\Delta M_s = {(4.33\pm 0.93)\cdot 10^{-3}}$ if expressed in
  terms of the bottom mass in the $\overline{\rm MS}$ scheme and
  $\Delta\Gamma_s/\Delta M_s = {(4.20\pm 0.95)\cdot 10^{-3}}$ for the use of the
  potential-subtracted mass. While the controversy on $|V_{cb}|$ affects both
  $\Delta\Gamma_s$ and $\Delta M_s$, the ratio $\Delta\Gamma_s/\Delta M_s$ is
  not affected by the uncertainty in $|V_{cb}|$. This is in the spirit
  of my strategy with Elena Venturini \cite{Buras:2021nns}, but we did not consider this ratio.
\item
  J. Brod, S. Kvedarait\'e, Z. Polonsky, and A. Youssef: Electroweak corrections to the Charm-Top-Quark Contribution to $\varepsilon_K$,
  \cite{Brod:2022har}.

The authors calculated the LO and NLO logarithmic
  electroweak corrections to the charm-top-quark contribution to the
  effective $|\Delta S| = 2$ Lagrangian, relevant for the parameter
  $\varepsilon_K$. They found that these corrections lead to a $-0.5\%$
  shift in the corresponding Wilson coefficient. Most importantly
  their 
  calculation removed an implicit ambiguity in the SM
  prediction of $\varepsilon_K$, by fixing the renormalization scheme of
  the electroweak input parameters.
\item
M. Gorbahn, S. J\"ager, F. Moretti, and E. van der Merwe: Semileptonic weak
Hamiltonian to $\ord(\alpha\alpha_s)$ in momentum-space subtraction schemes
\cite{Gorbahn:2022rgl}.

The CKM unitarity precision test of the SM requires a systematic treatment of electromagnetic and strong corrections for semi-leptonic decays.
  Electromagnetic corrections require the renormalization of a semileptonic four-fermion operator. 
  In this work the authors calculated the $\mathcal{O}(\alpha\alpha_s)$ perturbative scheme conversion between the $\overline{\rm MS}$ scheme and several momentum-space subtraction schemes,
  which can also be implemented on the lattice. Applications of these results
  relevant for lattice calculations were presented.
\item
M. Gorbahn, S. J\"ager, and S. Kvedarait\'e: RI-(S)MOM to $\overline{\text{MS}}$ conversion for $B_K$ at two-loop order, 
\cite{Gorbahn:2024qpe}.

The parameter \( \hat{B}_K \) entering $\varepsilon_K$ and $\Delta M_K$ plays a critical role in constraining the parameters of the CKM matrix and in probing physics beyond the SM.
In this work, the authors improved the precision of \( \hat{B}_K \) to NNLO and provided
world averages for both $3$- and $4$-flavour theories.
In the course of this, they carried out the two-loop matching between the RI-(S)MOM and $\overline{\mbox{MS}}$ schemes. 
They found $\hat B_{K}^{(f=3)} = 0.7627(60)$, which comprises the complete set of $3$- and $4$-flavour
lattice results and can be used directly in phenomenological applications.
They found
$|\varepsilon_K|=2.171(65)_\text{pert.}(71)_\text{non-pert.}(153)_\text{param.} \times 10^{-3}$. The large parametric uncertainty related primaly to strong
dependence of $\varepsilon_K$ ($\vcb^{3.4}$) should be noticed.
\item
  M. Egner, M. Fael, A.  Lenz, M. L. Piscopo, A. V. Rusov,  K. Sch{\"o}nwald and
  M. Steinhauser: Total decay rates of B mesons at NNLO-QCD \cite{Egner:2024lay}.

  The authors updated the SM predictions for the lifetimes of the $B^+$, $B_d$ and $B_s$ mesons 
within the heavy quark expansion (HQE),
including the recently determined NNLO-QCD corrections to non-leptonic decays of the free $b$-quark. In addition, they updated the HQE predictions for the lifetime ratios $\tau (B^+)/\tau (B_d)$ and $\tau (B_s)/\tau (B_d)$, and provided new results for the semileptonic branching fractions of the three mesons entirely within the HQE. They obtained a considerable improvement of the theoretical uncertainties, mostly due to the reduction of the renormalisation scale dependence when going from LO to NNLO, and for all the observables considered, they found good agreement, within uncertainties, between the HQE predictions and the corresponding experimental data. Their results can be found already in the abstract of 
their paper. An outlook for further improvements has been  presented.
\item
M. Gerlach, U. Nierste, P. Reeck, V. Shtabovenko and M. Steinhauser:
Current-current operator contribution to the decay matrix in $B$-meson mixing at next-to-next-to-leading order of QCD \cite{Gerlach:2025tcx}.

The authors computed NNLO perturbative corrections to the
  decay width difference of mass eigenstates and the charge-parity asymmetry
  $a_{\rm fs}$ in flavour-specific decays of neutral $B$ mesons.  They 
  took into account the full dependence on the charm and bottom quark masses
  for the current-current operator contributions up to three-loop
  order. Special emphasis was put on the proper construction of the so-called
  $|\Delta B|=2$ theory such that Fierz symmetry is preserved. They provided
  updated phenomenological predictions, for $\Delta\Gamma$,
  $\Delta\Gamma/\Delta M$ and $a_{\rm fs}$ for the $B_d$ and $B_s$ system,
  including a detailed analysis of the uncertainties of their predictions.  
  The calculated NNLO correction reduced the perturbative uncertainties
  present at NLO level.
\item
  M. Gorbahn, F. Moretti and S. J{\"a}ger: Beyond Leading Logarithms in $g_V$:
  The Semileptonic Weak Hamiltonian at $\ord(\alpha\alpha_s^2)$ \cite{Gorbahn:2025ssv}. 

  The authors presented the first NLO QCD analysis of the electromagnetic corrections
  to the semileptonic weak Hamiltonian, including the mixed
  $\ord(\alpha\alpha_s^2)$ corrections to the vector coupling $g_V$ . The analysis combines the evaluation of three-loop anomalous dimensions and two-loop
matching corrections with a consistent factorization of short-distance QCD effects. The resulting renormalization-group–improved expression for the
radiative correction can be systematically refined using input from lattice QCD
and perturbation theory and improves the consistency of first-row CKM
unitarity tests.
\item
  U. Nierste,  P. Reeck,  V. Shtabovenko and M. Steinhauser:
Complete NNLO QCD corrections to the decay matrix in B-meson mixing at leading power \cite{Nierste:2025muk}.

  The authors computed NNLO corrections to the decay width
difference of mass eigenstates and the charge-parity (CP) asymmetry $a_{fs}$ in flavour-
specific decays of neutral B mesons. They included both current-current and penguin
operators at three-loop order. All input integrals in the transition amplitude are reduced to a small set of master integrals which depend on the ratio of the charm and
bottom quark masses. The latter are computed using semi-analytic methods which
provide expansions around properly selected values of $m_c/m_b$. They provided numerical results for $\Delta\Gamma$  and $\Delta\Gamma/\Delta M$ , both for the $B_d$ and $B_s$ system. Many useful results on these quantities can be
found in this paper.
\item
M. Lang, A. Lenz, A. Mohamed, M. L. Piscopo and A. V. Rusov:
B-meson decay width up to $1/m^3$  corrections
within and beyond the Standard Model \cite{Lang:2025ios}.

Starting from the most general effective Hamiltonian describing non-leptonic
b-quark decays $b\to q_1\bar q_2q_3$, the authors computed analytic expressions for all matching coefficients of the two-quark
operator contributions in the heavy quark expansion (HQE) of a B meson, up to mass-dimension-six.
In addition, they calculated the weak-annihilation contributions, which enter the matching of four-quark
operators in the HQE at dimension-six and were previously missing. Their results complete the calculation of beyond Standard Model effects in non-leptonic, tree-level, b-quark decays relevant for B
meson lifetimes and lifetime ratios such as $\tau(B^0_s)/\tau(B_d^0)$. Such effects naturally arise in extensions of
the SM that aim to address the observed tensions between experimental
measurements and theoretical predictions based on QCD factorisation in several colour-allowed non-leptonic B-meson decays. As a by-product of their calculation, they  also determined the matching coefficients
in the HQE induced by the QCD-penguin operators within the SM, including both the interference
between current-current and penguin operators and the contributions quadratic in the penguin operators.
 Owing to the suppression of the QCD-penguin Wilson coefficients within the SM, these effects
are typically regarded as corrections of order $\alpha_s$ and $\alpha_s^2$,
respectively. Their  results reproduce the known expressions at dimension-three and provide new results for the coefficients of the chromomagnetic operator at dimension-five and of the Darwin operator at dimension-six.
\item
  In 2016 the first message on NNLO QCD contribution of QCD-penguins to $\epe$ has been
  presented in \cite{Cerda-Sevilla:2016yzo} and followed by two contributions
  in 2016 and 2018 \cite{Cerda-Sevilla:2016tpw,Cerda-Sevilla:2018hjk}. These
  papers written in the group of Martin Gorbahn indicated significant
  suppression   of $\epe$ by these contributions. Unfortunately, the final paper did not appear until today.
\item
  Similarly,  in 2019 Maria Cerd{\'a}-Sevilla reported on the status of
  NNLO QCD corrections to $X(x_t)$ obtained  in the group of Martin Gorbahn
\cite{Cerda-Sevilla:2019use}. Next at Kaon 2023  Martin Gorbahn himself presented preliminary results on these NNLO QCD corrections relevant for $K\to\pi\nu\bar\nu$ decays  and
  Emmanuel Stamou in his talk at Kaon 2025 promised, reporting on the final result of their paper, its   submission 
  to hep-arXiv  in September 2025. Unfortunately, this did not happen until today. Let us hope that the final version of this paper will appear in hep-arXiv in 2026.
\end{itemize}

I am extremely happy about all these calculations and also about the ones I was involved in. In October 1988 when I got the idea for this grand project some of my colleagues told me 
that I was crazy starting it because of various non-perturbative uncertainties.
But 37 later I think it was probably my best strategic idea I had during my career as a
particle physicist. Yet, I want to emphazise that I was very lucky to find
strong collaborators and convince younger generation to continue this
project once I got older. Indeed without the full army
of excellent multi-loop calculators this project could not be realized.
Therefore it is justified to devote one section to a table that lists
the NLO and NNLO Summits Conquered by the Members of the MNLC. No additional
words are required. The description of these expeditions are given above and of
course in original papers.


\section{Summary of NLO and NNLO Summits Conquered by the Members of the MNLC.}\label{TUMsumitts}
   \begin{table}[thb]
     \setlength{\tabcolsep}{2pt}
\centering
  \begin{adjustbox}{width=0.9\textwidth}
\begin{tabular}{|l|l|l|}
\hline
\bf \phantom{XXXX} Decay &  {\bf NLO} & {\bf NNLO}  \\
\hline
\hline
Current-Current $(Q_1,Q_2)$      &Buras, Weisz& Gorbahn, Haisch\\
QCD P $(Q_3,Q_4,Q_5,Q_6)$  &Buras, Jamin, Lautenbacher, Weisz & Gorbahn, Haisch\\
     & Fleischer & \\
EW P $(Q_7,Q_8,Q_9,Q_{10})$  & Buras, Jamin, Lautenbacher & 
Buras, Gambino, Haisch\\
$\mathcal{B}(B)_{SL}$   &Ball, Lenz, Nierste, Ostermaier & \\
inclusive non-leptonic decays       & Ball, Buchalla, Jamin   & \\
B-Decays in QCDF   & 
Bartsch, Buchalla &
 Bell, Pilipp \\
Current-Current (BSM) & Buras, Misiak, Urban    & \\
Penguins (BSM)  & Buras, Misiak, Urban   & \\
Semi-Leptonic ($\vcb$, $\vub$) & Ball, Czarnecki, Gambino & 
Czarnecki, Gambino \\
Semi-leptoni $\ord(\alpha\alpha_s^2)$ & Gorban, J{\"a}ger &\\
\hline
$\eta_1$                   & Herrlich, Nierste& Brod, Gorbahn, Stamou \\
$\eta_2,~\eta_B$           & Buras, Jamin, Weisz; Urban & Brod, Gorbahn, Stamou\\
$\eta_3$                   & Herrlich, Nierste& 
Brod, Gorbahn, Stamou\\
Tree-Level FCNC & Buras, Girrbach-Noe &\\
$\Delta\Gamma_{B_s}$, $\Delta\Gamma_{B_d}$       & Buchalla, Lenz, Nierste & Lenz, Nierste\\
\hline
$K^0_L \rightarrow \pi^0\nu\bar{\nu}$, $B_{s,d}\rightarrow l^+l^-,~
X_{\rm s}\nu\bar{\nu}$  & Buchalla, Buras ; Misiak, Urban &  Bobeth, Gorbahn, Misiak \\ 
$K^+   \rightarrow \pi^+\nu\bar{\nu}$, $K_{\rm L} \rightarrow \mu^+\mu^-$
                                      & Buchalla, Buras & Buras, Gorbahn, Haisch\\
&& Nierste, Stamou \\
$K^+\to\pi^+\mu\bar\mu$               & Buchalla, Buras & \\
EW to Charm in $K^+   \rightarrow \pi^+\nu\bar{\nu}$ & Brod, Gorbahn  &    \\
EW to Top in $K\to\pi\nu\bar\nu$ & Buchalla, Buras; Brod, Gorbahn  &  \\
and $B_{s,d}\to\mu^+\mu^-$ & Stamou, Bobeth &\\
\hline
$B\to X_s\gamma$ &  &\\
\hline
$C_i(M_W)$ & Buras, Kwiatkowski, Pott; Gambino  & Bobeth, Misiak, Urban 
\\
$\hat\gamma(Q_{7\gamma},Q_{8G})$  & Misiak, M\"unz; & Gorbahn, Haisch, Misiak\\
$\hat\gamma(\rm{Mixing})$     &
  Chetyrkin, Misiak, M\"unz & Haisch, Misiak
\\
Matrix Elements & 
     Buras, Czarnecki, Misiak, Pott, Urban  & 
Ewerth; Haisch; Misiak\\
$B\to K^*(\rho)\gamma$ & Bosch, Buchalla & \\
\hline
$K_{\rm L} \rightarrow \pi^0e^+e^-$         & Buras, Lautenbacher, Misiak, 
M\"unz & \\
$B\rightarrow X_s \ell^+\ell^-$:
$C_i(M_W)$   & Misiak; Buras, M\"unz & Bobeth, Misiak, Urban\\
$\hat\gamma({\rm Mixing})$ & Buras, Lautenbacher, Misiak, M\"unz & 
Gambino, Gorbahn, Haisch  \\
Matrix Elements & Misiak; Buras, M\"unz &  
Bobeth, Gambino, Haisch \\
& & Bell, Pilipp \\
Electric Dipole Moment & Brod, Stamou & \\
\hline
\end{tabular}
\end{adjustbox}
  \caption{Summary of NLO and NNLO summits conquered by the members of MNLC.}
  \label{TAB11}
\end{table}

\boldmath
\part{The CKM Matrix and The Unitarity Triangle}\label{Rare}
\unboldmath
New ideas in the context
of the UT and the parametrization of the CKM matrix as well as
strategies for the determination of the CKM parameters will be discussed next.

  \boldmath
  \section{Improved Wolfenstein Parametrization: $(\bar\varrho,\bar\eta)$ (1994)}
  \unboldmath
  The top quark has been  discovered in 1995, although some indication for this
  discovery came already in 1994.  Anticipating this discovery and some
  progress on the determination of the CKM parameters, of $\hat B_K$ and $F_B\sqrt{\hat B_B}$ in the following five years, I made an excursion into the future in order to find a possible picture of the unitarity triangle (UT), of quark mixing and of CP violation around the year 2000. I was accompanied by Markus Lautenbacher and my new PhD student  Gaby Ostermaier. In this context we have generalized the
  Wolfenstein parametrization \cite{Wolfenstein:1983yz}  of the CKM matrix beyond the leading order in $\lambda$ and replaced his parameters $(\varrho,\eta)$ used by everybody in the
  literature until then by $(\bar\varrho,\bar\eta)$ \cite{Buras:1994ec}
  \be\label{2.88d}
\bar\varrho=\varrho\left(1-\frac{\lambda^2}{2}\right)   \qquad \bar\eta= \eta\left(1-\frac{\lambda^2}{2}\right),
\ee
where $\lambda=\vus$. These parameters became the standard ones   since then. We have also extended Wolfenstein's parametrization of the  full CKM matrix by including $\lambda^4$ corrections.

  While this result had an important impact on the future phenomenology of FCNCs and of the UT, reading this paper recently again, I realized that
  many useful expressions on the UT used in our subsequent papers until now were
  in fact introduced in this paper. These expressions are collected in Section 2.5.6 of my book \cite{Buras:2020xsm}.

While these expressions are useful for the derivation of analytic formulae 
for various observables, in numerical calculations these days it is better 
to use exact standard parametrization of the CKM matrix. To 
this end, as suggested to my knowledge by CKMfitters \cite{Charles:2004jd},  $\bar\varrho$ and $\bar\eta$ can be defined in a phase convention-independent manner through
\begin{equation}\label{CKM2.88d}
\boxed{\bar\varrho+ i\bar\eta =-\frac{V_{ud}V^*_{ub}}{V_{cd}V^*_{cb}}\,.}
\end{equation}

One can check that  this definition agrees to an excellent accuracy 
with the original one in (\ref{2.88d}).

  Our 1994 numerical analysis of $\kpn$, of
  $B^0_s-\bar B^0_s$ mixing and of CP asymmetries in B-decays is clearly out of
  date but it demonstrates the impressive progress made in this field in the
  last 30 years.

 \begin{figure}
\centering
\includegraphics[width = 0.55\textwidth]{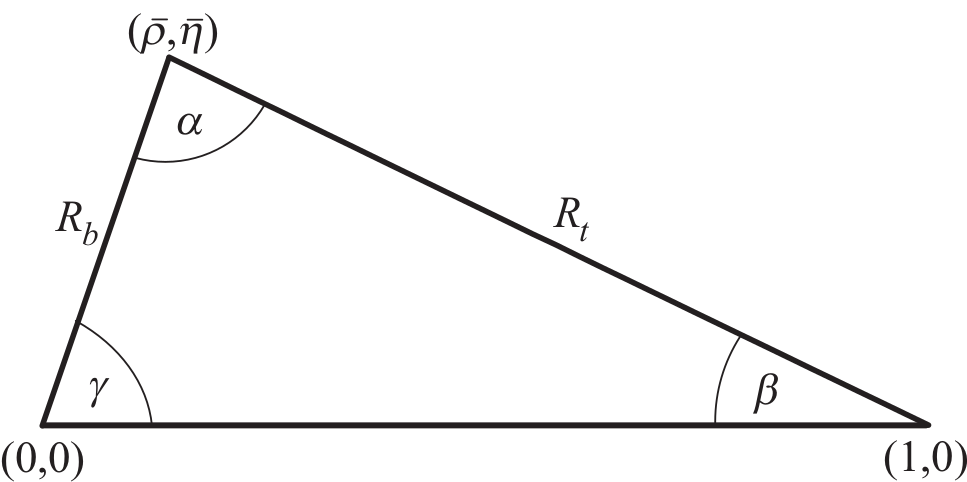}
 \caption{\it The Unitarity Triangle. }\label{UUTa}
\end{figure}

\section{Another Look at CKM and UT (2002, 2022, 2023)}\label{BPS}
During the first CKM Workshop in 2001 at CERN two experimentalists,
Fabrizio Parodi and Achille Stocchi, and I got the following idea
that resulted in the paper \cite{Buras:2002yj}.

As the unitarity triangle (see Fig.~\ref{UUTa}) can be determined by means of two measurements 
of its sides or angles, the question arose which two measurements would
be most efficient to determine UT or equivalently $(\bar\varrho,\bar\eta)$.
 Assuming then the same relative errors on the
angles $(\alpha,\beta,\gamma)$ and the sides $(R_b,R_t)$, we found that the
pairs $(\beta,\gamma)$ and $(\gamma,R_b)$ are most efficient in this
respect. 
They are followed by  
$(\alpha,\beta)$, $(\alpha,R_b)$, $(R_t,\beta)$, $(R_t,R_b)$ and 
$(R_b,\beta)$.
As the set $\vus$, $\vcb$, $R_t$ and $\beta$ appeared to be the best 
candidate for the fundamental set of flavour violating parameters in
the following years, we showed various constraints on the CKM matrix in the
$(R_t,\beta)$ plane.

20 years later the situation with the determination of $\gamma$ improved
by much while $R_b$ has problems with the unique determination of $\vub$.
Therefore the $(\beta,\gamma)$ strategy is a clear winner and in my view
the optimal set of CKM variables is the following one
\be
\vus,\qquad \vcb, \qquad \beta, \qquad \gamma\,.
\ee

In this context let me present two simple formulae that are central in the
$(\beta,\gamma)$ strategy as they allow to calculate the appex of the UT in no
time, but 
to my knowledge they have been presented  only recently for the first time
\cite{Buras:2023ric,Buras:2023qaf}.  
They read
\be\label{AJB23}
\boxed{\bar\varrho=\frac{\sin\beta\cos\gamma}{\sin(\beta+\gamma)},\qquad
  \bar\eta=\frac{\sin\beta\sin\gamma}{\sin(\beta+\gamma)}.}
\ee
Evidently they can be derived by high-school students, but the UT is unknown to them and somehow no flavour physicist got the idea to present them in print so far. Numerical tables for $\bar\varrho$ and $\bar\eta$ for different values of
$\beta$ and $\gamma$ are given in
\cite{Buras:2023ric,Buras:2023qaf}.

\begin{figure}[t!]
  \centering%
  \includegraphics[width=0.56\textwidth]{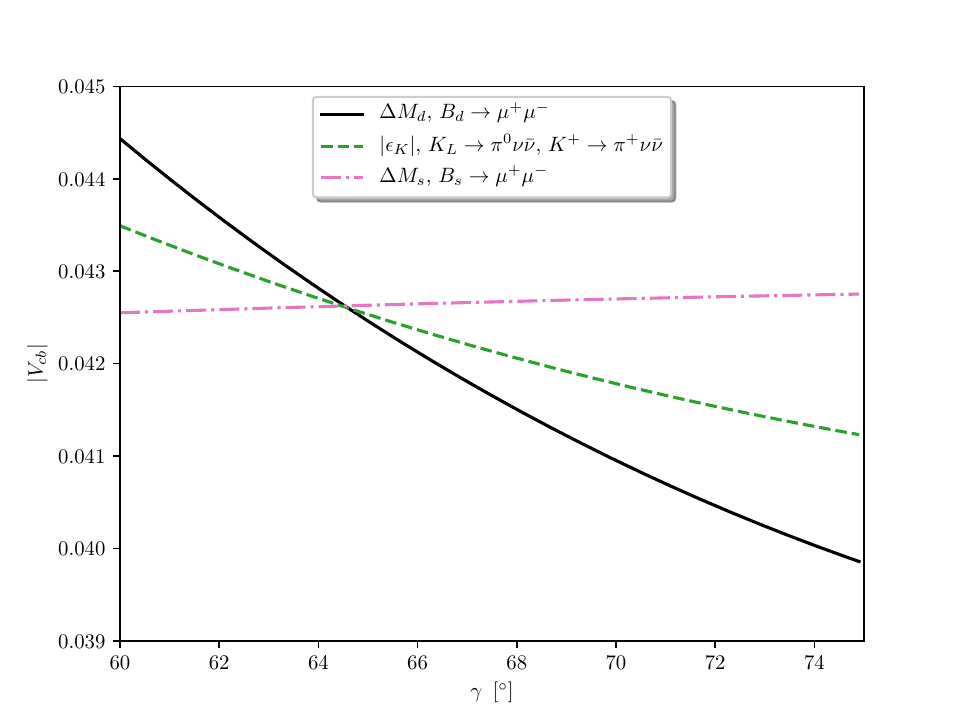}
  \includegraphics[width=0.42\textwidth]{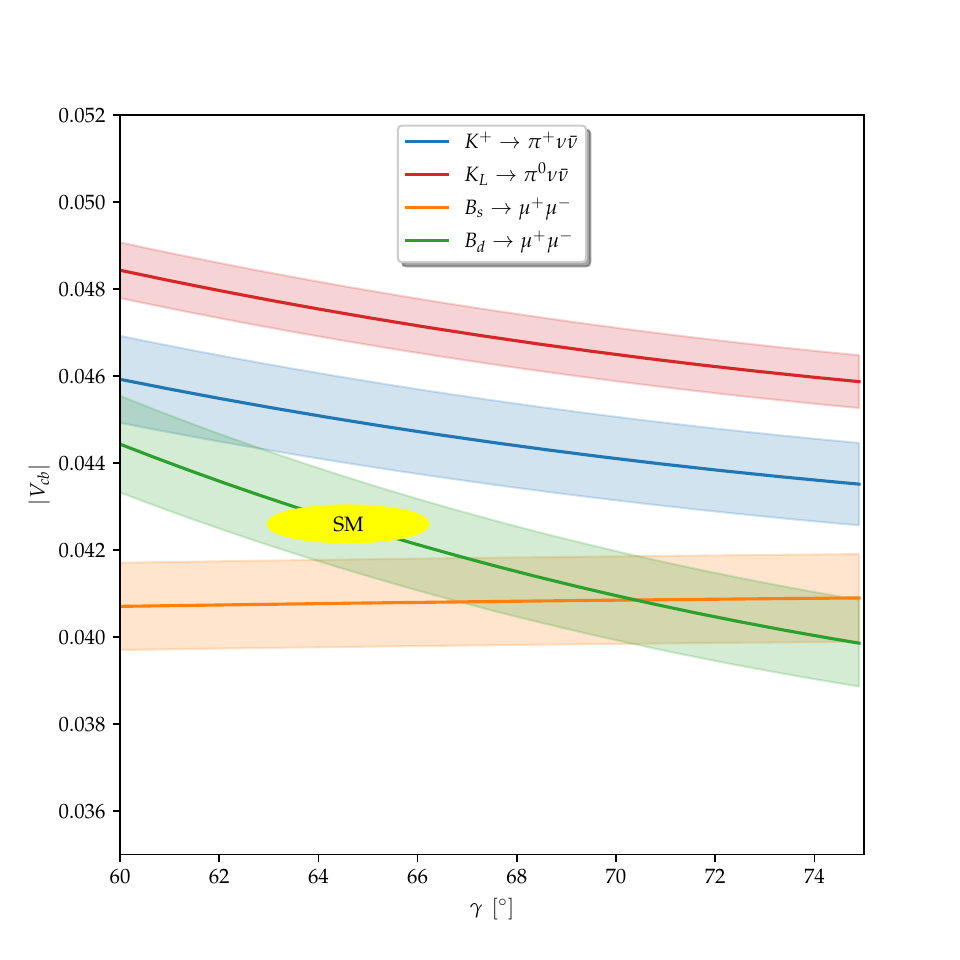}
  \caption{\it {Left:Schematic illustration of the action of the seven observables in the $\vcb-\gamma$ plane in the context of the SM. We set $\beta=22.2^\circ$ and all uncertainties to zero. Right:The impact of hypothetical future measurements of the branching ratios for $\kpn$, $\klpn$,  $B_d\to\mu^+\mu^-$ and $B_s\to\mu^+\mu^-$  on the $\vcb-\gamma$ plane. All uncertainties  are included. The yellow disc  represents the SM.                  From \cite{Buras:2022nfn}.}
\label{fig:XX}}
\end{figure}

Having $\vcb$ and $\gamma$ among the optimal quartet and being presently
not as well determined as $\vus$ and $\beta$ 
the $\vcb-\gamma$ plots for fixed $\beta$ have been proposed in  \cite{Buras:2022wpw,Buras:2021nns}. They are, as emphasized in \cite{Buras:2022nfn}, useful companions to   common unitarity triangle fits because they exhibit better possible inconsistences between $\vcb$ and $(\beta,\gamma)$ determinations than the latter fits.
Indeed,
\begin{itemize}
\item
  They exhibit $\vcb$ and its correlation with $\gamma$ determined through a given observable in the SM, allowing thereby monitoring the progress 
  on both parameters expected in the coming years. Violation of this correlation in experiment will clearly indicate NP at work.
\item
  They utilize the strong sensitivity of rare $K$ decay  processes to $\vcb$ thereby providing
  precise determination of $\vcb$
  even with modest experimental precision on their branching ratios.
\item
  They exhibit, as shown in Fig.~\ref{fig:XX}, the action of $\Delta M_{s}$,
  and of $B_s$ decays,
like $B_s\to\mu^+\mu^-$
which is not possible in the common UT-plot.
\item
  Once the accuracy of $\gamma$ measurements  will approach $1^\circ$ it will be easier to monitor this progress on the $\vcb-\gamma$ plot than on the UT-plot.
\item
  It should be noticed that the lines corresponding to Kaon observables in the plots in Fig.~\ref{fig:XX} depend on $\beta$ so that $\beta$ enters these plots
  indirectly.
\end{itemize}

This is the right place to mention the recent idea of Avital Dery \cite{Dery:2025pcx}, who proposed a new complex plane for Kaon CKM data, that stresses
the determination of four parameters of the CKM from Kaon data alone. Instead
of $(\bar\varrho,\bar\eta)$ plane dominated presently by B-observables, 
Avital proposes to consider the $(x,y)$ plane where
\be
x=A^2(1-\hat\varrho), \qquad y= A^2\hat\eta
\ee
with $\hat\varrho$ and $\hat\eta$ defined through
\be
A^2\lambda^4(1-\hat\varrho+i\hat\eta)\equiv-R_{ct}e^{-i\theta_{ct}}.
\ee
Here the Wolfenstein parameter $A$ is related to $\vcb$ through $\vcb=A\lambda^2$
and
\be
R_{ct}=\left|\frac{V_{td}V^*_{ts}}{V_{cd}V^*_{cs}}\right|,\qquad
\theta_{ct}=\text{arg}\left(-\frac{V_{td}V^*_{ts}}{V_{cd}V^*_{cs}}\right).
\ee

While I find this idea nice and useful, in my view the strength of the CKM
framework is to relate flavour violating effects in all meson systems and
to check whether they are consistent with each other.

In this sense among the three proposals: standard UT, the $\vcb-\gamma$ plane and the one just discussed, I claim that the $\vcb-\gamma$ plane proposed by
Elena and myself \cite{Buras:2022wpw,Buras:2021nns} is most useful and transparent
in presenting the flavour data in the coming years. But of course I am biased here. Additional
$\vcb-\gamma$ plots will appear in Part~\ref{BV}.

As Avital just entered the scene, I should emphasize that she is not only a very good
physicist and speaker but also a very good singer. Let me just advertise a video
in YouTube where she sings in {\em Vivaldi's Stabat Mater}. I suggest
that at some future conferences she should be invited not only to give a talk
but also to sing at the conference dinner.

  \section{Futuristic Proposals for the Determination of the CKM  Matrix (1994)}
  Let me jump back to 1994 in which the first ideas to determine some of the elements of the CKM
  matrix from rare Kaon decays have been put forward.
  After the calculation of NLO QCD corrections to theoretically clean decays
  $\kpn$ and $\klpn$ \cite{Buchalla:1993bv,Buchalla:1993wq}
  I have written in May 1994 two papers pointing out that these two
  decays, being theoretically very clean, could be used to determine  some elements of the CKM matrix, in particular $\vcb$ and the angle  $\beta$ of the UT.
  \boldmath
  \subsection{$\sin 2\beta$ from $K\to\pi\nu\bar\nu$}
  \unboldmath
  In the first paper, written with Gerhard Buchalla \cite{Buchalla:1994tr}, we have pointed out that the measurement of just the branching ratios for $\kpn$ and $\klpn$ decays could determine $\sin 2\beta$ almost independently of $m_t$ and $\vcb$ that
  were both poorly known at that time. In 2026, 32 years later, $m_t$ is very
  precisely known while the determination of $\vcb$ is still subject to significant uncertainties due to differences in inclusive and exclusive determinations. While $\mathcal{B}(\klpn)$ has not been measured yet, $\sin 2\beta$ has
  been measured with high precision. Once the final result for  $\mathcal{B}(\kpn)$ from the NA62 collaboration will be known one will be able to see what it implies for   $\mathcal{B}(\klpn)$.
  Defining 
\be\label{b1b2}
B_1=\frac{B(\kpn)}{4.64\cdot 10^{-11}}\quad
B_2=\frac{B(\klpn)}{1.94\cdot 10^{-10}},\quad \sigma=\frac{1}{(1-\lambda^2/2)^2}   \ee
we found \cite{Buchalla:1994tr}.
\be\label{cbb}
\cot\beta=
\sqrt{\sigma}\frac{\sqrt{\sigma(B_1-B_2)}-P_0(K^+)}{\sqrt{B_2}}, \ee
with $P_0(K^+)$ parametrizing charm contribution to $\kpn$.

\begin{figure}[t]
\centering%
\includegraphics[width=0.85\textwidth]{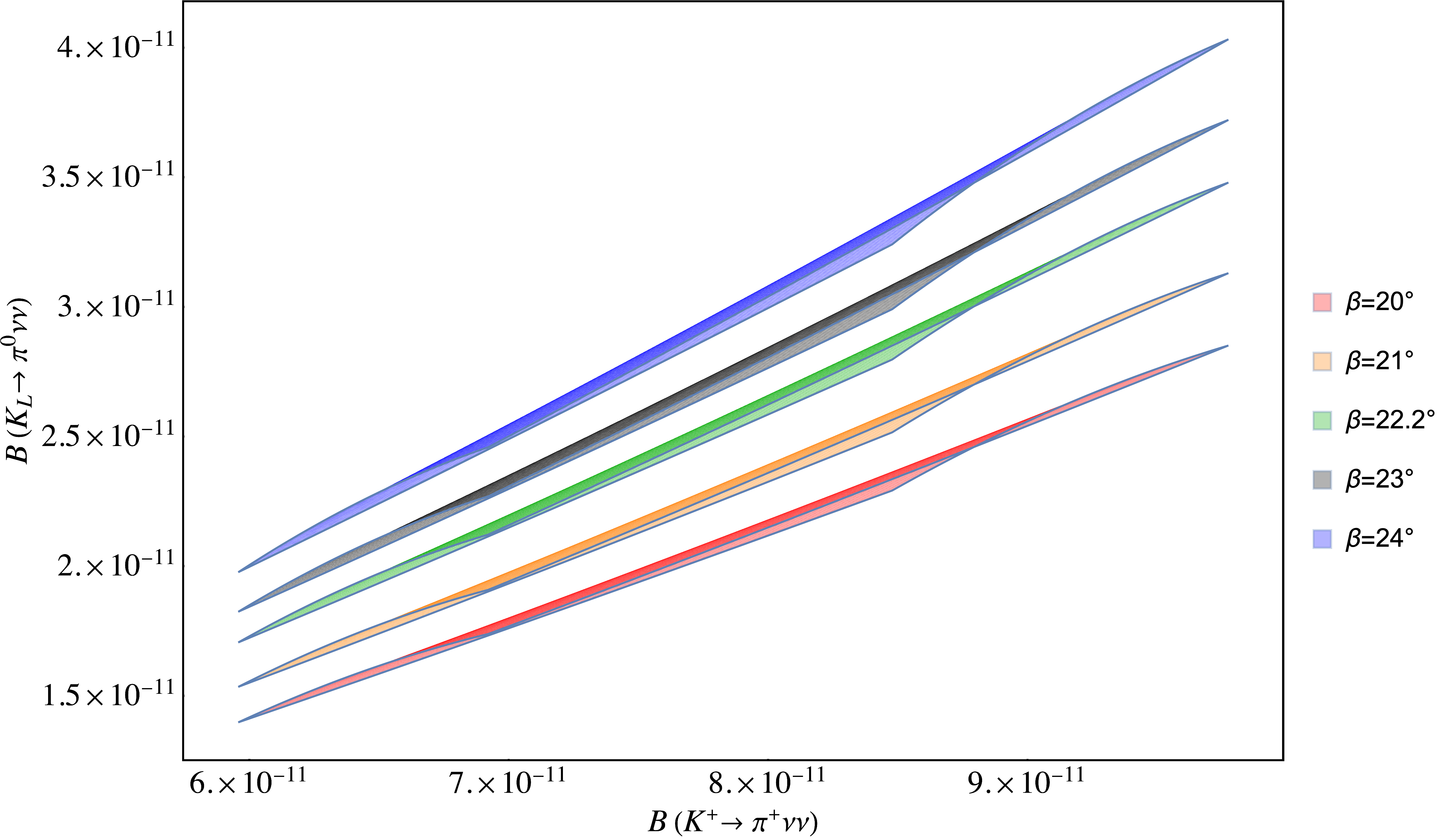}%
\caption{\it The correlation between branching ratios for $\klpn$ and $\kpn$
  as given in (\ref{cbb}) for different  values of $\beta$ within the SM.
  The ranges of branching ratios correspond to $38 \leq |V_{cb}|\times 10^{3} \leq 43$ {and $60^\circ\leq\gamma \leq 75^\circ$}. From \cite{Buras:2021nns}.\label{fig:3}}
\end{figure}

However, it should be emphasized that although this formula applies not
only to the SM but also to models with constrained MFV (see Section~\ref{CMFV}) it does not apply
to models with new sources of flavour and CP violation beyond the CKM matrix
and in which new operators contribute to $\kpn$ and $\klpn$.
In Fig.~\ref{fig:3} we illustrate this correlation between $\kpn$, $\klpn$
and $\beta$ that has been updated in 2021 in a paper with Elena Venturini \cite{Buras:2021nns}.

\boldmath
 \subsection{ 
Precise Determinations of the CKM Matrix from CP Asymmetries  in
B Decays and $\klpn$}
 \unboldmath
  The second  paper, this time  by myself alone \cite{Buras:1994rj}, was even more futuristic.
  I have pointed out that the measurements of CP asymmetries in neutral B decays together with the measurement of $\mathcal{B}(\klpn)$ and the known value of
  $\vus$ can determine {\em all} elements of the CKM matrix without essentially any hadronic uncertainties. In particular the strong dependence of $\mathcal{B}(\klpn)$ on $\vcb$ ($\vcb^4$) can be used to determine this element precisely.
  Consequently $\vub$, $\vtd$ and $\vts$ can also be determined. In this context
  I derived a number of formulae which would allow to realize this strategy.

  I think this was a good idea provided $\mathcal{B}(\klpn)$ could be measured.
  However,  I received an email from a Brookhaven experimentalist who
  essentially
  wrote that I must be crazy to propose such an idea. As after 32 years this branching ratio has not been measured he was partly right. But one day it will be measured. Even if one should hope that the measured $\mathcal{B}(\klpn)$ will
  be affected by NP contributions, the formulae presented in this paper can turn
  out to be useful for various tests of the SM.

  These ideas have been presented
  in the proceedings of the First International Conference on Phenomenology of Unification from Present to Future that took place in Rome \cite{Buras:1994sd}
  and also at the conference in Cracow dedicated to the 60th Birthday of Kacper Zalewski \cite{Buras:1995pp}. It took place in January 1995. In fact it was my
  first visit I made to Poland after my emigration to Denmark in May 1971.

\part{Strategies Beyond the Standard Model}\label{SBSM}
    
  \section{Most Important Strategies}
  \subsection{Preface}
  In 1983 the  $W^\pm$ bosons have been discovered at CERN. The world celebrated this discovery. Yet, at a colloquium in MPI for Physics Victor
  Weisskopf (1908 -2002) stated
  
\begin{figure}
\centering
\includegraphics[width = 0.75\textwidth]{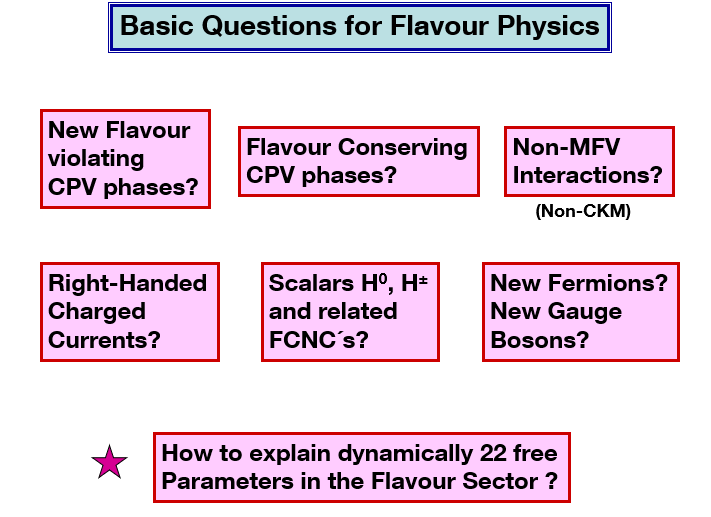}
 \caption{\it Basic Questions for Flavour Physics. }\label{BasicQ}
\end{figure}

\vspace{0.1cm}
 {\bf If a theory agrees with data there is no progress!}
 \vspace{0.1cm}

 \noindent
   which surprised the audience and some said that he is getting really old. He
  was 75 at that time. By the theory he meant  the SM and he was
  of course right. Let me just list the main reasons why the SM cannot be the
  whole story and NP must exist. Moreover there are many questions with possible answers requiring the presence of NP.  Here we go:
  \begin{itemize}
  \item
    Why is our universe dominated by matter? This is related to the violation of CP-symmetry soon after the BIG BANG and is crucial for our existence. The size
    of CP violation in the SM is by far insufficient to explain this.
  \item
    Why  the Higgs boson remains light? Is it a fundamental particle or a composite particle arising from some new strong dynamics responsible for spontaneous
    breakdown of electroweak symmetry of the SM? My excellent PhD student
    and collaborator, Gerhard Buchalla contributed in an important manner to this field. One
    day he will summarize his activities in this field  in his autobiography.
  \item
    Is the form of Higgs potential affected by NP? Another autobiography. This time by Gino Isidori, another great collaborator of mine.
  \item
    Strong CP problem and related solutions involving axions.
  \item
    What is the origin of neutrino masses that are vanishing in the SM?
  \item
    Are neutrinos Majorana fermions, i.e. their own antiparticles? This
    would indicate that lepton number is not a fundamental symmetry of nature.
    Neutrinoless double-beta decay would establish this property.
  \item
    Is there CP violation in the lepton sector? If established, it would strengthen the case of leptogenesis related to the first point on our list.
  \item
    Why there are three generations of quarks and leptons?
  \item
    What is the origin of vast differences between particle masses with
    the top quark mass larger by five orders of magnitude than the electron
    mass and eleven orders of magnitude larger than neutrino masses.
  \item
    What is the origin of the pattern of their interactions summarized by
    CKM and PMNS matrices?
  \item
    What is Dark Matter which amounts to $27\%$ of the matter of the Universe
    and similar question is related to Dark Energy.
  \end{itemize}
  In view of these questions it is obvious that new forces and new particles
  beyond those shown in Fig.~\ref{SM} must exist and our duty is to find them
  somehow. As far as flavour physics is concerned the basic questions are listed in Fig.~\ref{BasicQ}.

  In this rather long section I want to describe a number of strategies and analyses for searching for NP 
  which in my view belong to the best among those I alone 
  and very often with my collaborators   suggested until now.
  Some deal only with the SM with the goal to obtain very precise
  predictions in this model with the hope to find anomalies in an efficient
  manner. Other, illustrate how one could test  different 
  simplified models or more involved models and distinguish between them on the basis of rare processes. Those which we analyzed
  in detail separately will be presented in Part~\ref{NPModels}.

  Let me stress first that in the indirect search for NP with the help of
  weak decays, particle-antiparticle mixing observables, EDMs and other observables,  the first very important step is to obtain SM predictions as accurate as possible. Simply because  the hints for NP will first come from deviations from SM expectations.
  As the masses of the involved SM quarks and leptons, gauge bosons and of the Higgs
  are by now very well known and this also applies to gauge couplings, the
  main challenge in obtaining precise SM predictions for the processes in
  question are:
  \begin{itemize}
  \item
    Calculation of short distance QCD, QED and electroweak corrections.
  \item
    Calculation of non-perturbative QCD effects.
  \item
    Determination of four CKM parameters.
  \item
    Precise experimental measurements of the observables involved.
    \end{itemize}

  In my view the first challenge on this list has been overcome to a large extent as
  described in Part~\ref{NLOQCD}. Impressive progress has also been made
  by LQCD in the evaluation of non-perturbative QCD effects with regular
  reports from FLAG \cite{FlavourLatticeAveragingGroupFLAG:2021npn}. This is in particular the case of particle-antiparticle mixing, weak decay constants,
  formfactors etc. Similar ChPT made important contributions here
\cite{Cirigliano:2011ny}.
In this
  manner leptonic  meson decays contain
  by now only small non-perturbative uncertainties. This is also the case
  of semileptonic FCNCs with neutrinos in the final state, where the main
  challenge are formfactors. Significant progress on them has been made
  in the last ten years \cite{Parrott:2022rgu,Parrott:2022zte,Parrott:2022smq,Khodjamirian:2023wol}.

  The case of semileptonic decays with charged
  leptons in the final state is another story because of non-factorizable
  long distance contributions that require detail studies. Presently
  they do not allow unique agreement between various groups on whether
  the anomalies observed in $b\to s \ell^+\ell^-$ transitions result
  from NP or non-perturbative effects. There is a very reach literature
  on this subject and I list here only few papers in which further references
  can be found \cite{Jager:2012uw,Jager:2014rwa,Gubernari:2020eft,Ciuchini:2022wbq,Gubernari:2022hxn,Gubernari:2023puw,Isidori:2024lng,Frezzotti:2025hif}.
  It should be noticed that my PhD student Sebastian J{\"a}ger and my two
  collaborators Gino Isidori and Luca Silvestrini were contributing in
  an important manner to these discussions.

  Unfortunately the situation is even worse in the case of
  non-leptonc meson decays where the non-perturbative uncertainties
  remain large. This was already clear from Section~\ref{DQCD} and will
  be evident as we proceed.

  Another issue are CKM parameters that are free parameters in the SM and
  have to be determined from experiment, preferably from tree-level decays
  that are believed to be less affected by NP than loop induced processes like
  FCNC processes within the SM. Unfortunately there are inconsistencies between
  different  determinations of some CKM parameters, in particular $\vcb$. This
  introduces significant uncertainties in decays like $\kpn$ and $\klpn$ that
  are sensitive functions of $\vcb$.

  Finally, crucial are the measurements of various observables, like branching
  ratios, CP asymmetries, mass differences $\Delta M_{s,d}$ etc. Some are
  already measured with respectable precision, other should be measured in the coming years
  but for several we will have to wait still for a decade or more.

  I have described already my activities in some of these  challenges in Part~\ref{NLOQCD} and Section~\ref{DQCD}. In the latter case, our approach of DQCD applies only to Kaons. For Non-leptonic Charm and B mesons  decays  one has to use other methods like   LQCD, Heavy Quark Expansions and QCD-Factorization that are beyond
  the scope of this writing as I did not make any contributions there.
  I will only indicate where still an important progress has to be made.

  In the following I  will discuss 
  strategies related to particular processes that have been developed in
  my group postponing the strategies for the elimination of the $\vcb$
  dependence to Part~\ref{BV}.  The listing of NP models
  investigated in my group and specific strategies developed there will be discussed in several subsequent sections. Finally at the end of this part I will
  present a Dual Picture of short distance scales which hopefully is not only
  amusing but also useful.

  \boldmath
 \subsection{Upper Bounds on $K \to \pi \nu \bar \nu$ and
   $K_L\to\pi^0e^+e^-$ from $\epe$ and $K_L \to \mu^+ \mu^-$ (1998)}
 \unboldmath
 This was one of the first among ten papers I have written with Luca Silvestrini
 \cite{Buras:1998ed}. Luca was a postdoc in my group and it was real fun to
 collaborate with him. 

 We analyzed rare kaon decays in models in which the dominant new effect
is an enhanced $\bar s d Z$ vertex $Z_{ds}$. We pointed out that in
spite of large theoretical uncertainties the CP-violating ratio $\epe$
provided in 1998 the strongest constraint on $\IM Z_{ds}$. Assuming
$0 \le \epe \le 2 \cdot 10^{-3}$ and SM values for the CKM
parameters at that time, we obtained the bounds
$\mathcal{B}(\klpn)\le 2.4 \cdot10^{-10}$ and $\mathcal{B}(K_L\to\pi^0e^+e^-) \le 3.6 \cdot 10^{-11}$ which
turned out to be 
substantially stronger than the bounds found previously  by Colangelo and
Isidori \cite{Colangelo:1998pm}, using $\varepsilon_K$ instead of $\epe$. We illustrated how these bounds can be improved with the help of the forthcoming data on $\epe$. Using the bound on $\RE Z_{ds}$ from $K_L \to \mu^+ \mu^-$ we found
$\mathcal{B}(\kpn) \le 2.3 \cdot 10^{-10}$ which is satisfied in 2026.

In this context we derived an
analytic upper bound on $\mathcal{B}(\kpn)$ as a function of
$\mathcal{B}(\klpn)$ and the short distance contribution to
$\mathcal{B}(K_L \to \mu^+ \mu^-)$.
We also discussed NP scenarios in which in addition to an
enhanced $\bar s d Z$ vertex also neutral meson mixing receives
important new contributions. In this case larger values of the
branching ratios in question cannot be excluded.

\boldmath
\subsection{Waiting for Precise Measurements of $\kpn$ and $\klpn$ (2008)}
\unboldmath
After the first measurements of the $\kpn$ branching ratio at Brookhaven in
1997-2004 by E947 collaboration \cite{E949:2008btt}
and no signs of progress to measure it better, I thought it would be good
to review the importance of this decay and of $\klpn$ which I investigated
in detail with Gerhard Buchalla in the 1990s as already described before.
This time two very good  PhD students, Felix Schwab and Selma Uhlig, joined me.

The first version of our review appeared in arXiv already in May 2004 but in view of our papers on minimal flavour violation (MFV) and the LHT model  that we wanted to discuss in this review its
final version appeared in Reviews of Modern Physics  \cite{Buras:2004uu} only
in 2008. In this first long review on these decays in my group
we collected the relevant formulae for quantities of
interest and analyzed their theoretical and parametric uncertainties. 
We pointed out that in addition to the angle $\beta$ 
in the UT \cite{Buchalla:1994tr}, also the angle $\gamma$ can in principle
be determined from these decays with respectable 
precision and emphasized in this context the importance of the recent NNLO QCD calculation of 
the charm contribution to $\kpn$ \cite{Buras:2005gr,Buras:2006gb} and of the improved estimate of the long distance contribution by
means of chiral perturbation theory \cite{Isidori:2005xm}.
In addition to known expressions we presented several new ones that should
allow 
transparent tests of the SM and of its extensions.

While our 
presentation was centered around the SM, we also discussed  models with
 MFV and scenarios with new complex phases in 
decay amplitudes and meson mixing. 
We gave a brief review of existing results within
specific extensions of the SM, in particular the LHT Model, $Z^\prime$ models, the MSSM and a model with one universal extra
dimension. The description of the analyses of these models in my group
will be given in Part~\ref{NPModels}  of this write-up.

Moreover, we investigated the virtues of 
$(R_t,\beta)$, $(R_b,\gamma)$, $(\beta,\gamma)$ and $(\bar\eta,\gamma)$ 
strategies for the UT \cite{Buras:2002yj} described in Section~\ref{BPS}
in the context of $K\to\pi\nu\bar\nu$ decays with the goal of testing the SM and its extensions.

In another paper in order to stress the collaboration between theorists and experimentalists in the context of these decays, Douglas Bryman, Gino Isidori, Laurence Littenberg and myself
summarized the theoretical virtues of both  decays 
and emphasized the unique role of
$\klpn$ in probing the nature of physics beyond the SM, 
in particular concerning possible new sources of 
CP violation and flavor-symmetry breaking. A brief summary of the 
prospects for the measurement of the $\klpn$ rate 
was also given \cite{Bryman:2005xp}.

These two papers written twenty years ago 
contain useful material on these decays and in particular the
review with Felix and Selma earned many citations but
it cannot compete with the 2015 review on these decays written in my group which I will describe
soon. This is partly due to
the fact that much progress on these two decays has been made both by experiment and theory since our 2008 review. 

\boldmath
\subsection{On the Correlations between Flavour Observables in  Minimal 
  $\text{U(2)}^3$ Models (2012)}
\unboldmath
In one of the first papers with Jennifer \cite{Buras:2012sd}
we pointed out a number of correlations between flavour observables present 
in a special class of models, to be called $\text{MU(2)}^3$, with an approximate global
 $\text{U(2)}^ 3$ flavour symmetry, constrained by a minimal set of spurions 
governing the breakdown of this symmetry. Riccardo Barbieri and Gino Isidori
are the masters of this symmetry \cite{Barbieri:2011fc} and this paper was probably motivated by Gino's presence at the TUM-IAS where I had an office since
the end of 2010.

In the  $\text{MU(2)}^3$ framework only
 SM operators are relevant in $\Delta F=2$ transitions. 
While the NP contributions to $\varepsilon_K$ have the same 
pattern as in
models with constrained Minimal Flavour 
Violation (CMFV), the CP-violation induced by
$B^0_{s,d}-\bar B^0_{s,d}$ mixings can deviate from the one in the SM and CMFV models. 

But these deviations in the $B^0_d$ and $B^0_s$ systems are strictly correlated 
by the $\text{U(2)}^3$ symmetry with each other. The most important result of 
our paper was the identification of a stringent triple 
$S_{\psi K_S}-S_{\psi\phi}-\vub$ correlation in this class of models that 
constitutes an important test for them 
and in  this context allows to determine $\vub$ by means 
of precise measurements of $S_{\psi K_S}$ and $S_{\psi\phi}$ with only small 
hadronic uncertainties.

We summarized
briefly the pattern of flavour violation 
 in rare $K$ and $B_{s,d}$ decays in $\text{MU(2)}^3$ and compared it with the 
 one found in CMFV models.

 \boldmath
\subsection{On the SM Prediction for $\mathcal{B}(B_{s,d} \to \mu^+ \mu^-)$ (2012)}
\unboldmath

As the decay $B_{s} \to \mu^+ \mu^-$ was in 2012 one of the milestones of the flavour program at the LHC I thought with Jennifer that one could
have still another look at its SM prediction. Also Diego Guadagnoli and Gino
Isidori had similar thoughts and eventually we joined forces which resulted
in a useful paper, well accepted by the community \cite{Buras:2012ru}.

First, following the suggestion of Jennifer and me, by analyzing the theoretical rate in the light of its main 
parametric dependence, we highlighted the importance of a complete evaluation of higher-order 
electroweak corrections, in 2012 known only in the large-$m_t$ limit, and leaving sizable 
dependence on the definition of electroweak parameters. Using insights from a complete 
calculation of such corrections for $K\to\pi\nu\bar\nu$ decays, we found a scheme in which 
NLO electroweak corrections were likely to be negligible.

Second, following the idea of Diego and Gino, we addressed the issue of the correspondence between the initial and the final state 
detected by the experiments, and those used in the theoretical prediction.
This was even more important than the issue of higher-order electroweak corrections. 
Particular attention 
was  devoted to the effect of the soft radiation, that has not been discussed 
for this mode in the previous literature, and that can lead to $O(10\%)$ corrections to the decay 
rate. The ``non-radiative'' branching ratio (that is equivalent to the branching ratio
fully inclusive of bremsstrahlung radiation) was estimated to be 
$(3.23 \pm 0.27) \times 10^{-9}$ for the flavour eigenstate, with the main uncertainty resulting 
from the value of $F_{B_s}$, followed by the uncertainty due to higher order electroweak corrections. 
Applying the same strategy to $B_d\to\mu^+\mu^-$, we found for its non-radiative branching ratio 
$(1.07 \pm 0.10) \times 10^{-10}$.

It should be stressed that our prediction for $\mathcal{B}(B_s \to \mu^+ \mu^-)$           did not include the width effect pointed out later by Robert
Fleischer and collaborators. Including this $10\%$ upward effect would
bring our result close to what is known in 2026.

 A complete
calculation of electroweak corrections has been performed one year later
by my three physics sons Bobeth, Gorbahn and Stamou so that residual perturbative and QED uncertainties have been estimated to be less than $\pm 1\%$ \cite{Bobeth:2013tba}. Including the NNLO QCD corrections from Hermann, Misiak and Steinhauser \cite{Hermann:2013kca} and the width effects these six loop masters found
$\overline{\mathcal{B}}(B_s\to\mu^+\mu^-)=(3.65 \pm 0.23)\times 10^{-9}$
\cite{Bobeth:2013uxa}.

But this is not the end of the story. In 2017 and 2019 still some improvements
by Beneke, Bobeth and Szafron regarding electromagnetic corrections
have been made \cite{Beneke:2017vpq,Beneke:2019slt} which
practically did not modify the above result. Still an important calculation
reducing the uncertainties in this decay.

In view of all these improvements the main uncertainty in this decay comes from the value
of $\vts$ or equivalently $\vcb$.
Indeed, the final result depends quadratically on $\vcb$ and the one quoted above uses its inclusive value.
For the exclusive  value one would get a significantly lower branching ratio. We will return to
this issue in Part~\ref{BV}.

In any case it is very important to have one day  very precise experimental values
not only for the $B_s\to\mu^+\mu^-$ branching ratio but also for the
 $B_d\to\mu^+\mu^-$ one.

\boldmath
\subsection{$\kpn$ and $\klpn$ in the Standard Model:\\
  Status and Perspectives (2015)}\label{Per2015}
\unboldmath
In 2015 Dario Buttazzo,  Jennifer Girrbach-Noe and Robert Knegjens joined me in this new adventure on these two decays \cite{Buras:2015qea}.
It became  a true
bestseller and was cited more than 450 times by now. Dario was a postdoc
at IAS in the context of my Advanced ERC Grant (2011-2016) and Robert was a
PhD student of Robert Fleischer. He spent three months at IAS. Both Dario
and Robert were very efficient and pleasant collaborators so that having
Jennifer in addition it is not surprising that our project was successful.

One of the reasons to write this review in 2015 was 
the recent start of the NA62 experiment at CERN that was expected to measure the $\kpn$ branching ratio with a precision of $10\%$. We summarised the present status of this promising decay within the SM.
We did likewise for $\klpn$, which was expected to be measured by the KOTO experiment around 2020. Both expectations turned out to be too optimistic.
As the perturbative QCD and electroweak corrections in both decays were already in 2015 almost under full control\footnote{Only NNLO QCD contribution to the function $X(x_t)$ was missing.}, the dominant uncertainties within the SM  originated  as today from the CKM parameters $\vcb$, $\vub$ and $\gamma$. 
We showed this dependence with the help of analytic expressions as well as accurate interpolating formulae. These formulae turned out to be useful for
the BV-Strategy in 2021 \cite{Buras:2021nns,Buras:2022wpw}.

Not expecting a soon  clarification of the discrepancies between inclusive and exclusive determinations of $\vcb$ and $\vub$ from tree-level decays,
we investigated whether higher precision on both branching ratios  is achievable by  determining $\vcb$, $\vub$ and $\gamma$ by means of other observables that are already precisely measured. 
In this context $\varepsilon_K$ and  $\Delta M_{s,d}$, together with the expected progress in QCD lattice calculations  was expected to play a prominent 
role. We found $\mathcal{B}(\kpn)= \left(9.11\pm 0.72\right) \times 10^{-11}$ and $\mathcal{B}(\klpn)= \left(3.00\pm 0.30 \right) \times 10^{-11}$, that turned out to
 more precise than using averages of the 2015 tree-level values of 
 $\vcb$, $\vub$ and $\gamma$:
$\mathcal{B}(\kpn)= \left(8.4\pm 1.0\right) \times 10^{-11}$ and $\mathcal{B}(\klpn)= \left(3.4\pm 0.6 \right) \times 10^{-11}$.

 Furthermore, we pointed out  correlations
 between $\mathcal{B}(\kpn)$, $\overline{\mathcal{B}}(B_{s,d}\to\mu^+\mu^-)$  and $\gamma$ within the SM, that are only very weakly dependent on  other CKM parameters. They are  given as follows
\begin{align}
\mathcal{B}(\kpn) &= (8.39\pm 0.58)\times 10^{-11} \cdot  \left[\frac{\gamma}{73.2^\circ}\right]^{0.81}\notag\\
&\qquad\qquad\qquad\quad\times\left[\frac{\overline{\mathcal{B}}(B_s\to\mu^+\mu^-)}{3.4\times 10^{-9}}\right]^{1.42}\left[\frac{227.7\mev}{F_{B_s}}\right]^{2.84},\label{master1a}
\end{align}
\begin{align}
\mathcal{B}(\kpn) &= (8.41 \pm 0.77)\times 10^{-11} \cdot \left[\frac{\overline{\mathcal{B}}(B_s\to\mu^+\mu^-)}{3.4\times 10^{-9}}\right]^{0.74}\left[\frac{227.7\mev}{F_{B_s}}\right]^{1.48}\notag\\
&\qquad\qquad\qquad\quad\times\left[\frac{\mathcal{B}(B_d\to\mu^+\mu^-)}{1.08\times 10^{-10}}\right]^{0.72}\left[\frac{190.5\mev}{F_{B_d}}\right]^{1.44}.\label{master2a}
\end{align}
Note that both relations are independent of $\vcb$ and (\ref{master1a}) depends only on $\gamma$. In (\ref{master2a}) the $\gamma$ dependence has been
eliminated with the help of $\mathcal{B}(B_d\to\mu^+\mu^-)$.

These formulae
are consistent with the ones listed in App.~\ref{UsefulEX} but one should
note that various input parameters changed since 2015 and recent NNLO QCD corrections to $X(x_t)$ have been taken into account there so that for the
phenomenology of 2026 it is better to use the expressions in App.~\ref{UsefulEX}.

In particular the correlation (\ref{master1a}) should be of interest in 
the coming years due to  
the improved measurements of $\kpn$ by NA62, of $B_s\to\mu^+\mu^-$ by LHCb and CMS and of $\gamma$ by LHCb. Moreover the last factor can  still  be improved by LQCD.

 There is no question about that some ideas presented in this paper led 6 years
 later to the BV-Strategy which also included other $K$ and $B$ decays as reported on in Part~\ref{BV}.

Unfortunately, soon after our 2015 paper was completed, Jennifer decided to stop the research which was a shock for
me as she was the one with whom I have written 19 papers within 4  years
she was in my group. She
was one of the best and possibly the most efficient collaborator, together with
Monika Blanke and Fulvia De Fazio, I ever had.

In this context I want to mention one short story. One day I got an idea
on some correlations between observables and went to Jennifer's office to
share with her my new idea. She was busy with typing something but stopped
and discussed with me my new idea. Suddenly in the middle of our discussion
she continued the typing which I found rather impolite but as she still
simultaneously was listening to my comments and was reacting to them I did
not protest. After returning to my office I opened my laptop to find
an email from Jennifer sending me the plots proposed by me.
It is then not surprising that within 4 years we could write so many
papers which as one can see in the right column in  Table~\ref{tab:ranking}
were very well accepted by the community. It is really a pity she left research but
fortunately I have still contact with her and her two wonderful  daughters,
 Lorena and Romy.

The next paper on $K\to\pi\nu\bar\nu$ decays has been then written
only with Dario and Robert \cite{Buras:2015yca}. It studied the correlations
between these two decays and the ratio $\epe$ in simplified NP models.  Let me describe briefly what can be found in this paper.

\boldmath
\subsection{$K\to\pi\nu\bar\nu$  and $\epe$ in Simplified New Physics Models (2015)}\label{Dario}
\unboldmath
The decays  $\kpn$ and $\klpn$ have been studied over many years in various 
concrete extensions of the SM. A review of the analyses performed until 
August 2007 can be found in \cite{Buras:2004uu}. More recent reviews until 2015 can be 
found in \cite{Buras:2010wr,Buras:2012ts,Buras:2013ooa,Blanke:2013goa,Smith:2014mla}. Most extensive analyses have been performed in supersymmetric models \cite{Buras:1997ij,Colangelo:1998pm,Buras:1999da,Buras:2004qb,Crivellin:2011sj}, 
the Littlest Higgs (LH) model without T-parity \cite{Buras:2006wk}, the LH model with T-parity (LHT) \cite{Blanke:2009am,Blanke:2015wba}, Randall-Sundrum models \cite{Blanke:2008yr,Bauer:2009cf}, models with partial compositeness \cite{Straub:2013zca} and  331 models \cite{Buras:2012dp,Buras:2013dea,Buras:2014yna}. All these models 
contain several new parameters related to couplings and masses of new fermions,
 gauge bosons and scalars and the analysis of  $\kpn$ and $\klpn$  requires 
the inclusion of all constraints on couplings and masses of these particles and 
consequently is rather involved. Moreover, the large number of parameters 
present in these models did not allow in 2015 and similarly does not allow presently for clear cut conclusions beyond rough bounds on the size of NP contributions to  $\kpn$ and $\klpn$. 

Therefore, we thought that  in order to get a better insight into the structure of the possible impact of NP on  $\kpn$ and $\klpn$ decays, and in particular on the correlation between them and other observables, it is useful to consider models with a small number of parameters.

With this idea in mind 
we studied $\kpn$ and $\klpn$ decays in the simplest extensions of the SM in which 
 stringent correlations between these two decays and other flavour observables 
are present. We first considered simple models with tree-level $Z$ 
and $Z^\prime$ contributions in which either MFV or the $\text{U(2)}^3$ symmetry is imposed on 
the quark flavour-violating couplings. We then compared the resulting correlations with 
those present in generic models in which the latter couplings are arbitrary, subject 
to the constraints from $\Delta F=2$ processes, electroweak and collider 
data. Of particular interest are the correlations with $\epe$ and $K_L\to \mu^+\mu^-$ which limit the size of NP contributions 
to $\kpn$ and $\klpn$, depending on the Dirac structure of couplings and 
the relevant operators. But in MFV models also the constraint from 
$B_s\to\mu^+\mu^-$ turns out to be important. We took into account the 
recent results from large $N$ approach that indicated $\epe$ 
in the SM to be significantly below the data.
While in many models the enhancement of $\epe$ implies the suppression of $\klpn$, we presented two models in which 
 $\epe$  and $\klpn$ can be simultaneously enhanced relative to 
SM predictions (see Fig.~\ref{ZepsilonKL}).  A correlation between $\kpn$ and  
$B\to K(K^*)\mu^+\mu^-$, found by us in the simple models considered here, as expected by us, turned out ten years later to be of interest 
for NA62 and LHCb experimentalists at CERN. The one 
with $B\to K(K^*)\nu\bar\nu$ for Belle II.

\begin{figure}[t]
\centering%
\includegraphics[width=0.6\textwidth]{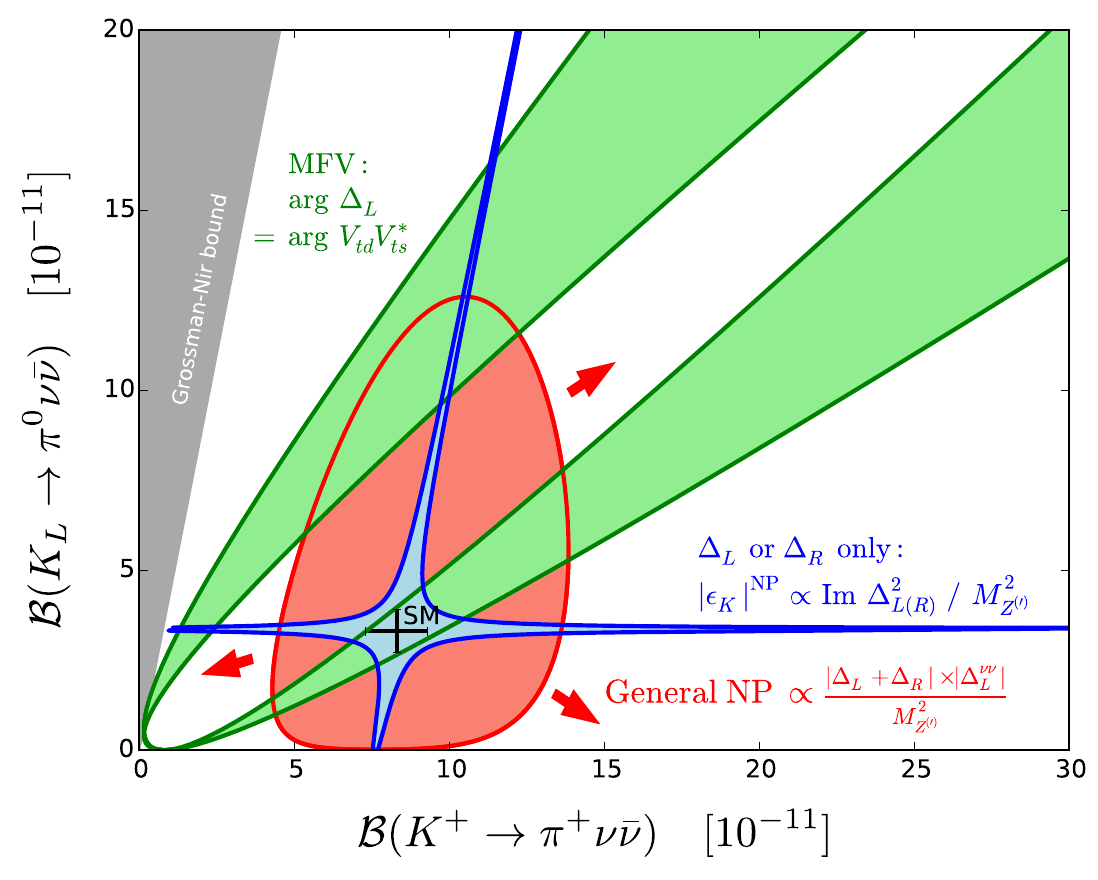}%
\caption{\it Illustrations of common correlations in the $\mathcal{B}(\kpn)$ versus $\mathcal{B}(\klpn)$ plane. The expanding red region illustrates the lack of correlation for models with general LH and RH NP couplings. The green region shows the correlation present in models obeying CMFV. The blue region shows the correlation induced by the constraint from $\varepsilon_K$ if only LH or RH couplings  are present. From \cite{Buras:2015yca}.\label{fig:illustrateEpsK}}
\end{figure}

\begin{figure}
\centering%
\includegraphics[width=0.32\textwidth]{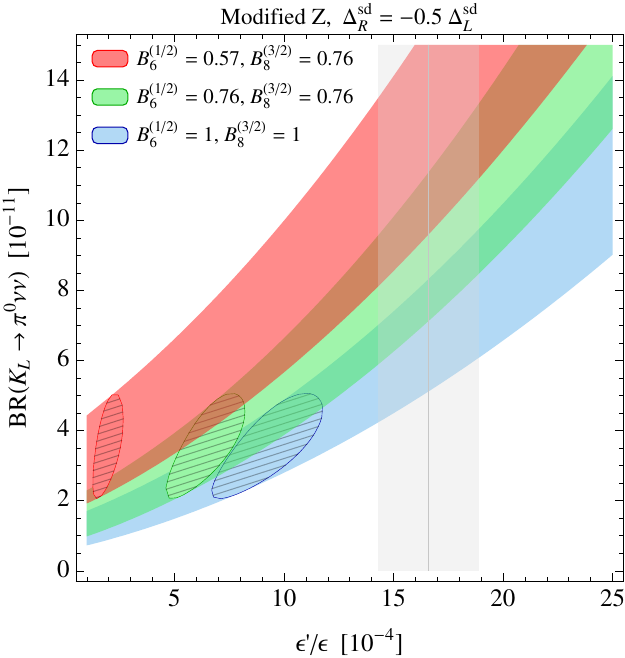}\hfill%
\includegraphics[width=0.32\textwidth]{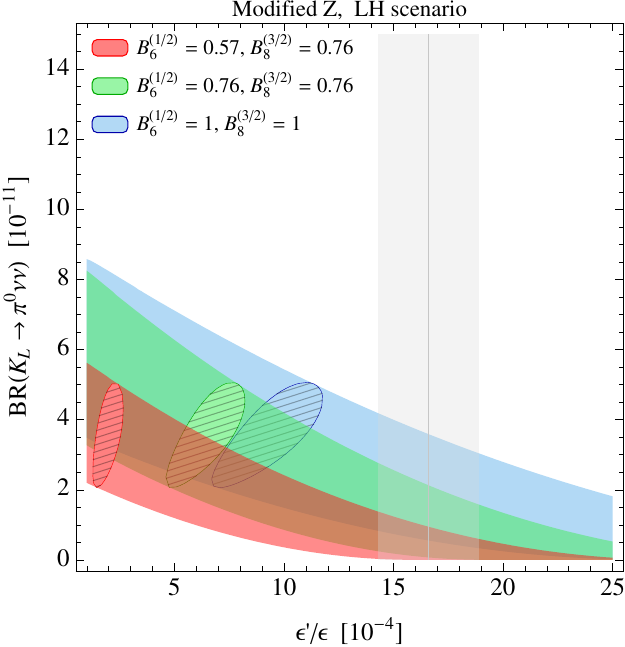}\hfill%
\includegraphics[width=0.32\textwidth]{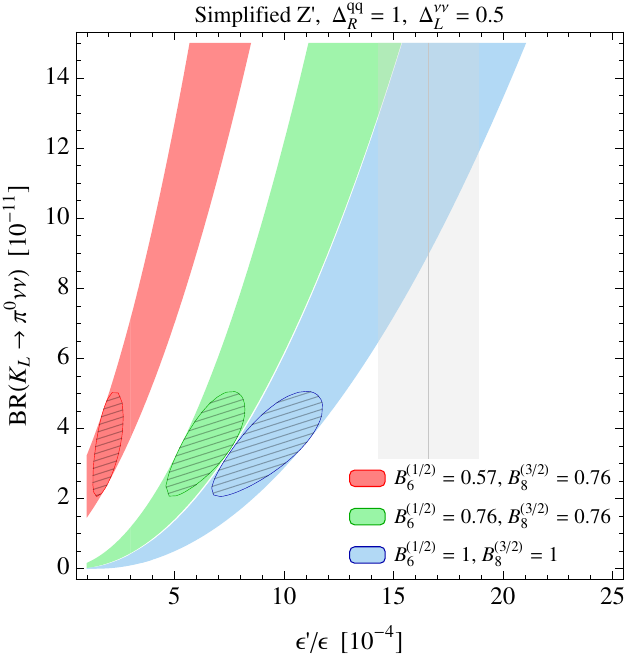}
\caption{\it 95\% C.L. allowed regions for $\epe$ and $\klpn$. Left: model with flavour-changing Z boson couplings $\Delta_R^{sd} = -0.5 \Delta_L^{sd}$. Center: modified Z, LH scenario $\Delta_R^{sd} = 0$. Right: 5 TeV Z' with $\Delta_R^{qq} = 1$ and $\Delta_L^{\nu\nu} = 0.5$. The plots are for different values of $\bsi$ and $\bei$ as indicated by different colours.The hatched regions are the SM predictions at $2\sigma$. The gray band shows the experimental result for $\epe$. From \cite{Buras:2015yca}.\label{ZepsilonKL}}
\end{figure}

Many plots showing correlations between various observables can be found in
our paper. I show here only two. 
Fig.~\ref{fig:illustrateEpsK}   is shown in all papers that study $K\to\pi\nu\bar\nu$ decays. It has been motivated by an important paper by Monika Blanke who
in 2009 studied the impact of NP physics in $\varepsilon_K$ on the correlation of $\kpn$ and  $\klpn$ \cite{Blanke:2009pq}. Monika found that in the limit of no NP in $\varepsilon_K$ these two branching ratios are correlated along the two  branches centered
in the {\em blue} region shown in this figure. Clearly, the branch parallel to the Grossman-Nir bound is more interesting as it implies NP both in $\kpn$ and $\klpn$. The blue region itself corresponds to the case if only LH or RH couplings
are present. Otherwise as shown in Fig.~\ref{fig:illustrateEpsK} the correlation is more involved.
In  Fig.~\ref{ZepsilonKL} correlation between $\epe$ and $\klpn$
is shown.

One day these plots could turn out to be very useful when both $\kpn$
and $\klpn$ decays will be measured and the situation of $\epe$ in the SM
will be clarified. Equally exciting will be the study that also includes 
$B\to K(K^*)\mu^+\mu^-$, $B\to K(K^*)\nu\bar\nu$ and $B_{s,d}\to\mu^+\mu^-$ decays. The study of the last decays and of correlations between their branching
ratios in a $Z^\prime$ model has been studied by us in two papers in 2014
\cite{Buras:2014fpa} and 2024 \cite{Buras:2024mnq} and I will describe these papers as we proceed.

\boldmath
\subsection{Symmetries and Asymmetries of $B\to K^*\mu^+\mu^-$ in the SM and Beyond (2008)}
\unboldmath
Another very successful paper turned out to be the one on Symmetries and Asymmetries
in the $B\to K^*\mu^+\mu^-$ decay \cite{Altmannshofer:2008dz} which became a standard
reference for this decay with more than 700 citations by now. This was the only paper I wrote with Patricia Ball. Her
presence in this collaboration was very important as she had much
larger knowledge of this decay than anybody in our team including myself.
But the contributions of other members of the team, Wolfgang Altmannshofer, Aoife Barucha, David Straub and Michael Wick were also very important.
Let me describe briefly what can be found in this paper.

First, let us stress that the rare decay $B\to K^*(\to K\pi)\mu^+\mu^-$ is regarded as one of
the crucial channels for $B$ physics as the polarization of the $K^*$ 
allows a precise angular reconstruction resulting in many observables
that offer new important tests of the SM and its
extensions. These angular observables can be expressed in terms of 
CP-conserving and CP-violating quantities which we studied in terms of
the full form factors calculated from QCD sum rules on the light-cone, 
including QCD factorization corrections. In this paper we investigated all 
observables in the context of the SM and various NP
models, in particular the LHT model
and various MSSM scenarios, identifying those observables with  
small to moderate 
dependence on hadronic quantities and large impact of NP. 
One important result of our study was that new CP-violating phases
will produce clean signals in CP-violating asymmetries. We also identified a 
number of correlations between various observables which will allow a 
clear distinction between different NP scenarios.

\begin{figure}
\centering
\includegraphics[width=0.6\textwidth]{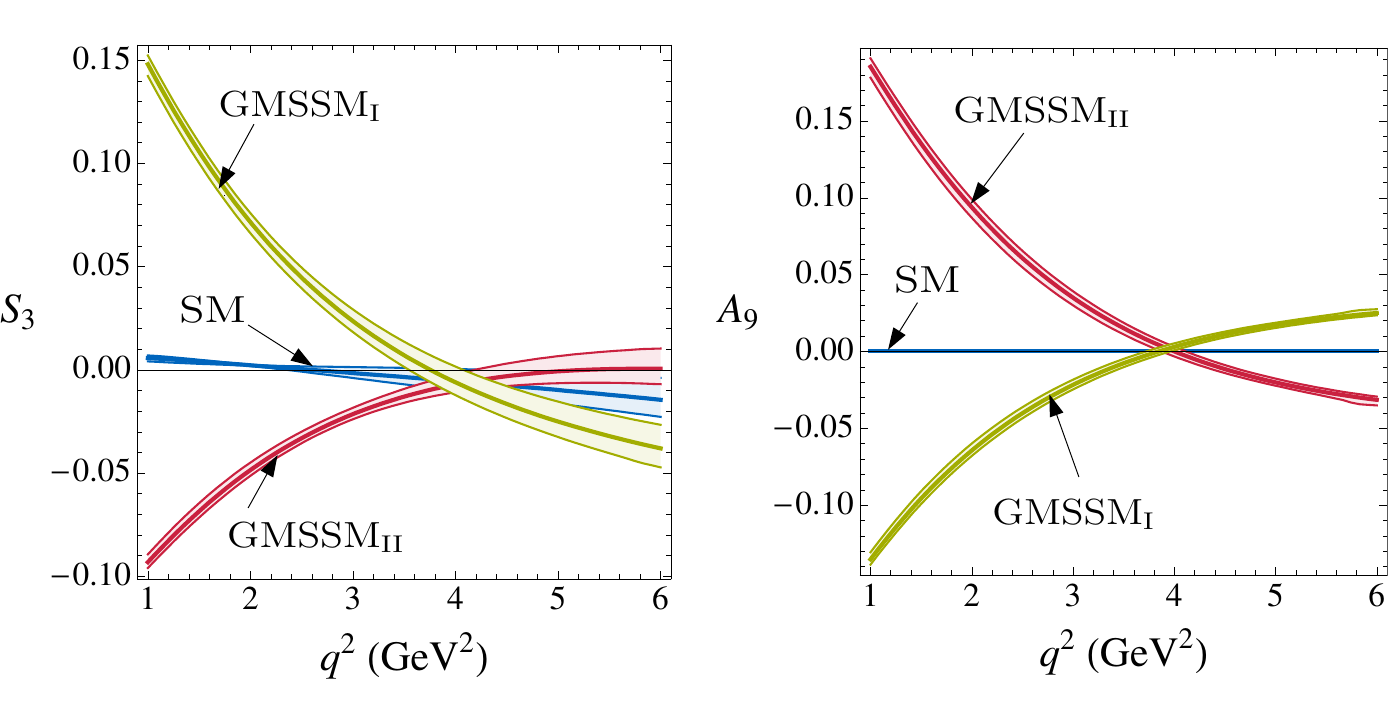}
\vskip-10pt
\caption[]{\small The observables $S_3$ and $A_9$ in the SM (blue
  band) and the two GMSSM scenarios GMSSM$_{\rm I,II}$ 
with large complex contributions to $C_7^\prime$ as described in \cite{Altmannshofer:2008dz}.}
\label{fig:GMSSMA9}
\hskip10pt
\centering
\includegraphics[width=0.9\textwidth]{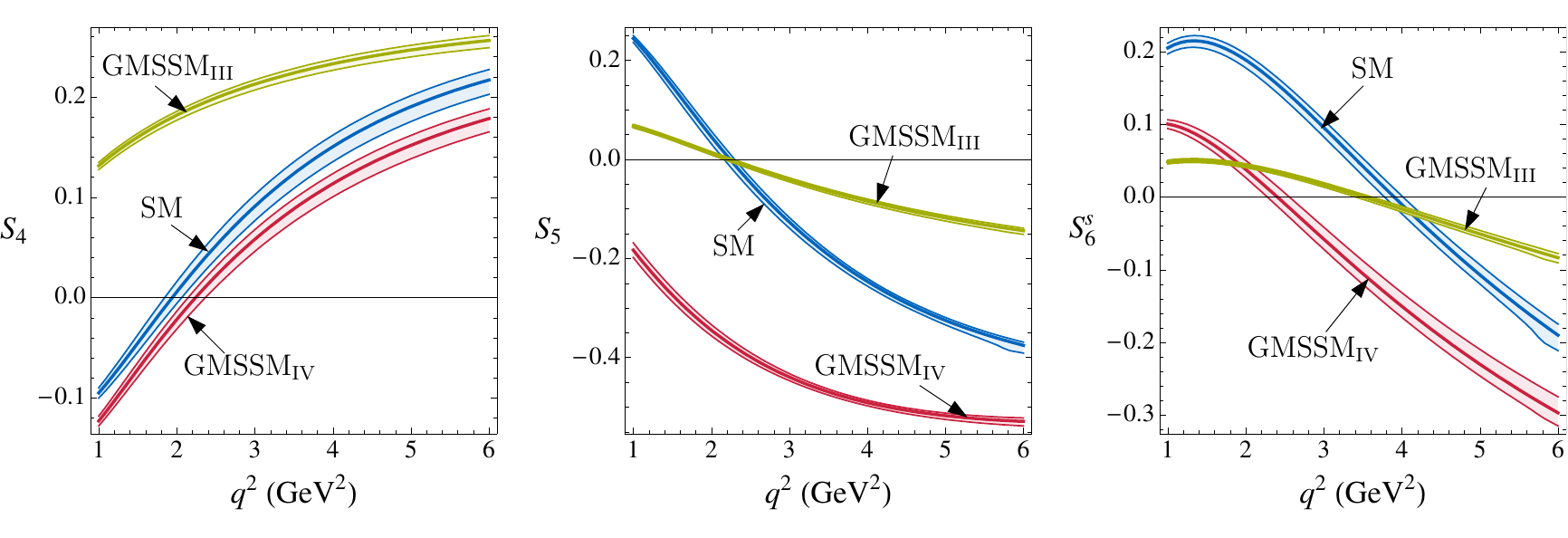}
\includegraphics[width=0.6\textwidth]{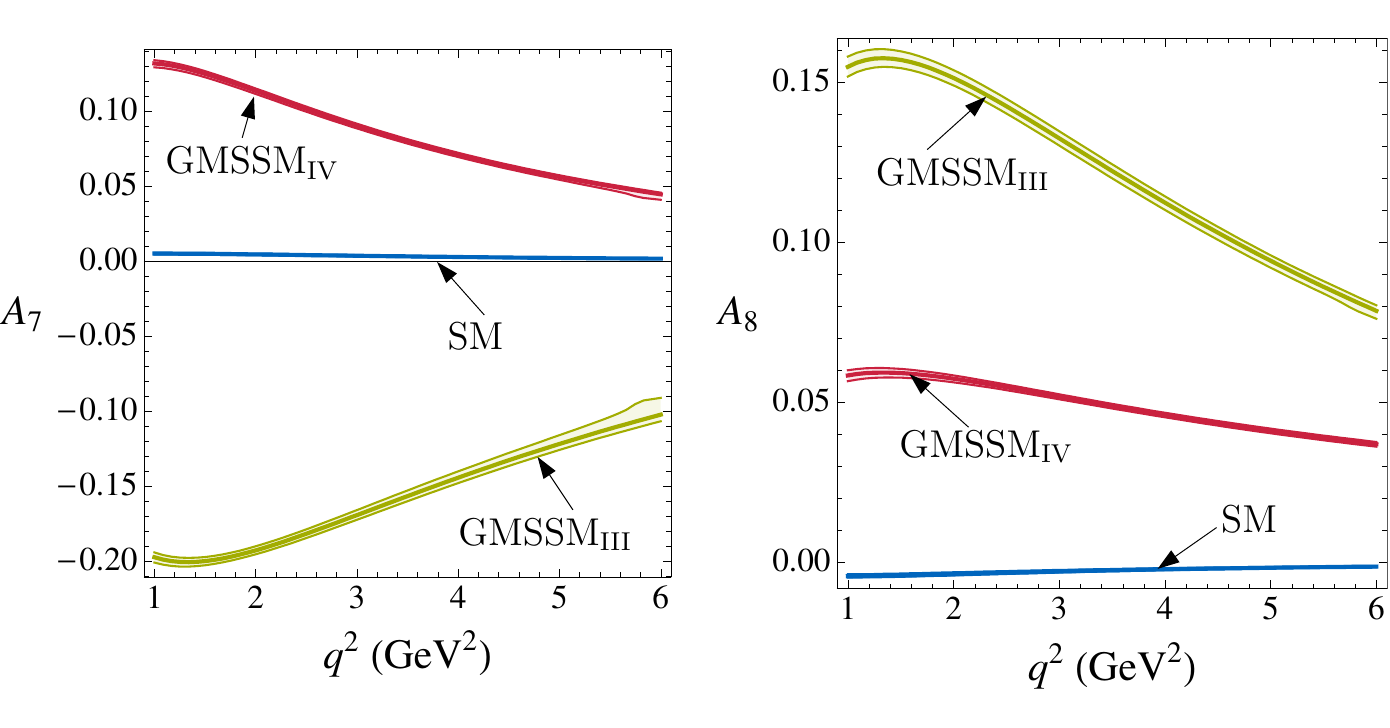}
\includegraphics[width=0.6\textwidth]{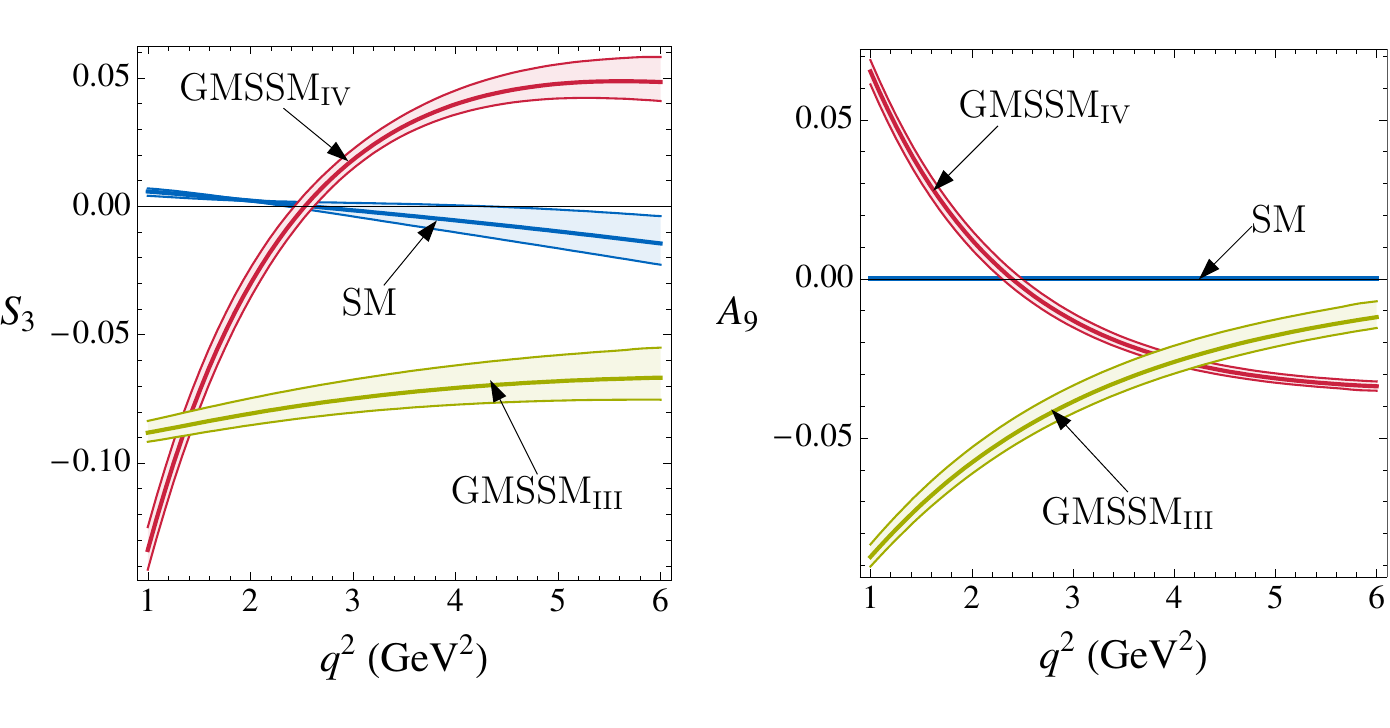}
\vskip-10pt
\caption[]{\small Several observables in the SM (blue band) and two selected GMSSM scenarios that show large non-standard behaviour. See \cite{Altmannshofer:2008dz}for details.}
\label{fig:GMSSM}
\end{figure}

Among the observables an important role in the phenomenology is played by the CP symmetries $S_i$ and the CP-asymmetries $A_i$ with $i=1,..12.$. Few examples are shown in Figs.~\ref{fig:GMSSMA9} and \ref{fig:GMSSM}. In addition to the SM
the results for two supersymmetric scenarios are shown. These plots show
that indeed a distinction between various NP scenarios can be made in this
manner. They show also that in 2008 the supersymmetric models were very popular
which is not the case now.

  \boldmath
\subsection{$B\to K\nu\bar\nu$ and $B\to K^*\nu\bar\nu$ in the SM and Beyond}
\unboldmath

\boldmath
\subsubsection{New Strategies for NP Search in $B\to K\nu\bar\nu$,  $B\to K^*\nu\bar\nu$ and\\ $B\to X_s\nu\bar\nu$ (2009)}
\unboldmath

Soon after the completion of the $B\to K^*\mu^+\mu^-$ paper we realized
that we could also look at $B\to K\nu\bar\nu$ and $B\to K^*\nu\bar\nu$ decays
that were expected to be measured by Belle and Belle II one day.
As Patricia and Aoife returned to Durham, only the remaining quartet worked
on these decays \cite{Altmannshofer:2009ma}.

Let me stress first that the rare decay $B\to K^*\nu \bar \nu$ is regarded as one of the important channels in $B$ physics as it allows a transparent study of $Z$ penguin and other electroweak penguin effects in NP scenarios in the absence of dipole operator contributions and Higgs (scalar) penguin contributions that are often more important than $Z$ contributions in $B\to K^*\ell^+\ell^-$ and $B_s\to \ell^+\ell^-$ decays. We presented a new analysis of $B\to K^*\nu\bar\nu$ with improved form factors and of the decays $B\to K\nu\bar\nu$ and $B\to X_s\nu\bar\nu$ in the SM and in a number of NP scenarios like the general MSSM, general scenarios with modified $Z$/$Z'$ penguins and in a singlet scalar extension of the SM. We also summarized the results in the LHT  and a Randall-Sundrum (RS) model with custodial protection of left-handed $Z d_i \bar d_j$ couplings
which have been studied separately in my group as described in Sections~\ref{LHTG} and \ref{RSM}, respectively.

While the decay $B\to K^*\nu \bar \nu$ is theoretically not as clean as $\kpn$ and $\klpn$ decays  because of the presence of form factors that have to be calculated by non-perturbative
methods, it should be emphasized that the existence of angular observables in
$B\to K^*\nu \bar \nu$ allows a deeper insight into the issue of {\em right-handed currents} than it is possible in the two
rare K decays in question. Indeed the latter decays are only sensitive to the sum of the Wilson coefficients of left-handed and right-handed couplings, whereas $B\to K^*\nu \bar \nu$ is also
sensitive to their difference.

In this context the results for the SM and NP scenarios can be transparently summarized in a $(\epsilon,\eta)$ plane \cite{Grossman:1995gt,Melikhov:1998ug}, analogous to the known $(\bar\varrho,\bar\eta)$ plane, with a non-vanishing $\eta$ signalling this time not CP violation but the presence of new right-handed down-quark flavour violating couplings which can be ideally probed by the decays in question. Measuring the three branching ratios and one additional polarization observable in $B\to K^*\nu \bar \nu$ $(F_L)$ allows to overconstrain the resulting point in the $(\epsilon,\eta)$ plane with $(\epsilon,\eta)=(1,0)$ corresponding to the SM.

Explicitly, defining $\epsilon$ and $\eta$ in terms of WCs $C^\nu_{L,R}$ as follows
\begin{equation}  \label{eq:epsetadef}
 \epsilon = \frac{\sqrt{ |C^\nu_L|^2 + |C^\nu_R|^2}}{|(C^\nu_L)^\text{SM}|}~, \qquad
 \eta = \frac{-\text{Re}\left(C^\nu_L C_R^{\nu *}\right)}{|C^\nu_L|^2 + |C^\nu_R|^2}~,
\end{equation}
the four observables in question can be expressed in terms of $\epsilon$ and $\eta$ as follows\footnote{For the most recent 2025 estimate of $\mathcal{B}(B \to X_s \nu\bar\nu)$ in the SM see \cite{Fael:2025xmi}.}
\begin{align}
\label{eq:epseta-BKsnn}
 \mathcal{B}(B \to K^* \nu\bar\nu) & = 6.8 \times 10^{-6} \, (1 + 1.31 \,\eta)\epsilon^2~, \\
\label{eq:epseta-BKnn}
 \mathcal{B}(B \to K \nu\bar\nu)   & = 4.5 \times 10^{-6} \, (1 - 2\,\eta)\epsilon^2~, \\
\label{eq:epseta-BXsnn}
 \mathcal{B}(B \to X_s \nu\bar\nu) & = 2.7 \times 10^{-5} \, (1 + 0.09 \,\eta)\epsilon^2~, \\
\label{eq:epseta-FL}
 \langle F_L \rangle             & = 0.54 \, \frac{(1 + 2 \,\eta)}{(1 + 1.31 \,\eta)}~.
\end{align}
As $\epsilon$ and $\eta$ can be calculated in any model by means of  (\ref{eq:epsetadef}), these four expressions can be considered as fundamental formulae for any phenomenological analysis of the decays in question.

\begin{figure}[tb]
\centering
\includegraphics[width=0.45\textwidth]{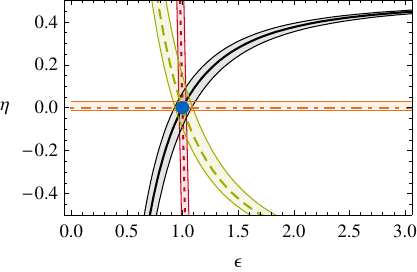}\qquad
\includegraphics[width=0.45\textwidth]{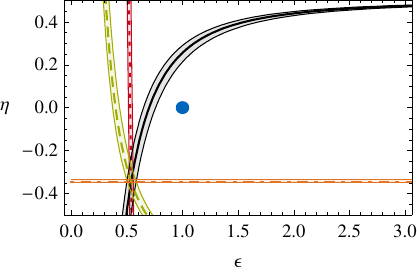}
\caption[]{\small Hypothetical constraints on the $\epsilon$-$\eta$-plane, assuming all four observables have been measured with infinite precision. The error bands reflect the theoretical uncertainty in 2009.
The green band (dashed line) represents $\mathcal{B}(B \to K^* \nu\bar\nu)$, the black band (solid line) $\mathcal{B}(B \to K \nu\bar\nu)$, the red band (dotted line) $\mathcal{B}(B \to X_s \nu\bar\nu)$ and the orange band (dot-dashed line) $\langle F_L \rangle$.
Left: SM values for the Wilson coefficients, right: assuming $C^\nu_L=0.5(C^\nu_L)^\text{SM}$ and $C^\nu_R=0.2(C^\nu_L)^\text{SM}$. The blue circle represents the SM point. From \cite{Altmannshofer:2009ma}.}
\label{fig:exp-hypo}
\end{figure}

Since the four observables depend on only two parameters, a measurement of all of them would overconstrain the resulting $(\epsilon,\eta)$ point. To illustrate the theoretical cleanliness of the various observables, we show in Fig.~\ref{fig:exp-hypo} the combined constraints after hypothetical measurements with infinite precision, first assuming the SM and then for a toy NP example.

We also pointed out that the correlations of these three channels with the rare decays $K^+\to\pi^+\nu\bar\nu$, $K_L\to\pi^0\nu\bar\nu$, $B\to X_s \ell^+\ell^-$ and $B_s\to \mu^+\mu^-$ offer powerful tests of NP with new right-handed couplings and non-MFV interactions.

While we improved on the SM predictions for decays in question, I do not
present these results because they are out of date due to changed CKM
parameters and formfactors in the last 15 years. For an updated analysis of this
strategy see my 2024 paper with Julia Harz and Martin Mojahed \cite{Buras:2024ewl} that I will discuss soon. In particular we have generalized the expression
in (\ref{eq:epseta-BKsnn})-(\ref{eq:epseta-FL}) to the cases of lepton flavour
and lepton universality violations.

 \boldmath
\subsubsection{$B\to K\nu\bar\nu$ and $B\to K^*\nu\bar\nu$ in the SM and Beyond (2014)}\label{2014nunubar}
\unboldmath

Almost six years later
David Straub suggested that we could return to these decays in view
of the progress being made at Belle. David, my excellent PhD student several years before, being a fellow at our excellence
cluster Universe,  supervised his PhD student Christoph Niehoff. On the other
hand Jennifer Girrbach-Noe, my outstanding collaborator, was still working
at IAS and our new quartet published in 2014 a new paper on these decays \cite{Buras:2014fpa}
that turned out to be even more popular than our 2009 paper described above.

\begin{figure}[ptb]
\centering
\includegraphics[height=0.8\textheight]{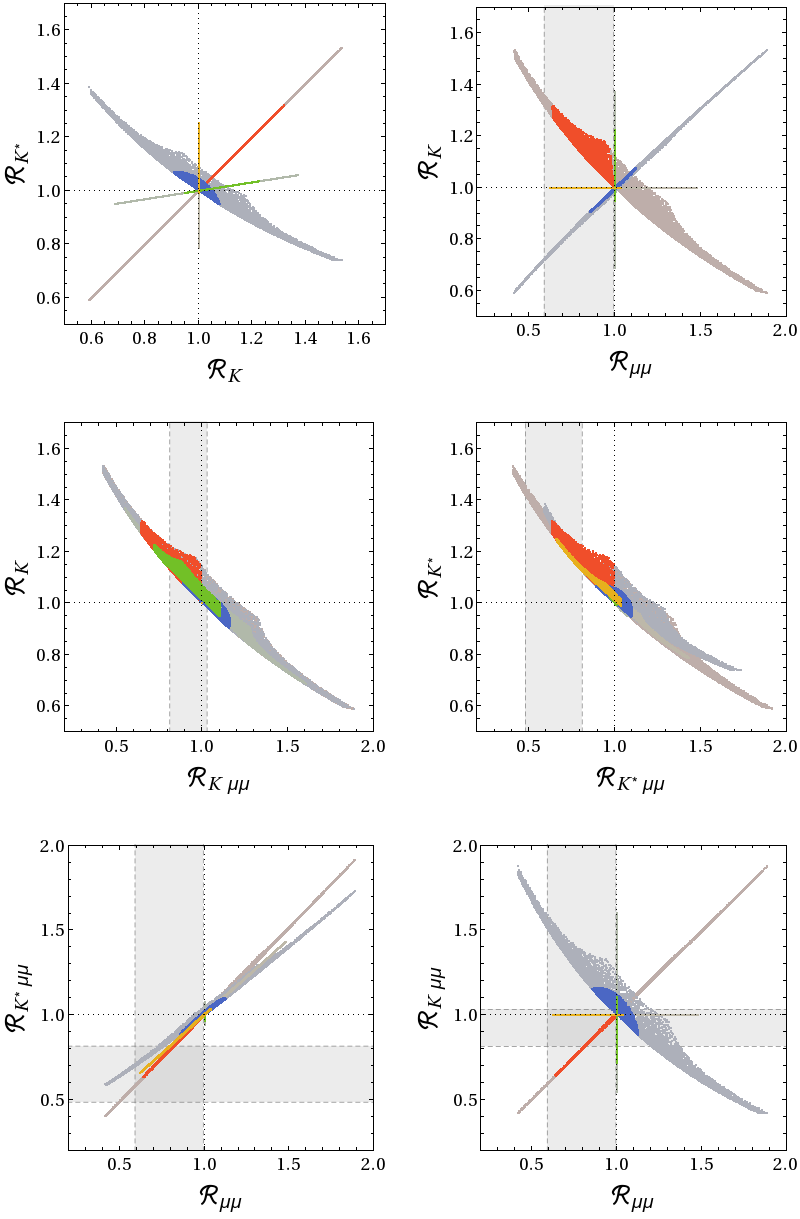}
\caption{Various correlations between observables in  LHS (red), RHS (blue), LRS (green), ALRS (yellow), assuming LFU and $\Delta_R^{\nu\nu} = \Delta_R^{\ell\ell}=0$. All points satisfy  $0.9\leq \Delta M_s/(\Delta M_s)_{\text{SM}}\leq 1.1$, $-0.14\leq S_{\psi\phi}\leq 0.14$.
Grey  regions are disfavoured at $2\sigma$ by $b\to s\mu^+\mu^-$ constraints in 2014. From \cite{Buras:2014fpa}.}
\label{fig:Zprime}
\end{figure}

We presented an analysis of $B\to K\nu\bar\nu$ and $B\to K^{*}\nu\bar\nu$
within the SM, in a model-independent manner, and in a number of NP models.
Combining new form factor determinations from lattice QCD with light-cone sum rule results and including complete two-loop electroweak corrections to the SM Wilson coefficient \cite{Brod:2010hi}, calculated again by my physics family, we obtained the following SM predictions
\be
\mathcal{B}(B^+\to K^+\nu\bar\nu) = (4.0 \pm 0.5) \times 10^{-6},\qquad
\mathcal{B}(B^0\to K^{* 0}\nu\bar\nu)  =  (9.2\pm1.0) \times 10^{-6},
\ee
more precise and more robust than previous estimates. Yet as we will see in
the recent analysis in 2024  with Peter Stangl \cite{Buras:2024mnq} both predictions went slightly up
in the last ten years. This is in particular due to new formfactors from HPQCD collaboration that became available only in 2022 \cite{Parrott:2022zte}, and modified CKM parameters. 
 I will describe this analysis in Section~\ref{AJBPS}.

Beyond the SM, we made use of the unbroken $\text{SU(2)}_L$ symmetry of the SMEFT at dim-6  to relate NP effects in $b\to s\nu\bar\nu$ transitions to $b\to s\ell^+\ell^-$ transitions and used the wealth of experimental data on $B\to K^{(*)}\ell^+\ell^-$ and related modes to constrain
NP effects in $B\to K^{(*)}\nu\bar\nu$.
We then considered several specific NP models, including $Z'$ models, the MSSM, models with partial compositeness, and leptoquark models, 
demonstrating that 
 the correlations between $b\to s\nu\bar\nu$ 
observables among themselves and with 
$B_s\to\mu^+\mu^-$ and 
$b\to s\ell^+\ell^-$ transitions offer 
powerful tests of NP with new right-handed couplings and non-MFV interactions.

Such correlations are shown in Fig.~\ref{fig:Zprime}.  All ratios are defined
by
\be\label{nunuratios}
\mathcal{R}_{K(K^*)}=\frac{\mathcal{B}(B\to K(K^*)\nu\bar\nu)}{\mathcal{B}(B\to K(K^*)\nu\bar\nu)_{\text{SM}}},\quad
\mathcal{R}_{K(K^*)\mu\mu}=\frac{\mathcal{B}(B\to K(K^*)\mu\bar\mu)}{\mathcal{B}(B\to K(K^*)\mu\bar\mu)_{\text{SM}}},\,
\ee
and $\mathcal{R}_{\mu\mu}$ for $B_s\to\mu\bar\mu$.
 Moreover the four scenarios shown have left-handed quark couplings (LHS), right-handed (RHS) ones, left-right symmetric (RHS) ones and anti-left-right symmetric (ALRS) ones. $\mathcal{R}_i=1$ in the SM.

I am very satisfied  with this paper. I am sure it will play an important role in phenomenological analyses when the data for all decays considered
in our paper will improve in this decade.

We will see the 2024 version of such correlations in Section~\ref{AJBPS}
obtained in an analysis with Peter Stangl  \cite{Buras:2024mnq}. As seen
in Fig.~\ref{fig:Zprime} the LHS (red) scenario implies significant suppression
of $\mathcal{R}_{\mu\mu}$ which is not seen in the present data and scenario close
to the LRS (green) one is favoured presently.

\begin{figure}[th]
\centering
\includegraphics[width=0.75\textwidth]{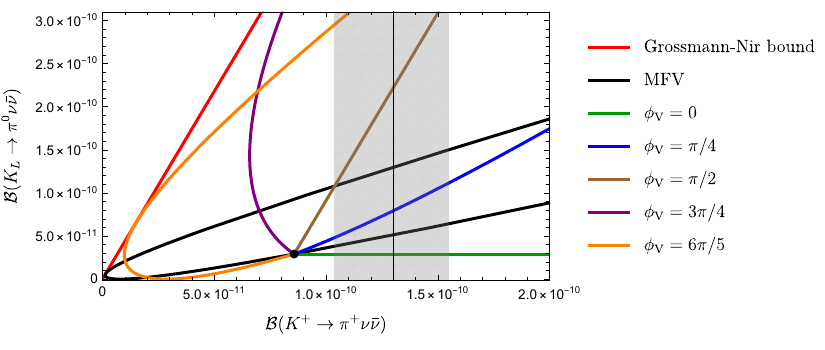}\\
\includegraphics[width=0.75\textwidth]{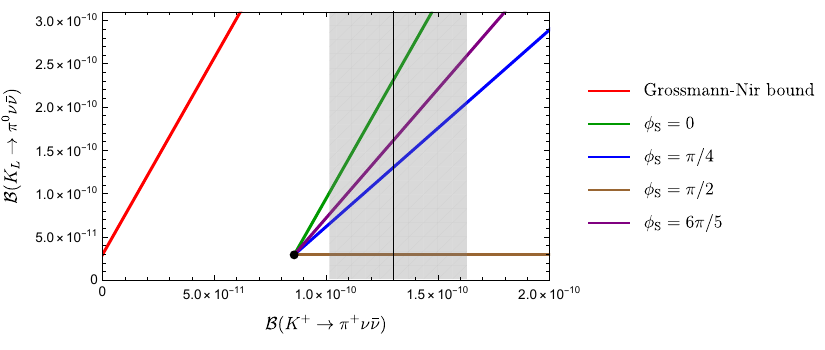}
\caption{$\mathcal{B}(\kpn)-\mathcal{B}(\klpn)$-plane for
  vector-current contributions with $C_S=0$ (top) and scalar-current contributions with $C_V=0$ (bottom) assuming lepton-flavor universality. $\phi_V$ $(\phi_S)$ is fixed to different values (see legend) and $C_V$ $(C_S)$, the size of
  vector (scalar) contributions,varied.
    The red line indicates the Grossmann-Nir bound. The SM contribution is represented by a black point. The grey region represents the present experimental $1\sigma$ range. See \cite{Buras:2024ewl} for details.}
\label{fig:KaonVectorScalar}
\end{figure}

\boldmath
\subsection{Disentangling New Physics in $K\to\pi\nu\Bar{\nu}$ and $B\rightarrow K(K^*)\nu\Bar{\nu}$ Observables (2024)}\label{Harz}
\unboldmath
Until 2022 all my papers on $K\to\pi\nu\Bar{\nu}$ and $B\rightarrow K(K^*)\nu\Bar{\nu}$ decays assumed that neutrinos are Dirac particles. As there were
already few papers studying $B\rightarrow K(K^*)\nu\Bar{\nu}$ decays with
Majorana neutrinos \cite{Li:2019fhz,Deppisch:2020oyx,Felkl:2021uxi},
I thought that it would be good to present a general analysis with Majorana neutrinos not only for the latter decays but also for 
$K\to\pi\nu\Bar{\nu}$. For this I needed a neutrino expert and I was lucky to
find one that wanted to collaborate with me. It was Julia Harz, who in 2022
was still in the Physics Department of TUM.

Our main goal was to investigate the possibility of disentangling different NP physics contributions to the rare meson decays $K\to\pi+\slashed{E}$ and $B\to K(K^*)+\slashed{E}$ through kinematic distributions in the missing energy $\slashed{E}$. We employed dimension-$6$ operators within the Weak Effective Field Theory (WET), identifying the invisible part of the final state as either active or sterile neutrinos. Special emphasis was given to lepton-number violating (LNV) operators with scalar and tensor currents. By February 2023 we showed analytically that contributions from vector, scalar, and tensor quark currents can be uniquely determined from experimental data of kinematic distributions. In addition, we presented new correlations of branching ratios for $K$ and $B$-decays involving scalar and tensor currents.

As already mentioned above we have generalized the expression
in (\ref{eq:epseta-BKsnn})-(\ref{eq:epseta-FL}) to the cases of lepton flavour
and lepton universality violations. These are the expressions (4.91)-(4.94) in
\cite{Buras:2024ewl}.

However, in March 2023 Julia moved to Mainz University where she
became  Associate Professor for Theoretical Physics. This slowed down
our project for at least six months. But this turned out to be fortunate
because in October 2023 an excellent Norwegian PhD student of Julia, Martin Mojahed, joined our collaboration which allowed to improve our analysis significantly and also to extend it to new invisible particles. The latter extension
was dominantly due to Martin.

\begin{figure}[th]
\centering
\includegraphics[width=0.75\textwidth]{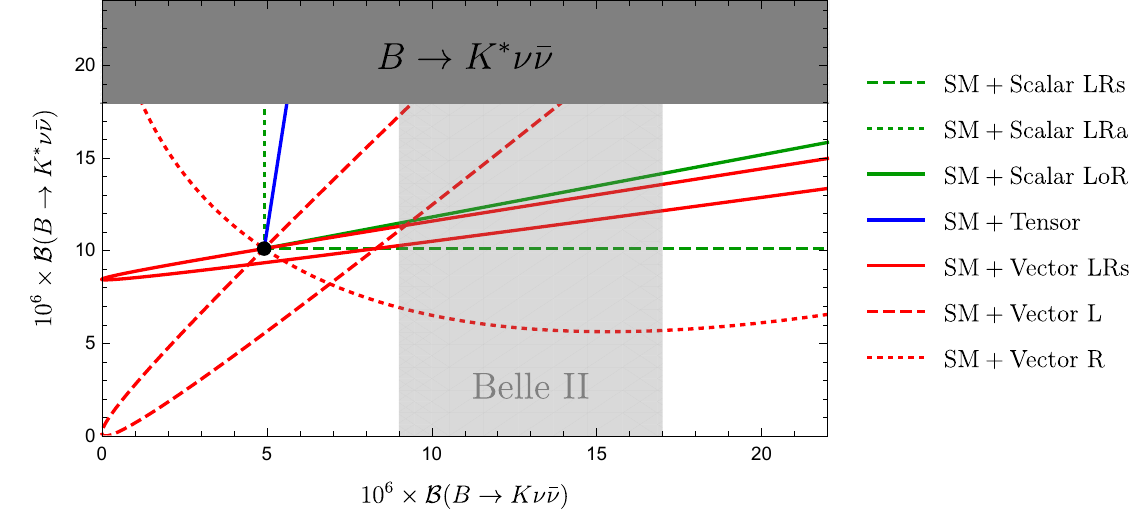}\\
\includegraphics[width=0.75\textwidth]{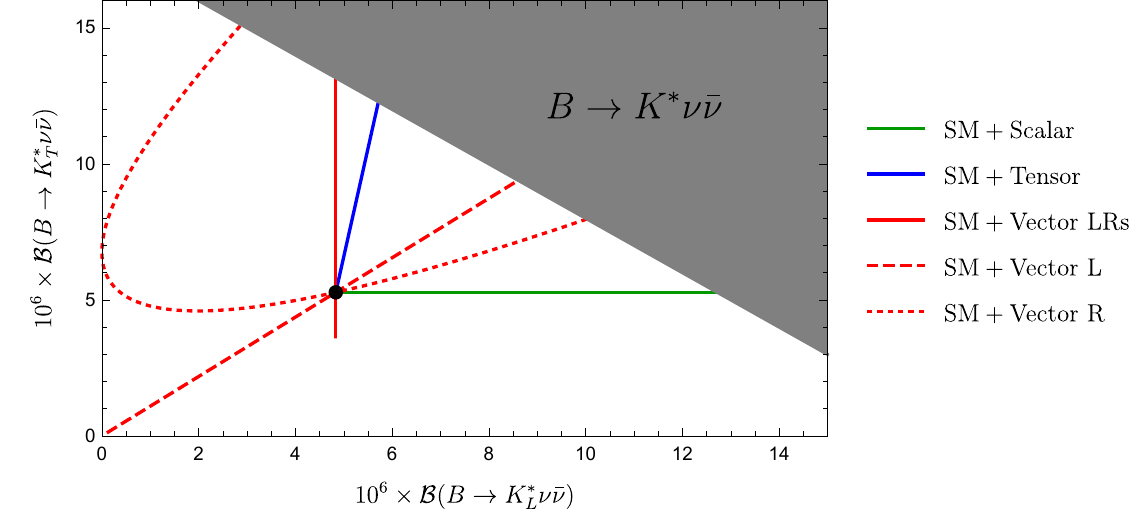}
\caption{The figure displays the $\mathcal{B}(B^+\rightarrow K^+\nu\widehat{\nu})-\mathcal{B}(B\rightarrow K^*\nu\widehat{\nu})$-plane (top) and $\mathcal{B}(B\rightarrow K^*_L\nu\widehat{\nu})-\mathcal{B}(B\rightarrow K^*_T\nu\widehat{\nu})$-plane (bottom) for different NP scenarios. The SM predictions are represented by black points. The light gray region in the upper plot indicates the 2023 experimental range quoted from Belle II~\cite{Belle-II:2023esi}, 
and the dark gray regions are excluded by the experimental limit on $B\rightarrow K^*\widehat{\nu}\nu$. The green lines show NP scenarios with scalar currents, the blue lines NP scenarios with tensor currents, and the red curves NP scenarios with vector currents (see \cite{Buras:2024ewl} for details).}
\label{fig:BPlanes}
\end{figure}

Indeed, as new invisible particles could \textit{a priori} appear in the final state, we included dark-sector operators leading to two dark scalars,
fermions, or vectors in the final state. In this context, we presented new calculations of the inclusive decay rate $B\to X_s+\slashed{E}$ for dark operators.  We showed that careful measurements of kinematic distributions make it theoretically possible to disentangle the contribution from WET operators from most of the dark-sector operators, even when multiple operators are contributing. We revisited sum rules for vector currents in WET and showed that the latter are also satisfied in some new dark-physics scenarios that could mimic WET. Finally, we pointed out that an excess in rare meson decays consistent with a LNV  hypothesis would point towards highly flavor non-democratic physics in the UV, and could put high-scale leptogenesis under tension. There is a very rich literature on
the topics considered in our paper. I just refer to \cite{Fridell:2023rtr,Kim:2025zaf} where many useful references can be found.

Our paper appeared finally in May 2024 \cite{Buras:2024ewl}. I like it
very much not only because in my view it is a useful paper but also
because I learned a lot participating in it. The most interesting plots are
shown in Figs.~\ref{fig:KaonVectorScalar} and \ref{fig:BPlanes}.
But in addition there are numerous expressions and tables that could
play a significant role in phenomenology when experimental data improves.

\subsection{Towards the Identification of NP through Quark Flavour Violating Processes: DNA Charts of NP (2013)}\label{DNAS}
\subsubsection{Preface}
Another strategy for searching for NP has been developed in a great collaboration with Jennifer Girrbach-Noe \cite{Buras:2013ooa}. In principle I could
 begin with this paper in this strategy-section. But seeing first other strategies this strategy can be explained better and appreciated even more than
otherwise. In my view it is the fastest route to select the
candidates for NP models before doing global fits.
The main idea 
is to study first the patterns of anomalies observed in 
the data and compare them  with the patterns of deviation from  SM
predictions in a given NP scenario. Such patterns, that expose suppressions and enhancements of various observables relative to SM predictions, can be considered as DNAs of the animalcula hunted by us.
In particular the
correlations between various enhancements and suppressions can rule out
some NP scenarios already before any global fit is performed. This
strategy proposed by us  in
2013 \cite{Buras:2013ooa} has been documented in several subsequent
papers, in particular in Chapter 19.4 of my book \cite{Buras:2020xsm} and recently in \cite{Buras:2024mhy} with many colourful plots. Still I think I should
describe it here in some detail.

As reviewed in \cite{Buras:2010wr,Buras:2012ts} extensive studies of many 
models  allowed to construct various classifications of NP contributions 
in the form of ``DNA'' tables \cite{Altmannshofer:2009ne} and {\it flavour codes}  \cite{Buras:2010wr}. The  ``DNA'' tables in 
\cite{Altmannshofer:2009ne} 
had as a goal to indicate whether in a given theory a value of a 
given observable can differ by a large, moderate or only tiny amount 
from the prediction of the SM. We will see them in Section~\ref{MSSM}.
The {\it flavour codes}  \cite{Buras:2010wr} were more a description of a given model in terms of the presence or absence of left- or right-handed  currents in it and the presence or the absence of new CP violating  phases, flavour violating and/or flavour conserving.

\begin{figure}[!tb]
\centering
\includegraphics[width = 0.98\textwidth]{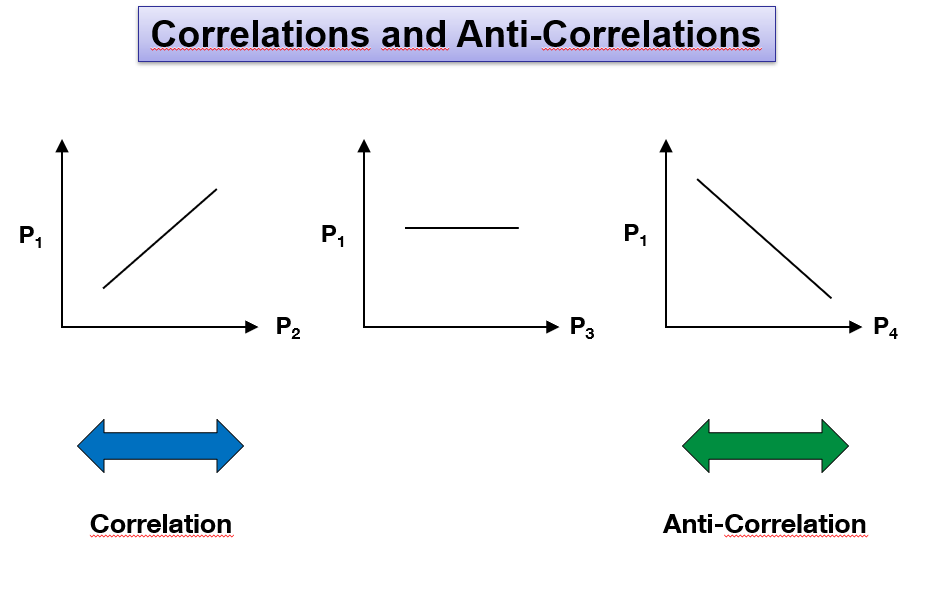}
\caption{\it Correlations, no correlations and anti-correlation. $P_i$ is
  probability representing in public colloquia the branching ratio}
 \label{Picorr}~\\[-2mm]\hrule
\end{figure}

Certainly in both cases there is a room for improvements. In particular in the 
case of the  ``DNA'' tables in  \cite{Altmannshofer:2009ne} we know now that 
in most quark  flavour observables considered there NP effects can be at most by  a factor of $2$ larger than the SM 
contributions. Exceptions are the cases in which 
some branching ratios or asymmetries vanish in the SM. But the particular 
weakness of this approach is the difficulty in depicting the correlations 
between various observables that could be characteristic for a given theory. 
Such correlations are much easier to show on a circle and to this end in
\cite{Buras:2013ooa} this new idea has been formulated in three steps and illustrated  with a few examples. Here comes the express version of this DNA-Strategy.
\subsubsection{DNA-Strategy}

{\bf Step 1}

We construct a chart showing different observables, typically 
a branching ratio for a given decay or an asymmetry, like CP-asymmetries 
$S_{\psi K_S}$ and $S_{\psi\phi}$ and quantities $\Delta M_s$, $\Delta M_d$, 
$\varepsilon_K$ and $\epe$. The important point is to select 
the optimal set of observables which are simple enough so that definite 
predictions in a given theory can be made and can be measured in the
coming decades. Following \cite{Buras:2013ooa} one could choose the following observables.

\be\label{Observables1} 
\varepsilon_K, \quad \Delta M_{s,d}, \quad S_{\psi K_S}, \quad S_{\psi \phi},
\ee

\be\label{Observables2} 
 \kpn, \quad \klpn, \quad K_S\to\mu^+\mu^-,\quad K_L\to\mu^+\mu^-,\quad \epe,
 \ee

\be\label{Observables3} 
 B_{s,d}\to\mu^+\mu^-, \qquad B\to X_{s}\nu\bar\nu, \quad B\to K^*(K)\nu\bar\nu,
\ee

\be \label{Observables4} 
B\to X_s\gamma, \quad B^+\to \tau^+\nu_\tau~, \quad B\to K^*(K)\mu^+\mu^-.
\ee
I assume here that the issues related to non-perturbative QCD uncertainties
in $\epe$ and $ B\to K^*(K)\mu^+\mu^-$ will be resolved one day.

The remaining observables not listed here but discussed in \cite{Buras:2013ooa}
will then serve as constraints on the model and if measured could also be chosen.

{\bf Step 2}

In a given theory we calculate the selected observables and investigate 
whether a given observable is {\em enhanced or suppressed} relative to the SM 
prediction or is 
basically unchanged. What this means requires a measure, like two 
$\sigma$. In the case of asymmetries one can proceed in the same manner if 
its sign remains unchanged relative to the one in the SM but otherwise one can define the change of its 
sign from $+$ to $-$ as a suppression and the change from  $-$ to $+$ as an
enhancement. For these three situations one can use the following colour 
coding:
\be
{\rm  \colorbox{yellow}{enhancement}}~=~{\rm yellow}, \qquad {\rm \framebox{no~change}}~=~{\rm white},  \qquad {\rm 
\colorbox{black}{\textcolor{white}{\bf suppression}}}~=~{\rm black}.
\ee
To this end the predictions within the SM have to be known precisely.

{\bf Step 3}

It is only seldom that a given observable in  a given theory is uniquely 
suppressed or enhanced but frequently two observables are correlated or
uncorrelated with each other. That is the enhancement of one observable implies uniquely
an enhancement (correlation) or suppression (anti-correlation) of another 
observable. It can also happen that no change in the value of a given 
observable implies no change in another observable. This is illustrated in
Fig.~\ref{Picorr}.

 The idea then is to connect in our DNA-chart 
a given pair of observables that are correlated with each other
by a line. Absence of a line means that two given observables are 
uncorrelated. In order to distinguish the correlation from anti-correlation 
one can use the following colour coding for the lines in question:
\be
{\rm correlation}~\textcolor{blue}{\Leftrightarrow}~{\rm blue} , \qquad {\rm anti-correlation}~\textcolor{green}{\Leftrightarrow}~{\rm
green}.
\ee

Let us consider a number of prominent rare decays and divide them 
into two classes:

{\bf Class A:} Decays that are governed by vector ($V=\gamma_\mu$) quark couplings. These are for instance
\be
\kpn, \qquad \klpn, \qquad B\to K\nu\bar\nu, \qquad B\to K \mu^+\mu^-.
\ee
In this case the change from left-handed to right-handed quark couplings does not 
introduce any change of the sign of NP contribution relatively to the SM one.

{\bf Class B:} Decays that are governed by axial-vector ($A=\gamma_\mu\gamma_5$) quark couplings. These are for instance
\be
K_L\to\mu^+\mu^-, \qquad B\to K^*\nu\bar\nu, \qquad B_{s,d}\to \mu^+\mu^-, \qquad B_d\to K^* \mu^+\mu^-.
\ee

In this case the change from left-handed to right-handed couplings implies
 the sign flip of NP contribution relatively to the SM one. Strictly speaking 
in the case of  $B\to K^*\nu\bar\nu$ and $B_d\to K^* \mu^+\mu^-$ this rule only  applies if the contributions from the longitudinal and parallel transversity components dominate. For perpendicular component there is no sign flip.

Thus if there is a correlation between two observables belonging to class A 
and B in the presence of left-handed couplings, it is changed into anti-correlation when right-handed couplings are at work. This difference allows then 
to probe whether one deals with left-handed or right-handed couplings. Of 
course if both left-handed and right-handed couplings are involved the 
structure of correlations is modified, but still studying it one can in 
principle extract the relative size of these couplings from the data. Moreover, 
if there is a correlation or anti-correlation of two observables belonging to 
one class, the flip of sign of $\gamma_5$ will not have an impact on these 
relations, but can of course have an impact on whether a given observable is 
suppressed or enhanced relative to the SM prediction.

A graphical representation of these properties are the DNA charts \cite{Buras:2013ooa} which we will briefly discuss now.
Let me then end this description by four charts presented in
\cite{Buras:2013ooa} where further details can be found.

\begin{figure}[!tb]
\centering
\includegraphics[width = 0.49\textwidth]{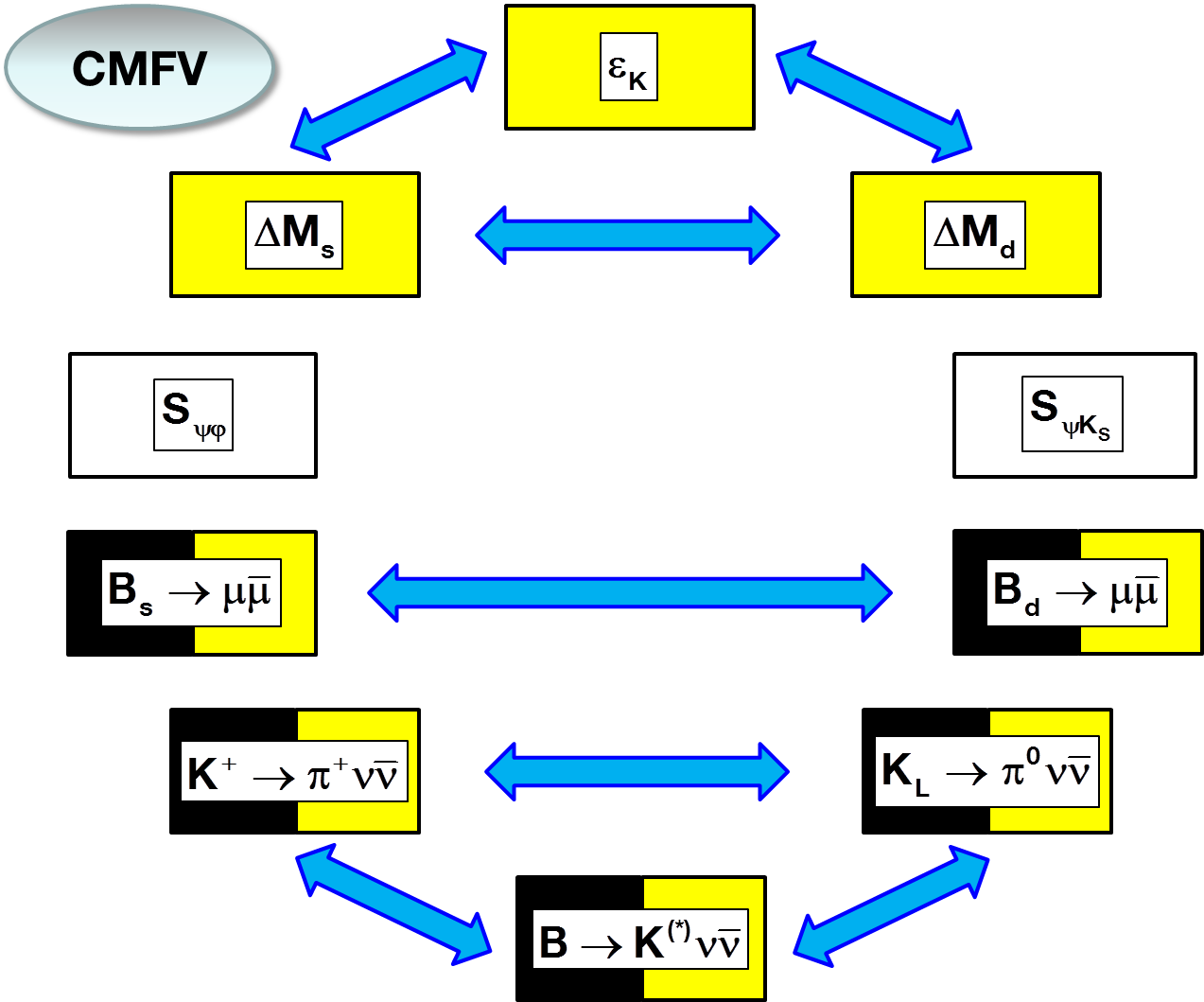}
\includegraphics[width = 0.49\textwidth]{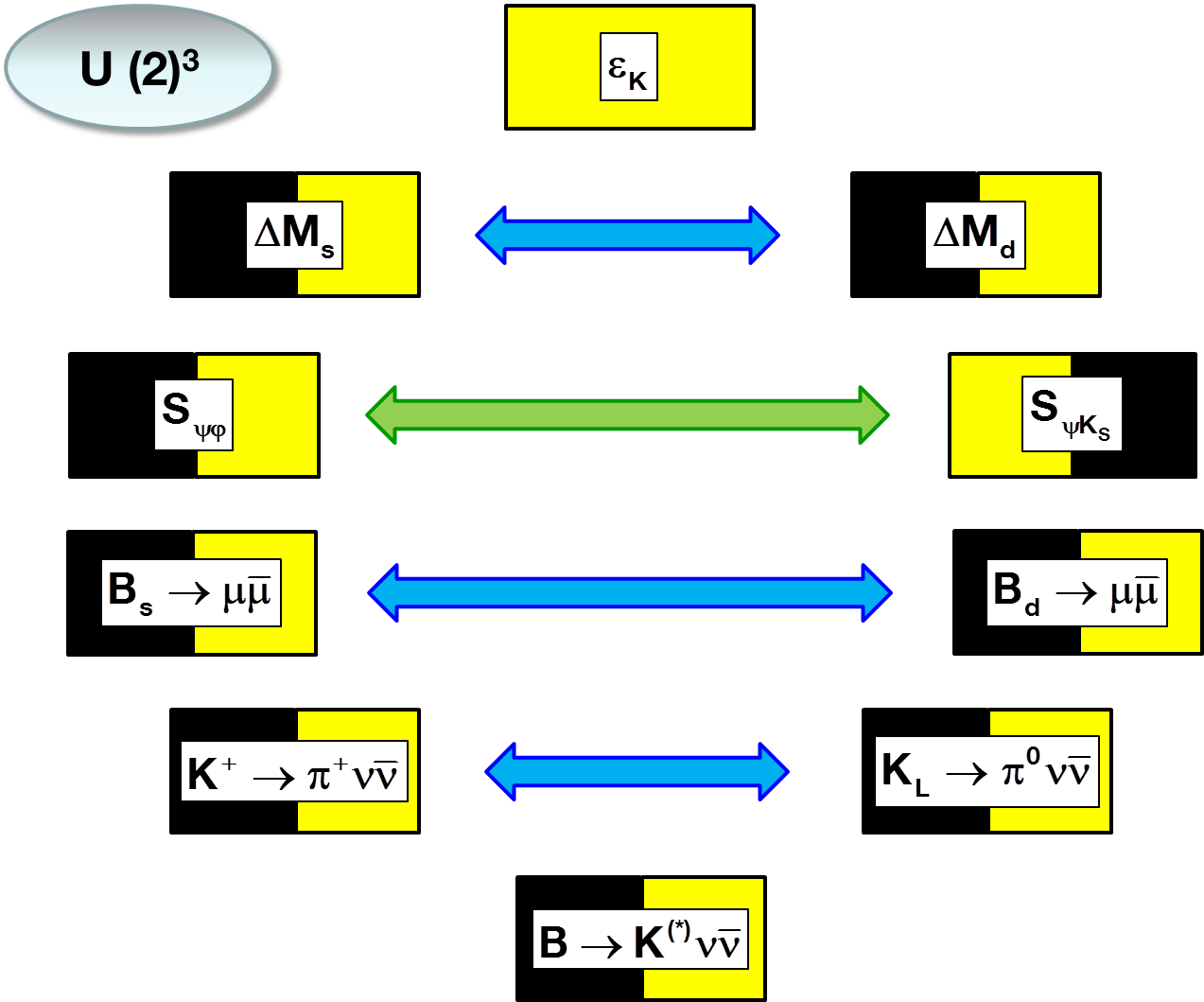}
\caption{\it DNA-chart of MFV  models (left) and of  $\text{U(2)}^3$ models (right). Yellow means   \colorbox{yellow}{enhancement}, black means
\colorbox{black}{\textcolor{white}{\bf suppression}} and white means \protect\framebox{no change}. Blue arrows
\textcolor{blue}{$\Leftrightarrow$}
indicate correlation and green arrows \textcolor{green}{$\Leftrightarrow$} indicate anti-correlation. From \cite{Buras:2013ooa}.}
 \label{fig:CMFVchart}~\\[-2mm]\hrule
\end{figure}

\begin{figure}[!tb]
\centering
\includegraphics[width = 0.49\textwidth]{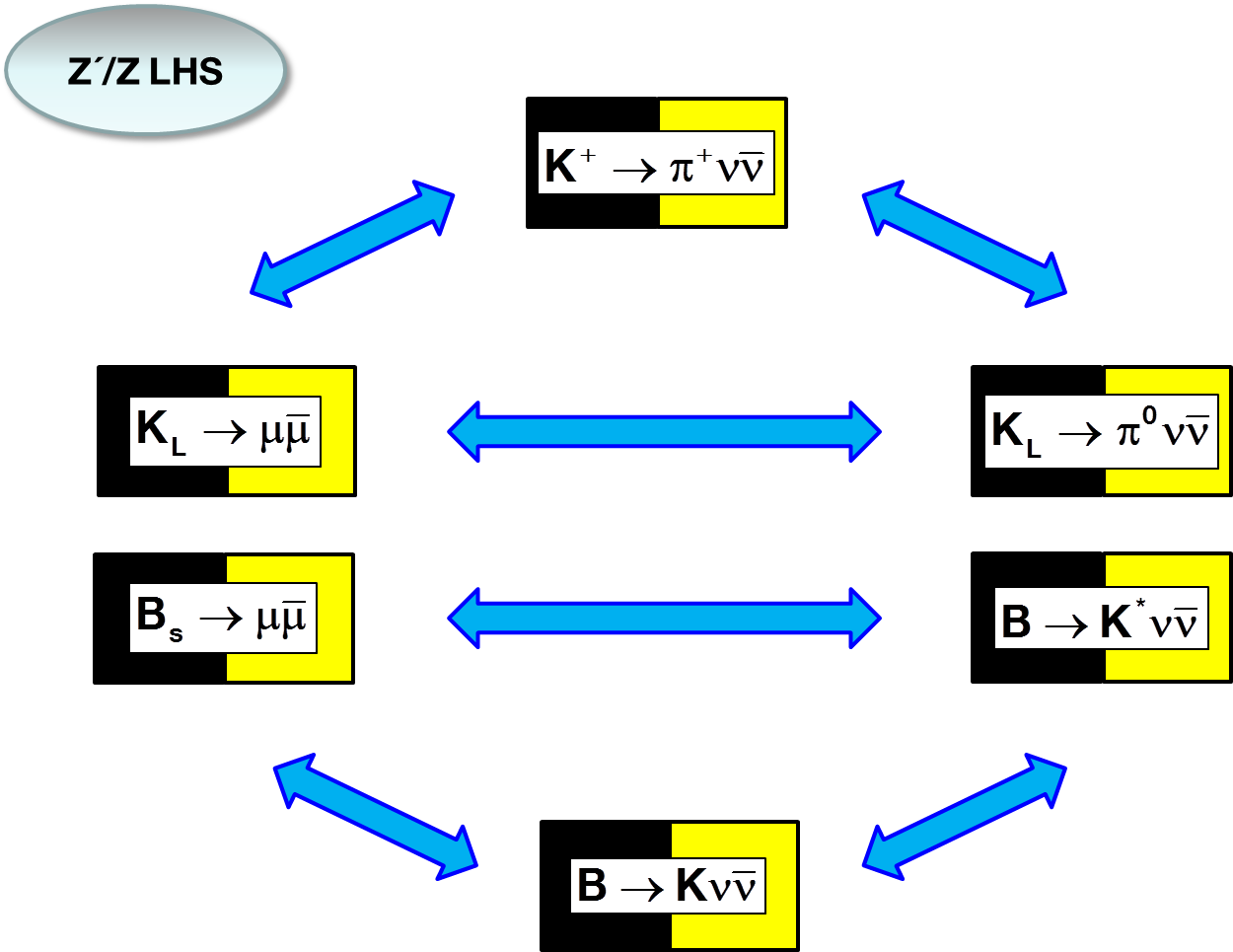}
\includegraphics[width = 0.49\textwidth]{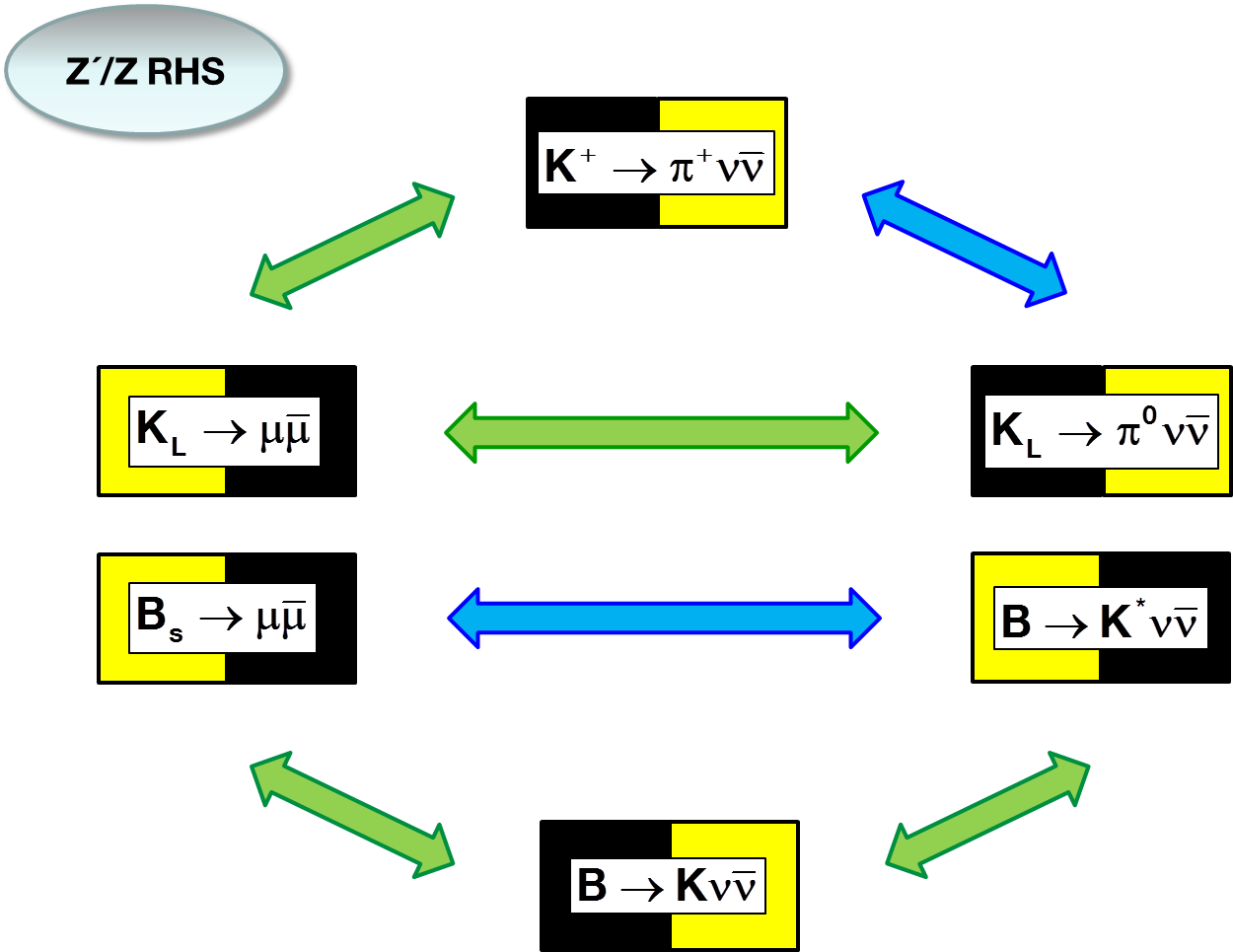}
\caption{\it DNA-charts of $Z^\prime$ models with LH and RH currents.  Yellow means   \colorbox{yellow}{enhancement}, black means
\colorbox{black}{\textcolor{white}{\bf suppression}} and white means \protect\framebox{no change}. Blue arrows
\textcolor{blue}{$\Leftrightarrow$}
indicate correlation and green arrows \textcolor{green}{$\Leftrightarrow$} indicate anti-correlation.  From \cite{Buras:2013ooa}.}
 \label{fig:ZPrimechart}~\\[-2mm]\hrule
\end{figure}

One can check that these charts summarize compactly the 
(anti-) correlations between processes of class A and B that we discussed before and also the correlations and anti-correlations within each class. In particular, the change of a correlation into an anti-correlation between two observables belonging to two different classes, when left-handed couplings 
are changed to right-handed ones, are clearly visible in these charts. We observe the following features:
\begin{itemize}
\item
Comparing   the DNA-charts of CMFV and $\text{U(2)}^3$ models 
in  Fig.~\ref{fig:CMFVchart}  we observe that the correlations 
between $K$ and $B_{s,d}$ systems are broken in the $\text{U(2)}^3$ case as the flavour symmetry is reduced from 
$\text{U(3)}^3$ down to $\text{U(2)}^3$. 
\item
As the decays $\kpn$, $\klpn$ and $B\to K\nu\bar\nu$  belonging to class A are only sensitive
to the vector quark currents, they do not change when the couplings are changed from  left-handed to right-handed ones. On the other hand, the remaining 
three decays in   Fig.~\ref{fig:ZPrimechart} belonging to class B are sensitive to axial-vector 
couplings implying interchange of enhancements and suppressions when going from 
$L$ to $R$ and also change of correlations to anti-correlations between the 
latter three and the former three decays. Note that the correlation between 
$B_s\to\mu^+\mu^-$  and $B\to K^*\mu^+\mu^-$ does not change as both decays are  sensitive only to axial-vector coupling if in the latter case the contribution 
from the longitudinal and parallel transversity components dominate.
\item
However, it should be remarked that in order to obtain the correlations or 
anti-correlations in LHS and RHS scenarios it was assumed in the DNA charts 
presented here that the signs 
of the left-handed couplings to neutrinos and the axial-vector couplings 
to muons are the same which does not have to be the case. If they are 
opposite the correlations between the decays with neutrinos and muons in 
the final state change to anti-correlations and vice versa. 
\item
On the other hand, due to $\text{SU(2)}_L$ symmetry the left-handed $Z^\prime$
 couplings to muons and neutrinos are equal and this implies the relation
\be\label{SU2}
\Delta_{L}^{\nu\bar\nu}(Z')=\frac{\Delta_V^{\mu\bar\mu}(Z')-\Delta_A^{\mu\bar\mu}(Z')}{2}. 
\ee
Therefore, once two of these couplings are determined, the third follows uniquely without the freedom mentioned in the previous item.
\item
In the context of the DNA-charts in  Fig.~\ref{fig:ZPrimechart}, the correlations involving $\klpn$ apply only if NP contributions carry some CP-phases. If this is not the case the branching ratio for $\klpn$ will remain unchanged relative to the SM one.
\end{itemize}

If in the case of tree-level $Z^\prime$ and $Z$ exchanges 
both LH and RH quark couplings are present and are equal to each 
other (LRS scenario) or differ by sign (ALRS scenario), then one finds 
\cite{Buras:2012jb}
\begin{itemize}
\item
In LRS NP contributions to $B_{s,d}\to\mu^+\mu^-$ vanish, but they are present 
in  $\klpn$, 
$\kpn$, $B_d\to K\mu^+\mu^-$ and $B\to K\nu\bar\nu$.
\item
In ALRS NP contributions to $B_{s,d}\to\mu^+\mu^-$ are non-vanishing. 
 On the other hand 
they are absent in the case of  $\klpn$, $\kpn$, $B_d\to K\mu^+\mu^-$ and 
 $B\to K\nu\bar\nu$.
\item
In  $B_d\to K^*\mu^+\mu^-$ and  $B\to K^*\nu\bar\nu$ this rule is more complicated as already stated above, but generally the LH and RH contributions interfere
destructively in LRS and constructively in ALRS. The details depend on form factors.
\end{itemize}

I read this paper after 13 years again and was very satisfied with it. In addition to the strategy just presented, it reviews 11 models for which such DNA-Charts could be constructed and provides useful information for flavour phenomenology.

\subsection{Flavour Expedition to the Zeptouniverse (2015)}
In October 2014 I gave two lectures at the school ``Flavour Ways to New Physics'' organized by the Karlsruhe University. I have written them up first in May
2015 so that I could include many of the findings we made in the years before
that I described above \cite{Buras:2015nta}.

In particular I emphasized again the power of correlations between flavour observables in the search for NP and  identified
a number of correlations that could allow to discover NP even if it would appear at the level of $20\%$ of the SM contributions.  The correlation between $\mathcal{B}(\kpn)$, $\overline{\mathcal{B}}(B_s\to\mu^+\mu^-)$  and $\gamma$ extracted from tree-level decays within the SM is one of them. It is shown in
(\ref{master1a}) and its updated version will be given in Part~\ref{BV}.

\begin{figure}[!tb]
  \includegraphics[width=0.49\textwidth]{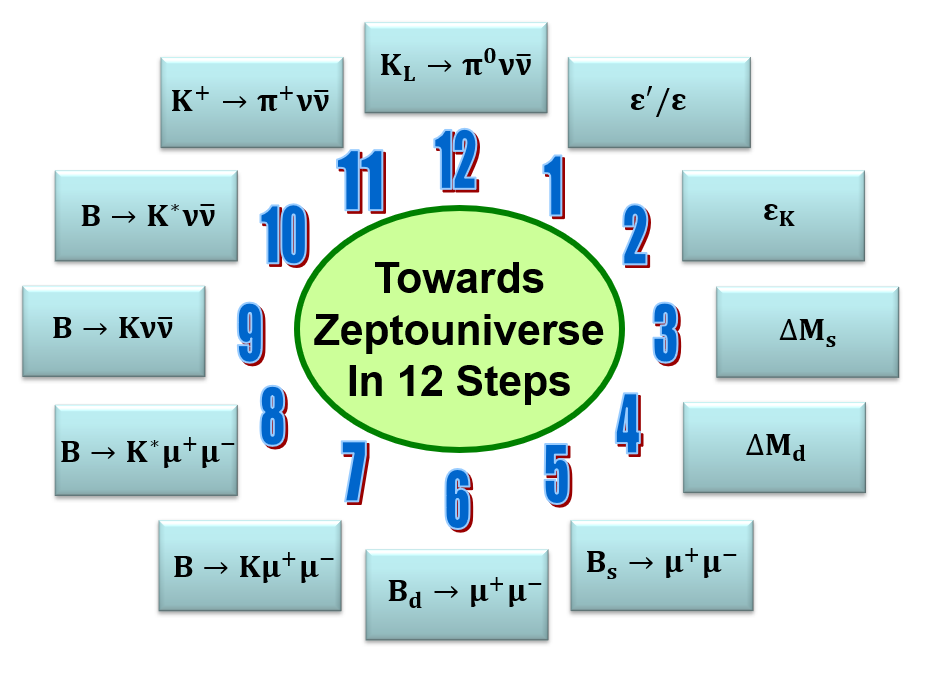}
  \includegraphics[width=0.49\textwidth]{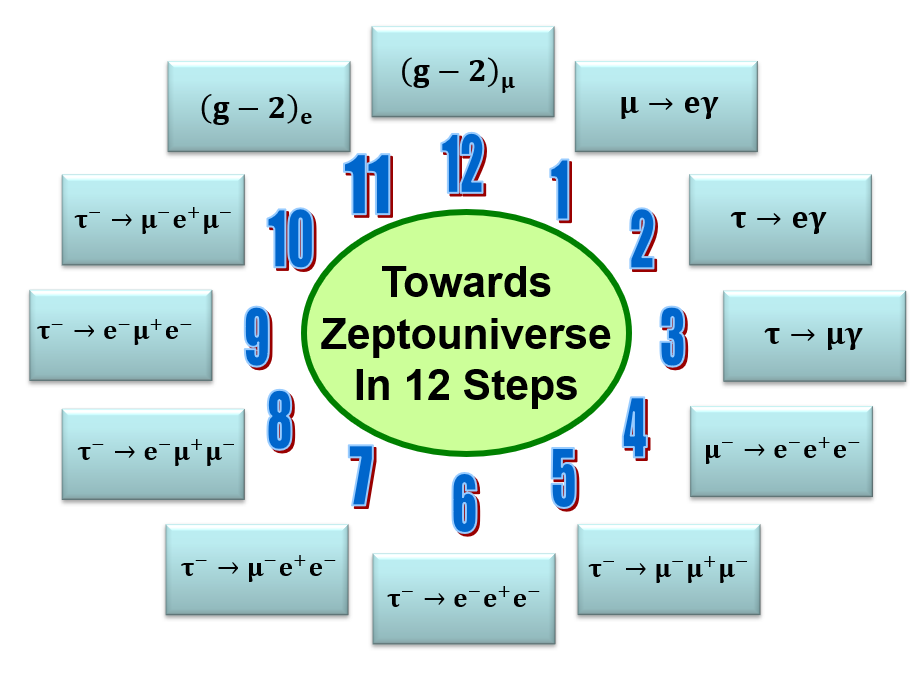}
  \includegraphics[width=0.49\textwidth]{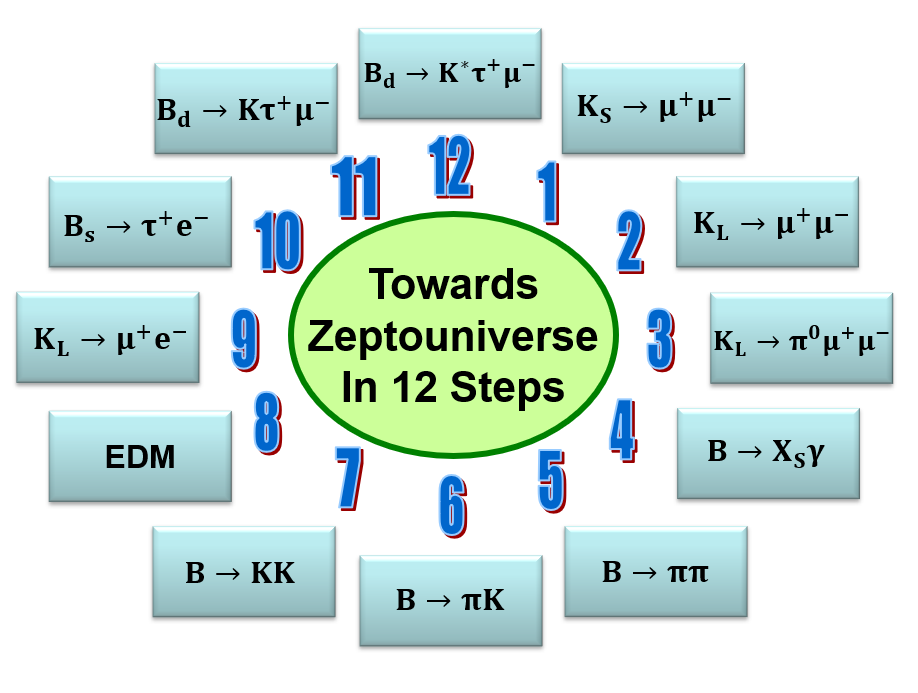}
  \includegraphics[width=0.49\textwidth]{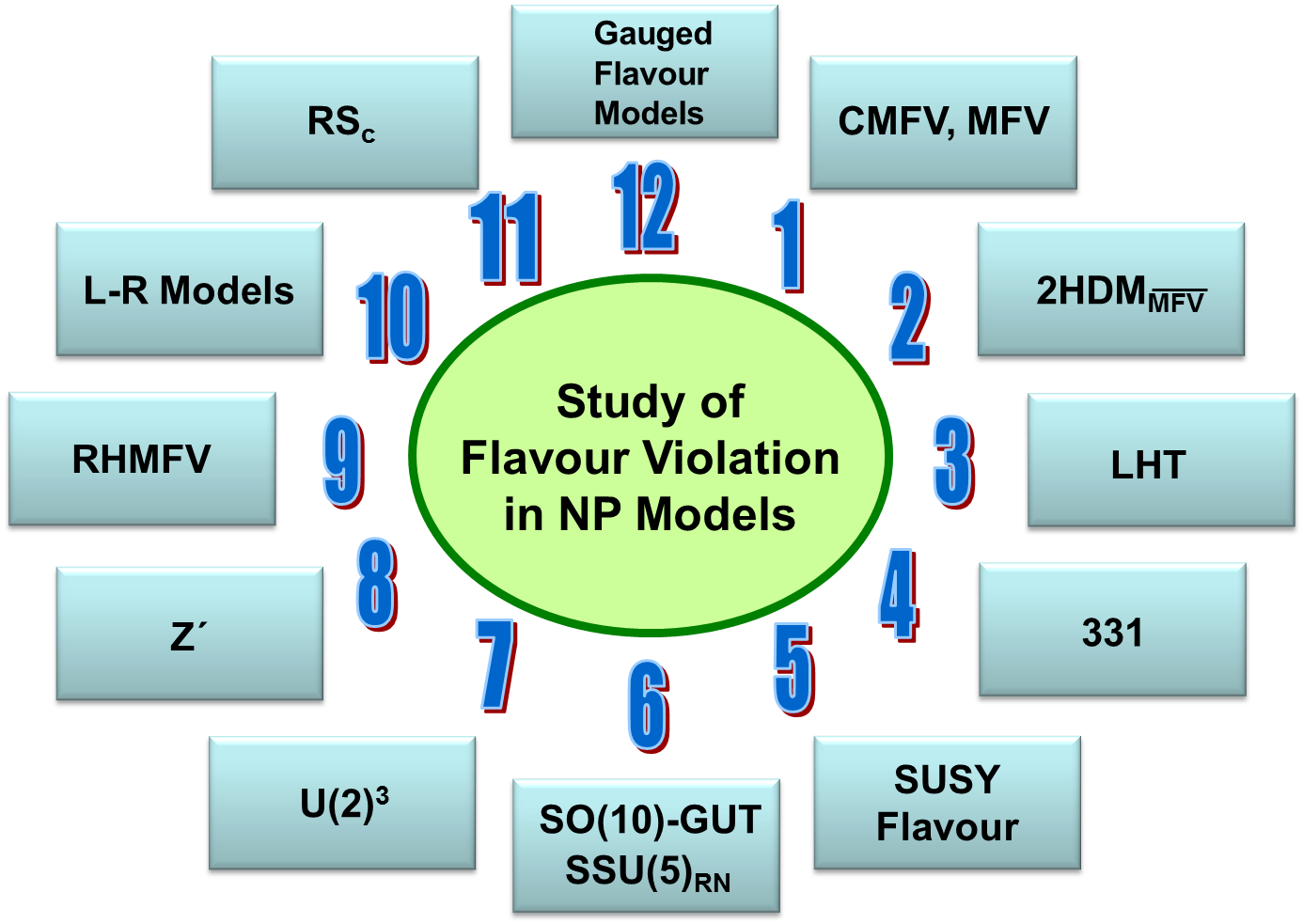}
\caption{\it Towards the Zeptouniverse in 12 Steps.}\label{Fig:1}~\\[-2mm]\hrule
\end{figure}

After presenting the simplest correlations in CMFV and $\text{U(2)}^3$ models 
I addressed the existing data on $B_{s,d}\to\mu^+\mu^-$ and the 
anomalies in $B_d\to K(K^*)\mu^+\mu^-$ in the context of $Z^\prime$-models  
 with quark flavour violating neutral couplings. A brief discussion of leptoquark models was also given. I emphasized the correlations of  $B_d\to K(K^*)\mu^+\mu^-$ with $B_d\to K(K^*)\nu\bar\nu$ 
that allows to distinguish between various explanations of the 
anomalies in question. Finally, I summarized our studies of $K\to\pi\nu\bar\nu$ and $B_{d,s}\to\mu^+\mu^-$ decays which demonstrates that these decays play important roles in finding out what happens in the Zeptouniverse. The clocks
presented in Figs. 1 and 4 of these lectures and in several of my talks are called by some
``Buras-Clocks''. Here I present more clocks of that type.
In Fig.~\ref{Fig:1}  various observables of interest are
exposed around the first three clocks. In the last clock several NP models (theories) 
are list for which the first three clocks can be one day constructed.

\begin{figure}[!tb]
  \centerline{\includegraphics[width=0.98\textwidth]{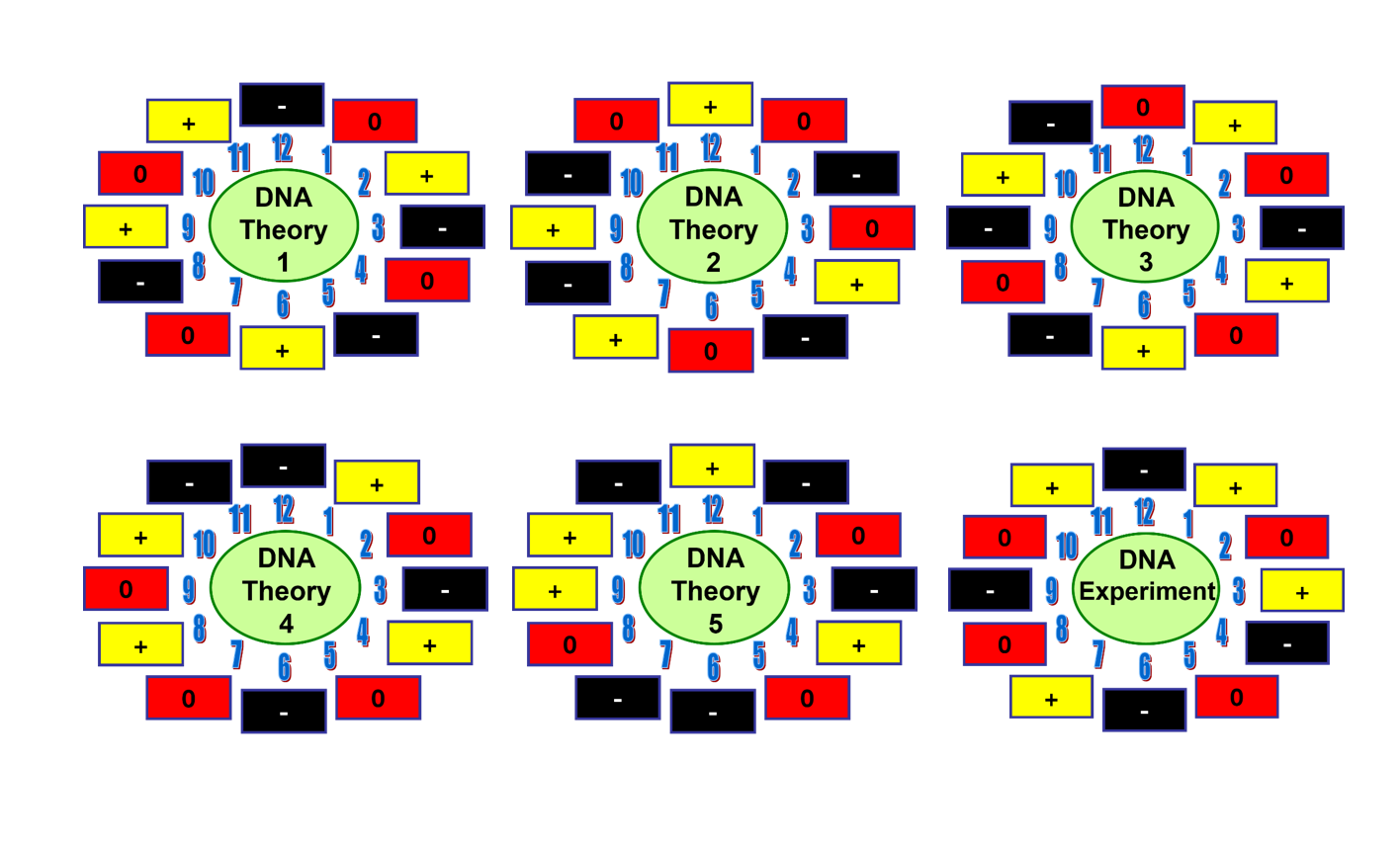}}
  \vspace{-1.0cm}
\caption{\it DNA Tests of several theories with the last one being experimental DNA.}
\label{Fig:2}~\\[-2mm]\hrule
\end{figure}

This is ilustrated in  Fig.~\ref{Fig:2} which shows DNA tests of different
theories with enhancements, suppressions and no changes for specific observables
in a given clock of Fig.~\ref{Fig:1}. The last clock in Fig.~\ref{Fig:2}
shows the experimental result for the chosen observables. As one can see
all five theories are ruled out.

Of course because of free parameters in a given theory, it is seldom that
the signs in these clocks are unique. But as in the special NP scenarios
presented in Figs.~\ref{fig:CMFVchart} and \ref{fig:ZPrimechart} the signs
in question could be correlated. In the future when many observables
will be measured, the clocks presented here together with the DNA charts
could help to present results in a given theory in an artistic manner.

I think it was a good summary of our strategies for finding NP but somehow
it did not catch the attention of the community until in May 2025 I have received
the following Email from Harry Cliff from the  LHCb collaboration:

``Dear Andrzej,

I’ve been commissioned by New Scientist magazine to write a feature article on accessing the Zeptouniverse, inspired by the article you wrote on exploring the
Zeptouniverse using flavour in 2015. Would you be willing to be interviewed for the piece? I would really appreciate your perspective on where we stand on this goal ten years on, particularly given the changed picture in terms of the flavour anomalies, and new results etc.''

I didn't know him personally but I checked that in addition to being involved in the LHCb experiment he is vigorously popularizing high energy particle physics so
that I was not surprised that New Scientist contacted him. We had a very nice
zoom discussion in which I told him about my 2021/2022 strategies with Elena Venturini \cite{Buras:2021nns,Buras:2022wpw} and their summary in my 2024 paper \cite{Buras:2024per}. I will present them in Part~\ref{BV}.

Harry wrote a very nice article in New Scientist which can be found in its
26th July 2025 edition.
There also my 2024 article for CERN Courier(July/August) is mentioned.

But this is not the end of this story. A Tunisian student, Mohamed Zied Jaber, whom I mentored in 2023 and whom I told  few things about the Zeptouniverse  during his bachelor studies, contacted me to inform me that he read Harry's article and that there are two french articles in GEO.fr and SLATE.fr describing roughly what Harry wrote in New Scientist. Zeptouniverse
becomes popular. What follows could possibly make it even more popular.

\section{Dual Picture of Short Distance Scales}
Our colleagues in Astrophysics and Cosmology can easily impress the community
by showing impressive pictures of the Universe at large. One can look
at them with naked eyes or powerful telescopes. This is much harder in particle physics as with
good eyes one can sense $10^{-3}$ m and using microscopes that are more powerful
can only give us a picture of nanoscience but not really of short distance scales explored in particle physics.

While preparing my talk at the EPS-2009 in Cracow I got the idea to improve
this situation by simply removing the sign from the exponential, that is
instead of using $10^{-x}$ m I proposed to use $10^{x}$ m. After all what matters
is the relative size to our hight which for adults is roughly $1.75\pm0.20$.

In fact in this manner one can really get impressed by the shortness of the scales
resolved in particle physics. In Figs.\ref{Micro} and \ref{Zepto} I show the dual pictures of
the following Universes:

\begin{figure}[t]
\centering%
\includegraphics[width=0.45\textwidth]{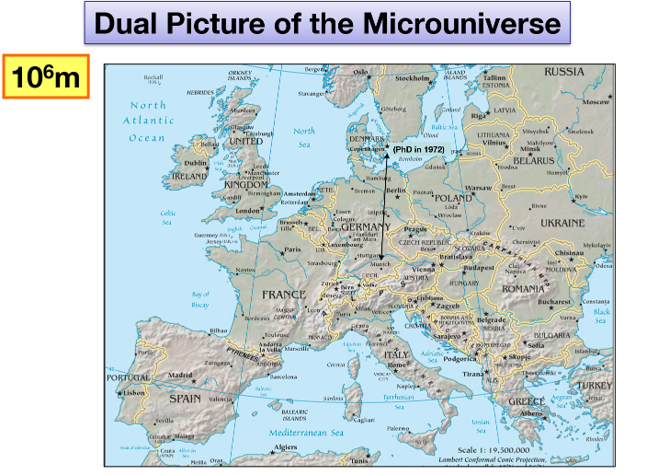}
\includegraphics[width=0.45\textwidth]{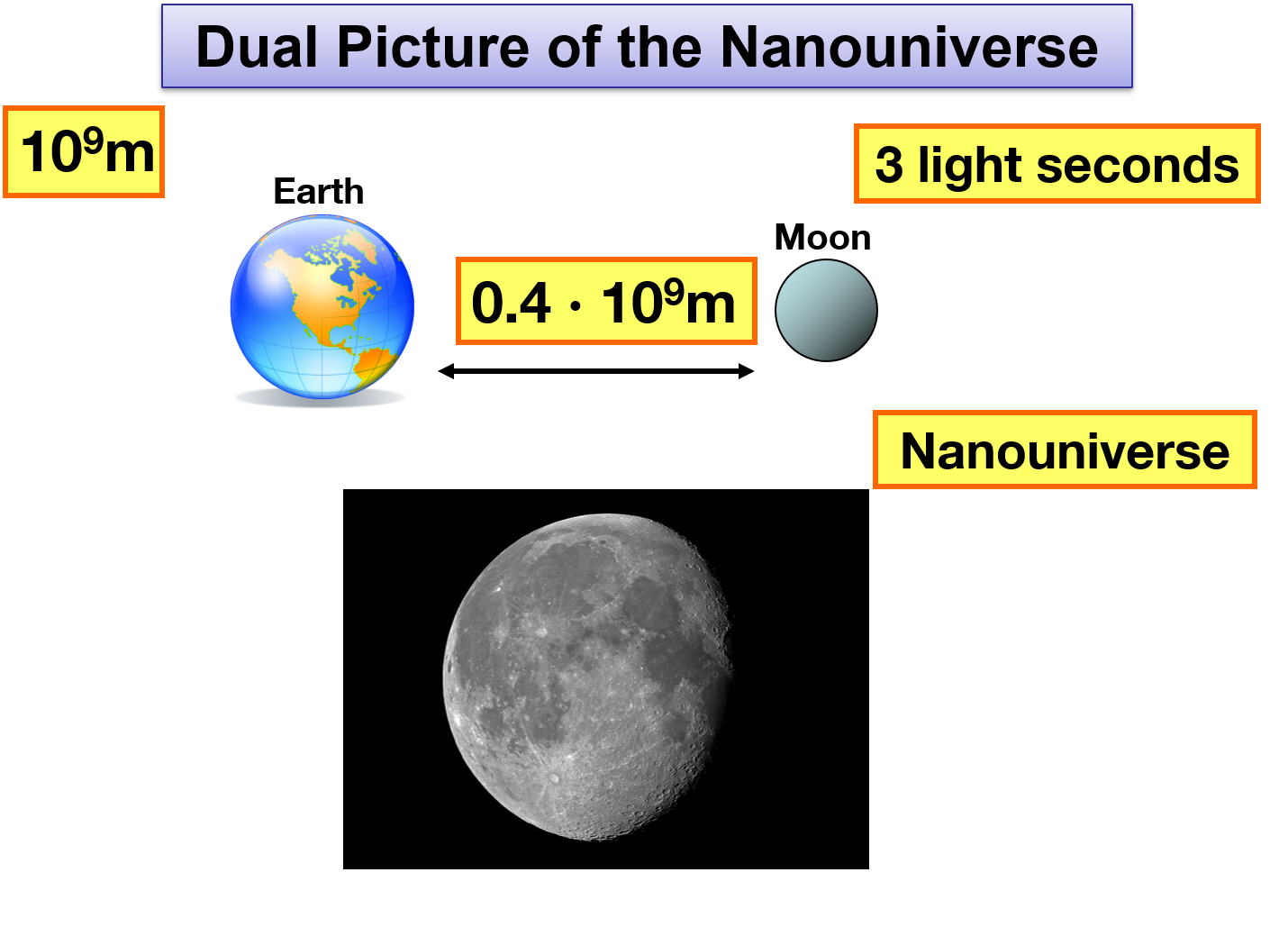}%
\caption{\it Dual Picture of the Microuniverse (Left) and of the Nanouniverse (right).\label{Micro}}
\end{figure}

\begin{figure}[t]
\centering%
\includegraphics[width=0.45\textwidth]{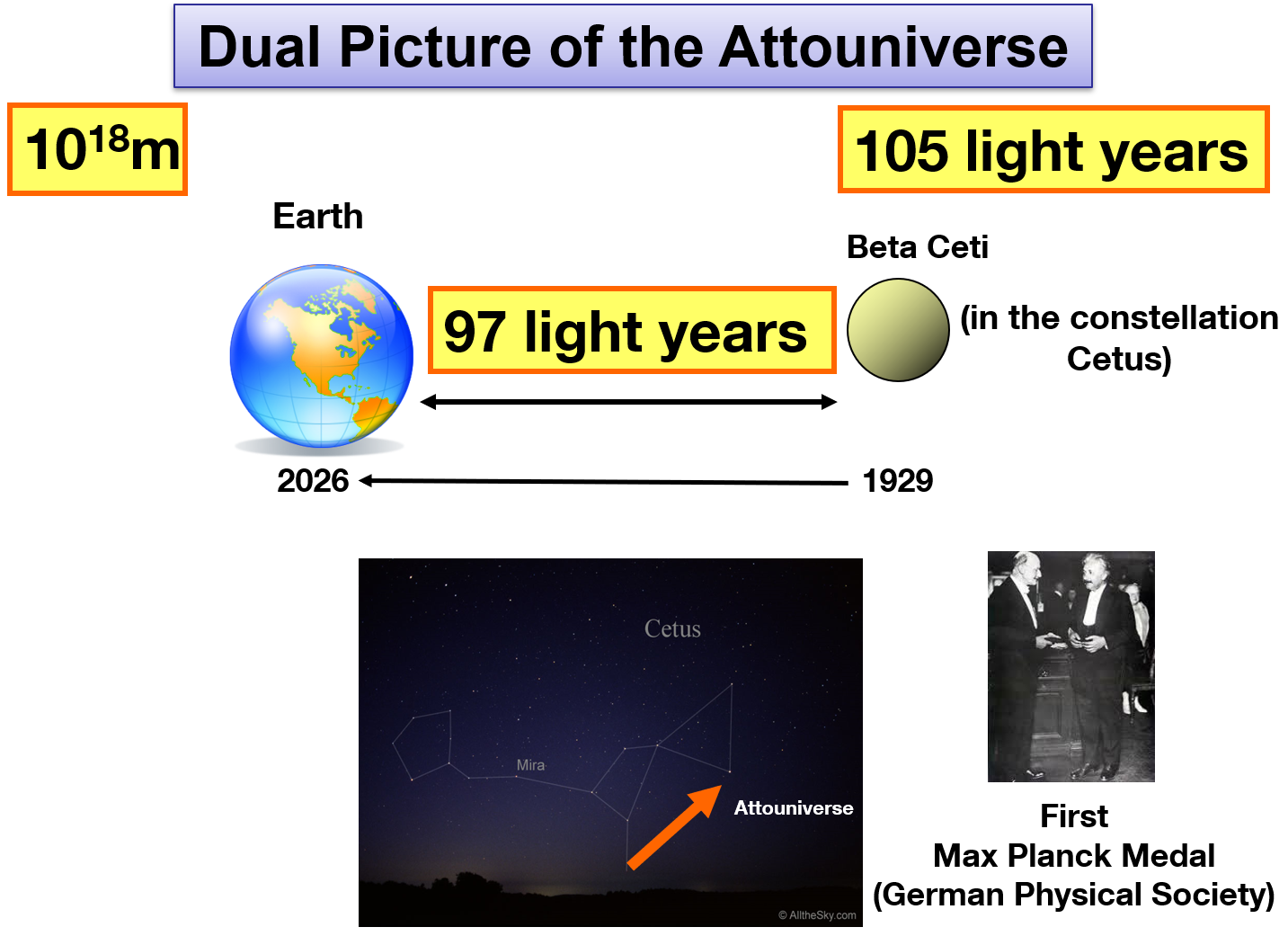}
\includegraphics[width=0.45\textwidth]{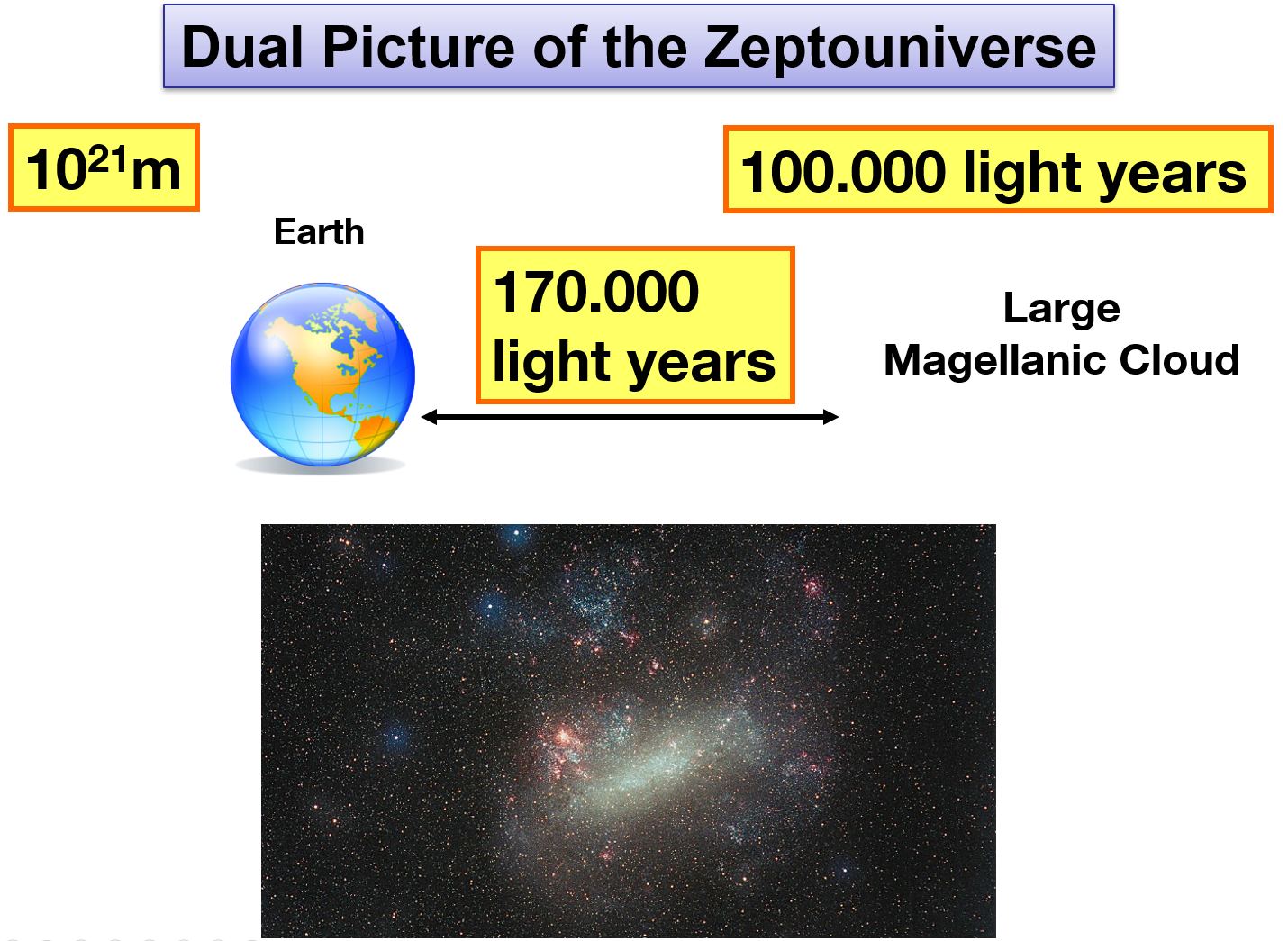}%
\caption{\it Dual Picture of the Attouniverse (left) and of the Zeptouniverse (right).\label{Zepto}}
\end{figure}

\begin{figure}[t]
\centering%
\includegraphics[width=0.45\textwidth]{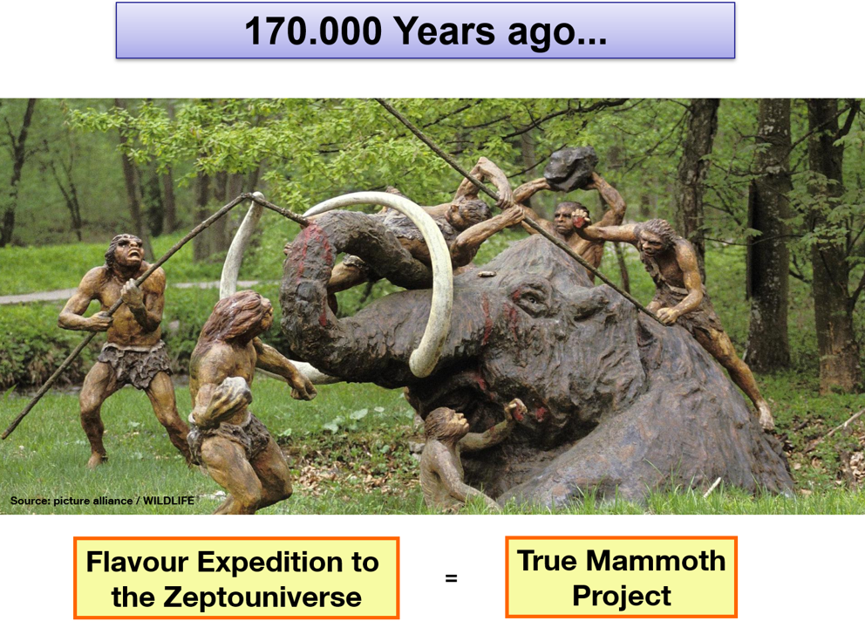}
\includegraphics[width=0.45\textwidth]{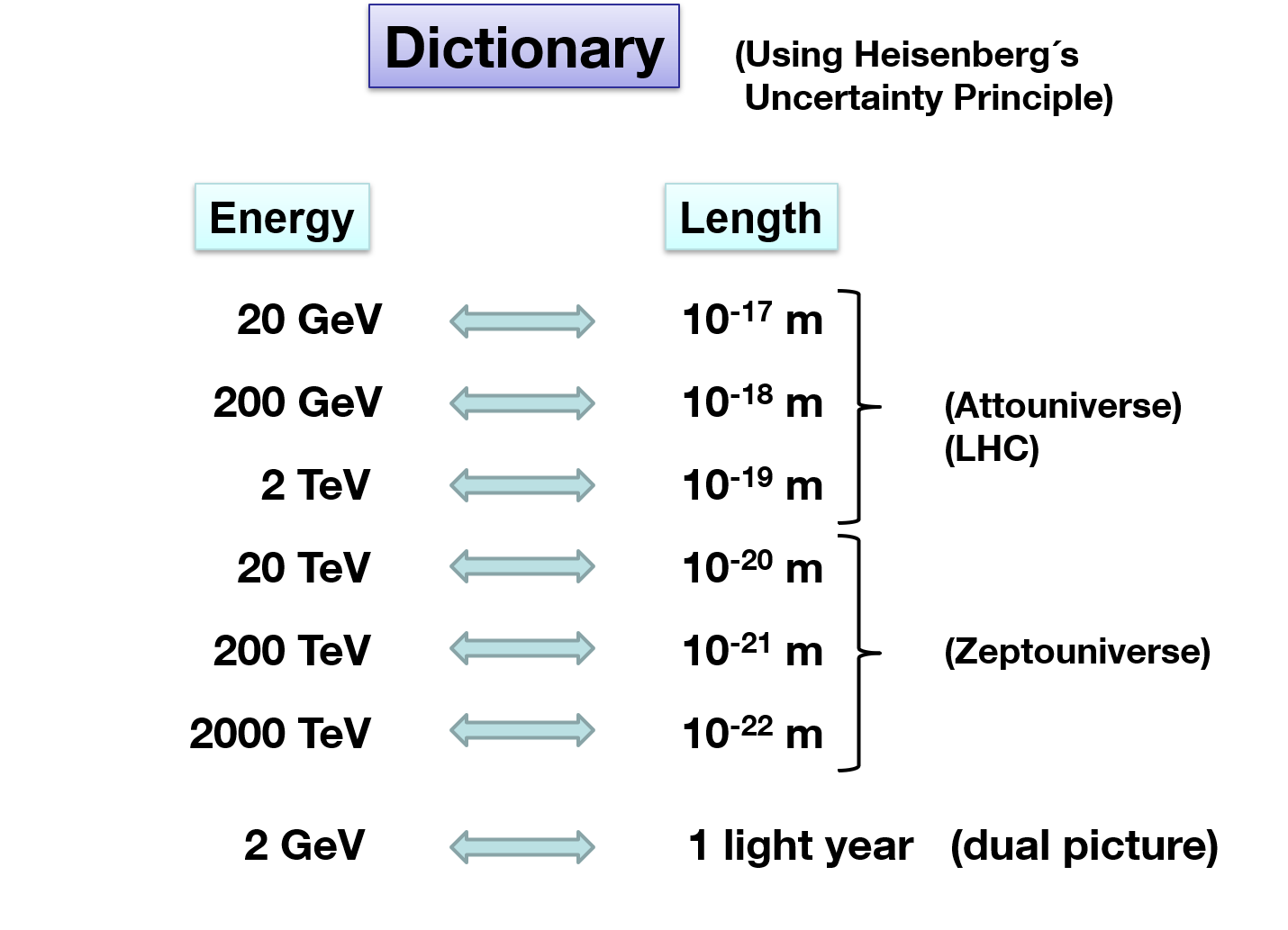}%
\caption{\it  Old Collaboration at Work (left) and the Dictionary (right).\label{Mamoth}}
\end{figure}

{\bf Dual Picture of the Microuniverse}

This corresponds to the distance from Munich to Copenhagen, where in 1972
I completed my PhD studies. See left picture in Fig.~\ref{Micro}.

{\bf Dual Picture of the Nanouniverse}

Looking at the moon one can get impressed by the distance scales explored
by the nanotechnologists. See right picture in Fig.~\ref{Micro}. Indeed, if we were as toll as the distance from
the earth to the moon, they would explore the distance scales of a few meters.
At first sight very impressive but cannot really compare with what particle
physicists can achieve. 

{\bf Dual Picture of the Attouniverse}

Indeed, already the Attouniverse is much more impressive than the Nanouniverse.
Let us assume that during the celebration of the first Max Planck Medal (1929)
given to Albert Einstein and Max Planck himself, a signal has been sent from
Beta Ceti (an object in the constellation Cetus, see Fig.~\ref{Zepto}) (left) has been sent to the earth. It will arrive only this year. 

{\bf Dual Picture of the Zeptouniverse}

We make next step to the Large Magellanic Cloud, see Fig.~\ref{Zepto} (right).
Let us then assume that
a signal from there will arrive on the earth this year. One can then ask what
was on the earth when this signal was sent to us. Certainly there was
no LHCb, CDF, ATLAS or Belle II collaboration but another one with different
goals as seen  on the left in Fig.~\ref{Mamoth}. The first guy on the left was probably the spokesman
of this collaboration. I doubt he expected to be cited by anybody in 2026.

As it is already evident  from the text until now and will be
even more evident as we continue, this picture fits very well to my story:
Flavour Expedition to the Zeptouniverse is indeed the true Mammoth Project.

The dictionary between energy scales explored by present and future colliders and distance scales used to construct
this dual picture of short distances is shown on the right in Fig.~\ref{Mamoth}.

\boldmath
\part{$\varepsilon_K$ (2008-2010, 2019)}\label{eKunitarity}
\unboldmath
\section{Preface}
Until October 2008 the formula for $\varepsilon_K$ in the SM used
in the literature was
\be
\eps_K = e^{i \phi_\eps} \sin \phi_\eps \left( \frac{{\rm Im}(M^K_{12})}{\Delta M_K} + \xi \right)~,\qquad \xi = \frac{{\rm Im} A_0}{{\rm Re} A_0}~,
\label{epsexact}
\ee
with $A_0$ the 0-isospin amplitude in $K \to \pi \pi$ decays, $M^K_{12} = \langle K^0 | \mc{H}^{\rm full}_{\D F = 2} | \bar K^0 \rangle$ and 
$\Delta M_K$ the $K^0 - \bar K^0$ mass difference. The phase $\phi_\eps$ is measured to be 
\be
\phi_\eps = (43.51 \pm 0.05)^\circ~.
\ee
In view of other uncertainties the term $\xi$ has been neglected in the literature.

In the fall of 2008, Diego Gaudagnoli and me decided to look at this term and
possible long-distance contributions to $\varepsilon_K$. Eventually
we introduced the quantity
\be
\keps = \sqrt 2 \sin\phi_\eps \bar\keps,
\ee
with $\bar\keps$ parameterizing the effect of $\xi \neq 0$. Estimating
$\bar\keps$ to be $0.94\pm 0.02$ we finally found \cite{Buras:2008nn}
\be
\keps = 0.92 \pm 0.02~.
\label{keps0}
\ee

In obtaining this result for $\keps$ we included only LD contributions to $\im \Gamma_{12}$. In 2010 we included in collaboration with Gino Isidori also the LD
contributions to $\im M_{12}$ so that the full result has been found to be
\cite{Buras:2010pza}
\be
\keps = 0.94 \pm 0.02~.
\label{keps}
\ee
For RBC-UKQCD calculations of LD effects to $\varepsilon_K$ see \cite{Bai:2023lkr}. As far as I understand they agree with our estimate within uncertainties of both approaches.

The formula (\ref{epsexact}) can now be written as
\begin{equation}
\varepsilon_K=C_{\eps} \kappa_{\eps}\hat B_K \IM\lambda_t \left\{
\RE\lambda_c \left[ \eta_1 S_0(x_c) - \eta_3 S_0(x_c, x_t) \right] -
\RE\lambda_t \eta_2 S_0(x_t) \right\} e^{i\phi_{\varepsilon}}\,,
\label{eq:epsformula}
\end{equation}
\noindent
where $\lambda_i=V_{id}V_{is}^*$, $\eta_i$ are QCD factors and the numerical constant $C_\eps$ is given in 2026 by
\begin{equation}
C_\eps = \frac{G_{\rm F}^2 F_K^2 m_K M_W^2}{6 \sqrt{2} \pi^2 \Delta M_K}
       = 3.635 \cdot 10^4 \, .
\label{eq:Ceps}
\end{equation}
$S_0(x_i)$ and $S_0(x_i,x_j)$ are the standard Inami-Lim functions \cite{Inami:1980fz,Buras:1983ap}.

In 2019  a more accurate formula for $\varepsilon_K$ has been presented by three members of my physics family \cite{Brod:2019rzc}.
It uses
the unitarity relation $\lambda_c=-\lambda_u-\lambda_t$ instead of
$\lambda_u=-\lambda_c-\lambda_t$ as done in the previous literature. This allows
to remove significant theoretical uncertainties from charm contribution to $\varepsilon_K$. The new SM expression for  $\varepsilon_K$ reads \cite{Brod:2019rzc}
\begin{equation}\label{BGS}
    |\varepsilon_K|
=  \kappa_\epsilon C_\varepsilon \hat{B}_K
|V_{cb}|^2 \lambda^2 \bar \eta 
 \times \Big[|V_{cb}|^2(1-\bar\rho)
\eta_{tt} S_{tt}(x_t) - \eta_{ut} S_{ut}(x_c, x_t) \Big]\,.
\end{equation}
It replaces the phenomenological expression given in (\ref{eq:epsformula}) \cite{Buras:2008nn}. Here
\be
S_{tt}(x_t)=S_0(x_t)+S_0(x_c)-2 S_0(x_c,x_t),\qquad  S_{ut}(x_c, x_t)=S_0(x_c)-S_0(x_c,x_t).
\ee

The QCD factors in (\ref{BGS}) at the NLO and NNLO level, respectively, read \cite{Brod:2019rzc}
\be
\eta_{tt}=0.55(2),\qquad \eta_{ut}=0.402(5).
\ee

The new expression in (\ref{BGS}) is significantly more accurate than the old one as far as theoretical uncertainties are concerned but  it is still subject
to large uncertainty due to $\vcb$. We will return to this issue in Part~\ref{BV}.

But let me go back to 2008 and 2010 and 
describe what can be found in \cite{Buras:2008nn,Buras:2010pza} and also in an additional 2009 paper with Diego \cite{Buras:2009pj} that had significant smaller impact on the literature  than the two papers just mentioned.

\section{Correlations among New CP Violating Effects in $\Delta F = 2$ Observables (2008)}

  In \cite{Buras:2008nn} we pointed out that the observed CP violation in $B_d - \bar B_d$ mixing, taking into account the measured ratio 
$\Delta M_d / \Delta M_s$, the recently decreased lattice value of the non-perturbative parameter $\hat B_K$ and an additional 
effective suppression factor $\keps \simeq 0.92$ in $\eps_K$, neglected sofar in most analyses, may be insufficient 
to describe the measured value of $\eps_K$ within the SM, thus hinting at new CP violating contributions
to the $K^0 - \bar K^0$ and/or $B^0_d - \bar B^0_d$ systems. Furthermore, assuming $\Delta M_d / \Delta M_s$ to be SM-like,
the signs and the magnitudes of NP effects in $\eps_K$ and in the CP asymmetries $S_{\psi K_s}$ and $S_{\psi \phi}$
may turn out to be correlated. For example, in a scenario with new CP-phases in $B_d$ and $B_s$ mixings being approximately 
equal and negative, a common new phase $\approx - 5^\circ$ could remove the tension between $\eps_K$ and $S_{\psi K_s}$
present in the SM and simultaneously accommodate, at least partly, the 2008 claim of $S_{\psi \phi}$ being much larger 
than the SM expectation.
We emphasized the importance of precise determinations of $V_{cb}$, $\hat B_K$, $F_K$ and $\xi_s$,~to which the parameter 
$\eps_K$ and its correlation with the CP violation in the $B_d - \ov B_d$ system are very sensitive. $\xi_s=F_{B_s}\sqrt{\hat B_s}/ F_{B_d}\sqrt{\hat B_d} $.

I do not think such tension is still valid as we have used
$\vcb=41.2(1.1)\cdot 10^{-3}$ instead of the higher inclusive value $\vcb\simeq 0.0425$ favoured in 2026. Moreover,
$\hat B_K\simeq 0.76$ is presently the best value for this parameter instead of
$\hat B_K = 0.720(13)(37)$ used by us in 2008 and  in our 2010 paper described next 
$\kappa_\epsilon =0.94\pm 0.02$ instead of $\kappa_\epsilon =0.92\pm 0.02$ used
in our 2008 paper. Yet, some ideas presented in this paper could turn out
to be useful when the community agrees on the best value of $\vcb$ one day.

\section{ On $\epsK$ Beyond Lowest Order in the Operator Product Expansion (2010)}
 In   \cite{Buras:2010pza}   we analyzed the structure of long distance (LD) contributions to $\varepsilon_K$, that generally affect both the {\it absorptive} 
($\Gamma_{12}$) and the {\it dispersive} ($M_{12}$) parts of the $K^0-\bar K^0$ mixing amplitude. 
We pointed out that, in a consistent framework, in addition to 
LD contributions to $\im \Gamma_{12}$, estimated by Diego and me in \cite{Buras:2008nn}, also LD 
contributions to $\im M_{12}$ have to be taken into account. Estimating the latter 
contributions the impact of LD effects on $\varepsilon_K$ is significantly
reduced (from $-6.0\%$ to $-3.6\%$). 
The overall effect of LD corrections and of the superweak 
phase being different from $45^\circ$ is summarised by the multiplicative factor 
$\kappa_\epsilon =0.94\pm 0.02$ as given already above. As LQCD did not yet include the dispersive part, this is the value used in all analyses to date resulting
in more than 200 citations for this analysis.

As a side remark, this paper was written in Florence in the Spring of 2010
during a Galileo Workshop organized by Paolo Gambino, Cecilia Tarentino
and myself. I enjoyed the four weeks I spent in Florence very much.

\section{On the Consistency between the Observed Amount of CP violation in the $K$- and $B_d$-systems within MFV (2009)}

In between these two papers, Diego suggested 
reappraising  whether the SM and MFV, could simultaneously describe the observed CP violation in the $K$- and $B_d$-systems.
In 2009 we found that CP violation in the $B_d$-system, measured most precisely through $(\sin 2 \beta)_{J/\psi K_s}$, implied $|\eps_K^{\rm SM}| = 1.78(25) \times 10^{-3}$ 
for the parameter $\eps_K$, measuring indirect CP violation in the $K$-system, to be compared with the experimental value $|\eps_K^{\rm exp}| = 2.23(1) \times 10^{-3}$. 
To bring this prediction to $1\sigma$ agreement with experiment, we explored then the simplest NP  possibility, that is the one of CMFV.
In particular we pointed out that this tension could be removed in this framework, with interesting signatures, e.g. correlated 
suppression patterns for rare K decay branching ratios. 

As I already mentioned above I do not think such tension is still valid
when the parameters of 2026 are used. Yet, some ideas presented in this paper could also turn out
to be useful when the community agrees on the best value of $\vcb$ one day.

  \part{Non-Leptonic B Decays}

\boldmath
\section{Non-leptonic Two-Body B Meson Decays}\label{NonL}
\unboldmath
After some success of our 1985 paper on two-body charm decays \cite{Buras:1985xv}, I made in 1994 another
look at theses decays but this time including NLO QCD corrections
 \cite{Buras:1994ij}. I still remember working on it during our vacation in Austria near Innsbruck.
The paper was finished in September 1994 just after we returned to Ottobrunn.
I did not expect that this paper would be cited by many but to my surprise it
was cited more than 100 times by now.

But 1994 was in any case the year when I started to study non-leptonic decays
of B mesons. Not alone but   with my excellent PhD student Robert Fleischer, who by now definitely
belongs to the world experts on these decays.
Most of my papers on non-leptonic two-body B decays have 
been written between 1994 and 2005 with Robert Fleischer. Two papers have been written with Luca
Silvestrini but let me first describe the ones with Robert.
Basically, they can be divided into two groups.

\begin{figure}[!bt]
 \centering
\includegraphics[width = 0.90\textwidth]{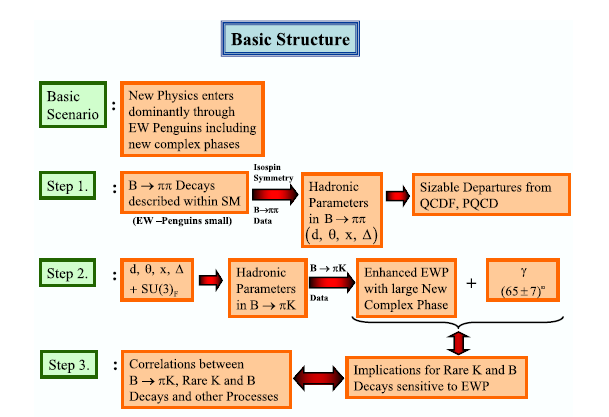}
\caption{Outline of our strategy. From \cite{Buras:2004dc}.
}\label{fig:BFRS}~\\[-2mm]\hrule
\end{figure}

\begin{enumerate}
\item
  In \cite{Buras:1994pb,Buras:1995pz,Buras:1995xb,Buras:1997cv,Buras:1998rb,Buras:2000gc} we have studied dominantly $B\to\pi K$ decays but also $B\to \pi\pi$ and $B\to KK$. The main goal was the determination of the angle $\gamma$ and to
  get control over penguin and rescattering effects. All these papers have been
  noticed by the community, in particular our 1998 paper \cite{Buras:1998rb}.
  I want to stress that although I definitely contributed to these papers
  in several ways, Robert was leading all these projects. I was often
  truly impressed by his way of using $\text{SU(3)}$ flavour symmetry.
\item
  In 2001 Robert also noticed that one could improve the bounds on $\sin 2\beta$
  and $K\to\pi\nu\bar\nu$ from MFV and this resulted in a new paper                  \cite{Buras:2001af}. While I was involved in studying these decays already
  for a decade, we realized while writing this paper that one could perform a general analysis correlating  non-leptonic B decays and rare B and K decays. This was in particular motivated by anomalies found in $B\to\pi K$ decays so that it was interesting
  to find out what possibe NP responsible for these anomalies would imply
  for rare B and K decays. We have involved Stefan Recksiegel, my assistant at that time and a PhD student Felix Schwab. Five paper written between 2003 and 2005 resulted from this collaboration \cite{Buras:2003yc,Buras:2003dj,Buras:2004ub,Buras:2004th,Buras:2005cv}. All these papers are well cited, in particular
  \cite{Buras:2003dj,Buras:2004ub}. Our strategy is summarized in the chart in
  Fig.~\ref{fig:BFRS}. It is interesting that the central value of $\gamma$
  obtained by us agrees very well with its best values known now, although
  the quoted errors these days are smaller, typically in the ballpark of
  $2^\circ-3^\circ$.
\end{enumerate}

As seen above I have written several papers with Robert on non-leptonic B
decays that were very well cited as seen in Table~\ref{tab:ranking}. As I
did not work on these decays for twenty years I prefer not to describe
them in details here but rather refer the reader to my description on
pages 317-328 in my book and to numerous papers by Robert and his collaborators
that were published in the last twenty years. They are reviewed in
an excellent paper by Robert \cite{Fleischer:2024uru}. I strongly recommend
this paper in which the most important issues related not only to non-leptonic
decays but also to $B_{s,d}\to\mu^+\mu^-$ and semi-leptonic decays $B\to K(K^*)\mu^+\mu^-$ are discussed. In particular the power of these decays in the
search for new CP violating phases is emphasized. In order to get an idea
on the complexity of such analyses it is advisable to have a look
at several papers from his group that are listed there. In
particular the analyses in \cite{Fleischer:2018bld,Barel:2020jvf}. In
these papers new strategies have been developed, in particular in order
to determine the parameters of electroweak penguins.
While I leave this to Robert's autobiograhy one day, his main collaborators
in the last ten years were Kristof De Bruyen, Ruben Jaarsma,  Eleftheria Malami and Keri Vos. While I did not meet Robert already for almost a decade,
I listened to a few nice presentations of this work by  Eleftheria and Keri.
See also the last paper of Robert and his collaborators on non-leptonic $D^0$ decays \cite{Fleischer:2025zhl}.

The $B\to\pi K$ anomalies studied by us 20 years ago did not disapear
but somehow the flavour community was not as much involved in studying them as
in other decays
because of hadronic uncertainties. But with much more data collected in the
last 20 years for $B\to PP$ decays with $B=B_{d,s}^0,B^+$ and $P=\pi,K,K^*$, one can perform fits to test the SM and in particular to test $\text{SU(3)}_F$ flavour symmetry. In addition to Robert  and his collaborators
also David London and his collaborators have written a number of  papers on this topic 
\cite{Berthiaume:2023kmp,Datta:2024zrl,Bhattacharya:2025wcq,Bhattacharya:2025rrv}. See also \cite{Grossman:2024amc}. Further literature to papers dealing with this topic can be found there as
well as numerous papers by Eleftheria, Keri and Robert.
Definitely it looks like the anomalies are present in hadronic $B$ decays
and I hope to return to them one day after the 20 years break.

In order to complete my story about non-leptonic $B$ decays, in 1998 I have written two papers on non-leptonic decays with Luca Silvestrini
\cite{Buras:1998us,Buras:1998ra}. They dealt with effects beyond factorization
in these decays. I think these were good papers, in particular the second one,
but they were soon shadowed by the papers by Beneke, Buchalla, Neubert and
Sachrajda who proposed QCD factorization in non-leptonic decays. It is very nicely summarized  by Gerhard Buchalla in Section 9 of my review on NLO and NNLO
corrections \cite{Buras:2011we}.

In this context I have to mention a strategical mistake I made in 1998. I have
been approached by  Martin Beneke asking me whether
I would be interested in joining them in developing QCD factorization for
non-leptonic B decays. I still remember how Martin was explaining this idea
to me on the blackboard in my office in the Physics Department. But I was very
busy with studying $\epe$ and rare decays at that time and I did not join them.
I think it was a mistake as their project was a great success and I would learn
more about these decays from a different perspective. On the other hand
several of their papers were written around the year 2000 and this was the
time in which new measurements of $\epe$ were performed and it was natural
to perform new analyses on $\epe$ \cite{Bosch:1999wr,Buras:2000qz}.

\part{An Atlas of New Physics Models}\label{NPModels}
\section{Preface}
\begin{figure}[!tb]
  \centerline{\includegraphics[width=0.98\textwidth]{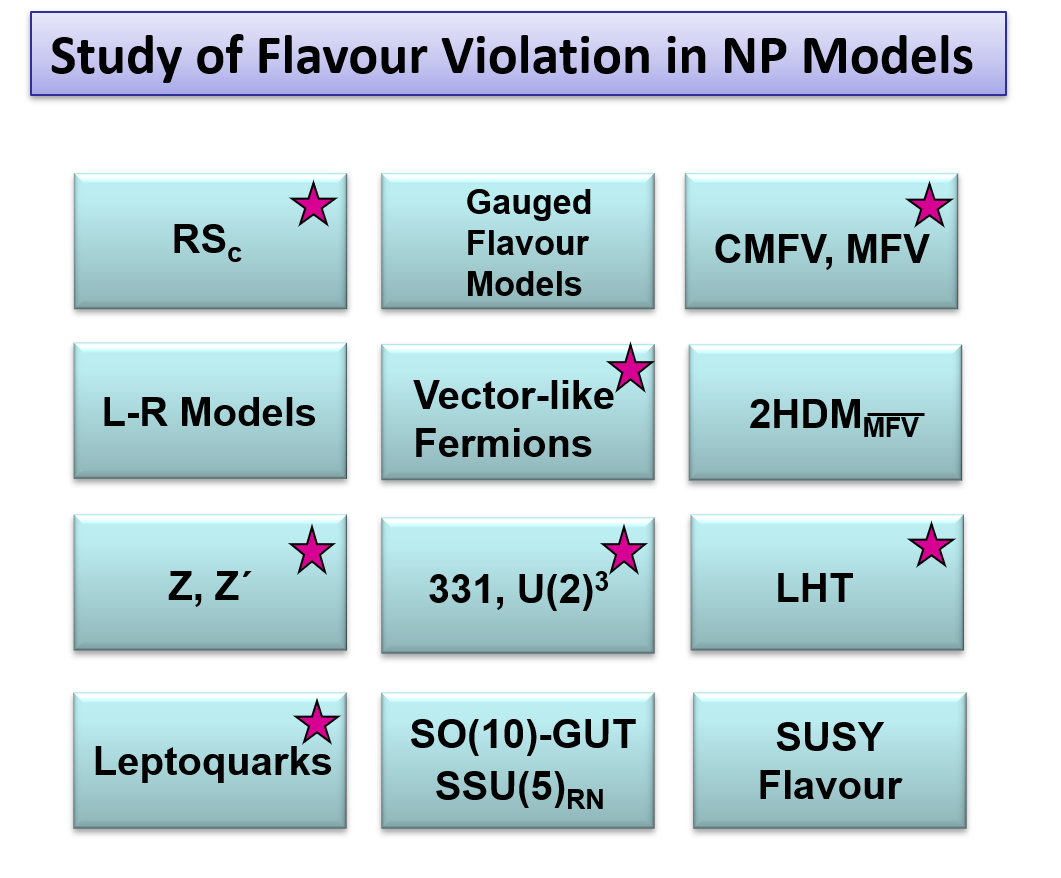}}
  \vspace{-0.5cm}
 \caption{\it NP Models investigated in my group at TUM.}
\label{Fig:NP}~\\[-2mm]\hrule
\end{figure}

As already stated in the outline, this part describes possibly my main contributions to the search for NP through FCNC
processes except for Part VII that I rate higher. It deals with detailed
analyses of specific NP models. It contains 15 sections and can be considered as an Atlas of these models. My PhD students and
postdocs contributed in a crucial manner to these analyses. To my knowledge there is no other
group worldwide that analyzed so many NP models in details.
This means deriving Feynman Rules, calculating flavour observables and
studying phenomenological implications of these models. I list in
Fig.~\ref{Fig:NP} the models we have investigated in detail, in
particular the ones indicated by red stars. Additional models investigated by us
will be briefly described below.

Several of these models lost on popularity during last ten years because no convincing signs of NP have been found in collider data. Yet, when the luminosity
at the LHC will be increased in this and the coming decade the situation could change. Yet, possibly the best hope to see the dynamics of these various
extensions is to see them in rare processes discussed in this Flavour Autobiography. Then many formulae and strategies developed in this part and also
in Part~\ref{SBSM} will turn out to be useful.

\section{Minimal Flavour Violation}\label{CMFV}
In the spring of 2000 standing with Luca Silvestrini in a queue in the TUM Mensa we got
the idea of possibly the simplest extension of the SM which we called
minimal flavour violation\footnote{It is the second time I got an idea in the TUM Mensa which brought over 600 citations. The first one was with Manfred M\"unz \cite{Buras:1994dj} presented in Part~\ref{NLOQCD}. Unfortunately, I stopped going there after I retired in 2012 and have to get ideas somewhere else.}. It consists of two ingredients \cite{Buras:2000dm}
\begin{itemize}
\item
  The CKM matrix is the only source of flavour violation.
\item
  The only relevant operators are the ones present in the SM.
\end{itemize}

Relatively to the MFV formulated two years later by D'Ambrosio, Jan Giudice, Gino Isidori and A. Strumia  \cite{DAmbrosio:2002vsn} this formulation of MFV was not as profound as the one of the Italian Masters and Gino used to call it {\em pragmatic} definition of MFV. I prefer
to call it constraint MFV, that is CMFV. It is a very simple framework in which
the one-loop functions are modified by NP in a universal manner for
all meson systems. Therefore various
ratios of observables in which these functions cancel out, remain unchanged
relative to the SM.

I have written a number of papers alone and with other collaborators calculating
implications of this idea which clearly is expected to be violated as new
CP-violating phases are expected to be present in order to explain our existence.
Let me just describe briefly some of them.
\begin{enumerate}
\item
  Basically the first application was to study with our son Robert the lower bound on $\sin 2\beta$ from CMFV  \cite{Buras:2000xq}.
\item
  Next, as reported already, in collaboration with Robert Fleischer, bounds on the unitarity triangle
  and $K\to\pi\nu\bar\nu$ decays have been derived   \cite{Buras:2001af}.
\item
  As $\sin 2\beta$, the angle $\gamma$ and the ratio $\Delta M_s/\Delta M_d$ are in CMFV the same as in the   SM, in   \cite{Buras:2001mb} we have studied
  how they could be affected by new $\Delta F=2$ operators. This paper written
  with Piotr Chankowski, Janusz Rosiek and Lucja Slawianowska was only a warming
  up for papers on the MSSM which I describe  later on.
\item
  In 2003 I  got the idea to study the ratios  $\mathcal{B}(B_{s,d}\to\mu^+\mu^-)/\Delta M_{s,d}$ which are free of CKM parameters and weak decay constants
  \cite{Buras:2003td}. As described later in Part~\ref{BV} this idea
  has been generalized in 2021 in  collaboration with Elena Venturini to rare B and K decays. I am puzzled why
  I did not get this idea earlier and why nobody else got it in 18 years.
\item
  Next in 2003 I gave lectures on CMFV at the Zakopane school \cite{Buras:2003jf}
\item
  In \cite{Bobeth:2005ck} and   \cite{Blanke:2006ig} various implications
  of CMFV for rare decays and observables in mixing have been studied.
  Both papers were well accepted by the community, in particular the second one
  which was the first paper of Monika Blanke, my excellent Diploma student and subsequently PhD student. Certainly, I could not know at that time that
  I would write additional 19 papers with her until now. But this
  paper was also the first one of several papers with my two Italian friends
  Diego Guadagnoli and Cecilia Tarantino who were postdocs in my group at that
  time. I  wrote still 8 papers with Diego and 9 with Cecilia until today and
  they are mentioned at various places in this write-up. These were some of the most   pleasant collaborations I had.

\item
  In 2006 Gustavo Branco visited Munich for the second time and this allowed
  me, Sebastian  J\"ager, Selma Uhlig and Andi Weiler to learn something about
  the   Leptogenesis \cite{Branco:2006hz}.

We analyzed lepton flavour violation (LFV), as well as generation of the observed baryon-antibaryon asymmetry of the Universe (BAU) within a generalized 
minimal lepton flavour violation (MLFV) framework \cite{Cirigliano:2005ck},
where we allowed for CP 
violation both at low and high energies. The generation of BAU is obtained 
through  {\it radiative resonant leptogenesis} (RRL), where starting with 
three exactly degenerate right--handed neutrinos at $\Lambda_{\rm GUT}$, we demonstrated explicitly within the SM and the MSSM that 
the splittings between their masses at the see-saw scale $M_\nu$, generated by
renormalization group effects, are sufficient for a successful leptogenesis
for $M_\nu$ even as low as $10^6\gev$.

The inclusion of flavour effects plays an important role in this
result and can lead to the observed BAU even in the absence
of CP violation beyond the PMNS phases.
The absence of a stringent lower bound on $M_\nu$ in this type of leptogenesis
allowed to easily satisfy 2006  
upper bounds on 
$\mu\to e\gamma$ and other charged lepton flavour violating (LFV) processes
even for $\Lambda_{\rm LFV}=\ord(1\tev)$.

We found, that the 
MLFV framework in the presence of heavy right-handed neutrinos and 
leptogenesis is  not as predictive 
as MFV in the quark sector and  
pointed out that without a specific MLFV model, there is a 
rich spectrum of possibilities for charged LFV processes and 
for their correlation 
with low energy neutrino physics and the LHC physics, 
even if the constraint from the observed BAU is taken into 
account.  While certain qualitative features of our analysis confirm findings of Cirigliano et al., \cite{Cirigliano:2005ck} at the quantitative level we found phenomenologically important differences. We explained the origin of these differences but there
is no space to describe them here.

I did not follow this field since then except for one paper with Julia Harz and
Martin Mojahed \cite{Buras:2024ewl} described above
and was curious how much of our findings are still 
relevant after 20 years. Therefore I asked Alejandro Ibarra, an expert in this field, to have a quick look at it. In his opinion our
paper is still relevant, since there has been no experimental breakthrough in the last 20 years (no $\mu\to  e \gamma$, no EDM, no signals of NP at the LHC).
There has been quite some progress in technical aspects of leptogenesis (mostly concerning QFT out of equilibrium), but this does  not change qualitatively the picture. According to him our paper still provides a valid overview about what to expect in scenarios with MLFV. Definitely good news but this shows that similar to rare K decays one has to wait long to know how ones predictions confront
the experimental data. 
\item
  In 2006 an interesting observation has been made by Monika Blanke: in CMFV the
  mass differences $\Delta M_s$ and $\Delta M_d$ can only be
  enhanced. This led to another joint paper with her \cite{Blanke:2006yh}. Actually, today
  I do not understand why Monika did not write this paper alone, but as she was the first author anyway it is not a problem. In any case in 2009 she wrote alone
   an important paper \cite{Blanke:2009pq} in which she studied the
  impact of NP physics in $\varepsilon_K$ on the correlation of $\kpn$ and
  $\klpn$. Her analysis motivated in 2015 Dario, Robert and myself to
  present Fig.~\ref{fig:illustrateEpsK} in the context of simplified models \cite{Buras:2015yca} as described in Section~\ref{Dario}.
\item
  In 2010 I gave again lectures at the Zakopane school, this time describing CMFV and how it could be broken by new operators \cite{Buras:2010wr}. In these
  lectures I got some ideas for classifying differnt properties of the extensions
  of the SM. But they cannot compete with the DNA charts developed with Jennifer
  three years later \cite{Buras:2013ooa} that I described in Section~\ref{DNAS}.
\item
  Then still a number of papers \cite{Buras:2009pj,Buras:2013raa,Blanke:2016bhf,Blanke:2018cya} have been written in the last decade which mainly tried to test
  CMFV as the experimental data where evolving.
\end{enumerate}

\section{Universal Extra Dimensions (2002, 2003)}
In the Summer of 2002 Andi Weiler told me that it would be good to investigate models with extra dimensions. Somehow I was not excited about it but as this was the
first time my PhD student suggested to me a project I agreed and we involved in this project
Michael Spranger, another PhD student of mine. Two papers resulted from this
collaboration. In the second paper also a very good PhD student Anton
Poschenrieder joined us. Somewhat surprisingly for me both papers have
been cited more than 200 times each.

\subsection{The Impact of Universal Extra Dimensions on the UT and Rare K and B Decays}

In the first paper \cite{Buras:2002ej}, that was fundamental for our project, we derived
Feynman rules for the model with one universal extra 
dimension that was proposed by Appelquist, Cheng and Dobrescu (ACD)
\cite{Appelquist:2000nn}. Subsequently, we calculated the contributions of the Kaluza-Klein (KK) modes  to 
the $K_L-K_S$ mass difference $\Delta M_K$, the parameter $\varepsilon_K$, 
the $B^0_{d,s}-\bar B^0_{d,s}$ mixing mass differences $\Delta M_{d,s}$ and 
rare decays  
$\kpn$, $\klpn$, $\klm$, $B\to X_{s,d}\nu\bar\nu$ and $B_{s,d}\to\mu\bar\mu$ 
in this model.

This model belongs to the class of CMFV models so that NP enters only through
universal one-loop functions in the   meson system considered by us and thereby implies many correlations known from other CMFV analyses.
For the compactification scale $1/R= 200~\gev$ the KK effects in these 
processes are governed by a $17\%$ enhancement of the $\Delta F=2$ box 
diagram function $S(x_t,1/R)$ and by a $37\%$ enhancement of the $Z^0$ 
penguin diagram function $C(x_t/1/R)$ relative to their 
 SM values. This implies 
the suppressions of $\vtd$ by $8\%$, of $\bar\eta$ by $11\%$ and 
of the  angle $\gamma$  in the unitarity triangle by $10^\circ$. 
$\Delta M_s$ is increased by $17\%$. $\Delta M_K$ is essentially 
uneffected.
All branching ratios considered in this paper are increased with 
a hierarchical structure of enhancements: $\kpn~(16\%)$, $\klpn~(17\%)$, 
$B\to X_{d}\nu\bar\nu~(22\%)$, $(K_L\to\mu\bar\mu)_{\rm SD}~(38\%)$,
 $B\to X_{s}\nu\bar\nu~(44\%)$,  $B_{d}\to\mu\bar\mu~(46\%)$ and 
$B_{s}\to\mu\bar\mu~(72\%)$.
For $1/R= 250~(300)\gev$ all these
effects are decreased roughly by a  factor of $1.5~(2.0)$.
We emphasized that the GIM mechanism
assures the convergence of the sum over the KK modes in the 
case of $Z^0$ penguin diagrams. 
We also emphasized that a consistent calculation of branching ratios has 
to take into account the modifications in the values of the CKM parameters.

\boldmath
\subsection{The Impact of Universal Extra Dimensions on
 $B\to X_s\gamma$, $B\to X_s~ {\rm gluon}$, $B\to X_s\mu^+\mu^-$,
  $K_L\to \pi^0e^+e^-$ and $\epe$}
\unboldmath

While in \cite{Buras:2002ej} we have calculated the KK effects on box diagrams and $Z^0$ penguins, in the second paper we calculated 
the contributions of the KK modes  to 
the $\gamma$--penguins, gluon--penguins, $\gamma$--magnetic penguins and 
chromomagnetic penguins \cite{Buras:2003mk}.
Together with our previous calculation of $Z^0$ penguin and box diagrams 
this allowed to study the impact of the KK modes on the decays 
$B\to X_s\gamma$, $B\to X_s~{\rm gluon}$, $B\to X_s\mu^+\mu^-$ and 
$K_L\to \pi^0e^+e^-$ and on the CP-violating ratio $\epe$.
For the compactification 
scale $1/R= 300\gev$ the perturbative part of the branching ratio for 
$B\to X_s\mu^+\mu^-$ is
enhanced by $12\%$ while the zero in the $A_{\rm FB}$ asymmetry is shifted 
from $\hat s_0=0.162$ to $\hat s_0=0.142$. The sizable suppressions of 
$\mathcal{B}(B\to X_s\gamma)~(20\%)$ and $\mathcal{B}(B\to X_s~{\rm gluon})~(40\%)$ implied several years later an improved lower bound on 
$1/R$. 
Similar comments apply to $\epe$ that is suppressed relative to the 
SM expectations with the size of the suppression depending 
sensitively on the hadronic matrix elements. The impact on 
$K_L\to \pi^0e^+e^-$ is below $10\%$. We pointed out the correlation between
the zero $\hat s_0$ in  the $A_{\rm FB}$ asymmetry and 
$\mathcal{B}(B\to X_s\gamma)$ that should be valid in most models with minimal 
flavour violation.  

Certainly, these two papers provided a good training in simple one-loop calculations for young generation and learning about FCNC processes.
But in the spirit of my usual procedure in supervising my students 
I did all these calculations as well. It was a real fun. One day when the data for all
these processes will be available it would be interesting to repeat the phenomenology of the ACD model.

The results of these two papers are summarized by me in the proceedings of the CKM 2003
workshop that took place in April 2003 in  Durham \cite{Buras:2003wc}.

In 2007 Uli Haisch and Andi Weiler following my suggestion
calculated the impact of the new measurement of $B\to X_s\gamma$ on this model
\cite{Haisch:2007vb}. They found an improved lower bound on the the inverse compactification radius $1/R\ge 600\gev$ at $95\%$ confidence level. This bound
suppressed the effects of KK modes by at least a factor of two. I am sure this
suppression is even stronger today.

\boldmath
\section{Littlest Higgs Model (LH)}\label{LHTG}
\unboldmath
\subsection{Preface}
An attractive idea to solve the gauge hierarchy problem is to regard the electroweak Higgs boson as a pseudo-goldstone boson of a certain global symmetry that is broken spontaneously at a scale $\Lambda=4\pi f=\ord(10\tev)$), much higher than the vacuum expectation value $v$ of the standard Higgs doublet. Concrete realizations of this idea are the “Littlest Higgs” models with the first one
proposed in \cite{Arkani-Hamed:2002ikv}
in which the Higgs field remains light, being protected by
the approximate global symmetry from acquiring quadratically divergent contributions to its mass at the one-loop level. In models of this type new heavy particles are present, that analogously to supersymmetric particles allow to cancel the quadratic divergences in question.

In addition to the SM particles, new charged heavy vector
bosons ($W_H^\pm$), a neutral heavy vector boson ($Z_H$), a heavy photon ($A_H$), a heavy top quark ($T$ ) and a triplet of heavy Higgs scalars $\Phi^{++},\Phi^+,\Phi^0$) are present. The details of
this model including the Feynman rules have been worked out in \cite{Han:2003wu}.
An important role in phenomenology plays the $t-T$ mixing parameter $0\le x_L \le 1$.

\boldmath
\subsection{Particle-Antiparticle Mixing, $\varepsilon_K$ and 
the UT in the LH Model (2004)}
\unboldmath
In the first paper \cite{Buras:2004kq}, in collaboration with my two diploma students Anton Poschenrieder and Selma Uhlig,  we calculated
the $K^{0}-\bar{K}^{0}$, $B_{d,s}^{0}-\bar{B}_{d,s}^{0}$
mixing mass differences $\Delta M_K$, $\Delta M_{d,s}$ and the 
CP-violating parameter $\varepsilon_{K}$ in the Littlest Higgs (LH) model.
It should be stressed that this model did not have the imposition of T-parity which was introduced later and studied by my group in details. I will report
on it in the next section.

We found generally 
the enhancements of 
$\Delta M_{d}$, $\Delta M_s$  and $\varepsilon_{K}$ by at most $20\%$ with negligible correction to $\Delta M_{K}$. 
 The dominant new contribution in this parameter range, 
calculated by us for the first time, 
comes from the box diagrams with ($W_L^\pm,W_H^\pm$) exchanges and 
ordinary quarks that are only suppressed by the mass of $W_H^\pm$ 
but do not involve explicit $\mathcal{O}(v^2/f^2)$ factors. 
This contribution is strictly positive.
The explicit $\mathcal{O}(v^2/f^2)$ corrections to the SM diagrams
with ordinary quarks and two $W_L^\pm$ exchanges have to be combined with 
the box diagrams with a single heavy $T$ quark exchange 
for the GIM mechanism to work. 
These $\mathcal{O}(v^2/f^2)$ corrections where found to be of 
the same order of magnitude as the ($W_L^\pm,W_H^\pm$) contribution but 
only for $x_L$ approaching $0.8$ they can compete with it. 
We pointed out that for $x_L>0.85$ box diagrams with two $T$ exchanges have to
be included. Although formally $\mathcal{O}(v^4/f^4)$, 
this contribution is dominant for $x_L \approx 1$ due to non-decoupling of $T$
that becomes fully effective only at this order.
We emphasized, that the concept of the unitarity triangle is still useful
in the LH model, in spite of the $\mathcal{O}(v^2/f^2)$ corrections
to the CKM unitarity involving only ordinary quarks. We demonstrated
the cancellation of the divergences in box diagrams that appear when one 
uses the unitary gauge for $W_L^\pm$ and $W_H^\pm$.

\boldmath
\subsection{Non-Decoupling Effects of the Heavy T in the $B_{d,s}^{0}-\bar{B}_{d,s}^{0}$ Mixing and Rare K and B Decays (2005)}
\unboldmath
In the second paper \cite{Buras:2005iv}, in which both Anton and Selma started
their PhD studies, we investigated in more details 
the case of the t-T mixing parameter $x_L > 0.90$. In this case the contribution to $B^0_{d,s}-\bar B^0_{d,s}$ mixing from box diagrams with two T exchanges cannot be neglected as already found in our first paper. It increases linearly with $x_T = m_T^2/M^2_W$ and with $x_T = O(f^2/v^2)$ constitutes effectively an $O(v^2/f^2)$ correction. For $x_L\approx 1$, this contribution turns out to be more important than the genuine $O(v^2/f^2)$ corrections. In particular it is larger than the $O(v^2/f^2)$ contribution of box diagrams with a single T exchange that increases only logarithmically with $x_T$. For $x_L = 0.95$ and $f/v = 5,10,15,$ the short distance function $S$ governing the $B^0_{d,s}-\bar B^0_{d,s}$ mixing mass differences $\Delta M_{d,s}$ receives $56\%, 15\%$ and $7\%$ enhancements relative to its SM value, implying a suppression of the CKM element $\vtd$ and an enhancement of $\Delta M_s$. The short distance functions $X$ and $Y$, relevant for rare $K$ and $B$ decays, increase only logarithmically with
$x_T$. With the suppressed $V_{td}$, $K\to\pi\nu\bar\nu$ and $B_d\to\mu^+\mu^-$ decays are only insignificantly modified with respect to the SM, while the branching ratio $\mathcal{B}(B_s\to\mu^+\mu^-)$ receives $66\%$, $19\%$ and $9\%$ enhancements for $ x_L = 0.95$ and $f/v = 5,10,15$, respectively. This enhancement
is excluded by the present data.
\boldmath
\subsection{ Rare $K$ and $B$ Decays in the LH Model  (2006)}
\unboldmath
Bill Bardeen joined us in the third paper on the LH model \cite{Buras:2006wk}.
He found  that the final result for the
$Z^0$-penguin contribution contains a divergence that is generated
by the one-loop radiative corrections to the currents corresponding
to the dynamically broken generators. Including an estimate of these
logarithmically enhanced terms, we calculated the branching ratios
for the decays $\kpn$, $\klpn$,
$B_{s,d}\to\mu^+\mu^-$ and $B\to X_{s,d}\nu\bar\nu$. We found that for the high
energy scale $f=\mathcal{O}\left(2-3\right) \text{TeV}$, as required by the
electroweak precision studies, the enhancement of all branching ratios
amounts to at most $15 \%$ over the SM values. On the technical side we
identified a number of errors in the existing Feynman rules in the LH model
without T-parity that could have some impact on other analyses
present in the literature. Calculating penguin and box diagrams in the
unitary gauge, we found divergences in both contributions that are cancelled
in the sum except for the divergence mentioned above.

On the whole these three papers were a good training for Anton and Selma in
calculating one-loop diagrams but the lesson from our study, as seen today, is that the LH
model does not seem to be  chosen by nature.

  \boldmath
  \section{Littlest Higgs Model with T-Parity (LHT)}
  \unboldmath
  \subsection{Preface}
 
In the original LH model just discussed, the custodial $\text{SU(2)}$ symmetry, of fundamental importance for electroweak precision studies, is unfortunately broken already at
tree level, implying that the relevant scale of NP, f , must be at least $2-3\tev$
in order to be consistent with electroweak precision data. As a consequence the
contributions of the new particles to FCNC processes turn out to be at most
$10-20\%$, which will not be easy to distinguish from the SM in view of experimental and
theoretical uncertainties.
More promising and more interesting from the point of view of FCNC processes is
the Littlest Higgs model with a discrete symmetry (T-parity) \cite{Cheng:2004yc}
, called LHT, under which all new
particles in the LH model, except $T_+$, are odd and do not contribute to processes with external
SM quarks (even under T-parity) at tree level. As a consequence the NP scale f
can be lowered down to $1~\tev$ and even below it, without violating electroweak precision constraints.

A consistent and phenomenologically viable LHT  model 
requires the introduction of three doublets of ``mirror quarks'' and three doublets of
``mirror leptons'' which are odd under T-parity, transform vectorially under
$\text{SU(2)}_L$ and
can be given a large Dirac mass. Moreover, there is an additional heavy $T_-$ quark that is also odd under T-parity.

Soon after completing the third paper on the LH model I decided to look at this model. Our team has been increased, by Monika Blanke,  Cecilia Tarantino, Stefan Recksiegel and Andreas Weiler. But it will be seen that not all members of
this team participated in all our LHT papers.

\boldmath
\subsection{Particle-Antiparticle Mixing, $\varepsilon_K$ and $B\rightarrow X_{s,d}\gamma$\\ in the LHT model}\label{LHT1}
    \unboldmath

    In our first LHT paper \cite{Blanke:2006sb}
Monika described the model in some details.
This is Section 2 of this paper\footnote{In fact, in basically all papers I wrote with Monika, she was responsible for section 2. But also contributed in an important manner to other sections.}.
Subsequently,
we calculated a number of observables related to particle-antiparticle 
mixing in LHT model going far beyond the previous phenomenological analyses
of other authors. In particular we studied $\Delta \Gamma_q$, $A_{\rm SL}^q$,
the CP asymmetries $S_{\psi K_S}$ and $S_{\psi \phi}$ and 
$B\rightarrow X_{s,d}\gamma$ decays.

In particular, we pointed out the role of mirror fermions 
with new flavour and CP-violating interactions in these quantities that in
2006 could solve some anomalies found by CDF collaboration. As these
anomalies disappeared I will not describe our numerical results.

We presented a number of correlations between the observables in question
which could be useful in testing this model better than it was possible in
2006. In particular one could learn in this manner about
the mass spectrum and the weak mixing 
matrix of mirror fermions that could have  a hierarchical structure that differs
significantly from  the CKM one.

\subsection{Another Look at the Flavour Structure of the Littlest Higgs Model with T-Parity}
In our second paper on the LHT model \cite{Blanke:2006xr}
we discussed the mixing matrix $V_{Hd}$ that describes the charged and neutral current
interactions between ordinary down-quarks and up- and down-mirror quarks in
the LHT model. We pointed out that this matrix
in addition to three mixing angles contains three physical complex phases and not
only one as used in the literature at that time. We explained the reason for the presence of
two additional phases, proposed a new standard parameterization of $V_{Hd}$ and briefly
commented on the relevance of these new phases for the phenomenology of FCNC
processes in the LHT model. A detailed numerical
analysis, including these new phases, of $K$ and $B$ physics, with particular attention to the most interesting rare decays has been presented in our third
LHT paper to be described next.

 \boldmath
\subsection{Rare and CP-Violating $K$ and $B$ Decays in
    the LHT Model (2006)}
\unboldmath
In our third LHT paper \cite{Blanke:2006eb}, which dealt with $K$ and
$B$ decays,
we gave a collection of Feynman rules including $v^2/f^2$
contributions that were presented in this paper for the first time and 
turned out to be useful also for applications outside flavour physics. We
adopted a model-independent parameterization of rare decays in
terms of the gauge independent functions $X_i,\,Y_i,\,Z_i\; (i=K,d,s)$,
which is in particular useful for the study of the breaking of the
universality between $K$, $B_d$ and $B_s$ systems through non-MFV
interactions.
Performing the calculation in the unitary and 't Hooft-Feynman
gauge, we found that the final result in the Z-penguin contained a
logarithmic UV cutoff dependence
which appeared
to signal some sensitivity to the ultraviolet completion of the
theory. We were not particularly worried about it as similar feature
was found in the LH model as described above. Unfortunately, as stated
below this logarithmic divergence was in fact cancelled by one contribution that we
missed.

In spite of this I continue to describe this and following  LHT papers to indicate which processes have been analyzed but omitting size of effects found by us
in these papers because finally they turned out to be smaller in some decays
than found by us first.

Including an estimate of this contribution, we calculated
the branching ratios for the decays $K^+\to\pi^+\nu\bar\nu$,
$K_L\to\pi^0\nu\bar\nu$, $B_{s,d}\to \mu^+\mu^-$, $B\to
X_{s,d}\nu\bar\nu$, $K_L \to\pi^0\ell^+\ell^-$ and $B\to
X_{s,d}\ell^+\ell^-$, paying particular attention to
non-MFV contributions present in the model. 

The main feature of mirror fermion effects is the possibility of large
modifications in rare $K$ decay branching ratios and in those $B$ decay
observables, like $S_{\psi \phi}$ and $A^s_\text{SL}$, that are very small in
the SM.
Imposing all available constraints we found that the decay rates for
$B_{s,d}\to \mu^+ \mu^-$, $B \to X_{s,d} \nu \bar \nu$, 
 $K^+\to\pi^+\nu\bar\nu$ and $\klpn$ and $K_L \to\pi^0\ell^+\ell^-$  could be at that time significantly enhanced.

\boldmath
\subsection{Charged Lepton Flavour Violation and $(g-2)_\mu$ in the LHT Model:
   a Clear Distinction from Supersymmetry (2007)}
\unboldmath
In our fourth LHT paper \cite{Blanke:2007db}
we calculated the rates for the charged lepton flavour violating decays $\ell_i\to\ell_j\gamma$, $\tau\to\ell\pi$, $\tau\to\ell\eta$, $\tau\to\ell\eta'$, $\mu^-\to e^- e^+ e^-$, the six three body leptonic decays $\tau^-\to \ell_i^-\ell_j^+\ell_k^-$ and the rate for $\mu-e$ conversion in nuclei. 

We also calculated the rates for $K_{L,S}\to\mu e$, $K_{L,S}\to\pi^0\mu e$ and $B_{d,s}\to\ell_i\ell_j$. We found that the relative effects of mirror leptons in these transitions are by many orders of magnitude larger than analogous mirror quark effects  in rare $K$ and $B$ decays analyzed in \cite{Blanke:2006eb}.
In particular, in order to suppress the $\mu\to e \gamma$ and $\mu^-\to e^-e^+e^-$ decay rates and the $\mu-e$ conversion rate below the experimental upper bounds, the relevant mixing matrix in the mirror lepton sector $V_{H\ell}$ must be rather hierarchical, unless the spectrum of mirror leptons is quasi-degenerate. 
We found that the pattern of the LFV branching ratios in the LHT model differs significantly from the one encountered in the MSSM, allowing in a transparent manner to distinguish these two models with the help of LFV processes.
 We also calculated $(g-2)_\mu$ and found the new contributions to $a_\mu$ below $1\cdot 10^{-10}$ and consequently negligible. 

 Independently of whether the LHT model will be chosen by nature one day,
 this paper was very useful for learning about the general structure of LFV processes.

 \boldmath
 \subsection{LHT Model Confronting the New Data on $D^0-\bar D^0$ Mixing (2007)}
   \unboldmath
  In our fifth LHT paper \cite{Blanke:2007ee},
motivated by the first experimental evidence of meson oscillations in the $D$
system, we studied $D^0 - \bar D^0$ mixing, investigating its role in constraining the model parameters
and its impact on the most interesting flavour observables.
We found that the experimental data were potentially strongly constraining but limited by large theoretical uncertainties in the long-distance SM
contribution to $D^0 - \bar D^0$ mixing.

\boldmath
 \subsection{  Correlations between $\epe$ and Rare $K$ Decays
in the LHT Model (2007)}                        
   \unboldmath
   In our sixth LHT paper \cite{Blanke:2007wr}, not counting the review with Monika \cite{Blanke:2007ww},   
we calculated the CP-violating ratio $\epe$ and investigated its correlations with the branching ratios for
$\klpn$, $K_L\to\pi^0\ell^+\ell^-$ and $\kpn$. The resulting correlations turned out to be
rather strong in the case of $K_L$ decays, but less pronounced in the case of
$\kpn$. Unfortunately, they are subject to large hadronic uncertainties present
in $\epe$. As the theoretical situation of $\epe$ changed significantly since
our analysis I will not describe the numerical analysis in this paper.

\boldmath
 \subsection{The LHT Model Facing CP-Violation in $B_s - \bar B_s$ Mixing (2008)}                      
 \unboldmath
 
 In our seventh LHT paper \cite{Blanke:2008ac} we updated our 2006 analysis
 in view of a large new CP phase in $B_s - \bar B_s$ mixing hinted by the CDF and D{\O} data. As this data was not supported by subsequent measurements
 and presently this phase is in agreement with its SM predictions I
 will not describe the results of this analysis here.

\subsection{FCNC Processes in the LHT Model: an Update (2008)}
Then in 2008 came a real shock.
An additional contribution to the $Z$ penguin in the LHT has been pointed out in \cite{Goto:2008fj,delAguila:2008zu}, which has been overlooked in our papers. This contribution leads to the cancellation of the left-over
logarithmic UV cutoff dependence in the calculation of some rare decay amplitudes. Instead of presenting separate errata to  our papers \cite{Blanke:2007db,Blanke:2007ee,Blanke:2007wr,Blanke:2008ac} partially affected by this omission, we have presented a corrected and updated analysis of FCNC processes in \cite{Blanke:2009am}.

Although this was indeed a shock, one should stress that only $Z$-penguin
contributions were affected. Moreover, while the removal of the logarithmic UV cutoff dependence in
our previous results made the deviations from
the SM expectations in the quark sector less spectacular, it still allowed 
for sizable NP effects  in $K\to\pi\nu\bar\nu$ and $K_L\to \pi^0 \ell^+\ell^-$ decays and in 
the CP-asymmetry $S_{\psi\phi}$ with the latter unaffected by the new contribution. We extended our analysis by a study of the fine-tuning required to fit the data on $\varepsilon_K$ and by the inclusion of the decay $K_L\to\mu^+\mu^-$. A number of correlations can distinguish this model from the custodially protected {Randall-Sundrum model  analyzed by us as described in Section~\ref{RSM}.

We also reconsidered lepton flavour violating decays, including now a discussion of fine-tuning. 
While the $\ell_i\to \ell_j\gamma$ decays are 
unaffected by {the removal of the logarithmic cutoff dependence}, the branching ratios for decays with three
leptons in the final state, like $\mu\to 3 e$ are lowered by almost an
order of magnitude. In spite of this, the pattern of lepton flavour violation in
the LHT model can still be distinguished from the one in supersymmetric
models.

\boldmath
\subsection{CP~Violation in $D^0 -\bar D^0$ Oscillations: General\\ Considerations and  {Applications} to the LHT Model (2009)}
\unboldmath
In 2009  I returned to look at the charm system having the opportunity
to work with my friend and leading world expert in charm physics Ikaros Bigi
\cite{Bigi:2009df}.

Indeed, the observed $D^0-\bar D^0$ oscillations provided
a new stage in our search for NP in heavy flavour dynamics. The theoretical verdict on the observed values of the relevant parameters $x_D$ and $y_D$ remained still ambiguous: while they could be totally generated by SM dynamics, they could also contain a sizable or even leading contribution from NP.  Those  oscillations are likely to 
enhance the observability of CP~violation as clear manifestations of NP. 
In our paper with Ikaros we presented general formulae for $D^0-\bar D^0$ oscillations,  concentrating on the case of negligible direct CP violation. In particular we derived a general formula for the time-dependent mixing-induced CP asymmetry in decays to a CP eigenstate and its correlation with the semileptonic CP asymmetry $a_\text{SL}(D^0)$ in $D^0(t)\to \ell\nu K$.
We applied our formalism to the LHT model, using the time-dependent CP asymmetry in {$D^0\to K_S\phi$} as an example. We found observable effects at a level well beyond anything possible with CKM dynamics. Comparisons with CP violation in the $K$ and $B$ systems offer an excellent test of this scenario and reveal the specific pattern of flavour and CP violation in the $D^0-\bar D^0$ system predicted by this model. 
We discussed a number of charm decays that could potentially offer an insight in the dynamics of CP violation in $D$ decays.
We also applied our formalism to $B^0_s-\bar B^0_s$ mixing.

\boldmath
\subsection{Quark Flavour Observables in the LHT Model After LHC Run 1 (2015)}
\unboldmath

Several years later we have returned to the LHT model
with the goal to present a new analysis of quark flavour observables in this model  in view of the oncoming flavour precision era. This time only Monika and Stefan joined me in this new analysis \cite{Blanke:2015wba}. The remaining
members of our LHT-Team were not working anymore at TUM or left research.
Monika was working at that time at CERN and in Karlruhe  but clearly I could not work
without her on the LHT model and indeed she contributed in an important
manner to our last LHT paper.

We used all available information on the CKM parameters, lattice QCD input and experimental data on quark flavour observables and corresponding theoretical calculations, 
taking into account new lower bounds on the symmetry breaking scale and the mirror quark masses from the LHC. We investigated  by how much the branching ratios for a number of rare $K$ and $B$ decays are still allowed to depart from their SM values. This included $\kpn$, $\klpn$, $K_L\to \mu^+\mu^-$,  
$B\to X_s\gamma$, $B_{s,d}\to\mu^+\mu^-$, $B\to K^{(*)}\ell^+\ell^-$, $B\to K^{(*)}\nu\bar\nu$, and $\epe$.

Taking into account the constraints from $\Delta F=2$ processes, significant departures from the SM predictions for 
$\kpn$ and $\klpn$ were still possible, while the effects in $B$ decays turned out to be much smaller.
In particular, the LHT model similar to LH model, investigated 10 years earlier
and reported on in Section~\ref{LHTG},
favours  $\mathcal{B}(B_{s}\to\mu^+\mu^-) \ge  \mathcal{B}(B_{s}\to\mu^+\mu^-)_{\rm SM}$, which is not supported by the data, and the anomalies in  $B\to K^{(*)}\ell^+\ell^-$ decays could not be explained in this model. With the recent  large $N$ input the requirement of the significant enhancement of the ratio 
$\epe$  relative to its SM value implied a significant suppression of the branching 
ratio for $\klpn$ with respect to its SM value while  allowing only for 
small modifications of $\kpn$.
Finally, we investigated how the LHT physics could 
be distinguished from other models by means of indirect measurements and
discussed the consequences for quark flavour observables of not finding any LHT state in the coming years.

On the whole it was interesting to study this model but the prediction
for $B_s\to \mu^+\mu^-$ worries me that, similar to the LH model, the LHT
model is likely not to be chosen by nature. With all these formulae at hand, possibly one
day, when many observables are precisely measured, one could still
perform new analysis and possibly introduce some modifications if necessary.

\section{Randal-Sundrum Scenario}\label{RSM}
\subsection{Preface}
Several members of my group contributed in various combinations to the study of phenomenological
implications of the fameous Randal-Sandrum scenario \cite{Randall:1999ee}.
Let me just list their names in alphabetic order: Michaela Albrecht, Monika
Blanke, Bj\"orn Duling, Katrin Gemmler, Stefania Gori and  Andi Weiler.

\subsection{Electroweak and Flavour Structure of a Warped Extra\\ Dimension  
with Custodial Protection (2009)}
Let me begin by recalling the basic electroweak and flavour structure of a model with a warped
extra dimension and the bulk gauge group 
$\text{SU(3)}_c\times \text{SU(2)}_L\times \text{SU(2)}_R\times\text{P}_{LR}\times \text{U(1)}_X$
as  described by us in \cite{Albrecht:2009xr}.
The presence of $\text{SU(2)}_R$ implies an unbroken custodial symmetry in the Higgs system
allowing to eliminate large contributions to the T-parameter in the electroweak precision tests (EWT), whereas the
$\text{P}_{LR}$ symmetry and the enlarged fermion representations provide a custodial 
symmetry for flavour diagonal and flavour changing couplings of the SM $Z$
boson
to left-handed {down-type} quarks. We diagonalised analytically  the mass matrices of
charged and neutral gauge bosons including the first KK modes. We presented the
mass matrices for quarks including heavy KK modes and discussed the neutral and
charged currents involving light and heavy fields. We provided
the corresponding 
complete set of Feynman rules in the unitary gauge.
This allowed to perform three phenomenological analyses that I describe in the
following. These analyses appeared in fact before this basic paper of our
studies for reasons which I do not remember now.

\boldmath
\subsection{$\Delta F=2$ Observables and Fine-Tuning in a Warped Extra Dimension with Custodial Protection (2008)}
\unboldmath

In the first phenomenological paper on this NP scenario \cite{Blanke:2008zb}
we presented a complete analysis of $\Delta S=2$ and $\Delta B=2$ processes 
in  a warped extra dimensional model with a custodial protection of $Zb_L\bar b_L$,
 including $\varepsilon_K$, $\Delta M_K$,  $\Delta M_s$, $\Delta M_d$, $A_{\rm
   SL}^q$, $\Delta\Gamma_q$, $A_{\rm CP}(B_d \rightarrow \psi K_S)$ and
 $A_{\rm CP}(B_s \rightarrow \psi \phi)$. These processes are affected by tree
 level contributions from Kaluza-Klein gluons, the heavy KK photon,
   new heavy electroweak gauge bosons $Z_H$ and $Z'$, and in principle by
     tree level $Z$ contributions.

We confirmed earlier findings \cite{Csaki:2008zd}  that the fully anarchic approach where all the hierarchies in quark masses and
weak mixing angles are geometrically explained seems implausible and we
confirmed that the KK mass scale $M_\text{KK}$ generically has to be at least {$\sim 20\tev$} to satisfy the $\varepsilon_K$ constraint. We pointed out, however, that there exist regions in parameter space with only modest fine-tuning in the 5D Yukawa couplings which satisfy all existing $\Delta F=2$ and electroweak precision constraints for scales $M_\text{KK}\simeq 3\tev$ in the reach of the LHC.  Simultaneously we found that $A_\text{CP}(B_s\to\psi\phi)$ and $A^s_\text{SL}$ could  be much larger than in the SM as indicated by 2008 results from CDF and D{\O} data. This was exciting but future measurements disproved these experimental findings.

We also pointed out that, for $B_{d,s}$ physics $\Delta F=2$ observables, the
  complex $(Z_H,Z')$ can 
  compete with KK gluons, while the 
    tree level $Z$ and KK photon contributions are
very small. In particular we pointed out that the $Zd^i_L\bar d^j_L$ couplings
are protected by the
 custodial symmetry.
As a by-product we showed the relation of the RS flavour model to the Froggatt-Nielsen mechanism \cite{Froggatt:1978nt} and we provided analytic formulae for the effective flavour mixing matrices in terms of the fundamental 5D parameters.

\boldmath
\subsection{Rare $K$ and $B$ Decays in a Warped Extra Dimension with Custodial Protection (2008)}
\unboldmath

Subsequently in 
\cite{Blanke:2008yr} we extended our studies to $\Delta F=1$ processes
presenting a complete study of rare $K$ and $B$ meson decays 
in a warped extra dimensional model with a custodial protection 
of (both diagonal and non-diagonal) $Z d_L^i \bar d_L^j$ couplings,
including $K^+\to \pi^+\nu\bar\nu$, $K_L\to\pi^0\nu\bar\nu$, $K_L\to\pi^0 \ell^+\ell^-$,
$K_L\to \mu^+\mu^-$, $B_{s,d}\to \mu^+\mu^-$, $B\to K\nu\bar\nu$, 
$B\to K^*\nu\bar\nu$  and $B\to X_{s,d}\nu\bar\nu$.
 In this model in addition to SM one loop contributions
 these processes receive  tree level contributions from the $Z$ boson and the new heavy electroweak gauge bosons.

 We analyzed all these contributions that turn out to be dominated by tree level $Z$ boson exchanges governed by right-handed couplings to down-type quarks. Imposing
all existing constraints from $\Delta F=2$ transitions analyzed by us
as described above and fitting all quark masses and CKM mixing parameters we 
found that a number of branching ratios for rare $K$ decays 
can differ significantly from the SM predictions, while the corresponding effects in rare $B$ decays are modest, dominantly due to the custodial protection being more effective in
$B$ decays than in $K$ decays.

In order to reduce the parameter dependence we
investigated  correlations between various observables within the $K$ system, within the $B$ system and in 
particular between $K$ and $B$ systems, and also between $\Delta F=2$ and $\Delta F=1$ observables. These correlations allow for a clear
distinction between this NP scenarios and models with
minimal flavour violation or the LHT model, and could give an opportunity to future experiments to confirm or rule out this model.
  We  showed how our results would change if the custodial protection of 
$Z d_L^i \bar d^j_L$ couplings was absent. In the case of rare $B$ decays the modifications are spectacular.

\subsection{The Impact of KK Fermions on SM Fermion Couplings in a RS Model with Custodial Protection (2009)}\label{LastRS}

The last paper of this series was more general but as an explicit
example we used again the RS model studied above \cite{Buras:2009ka}. Only
Bj\"orn Duling, Stefania Gori and I were involved in this project.

Namely, we reconsidered  the impact of heavy vector-like fermions on the couplings of
SM quarks to the SM gauge bosons $W^\pm$ and $Z$ and to the
SM Higgs boson $H$. Integrating out these fermions at tree level we derived general
formulae that can be used in any model containing such particles. We applied
these formulae to the case of the lightest KK fermions in a RS model with a custodial protection of flavour conserving $Zd_L^i\bar d_L^i$
and flavour violating $Zd_L^i\bar d_L^j$ couplings. We pointed out that in this
model also the couplings of $Zu_R^i\bar u_R^i$ and $Zu_R^i\bar u_R^j$ are
protected.

In particular we demonstrated explicitly that this protection is not
spoiled by the mixing of the SM quarks with the KK fermions, which is
guaranteed by the underlying $\text{P}_{LR}$ symmetry. We found that the impact of KK
fermions on the $Z$-couplings, while not negligible, is significantly smaller than the one coming from the mixing of $Z$ with
the heavy KK gauge bosons $Z_H$ and $Z'$. The situation is similar for the $W^\pm$ couplings with the only exception of the $tb$ coupling where the impact of KK fermions can in principle be larger than the effects that are induced by gauge boson mixing.

We also showed explicitly that at $\mathcal{O}(v^2/M_\text{KK}^2)$ the
fermion--Higgs couplings of a Higgs placed on the infrared (IR) brane is not
affected by the KK contributions up to strongly chirally suppressed
contributions.
 The corrections to the CKM matrix, in particular the breakdown of its unitarity,  were found to be small. We also investigated the
 right-handed couplings of $W^\pm$ that are generated through the mixing with
 KK-fermions and introduced a $3\times 3$ matrix that describes the pattern of flavour
 violation in these new interactions of the standard $W^\pm$.

 As these four papers received on average  200 citations by now, they had 
 significant impact on the literature. As no KK-fermions have been found
 by the LHC so far, once many flavour observables will be precisely measured
 it will be interesting to investigate whether this NP scenario is still viable.

\section{4th-Generation of Quarks and Leptons (2010)}
In 2010 the fourth generation of quarks and leptons was still a realistic possibility and the team consisting of Thorsten Feldmann, the substitute professor during my Carl von Linde Senior Fellowship at TUM-IAS, of three of
my PhD students Bj\"orn Duling, Tillmann Heidsieck and Christoph Promberger and myself decided to
analyse this NP scenario from the point of view of flavour physics.
We have written three detailed papers. The first one on the impact of fourth generation of quarks and leptons on rare K and B decays and particle-antiparticle mixing \cite{Buras:2010pi}. 
In the second paper this analysis has been extended to the impact on mixing and CP Violation in the charm system \cite{Buras:2010nd}. In these two papers also
Stefan Recksiegel participated.
Finally in the third paper lepton flavour violation in the presence of fourth generation has been analyzed \cite{Buras:2010cp}.

I will not describe any results from these papers because two years later, after the Higgs discovery, this scenario of NP has been ruled out. In particular
as demonstrated in \cite{Eberhardt:2012gv} the presence of a sequential fourth generation
of quarks and leptons, coupling to the Higgs, would increase its width by much
in disagreement with experiment. Thus while it was fun to analyse this scenario,
it will remain as a mathematical exercise: a good training for PhD students.
On the other hand the exclusion of fourth sequential generation of quark and leptons was important because if new fermions exist in nature they must be vector-like, which allows them to decouple from Higgs and have masses much higher than
the top quarks mass. We will return to them in Section~\ref{VLQ}.

This was the last paper written with Bj\"orn Duling who was one of the best
PhD students I had. We worked together on the LHT and RSc models and also on
the 4G model. One can check that all papers written with him earned many citations
and he contributed to all of them in an important manner. Therefore,
it was a big disappointment for me when after completing his excellent thesis
he decided to leave research. But this was his choice and I had to accept it.

\section{Two-Higgs Doublet Models, Heavy Scalars}
\subsection{Preface}
Somehow my interest in working on heavy scalars was lower than for gauge bosons, like $Z^\prime$ and fermions.  Still I have written eventually several papers
on heavy scalars.
Some of them were
well accepted by the community.

Already during my days at MPI, in 1985, I have written two papers with Gustavo Branco and Jean-Marc G\'erard, my great collaborator over many years. Gustavo visited MPI for a
few months and as he was an expert on spontaneous CP violation, I thought it would be good to look at this issue with him. 

Weinberg proposed three scalar doublet model with spontaneous CP violation
\cite{Weinberg:1976hu} that had some phenomenological problems. In the first paper we have shown
how these problems could be solved \cite{Branco:1985ch}. This paper was practically ignored by the community. The second paper on CP Violation in models with two and three scalar doublets, written soon after the first paper was more successful \cite{Branco:1985pf}. Several formulae for $\epe$, $\varepsilon_K$, $\Delta M_K$ and electric dipole moment of the neutron with neutral and charged scalar
exchanges have been derived in this paper for the first time. {\bf Jean-Marc
  presented our work at  the 1985 Moriond Workshop on Heavy Quarks, Flavour Mixing and CP-Violation \cite{Gerard:1985pa}and had a long discussion with Bjorken who showed great interest in our paper, in particular in connection with possible light scalars.}

My second encounter with scalars was in the Summer of 1989. Together with
my PhD student Claudia Salazar, my diploma student at that time Markus Lautenbacher  and Pawel Krawczyk from Warsaw, we have studied $B^0-\bar B^0$ mixing,
CP violation and $\kpn$ in a two Higgs doublet model \cite{Buras:1989ui}.
In this analysis charged Higgs effects played an important role. While the
numerical analysis of this paper is out of date, this paper is still cited
almost 30 years later.

One year later  we have performed a renormalization group analysis of charged Higgs effects in $\epe$, $\varepsilon_K$ and $B_d^0-\bar B_d^0$ mixing in a two Higgs doublet model, this time including the effects of electroweak penguins in the presence of a heavy top quark. This analysis was motivated by our $\epe$ anatomy
within the SM for a heavy $m_t$ \cite{Buchalla:1989we} and this was natural
that my Diploma students Gerhard Buchalla and Michaela Harlander
took part in this collaboration\cite{Buchalla:1990fu}. Possibly, the main
result of this paper was additional suppression of $\epe$ by $Z^0$-penguins
with charged Higgs exchanges in addition to the suppression of $\epe$
by electroweak penguins. While after 25 years one would have to make a closer
look at this analysis, it appears that charged Higgs cannot help
in reducing the present $\epe$ anomaly. See Section~\ref{Statusepe}.

I returned to Two-Higgs Doublet models 20 years later when Gino Isidori spent several months in my group at TUM-IAS as a Hans Fischer Senior Fellow of this institute.

\subsection{Higgs-mediated FCNCs: Natural Flavour  Conservation\\ vs.~Minimal Flavour Violation (2010)}

The first paper written with him and  two of my PhD students Maria Valentina Carlucci and Stefania Gori dealt with Higgs-mediated FCNCs in models with more than one-Higgs doublet \cite{Buras:2010mh}. 

In this paper 
we compared the effectiveness of two hypotheses,  
Natural Flavour Conservation (NFC) and MFV, 
in suppressing the strength of  FCNCs
in models with more than one-Higgs doublet. We showed that the MFV 
hypothesis, in its general formulation, is more stable 
in suppressing FCNCs than the hypothesis of NFC alone
when quantum corrections are taken into account.
We discussed the phenomenological implications of these two scenarios 
analyzing meson-antimeson mixing observables and the rare decays $B_{s,d}\to \mu^+\mu^-$. 
We demonstrated that, introducing {\it flavour-blind} CP phases,
two-Higgs doublet models respecting the MFV hypothesis can accommodate a  
large CP-violating phase in $B_s$ mixing, as hinted by CDF and D0 data at that time and, without extra free parameters, soften significantly in a correlated manner  the observed anomaly in the relation between $\varepsilon_K$ and $S_{\psi K_S}$.

Evidently, the situation in 2026 is different. Both $\varepsilon_K$ and $S_{\psi K_S}$ are SM-like and no $H^+$ contributions are required to describe them.
But it is not excluded that they will play some role in other observables.
Then the formulae derived by us could turn out to be useful.

\boldmath
\subsection{EDMs vs.~CPV in $B_{s,d}$ Mixing in Two Higgs Doublet \\ Models with MFV (2010)}
\unboldmath

In the second paper with Gino and another great Italian collaborator Paride Paradisi \cite{Buras:2010zm},
we analyzed the correlations between electric dipole moments (EDMs)
of the neutron and heavy atoms and CP violation in $B_{s,d}$
mixing in two Higgs doublet models respecting the MFV hypothesis, with flavour-blind CP-violating (CPV) phases
that Gino, Stefania, Valentina and I analyzed in \cite{Buras:2010mh}.

In particular, we considered the case of flavour-blind CPV phases from
i) the Yukawa interactions and  ii) the Higgs potential.
We showed that in both cases the upper bounds on the above EDMs do not
forbid sizable non-standard CPV effects in $B_{s}$ mixing.
However, if a large CPV phase in $B_s$ mixing as hinted by CDF and D0 data in 2010 would  be confirmed, this
would imply EDMs very close to their 2010 experimental bounds,
within the reach of the next generation of experiments, as well as
$\mathcal{B}(B_{s,d}\to\mu^+\mu^-)$ typically largely enhanced over its
SM expectation. The two flavour-blind CPV mechanisms could then  be distinguished
through the correlation between $S_{\psi K_S}$ and $S_{\psi\phi}$ that is
strikingly different if only one of them is relevant. Which of these two
CPV mechanisms dominates depends on the precise values of $S_{\psi\phi}$
and $S_{\psi K_S}$, as well as on the CKM phase (as determined by tree-level
processes). The 2010  data seemed to show a mild preference for a {\it hybrid}
scenario where both these mechanisms are at work.

This was a nice analysis but similarly to previous paper the 
large CPV phase in $B_s$ mixing as hinted by CDF and D0 data in 2010
has not been confirmed by later experiments. Consequently also
the enhancements of $\mathcal{B}(B_{s,d}\to\mu^+\mu^-)$ in these scenarios
are not necessary which at least in the case of $\mathcal{B}(B_{s}\to\mu^+\mu^-)$
is fortunate for these scenarios as the present data require slight suppression
of this branching ratio and not the enhancement relative to the SM.

\subsection{The Anatomy of Neutral Scalars with FCNCs in the Flavour Precision Era (2013)}
Three years later I have written still two papers with scalars. The first one with my two
great frequent collaborators  Fulvia and Jennifer, a postdoc Minoru Nagai  and a PhD student of
Robert Fleischer, Robert Knegjens  \cite{Buras:2013rqa}. This was the anatomy
of neutral scalars with FCNCs in the flavour precision era and in contrast
to the two 2010 papers with Gino, it was more general.

In fact in many extensions of the SM FCNC processes can be mediated by tree-level heavy neutral scalars and/or  pseudo-scalars $H^0(A^0)$. This
generally introduces new sources of flavour violation and CP violation
as well as left-handed (LH) and  right-handed (RH) {\it scalar}
($1\mp\gamma_5$) currents.
These NP contributions imply then 
a pattern of deviations from SM expectations for FCNC processes that depends
only on the couplings of $H^0(A^0)$ to fermions  and on their  masses. In
situations in which a single $H^0$ or $A^0$ dominates NP contributions
stringent correlations between $\Delta F=2$ and $\Delta F=1$ observables exist.
 Anticipating the Flavour Precision Era (FPE) ahead
of us we illustrated this by searching for allowed oases in the
landscape of a given model assuming significantly smaller uncertainties in CKM and hadronic
parameters than  available in 2013.

To this end we  analyzed
$\Delta F=2$ observables in
$B^0_{s,d}-\bar B^0_{s,d}$ and $K^0-\bar K^0$ systems and rare $B$ and $K$ decays with charged leptons in the final state including
both left-handed and right-handed scalar couplings of $H^0$ and $A^0$
to quarks in various
combinations.  
We identified a number of correlations
between various flavour observables that could test and distinguish these different scenarios. The prominent
role of the decays $B_{s,d}\to \mu^+\mu^-$ in these
studies was emphasized. Imposing the existing flavour constraints, a rich
pattern of deviations from the SM expectations in rare $B_{s,d}$ decays emerges provided $M_{H}\le 1\tev$. NP effects in rare $K$ decays,
except for  $K_L \to\mu^+\mu^-$, turn out 
 to be very small. In $K_L \to\mu^+\mu^-$ they can be as large as the SM contributions but due to hadronic uncertainties this was in 2013 still
insufficient to 
learn much about new scalars from this decay in the context of models considered by us. Many years later the theoretical situation improved through the
analyses in \cite{DAmbrosio:2017klp,Dery:2021mct,Dery:2022yqc,DAmbrosio:2025mxa}.

Flavour violating SM Higgs contributions to rare $B_d$ and $K$ decays turned out to be negligible once the constraints from $\Delta F=2$ processes were taken into
account. But $\mathcal{B}(B_s\to\mu^+\mu^-)$  could still be enhanced 
up to $8\%$. Finally, we pointed out striking differences between the correlations found in this paper and in  scenarios in which tree-level 
FCNC are mediated by a new neutral gauge boson $Z'$.

\boldmath
\subsection{Probing New Physics with the $B_s\to \mu^+\mu^-$ Time-Dependent Rate  (2013)}
\unboldmath

The second paper written basically at the same time dealt with testing NP
with the $B_s\to\mu^+\mu^-$ time dependent rate \cite{Buras:2013uqa}.
It was written with Jennifer, Robert Knegjens and Robert Fleicher, a world
expert on time dependent rate in this decay. 

This paper was motivated by the 2013 results of the LHCb collaboration which
reported the first evidence for $B_s\to\mu^+\mu^-$
at the $3.5\,\sigma$ level, with a 
branching ratio in the ballpark of the SM prediction. Thanks to the recently (2012) established sizable decay width difference of the $B_s$ system, another observable, ${\cal A}^{\mu\mu}_{\Delta\Gamma}$, was available, which can be extracted from the time-dependent untagged  $B_s\to\mu^+\mu^-$ rate. If tagging information is available, a CP-violating asymmetry, ${\cal S}_{\mu\mu}$, can also be determined. These two observables exhibit sensitivity to NP that is complementary 
to the branching ratio. In our paper we defined and analyzed scenarios in which these quantities allow us to  
discriminate between model-independent effective operators and their CP-violating phases.
In this context we classified a selection of popular NP models into the considered scenarios. This is Section 3 of our paper in which various scenarios for scalar
and pseudoscalar couplings have been considered and each one illustrated
with specific models like CMFV, LHT, RSc and $Z^\prime$.

Furthermore, we considered specific models with tree-level FCNCs mediated 
by a heavy neutral gauge boson, pseudoscalar or scalar,
finding striking differences in the predictions of these scenarios for the observables considered and the correlations among them.
We updated the SM prediction for the time-integrated branching ratio taking 
the subtle decay width difference effects into account. We found $(3.56\pm0.18)\times 10^{-9}$ that is fully consistent with the 2026 SM prediction for this rate.

There was no question about already in 2013 that $B_s\to \mu^+\mu^-$ and in particular its time-dependent rate will play an important role in the
search for NP and the subsequent analyses by Robert Fleischer and his collaborators confirmed these expectations. See \cite{Fleischer:2024uru,Fleischer:2024fkm} and references
therein.

\section{MSSM}\label{MSSM}
\subsection{General View}
  Despite the fact that Supersymmetry became very popular already in the early 1980s
  and I learned its basic futures already in 1982, my activities in SUSY began
  first in 1997 when Andrea Romanino and Luca Silvestrini were visiting
  my group at TUM. Eventually I have written 16 papers between 1997 and 2010 on SUSY, most of them well cited. While supersymmetric particles have not been
  discovered, it is not excluded that one day they will be and then these
  papers could turn out to be useful. Moreover, if rare decays considered by us will be measured with sufficient precision and the situation with $\epe$ within the SM will be clarified, the strategies developed in these papers could help to discover SUSY indirectly. Therefore let me briefly describe what can be found in these papers. Here we go.

  \begin{enumerate}
  \item
    As already reported above, Gerhard Buchalla and I derived a relation between the branching ratios for $\kpn$ and $\klpn$ and $\sin 2\beta$ that can be measured through the CP asymmetry $S_{\psi K_S}$ \cite{Buchalla:1994tr}. It is given in (\ref{cbb}). The main
    goal of the paper with Andrea and Luca was to investigate how this
    {\em Golden Relation} is violated beyond the SM, first model independently
    and then specifically within SUSY model  
    \cite{Buras:1997ij}.
  \item
    In 1999 our trio has been extended by Gilberto Colangelo and Gino Isidori.
    Here we concentrated on SUSY and studied correlations between $\kpn$, $\klpn$, $K_L\to\pi^0 e^+e^-$, $K_L\to\mu^+\mu^-$ and the ratio $\epe$ in the SUSY framework     \cite{Buras:1999da}. While this was the only paper I wrote with Gilberto, I wrote several with Gino and Luca.
  \item
    My next paper on SUSY deals with the ratio $\epe$ and rare $K$ and $B$ decays in the
    MSSM with minimal flavour and CP violation, including NLO QCD corrections
    and imposing constraints on supersymmetric parameters coming from
    $\varepsilon_K$, $B^0_{d,s}-\bar B^0_{d,s}$ mixings and few electroweak observables. A compendium of phenomenologically relevant formulae in the MSSM presented by us could
    be useful one day if MSSM will turn out to be popular  again
    \cite{Buras:2000qz}. In addition to Luca and Paolo Gambino,  two of my excellent PhD students, Martin Gorbahn and Sebastian Jäger, were involved in this project. Note that this is the same team which proposed CMFV \cite{Buras:2000dm}.
  \item
    Next Christoph Bobeth, Frank Kr\"uger, J\"org Urban and myself calculated
    for the first time  NLO QCD corrections to $\bar B\to X_{d,s}\nu\bar\nu$, $\bar B_{d,s}\to\mu^+\mu^-$, $\kpn$, $\klpn$ and $K_L\to \mu^+\mu^-$ in the MSSM with MFV
    \cite{Bobeth:2001jm}. In particular new operators relevant for
    large $\tan\beta$ have been considered. From the present perspective it was an overkill but in 2001 the community was strongly beliving in the existence of SUSY and expecting to discover it at the LHC even before the SM Higgs.
\item
  In 2002 Janusz Rosiek spent few months in my group and together with
Piotr Chankowski and Lucja Slawianowska we have written two papers on $\Delta M_s$, $B_s\to\mu^+\mu^-$ and $B\to X_s\gamma$ in the MSSM with large $\tan\beta$
  \cite{Buras:2002wq,Buras:2002vd}. The fact that these two papers, in particular the second one, are very well cited shows the popularity of the MSSM in the first decade of this millenium.
\item
  For this reason having Janusz in my group we have calculated $\kpn$ and $\klpn$ in general SUSY  \cite{Buras:2004qb}. Thorsten Ewerth and Sebastain J\"ager
  contributed in an important manner to this paper.
\item
  The next paper \cite{Bobeth:2004jz} was in my view an overkill. It is clearly important to
  calculate QCD NNLO  corrections to $B\to X_s \ell^+\ell^-$ in the SM
    but to calculate them for the MSSM before SUSY has been discovered is really questionable and when Christoph and Thorsten suggested it to me I
    told them that I will not be involved and hoped that they could do it fast
    because their stay in Munich ended that year.
  They claimed it would take only few months but it took over a year. They did
  all calculations and I was only discussing this project with them helping a bit   writing up their paper. If the paper was finished during their stay in my
  group I would not sign it but as they left my group several months before and
  they were payed by the DFG I had to sign it. The only time in my career I did it. When
  I stated this in one of my reviews Thorsten told me that I contributed a lot.
  Nice guy!
\item
  I have written a number of papers on various aspects of SUSY between 2007 and
  2010. In most of these papers Wolfgang Altmannshofer, Diego Guadagnoli and Paride Paradisi were involved. In a few my PhD students Michaela Albrecht, Stefania Gori, David Straub and Michael Wiik. Also postdocs Minoru Nagai and Lorenzo Calibi. 
  These are \cite{Altmannshofer:2007rj,Albrecht:2007ii,Altmannshofer:2007cs,Altmannshofer:2008hc,Altmannshofer:2009ne,Buras:2009sg,Buras:2010pm,Altmannshofer:2010ad}. Except for \cite{Altmannshofer:2009ne} none of these papers was
  cited more than 100 times and I will not describe them except this one which reached more than 250 citations and the results of this paper have been presented by us at several conferences. Let me then decribe what can be found in this paper.
\item
  The paper\cite{Altmannshofer:2009ne} written in collaboration with Wolfgan Altmannshofer, Stefania Gori, Paride Paradisi and David Straub was also the one
  which I liked most among my MSSM papers because it involved many correlations between various
  observables and the one to which I have contributed more than to other papers
  listed in the previous item.

In this paper we  performed an extensive study of FCNC and CP Violation within SUSY theories with
particular emphasis put on processes governed by $b\to s$ transitions and of their correlations 
with processes governed by $b\to d$ transitions, $s\to d$ transitions, $D^0-\bar D^0$ oscillations,
lepton flavour violating decays, electric dipole moments and $(g-2)_{\mu}$.
We first performed a comprehensive model-independent analysis of $\Delta F=2$ observables and we 
emphasized the usefulness of the $\vub/\vcb-\gamma$ plane in exhibiting transparently various tensions present in the UT analyses of 2009.

Secondly, we considered a number of SUSY models: the general MSSM, a flavour blind MSSM, the MSSM
with Minimal Flavour Violation as well as SUSY flavour models based on abelian and non-abelian 
flavour symmetries that show representative flavour structures in the soft SUSY breaking terms.
We showed how the characteristic patterns of correlations among the considered flavour observables
allow to distinguish between these different SUSY scenarios.
Of particular importance are the
correlations between the CP asymmetry $S_{\psi\phi}$ and $B_s\rightarrow\mu^+\mu^-$, between
the 2009 anomalies in $S_{\phi K_S}$ and $S_{\psi\phi}$, between $S_{\phi K_S}$ and $d_e$, between
$S_{\psi\phi}$ and $(g-2)_{\mu}$ and also those involving lepton flavour violating decays.

In our analysis, the presence of right-handed currents and of the double Higgs penguin 
contributions to $B_s$ mixing played a very important role.
We proposed a ``DNA-Flavour Test'' of NP models including Supersymmetry, the LHT model
and the Randall-Sundrum model with custodial protection, with the aim of showing a
tool to distinguish between these NP scenarios, once additional data on flavour changing processes
become available. Such DNA Tables are discussed below.

As a byproduct, we presented the SM prediction for
$\mathcal{B}(B^+\to\tau^+\nu)=(0.80\pm 0.12)\times 10^{-4}$ that follows solely from an analytical formula for this branching ratio in terms of $\Delta M _{s,d}$ and $S_{\psi K_S}$ asymmetry and which does not involve $V_{ub}$ and $F_B$ uncertainties. This formula reads
\begin{eqnarray} \label{eq:Btaunu_DMd}
\mathcal{B}(B^+ \to \tau^+ \nu)_{\rm SM} &=& \frac{3 \pi}{4
    \, \eta_B \, S_0(x_t) \, \hat B_{B_d}} \frac{m_\tau^2}{M_W^2} \left(1 -
  \frac{m_\tau^2}{m_{B^+}^2} \right)^2 \left\vert \frac{V_{ub}}{V_{td}}
  \right\vert^2 \tau_{B^+}~\Delta M_d ~,
\end{eqnarray}
with 
$\left\vert V_{ub}/V_{td}\right\vert^2$ found using 
\begin{equation}
\left\vert \frac{V_{ub}}{V_{td}}\right\vert^2 =
\left( \frac{1}{1-\lambda^2/2} \right)^2
~\frac{1+R_t^2-2 R_t\cos\beta}{R^2_t} ~.
\end{equation}
Here $R_t$ and the angle $\beta$ (see Fig.~\ref{UUTa})
can be obtained from $\Delta M_d/\Delta M_s$ and $S_{\psi K_S}$, respectively. $\eta_B$ is QCD correction in $\Delta M_{d,s}$.

While some of the tensions considered in our paper disappeared by now, these correlations, in particular the ones that distinguish between SUSY, LHT model and RS-model with custodial protection could indeed be very useful in the present decade when hopefully the experimental data improve. 
\end{enumerate}

\subsection{DNA-Flavour Test of New Physics Models (2009,2010)}\label{sec:dna}

We have seen in the previous sections that the patterns of flavour
violation found in various extensions of the SM differed from model to model, thereby allowing
in the future to find out which of the models considered by us, if any, can survive the future
measurements. Undoubtedly, the correlations between various observables that are often
characteristic for a given model will be of the utmost importance in these tests.

In Tab.~\ref{tab:DNASUSY} from \cite{Altmannshofer:2009ne} and Tab~\ref{tab:DNA2} from \cite{Buras:2010wr}
 I show a summary of the potential size of deviations from the SM results
 allowed for a large number of observables considered in several models
when all existing constraints as known in 2009 (2010)
from other observables not listed were taken into account. We distinguished among:
\begin{itemize}
\item large effects (three {\it red} stars),
\item moderate but still visible effects (two {\it blue} stars),
\item vanishingly small effects (one {\it black} star).
\end{itemize}
%

\newcommand{\three}{{\color{red}$\bigstar\bigstar\bigstar$}}
\newcommand{\two}{{\color{blue}$\bigstar\bigstar$}}
\newcommand{\one}{{\color{black}$\bigstar$}}

\begin{table}[t]
\addtolength{\arraycolsep}{4pt}
\renewcommand{\arraystretch}{1.5}
\centering
\begin{tabular}{|l|c|c|c|c|c|c|}
\hline
&  AC & RVV2 & AKM  & $\delta$LL & FBMSSM & $\text{SSU(5)}_{\rm RN}$
\\
\hline\hline
$D^0-\bar D^0$& \three & \one & \one & \one & \one & \one
\\ \hline
$\epsilon_K$& \one & \three & \three & \one & \one & \three
\\ \hline
$ S_{\psi\phi}$ & \three & \three & \three & \one & \one & \three
\\ \hline\hline
$S_{\phi K_S}$ & \three & \two & \one & \three & \three & \two
\\ \hline
$A_{\rm CP}\left(B\rightarrow X_s\gamma\right)$ & \one & \one & \one & \three & \three & \one
\\ \hline
$A_{7,8}(K^*\mu^+\mu^-)$ & \one & \one & \one & \three & \three & \one
\\ \hline
$B_s\rightarrow\mu^+\mu^-$ & \three & \three & \three & \three & \three & \three
\\ \hline
$K^+\rightarrow\pi^+\nu\bar\nu$ & \one & \one & \one & \one & \one & \one
\\ \hline
$K_L\rightarrow\pi^0\nu\bar\nu$ & \one & \one & \one & \one & \one & \one
\\ \hline
$\mu\rightarrow e\gamma$& \three & \three & \three & \three & \three & \three
\\ \hline
$\tau\rightarrow \mu\gamma$ & \three & \three & \one & \three & \three & \three 
\\ \hline
$\mu + N\rightarrow e + N$& \three & \three & \three & \three & \three & \three
\\ \hline\hline
$d_n$& \three & \three & \three & \two & \three & \three
\\ \hline
$d_e$& \three & \three & \two & \one & \three & \three
\\ \hline
$\left(g-2\right)_\mu$& \three & \three & \two & \three & \three & \three
\\ \hline
\end{tabular}
\renewcommand{\arraystretch}{1}
\caption{\small
``DNA'' of flavour physics effects for the most interesting observables in a selection of SUSY
models. \three\ signals large NP effects, \two\ moderate to small NP effects and \one\
implies that the given model does not predict visible NP effects in that observable. From \cite{Altmannshofer:2009ne} and \cite{Buras:2010pm}.}
\label{tab:DNASUSY}
\end{table}

These tables can be considered as the collection of the DNAs for various models.
These DNAs will be modified as new experimental data will be availabe and in certain
cases we will be able to declare certain models to be disfavoured or even ruled out.

In constructing these tables  we did not take into account possible correlations among
the observables listed there. We have seen that in some models, it is not possible to
obtain large effects simultaneously for certain pairs or sets of observables and
consequently future measurements of a few observables considered in these tables 
will have an impact on the patterns shown in these DNA tables. In this context
the DNA-Charts proposed with Jennifer in 2013 \cite{Buras:2013ooa} and discussed in Section~\ref{DNAS} are superior to these DNA tables.

Yet, I wanted to show them not only for historical reasons.
It will be interesting to
monitor the changes in these tables when the future experiments will provide new results and then it
will be fun to update these tables after so many years.

\begin{table}[t]
\addtolength{\arraycolsep}{4pt}
\renewcommand{\arraystretch}{1.5}
\centering
\begin{tabular}{|l|c|c|c|c|c|}
\hline
& LHT & RSc & 4G & 2HDM  & RHMFV
\\
\hline\hline
$D^0-\bar D^0$ (CPV)&  \three & \three & \two & \two & 
\\ \hline
$\epsilon_K$& \two & \three & \two &\two & \two
\\ \hline
$ S_{\psi\phi}$ & \three & \three & \three & \three  &  \three
\\ \hline\hline
$S_{\phi K_S}$ &  \one & \one  & \two & &
\\ \hline
$A_{\rm CP}\left(B\rightarrow X_s\gamma\right)$ & \one &  & \one & &
\\ \hline
$A_{7,8}(K^*\mu^+\mu^-)$ & \two &\one  & \two & &
\\ \hline
$B_s\rightarrow\mu^+\mu^-$ & \one & \one & \three & \three & \two
\\ \hline
$K^+\rightarrow\pi^+\nu\bar\nu$ & \three & \three & \three & & \two
\\ \hline
$K_L\rightarrow\pi^0\nu\bar\nu$ &  \three & \three & \three & & \two
\\ \hline
$\mu\rightarrow e\gamma$& \three & \three & \three & &
\\ \hline
$\tau\rightarrow \mu\gamma$ & \three & \three & \three & &
\\ \hline
$\mu + N\rightarrow e + N$& \three & \three & \three & &
\\ \hline\hline
$d_n$& \one & \three & \one & \three &
\\ \hline
$d_e$& \one & \three & \one & \three&
\\ \hline
$\left(g-2\right)_\mu$& \one & \two & \one & &
\\ \hline

\end{tabular}
\renewcommand{\arraystretch}{1}
\caption{\small
``DNA'' of flavour physics effects for the most interesting observables in a selection of non-SUSY models. \three\ signals large NP effects, \two\ moderate to small NP effects and \one\
  implies that the given model does not predict visible NP effects in that observable. Empty spaces reflect my present ignorance about the given entry. From
\cite{Buras:2010pm}}
\label{tab:DNA2}
\end{table}

\section{Right-Handed Currents}
Possibly motivated by our analyses in the RS-scenario, in which right-handed
currents play a role, I thought we could have a closer look at right-handed
currents more generally. The two papers resulting from this work turned out
to be the basis of the PhD thesis of Katrin Gemmler.

\subsection{ Quark Flavour Mixing with Right-handed Currents:\\ an Effective Theory Approach (2010)}
In 2010 we were lucky to have Gino Isidori as a Hans Fischer Fellow at IAS.
When I mentioned to Gino my plans to look at right-handed currents he suggested
to discuss right-handed currents in FCNC processes in the frameork of an
effective theory approach \cite{Buras:2010pz}. To this end 
we analyzed 
the structure of dimension-six operators assuming a left-right symmetric flavour group, commuting with an underlying $\text{SU}(2)_L\times\text{SU}(2)_R\times\text{U(1)}_{\rm B-L}$ global symmetry, broken only by two
Yukawa couplings. This model contains a new unitary matrix $V^R$ controlling flavour-mixing in the right-handed sector.

We determined the structure of this matrix by charged-current data, 
where the tension between inclusive and exclusive determinations of 
$|V_{ub}|$ can be solved. Having determined the size and the flavour structure of right-handed currents, we investigated how they would manifest themselves in 
neutral current processes, including particle-antiparticle mixing,
$Z\to b\bar b$, $B_{s,d}\to\mu^+\mu^-$, $B\to (X_s,K,K^*)\nu\bar\nu$
  and $K\to\pi\nu\bar\nu$ decays. I think it is a useful paper and once the data
  for these processes improves one could have another look at this model.

  \boldmath
  \subsection{$\Delta F=2$ Observables and $B\to X_q\gamma$ Decays in the Left-Right Model: Higgs Particles Striking Back (2011)}
  \unboldmath

  In the second paper again with Katrin but this time with Monika Blanke and
  another PhD student of mine, Tillmann Heidsieck, we extended the previous
  analysis (\cite{Buras:2010pz}) to various $\Delta S=2$ and $\Delta B=2$ observables \cite{Blanke:2011ry}. This included 
$\varepsilon_K$, $\Delta M_K$,  $\Delta M_q$, $A_{\rm SL}^q$, $\Delta\Gamma_q$ with $q=d,s$
and the mixing induced CP asymmetries $S_{\psi K_S}$ and  $S_{\psi \phi}$.
Compared to the SM these observables 
are affected by tree level
contributions from heavy neutral Higgs particles ($H^0$) as well as new box diagrams with $W_R$ gauge boson and charged Higgs 
$(H^\pm)$  exchanges. We also analyzed the 
$B\to X_{s,d}\gamma$ decays that receive important new contributions from the 
$W_L-W_R$ mixing and $H^\pm$  exchanges.

Compared to the literature existing in 2011, the novel feature of our analysis
was the search for correlations between various observables that could help us to distinguish this model from other extensions of the SM and to obtain an insight into the structure of 
the right-handed mixing matrix $V^R$, that governs right-handed currents.
An important part of this analysis was the study of neutral
and charged heavy scalar particles with $H^0$ playing an important role
in $\Delta F=2$ processes through its contribution already at tree-level. On the other hand  $H^\pm$ plays an important role in $B\to X_q \gamma$ decay.
Finally we presented a rather complete list of Feynman rules involving quarks, gauge bosons and Higgs particles.

I hope that one day right-handed currents will show up and then it will be fun
to update these two analyses.

As this is the last appearance of Katrin in this write-up I want to stress
that she was one of the nicest PhD students I had and the only one I could
discuss classical music with. That  this was important for me will be clear from the
last entry of my autobiography. Unfortunately after a postdoc position
at Fermilab where she wrote a number of interesting papers \cite{Agrawal:2014aoa,Bauer:2015fxa,Bauer:2015kzy}, the first one with her husband Prateek Agrawal and Monika Blanke, she left research. Fortunately, she is one of the few of my
PhD students who left physics,  with whom I still have some contact. In particular I meet from
time to time Prateek, an excellent physicist, so that he can report to me how Katrin
and their daughter Anjali are doing.

\section{Gauge Flavour Models}

\boldmath
\subsection{The Impact of Flavour Changing Neutral Gauge Bosons on $B\to X_s\gamma$ (2011)}
\unboldmath
The branching ratio of $B\to X_s\gamma$ decay provides
potentially strong constraints on models beyond the SM because it is already
precisely measured. In this project
I have been joined by Luca Merlo, a very good postdoc in my group and Emmanuel
Stamou, my excellent PhD student, in fact the last one I had \cite{Buras:2011zb}.

Considering a general scenario with new heavy neutral gauge bosons, present in particular
in $Z'$ and gauge flavour models, that were popular at that time, we pointed out two new contributions to
the ${B}\to X_s\gamma$ decay. The first one originates from one-loop diagrams mediated by gauge bosons and heavy exotic quarks with electric charge $-1/3$. The second contribution stems from the QCD mixing of neutral current-current operators generated by heavy neutral gauge bosons and the dipole operators responsible for the $B\to X_s\gamma$ decay. The latter mixing was calculated in our paper for the 
first time.
We discussed general sum rules which have to be satisfied in any
model of this type. We emphasised that the 
neutral gauge bosons in question could also significantly affect other fermion radiative decays as well as non-leptonic two-body $B$ decays, $\epe$, anomalous $(g-2)_\mu$ and electric dipole moments.

\boldmath
\subsection{Phenomenology of a Gauged $\text{SU(3)}^3$ Flavour Model (2011)}
\unboldmath
In the second paper also Valentina Carlucci took part \cite{Buras:2011wi}.
We presented an extensive analysis of $\Delta F=2$ observables 
and of $B\to X_s\gamma$ in the framework of a specific 
Maximally Gauged Flavour (MGF) model of 
Grinstein, Redi and Villadoro \cite{Grinstein:2010ve} including all relevant contributions, in particular tree-level heavy gauge boson exchanges whose effects we studied in detail in this paper for the first time.

The model allows in principle for significant deviations from the SM 
predictions for $\varepsilon_K$, $\Delta M_{B_{d,s}}$, mixing induced $CP$-asymmetries $S_{\psi K_S}$
and $S_{\psi \phi}$  and $B\to X_s\gamma$ decay. 
The tension between $\varepsilon_K$ and $S_{\psi K_S}$ present in the SM in 2011 could be removed by enhancing $|\varepsilon_K|$ without modifying $S_{\psi K_S}$. In this case, we found that in this model i) the results 
for $S_{\psi \phi}$  and $B\to X_s\gamma$ turned out to be 
SM-like, ii) the {\it exclusive determination} of $\vub$ is favoured and most 
importantly iii) the values of $\Delta M_{B_d}$ and $\Delta M_{B_s}$ being strongly
correlated in this model with $\varepsilon_K$ turn  out to be much larger than the data for the central
values of input parameters: $\Delta M_{B_d}\approx0.75/ps$ and $\Delta M_{B_s}\approx27/ps$.
Therefore, from the present perspective, the model suffers from a serious 
$\varepsilon_K$--$\Delta M_{B_{d,s}}$ tension. However, this tension can be softened considering theoretical and parametric uncertainties and in particular the decrease of the weak decay constants. On the other side, the model can be strongly constrained considering the theoretically cleaner ratios $\Delta M_{B_d}/\Delta M_{B_s}$ and $\mathcal{B}(B^+\to\tau^+\nu)/\Delta M_{B_d}$ and we found that it is unable to remove simultaneously all the SM tensions on the data. Finally, we compared the pattern of flavour violation in MGF with selected extensions of the SM.

This was the picture in 2011. In the meantime both the data and the hadronic matrix elements relevant for $\Delta M_{s,d}$ changed. Therefore in order to assess the validity of this model 15
years later one would have to update this analysis. But possibly it is better
to wait a couple of years to see how the experimental data evolve.

\section{Minimal Theory of Fermion Masses}
\subsection{Preface}
It is evident from my list of publications that I am not a real model builder.
In the case of flavour physics there are only two papers \cite{Buras:2011ph,Buras:2013td} that are related to model building. More important are
three papers on the so-called {\em Petite Unification} \cite{Hung:1981pd,Buras:2003du,Buras:2004ij} but they almost do not deal with flavour and I will
not discuss them here. I will describe them in App.~\ref{Petite}. The sad thing is that both Bjorken and Hung passed
away in 2024 so that I am the only survivor of this idea.

Let me therfore
describe here only the two papers in which flavour played a role.

\subsection{FCNC Effects in a Minimal Theory of Fermion Masses (2011)}			
The first one is with Christophe Grojean, Stefan Pokorski and Robert Ziegler.
Robert was a postdoc in my group at IAS at that time and Stefan Hans Fischer
Fellow at IAS. The idea was not mine but I was asked to join them because of my expertise in FCNCs. I worked
dominantly with Robert and had several discussions with Stefan but no
discussions
with Christophe.

The main goal of \cite{Buras:2011ph} was to construct  a minimal theory of fermion masses. To this end  we extended the SM by heavy vectorlike fermions, with flavour-anarchical Yukawa couplings, that mix with chiral fermions such that small SM Yukawa couplings arise from small mixing angles. This model can be regarded as an effective description of the fermionic sector of a large class of existing flavour models and thus might serve as a useful reference frame for a further understanding of flavour hierarchies in the SM. Already such a minimal framework gives rise to FCNC effects through exchange of massive SM bosons whose couplings to the light fermions get modified by the mixing. We derived general formulae for these corrections and discussed the bounds on the heavy fermion masses. Particularly stringent bounds, in a few TeV range, come from the corrections to the $Z^0$ couplings. 

\boldmath
\subsection{Particle-Antiparticle Mixing, CP Violation and Rare $K$ and $B$ Decays in a Minimal Theory of Fermion Masses (2012)}
\unboldmath
The real phenomenology in \cite{Buras:2011ph} was rather limited and in 2012
I discussed with Robert whether we could perform an extensive phenomenological
analysis of our ``Minimal Theory of Fermion Masses" (MTFM). Jennifer joined us in these efforts.

In \cite{Buras:2013td}
we presented a detailed study of $\Delta F=2$ observables and of
rare $K^+(K_L)$ 
and $B_{s,d}$ meson decays in the MTFM. 
As stated above, in this model
Yukawa couplings are generated through the mixing with heavy vectorlike (VF)
fermions. This implies corrections to the SM quark 
couplings to $W^\pm$, 
$Z^0$ and Higgs so that FCNC processes receive contributions from tree level 
$Z^0$ and Higgs exchanges and  $W^\pm$ bosons couple to right-handed quarks.
In a particular version of this model in which  the Yukawa matrix $\lambda^D$
in the heavy {\it down} fermion sector is {\it unitary}, 
$\lambda^U = \mathds{1}$ and  $M = M_{\rm VF}$ 
is fixed,
only three real and positive definite parameters describe NP
contributions to all $\Delta F=2$ and $\Delta F=1$ observables  in 
$K$ and $B_{s,d}$ systems  once the known quark masses and 
the CKM matrix are correctly reproduced.  For $M\ge 1\tev$ NP 
 contributions to $B_{s,d}^0-\bar B_{s,d}^0$ mixings are found to be very small.

 While in principle NP contributions to $\varepsilon_K$ and $\Delta F=1$ processes 
could be large, the correlation between $\varepsilon_K$ and $K_L\to\mu^+\mu^-$
eliminates basically NP contributions to $\varepsilon_K$  and  right-handed 
current contributions to $\Delta F=1$ FCNC observables.
We found CMFV structure in $B_{s,d}$ decays with 
$\mathcal{B}(B_{s,d}\to\mu^+\mu^-)$ {\em uniquely enhanced} for  $M=3~\tev$ 
by at least $35\%$ over 
their SM values. Also $\mathcal{B}(\kpn)$ and $\mathcal{B}(\klpn)$ are 
uniquely enhanced 
by similar amount but the correlation between them differs from the CMFV one.
We emphasized various correlations between $K$ and $B_{s,d}$ decays that could 
test this scenario. The model favours $\gamma\approx 68^\circ$, $\vub\approx 0.0037$, $S_{\psi K_S}\approx 0.72$, 
$S_{\psi\phi}\approx0.04$ and $ 4.2 \times 10^{-9}\le\mathcal{B}(B_s\to\mu^+\mu^-)\le 5.0 \times 10^{-9}$ for $M=3-4\tev$.

I described these results in some detail because several of the implications
of our model are correct. But the enhancement of $\mathcal{B}(B_s\to\mu^+\mu^-)$
is ruled out by now as some slight suppression of this branching ratio by NP is required to fit the data. While this was fun to do this analysis, our MTFM does not
seem to be chosen by nature unless some modifications are made. But I expect
that this could be the case of many models for fermion masses proposed
in last decades if the authors of these models performed  extensive
FCNC analyses as ours. Just to reproduce the mass spectrum of quark and lepton
masses is insufficient.

  \boldmath
  \section{$Z^\prime$ Models}\label{ZprimeModels}
  \unboldmath

  Among the simplified scenarios of NP that I have analyzed, many contained a heavy  $Z^\prime$ gauge boson. Several of these analyses have been done either with Jennifer Girrbach-Noe or Fulvia De Fazio and also some with both of them. I enjoyed these collaborations very much. Let me describe all of them
  in what follows leaving out two papers on NLO QCD corrections with Jennifer 
  \cite{Buras:2012fs,Buras:2012gm} that are described in Section~\ref{BG12}.

  \boldmath
  \subsection{The Anatomy of $Z^\prime$ and Z with Flavour Changing Neutral Currents in the Flavour Precision Era (2012)}
  \unboldmath

The simplest extension of the SM that generally introduces new sources
of flavour violation and CP violation as well as right-handed (RH) currents is the addition
of a $\text{U(1)}$ gauge symmetry to the SM gauge group. If the corresponding heavy gauge boson $Z^\prime$
 mediates FCNC processes in the quark sector at tree-level, these NP
contributions imply a pattern of deviations from SM expectations for FCNC processes that
depends only on the couplings of $Z^\prime$ to fermions and on its mass. This implies stringent
correlations between $\Delta F=2$  and $\Delta F = 1$ observables which govern the landscape of the
allowed parameter space for $Z^\prime$-models. Anticipating the Flavour Precision Era (FPE)  ahead
of us we illustrated this in \cite{Buras:2012jb}
by searching for allowed oases in this landscape assuming significantly
smaller uncertainties in CKM and hadronic parameters than available in 2012.

To this end we analyzed $\Delta F=2$ observables in $K^0-\bar K^0$ and
$B^0_{s,d}-\bar B^0_{s,d}$ systems and rare K and B decays
including both left-handed and right-handed $Z^\prime$ couplings to quarks in various combinations.
We identified a number of correlations between various flavour observables that could test and
distinguish these different $Z^\prime$ scenarios. We emphasized the important role of $b\to s\ell^+\ell^-$ and $b\to s\nu\bar\nu$ 
transitions in these studies that has been confirmed ten years later. Imposing the flavour constraints of 2012, a rich
pattern of deviations from the SM expectations in $B_{s,d}$ and K meson systems emerges provided $M_{Z^\prime}\le 3 \tev$.

While for $M_{Z^\prime}\ge 5 \tev$ $Z^\prime$ 
effects in rare $B_{s,d}$ decays were found typically below
$10\%$ and hard to measure even in the FPE, $\kpn$, $\klpn$ and $K_L\to \pi^0 \ell^+\ell^-$  decays provided an 
important portal to scales beyond those explored by the LHC. We applied our formalism to NP
scenarios with induced flavour changing neutral $Z$-couplings to quarks. We found that in the
case of $B_d$ and K decays such Z-couplings still allowed for sizable departures from the SM. On
the other hand in the $B_s$ system, constraints on $b\to s\ell^+\ell^-$
transitions basically eliminated NP
effects from such couplings.

Our paper contains a number of useful formulae that we used in subsequent
$Z^\prime$ papers to be described below. Once the data on the observables
considered in our paper will further improve, it would be interesting to repeat
this analysis.

In fact I will return to an analysis of 2024 with Peter Stangl\cite{Buras:2024mnq} in Section~\ref{AJBPS} that discusses only B decays but still shows
what is the status of such models now.

\boldmath
  \subsection{Left-handed $Z^\prime$ and Z Quark Couplings Facing New $b\to s\mu^+\mu^-$ Data (2013)}
  \unboldmath
  In view of the improved data on $B_{s,d}\to\mu^+\mu^-$ and
  $B_d\to K^*\mu^+\mu^-$ presented at Summer conferences 2013 Jennifer and I revisited in \cite{Buras:2013qja}
  two simple NP scenarios among those analyzed by us with Fulvia in 2012
  and described above \cite{Buras:2012jb}. The ones 
in which new FCNC
currents in $b\to s \mu^+\mu^-$ are mediated either entirely by a neutral heavy
$Z^\prime$ gauge boson with purely left-handed complex couplings $\Delta_L^{qb}(Z^\prime)$  and real couplings to muons $\Delta_A^{\mu\bar\mu}(Z^\prime)$ and
$\Delta_V^{\mu\bar\mu}(Z^\prime)$ or the SM Z boson with left-handed
complex couplings $\Delta_L^{qb}(Z)$ . We demonstrated how the 
{\it reduced couplings}, the couplings in question divided by $M_{Z^\prime}$ or 
$M_Z$,  can be determined by future  $\Delta F=2$ and $b\to s\mu^+\mu^-$ 
observables up to sign ambiguities.

We calculated the violation of the CMFV relation between ${\mathcal{B}}(B_{s,d}\to\mu^+\mu^-)$ and $\Delta M_{s,d}$  in these scenarios. Correlations between $b\to s\mu^+\mu^-$ and $b\to s\nu\bar\nu$ observables in these scenarios were emphasized. 
We also discussed briefly scenarios in which the $Z^\prime$ boson
has right-handed FCNC couplings.   In this context we  pointed out 
a number of correlations between angular observables measured in $B_d\to K^*\mu^+\mu^-$ that arise in the absence of new CP-violating phases in scenarios with only left-handed or right-handed couplings or scenarios in which 
left-handed and right-handed couplings are equal to each other or differ by sign.

Evidently, this analysis has been motivated not only by the new data from CMS
and LHCb but also by our DNA review \cite{Buras:2013ooa} that appeared in June
2013 which I described it in Section~~\ref{DNAS}. Several correlations
described in both papers should be very useful for phenomenology
in this decade when the data from CDF, ATLAS, LHCb and Belle II improves.
I do not describe our numerical results of 2013 because the data changed since then 
in a significant manner. While in 2013 the CMS data for $B_{s}\to\mu^+\mu^-$ where roughly
by a factor of three higher than the ones of the SM, the present values
from CDF, ATLAS, LHCb are presently by $1\sigma$ below the SM prediction.
Also various non-parturbative parameters, like the ones related to $\Delta M_{s,d}$ changed in the last 13 years.

\boldmath
\subsection{$\Delta I=1/2$ Rule, $\epe$ and $K\to\pi\nu\bar\nu$ in $Z^\prime(Z)$ and $G^\prime$ Models with FCNC Quark Couplings (2014)}
  \unboldmath
The experimental value for the isospin amplitude ${\rm Re}A_2$ in  $K\to\pi\pi$ decays 
has been by 2014 successfully explained within the SM  both within 
DQCD \cite{Bardeen:1986vz,Buras:2014maa} and 
by RBC-UKQCD Lattice collaboration \cite{Blum:2015ywa}.
On the other hand within DQCD
the value of ${\rm Re}A_0$ turned out to be by roughly  $30\%$ below the data.
In 2014 the value of ${\rm Re}A_0$ from RBC-UKQCD collaboration was very poorly
known with the central value by a factor of two below the data and the error
approaching $100\%$ compared to the $15\%$ error from DQCD.
In both cases the current-current operators dominated the dynamics of this rule.
See Section~\ref{DQCD} for details.

While this deficit found in 2014 could be the result of theoretical uncertainties in the DQCD
approach and as we expected could be removed by future precise QCD lattice calculations, it could not be excluded that the missing piece in  ${\rm Re}A_0$
comes from NP. In fact, even in 2026, the
result for ${\rm Re}A_0$, obtained in 2020 by the RBC-UKQCD collaboration
\cite{Abbott:2020hxn}, although consistent with experiment, has an error of about $25\%$. Therefore, even
now NP could be at work here.

Already in 2014, Fulvia, Jennifer and I
demonstrated in \cite{Buras:2014sba} that the deficit in  ${\rm Re}A_0$
as seen in 2014, can be significantly softened by 
tree-level FCNC transitions  mediated by a heavy colourless $Z^\prime$ gauge boson with flavour violating {\it left-handed} coupling  $\Delta^{sd}_L(Z^\prime)$ and approximately 
universal flavour diagonal {\it right-handed} coupling  $\Delta^{qq}_R(Z^\prime)$ to quarks. The approximate flavour universality of the latter coupling assures negligible NP contributions to  ${\rm Re}A_2$.
 This property together with the breakdown of GIM mechanisms at
tree-level allows to enhance significantly the contribution of the leading 
QCD penguin operator $Q_6$ to ${\rm Re}A_0$.

A large fraction of the missing piece in the $\Delta I=1/2$ rule can be
explained in this manner for $M_{Z^\prime}$ in the reach of the LHC, while satisfying constraints from $\varepsilon_K$, $\epe$, $\Delta M_K$, LEP-II and the LHC. 
The presence of a small right-handed flavour violating coupling $\Delta^{sd}_R(Z^\prime)\ll\Delta^{sd}_L(Z^\prime)$ and of enhanced matrix elements 
of $\Delta S=2$ left-right operators allows to satisfy simultaneously the constraints from   ${\rm Re}A_0$ and $\Delta M_K$, although this requires some 
fine-tuning. We identified {\it quartic} correlation 
between  $Z^\prime$ contributions to ${\rm Re}A_0$, $\epe$, $\varepsilon_K$ and 
$\Delta M_K$. See Fig.~4 in our paper.

The tests of this proposal will require 
much improved evaluations of  ${\rm Re}A_0$ and $\Delta M_K$ within the SM, of the hadronic matrix element $\langle Q_6 \rangle_0$ as well as precise tree level determinations of $\vub$ and $\vcb$. We presented correlations between $\epe$, $\kpn$ and $\klpn$ 
with and without the $\Delta I=1/2$ rule constraint and generalized the 
whole analysis to $Z^\prime$ with colour ($G^\prime$) and $Z$ with FCNC couplings. In the 
latter case no improvement on  ${\rm Re}A_0$  can be achieved without 
destroying the agreement of the SM with the data on ${\rm Re}A_2$. Moreover, this scenario is 
very tightly constrained by $\epe$.  On the other hand in the context of the 
$\Delta I=1/2$ rule  $G^\prime$ is even more effective than 
$Z^\prime$: it provides the missing piece in  ${\rm Re}A_0$, as seen from the point of view of DQCD in 2014, for $M_{G^\prime}=(3.5-4.0)\tev$.

While in 2026 the status of the $\Delta I=1/2$ changed, our paper was the
first one which pointed out that the presence of both left-handed and
right-handed $Z^\prime$ and $G^\prime$ couplings is necessary to satisfy
simultaneously the constraints from ${\rm Re}A_0$ and $\Delta M_K$.

\boldmath
\subsection{Can We Reach the Zeptouniverse with Rare $K$ and $B_{s,d}$ Decays? (2014)}
\unboldmath
The idea of using both left-handed and right-handed couplings of new gauge
bosons to satisfy experimental constraints suggested in \cite{Buras:2014sba}
in connection with the $\Delta I=1/2$ rule has been extended in \cite{Buras:2014zga} to $B_{s,d}$ decays. In this paper, written in collaboration with Jennifer,
Dario Buttazzo and Robert Knegjens, we have addressed the following question.

Could NP be probed up to energy scales as high as $200\tev$, corresponding to distances as small as $\ord( 10^{-21})$m  -- {\it the Zeptouniverse} --  by means of  rare $K$ and $B_{s,d}$ decays, subject to present $\Delta F=2$ constraints and perturbativity?

Evidently, such short distance scales cannot be probed by the LHC which can probe at best distance scales as short as $5\times 10^{-20}~{\rm m}$, corresponding to energy scales at the level of a $(5-10)\tev$.
In order to reach even higher resolutions before the advent of future high-energy colliders, it is necessary to consider indirect probes of NP, a prime example being $\Delta F=2$ neutral meson mixing processes, which are sensitive to much shorter distance scales 
\cite{UTfit:2007eik,Isidori:2010kg,Charles:2013aka}.
These analyses, which dealt dominantly with $\Delta F=2$ observables, have 
already shown that in the presence of four-quark left-right operators one could be sensitive to scales as high as $10^3\tev$, or even higher scales.

However, as pointed out by us in \cite{Buras:2014zga},
$\Delta F=2$ processes alone cannot tell us much about the structure of NP  beyond the LHC scales.
To identify for instance the presence of new quark flavour-changing dynamics of a left-handed (LH) or right-handed (RH) nature, complementary results from $\Delta F=1$ rare decay processes are vital.
Focusing in particular on a heavy $Z^\prime$ gauge boson we found the following.

If restricted to purely LH or RH $Z^\prime$ couplings to quarks, we found that 
rare $K$ decays, in particular $\kpn$ and $\klpn$, allow us to probe the Zeptouniverse. 
On the other hand rare $B_s$ and $B_d$ decays, which receive stronger $\Delta F=2$ constraints, allow us to reach about $15\tev$.
Allowing for both LH and RH couplings a loosening of the $\Delta F=2$ constraints is possible, and we found that the  maximal values of $M_{Z^\prime}$ at which NP effects could be found that are consistent with perturbative couplings are approximately $2000\tev$ for $K$ decays and $160\tev$ for rare $B_{s,d}$ decays.
Let me just show the two plots in Fig.~\ref{KLKP}

\begin{figure}[!tb]
\centering%
\includegraphics[width = 0.45\textwidth]{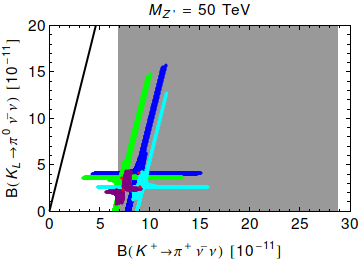}
\includegraphics[width = 0.45\textwidth]{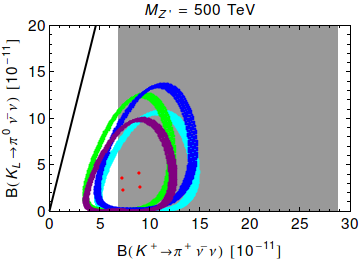}
\caption{\it $\mathcal{B}(\klpn)$ versus
$\mathcal{B}(\kpn)$ for $M_{Z^\prime} = 50~{\rm TeV}$ in the LHS (left) and for 
$M_{Z^\prime} = 500~{\rm TeV}$ in L+R scenario. The colours distinguish between different CKM input considered by us which also implies the four red points corresponding to
the SM central values for these four CKM scenarios. The black line corresponds to the Grossman-Nir bound \cite{Grossman:1997sk}. The gray region shows the
experimental range of $\mathcal{B}(\kpn))_\text{exp}=(17.3^{+11.5}_{-10.5})\times 10^{-11}$. From \cite{Buras:2014zga}.}\label{KLKP}
\end{figure}

Because $Z^\prime$ exchanges in the $B_{s,d}\to \mu^+\mu^-$ rare decays are helicity suppressed, we also considered tree-level scalar exchanges for these decays, for which we find that scales close to $1000\tev$ can be probed for the analogous pure and combined LH and RH scenarios.
We further presented a simple idea for an indirect determination of  $M_{Z^\prime}$ that could be realised at  the next linear $e^+e^-$ or $\mu^+\mu^-$ collider and with future precise flavour data.

As evident from these results NP at scales well above 1000 TeV cannot be probed by rare meson decays in contrast to  $\Delta F=2$ processes. The main reason is the absence of left-right operators involving four quarks that are 
the driving force in probing high energy scales with the help of $\Delta F=2$
processes. Rare decays are dominated 
by
operators built out of one quark current and one lepton current and the QCD
enhancement of left-right four quark operators is absent. But, as we have seen above scalar operators could provide significant enhancements.

On the other hand high scales beyond 1000 TeV can 
in principle  be reached in the future with the help of lepton flavour 
violating decays such as $\mu\to e \gamma$ and $\mu\to 3e$, $\mu\to e$ 
conversion in nuclei, and electric dipole moments as stressed already in 2013.
See refs. 6-15 in our paper.

\boldmath
\subsection{New Physics Patterns  in $\epe$ and $\varepsilon_K$ with Implications for Rare Kaon Decays and $\Delta M_K$ (2015)}
  \unboldmath
  The results for the ratio $\epe$ in the SM presented by the RBC-UKQCD collaboration   in May 2015 \cite{RBC:2015gro} turned out to 
  be significantly below the experimental data. Also $\varepsilon_K$ in the SM
  appeared to be below the data. Any NP removing 
these possible anomalies would first of all have impact on flavour observables in the $K$ 
meson system, in particular on rare decays $\kpn$, $\klpn$,
$K_L\to \mu^+\mu^-$ and $K_L\to \pi^0\ell^+\ell^-$ and $\Delta M_K$. Restricting the operators 
contributing to $\epe$ to the SM ones and to the corresponding primed operators, NP contributions to 
$\epe$ are quite generally dominated either by  QCD penguin (QCDP) operators 
$Q_6(Q_6^\prime)$ or  electroweak penguin (EWP) operators $Q_8(Q_8^\prime)$ 
with rather different implications for other flavour observables.

In October 2015 I thought it would be interesting to look at this issue
again in general models with tree-level $Z$ and $Z^\prime$ flavour violating exchanges \cite{Buras:2015jaq}. In addition to summarizing  known results from my work with Fulvia and Jennifer, I  added several new ones and also briefly
discussed few  specific models. The correlations of $\epe$ with 
flavour observables listed above allow to  differentiate between  models in which $\epe$ can be enhanced.  Various DNA-tables are helpful in this respect. I found  that simultaneous enhancements of $\epe$, $\varepsilon_K$, $\mathcal{B}(\klpn)$ and $\mathcal{B}(\kpn)$ in $Z$ scenarios are only possible in the presence of both left-handed  and right-handed flavour-violating couplings.

In $Z^\prime$ scenarios  this is not required but the size of NP effects 
 and the correlation between  $\mathcal{B}(\klpn)$ and $\mathcal{B}(\kpn)$ 
depends strongly on whether QCDP or EWP dominate NP 
contributions to $\epe$. In the QCDP case possible enhancements of both
branching ratios are much larger than for EWP scenario and take place only on the branch parallel to the Grossman-Nir bound, which is in the case of EWP dominance  only possible in the absence of NP in $\varepsilon_K$. In addition I pointed  out that 
 QCDP and EWP scenarios of NP in $\epe$ 
 can also be uniquely distinguished by the size and the sign of NP contribution to $\Delta M_K$, elevating the importance of the precise calculation of $\Delta M_K$ in the SM.

 I also emphasized the importance of the theoretical improvements
not only on $\epe$,  $\varepsilon_K$  and $\Delta M_K$   but also on   $K_L\to\mu^+\mu^-$,  $K_L\to \pi^0\ell^+\ell^-$, and the $K\to\pi\pi$ isospin amplitudes ${\rm Re} A_0$ and  ${\rm Re} A_2$ which would in the future enrich my analysis.

This paper contains many useful expressions and plots which have been
made by our son Robert. With 79 citations in ten years, this paper cannot
be considered as a big success, but certain observations made there could
still be useful when all data for rare decays improves and the status of
$\epe$ will be clarified. See the summary of the status of $\epe$  as of 2026
in Section~\ref{epe}. 

One change can already  be mentioned. The QCD penguin role in enhancing
$\epe$ seems to be ruled out as demonstrated by Jason Aebischer, Jacky
Kumar and myself in a SMEFT analysis five years later \cite{Aebischer:2020mkv}
so that either electroweak penguins
$Q_8(Q_8^\prime)$ or other operators could provide the necessary enhancement
of $\epe$ if this will turn out one day to be necessary. See Section~\ref{JJ1}
for more details.

\boldmath
\subsection{Quark-Lepton Connections in $Z^\prime$ Mediated FCNC\\ Processes: Gauge Anomaly Cancellations at Work (2019)}
\unboldmath

Until now in this section on simplified $Z^\prime$ models only models with quark flavour violation have been
considered. Let me then finish this section with two papers in which also
lepton flavour violation has been addressed.

It is well known that in the SM the cancellation of gauge anomalies
takes place because of the same number of lepton and quark generations.
Discussing one day with Fulvia, we asked ourselves whether in simple $Z^\prime$ models some correlations between quark flavour violation and lepton flavour
violation could be caused by the the cancellation of gauge anomalies.
This led to the analysis in \cite{Aebischer:2019blw} in which also Jason
Aebischer and my postdoc Maria Cerd{\'a}-Sevilla participated.

We considered then scenarios with a heavy $Z^\prime$ gauge boson with flavour
non-universal quark and lepton couplings with the goal to illustrate
how the cancellation of gauge anomalies generated by the presence of an additional
 $\text{U(1)}^\prime$  gauge symmetry would imply correlations between FCNC processes within
the quark sector, within the lepton sector and most interestingly between quark flavour
and lepton flavour violating processes.

To this end we presented  simple
scenarios with only left-handed flavour-violating $Z^\prime$ couplings and those
in which also right-handed flavour-violating couplings were present. The considered scenarios were characterized by a small number of free parameters but in contrast to gauge anomaly cancellation in the SM, in which it takes
place separately within each generation, in our scenarios anomaly cancellation
involved simultaneously quarks and leptons of all three generations.
Our models involved, beyond the ordinary quarks and leptons, three heavy right-handed neutrinos. The
models with only left-handed FCNCs of $Z^\prime$ involved beyond
$g_{Z^\prime}$ and $M_{Z^\prime}$  two real parameters characterizing the charges of all fermions under the $\text{U(1)}^\prime$ gauge symmetry and the CKM and
PMNS ones in the quark and lepton sectors, respectively. The models with the right-handed FCNCs of $Z^\prime$ involved few additional parameters.

Imposing constraints
from well measured $\Delta F=2$ observables we identified a number of interesting
correlations that involve e.g. $\epe$, $B_{s,d}\to\mu^+\mu^-$, $B\to K(K^*) \ell^+\ell^-$, $\kpn$, $\klpn$ and purely lepton flavour violating
decays like $\mu\to e\gamma$, $\mu\to 3 e$, $\tau\to 3\mu$ and $\mu-e$ conversion among others. Also $(g-2)_{\mu,e}$ were considered.  The impact of the experimental  $\mu\to e\gamma$,  $\mu\to 3 e$ and in particular $\mu-e$ conversion bounds on rare $K$ and $B$ decays has been emphasized.

I think this is an interesting paper even if I expected the correlations to be
stronger than we have found. I should also stress that although we all
contributed to this paper the true leader in this collaboration was Fulvia.

This was the only paper I wrote with Maria Cerd{\'a}-Sevilla. Unfortunately she left physics research one year later. She is a very nice person
and from time to time we contact each other.

  \boldmath
  \subsection{Global Analysis of Leptophilic $Z^\prime$ Bosons (2021)}
  \unboldmath
  
  In January 2021 I have been contacted by my physics grandson, Andi Crivellin,
  who  expressed again the wish to write a paper with me. We have discussed
  possible collaboration for years but it was not realized until 2021 when
  we completed our only
  paper by now \cite{Buras:2021btx}. Three young physicists from his group
  took place in this collaboration. The main idea came from Andi but I mainly
  collaborated with Fiona Kirk, my first physics great granddaugther  I have collaborated
  with. I had  several zoom discussions with Fiona and enjoyed them as well as
  collaboration with her very much. Claudio was another PhD student of Andi,
  my great grandson in physics and Marc his postdoc.

  The main idea was as follows.
  New neutral heavy gauge bosons ($Z^\prime$) are predicted within many extensions of the SM. While in case they couple to quarks the LHC bounds are very stringent, leptophilic $Z^\prime$ bosons (even with sizable couplings) can be much lighter and therefore can lead to interesting quantum effects in precision observables {(like $(g-2)_\mu$)} and generate flavour violating decays of charged leptons. In particular, $\ell\to\ell^\prime\nu\bar\nu$ decays, anomalous magnetic moments of charged leptons, $\ell\to\ell^\prime\gamma$ and $\ell\to3\ell^\prime$ decays place stringent limits on leptophilic $Z^\prime$ bosons. Furthermore, in case of the mixing of $Z^\prime$ with the SM $Z$, $Z$-pole observables are affected.

  In light of these many observables we performed a global fit to leptophilic $Z^\prime$ models with the main goal of finding the bounds for the $Z^\prime$ couplings to leptons. To this end we considered a number of scenarios for these couplings.  While in generic scenarios correlations turned out to be weak, this changes once additional constraints on the couplings are imposed. In particular, if one considers an $L_\mu-L_\tau$ symmetry broken only by left-handed rotations, or considers the case of $\tau-\mu$ couplings only. In the latter setup, one could explain  the $(g-2)_\mu$ anomaly and the hint for lepton flavour universality violation in $\tau\to\mu\nu\bar\nu/\tau\to e\nu\bar\nu$ without violating bounds from electroweak precision observables. As in 2025  the $(g-2)_\mu$ anomaly
  practically disappeared, it would be interesting to repeat this analysis.

Now, as evident from my first statements in this autobiography I was
not involved in the global fit my collaborators made. But Section 3.4 and  Appendix B of our paper, dealing with 
lepton flavour violation in $\ell\to 3\ell^\prime$ decays, were exclusively done by Fiona and myself.
Simply, all formulae could be derived entirely by hand.

Now, there is the following peculiarity of pure lepton flavour violation
as far as correlation between lepton flavour violating processes are
concerned, which makes the distinction between various NP models more
difficult than in the case of quark flavour violation.

As stressed in several papers on quark flavour violation with Fulvia and Jennifer, different NP models can be distinguish by correlations and anticorrelations
between different shifts in observables relative to SM expectations. This
is the basis of the DNA charts in \cite{Buras:2013ooa}.
But if only lepton flavour violating observables are considered they
can only be enhanced relative to the SM expectations that are vanishing
and at best only correlations can be present.

  \boldmath
  \section{331 Models}\label{331Models}
  \unboldmath
  \subsection{Preface}
The 331 models, based on the gauge group $\text{SU(3)}_C\times \text{SU(3)}_L\times \text{U(1)}_X$ \cite{Pisano:1992bxx,Frampton:1992wt}
are among the simplest NP scenarios with new sources of flavour and CP violation. The latter originate dominantly through the flavour
violating interactions of ordinary quarks and leptons with a new heavy $Z^\prime$ gauge boson. Due to the enlarged non-abelian gauge group the SM fermions
have to be put in the representations of $\text{SU(3)}_L$ in a special manner in order to avoid gauge anomalies.

One of the nice features of this model is that the requirement of anomaly cancellation, together with that of asymptotic freedom of QCD, constrains the number of generations to be equal to the number of colours, thus providing an explanation for the existence of 3 generations, long sought for in the SM. Requirement of anomaly cancellation has also consequences on the transformation properties of fermions. In fact, as a first consequence of the extension of the SM gauge group
$\text{SU(2)}_L$ to $\text{SU(3)}_L$,
 one has that left-handed
fermions transform as triplets (or antitriplets) under the action of $\text{SU(3)}_L$ . In order to
have an anomaly-free theory, the number of triplets should be equal to the number of antitriplets. A possible choice is to assume that the three lepton generations transform
as antitriplets, so that, taking into account the three colour possibilities for the quarks, the number of quark triplets should be equal to the number of antitriplets plus one. Hence two quark generations should transform as triplets, one as an antitriplet. The common choice of having the third generation with different transformation properties might be at the origin of the large mass of the top quark with respect to the other quarks. However, in contrast to the SM, where anomaly is canceled for each generation of fermions, in the 331 model the cancellation is fulfilled only when all the generations are considered.

There is a reach literature on these
models beyond the first papers \cite{Pisano:1992bxx,Frampton:1992wt}. The ones
that were published until 2019 are listed in my book. 
I will concentrate here on the ones I was involved in.

The first  three papers I have written on these models were
done again with Fulvia and Jennifer. Among them one also with Valentina. As Jennifer unfortunately left research
in January 2015, the remaining four papers where done only with Fulvia. In one of them also Pietro Collangelo and younger members of the Bari group
were participating.

Before continuing I should mention that in 2007 my PhD
students Christoph Promberger and Felix Schwab together with Sebastian Schacht
wrote a paper on a particular 331 model, ``$\beta=1/\sqrt{3}$'' model
(to be called $\overline{331}$ model) \cite{Promberger:2007py}. Christoph's diploma thesis on
this model was very useful for us five years later. The parameter $\beta$ in these models determines the charges of new heavy fermions and gauge bosons.

Here come short summaries what we achieved in these papers. Compact summary
of 331 models can be found in Section 16.2 of my book \cite{Buras:2020xsm}.
  \boldmath
  \subsection{The Anatomy of Quark Flavour Observables in 331 Models in the Flavour Precision Era (2012)}
  \unboldmath
In our first paper \cite{Buras:2012dp}
  after presenting first these models in some generality,
we presented a detailed study of $\Delta F=2$ observables and of rare $K$ and $B$ meson decays in the  $\overline{331}$ model assuming significantly smaller
uncertainties in CKM and hadronic parameters than  available in 2012.
This was the same strategy as in our first $Z^\prime$ paper \cite{Buras:2012jb}.

The most prominent roles in our analysis played $\varepsilon_K$, $\Delta M_q$ $(q=d,s)$, the mixing induced CP asymmetries $S_{\psi K_S}$ and $S_{\psi\phi}$,
rare decays $B_{s,d}\to \mu^+\mu^-$ and in particular the CP-asymmetry
$S_{\mu^+\mu^-}^s$ in $B_{s}\to\mu^+\mu^-$.
As the phenomenology of $Z'$
contributions in this particular model with fixed $\beta$ is governed only by $M_{Z'}$ and four new mixing parameters
$\tilde s_{13}$, $\tilde s_{23}$, $\delta_1$ and $\delta_2$,
we identified a number of correlations between various observables that
differ from those known from CMFV models.
While, the $\Delta F=2$ observables allowed in 2012 still
for four oases in the new parameter space, we demonstrated how the inclusion
of $\Delta F=1$ observables could in the future identify the optimal
oasis for this model. The numerical analysis of this model is in 2026
out of date as it uses the inclusive value of $\vub$, which from present
perspective is not favoured by experiment. But this paper was the warming up
for the subsequent more general analyses of these models to be described next.
In fact in 2013, 2014, 2015 and 2016 we have written four papers which significantly extended our analysis of 2012. They were motivated by the developement
of the experimental data, mostly from CMS, ATLAS and LHCb.

\boldmath
\subsection{331 Models Facing New $b\to s\mu^+\mu^-$ Data (2013)}
\unboldmath
In our 2013 paper \cite{Buras:2013dea}
we investigated how 
the 331 models face  new data on $B_{s,d}\to \mu^+\mu^-$ and $B_d\to K^*(K)\mu^+\mu^-$
taking into account 2013 constraints from $\Delta F=2$ observables,
low energy precision measurements, LEP-II and the LHC data.
The strength of the relevant couplings is governed by four new mixing parameters  in the quark
sector listed above and the parameter $\beta$ which as already stated above determines in these models  the charges of new heavy fermions and gauge bosons. We studied the
implications of these models  for $\beta=\pm n/\sqrt{3}$ with $n=1,2,3$.  The case $\beta=-\sqrt{3}$
leading to Landau singularities for  $M_{Z^\prime}\approx 4\tev$ could be ruled out when the  constraints on
$Z^\prime$ couplings, in particular from LEP-II, were taken into account.

For $n=1,2$ interesting results are found for  $M_{Z^\prime}< 4\tev$ with
largest NP effects for $\beta <0$ in   $B_d\to K^*\mu^+\mu^-$ and the ones
in  $B_{s,d}\to\mu^+\mu^-$ for $\beta>0$. As $\RE(C_9^{\rm NP})$  could
reach the values $-0.8$ and  $-0.4$  for $n=2$ and $n=1$, respectively, the $B_d\to K^*\mu^+\mu^-$ anomalies found by the LHCb collaboration could be
softened with
the size depending on
$\Delta M_{s}/(\Delta M_{s})_{\rm SM}$ and the CP-asymmetry $S_{\psi\phi}$.
A correlation between  $\RE(C^{\rm NP}_{9})$ and $\overline{\mathcal{B}}(B_{s}\to\mu^+\mu^-)$, identified for $\beta<0$, implies for {\it 
negative}
$\RE(C^{\rm NP}_{9})$ uniquely suppression of $\overline{\mathcal{B}}(B_{s}\to\mu^+\mu^-)$  relative to its SM value which is favoured by 
the data. In
turn also
$S_{\psi\phi}< S_{\psi\phi}^{\rm SM}$ was favoured with $S_{\psi\phi}$
having dominantly opposite sign to $S_{\psi\phi}^{\rm SM}$
and at that time closer to its central experimental value. This
is no longer the case in 2026. The $S_{\psi\phi}^{\rm SM}$ agrees fully with experiment. 

 Another triple correlation is the one between $\RE(C^{\rm NP}_9)$, $\overline{\mathcal{B}}(B_{s}\to\mu^+\mu^-)$ and  $\mathcal{B}(B_d\to K\mu^+\mu^-)$.  NP effects in $b\to s\nu\bar\nu$ transitions, $\kpn$
and $\klpn$ turn out  to be small. We
found that the absence of $B_d\to K^*\mu^+\mu^-$ anomalies in the future
data and the confirmation of the suppression of $\overline{\mathcal{B}}(B_{s}\to\mu^+\mu^-)$
relative to its SM value would favour $\beta=1/\sqrt{3}$ and $M_{Z^\prime}\approx
3\tev$. Assuming lepton universality,
 we found an upper bound  $|C^{\rm NP}_{9}|\le 1.1 (1.4)$ from
LEP-II data for {\it all}  $Z^\prime$ models with only left-handed flavour
violating couplings to quarks when NP contributions to
$\Delta M_s$ at the level of  $10\%(15\%)$  are allowed.

\boldmath
  \subsection{ $Z$-$Z^\prime$ Mixing and $Z$-Mediated FCNCs in 
 $\text{SU(3)}_C\times \text{SU(3)}_L\times \text{U(1)}_X$ Models (2014)}                \unboldmath
  Until our 2014 paper \cite{Buras:2014yna}
most of the existing analyses of FCNC processes in 
the 331 models were fully dominated by tree-level $Z^\prime$ exchanges. However, due to the $Z-Z^\prime$ mixing also 
corresponding contributions from $Z$ boson are present. As the $Z-Z^\prime$ 
mixing is estimated generally in $Z^\prime$ models to be at most $\ord(10^{-3})$,  the latter contributions are usually neglected. The paucity of relevant 
parameters in 331 models allows to check whether this neglect is really 
justified in these concrete models.

In \cite{Buras:2014yna}
we calculated the impact of these contributions on $\Delta F=2$ processes and rare $K$, $B_s$ and $B_d$ decays for 
 different values of the parameter $\beta$, which distinguishes between various 331 models and for  different fermion representations under the $\text{SU(3)}_L$ group.
 We derived  a general expression for the $Z-Z^\prime$ mixing in terms 
 $\beta$, $M_Z$, $M_{Z^\prime}$ and $\tan\bar\beta$, familiar from 2 Higgs Doublet models, that differs from the one quoted in the 
literature.
 We studied in particular the models with $\beta=\pm n/\sqrt{3}$ with $n=1,2$ 
 investigated by us in our 2013 paper \cite{Buras:2013dea} as described above.
 We found that 
these new contributions can indeed be neglected in the case of 
$\Delta F=2$ transitions and decays, like  $B_d\to K^*\mu^+\mu^-$, where they are suppressed by the small vectorial $Z$ coupling to charged 
leptons.

However, 
the contributions of tree-level $Z$ exchanges to decays sensitive to 
axial-vector couplings, like $B_{s,d}\to \mu^+\mu^-$ and $B_d\to K\mu^+\mu^-$, 
and those with neutrinos in the final state, like  $b\to s\nu\bar\nu$ transitions, $\kpn$ and $\klpn$ cannot be  generally neglected with size of 
$Z$ contributions depending on $\beta$, $\tan\bar\beta$ and $M_{Z^\prime}$. We analyzed how our 2013 results 
on FCNCs in 331 models, in particular correlations between various observables, are modified by these new contributions.  As a byproduct we 
analyzed for the first time the ratio $\epe$ in 
these models including both $Z^\prime$ and $Z$ contributions.   Our 
analysis of electroweak precision  observables within 331 models demonstrates transparently 
that the interplay of NP effects in electroweak precision observables and those in flavour observables could allow in the future to identify the favourite 
331 model.

One can find numerous plots in this paper, all made by Fulvia and Jennifer.
Examples are shown in Fig.~\ref{fig:BsmuRnuF2}.

\begin{figure}[!tb]
 \centering
\includegraphics[width = 0.45\textwidth]{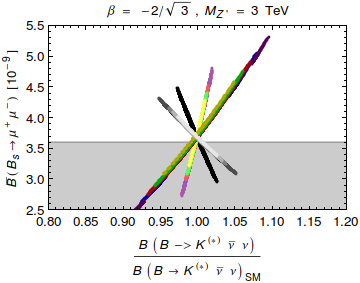}
\includegraphics[width = 0.45\textwidth]{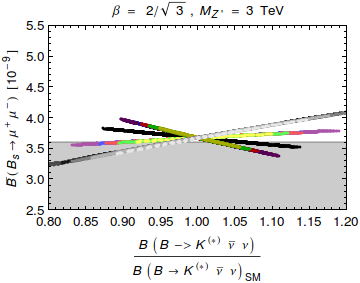}

\includegraphics[width = 0.45\textwidth]{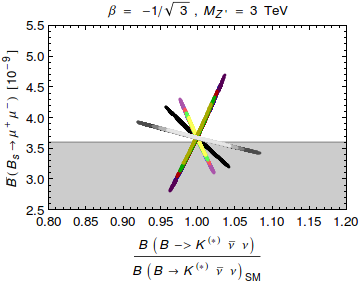}
\includegraphics[width = 0.45\textwidth]{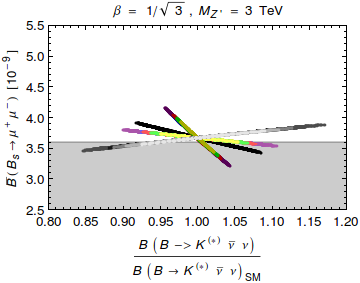}
\caption{ \it $\overline{\mathcal{B}}(B_s\to\mu^+\mu^-)$ versus the ratio ${\mathcal{B}(B\to K \nu \bar \nu)}/{ \mathcal{B}(B\to K \nu \bar 
\nu)_{\rm SM}}$  for all four $\beta=\pm\frac{2}{\sqrt{3}},
\pm
\frac{1}{\sqrt{3}}$. Colours describe different values of $\Delta M_s/(\Delta M_s)_{\rm SM}$. From \cite{Buras:2014yna}.
}\label{fig:BsmuRnuF2}~\\[-2mm]\hrule
\end{figure}

\boldmath
\subsection{$\epe$ in 331 Models (2015)}
\unboldmath
As already reported several times, in May 2015 RBC-UKQCD collaboration \cite{Abbott:2020hxn},
found  that the ratio $\epe$ in the SM  appeared to be significantly below
the data as expected earlier from DQCD. This was discussed vigorously at the
EPS-HEP conference in Vienna which both Fulvia and me attended. In the fall
we decided to investigate whether the necessary  enhancement of $\epe$
can be obtained in
331 models which resulted in our 2015 paper on these models \cite{Buras:2015kwd}.

NP contributions to $\epe$ in these models are governed by the electroweak penguin operator $Q_8$ for which the hadronic matrix element was already in 2015 rather well known both from DQCD and RBC-UKQCD collaboration so that our analysis of NP contributions
was subject to much smaller theoretical uncertainties than within the SM. In particular 
strong cancellations between different contributions do not take place.
The size of NP effects
depends not only on $M_{Z^\prime}$ but in particular on the parameter $\beta$, which distinguishes between various 331 models. Also the parameter $\tan\bar\beta$
present in the $Z-Z^\prime$ mixing plays a role.
We performed the 
analysis in {\em seven} 331 models characterized by different $\beta$, $\tan\bar\beta$  and fermion representations under the gauge group that have been selected 
in our 2014 analysis on the basis of electroweak precision data \cite{Buras:2014yna}.

Imposing the constraints from $\Delta F=2$ transitions we found that only {\em three} of these 
models can provide a significant positive shift in $\epe$ up to $6\times 10^{-4}$ for
 $M_{Z^\prime}=3\tev$. Two of them allow simultaneously a suppression of the 
rate for   $B_{s}\to \mu^+\mu^-$ by $20\%$ thereby bringing the  theory closer to the data without any significant impact 
on the Wilson coefficient $C_9$.  The third model provides simultaneous 
shift  $\Delta C_9=-0.6$, softening the anomalies in $B\to K^*\mu^+\mu^-$, without any significant impact on  $B_{s}\to \mu^+\mu^-$.
NP effects in rare $K$ decays, in particular in $\kpn$, turn out to be 
 small. This is also the case of $B\to K(K^*)\nu\bar\nu$. Both predictions 
could be challenged by NA62 experiment and Belle II in this decade.

The special flavour structure of 331 models implies that  even for  $M_{Z^\prime}=30\tev$ a shift of $\epe$ up to $8\times 10^{-4}$ and a significant shift in 
$\varepsilon_K$ can be obtained, while the effects in other flavour observables
are small. In this manner $\epe$ and $\varepsilon_K$ appear to be  unique flavour observables in these models which provide the possibility of accessing masses of $M_{Z^\prime}$ far beyond the LHC reach.

It should be stressed that although the RBC-UKQCD results for $\epe$ moved up in 2020 to be consistent with the data, the absence of isospin-breaking effects
in their calculations does not yet exclude the possibility that indeed NP
is required to bring $\epe$ to agree with data. I am repeating these sentences
in many of my papers. After all the data on $\epe$ are already more than 20 years
old and in this case experimentalists are waiting for theory as opposed
to most other observables. Another important exception is
 the $\Delta I=1/2$ rule which  still allows for significant NP contributions.

\boldmath
\subsection{ 331 Models Facing the Tensions in $\Delta F=2$ Processes with the Impact on  $\epe$, $B_s\to\mu^+\mu^-$ and $B\to K^*\mu^+\mu^-$ (2016)}                
\unboldmath
In our 2016 paper \cite{Buras:2016dxz}, motivated by the recently improved results from the Fermilab Lattice and MILC Collaborations \cite{FermilabLattice:2016ipl} on the hadronic matrix elements entering  $\Delta M_{s,d}$ in
 $B_{s,d}^0-\bar B_{s,d}^0$ mixings and the resulting increased tensions 
between $\Delta M_{s,d}$  and $\varepsilon_K$ in the SM and 
CMFV models, we demonstrated that these tensions can be removed in  331 models.

I will not describe our results because in my view the situation in 2026
is very different and it appears that there are no tensions between
the data and SM predictions for $\Delta F=2$ observables like $\Delta M_{s,d}$ 
and  $\varepsilon_K$. The main reason seems to be the following one. The Fermilab Lattice and MILC Collaborations  \cite{FermilabLattice:2016ipl} when calculating  the hadronic matrix elements entering  $\Delta M_{s,d}$  included only three quark flavours (u,d,s) although at B-meson scales
also charm should be included. With the four-quark results presented by HPQCD
collaboration \cite{Dowdall:2019bea} the tensions in the case of  $\Delta M_{s,d}$ disappear
if the inclusive value of $\vcb$ is used. It appears also that then the SM
prediction for $\varepsilon_K$ is also in good agreement with data.

Our analysis then shows that if HPQCD results will be confirmed by another
lattice collaboration that performs the calculations of the hadronic
matrix elements in question with four flavours, NP effects in 331 models will be severly constrained unless the exclusive value of $\vcb$ is used.
A similar problem appears in the 2016 paper with Monika Blanke in which
we analyzed CMFV models in view of the Fermilab and MILC results 
instead of 331 models \cite{Blanke:2016bhf}.

This example shows how important both non-perturbative calculations and the
experimental data are. It is also important that at least two LQCD collaborations and two experiments address the same calculations and measurements, respectively. Until the LQCD calculations in question are done for four flavours by a second collaboration we cannot be sure that the HPQCD values are the final ones.
Fig.~\ref{fig:10052} in Section~\ref{BVVisions} demonstrates clearly that it is an important issue.

\subsection{The Charm of 331 (2021)}
In the Spring of 2021 I have been contacted by Fulvia suggesting that I will
join her, Pietro Colangelo and her excellent PhD student Francesco Loparco  in the analysis of FCNCs in
the charm sector in the context of 331 models. As charm was never my favourite
meson system I was first not sure I would accept this invitation but eventually
I joined them for two reasons. First of all knowing Fulvia and Pietro already
for years I was sure that it will be a very nice and efficient collaboration.
The second reason was their recent paper \cite{Colangelo:2021myn} in which they pointed out that
in the case of $Z^\prime$ contributions in these models 
there are no new free parameters in the charm sector beyond those already present  in the
$B_{d,s}$ and $K$ meson systems   analyzed by Fulvia, Jennifer and myself 
in the past. As a result, definite ranges for NP effects
in various charm observables could be automatically obtained.

As we have shown in \cite{Buras:2021rdg},
while generally NP
effects turn out to be small, in a number of observables they are much larger
than the tiny effects predicted within the SM. In particular we found that the branching ratio of the mode $D^0 \to \mu^+ \mu^-$,  despite remaining tiny, can be enhanced by 6 orders of magnitude with respect to the SM value. We also worked out correlations between this mode and rare $B_{d,s}$ and $K$ decays. 
We also discussed neutral charm meson oscillations and CP violation in the charm system. In particular, we pointed out that 
  331 models provide new weak phases that are a necessary condition to have non-vanishing CP asymmetries.
In the case of  the anomalous $\Delta A_{CP}$,   the difference between the CP asymmetries in $D^0 \to K^+ K^-$ 
and $D^0 \to \pi^+ \pi^-$, we found that agreement with experiment  can be obtained provided that two conditions are verified: the phases  in the ranges predicted in 331 models and large hadronic matrix elements.

\boldmath
\subsection{331 Model Predictions for Rare $B$ and $K$ Decays and $\Delta F=2$ Processes: an Update (2023)}
\unboldmath
In January 2023 motivated by the improved results from the HPQCD lattice collaboration
  on the hadronic matrix elements entering  $\Delta M_{s,d}$ in
  $B_{s,d}^0-\bar B_{s,d}^0$ mixings and the increase of the experimental
  branching ratio for $B_s\to\mu^+\mu^-$, Fulvia and I updated
  in \cite{Buras:2023ldz}   our 2016 analysis \cite{Buras:2016dxz}
  of various flavour observables   in four  331 models, M1, M3, M13 and M16
  defined there.

    These four models, which are distinguished by the quantum numbers, have been  selected in  \cite{Buras:2016dxz}    among 24 331 models through their
  consistency with the electroweak precision tests and simultaneously
  by the relation  $C_9^\text{NP}=-b\, C_{10}^\text{NP}$ with $2\le b\le 5$.
After the  new 2022 result on  $B_s\to\mu^+\mu^-$ from CMS this relation is favoured over the
  popular relation $C_9^\text{NP}=- C_{10}^\text{NP}$  predicted by several
  leptoquark models.  In this context we investigated in particular
  the dependence of various observables on $\vcb$, varying  it in the broad range $[0.0386,\,0.043]$, that encompasses  both its inclusive and exclusive determinations.

  Imposing the experimental constraints from $\varepsilon_K$, $\Delta M_s$, $\Delta M_d$ and the mixing induced CP asymmetries $S_{\psi K_S}$ and $S_{\psi \phi}$, we investigated for which values of $\vcb$
    the four models can be made compatible with these data and what is the
    impact on $B$ and $K$ branching ratios. In particular we analyzed
    NP contributions to the Wilson coefficients $C_9$ and $C_{10}$ and
    the decays $B_{s,d}\to\mu^+\mu^-$, $\kpn$ and $\klpn$.
  This allowed us
  to illustrate how the value of $\vcb$ determined together with other
  parameters of these models is infected  by NP contributions and compare it
  with the one obtained recently under the assumption of the absence of NP
  in $\varepsilon_K$, $\Delta M_s$, $\Delta M_d$ and $ S_{\psi K_S}$
  \cite{Buras:2022wpw,Buras:2022qip}.

  In my view this paper summarizes the present status of 331 models as far as
  FCNC processes in $K$ and $B$ meson systems are concerned
  and one has to wait for new measurements of rare processes to see
  which of the four models analyzed by us, if any, survives the future tests.
  While this paper has only been signed by Fulvia and me, the contributions
  of other collaborators in the study of 331 models, in particular of Jennifer,
  should be emphasized.

\section{Vector-like Quarks}\label{VLQ}

\subsection{Patterns of Flavour Violation in Models with Vector-Like Quarks (2016)} \label{VLQ1}
In the Summer of 2016 I thought it would be good to have another  look at the
models with vector-like quarks (VLQs). In fact, as described in Section~\ref{LastRS},  in 2009 I made already a study
of VLQs in the context of the  RS model where they are represented by
KK-fermions \cite{Buras:2009ka}. 
I was joined in this new expedition by Christoph
Bobeth, Alejandro Celis, a postdoc in the group of Gerhard Buchalla at the LMU
and Martin Jung a postdoc at IAS and the Excellence Cluster Universe at TUM in
Garching.

There was already a nice classification of various VLQ models done by Ishiwata, Ligeti and Wise \cite{Ishiwata:2015cga}
but a general detailed study of the patterns of flavour violation in these
models was still missing. To this end we studied 
renormalisable extensions of the SM 
that contain VLQs in a single complex
representation of either the SM gauge group $\GSM$  or the one extended by
$\text{U(1)}_{L_\mu-L_\tau}$ that we denoted by  $\GSMUpr$ \cite{Bobeth:2016llm}. In fact we found that the patterns of flavour violation in $\GSM$-models and
$\GSMUpr$-models differ significantly from each other. In the latter case
we considered $\GSMUpr(S)$ and $\GSMUpr(\Phi)$ models with different scalar
sectors. While in $\GSM$ and  $\GSMUpr(S)$ models $\Delta F=1$ FCNCs are
mediated by $Z$ and $Z^\prime$, respectively, in $\GSMUpr(\Phi)$ models
both $Z$ and $Z^\prime$ are at work. Here $S$ is a singlet under $\text{SU(2)}_L$, while $\Phi$ is a doublet.

We first decoupled VLQs in the $M=(1 - 10)$~TeV range and then at the 
electroweak scale also $Z, Z'$ gauge bosons and additional scalars to study the
phenomenology. Our results depend on the relative size of $Z$- and $Z'$-induced
flavour-changing neutral currents, as well as the size of $|\Delta F|=2$
contributions including the effects of renormalisation group Yukawa evolution 
from $M$ to the electroweak scale that turned out to be very important for models
with right-handed currents through the generation of left-right operators.
In addition to rare decays like $P\to \ell\bar\ell$, $P\to P' \ell\bar\ell$,
$P\to P'\nu\bar\nu$ with $P=K, B_s, B_d$ and $|\Delta F|=2$ observables we
analyzed  the ratio $\epe$ which in 2016 not only in DQCD but also in LQCD
appeared in the SM to be significantly below
the data.

\begin{table}[t]
\addtolength{\arraycolsep}{4pt}
\renewcommand{\arraystretch}{1.5}
\centering
\begin{tabular}{|l|c|c|c|c|c||c|c||c|c|c|c|}
\hline
  & \multicolumn{5}{|c||}{$\GSM$}
  & \multicolumn{2}{c||}{$\GSMUpr(S)$}
  & \multicolumn{4}{c|}{$\GSMUpr(\Phi)$}
\\
& $D$    & $Q_V$  & $Q_d$  & $T_d$  & $T_u$  & $D$    & $Q_V$  & $D$     & $Q_d$   & $T_d$   & $T_u$
\\
\hline\hline
  $C_9^{ij,\ell}$
& \red   &  ---   &  ---   & \red   & \red   &  ---   & \green & \sred    &  ---    & \sred   & \sred
\\
\hline
  $C_9^{\prime\, ij,\ell}$ 
&  ---   & \red   & \red   &  ---   &  ---   & \green &  ---   &  ---    & \sred   &  ---    &  --- 
\\  
\hline
  $C_{10}^{ij,\ell}$ 
& \green &  ---   &  ---   & \green & \green &  ---   &  ---   & \sgreen &  ---    & \sgreen & \sgreen 
\\
\hline
  $C_{10}^{\prime\, ij,\ell}$ 
&  ---   & \green & \green &  ---   &  ---   &  ---   &  ---   &  ---    & \sgreen &  ---    &  ---  
\\
\hline\hline
  $C_L^{ij,\nu}$
& \green &  ---   &  ---   & \green & \green &  ---   & \green & \sgreen &  ---    & \sgreen & \sgreen
\\
\hline
  $C_R^{ij,\nu}$ 
&  ---   & \green & \green &  ---   &  ---   & \green &  ---   &  ---    & \sgreen  &  ---   &  --- 
\\  
\hline
\end{tabular}
\renewcommand{\arraystretch}{1.0}
\setlength{\tabcolsep}{2pt}
\caption{\small
  ``DNA'' table for VLQ contributions to the $b \to s \mu^+ \mu^-$ Wilson coefficients
  $C_{9,10}^{(\prime)}$ and to the $d_j \to d_i \nu\bar\nu$ ones $C_{L,R}^{\nu}$.
  \green~means that the NP contribution is potentially large, while \red~stands for a
  generically small contribution, due to the suppressed vector couplings of the $Z$
  to leptons compared to its axial-vector couplings. Smaller symbols in the $\GSMUpr(\Phi)$
  models indicate the general suppression by $c^2_\beta$ w.r.t. $\GSM$ models.
 From \cite{Bobeth:2016llm}.}
\label{tab:DNA:bsll-WC}
\end{table}

We studied patterns and correlations between these
observables which taken together should in the future allow
for differentiating between various VLQ models. In particular the patterns in models with
left-handed and right-handed currents are markedly different from each other.
Among the highlights of our paper are large $Z$-mediated NP effects in Kaon
observables  in some of the models and significant
effects in $B_{s,d}$-observables. $\epe$ can easily be made consistent
with the data, implying then {\em uniquely the suppression of}  $\klpn$ in contrast
to many $Z^\prime$ models in which the enhancement of $\epe$ implies the enhancement of $\klpn$ rate in a correlated manner. See for instance \cite{Aebischer:2023mbz}. But similar to $Z^\prime$ models, 
significant
enhancements of $\mathcal{B}(\kpn)$ in VLQ models are still possible.

Finally, we pointed out that the combination of NP contributions to $|\Delta F|=2$ and $|\Delta F|=1$
observables in a given meson system generally allows to determine the masses
of VLQs in a given representation independently of the size of VLQ couplings.

Actually, it is difficult to imagine that there are no additional fermions in nature and then VLQs enter the game. Therefore I devoted Section 16.3.2 in
my book \cite{Buras:2020xsm} to these new fermions.
This is also the belief of Gustavo
Branco who invested a lot of energy to study these fermions. I find
his recent review in Physics Reports \cite{Alves:2023ufm} very interesting.
Also the recent papers \cite{Albergaria:2023vls,Albergaria:2024pji}.
Many useful references to new papers that appeared after our paper can be found
in these papers.

\begin{table}[t]
\addtolength{\arraycolsep}{4pt}
\renewcommand{\arraystretch}{1.5}
\centering
\begin{tabular}{|l|c|c|c|c|c||c|c||c|c|c|c|}
\hline
  & \multicolumn{5}{|c||}{$\GSM$}
  & \multicolumn{2}{c||}{$\GSMUpr(S)$}
  & \multicolumn{4}{c|}{$\GSMUpr(\Phi)$}
\\
  & $D$ & $Q_V$ & $Q_d$ & $T_d$ & $T_u$& $D$ & $Q_V$ & $D$ & $Q_d$ & $T_d$ &$T_u$
\\
\hline\hline
$|\Delta F|=2$& \rSt & \bSt & \bSt & \rSt & \rSt & \bSt & \rSt & \rSt & \bSt & \rSt & \rSt
\\
\hline
$B_{s,d}\to\mu\bar\mu$ & \rSt &\bSt  & \bSt & \rSt & \rSt &  &  & \rst & \bst & \rst & \rst
\\
\hline
$B\to K\mu\bar\mu$ & \rSt &\bSt  & \bSt & \rSt & \rSt & \bSt & \rSt & \rst & \bst & \rst & \rst
\\
\hline
$B\to K^*\mu\bar\mu$ &  &  &  &  &  & \bSt & \rSt & \rst & \bst & \rst & \rst
\\
\hline
$B\to K(K^*)\nu\bar\nu$ & \rSt &\bSt  & \bSt & \rSt & \rSt &  &  & \rst & \bst & \rst & \rst
\\
\hline
$\kpn$ & \rSt &\bSt  & \bSt & \rSt & \rSt &  &  & \rst & \bst & \rst & \rst
\\
\hline
$\klpn$ & \rSt & \bSt & \bSt & \rSt & \rSt &  &  &  &  &  & 
\\
\hline
$\epe$ & \rSt & \bSt & \bSt & \rSt & \rSt &  &  & \rst & \bst & \rst & \rst
\\
\hline
\end{tabular}
\renewcommand{\arraystretch}{1.0}
\caption{\small
  ``DNA'' of flavour effects in VLQ models. A star indicates that significant
  effects in a given model and given process are in principle possible, but 
  could be reduced through correlation among several
  observables. Empty space means
  that the given model does not predict sizeable effects in that observable.
  The star \rSt{} indicates left-handed currents and the star \bSt{}
  right-handed ones, smaller stars indicate the suppression of $|\Delta F|=1$
  decays in $\GSMUpr(\Phi)$ models.  From \cite{Bobeth:2016llm}.
}
\label{tab:DNAVLQ}
\end{table}

\begin{figure}
  \centering
     \includegraphics[width=0.4\textwidth]{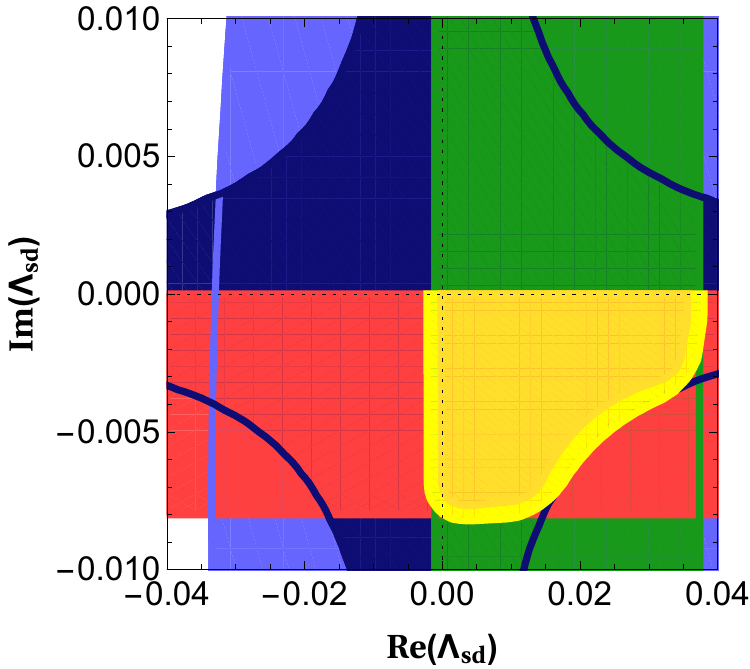} 
     \hskip 0.05\textwidth
     \includegraphics[width=0.4\textwidth]{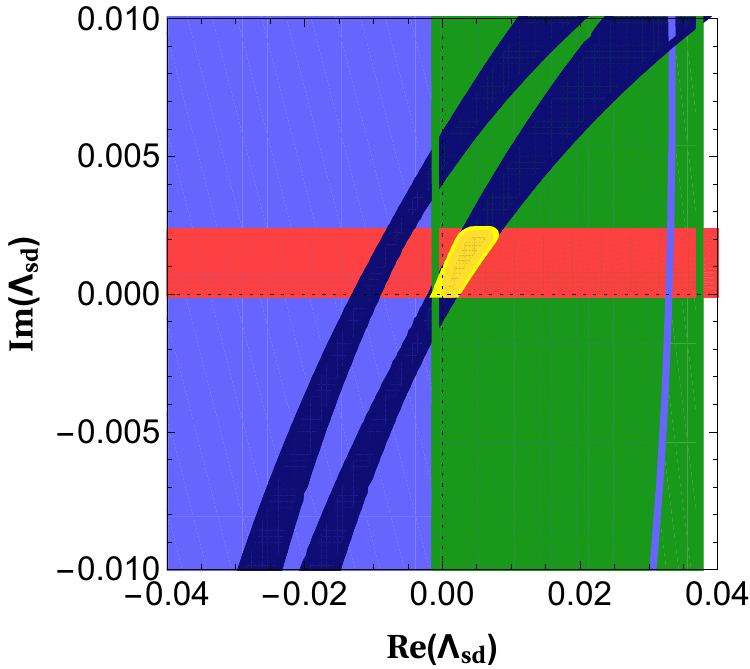}    
   \vskip 0.1cm
     \includegraphics[width=0.39\textwidth]{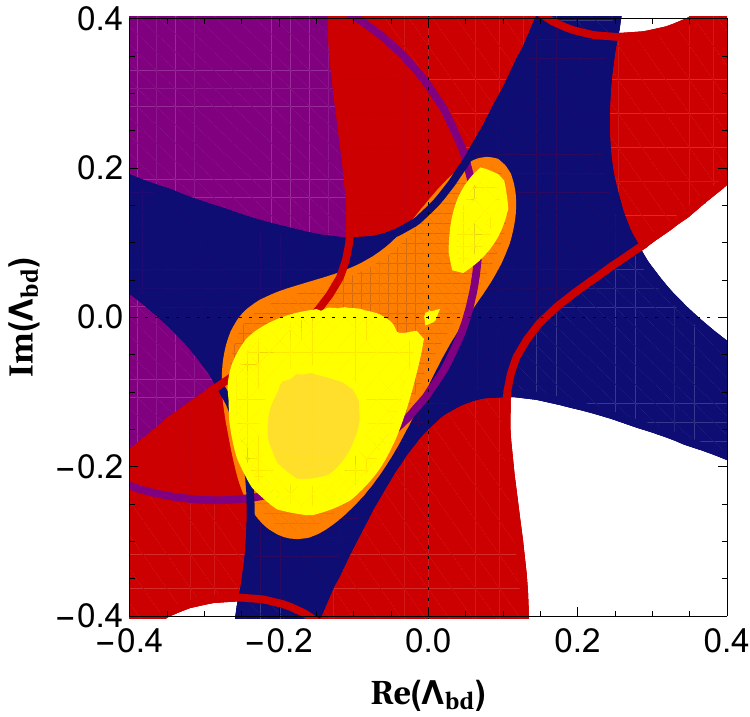} 
     \hskip 0.05\textwidth
     \includegraphics[width=0.39\textwidth]{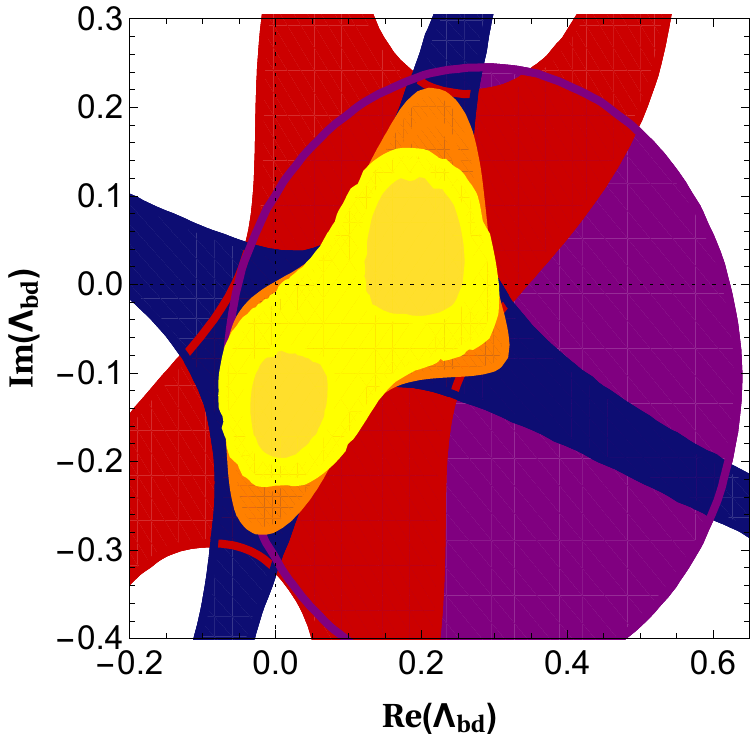}
   \vskip 0.1cm    
     \includegraphics[width=0.38\textwidth]{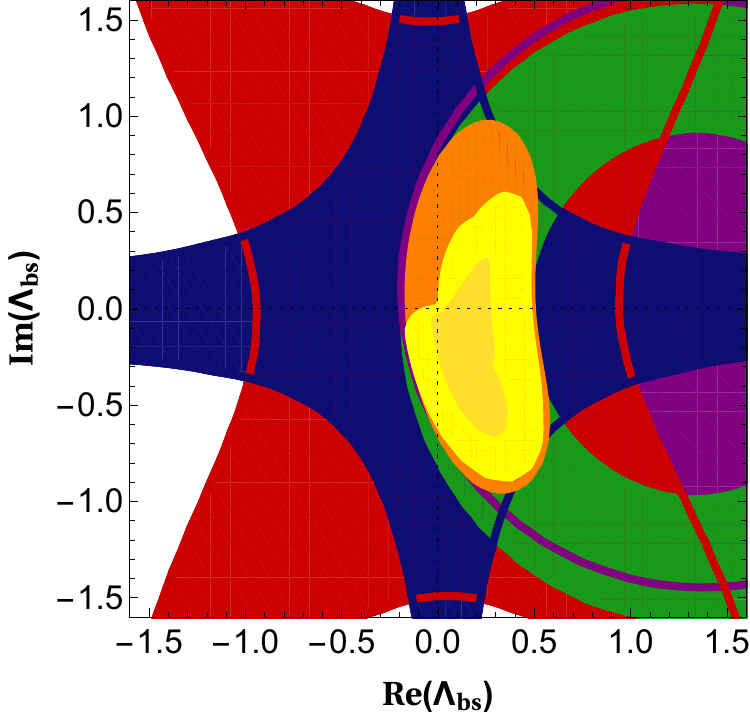} 
     \hskip 0.05\textwidth
     \includegraphics[width=0.38\textwidth]{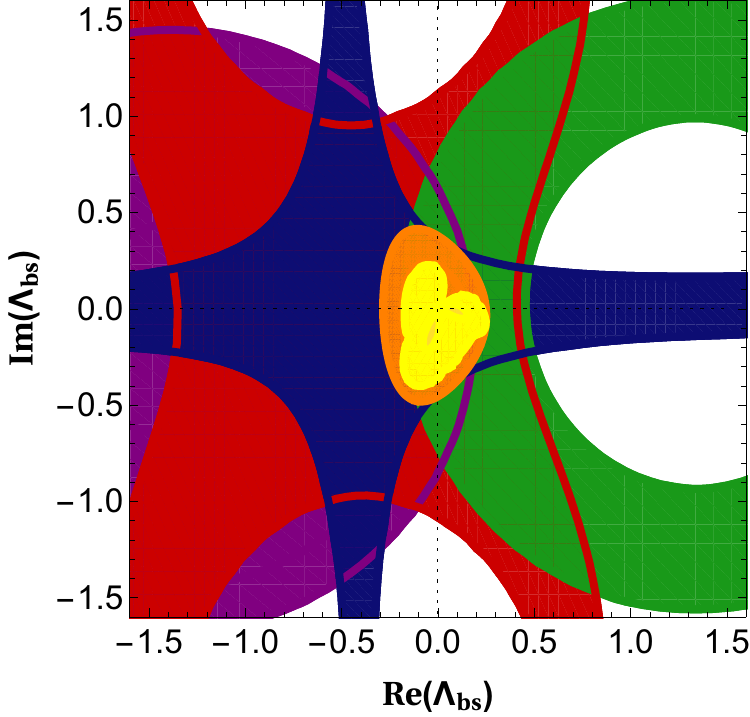}        
\caption{
  \small 
  \label{fig:GSM-A_ij-fits-10TeV}
  Martin Jung as Pablo Picasso. See \cite{Bobeth:2016llm} for explanations.
}
\end{figure}

DNA-tables~\ref{tab:DNA:bsll-WC} and \ref{tab:DNAVLQ} summarize
the patterns of NP effects in different models.  Fig.~\ref{fig:GSM-A_ij-fits-10TeV} is presented here for fun. There are several plots like this in this and
the next paper. Different colours represent constraints for couplings from
different observables.

\boldmath
\subsection{Yukawa Enhancement of $Z$-mediated New Physics in\\ $\Delta S = 2$ and $\Delta B = 2$ Processes (2017)}\label{YukawaE}
\unboldmath

Our study of VLQ models described above, in which the $Z$-mediated  NP
played a significant role, motivated our quartet to analyze more generally 
Yukawa-enhanced contributions from $Z$-mediated NP to
down-type quark $\Delta F=2$ processes in the framework of the SMEFT
\cite{Bobeth:2017xry}. In fact this was the first analysis of the induced
FCNCs mediated by $Z$ within the SMEFT framework. 

Besides RG
mixing of the $Z$-mediating $\psi^2 H^2 D$ operators into $\Delta F = 2$
operators, we included at the electroweak scale one-loop (NLO) matching
corrections consistently, necessary for the removal of the matching scale
dependence.  We pointed out that the right-handed $Z$-mediated interactions
generate through Yukawa RG mixing $\Delta F=2$ left-right operators, which are
further enhanced through QCD RG effects and chirally enhanced hadronic matrix
elements. We investigated the impact of these new effects on the known
correlations between $\Delta F=2$ and $\Delta F=1$ transitions in the SMEFT
framework and pointed out qualitative differences to previous parameterizations of
$Z$-mediated NP that arise for the left-handed case. The visualization of
all these effects is shown in Fig.~\ref{fig:visualization} .

In particular  we illustrated how
specific models fit into our model-independent framework by using four models
with VLQs.  To this end we carried out model-independent analyses of scenarios
with purely left-handed and purely right-handed NP $Z$ couplings for
each of the three sectors $s\to d$, $b\to s$ and $b\to d$. Specifically we discussed the correlations between $\epe$, $\varepsilon_K$, $K_L\to \mu^+\mu^-$
$\kpn$ and $\klpn$ in the Kaon sector, and $B_s\to\mu^+\mu^-$ and
$B\to K^{(*)} (\mu^+\mu^-, \nu\bar\nu)$ in the $b\to s$ sector and
$B_d\to\mu^+\mu^-$ in the $b\to d$ sector. See also related papers  \cite{Endo:2016tnu,Endo:2018gdn}.

\begin{figure}
  \includegraphics[width=\textwidth]{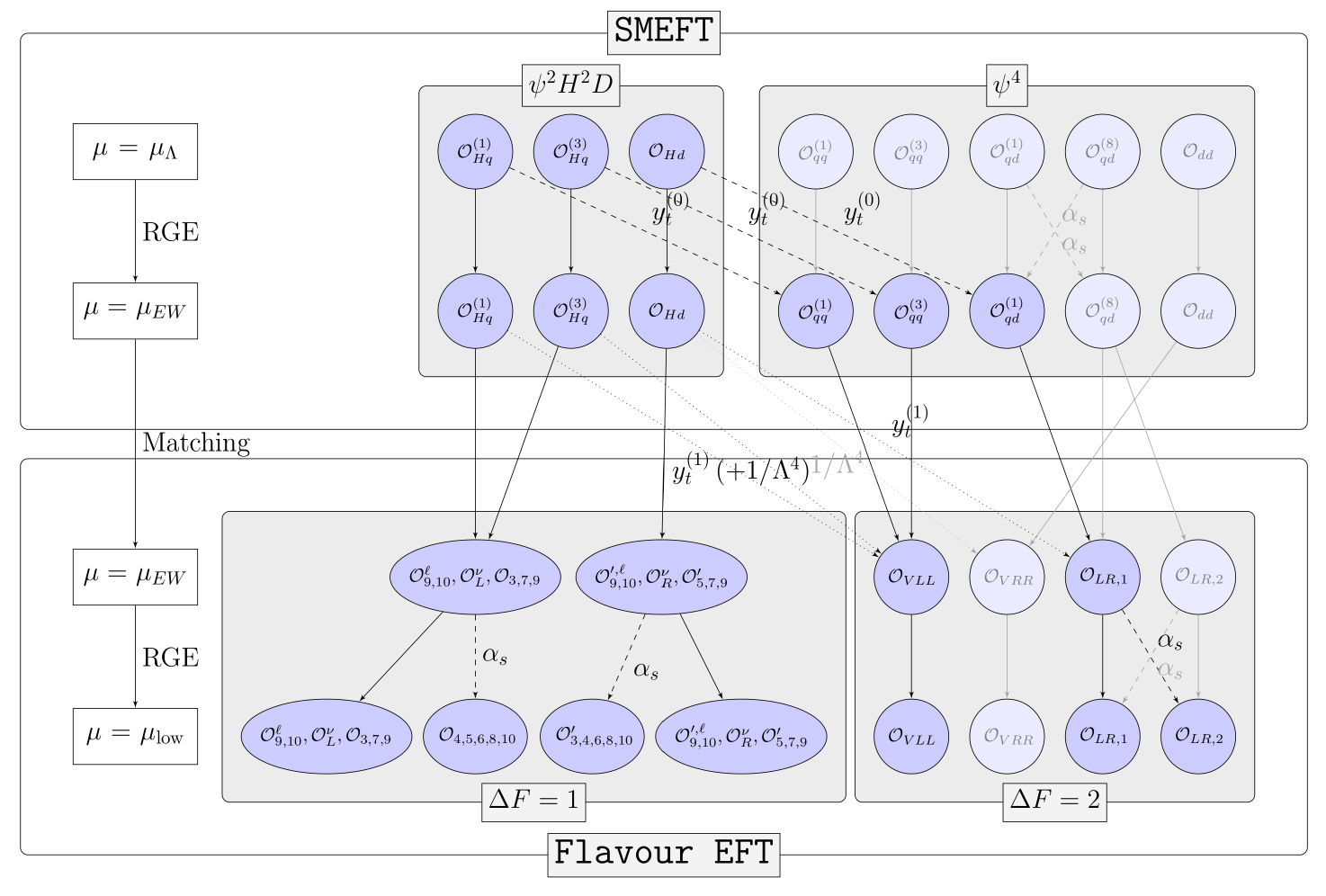}
\caption{\small \label{fig:visualization} 
  Visualization of matching and RG running for the operators under consideration
  in SMEFT and the $\Delta F=1,2$--EFTs. The darker nodes are those that are
  dominant within our approach. Solid lines indicate $\mathcal O(1)$ running/matching
  contributions, dashed lines RG mixing enhanced by a large logarithm and dotted
  lines NLO running/matching contributions.  Contributions via 1stLLA top-Yukawa
  RG mixing are denoted by $y_t^{(0)}$, 1-loop corrections to the matching at
  $\muEW$ by $y_t^{(1)}$. $1/\Lambda^4$ refers to contributions that appear at
  dimension eight, like double insertions of dim-6 operators. From \cite{Bobeth:2017xry}.
}
\end{figure}

Many lessons for FCNC processes within the VLQ models follow from this and the previous paper which I do not list here because they can be found on
pages 553-555 in my book: 14 lessons in total. Once more data will be available
it will be a real fun to go back to the analyses of these two papers with
the hope that the Picasso-like plots in Fig.~\ref{fig:GSM-A_ij-fits-10TeV} will
be more transparent than in our papers and successful VLQ models will be selected.

\boldmath
\section{ Leptoquarks Meet $\epe$ and Rare Kaon Processes (2017)}
\unboldmath

Leptoquark models gained on the popularity in 2013 due to the breakdown
of $\mu-e$ universality in $B\to K(K^*) \ell^+\ell^-$ decays claimed by the LHCb collaboration. While I found the suppression of $B\to K(K^*) \mu^+\mu^-$
relative to the SM predictions interesting, I did not write a single
paper on the so-called $R(K)$ and $R(K^*)$ anomalies, whose departure from unity
signalled violation of lepton flavour universality (LFU). I simply did not believe
in it and made a bet with Andreas Crivellin that this breakdown of LFU
will disappear one day. This indeed happened, but only in December 2022.

In spite of this, I found Leptoquark (LQ) models interesting and in the fall of 2017
analyzed with Christoph Bobeth  for the first time the ratio $\epe$ in $K\to \pi\pi$ decays in these models \cite{Bobeth:2017ecx}. We found the following.

Assuming a mass gap to the electroweak (EW)
scale, the main mechanism for LQs to contribute to $\epe$ is EW gauge-mixing of
semi-leptonic into non-leptonic operators, which we treated within the SMEFT. We performed also the one-loop decoupling for scalar
LQs, finding that in all models with both left-handed and right-handed LQ
couplings box-diagrams generate numerically strongly enhanced EW-penguin
operators $Q_{8,8'}$ already at the LQ scale.  We then investigated correlations
of $\epe$ with rare Kaon processes ($\klpn$, $\kpn$, $\klpll$, $\ksm$,
$\Delta M_K$ and $\eps_K$) and found that even imposing only a moderate
enhancement of $(\epe)_{\rm NP} = 5 \times 10^{-4}$ to explain the current
anomaly hinted by the Dual QCD approach 
leads to conflicts with experimental upper bounds on rare Kaon processes.

These bounds
exclude all LQ models with only a single coupling as an explanation of the
$\epe$ anomaly and put strong-to-serious constraints on parameter spaces of the
remaining models. Future results on $\kpn$ from the NA62 collaboration, $\klpn$
from the KOTO experiment and $\ksm$ from LHCb will even stronger exhibit the
difficulty of LQ models in explaining the measured $\epe$, in case the $\epe$
anomaly will be confirmed by improved lattice QCD calculations. Hopefully also
improved measurements of $\klpll$ decays will one day help in this context.
But if the $\epe$  anomaly will disappear one day, large
NP effects in rare K decays that are still consistent with present bounds will be allowed.

\section{Guide to Correlations between Observables}
We have seen that correlations between observables played a very important role
in the search for NP and the tests of the SM. This was in particular in this
part but also in previous parts, in particular in Part~\ref{SBSM}.
They will also be present in Parts~\ref{Anatomies}-\ref{BV}. But it
is useful already now to collect the references to most important papers
in these context.

In Table~\ref{tab:corr} the references to papers from my group that analyzed 
various correlations in all models discussed above have been collected for 
convenience. The more general strategies like the ones with Jennifer \cite{Buras:2013ooa} and Elena  \cite{Buras:2021nns,Buras:2022wpw,Venturini:2022sdf}  have not been listed here. While the last entry in this table is our recent SMEFT
review \cite{Aebischer:2025qhh}, presented in Section~\ref{SMEFT2025}, in which such correlations have been
described systematically, several analyses listed in this table were performed
 in the
framework of the SMEFT. This is in particular the case of \cite{Buras:2014fpa}
and \cite{Bobeth:2017xry}.

\begin{table}[thb]
\begin{center}
\begin{tabular}{|c|c|}
\hline
{\bf Model} & {\bf Reference}   \\
\hline
CMFV &   \cite{Buras:2000dm,Buras:2003jf,Blanke:2006ig,Buras:2003td}\\
$\text{U(2)}^3$& \cite{Buras:2012sd,Buras:2015yca}\\
${\rm 2HDM_{\overline{MFV}}}$ &  \cite{Buras:2010mh,Buras:2010zm} \\
ACD Model & \cite{Buras:2002ej,Buras:2003mk} \\
LH        &  \cite{Buras:2004kq,Buras:2006wk}\\
LHT &  \cite{Blanke:2009am,Bigi:2009df,Blanke:2015wba}\\
SM4 &  \cite{Buras:2010pi,Buras:2010nd,Buras:2010cp}\\
AC,~RVV2,~AMK,~$\delta$LL &  \cite{Altmannshofer:2009ne,Altmannshofer:2010ad}
 \\ 
$\text{SSU(5)}_{RN}$ & \cite{Buras:2010pm}\\
FBMSSM & \cite{Altmannshofer:2008hc}\\
RHMFV & \cite{Buras:2010pz}\\
RSc & \cite{Blanke:2008yr,Blanke:2008zb}\\
$Z^\prime$ &  \cite{Buras:2012jb,Buras:2014fpa,Aebischer:2023mbz,Buras:2024mnq}   \\
331 & \cite{Buras:2012dp,Buras:2014yna,Buras:2023ldz},  \\
Scalars & \cite{Buras:2013uqa,Buras:2013rqa}\\
Leptoquarks & \cite{Bobeth:2017ecx}   \\
VLQ  & \cite{Bobeth:2016llm}  \\
SMEFT & \cite{Aebischer:2025qhh} \\
\hline
\end{tabular}
\end{center}
\vspace{0.1cm}
\caption[]{References to correlations in various NP models studied in my group.
\label{tab:corr}}
\end{table}

\part{The Renaissance of Dual QCD (2013-2018)}\label{RDQCD}
\section{Preface}
In January 2013 I thought that it is time to have a look at DQCD again
and make some improvements. The reason was the progress in LQCD calculations.
I contacted Jean-Marc and he was excited about it although he warned me that
being the president of the Institute for Research in Mathematics and Physics (2012-2015)
at his University in Louvain\footnote{Prior to that, Jean-Marc founded and directed the Center for Particle Physics and Phenomenology (CP3) from 2003 to 2012. It is
amusing that this was also my case at TUM from 1988 to 2012.}
he cannot work as much as in previous
papers. Bill was not excited about my idea. Yet, after I found a 12 year
old talk by him, in which he made some progress already, he agreed to participate. The main goals of our new paper were:
\begin{itemize}
\item
  The inclusion of the vector mesons in the meson evolution which would
  make weaker the $\Lambda^2$ dependence of hadronic matrix elements and turning
  it into logarithmic one.
\item
  Taking care of the renormalization scheme dependence of hadronic matrix
  elements that would allow the consistent matching of these elements to
  the WCs evaluated in the NDR scheme.
\item
  Applying these improvements for the $\Delta I=1/2$ rule and the evaluation
  of the parameter $\hat B_K$ and the $K^0-\bar K^0$ mass difference $\Delta M_K$.
  \end{itemize}

In fact Jean-Marc started calculating the vector meson contributions already in
1988 \cite{Gerard:1988it}  and Bill looked at the issue of the RS dependence in his 2001 talk \cite{Bardeen:2001kd} which
unfortunately was ignored by the community.

It took almost one year in completing the paper due to Jean-Marc's obligations
but he was of course leading our calculation of vector meson contributions
which I checked. I tried to do as much as possible but was 
busy with other papers in 2013 which I described already. Finally, in January
2014 our paper appeared in arXiv. Bill, according to him, did not contribute much to this paper and did not want to sign it. Fortunately, I convinced him to be
at least the last author. After all his 2001 paper on the RS scheme
dependence in which he  introduced a momentum scheme for DQCD was important
\cite{Bardeen:2001kd}.
He also calculated there the one-loop shift to the NDR scheme and 't Hooft
scheme that I confirmed.

Let me then summarize what can be find in our paper
\cite{Buras:2014maa}.
This paper was indeed the beginning of the  Renaissance of DQCD. In the following  years we have written a number of papers that I will describe subsequently.
In some of them I had other collaborators.

\boldmath
\section{DQCD Approach 28 Years Later: $\Delta I=1/2$ Rule, $\hat B_K$ and $\Delta M_K$ (2014)}
\unboldmath

In  \cite{Buras:2014maa} we reviewed and updated our results for  $K\to\pi\pi$ decays and  $K^0-\bar K^0$ mixing obtained by us in the 1980s within DQCD. Let us recall that in this approach the SM  dynamics behind the enhancement of  ${\rm Re}A_0$ and suppression of  ${\rm Re}A_2$, the so-called
$\Delta I=1/2$ rule for $K\to\pi\pi$ decays,  has a simple 
structure: the usual octet enhancement through the long but slow quark-gluon 
 renormalization 
group evolution down to the scales $\ord(1\gev)$ is continued as a short but 
fast meson 
evolution down to zero momentum scales at which the factorization of 
hadronic matrix elements is at work.

The inclusion of lowest-lying vector meson contributions in addition to the
pseudoscalar ones and of Wilson coefficients in a momentum scheme in the 2014 paper improved significantly the matching between quark-gluon 
and meson evolutions. In particular, the anomalous dimension matrix governing the meson 
evolution exhibits the structure of the known anomalous dimension matrix in the quark-gluon evolution. While in 2014 this physical picture did not 
yet emerge 
from 
lattice simulations, the 2012 results on ${\rm Re}A_2$  and ${\rm Re}A_0$ 
from  the  RBC-UKQCD collaboration \cite{RBC:2012ynq}   gave  support for
its correctness. 
In particular, the signs of the two main contractions found numerically by 
these authors follow uniquely from our analytic approach. The anatomy of
various contributions both in DQCD and LQCD is presented in Sections 8 and
9 of our paper, respectively.

 Though the current-current 
operators dominate the $\Delta I=1/2$ rule, working with matching scales 
$\ord(1 \gev)$ we find that the presence of QCD penguin operator $Q_6$  
 is required to obtain satisfactory result for ${\rm Re}A_0$. At NLO in $1/N$ we obtained  
${\rm Re}A_0/{\rm Re}A_2= 16.0\pm 1.5$ which amounts to an order of magnitude 
 enhancement over the strict large $N$ limit value $\sqrt{2}$\footnote{In this context I received a nice email from Heinrich Leutwyler who found this
 explanation of the $\Delta I=1/2$ rule very nice.}. We also 
updated our results 
for the parameter $\hat B_K$, finding $\hat B_K=0.73\pm 0.02$. The smallness of $1/N$ corrections to the large $N$ value $\hat B_K=3/4$ 
results within our approach from an approximate cancellation between 
pseudoscalar  and vector meson one-loop contributions.

We 
also summarized the status of $\Delta M_K$ in this approach. We found
the SM model value slightly below the experimental value but
consistent with it. This is consistent with previous analyses of Jean-Marc
\cite{Gerard:1990dx,Bijnens:1990mz}.

  Interestingly the 2018 analysis 
of the RBC-UKQCD lattice QCD collaboration \cite{Bai:2018mdv} finds
$\Delta M_K$ in the SM by $2\sigma$ above the experimental value. 
Hopefully this will be clarified one day.

Unfortunately, this was the last paper we have written with Bill who
was,  as described in Section~\ref{DQCD}, the one who suggested this
approach. Already since 2015 Jean-Marc and me pushed DQCD without Bill and
I will describe it next. 

\boldmath
\section{Upper Bounds on $\epe$ Parameters $\bsi$ and $\bei$  from  
  Large $N$ QCD and other  News (2015)}\label{B6B8}
\unboldmath
In \cite{Buras:2015xba}
we demonstrated that in DQCD  the parameters $\bsi$ and $\bei$ parametrizing the $K\to\pi\pi$ matrix elements $\langle Q_6 \rangle_0$ and $\langle Q_8 \rangle_2$ of the dominant QCD and electroweak operators  receive both {\it negative} $\ord(1/N)$ corrections such that $\bsi \le \bei<1 $.
We also pointed out that the pattern of the size of  the hadronic matrix elements of all QCD and electroweak penguin operators $Q_i$ contributing to the $K\to \pi \pi$ amplitudes $A_0$ and $A_2$, obtained  in 2015 by the  RBC-UKQCD  collaboration \cite{Blum:2015ywa}, provided further  support to DQCD. As we will see later, while
in the case of $\bei$ this remained in 2020, their 2020 results violated
our bound $\bsi \le 1 $ \cite{Abbott:2020hxn}.

In any case with our values of $\bsi$ and $\bei$ we found 
$\epe$ in the SM to be  by a factor of two below its
experimental value $(16.6\pm 2.3)\times 10^{-4}$. We also briefly 
commented on the parameter $\hat B_K$ and the $\Delta I=1/2$ rule.

\boldmath
\section{Final State Interactions in $K\to\pi\pi$ Decays:\\ $\Delta I=1/2$ Rule vs.  $\epe$ (2016)}\label{39FSI}
\unboldmath
In \cite{Buras:2016fys} we revisited the 
dispersive effects from strong $\pi\pi$ rescattering in the final state (FSI) of weak $K\to\pi\pi$ decays with the goal to have a global view on their
{\it relative} importance for the $\Delta I=1/2$ rule and the ratio $\epe$  
in the SM.
 We pointed out that this goal cannot be reached within a pure effective (meson) field approach like {chiral perturbation theory}  in 
which the dominant current-current operators governing 
the $\Delta I=1/2$ rule and the dominant density-density (four-quark) 
operators governing $\epe$ cannot be disentangled from each other. 
But in the context of DQCD, which includes both long distance 
dynamics and the UV completion, that is QCD at short distance scales,
 such a distinction is possible. We found then that  beyond the strict large $N$ limit, FSI 
are likely to be important for the $\Delta I=1/2$ 
rule but much less relevant for $\epe$. In our view the latter finding diminished significantly hopes that improved calculations of $\epe$ would 
 bring its SM prediction to agree with the experimental data,
 opening  thereby an arena for important NP contributions to this ratio.

 While in 2025 the 2020 RBC-UKQCD result for $\epe$ has not been updated and agrees with experiment,
 the large error in their result and the omission of isospin breaking effects
 in their calculations, does not yet allow for any conclusions.

 \boldmath
 \section{$K\to\pi\pi$ and $K-\pi$ Matrix Elements of \\the Chromomagnetic Operators from  DQCD (2018)}
 \unboldmath
 In \cite{Buras:2018evv} Jean-Marc and I
performed for the {\it first time} a direct calculation of {\it on-shell} $K\to\pi\pi$ 
hadronic matrix elements of chromomagnetic operators (CMO) in the SM and
beyond. To this end, we used our DQCD approach in which we also 
considered {\it off-shell} $K-\pi$ matrix elements that allow the comparison with  
lattice QCD calculations presented in 2018 by the ETM collaboration \cite{Constantinou:2017sgv}. Working in the SU(3) chiral limit, we find
for the single $B$ parameter $B_{\rm CMO}=0.33$.  Using the numerical results provided by the ETM collaboration we argue that only {\it small} corrections beyond that limit are to be expected. Our results are relevant for NP scenarios in the context of  
the emerging $\epe$ anomaly strongly indicated within DQCD. I hope that this will be confirmed by the final result from RBC-UKQCD collaboration  or other
LQCD collaboration one day.

\boldmath
\section{DQCD Insight into  BSM Hadronic Matrix\\ Elements for $K^0-\bar K^0$ Mixing from LQCD  (2018)}
\unboldmath
In \cite{Buras:2018lgu}
we calculated BSM hadronic matrix elements  for  $K^0-\bar K^0$ mixing in the DQCD. 
The ETM, SWME and RBC-UKQCD lattice collaborations 
\cite{Carrasco:2015pra,SWME:2015oos,Garron:2016mva,Boyle:2017skn} found
  the matrix elements of the BSM density-density operators  $\mathcal{O}_i$ with
$i=2-5$ to be rather different from their vacuum insertion values (VIA) with 
$B_2\approx 0.5$, $B_3\approx B_5\approx 0.7$ and $B_4\approx 0.9$ 
at $\mu=3\gev$ to be compared with $B_i=1$ in the VIA.
We demonstrated that this pattern can 
be reconstructed within the DQCD through the non-perturbative 
meson evolution from very low scales, where factorization of matrix elements is valid, to scales $\ord(1\gev)$ with subsequent perturbative quark-gluon evolution to $\mu=3\gev$. This turns out to be possible in spite of a very different pattern displayed at  low scales with  $B_2=1.2$, $B_3=3.0$, $B_4=1.0$ and $B_5\approx 0.2$ in  the large $N$ limit. Our 
results imply that the inclusion of 
meson evolution in the phenomenology of any non-leptonic transition like $K^0-\bar K^0$ mixing  and $K\to\pi\pi$ decays is mandatory. While meson evolution, as demonstrated in our paper, is hidden in LQCD results, to our knowledge DQCD is the only analytic approach for non-leptonic transitions and decays which takes this important QCD dynamics into account.

In our view this analysis gave a very strong support for DQCD and gave the physical insight in LQCD calculations. To our surprise our paper
has been rejected by EPJC and Phys.Rev.D with the argument that the
LQCD results were more accurate than ours. Truly amazing. The referees
in both journals must have been some particular LQCD guys who are only interested in the
numbers coming out from their computers but not in the dynamics behind them.
But it is also amazing that the editors followed their advice. Fortunately,
Michal Praszalowich, one of the editors of Acta Physica Polonica, realized
right away the relevance of our paper.

 \boldmath
 \section{BSM Hadronic Matrix Elements for $\epe$ and \\$K\to\pi\pi$ Decays in the DQCD Approach (2018)}\label{42BSM}
 \unboldmath
 In 2018 I had a great pleasure to collaborate for the first time with Jason
 Aebischer, a young Swiss postdoc at our Excellence Cluster in the group
 of David Straub, my former excellent PhD student. He got interested in
 our paper just described and joined us in our next DQCD project
 \cite{Aebischer:2018rrz}.

 In this new analysis we calculated for the first time all four-quark  hadronic matrix elements  of local operators possibly contributing to $K\to\pi\pi$ decays and in particular to the ratio $\epe$ beyond the Standard Model (BSM). In addition to 7 new  mirror operators obtained from the SM ones by flipping the chirality, one has 13 BSM four-quark operators of a given chirality linearly independent of each other and of the other 14 operators, mentioned above, for which hadronic matrix elements were already known.

 We presented results in two bases for all these operators, one
termed DQCD basis useful for the calculation of the hadronic matrix elements in
 the DQCD approach and the other called SD basis suited to the short distance renormalization group  evolution above the $1\gev$ scale.
We demonstrated that the pattern of long
distance evolution (meson evolution) matches the one of short distance evolution (quark-gluon evolution), a property which to our knowledge cannot be presently achieved in any other analytical framework.  The highlights of our paper
are chirally enhanced matrix elements of tensor-tensor and scalar-scalar BSM
operators.
They could thereby explain the emerging $\epe$ anomaly which is strongly indicated within DQCD.  On the other hand
we did not expect the BSM operators to be relevant for the $\Delta I=1/2$ rule.

\boldmath
\part{Master Formulae and Anatomies of $\epe$ (2018-2020)}\label{Anatomies}
\unboldmath
\boldmath
\section{Preface }\label{epe}
  \unboldmath
  The ratio $\epe$ played an important role in my research until
  now. I have written many papers on this ratio both in the
  SM and in various NP models. Let me first just list some of them written until 2015. These are \cite{Buchalla:1989we,Buras:1993dy,Buras:1993bi,Buras:1996dq,Bosch:1999wr,Buras:1999if,Buras:2000kx,Buras:2001au,Buras:2003zz}.

  In 2015 the RBC-QCD collaboration \cite{RBC:2015gro} presented their first result with the central value in the ballpark of $1\times 10^{-4}$ which was even smaller than I expected. Their central value 20 times larger in 2020 demonstrated that they do not
  yet have their errors under control. The main obstacle in their calculations
  is the difficulty in including isospin breaking corrections which were know
  already in the 1980s through the work of Donoghue et al and G\'erard and myself   to suppress the QCD penguin contributions to $\epe$ by roughly $30\%$ bringing   it below $ 1\times 10^{-3}$. I described these papers in
  Section~\ref{NIB}. We have stressed this in \cite{Aebischer:2019mtr} and I stressed it in a number of talks  \cite{Buras:2018ozh,Buras:2019vik,Buras:2021ane}.
  My last 2020 paper on this issue with Jean-Marc  demonstrates it very well   \cite{Buras:2020pjp}.
  
  Our 2015 anatomy of $\epe$ \cite{Buras:2015yba} was based on the 2015 hadronic
  matrix elements of RBC-UKQCD which are out of date and I will not describe
  it here although some useful expressions can be found there.

Finally, there is the issue of final-state interactions. While these effects are $1/N$-suppressed in the DQCD approach, as shown in \cite{Buras:2016fys} they can be relevant for the $\Delta I=1/2$ rule, but are expected to be much less important for $\epe$. I discuss this in more detail in Section~\ref{39FSI}.

Useful general expressions for $\epe$, including NP contributions, can be found in
\cite{Aebischer:2018csl,Aebischer:2020jto,Aebischer:2021hws}, and I will present some of them shortly.

 \boldmath
 \section{Master Formula for $\epe$ beyond the SM (2018)}
 \unboldmath

 Having the matrix elements of new BSM operators, calculated using DQCD
 in \cite{Aebischer:2018rrz} (see Section~\ref{42BSM}), it was natural
 to find a master formula for $\epe$ beyond the SM. In this new project
 we were joined by my two excellent former PhD students,
 Christoph Bobeth and David Straub. Christoph was this time a postdoc in my group at IAS.
 
In \cite{Aebischer:2018quc}
we presented for the first time a master formula for $\epe$ valid in {\em any}
ultraviolet extension of the SM (BSM). This formula makes use
of hadronic matrix elements of
BSM operators calculated in the DQCD approach in \cite{Aebischer:2018rrz}
and the ones of the
SM operators from lattice QCD. We emphasized the large impact of several scalar
and tensor BSM operators in the context of the emerging $\epe$ anomaly. We have
implemented the results in the open source code flavio developed by David
\cite{Straub:2018kue}.

Writing $\epe$ as a sum of the SM and BSM contributions,
\begin{align}
  \frac{\varepsilon'}{\varepsilon} &
  = \left(\frac{\varepsilon'}{\varepsilon}\right)_\text{SM}
  + \left(\frac{\varepsilon'}{\varepsilon}\right)_\text{BSM} \,,
\end{align}
the master formula of \cite{Aebischer:2018quc} for the BSM part in the
version presented in our next project \cite{Aebischer:2018csl}
reads then as follows
\begin{align}
  \label{eq:master}
  \left(\frac{\varepsilon'}{\varepsilon}\right)_\text{BSM} &
  = \sum_i  P_i(\muEW) ~\text{Im}\left[ C_i(\muEW) - C^\prime_i(\muEW)\right]
  \times (1\,\text{TeV})^2,
\end{align}
where
\begin{align}
  \label{eq:master2}
  P_i(\muEW) & = \sum_{j} \sum_{I=0,2} p_{ij}^{(I)}(\muEW, \muLow)
  \,\left[\frac{\langle O_j (\muLow) \rangle_I}{\text{GeV}^3}\right],
\end{align}
with the sum over $i$ extending over the Wilson coefficients $C_i$ of all BSM
operators including their chirality-flipped counterparts, that is $36 + 36'$
linearly independent four-quark operators and $1 + 1'$ chromo-magnetic dipole operators.
The $C_i'$ are the Wilson coefficients of the corresponding
chirality-flipped operators obtained by interchanging $P_L\leftrightarrow P_R$.  The
relative minus sign accounts for the fact that their $K\to\pi\pi$ matrix
elements differ by a sign.  Among the contributing operators are also operators
present already in the SM but their Wilson coefficients in (\ref{eq:master})
include only BSM contributions.

The dimensionless coefficients $p_{ij}^{(I)}(\muEW,\muLow)$ include the QCD and
QED RG evolution from $\muEW$ to $\muLow$ for each
Wilson coefficient as well as the relative suppression of the contributions to
the $I=0$ amplitude due to ${\text{Re}A_2} / {\text{Re}A_0}\ll 1$ for the matrix
elements $\langle O_j (\muLow) \rangle_I$ of all the operators $O_j$ present at
the low-energy scale. The index $j$ includes also $i$ so that the effect of
self-mixing is included. The $P_i(\muEW)$ do not depend on $\muLow$ to the
considered order, because the $\muLow$-dependence cancels between matrix
elements and the RG evolution operator.  Moreover, it should be emphasized that
their values are {\em model-independent} and depend only on the SM dynamics
below the electroweak scale, which includes short distance contributions down to
$\muLow$ and the long distance contributions represented by the hadronic matrix
elements. The BSM dependence enters our master formula in (\ref{eq:master}) {\em
  only} through the Wilson coefficients $C_i(\muEW)$ and
$C^\prime_i(\muEW)$. That is, even if a given $P_i$ is non-zero, the fate of its
contribution depends on the difference of these two coefficients. In particular,
in models with exact left-right symmetry this contribution vanishes as first
pointed out by Gustavo Branco, Fr\'ere and Jean-Marc in \cite{Branco:1982wp}.

\boldmath
\section{Anatomy of $\epe$ beyond the SM (2018) }\label{Master1}
\unboldmath
Having the master formula just discussed, it was natural to present an anatomy
of it. 
While I have already written few papers in which SMEFT was involved before,
this second 2018 paper with Jason Aebischer, Christoph Bobeth and David Straub   was the first of a series of papers in which WET and SMEFT played the dominant role.

In \cite{Aebischer:2018csl} we have presented for the first time a model-independent anatomy of $\epe$ in the context of the $\Delta S = 1$ effective theory with operators invariant under QCD and
QED, that is WET, and in the context of the SMEFT with the operators invariant under the full SM gauge group. It should be emphasized that the formula
(\ref{eq:master}) can be applied to any BSM theory in which the WCs of all contributing
operators have been calculated at the electroweak scale. Due to the significant
uncertainties in hadronic matrix elements the values of  $P_i(\muEW)$ listed
in our paper will be modified when the calculations of the matrix elements
in question will improve with time.

This formula allows in principle to identify
the NP  scenarios that are probed by this ratio and which could help to
explain a possible deviation from the SM that is hinted by the data.
Within SMEFT, the constraints from $K^0-\bar K^0$  and $D^0-\bar D^0$  mixing as well
as electric dipole moments limit significantly potential NP contributions to
$\epe$. We have briefly discussed correlations of $\epe$  with $\kpn$ and $\klpn$ decays. We highlighted again
the role of the new scalar and tensor matrix elements in models with scalar mediators. These matrix elements have only been calculated in DQCD \cite{Aebischer:2018rrz} and
it would be very important if LQCD calculated them as well.

\boldmath
\section{On the Importance of NNLO QCD and Isospin-breaking Corrections
  in $\epe$ (2019)}
\unboldmath
In \cite{Aebischer:2019mtr}, following the 1999 analysis of Gambino, Haisch and myself \cite{Buras:1999st},
we stressed that
  all the recent (2019)  NLO analyses of $\epe$ in the SM did not include the NNLO QCD contributions to EW penguins and suffered from
  the {\em renormalization scheme} dependence present in the EW penguin
  contributions as well as from scale uncertainties in them related
  to the matching scale $\mu_W$ and in particular to $\mu_t$ in $m_t(\mu_t)$.
  We also reemphasized the important role of isospin-breaking  and QED effects
  in the evaluation of $\epe$.  We provided a table for $\epe$ for
  different values of $\bsi$ and the isospin-breaking parameter $\hat\Omega_{\rm eff}$,
  that should facilitate monitoring the values of $\epe$ in the SM when the
  RBC-UKQCD calculations of hadronic matrix elements including isospin-breaking
  corrections and QED effects will improve with time.
}
\boldmath
\section{$\epe$ in the SM at the Dawn of the 2020s}
\unboldmath

In view of new results for hadronic matrix elements from RBC-UKQCD collaboration  \cite{Abbott:2020hxn}   we have updated in May 2020 our SM analysis for $\epe$ in \cite{Aebischer:2020jto}. Only Jason, Christoph and I took part in this update. As a byproduct
we have also updated our BSM formula of \cite{Aebischer:2018csl}.

To this end we used the most important NNLO QCD corrections to electroweak penguin contributions
and the isospin-breaking corrections. We illustrated the
importance of the latter by using their latest estimate from chiral
perturbation theory (ChPT) based on the {\em octet} approximation for
lowest-lying mesons \cite{Cirigliano:2019cpi} and a very recent  estimate in the {\em nonet} scheme
that takes into account the contribution of $\eta_0$ \cite{Buras:2020pjp}. We found 
$(\epe)^{(8)}_\text{SM} = (17.4 \pm 6.1) \times 10^{-4}$ and
$(\epe)^{(9)}_\text{SM} = (13.9 \pm 5.2) \times 10^{-4}$, respectively.
Despite a very good agreement with the measured value $(\epe)_\text{exp}
= (16.6 \pm 2.3) \times 10^{-4}$, the large error in $(\epe)_\text{SM}$
still leaves room for significant new physics (BSM) contributions to this
ratio. We updated the 2018 master formula for $(\epe)_\text{BSM}$ valid in
any extension beyond the SM without additional light degrees of freedom.
We provided new values of the penguin parameters $\bsi(\mu)$ and $\bei(\mu)$
at the $\mu$-scales used by the RBC-UKQCD collaboration and at lower scales
$\ord(1\gev)$ used by ChPT and DQCD. We presented semi-analytic formulae for
$(\epe)_\text{SM}$ in terms of these parameters and $\hat\Omega_{\rm eff}$ that
summarizes isospin-breaking corrections to this ratio. We stressed the
importance of lattice calculations of the $\ord(\alpha_{em})$ contributions
to the hadronic matrix elements necessary for the  removal of renormalization
scheme dependence at $\ord(\alpha_{em})$ in the present analyses of $\epe$.

  \boldmath
  \section{The Present Status of $\epe$ within the SM}\label{Statusepe}
  \unboldmath

   Let me next 
  summarize the present status of $\epe$ within the SM.

  Probably, among my summaries of $\epe$ within the SM, the best is the one in \cite{Buras:2020wyv} where I described the history of this ratio from 1976 until 2021. It contains many references. It is dedicated to Jean-Marc on the occasion of his 64th birthday.
  A shorter description is given in \cite{Buras:2022cyc} and even shorter will be given now.
  
  The situation of $\epe$ as of March 2022 was as follows.

 The   experimental world average from NA48 \cite{Batley:2002gn} and
  KTeV \cite{AlaviHarati:2002ye, Worcester:2009qt} collaborations reads
  \be
    \label{EXP}
    \boxed{(\epe)_\text{exp} 
    = (16.6 \pm 2.3) \times 10^{-4}\,.}
  \ee

The most recent result from LQCD, that is from 
the RBC-UKQCD collaboration
\cite{Abbott:2020hxn}, reads
\be
  \label{RBCUKQCD}
  \boxed{(\epe)_{\rm SM} 
  = (21.7 \pm 8.4) \times 10^{-4} \,,\qquad (\text{RBC-UKQCD}-2020),}
\ee
where statistical, parametric and systematic uncertainties have been added in
quadrature. It is subject
to large systematic uncertainties which dominate the quoted error. It 
is based on the improved (relative to 2015) values of the hadronic matrix elements of QCDP,
includes the Wilson coefficients at the NLO level but does not
include  isospin breaking effects, charm contributions and NNLO QCD effects.

The most recent 
estimate of $\epe$ in the SM from ChPT \cite{Cirigliano:2019ani,Gisbert:2020wkb} reads
  \be
  \label{Pich}
  \boxed{(\epe)_\text{SM}   = (14 \pm 5) \times 10^{-4} \,,\qquad (\text{ChPT}-2019).}
\ee
  The large error is related to the problematic matching
  of LD and SD contributions in this approach which can be traced back
  to the absence of meson evolution in this approach. Here the isospin breaking corrections are included but only in the octet approximation and
  are subject to significant uncertainties
  again related to the problematic matching   of LD and SD contributions.

  Finally, based on the insight from DQCD obtained in collaboration
  with Jean-Marc G{\'e}rard  \cite{Buras:2015xba,Buras:2016fys} one finds
  \cite{Buras:2020wyv}
  \be\label{AJBFINAL}
 \boxed{(\epe)_{\rm SM}= (5\pm2)\cdot 10^{-4},\qquad (\text{DQCD}-2020).}
 \ee

 While the results in (\ref{RBCUKQCD}) and (\ref{Pich}) are fully consistent
 with the data in (\ref{EXP}), the DQCD result in (\ref{AJBFINAL}), if confirmed
 by LQCD one day, would imply a significant anomaly and NP at work.

 Interestingly, there is
a very good agreement between
LQCD and DQCD as far as EWP contribution to $\epe$ is concerned. This implies
that this contribution to $\epe$, that is unaffected by leading isospin breaking corrections, is already
known within the SM with acceptable accuracy:
\be
  \label{EWPSM}
  \boxed{(\epe)^{\text{EWP}}_\text{SM}   = - (7 \pm 1) \times 10^{-4} \,,\qquad (\text{LQCD~and~DQCD}).}
\ee
Because both LQCD and DQCD can perform much better in the case of EWPs than in the case of QCDPs I expect that this result will remain with us for the coming years.

On the other hand the value from ChPT \cite{Cirigliano:2019ani} is roughly by a factor of 2 below the result in (\ref{EWPSM}). This lower value originates from  the suppression due to FSI, which, however, does not appear to be  supported
by the results of the RBC-UKQCD collaboration. Yet, the uncertainty
in the ChPT estimate is large. It would be good to clarify this difference in the future.

The origin of the differences in $\epe$ between the three groups
are different values of
\be
\hat\Omega_{\rm eff}, \qquad \bsi, \qquad \bei,
\ee
that enter the SM formula
\cite{Aebischer:2020jto} 
\begin{equation}
\left(\frac{\varepsilon'}{\varepsilon}\right)_{\text{SM}} =  
\IM\lambda_{\rm t}\cdot \left[\,
\big(1-\hat\Omega_{\rm eff}\big) \big(-2.9 + 15.4\,\bsi(\mu^*)\big) + 2.0 -8.0\,\bei(\mu^*) \,\right].
\label{AN2020}
\end{equation}
It includes NLO QCD corrections to the QCD penguins  contributions and NNLO contributions to electroweak penguins (EWP). Interested readers are asked
to look at \cite{Buras:2022cyc} where the anatomy of this formula is presented.

It should be noted that $\bsi$ and $\bei$ are scale dependent but this dependence is cancelled by other factors and terms in this formula. Their values here
are for  $\mu^*=1~\gev$.

\begin{figure}[!bt]
 \centering
\includegraphics[width = 0.90\textwidth]{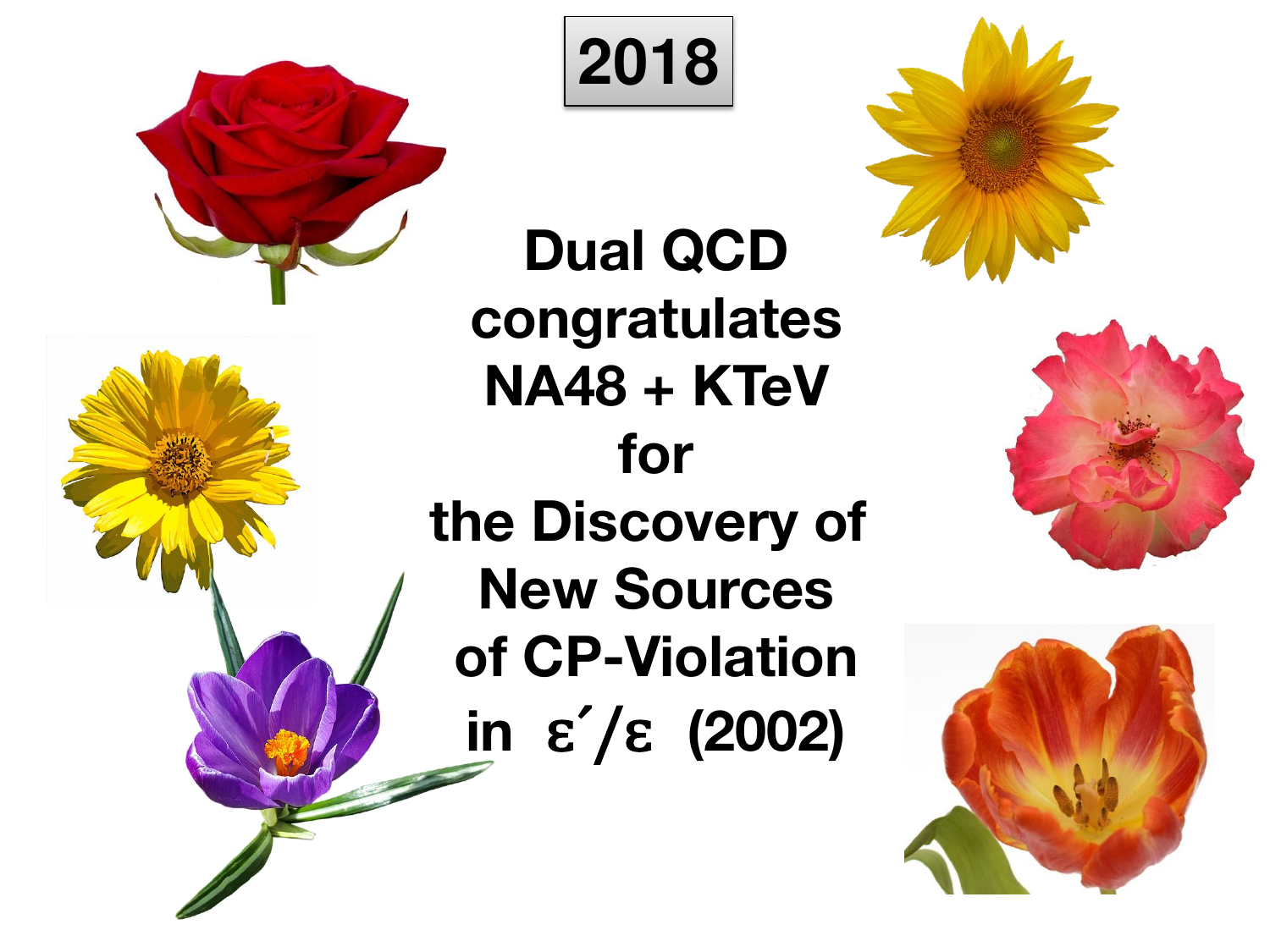}
\caption{Congratulations to NA48 and KTeV.
}\label{fig:flowers}~\\[-2mm]\hrule
\end{figure}

In order to stress my strong believe that NP has an important impact on the
value of $\epe$ I show Fig.~\ref{fig:flowers} from  \cite{Buras:2018ozh}
stressing that already in 2018 DQCD could congratulate both experiments
for discovering NP. I hope that LQCD will join DQCD in these congratulations 
  still in this decade or soon after.

\part{WET and SMEFT (2020-2025)}\label{SMEFT}

\section{Preface}
\begin{figure}[t]
\centering%
\includegraphics[width=0.90\textwidth]{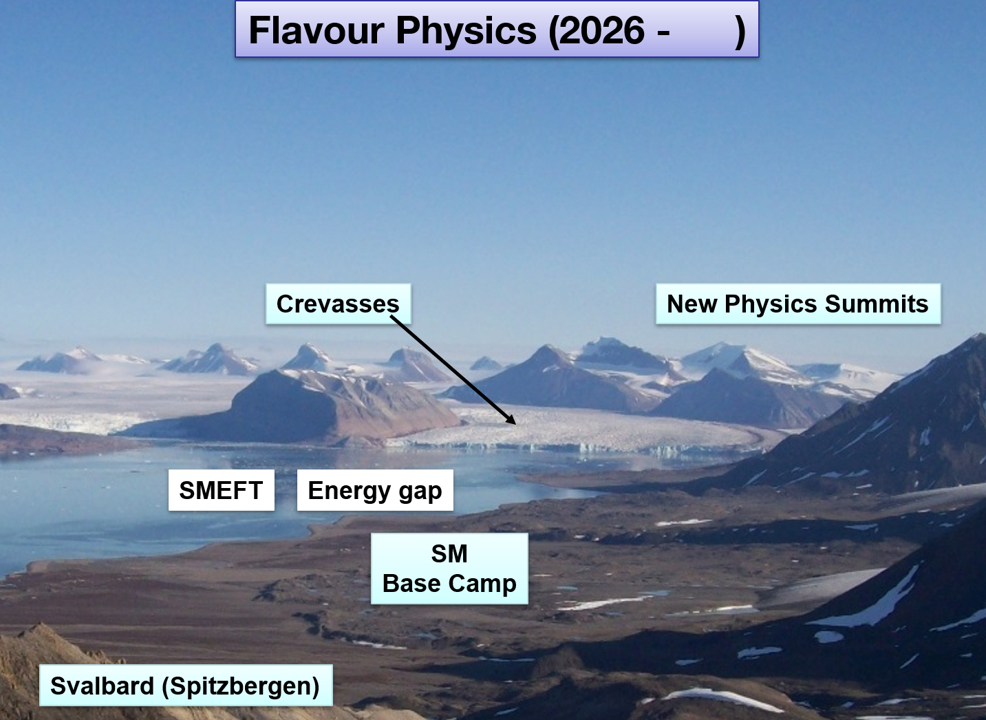}%
\caption{\it SMEFT and New Physics Summits.\label{Svalbard}}
\end{figure}

\begin{figure}[t]
\centering%
\includegraphics[width=0.8\textwidth]{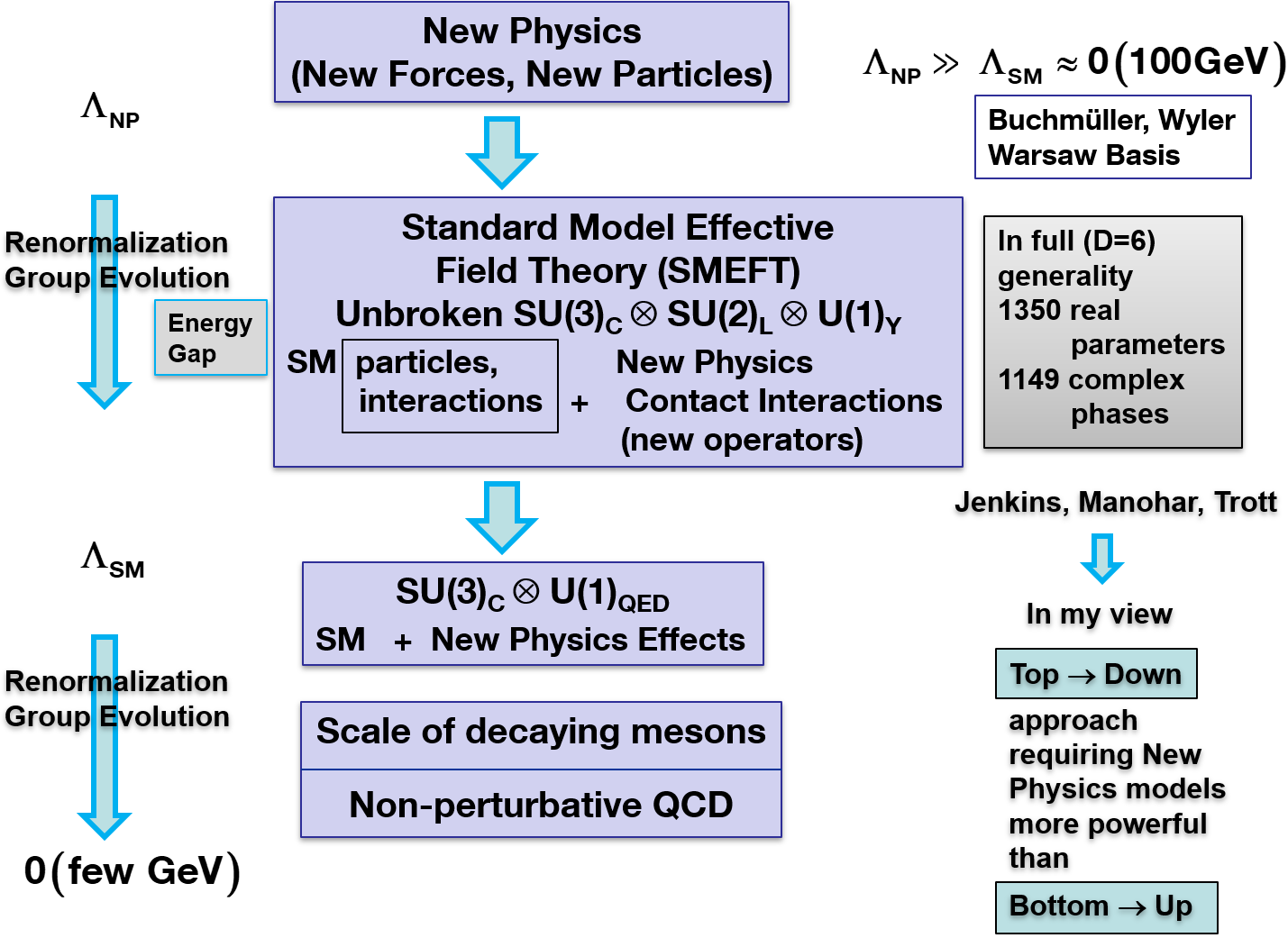}%
\caption{\it Grand View of the RG Evolution in the SMEFT.\label{GVSMEFT}}
\end{figure}

While WET and SMEFT appeared already previously I want to summarize them
by Figs.~\ref{Svalbard} and \ref{GVSMEFT}. The photo in Fig.~\ref{Svalbard}
has been taken by our younger son Allan in his expedition related to climate
rather than flavour.


\boldmath
\section{Another SMEFT Story: $Z^\prime$ Facing New Results on  $\epe$, $\Delta M_K$, $\kpn$ and $\klpn$\\ (2020)}\label{JJ1}
\unboldmath

In the summer of 2020 I gained another excellent collaborator, Jacky Kumar, who
was working at the Montreal University at that time. But he impressed me so
much that I arranged for him a Humboldt Fellowship so that he spent almost
two years in my group at IAS. He is the only Indian physicist I worked with
until now.

The first paper with him and Jason Aebischer \cite{Aebischer:2020mkv} was the first one of a series of
papers with Jason and Jacky written until now. I will call them JJ from now on.

In this first paper, in view of the new result on hadronic matrix elements
from RBC-UKQCD mentioned above \cite{Abbott:2020hxn} and their new result on $\Delta M_K$ \cite{Bai:2018mdv}, $2\sigma$
above the data, we have pointed out that the suppression of $\Delta M_K$
by NP requires the presence of new CP-violating phases with interesting implications for $\kpn$, $\klpn$, $K_S\to\mu^+\mu^-$ and $K_L\to \pi^0\ell^+\ell^-$
decays.

Considering a $Z^\prime$-scenario
within the SMEFT we analyzed the dependence of all these observables on the size of
NP still allowed by the data on $\epe$. The hinted $\Delta M_K$  anomaly together with the $\varepsilon_K$ 
constraint implies in the presence of only left-handed (LH) or right-handed (RH) flavour-violating $Z^\prime$  couplings strict correlation between $\kpn$ and $\klpn$ branching
ratios so that they are either simultaneously enhanced or suppressed relative to SM
predictions. An anticorrelation can only be obtained in the presence of both LH and RH
couplings. Interestingly, the NP QCD penguin scenario for the enhancement of $\epe$ considered by me five years earlier in \cite{Buras:2015jaq}
 has been  excluded by SMEFT
 renormalization group effects in $\varepsilon_K$  so that NP effects in $\epe$
  are governed by electroweak
  penguins, also considered by me in details in \cite{Buras:2015jaq}.

In \cite{Aebischer:2020mkv} we have 
 also investigated for the first time whether the presence of a heavy $Z^\prime$ 
with flavour violating couplings could generate, through top Yukawa renormalization group effects, FCNCs mediated by the SM Z-boson.
We have demonstrated that RG effects can in the presence of $Z^\prime$ contributions generate flavour-violating $Z$ contributions to $\epe$ and rare decays that have significant impact on the phenomenology. We also found  that in the presence of $\ord(1)$
diagonal $Z^\prime$ top-quark couplings, the $(V-A)\times (V+A)$ EWP operators can be generated
solely through the RG induced flavour-violating $Z$ couplings. This effect turns out to be sufficiently strong to provide significant BSM contributions to
$\epe$, if required, while simultaneously suppressing $\Delta M_K$.

\boldmath
\section{SMEFT ATLAS of $\Delta F=2$ Transitions (2020)}
\unboldmath
Our previous papers motivated us to 
present a model-independent anatomy of the  $\Delta F=2$ transitions
$K^0-\bar K^0$, $B^0_{s,d}-\bar B^0_{s,d}$   and $D^0-\bar D^0$  mixings 
in the context of the SMEFT \cite{Aebischer:2020dsw}.

We presented two master formulae for the mixing amplitude $[M_{12}]_{\rm  BSM}$:
\begin{itemize}
  \item
One in terms of the WCs of the WET
operators evaluated at the electroweak scale $\muEW$,
\item
One in terms of the
WCs of the SMEFT operators evaluated at the BSM scale $\Lambda$.
\end{itemize}

The coefficients $P_a^{ij}$
entering these formulae contain all the information below the scales $\muEW$ and $\Lambda$,
respectively. It turns out that RG effects from the top-quark Yukawa coupling
play the most important role.  The collection of the individual contributions of
the SMEFT operators to $[M_{12}]_{\rm  BSM}$ can be considered as the SMEFT ATLAS of
$\Delta F = 2$ transitions and constitutes a travel guide to such transitions far beyond the
scales explored by the LHC. We emphasized that this ATLAS depends on whether
the down-basis or the up-basis for SMEFT operators is considered. We illustrated
this technology with tree-level exchanges of heavy gauge bosons ($Z^\prime$, $G^\prime$)  and heavy scalars.

After our paper appeared in the hep-arXiv we received a critical comment
from one experimentalist that using ATLAS is misleading in view of the
ATLAS collaboration. A rather strange comment in view of many atlases outside
the particle physics. Moreover we specified that it is SMEFT ATLAS. Yet, possibly using ``Atlas'' instead is better.

\boldmath
\section{General Non-leptonic $\Delta F = 1$ WET at the NLO in QCD: (2021, 2022)}
\unboldmath
In the year 2000, as reported already in Section~\ref{BMU}, Misiak, Urban
and myself calculated two-loop ADMs for non-leptonic $\Delta F=1$ processes beyond the SM
\cite{Buras:2000if}. This calculation has been performed in a particular basis for operators that was useful for our calculations, the so-called BMU basis.
However, the JMS basis for WET is more useful for the matching of the WET to
the SMEFT. Therefore 21 years later I thought that it would be useful
to reconsider the complete set of four-quark operators in the WET relating
our old calculation done in the BMU basis to the JMS basis both in the SM
and beyond.

For this rather technical calculations we gained two NLO QCD experts
Christoph Bobeth and Mikolaj Misiak. One important issue was the treatment
of the evanescent operators in the process of relating operators in the JMS basis to the one in the BMU basis at NLO in QCD. These technical details
can be found in our paper \cite{Aebischer:2021raf}\footnote{Small errors in terms related to one operator $Q_{11}$ have been identified in \cite{Morell:2024aml} which in my view are of no phenomenological importance.}, and in a paper written in
2022 \cite{Aebischer:2022tvz}.

As we will see below our paper with Bobeth and Misiak \cite{Aebischer:2021raf}
was important for obtaining NLO QCD corrections in the SMEFT  for
the non-leptonic $\Delta F=2$ processes.

{\bf Meanwhile, as far as WET is concerned two-loop anomalous dimensions for all
WET operators, not only relevant for non-leptonic decays as calculated by us, have been calculated
in a series of papers by Luca Naterop and Peter Stoffer \cite{Naterop:2024cfx,Naterop:2025cwg,Naterop:2025lzc}.}

\boldmath
\section{BSM Master Formula for $\epe$ in the WET Basis at NLO in QCD (2021)}
\unboldmath
As reported already in Section~\ref{Master1} in 2018 we have presented
a master formula for $\epe$ in the WET and SMEFT at the LO in QCD
\cite{Aebischer:2018csl}. Having the results just obtained in \cite{Aebischer:2021raf} we were in position to generalize at least the WET part of our 2018 formula to NLO in QCD \cite{Aebischer:2021hws}.

In fact we presented
for the first time the NLO master formula for the BSM part of this ratio expressed
in terms of the WCs of all contributing operators evaluated at the electroweak scale. To this end we used the JMS  basis
for which tree-level \cite{Jenkins:2017jig} and one-loop matching \cite{Dekens:2019ept} to the SMEFT were 
already known at that time. This was 
 an important step towards a complete next-to-leading (NLO) QCD analysis of
$\epe$  within the SMEFT.

 \boldmath
 \section{NLO QCD RG Evolution for    Non-Leptonic\\ $\Delta F = 2$  Transitions in the SMEFT (2022)}
 \unboldmath
 In the Summer of 2022 JJ and I decided to generalize the non-leptonic
 $\Delta F=2$ processes within the SMEFT to NLO in QCD \cite{Aebischer:2022anv}.
 In fact we presented for the first time NLO QCD RG evolution matrices for these processes in the SMEFT.  To this end we transformed first the known two-loop QCD ADMs of the BSM operators in the BMU basis into the ones in the  JMS basis for which
tree-level \cite{Jenkins:2017jig} and one-loop matching \cite{Dekens:2019ept} to the SMEFT were already known. This allowed us
subsequently to find the two-loop QCD ADMs for the SMEFT non-leptonic
$\Delta F=2$ operators in the Warsaw basis without performing any two-loop
calculations.

Having all these ingredients we investigated the impact
of these NLO QCD effects on the QCD RG evolution of SMEFT Wilson coefficients for
non-leptonic $\Delta F=2$  transitions from the NP scale $\Lambda$ down to the electroweak scale $\muEW$. The main benefit of these new contributions is that they allow to remove
renormalization scheme dependences present both in the one-loop matchings between
the WET and SMEFT and also between SMEFT and a chosen UV completion. But
the NLO QCD effects, calculated here in the NDR-$\overline{\text{MS}}$ scheme, turn out to be small, in
the ballpark of a few percent but larger than one-loop Yukawa top effects when only the $\Delta F=2$  operators are considered.

{\bf Very recently the NLO running for all baryon-number-conserving sector of dimension-six SMEFT has bee computed by Banik, Crivellin, Naterop and Stoffer \cite{Banik:2025wpi} and independently by Born, Fuentes-Martin and Thomsen \cite{Born:2026xkr}.}
\boldmath
\section{Kaon Physics Without New Physics in $\varepsilon_K$ (2023)}
\unboldmath
Motivated by my papers with Elena Venturini \cite{Buras:2021nns,Buras:2022wpw}
and my own results in \cite{Buras:2022qip} that NP seemed to be suppressed
in $\Delta F=2$ processes, I thought it would be interesting to find out what would
this imply for $Z^\prime$ models in which $Z^\prime$ contributes to $\varepsilon_K$ already at tree-level. In this manner the new paper with JJ \cite{Aebischer:2023mbz}
was born.

Analyzing
this scenario, we pointed out, following the 2009 observations of Monika
Blanke \cite{Blanke:2009pq}   and ours of 2020 \cite{Aebischer:2020mkv}, that despite of the absence of NP contributions to $\varepsilon_K$, significant
NP contributions to $\kpn$, $\klpn$, $K_S\to\mu^+\mu^-$,  $K_L\to \pi^0\ell^+\ell^-$, $\epe$ and $\Delta M_K$ can be present. In the simplest scenario, this is guaranteed, as far as flavour
changes are concerned, by a single non-vanishing {\em imaginary} left-handed $Z^\prime$  coupling $g_L^{sd}$.

\begin{figure}[t]
  \begin{center}
    \includegraphics[width=0.65\textwidth]{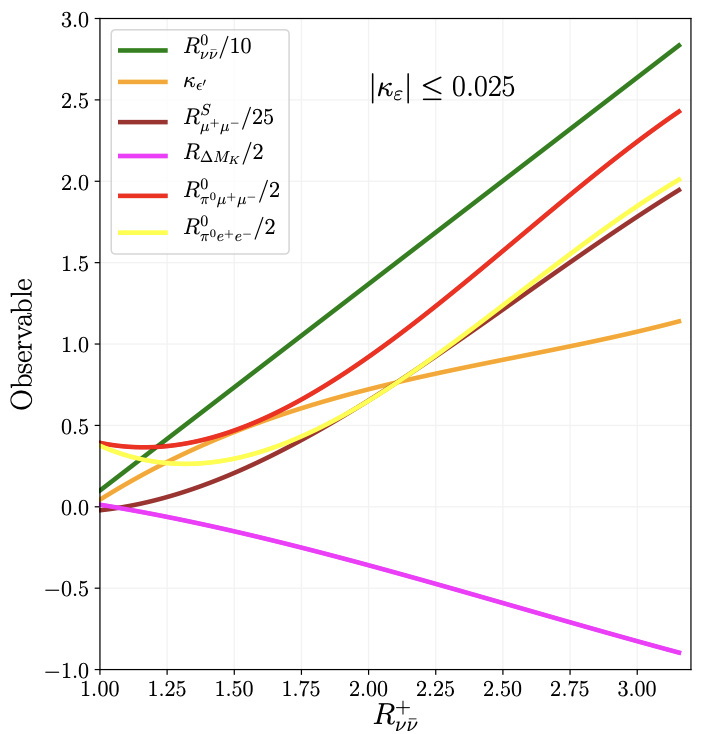}
   \caption{Correlations between the observable 
  $R_{\nu \bar \nu}^+$ and various other Kaon observables in a $Z^\prime$ model.
  All ratios $R_i=1$ in the SM. Note that for $\klpn$ and $K_S\to\mu^+\mu^-$ the
  ratios are divided by 10 and 25 respectively.  From \cite{Aebischer:2023mbz}.
  }
\label{Fig:3}
\end{center}
\end{figure}

As seen in Fig.~\ref{Fig:3} this scenario implies very stringent correlations between the Kaon observables
considered by us. In particular, the identification of NP in any of these observables implies automatically NP contributions to the remaining ones under the assumption of
non-vanishing flavour conserving $Z^\prime$ couplings to $q\bar q$, $\nu\bar\nu$ and $\mu^+\mu^-$. A characteristic
feature of this scenario is a strict correlation between $\kpn$ and $\klpn$ 
branching ratios on a branch parallel to the Grossman-Nir bound. Moreover, $\Delta M_K$  is
automatically suppressed as seems to be required by the results of the RBC-UKQCD
lattice QCD collaboration \cite{Bai:2018mdv}. Furthermore, there is no NP contribution to
$K_L\to\mu^+\mu^-$ 
which otherwise would bound NP effects in $\kpn$. Of particular interest are the
correlations of $\kpn$ and $\klpn$  branching ratios and of $\Delta M_K$  with
$\epe$. We investigated the impact of RG effects in the context of the SMEFT on this simple scenario.

\boldmath
\section{On the Interplay of Constraints from $B_{s}$, $D$, and $K$ Meson Mixing in $Z^\prime$ Models with Implications for $b\to s\nu\bar\nu$ Transitions (2024)}\label{AJBPS}
\unboldmath
I found  the scenario with strongly suppressed NP contributions to $\Delta F=2$, suggested in the papers with Elena, very interesting. As the paper with JJ, just
described,  was dealing with the $K$ system I thought it would be good to
extend this analysis to the $B$ meson system. In principle I could do this
analysis with JJ but we were very busy with completing our
long SMEFT-Review \cite{Aebischer:2025qhh} and I thought I should talk to somebody else who was
an expert in the SMEFT-codes. I found one rather quickly. It was one of
my many physics grandsons, Peter Stangl, former PhD student of David Straub.

Peter wrote already several very good papers and also developed SMEFT-codes,
in particular with JJ \cite{Aebischer:2018iyb,Proceedings:2019rnh,Aebischer:2023nnv}. Therefore I thought he would be an excellent
collaborator for this project and in fact his contribution to our
paper \cite{Buras:2024mnq} was far beyond my expections, although I knew already
from David that Peter was a very good physicist. I knew him from his
stay in our Excellence Cluster during his PhD studies but we did not collaborate at that time.

We studied again $Z^\prime$ models.
In the $B$ system in order to suppress NP contributions to $\Delta M_s$ and
$\Delta M_d$ from tree-level $Z^\prime$ exchanges one has to use the
strategy of Fulvia, Jennifer and myself \cite{Buras:2014sba} and invoke
some right-handed $Z^\prime$ couplings in order to suppress $Z^\prime$ contributions from the ones with only left-handed couplings involved.

Now, while my idea was to do a simple analysis, Peter extended it one day
to a general analysis of neutral meson mixing within the $Z^\prime$ models
in the context of the SMEFT and WET. He has rewritten to a large extent
the draft I have written at the beginning of our collaboration. I liked
this very much. Subsequently, he took over the leadership in our collaboration.
Now comes a short description what one can find in our paper which turned
out to be significantly longer than I planned.

 As evident from my previous $Z^\prime$ analyses, neutral meson mixing severely constrains NP  effects in FCNC processes.
  However, in certain regions of the $Z^\prime$ parameter space, the contributions to meson mixing observables become negligibly small even for large $Z^\prime$ couplings.
  While this a priori allows for significant NP effects in FCNC decays, we discussed how large $Z^\prime$ couplings in one neutral meson sector can generate effects in meson mixing observables of other neutral mesons, through correlations stemming from $\text{SU(2)}_L$ gauge invariance and RG effects in the SMEFT. We illustrated this with the example of $B_s^0-\bar B_s^0$ mixing, which in the presence of both left- and right-handed $Z^\prime bs$ couplings $\Delta_L^{bs}$ and $\Delta_R^{bs}$ remains SM-like for $\Delta_R^{bs}\approx 0.1\,\Delta_L^{bs}$. This is the strategy of \cite{Buras:2014sba}.
  
  We showed that in this case, large $Z^\prime bs$ couplings generate effects in $D$ and $K$ meson mixing observables, but that the $D$ and $K$ mixing constraints and the relation between $\Delta_R^{bs}$ and $\Delta_L^{bs}$ are fully compatible with a lepton flavour universality~(LFU) conserving explanation of the most recent $b\to s\ell^+\ell^-$ experimental data without violating other constraints like those from $e^+ e^-\to\ell^+\ell^-$ scattering.
  
    Assuming LFU, invariance under the $\text{SU(2)}_L$ gauge symmetry leads then to correlated effects in $b\to s\nu\bar\nu$ observables presently studied intensively by the Belle~II experiment, which allow to probe the $Z^\prime$ parameter space that is opened up by the vanishing NP contributions to $B_s^0-\bar B_s^0$ mixing. In this scenario the  suppression of $B\to K(K^*)\mu^+\mu^-$
      branching ratios implies {\em uniquely}  enhancements of
      $B\to K(K^*)\nu\bar\nu$ branching ratios up to $20\%$ and a small suppression of $B_s\to\mu^+\mu^-$, the pattern consistent with present experimental data.

\begin{figure}
\centering
\includegraphics[width=0.49\textwidth]{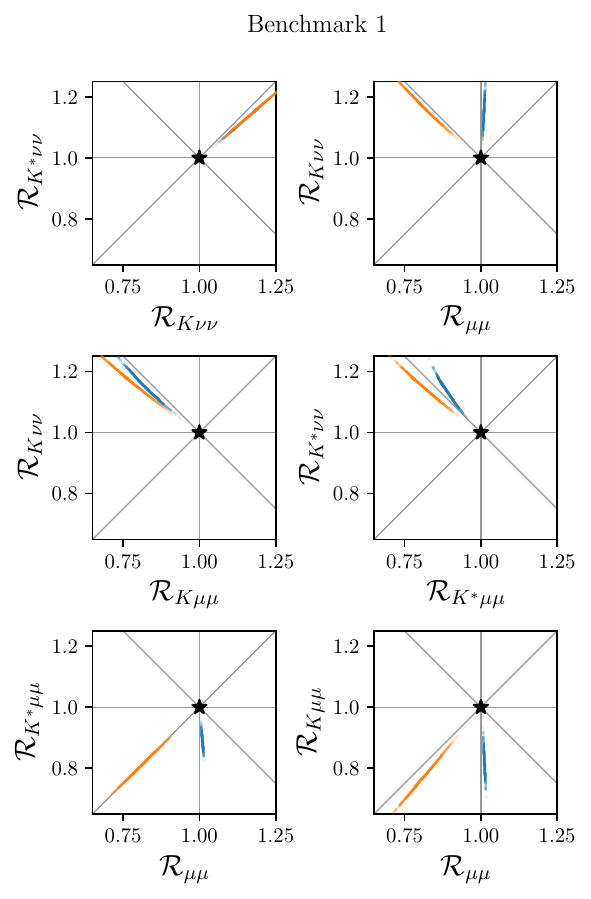}
\includegraphics[width=0.49\textwidth]{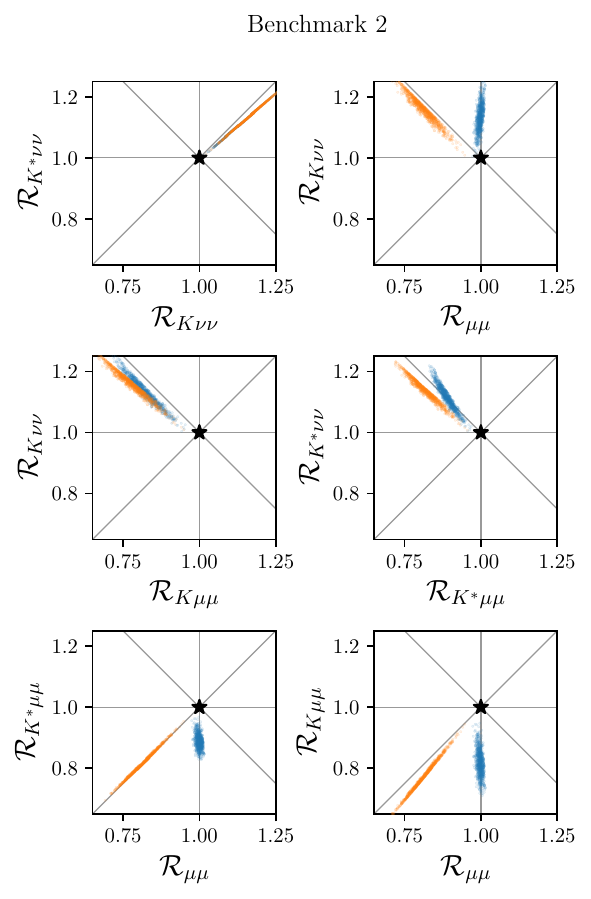}
\caption{%
  Predictions for various observables distributed according to the global fits
  performed in \cite{Buras:2024mnq}. The two left panels show the results for Benchmark~1, while the two right panels show the results Benchmark~2. The blue and orange points correspond to scenarios with vector and left-handed $Z^\prime$ lepton couplings, respectively.
}
\label{fig:predictions_R}
\end{figure}

      This can be seen in Fig.~\ref{fig:predictions_R} with the ratios
      already defined in (\ref{nunuratios}) in Section~\ref{2014nunubar}.
      
We considered the following scenarios:
\begin{enumerate}
 \item[\textbf{Scenario 1}] Benchmark~1 with left-handed $Z^\prime$-lepton couplings.
 \item[\textbf{Scenario 2}] Benchmark~2 with left-handed $Z^\prime$-lepton couplings.
 \item[\textbf{Scenario 3}] Benchmark~1 with vector $Z^\prime$-lepton couplings.
 \item[\textbf{Scenario 4}] Benchmark~2 with vector $Z^\prime$-lepton couplings.
\end{enumerate}
The Benchmarks~1 and 2 emphasize the importance of $D^0-\bar D^0$ and 
$K^0-\bar K^0$, respectively. I leave to the readers the comparison of 
Fig.~\ref{fig:predictions_R} with Fig~\ref{fig:Zprime} in Section~\ref{2014nunubar}.

{\bf The size of all effects is very similar in Benchmark~1 and Benchmark~2. The main difference between the two is that RG effects generating $C_{10}^{bs\ell_i\ell_i}(\mu_b)$ are allowed to be considerably larger in Benchmark~2, given that $Z^\prime$-quark couplings entering these RG effects are allowed to be larger.
This is in particular reflected by the slightly less strict (anti-)correlations between $R_{\mu\mu}$ and the other ratios.
The difference between Benchmark~1 and Benchmark~2 is more pronounced in the case of vector lepton couplings (blue dots), where $C_{10}^{bs\ell_i\ell_i}(\mu_b)$ is generated exclusively from RG effects.}

      \section{Analytic Inclusion of the Scale Dependence of the Anomalous Dimension Matrix in SMEFT (2018)}
      In the fall of 2017 during one of the MIAPP workshops at our Excellence Cluster Universe, Martin Jung, Emilie
      Passemar and myself got the idea to perform a detailed SMEFT analysis of  EDMs of heavy fermions.
 I have written two papers with Martin before, one on vector-like quarks 
 \cite{Bobeth:2016llm} and Yukawa enhancement of $Z$-mediated NP in
 $\Delta S=2$ and $\Delta B=2$ processes \cite{Bobeth:2017xry}
 described in Sections~\ref{VLQ1} and \ref{YukawaE}, respectively.
I expected therefore that now I will have a third successful paper with Martin.

The project went very well until the end of 2018 despite the fact that
we were separately busy with other projects. Basically what was left
to complete it was numerical evalua
tion which I admit I could not do.
 Unfortunately Martin and Emilie did not have time to perform this evaluation
 being busy with more important issues than EDMs. Emilie was expecting a daughter
 and both were applying for jobs.

 The paper did not appear in hep-arXiv until today but fortunately a part of it
 is described in chapter 17.3.6 of my book \cite{Buras:2020xsm}. I will
 not add anything to this paper here and hope that one day our paper
 will be completed. After all it takes also years to measure EDMs. In any
 case I have learned a lot from this study which allowed me to construct
 Section 17.3 in my book that is much more general than our unfinished paper.

 However, during these studies Martin and me playing with RGEs involving
 both gauge and Yukawa couplings derived a number of analytical expressions
 that I think are useful and fortunately we published them \cite{Buras:2018gto}.
The motivation for this analysis and what we derived by us were as follows.

The RGEs in the SMEFT are usually either solved analytically,
neglecting the scale dependence of gauge and Yukawa couplings, or numerically without such approximations. We presented
analytic solutions of RGEs that take into account the dominant scale dependence of the anomalous-dimension matrix due to
the running of the QCD coupling $\alpha_s$ and of the top-Yukawa coupling.
We considered first the 
case for which a given operator is generated directly through mixing with the {\it parent} operator whose Wilson coefficient is
non-vanishing at NP scale. Subsequently we considered the case of two-step running, in which two operators do not mix
directly, but only via a third {\it mediator operator}. We generalized  these solutions to an arbitrary number of operators and showed how
even in this case analytic solutions can be obtained.

An Express Review of this technology can be found in Appendix E of my
book but to appreciate the beauty of these analytic expressions one
has to read our paper. I like this paper but this does not seem to be
the case of the community which prefers to use numerical codes instead.
Until now our paper has been cited only 15 times of which 7 are self-citations.

\section{SMEFT Atlas: The Landscape Beyond the SM (2025)}\label{SMEFT2025}

In the Fall of 2020 I realized that the only existing long review of the SMEFT was the one
of Ilaria Brivio and Michael Trott \cite{Brivio:2017vri} which appeared in Physics Reports in 2017. It was a nice review, moreover written at Niels Bohr Institute where I received my PhD 48 years earlier. But it  discussed dominantly
electroweak precision tests and Higgs physics, moreover it was already four years
old. I got then the idea to write a Physics Reports article  on the SMEFT concentrating dominantly
on flavour physics, electric dipole moments but not fully ignoring electroweak
precision tests and high energy collisions.

As evident from this writing, I love writing reviews and after completing
my book in 2020 I thought it is time to write something new. This time
it was evident that I cannot write this review alone as I would not be
able to report properly on various numerical codes and the multitude of
papers which appeared in the last decade and were appearing almost every day on
arXiv. I told JJ about this idea and they got excited about this new project.

As I wrote already three reviews in Reviews of Modern Physics \cite{Buras:1979yt,Buchalla:1995vs,Buras:2004uu}, I thought it is time to write something for
Physics Reports. I contacted Patricia Hudson, Senior Editorial Manager
of Physics Reports and to my surprise received the answer not only from her but
also from Giulia Zanderighi, the wife of Ulrich Haisch, who in addition to being one of the directors of the MPI for physics and an excellent theorist, is an Editor of Physics Reports. I knew Giulia for decades and this simplified the full procedure.
On February 3rd 2021 I signed the formal Letter of Agreement for the submission of our review on 30th June 2022, subsequently changed to January 2023.

Meanwhile I got the idea to update my review on NLO and NNLO QCD corrections
to weak decays which appeared in arXiv in 2011 \cite{Buras:2011we}. Again
during the last decade several NLO and NNLO QCD corrections have been performed,
mostly by other authors, with the exception of my two papers with Jennifer
in 2012 \cite{Buras:2012gm,Buras:2012fs} described in Section~\ref{BG12}.
Again I decided to submit it to
Physics Reports and signed another Letter of Agreement on June 2, 2022 with the deadline
of May 2023. This agreement carried the name Buras2, the first one  Buras1.

But already in September 2022 it was evident that our SMEFT review will not
be finished in January 2023. On the other hand I made progress on Buras2 review
and I suggested to Physics Reports that we exchange the deadlines with Buras2
submitted in January 2023 and Buras1 in May 2023. I made this deadline
and Buras2 appeared in July 2023 in Physics Reports with an addition
to the 2011 title: 1988-2023.

To make the long story short we missed totally the deadline of May 2023 for
Buras1. It appeared in arXiv on July 9th 2025\footnote{Interestingly on the same day a very extensive SMEFT global fit made by Luca Silvestrini and collaborators appeared in the arXiv \cite{deBlas:2025xhe}.}, during the EPS-HEP 2025
Conference in Marsellie. Actually, this delay by two years was fortunate.
At least 300 papers on SMEFT have been published between the Summer of 2023
and the Summer of 2025 and we could include them in our review so that
our review \cite{Aebischer:2025qhh}, 298 pages long, refers to more than 1000 papers. Let me then briefly state what one can find in this review.
 A short summary of Kaon physics in the
SMEFT can be found in a talk by Jason at Kaon 2025 \cite{Aebischer:2025byg}.

We summarized the present
status of the SMEFT with a particular emphasize on its role in the indirect search for NP. While flavour physics of both quarks and leptons is the main topic
of our review, we also discussed electric dipole moments, anomalous magnetic moments $(g-2)_{\mu,e}$, $Z$-pole observables, Higgs observables and high-$p_T$ scattering processes within
the SMEFT. We grouped the observables into 10 classes:
  
{\bf Class 1: Meson-Antimeson Mixing}

\begin{center} 
{\bf 1A: Down type Mesons}
\end{center}
\be \label{Class1}
K^0-\bar K^0\,, \qquad  B_d^0-\bar B_d^0\,, \qquad  B_s^0-\bar B_s^0\,.
\ee

\begin{center}
{\bf 1B: Up type Meson}
\end{center}
\be
D^0-\bar D^0\,.
\ee         

{\bf Class 2: FCNC Decays of Mesons}

\begin{center}
{\bf 2A: Semileptonic Decays}
\end{center}
\be\label{Class2x}
B\to K (K^*)\ell^+\ell^-\,, \quad B\to K(K^*)\nu\bar\nu\,,\quad  B\to X_{s,d} \ell^+\ell^-\,,\quad  B\to X_{s,d}\nu\bar\nu \,, \nn
\ee
\be\label{Class2y} 
\kpn\,, \quad \klpn\,, \quad  K_L\to \pi^0 \mu^+\mu^-\,, \quad  K_L\to \pi^0 e^+e^-\,,
 \ee
\be\label{Class2z} 
B\to \pi\nu \bar \nu\,, \quad B\to \varrho \nu \bar \nu\,, \quad {D\to K \,\ell \,\nu_\ell\,, \quad D\to \pi \,e \,\nu_e\,.}\nn
\ee

\begin{center}
{\bf 2B: Leptonic Decays}
\end{center}
\be\label{Class2b}
B_{s,d}\to\mu^+\mu^-\,,\quad K_{L,S}\to\mu^+\mu^-\,,\quad D\to\mu^+\mu^-\,.
\ee

\begin{center}
{\bf 2C: Radiative Decays}
\end{center}
\be\label{Class2c}
B\to X_s\gamma\,, \quad  B\to X_d\gamma\,, \quad B\to K^*\gamma\,, \quad B\to \varrho\gamma\,, {\quad D\to \rho\gamma}\,.
\ee

{\bf Class 3: Non-Leptonic Decays of Mesons}

\be\label{Class3KB}
K_L\to \pi\pi\,, \quad \epe\,, \quad B\to\pi\pi\,,\quad B\to K\pi\,,\quad B\to KK\,.
\ee
\be\label{Class3D}
D\to \pi\pi\,,\quad D\to\pi K\,,\quad D\to KK\,.
\ee

{\bf Class 4: Electroweak Precision Observables }

\be \label{Class4}
{W, Z} \text{-Pole~Observables}\,.
\ee

{\bf Class 5:  Leptonic LFV Decays}

\begin{center}
{\bf 5A: $\Delta F=1$ Decays}
\end{center}
\be \label{Class5A}
\mu^-\to e^-e^+e^-\,, \quad \tau^-\to\mu^-\mu^+\mu^-\,, \quad\tau^-\to e^-e^+e^-\,,\quad
\tau^-\to\mu^-e^+e^-\,, \quad \tau^-\to e^-\mu^+\mu^-\,,
\ee
with the last two receiving also $\Delta F=2$ contributions.
\begin{center} 
{\bf 5B: $\Delta F=2$ Decays}
\end{center}
\be
\tau^-\to e^-\mu^+e^-\,, \qquad \tau^-\to\mu^- e^+\mu^-\,.
\ee

\begin{center} 
{\bf 5C: Radiative Decays}
\end{center}
\be
{{\mu\to e \gamma\,, \qquad \tau\to e\gamma\,, \qquad \tau\to \mu\gamma}\,.}
\ee

\begin{center}
{\bf 5D: LFV $Z$ Decays}
\end{center}
\be
Z \to f_i  \bar f_j.
\ee

{\bf Class 6: Semileptonic LFV Decays}

\begin{center}
{\bf 6A: Quark Flavour Conserving Decays}
\end{center}
\be
{\tau^-}\to \pi\ell^-\,,\qquad {\tau^-}\to \rho\ell^-\,,\qquad {\tau^-}\to\phi \ell^-\,, \quad \mu\to e\,~\text{conversion}\,.
\ee

\begin{center}
{\bf 6B: Quark Flavour Violating Decays}
\end{center}
\be
K_{L,S}\to\mu e\,, \qquad K_{L,S}\to\pi^0\mu e \,, \nn
\ee
\be
B_{d,s} \rightarrow \mu e\,, \qquad B_{d,s} \rightarrow \tau e\,, \qquad B_{d,s} \rightarrow \tau \mu\,,
\ee
\be
B_d\to K^{(*)}\tau^\pm\mu^\mp\,, \qquad B_d\to K^{(*)}\mu^\pm e^\mp\,, \nn
\ee
\be
D^0\to \mu^\pm e^\mp\,, \qquad D^0\to \pi^0\mu^\pm e^\mp\,. \nn
\ee

{\bf Class 7: Electric and Magnetic Dipole Moments}

\be
{\rm 7A: Electric~ Dipole~ Moments}\,, \qquad {\rm 7B:} (g-2)_{e,\mu}\,.
\ee

{\bf Class 8: Charged Current Processes}

\be
\bar B\to D \ell\bar\nu_l\,, \quad \bar B\to D^* \ell\bar\nu_l\,, \quad B_c\to J/\psi\tau\bar\nu\,,\quad   B_c^+\to \tau^+\nu_\tau\,, \quad  B^+\to \tau^+\nu_\tau\,,
\ee
\be
\pi^- \to e^- \bar \nu\,,\quad {\rm Beta-decay~ and ~neutron~ decay,}    \quad D^+\to e^+\nu\,. \nn
\ee

{\bf Class 9: Higgs Observables}

\be
\text{Higgs Signal Strengths}
\ee

\boldmath
{\bf Class 10: High $p_T$ Scattering Processes}
\unboldmath

\be
\begin{aligned}
&
{\rm 10A:} ~\ell^+ \ell^- \to \ell^+ \ell^- \,, \qquad 
{\rm 10B:} ~\ell^+ \ell^- \to q \bar q'  \,, \qquad  \\
&
{\rm 10C:} ~pp \to \ell^+ \ell^-\,, \quad  p p \to \ell \bar \nu  \,, \qquad 
{\rm 10D:} ~ pp \to q \bar q' .
\end{aligned}
\ee

\begin{figure}[htb]%
\centering
\includegraphics[width=0.7\textwidth]{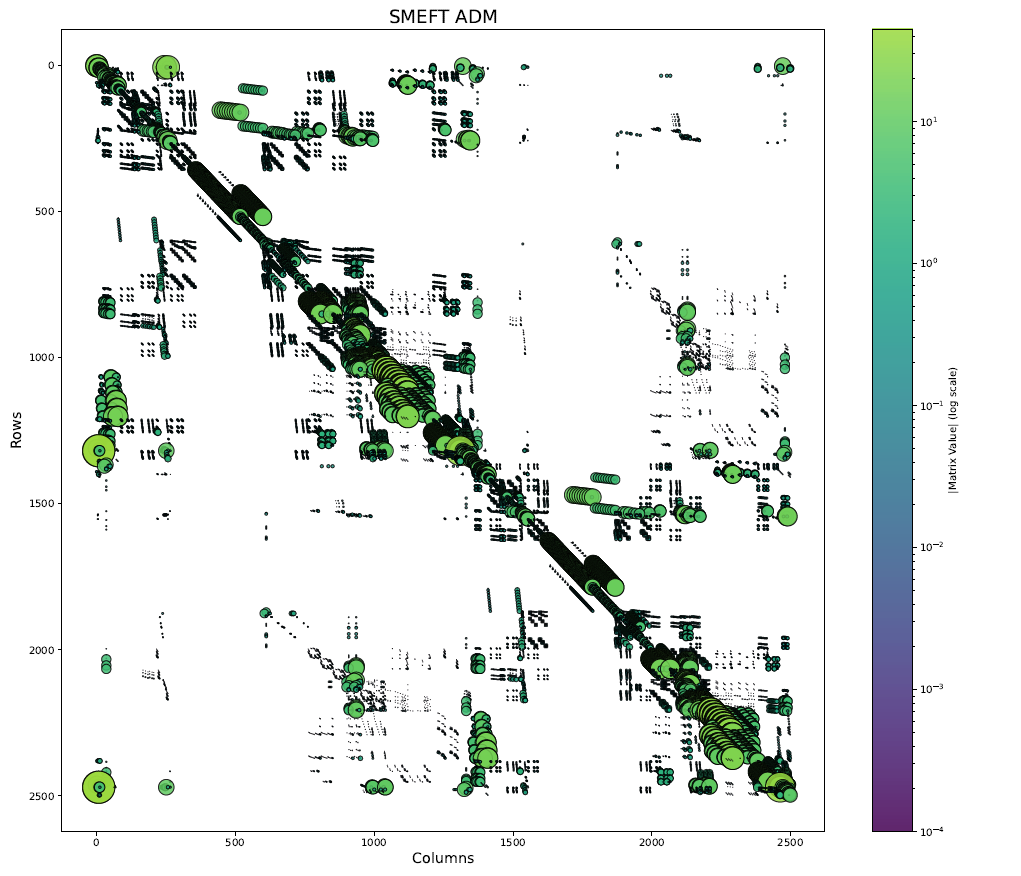}
\caption{ Jason Aebischer as Pablo Picasso: the non-zero entries of the ADM. From \cite{Aebischer:2025qhh}.}%
\label{fig:ADMplot}%
\end{figure}

For each class we listed  the most
relevant operators and the corresponding RG equations. Also tree-level matching
of SMEFT on to WET was given.
We exhibited the correlations between different classes implied both by the operator
mixing and the $\text{SU(2)}_L$ gauge symmetry. Our main goal was to provide an insight into the complicated operator structure of this framework which hopefully will facilitate the identification of valid ultraviolet completions behind possible anomalies observed in future data. Numerous colourful charts, and 85 tables, while representing rather
complicated RG evolution from the NP scale down to the electroweak scale, beautify the involved SMEFT landscape. The complexity of RG mixing between the operators is expressed with the plot of non-vanishing entries of the $2499\times 2499$
SMEFT ADM in Fig.~\ref{fig:ADMplot} constructed by Jason.

Over 1000 references to the literature underline the importance and the popularity of this field. We discussed both top-down and bottom-up approaches as well as their interplay. This allowed us eventually to present an atlas of different landscapes beyond the SM that includes heavy gauge bosons and scalars, vector-like quarks and leptons and leptoquarks. 

While in the present write-up  I  indicated specific NP effects characteristic
for a given NP model or scenario, in this SMEFT review we have listed 
tables of SMEFT operators most important for a given model.

In March 2023 we were surprised by an excellent SMEFT review of Gino Isidori,
Felix Wilsch and Daniel Wyler \cite{Isidori:2023pyp} that in 2024 was published
in Rev.Mod.Phys. Fortunately their and our review differ by much from each other.

Our review is divided into three parts and contains several appendices.

In {\bf Part I} after summarizing the Modern Effective Field Theories and the Grand View of
our SMEFT Expedition we have presented the following topics:
\begin{itemize}
\item SMEFT Lagrangian.
\item Choice of Weak Bases.
\item Effective Parameters.
\item SMEFT Renormalization Group Equations.
  \item SMEFT Beyond Leading Order.
  \item Flavour Symmetries in the SMEFT.
  \item SMEFT Tools. These include: Basis generation, Numerical basis transformations, Matching tools, running tools, Observable calculators and fitting tools, Monte-Carlo tools, Clustering tools. 
  \item A Guide to Existing SMEFT Analyses (Tables 2-14) with roughly 350 references. This includes tree level and NLO level analyses and global fits of SMEFT WCs.
\end{itemize}
This part contains many references to technology of the SMEFT. In particular
four of my collaborators, Jason Aebischer, Jacky Kumar, Peter Stangl and David
Straub developed already in 2018 very useful computer codes for RG running and phenomenology. These are {\em wilson} \cite{Aebischer:2018bkb}, {\em flavio} \cite{Straub:2018kue} and {\em smelli} \cite{Aebischer:2018iyb}. See also {\it RGESolver} with the participation of Luca Silvestrini \cite{DiNoi:2022ejg} and {\em Dsixtools} with the participation of Alejandro Celis \cite{Celis:2017hod}.

Moreover, some of them were active in the last two years in various applications
of these codes. These are \cite{Aebischer:2023djt,Aebischer:2023nnv,Aebischer:2024xnf,Aebischer:2024csk,Aebischer:2025hsx,Aebischer:2025zxg,Aebischer:2025ddl}. The titles tell us everything.

\begin{figure}[t]
\centering%
\includegraphics[width=0.8\textwidth]{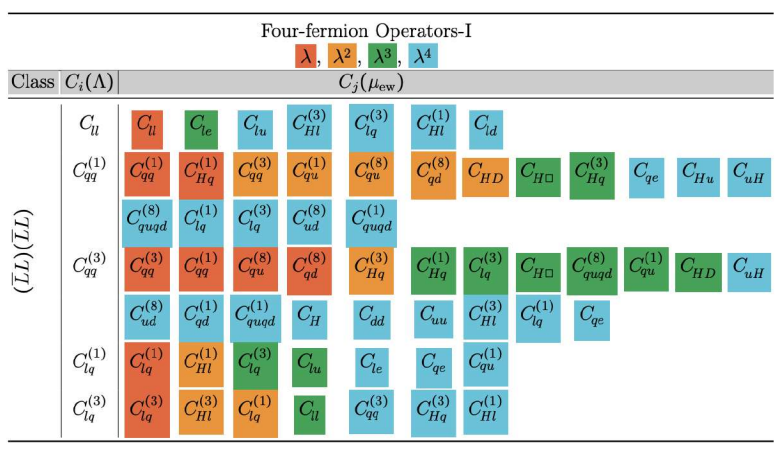}%
\caption{\small The pattern of operator mixing for the $(\overline LL) (\overline LL)$ four-fermion $\Delta B=\Delta L=0$ operators in the SMEFT. See the text for explanations. From \cite{Aebischer:2025qhh}. Constructed by Jacky Kumar.}
\label{tab:mixing}
\end{figure}

\begin{figure}[t]
\centering%
\includegraphics[width=0.8\textwidth]{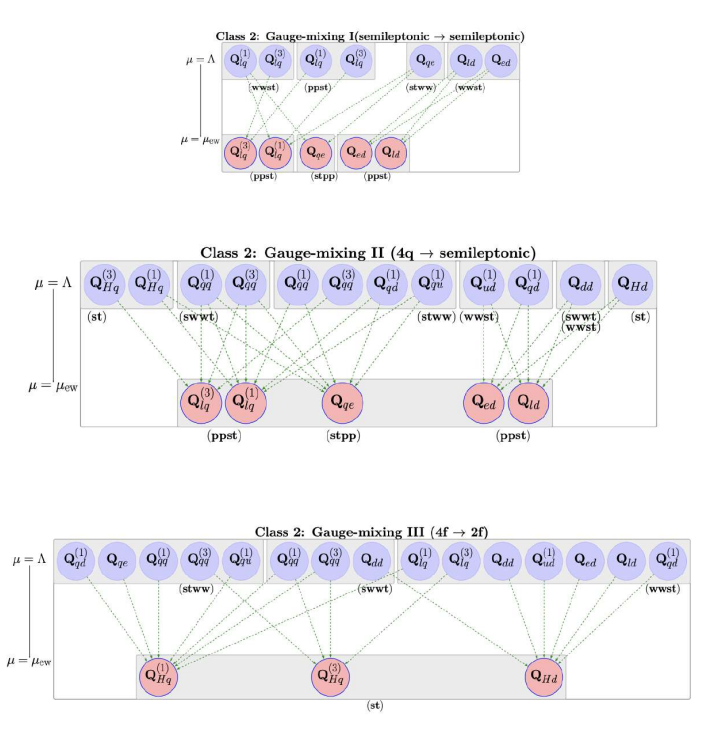}%
\caption{\small Class 2: Gauge coupling dependent operator mixing for $\Delta F=1$ transitions such as $b \to s\ell^+ \ell^-$, $b\to s\nu \bar \nu$ and $s\to d \nu \bar \nu$ in the Warsaw down-basis. The self-mixing is omitted. From \cite{Aebischer:2025qhh}. Constructed by Jacky Kumar.}%
\label{chart:df1-gauge}%
\end{figure}

    In {\bf Part II},  which can be considered as {\bf the SMEFT Atlas}, we have
    classified the processes in 10 classes as listed above and presented
      the following topics
    \begin{itemize}
    \item Operator Mixing in the SMEFT and Mapping Operators to Observables
      illustrating it with several colourful tables. An example is given in
      Fig.~\ref{tab:mixing}. It shows the impact of a coefficient $C_i(\Lambda)$
      on coefficients  $C_j(\mu_{ew})$ at the EW scale through RG operator mixing       with the strength of $\lambda^k$ ($\lambda=0.22$).
    \item
      As I stressed the importance of correlations between observables already in the abstract, Tables 17-19 in Section 12 and Tables 20-36 in Section 13 of our review are
      very important in this context.
    \item Description of all classes of observables listing for each one  the relevant  RGEs and matchings between WET and SMEFT. Colourfull charts help
      hopefully to digest the complicated RGEs. An example is given in
      Fig.~\ref{chart:df1-gauge}.    
      \end{itemize}

      In {\bf Part III} after the classification of BSM particles into gauge
      bosons, scalars, vector-like quarks, vector-like leptons and leptoquarks
      we have presented:
      \begin{itemize}
      \item
        Top-down Approach.
      \item
        Bottom-Up Approach
      \item Hunting New Particles using these two approaches.
      \end{itemize}


      It is not possible for me to describe more what can be found in this
      review but let me make here one point which hopefully will correct
      some statements by many authors present in the literature.

      In many papers it is stated that the SMEFT offers within the Bottom-Up
      approach {\em a model independent search for New Physics.} This is {\em incorrect} simply for at least two reasons.
      \begin{itemize}
      \item
        The outcome of the SMEFT analysis in  the absence of a $\text{U(3)}^5$  flavour symmetry in the gauge sector depends on how the Yukawa and other interaction matrices are oriented in flavour space. To know how they
        are oriented one has to know the UV completion. 
        This also explains why the bounds on various Wilson coefficients found
        in \cite{Silvestrini:2018dos} for the down-basis and up-basis differ from each other
        and as stressed in particular by Gino Isidori and collaborators
        (see e.g. \cite{Allwicher:2024ncl} )
        there could be other bases in between these two popular bases. They
        could be parametrized by a parameter $0\le\epsilon\le 1 $ which interpolates between these two bases. We have discussed this point in Section 5 of
        our review where references to other papers of Luca, Gino and Admir Greljo
        can be found.
      \item
        As discussed in Section 26 of our review it is not always possible to
        determine SMEFT Wilson Coefficients at the Electroweak Scale
        through the matching of them to the WET coefficients determined
        in experiment. In particular in the case of operators involving
        left-handed currents a WET coefficient is matched to a sum of two
        SMEFT coefficients and not to a single SMEFT coefficient. Consequently
        only their sum can be determined by experiment. We presented several
        examples with left-handed and right-handed currents demonstrating when
        a unique determination of SMEFT coefficients is possible and when not.
      \end{itemize}

      {\bf  As far as the issue of the model independence of the SMEFT itself is concerned some insight can be obtained from \cite{Datta:2025csr}.}
  
\boldmath
\part{Most Recent Strategy: Removing $\vcb$ Uncertainties}\label{BV}
\unboldmath

\section{Preface}
In order to motivate this strategy in explicit terms, let us recall  the values of $\vcb$
extracted from inclusive and exclusive tree-level semi-leptonic $b\to c$ decays
\cite{Finauri:2023kte,FlavourLatticeAveragingGroupFLAG:2021npn} 
\be
\vcb_\text{incl}=(41.97\pm0.48)\cdot 10^{-3},\qquad \vcb_\text{excl}=(39.21\pm0.62)\cdot 10^{-3}\,.
\ee
As rare K and B decays and mixing parameters are sensitive functions of $\vcb$, varying it from $39\cdot 10^{-3}$ to $43\cdot 10^{-3}$ changes $\Delta M_{s,d}$ and $B$-decay branching ratios
by roughly $21\%$, $\kpn$ branching ratio by $31\%$, $\varepsilon_K$ by $39\%$
and $\klpn$ and $K_S\to\mu^+\mu^-$ branching ratios by $48\%$.

Based on my report in Part~\ref{NLOQCD}, these uncertainties are clearly a disaster for those like me, my collaborators
and other experts in NLO and NNLO calculations who spent decades to reduce theoretical uncertainties in basically all important rare $K$ and $B$ decays and
quark mixing observables down to $(1-2)\%$.

It is also a disaster for lattice QCD experts who for quark mixing observables and in particular  meson weak decay constants achieved the accuracy at the level of a few percent. Also significant progress on formfactors from LQCD and Light-Cone Sum Rules has been done.

\begin{figure}[t!]
\centering%
\includegraphics[width=0.48\textwidth]{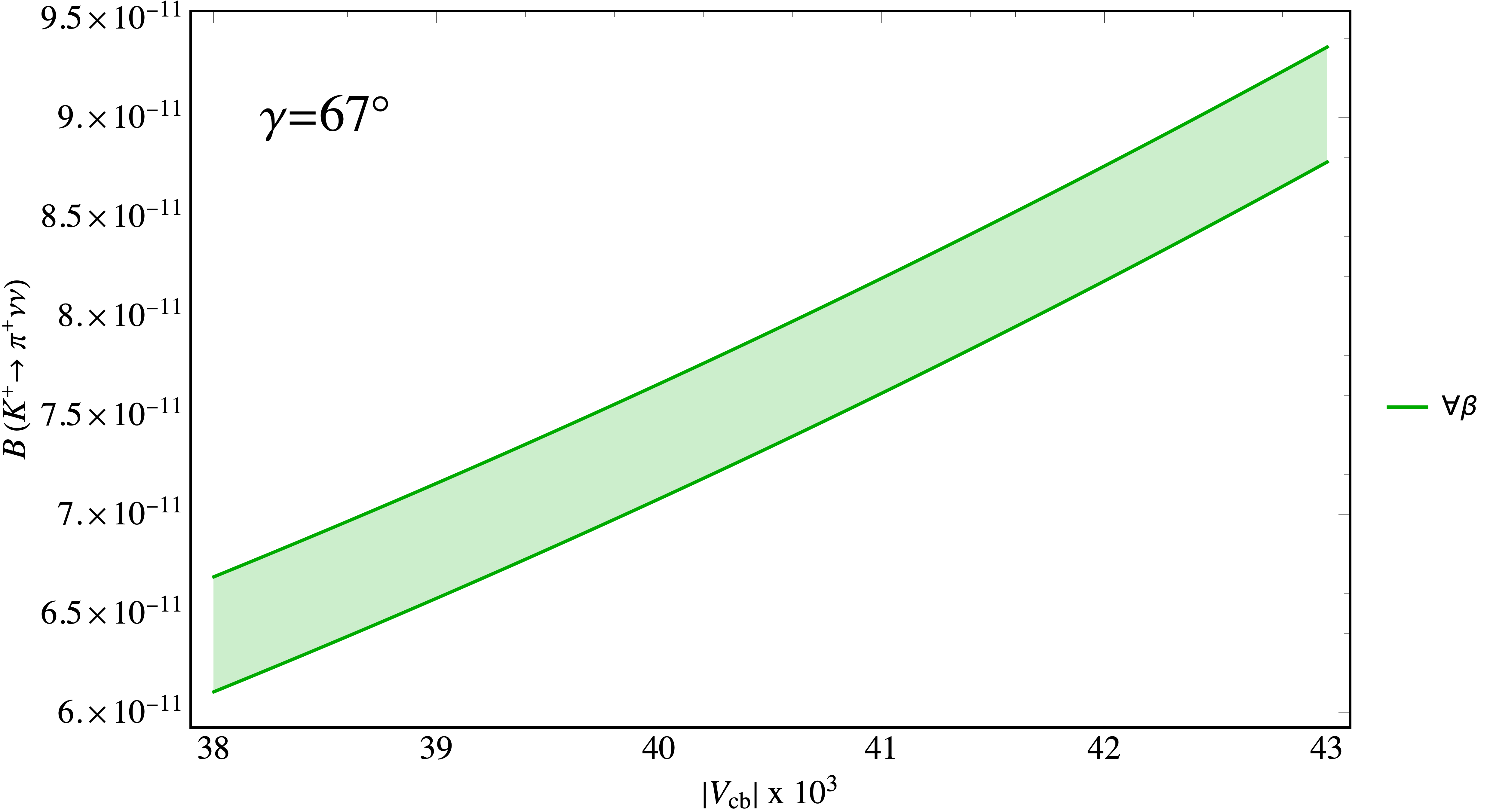}%
\hfill%
\includegraphics[width=0.48\textwidth]{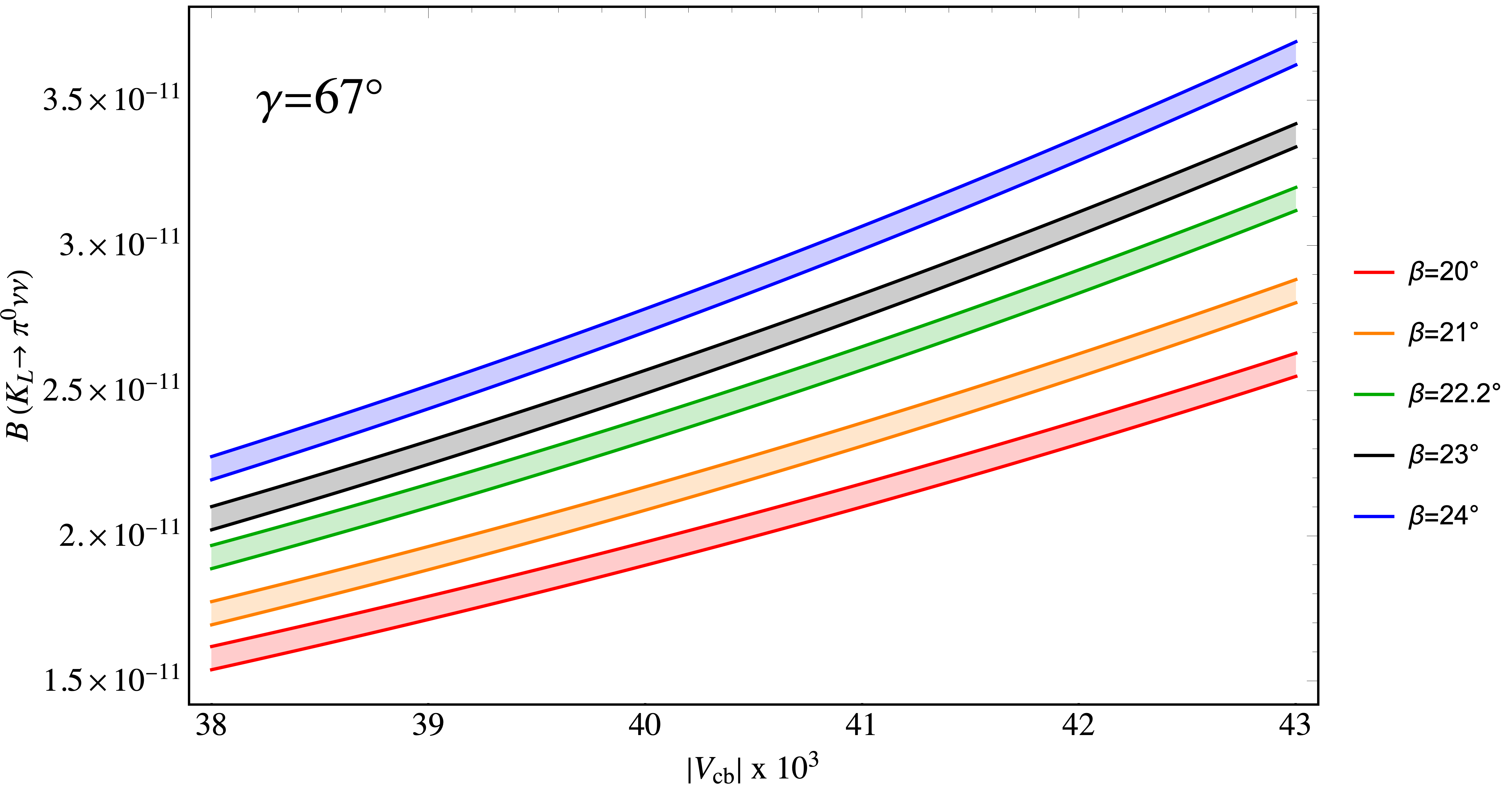}\\
\includegraphics[width=0.48\textwidth]{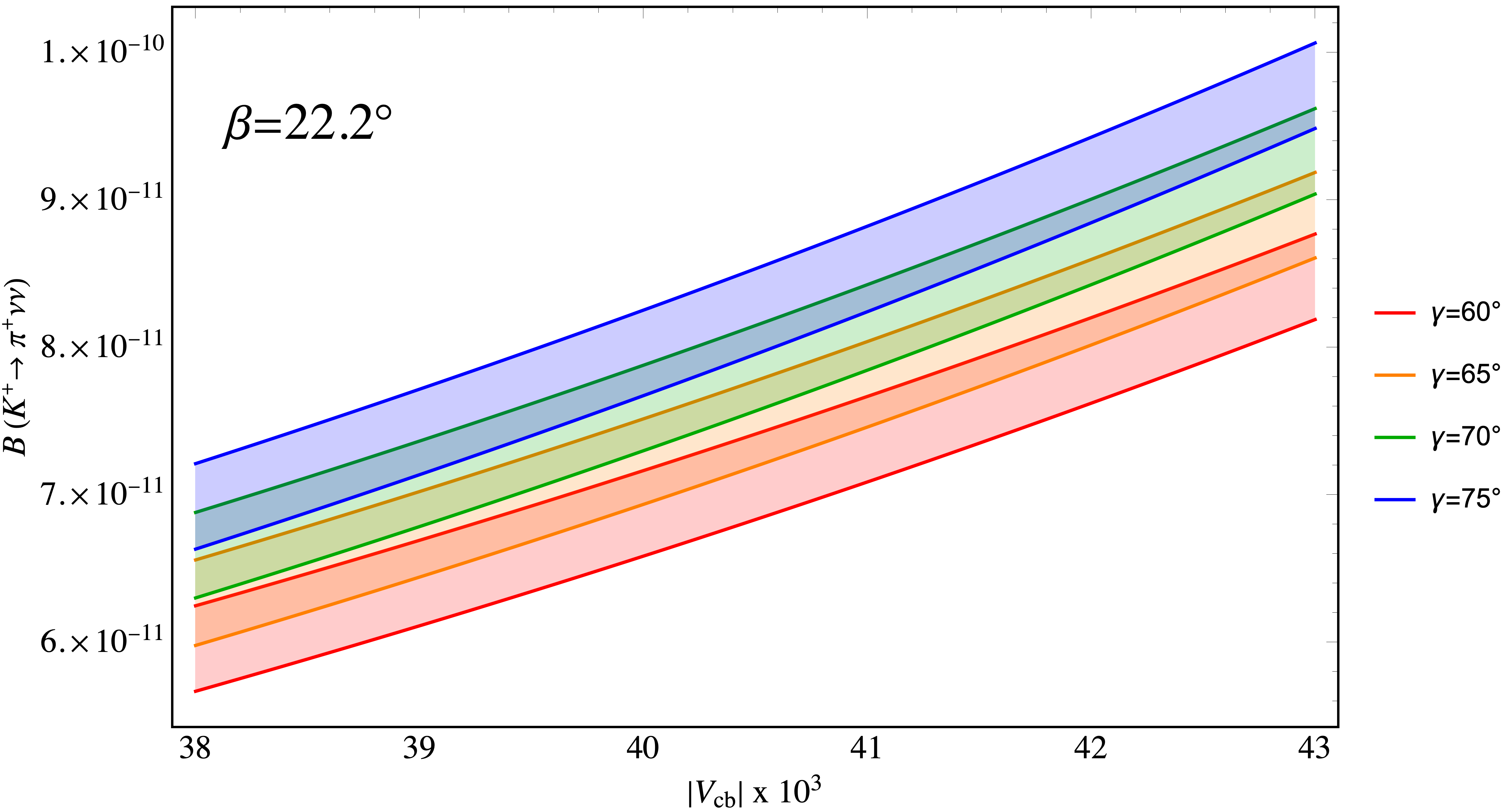}%
\hfill%
\includegraphics[width=0.48\textwidth]{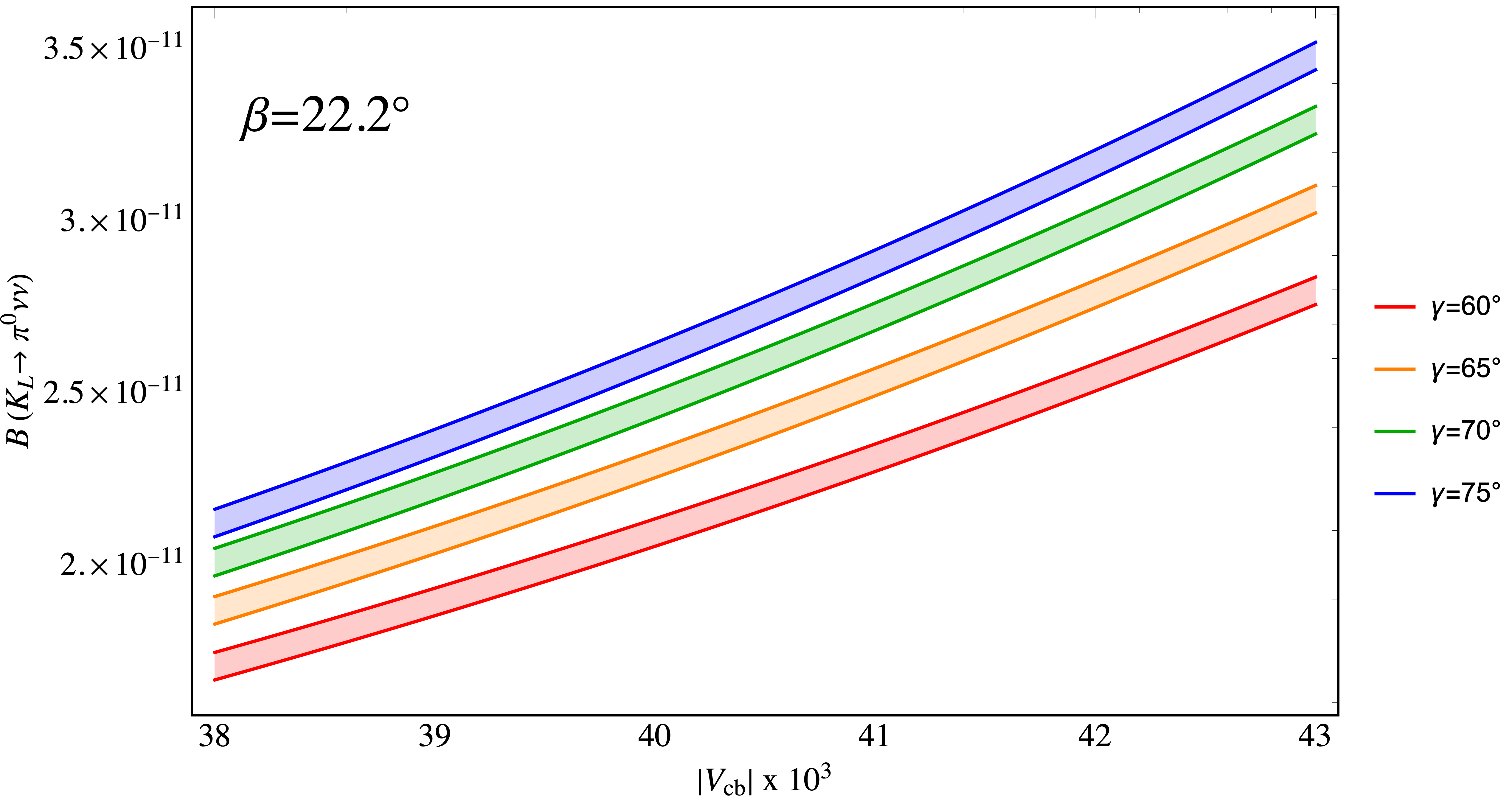}%
\caption{\it {The dependence of the branching ratios $\mathcal{B}(\kpn)$ (left panels) and $\mathcal{B}(\klpn)$ (right panels) on  $|V_{cb}|$ for different values of   $\beta=20.0^\circ,21.0^\circ, 22.0^\circ, 23.0^\circ, 24.0^\circ$  at fixed $\gamma=67^\circ$ and for different values of $\gamma=60.0^\circ, 65.0^\circ, 70.0^\circ, 75^\circ$  at fixed $\beta=22.2^\circ$ . {The width of the bands represents the uncertainties whose origin is not related to the $\gamma$, $\beta$ and $\vcb$ parameters. From \cite{Buras:2021nns}. }} 
\label{fig:CKMdependence}}
\end{figure}

In order to motivate this strategy further, I show in Fig.~\ref{fig:CKMdependence} the dependence of $\kpn$ and $\klpn$ branching ratios on $\vcb$, $\beta$ and
$\gamma$. This dependence has been already studied with Monika
in \cite{Blanke:2018cya} but this particular figure is from the first paper
with Elena that I will describe in details later \cite{Buras:2021nns}.
One can notice that the branching ratio for $\kpn$  is almost exactly independent {of} the $\beta$ angle, as already mentioned previously. One can also see that the largest uncertainties on these branching ratios {are}  due to the $\vcb$ parameter: the {elimination} of this source of error is the main focus of
\cite{Buras:2021nns}. But let us first go back to 2003 when this idea was born
in the context of $\Delta M_{s,d}$ and $B_{s,d}\to \mu\bar\mu$ \cite{Buras:2003td}.

  \unboldmath
  \boldmath
  \section{Relations between $\Delta M_{s,d}$ and $B_{s,d}\to \mu\bar\mu$
  in\\ Models with CMFV (2003, 2021)}
    \unboldmath
The predictions for the
$B_{s,d}-\bar B_{s,d}$ mixing mass differences $\Delta M_{s,d}$ and 
the branching ratios $\mathcal{B}(B_{s,d}\to\mu\bar\mu)$ 
within the SM and its extensions suffered in 2003, when I got this idea
\cite{Buras:2003td}, from considerable
hadronic uncertainties present in the $B_{s,d}$-meson 
decay constants $F_{B_{s,d}}$ and CKM parameters $V_{ts}$ and $V_{td}$ that all enter these quantities quadratically.
I pointed out that in the restricted class of 
 models with minimal flavour violation in which only the SM low 
energy operators are relevant (CMFV), the ratios 
$\mathcal{B}(B_{q}\to\mu\bar\mu)/\Delta M_q$ $(q=s,d)$
do not depend on $F_{B_{q}}$ and the CKM matrix elements.
They involve
in addition to the short distance functions and B meson lifetimes
only the non-perturbative parameters $\hat B_{q}$ entering $\Delta M_q$.

In 2003 several corrections to $B_s\to\mu^+\mu^-$ and $B_d\to\mu^+\mu^-$
branching ratios, 
that are included in the 2026 formulae given in  App.~\ref{UsefulEX}, were
unknown and I found
\be\label{CMFV61}
R_{q\mu}=\frac{\mathcal{B}(B_q\to\mu^+\mu^-)}{\Delta M_q}= 4.291\times 10^{-10}\ \frac{\tau_{B_q}}{\hat B_q}\frac{(Y(x_t))^2}{S(x_t)},\qquad q=d,s\,,
\ee
with $Y(m_t)$ and $S(x_t)$ known one loop $m_t$-dependent functions. In
particular the  $\Delta\Gamma_q$
effects that are only relevant for $B_s\to\mu^+\mu^-$ \cite{deBruyn:2012wk}
have not been included.

In the spring of 2021 in a paper with Christoph Bobeth the search for NP with the help of this ratio has been reemphasized \cite{Bobeth:2021cxm}.
    In particular we stressed again  that it does not suffer
from the persistent puzzle on the $|V_{cb}|$ determinations from inclusive
versus exclusive $b\to c\ell\bar\nu$ decays and which affects the
value of the CKM element $\vts$ that is crucial for the SM predictions of
both $\overline{\mathcal{B}}(B_s\to\mu\bar\mu)$ and $\Delta M_s$, but
cancels out in this ratio.

In our analysis we updated my analysis of 2003 by including higher order electroweak and QED corrections \cite{Bobeth:2013uxa} and also widths corrections  to
$\mathcal{B}(B_{s}\to\mu\bar\mu)$ \cite{deBruyn:2012wk}
which have been calculated in the previous decade. Moreover, there was a significant 
improvement for the parameters $\hat B_q$ that are known already with good precision from LQCD with 2+1+1 flavours \cite{Dowdall:2019bea}. In \cite{Bobeth:2021cxm} we have used  the average of 2+1+1 \cite{Dowdall:2019bea} and 2+1 results \cite{FermilabLattice:2016ipl} provided in \cite{Dowdall:2019bea}.
Using instead only the 2+1+1 flavour results of \cite{Dowdall:2019bea}
one finds \cite{Buras:2022wpw}
\be\label{F1}
  \boxed{R_{s\mu}=\frac{\overline{\mathcal{B}}(B_s\to\mu^+\mu^-)}{\Delta M_s}=
    (2.130^{+0.083}_{-0.053})\times 10^{-10}\text{ps}\,,}
\ee
   \be\label{F2}
   \boxed{R_{d\mu}=\frac{\mathcal{B}(B_d\to\mu^+\mu^-)}{\Delta M_d}=
(2.005^{+0.089}_{-0.066})\times 10^{-10}\text{ps}\,.}
\ee

Consequently in these models the predictions for $B_{q}\to\mu\bar\mu$
branching ratios are rather precise as  $\Delta M_q$ have already been precisely measured.  
Using then the relations (\ref{F1}) and (\ref{F2}) Elena Venturini and me found
\cite{Buras:2022wpw}
\be
\boxed{\mathcal{B}(B_{s}\to\mu\bar\mu)=(3.78^{+ 0.15}_{-0.10}) \cdot 10^{-9}, \qquad
\mathcal{B}(B_{d}\to\mu\bar\mu)=(1.02^{+ 0.05}_{-0.03})\cdot 10^{-10}}.
\ee
Present experimental values are given in Table~\ref{tab:SMBRs}.

\boldmath
\section{Searching for New Physics in Rare $K$ and\\ $B$ Decays without $\vcb$ and $\vub$ Uncertainties (2021)}\label{BVfirst}

\unboldmath
In the fall of 2021 my strategy of 2003 has been generalized to semi-leptonic
$B$ and $K$ decays. My new collaborator, Elena Venturini, was a postdoc in the
group of Andreas Weiler. We met first time in October 2020 but until we
started our collaboration in the Summer of 2021, we did not discuss physics
but classical music, practically every day exchanging emails. But finally
we thought we could do research together as well.

Only three papers resulted from this collaboration \cite{Buras:2021nns,Buras:2022wpw,Venturini:2022sdf}, but I expect that this strategy, which I call BV-strategy, will remain to
play a significant  role in the tests of the SM like my ratios of 2003 which
played this role in the last 20 years.

Let me then describe the first paper \cite{Buras:2021nns} which appeared
in September 2021. In this paper 
we reemphasized the strong dependence of the branching ratios
$\mathcal{B}(K^+\to\pi^+\nu\bar\nu)$  and  $\mathcal{B}(K_L\to\pi^0\nu\bar\nu)$ 
on $|V_{cb}|$ that is stronger than in rare $B$ decays, in particular for $K_L\to\pi^0\nu\bar\nu$ it is like $\vcb^4$.   Thereby
   the persistent tension between inclusive and exclusive determinations
  of $|V_{cb}|$ weakens the power of these theoretically
  clean decays   in the search for NP.
  We demonstrated how this uncertainty can be practically removed by
  considering within the SM suitable ratios of the two branching ratios  between each other and with other observables like the branching ratios for 
$K_S\to\mu^+\mu^-$, $B_{s,d}\to\mu^+\mu^-$ and $B\to K(K^*)\nu\bar\nu$.

  We used as basic CKM parameters $V_{us}$, $|V_{cb}|$ and the angles $\beta$ and $\gamma$ in the unitarity triangle (UT) with the latter two  determined through the measurements of tree-level $B$ decays. This avoids the use of the problematic $|V_{ub}|$. A ratio involving  $\mathcal{B}(K^+\to\pi^+\nu\bar\nu)$ and
  $\overline{\mathcal{B}}(B_{s}\to\mu^+\mu^-)$, given in (41) of \cite{Buras:2021nns}, 
  while being $|V_{cb}|$-independent exhibits sizable dependence on   the angle $\gamma$.  It should be of interest for   several experimental groups  in the coming years. I recall this formula and the one involving $B_d\to\mu^+\mu^-$ in
   Appendix~\ref{UsefulEX} updating some parameters after four years.
  
  We pointed out that the $|V_{cb}|$-independent ratio of
  $\mathcal{B}(B^+\to K^+\nu\bar\nu)$ and
  $ \overline{\mathcal{B}}(B_{s}\to\mu^+\mu^-)$  from Belle II and LHCb signals a $1.8\sigma$ tension with   its SM value. This was the case of the fall of 2021.
  I report on the present status of this ratio below.

  Let me list a few ratios of interest. In particular one has not including
  NNLO corrections to $X(x_t)$
\be\label{F3}
  \boxed{\frac{\mathcal{B}(\kpn)}{|\varepsilon_K|^{0.82}}=(1.27\pm0.06)\times 10^{-8}{\left(\frac{\sin\gamma}{\sin 64.6^\circ}\right)^{0.015}\left(\frac{\sin 22.6^\circ}{\sin \beta}\right)^{0.71},  }            }
  \ee
  \be\label{R12a}
\boxed{\frac{\mathcal{B}(\klpn)}{|\varepsilon_K|^{1.18}}=(4.03\pm 0.21)\times 10^{-8}
    {\left(\frac{\sin\gamma}{\sin 64.6^\circ}\right)^{0.03}\left(\frac{\sin\beta}{\sin 22.6^\circ}\right)^{0.9{8}}.}}
\ee
The negligible dependence on $\gamma$ should be noticed so that the angle
$\beta$ plays more important role than $\gamma$ in this context. Since the fall of 2021, when our analysis was performed,  the measurement of $\beta$
has been improved from  $\beta=22.2(7)^\circ$ to $\beta=22.62(45)^\circ$ \cite{HeavyFlavorAveragingGroupHFLAV:2024ctg}. Therefore, some numerical predictions listed below 
differ slightly from the ones presented by us. We plan to update our two papers
later this year but I present preliminary modified results already here.

Using the experimental values of $|\varepsilon_K|$
and $\beta$ these ratios 
implied already in 2021 the most accurate predictions for the two branching ratios in question in  the SM  to date \cite{Buras:2022wpw,Buras:2021nns}. Moreover, they are independent of the value of $\vcb$. Meanwhile in addition to $\beta$ the uncertainties in the function $X(x_t)$ in $\kpn$ and $\klpn$ and on the parameter $\hat B_K$ in
$\varepsilon_K$ have been decreased but this had only a minor impact
on the accuracy of predictions for branching ratios when using our strategy after adding all uncertainties in quadrature.

Now, as seen above the increase of $\beta$ decreased slightly the r.h.s of
(\ref{F3}) but increased the r.h.s of (\ref{R12a}). On the other hand the recently calculated  NNLO QCD corrections to the function $X(x_t)$\footnote{I extracted these modifications from a talk by Emannuel  Stamou at the Kaon 2025 conference that was based on his paper with Martin Gorbahn and Yu that unfortunately did not appear in arXiv until today.} that enters quadratically
on the r.h.s increased the r.h.s of these equations by  $1.8\%$.
Consequently our 2021 central value of the $\kpn$ branching ratio practically did not
change but the one for the $\klpn$ branching ratio increased by $3.5\%$.

Assuming then no NP in  $|\varepsilon_K|$ and $S_{\psi K_S}$, which is the case of the SM anyway,  and using the experimental value of $\varepsilon_K$,
the updated values of the two branching ratios determined independently of $\vcb$ and $\gamma$ read in 2026
\be\label{BV26}
\boxed{\mathcal{B}(\kpn)_\text{SM}= (8.65\pm0.42)\times 10^{-11}, \quad 
\mathcal{B}(\klpn)_\text{SM}=(3.05\pm 0.17)\times 10^{-11}}
\ee
with the uncertainties in $\beta$, $\hat B_K$ and $P_C$ taken into account.
 These are the most precise determinations to date. 
 Other results are listed in Table~\ref{tab:SMBRs}.

 These should be compared with the results quoted by Emmanuel Stamou at Kaon 25 where instead of eliminating
 $\vcb$ a global UT fit of CKM parameters has been used:
 \be\label{GSV}
\mathcal{B}(\kpn)_\text{SM}= (8.55\pm0.47)\times 10^{-11}, \quad 
\mathcal{B}(\klpn)_\text{SM}=(3.15\pm 0.23)\times 10^{-11}\,.
\ee
One observes that there is a good agreement between these two ways of
obtaining predictions for the $K\to\pi\nu\bar\nu$ branching ratios.
The significantly larger error in the $\klpn$ branching ratio in (\ref{GSV})
is related to the $\vcb^4$ dependence present in $\mathcal{B}(\klpn)$ that is absent
in the BV strategy.

Needless to say the differences between the results in (\ref{BV26}) and
(\ref{GSV}) are from the point of view of the search for NP in these decays
presently immaterial as they are far beyond the accuracy of the experimental data for both decays that one expects in the present and in the coming decade.

Next one finds: 
\be\label{R1}
\boxed{\frac{\mathcal{B}(\kpn)}{\left[{\overline{\mathcal{B}}}(B_s\to\mu^+\mu^-)\right]^{1.4}}= 53.46\pm2.75\,,}
\ee

\be\label{R5}
\boxed{\frac{\mathcal{B}(\kpn)}{\left[\mathcal{B}(B^+\to K^+\nu\bar\nu)\right]^{1.4}}={(2.28\pm0.13)\times 10^{-3}\,.}}
\ee
\be\label{R7}
\boxed{\frac{\mathcal{B}(B^+\to K^+\nu\bar\nu)}{{\overline{\mathcal{B}}}(B_s\to\mu^+\mu^-)}={(1.32\pm0.07)\times 10^{3}\,,}}
\ee

There is no question about that some ideas presented here where
influenced by the 2015 paper \cite{Buras:2015qea} presented in Section~\ref{Per2015}.
One important progress was the inclusion of 
 other $K$ and $B$ decays. Another one
 was
 analytical removal of the $\vcb$ dependence by taking suitable ratios of observables. By suitable I mean for instance that in order to eliminate the $\vcb$
 dependences in the branching ratios for $\kpn$ and $\klpn$ with the help
 of $\varepsilon_K$, this  parameter 
 had to be raised, as given in (\ref{F3}) and (\ref{R12a}) to the power $0.82$ and $1.18$, respectively. In finding these powers, the numerical analysis of Elena played a very important role. As an example she found that using $\sin\gamma$
 instead of $\gamma$ used in \cite{Buras:2015qea} improves the accuracy in numerics.

 As a complementary test of the SM we proposed to extract $|V_{cb}|$ from different observables as a function of $\beta$ and $\gamma$ and illustrated
 already in Section~\ref{BPS} in Fig.~\ref{fig:XX} and in Fig.~\ref{fig:10052}
 below.

 While the decay $K_L\to\mu^+\mu^-$ played already an important role in flavour
 physics, in particular in bounding possible enhancement of $\kpn$ branching
 ratio in NP models with new left-handed currents, it cannot be considered
 as a {\em golden mode} as $\kpn$ and in particular $\klpn$ as far as theoretical
 uncertainties are concerned. However, during the last years the short
 distance contribution to $K_S\to\mu^+\mu^-$ has been promoted to the third
 {\em golden model} that as shown in a number of papers can be extracted from
 the future LHCb data \cite{DAmbrosio:2017klp,Dery:2021mct,Dery:2021vql,Dery:2022bhj,Dery:2022yqc,DAmbrosio:2025mxa}. In this manner this decay, when measured, will play a very
 important role in testing the SM and NP models.

 Therefore, I present still another $\vcb$-independent SM ratio from our paper
 \be\label{SR1}
\boxed{R_{\rm SL}=\frac{\mathcal{B}(\ksm)_{\rm SD}}{\mathcal{B}(\klpn)}=1.55\times 10^{-2}\,\left[\frac{\lambda}{0.225}\right]^2
\left[\frac{Y(x_t)}{X(x_t)}\right]^2\,}
\ee
which is independent of any SM parameter except for $m_t$ and $\lambda$ which are both precisely known.

\boldmath
\section{The Exclusive Vision of Rare $K$ and $B$ Decays and of the Quark Mixing in the   Standard Model (2022) }\label{BVVisions}
\unboldmath
The most common predictions for rare $K$ and $B$ decay branching ratios  in the SM in the literature are based on   the CKM elements $|V_{cb}|$ and  $|V_{ub}|$  resulting from global fits, that are in the ballpark of their  inclusive and exclusive determinations, respectively. In our second paper \cite{Buras:2022wpw} we followed another route,  which to our knowledge has not been explored for
  $\Delta M_{s,d}$ and rare $K$ and $B$ decays by anybody to date. We assumed, in contrast to the prevailing {\em inclusive} expectations
  for  $|V_{cb}|$,
  that the future true  values of $|V_{cb}|$ and  $|V_{ub}|$  will be both from
  {\em exclusive} determinations. To this end we used the available averages from FLAG.

With the precisely known $\vus$ the
  resulting rare decay branching ratios, $\varepsilon_K$, $\Delta M_d$, $\Delta M_s$  and $S_{\psi K_S}$ depend then only on the angles $\beta$ and $\gamma$
  in the UT that moreover are correlated through the CKM unitarity.   An unusual pattern of SM predictions results from this study with
  some  existing tensions being dwarfed and new tensions being born.   In particular using HPQCD $B^0_{s,d}-\bar B^0_{s,d}$ hadronic matrix elements a $3.1\sigma$ tension in $\Delta M_s$ independently of $\gamma$ is found.
    For $60^\circ\le\gamma\le 75^\circ$  the tension in $\Delta M_d$ between $4.0\sigma$ and $1.1\sigma$ is found and in the case of $\varepsilon_K$
between $5.2\sigma$ and $2.1\sigma$.
  Moreover, the room for NP in $K^+\rightarrow\pi^+\nu\bar\nu$,
  $K_L\rightarrow\pi^0\nu\bar\nu$ and $B\to K(K^*)\nu\bar\nu$ decays is significantly increased.

   While I think this was an interesting exercise, I do not think this is the
  situation realized in nature. Fortunately, in this paper we have made
  as a byproduct the following analysis which in my view is the most important
  part of this paper. Namely, we compared the determination of $\vcb$
  from $\Delta M_s$, $\Delta M_d$, $\varepsilon_K$ and $S_{\psi K_S}$ 
  using $B^0_{s,d}-\bar B^0_{s,d}$ hadronic matrix elements from LQCD with
  $2+1+1$ flavours, $2+1$ flavours and their average. Only for the
  $2+1+1$ case values for $\beta$ and $\gamma$ exist for which the
  same value of $\vcb$ is found from these observables\footnote{Relative to
\cite{Buras:2022wpw} I have updated the value of $\beta$ from $\beta=22.2(7)$
to a more recent one, $\beta=22.62(45)^\circ$, from HFLAV \cite{HeavyFlavorAveragingGroupHFLAV:2024ctg}, which through $\varepsilon_K$ slightly decreased the
value of $\vcb$ that was previously found to be $\vcb=42.6(4)\times 10^{-3}$. This had also slight impact on other CKM parameters.} 
  :
  \be\label{CKMBV}
\boxed{\vcb=42.5(5)\times 10^{-3}, \quad 
  \gamma=64.6(16)^\circ, \quad \beta=22.62(45)^\circ, \quad \vub=3.76(11)\times 10^{-3}\,}
\ee
and consequently
\be\label{CKMoutput2}
\boxed{\vts=41.8(4)\times 10^{-3}, \qquad \vtd=8.64(14)\times 10^{-3}\,,\qquad
{\IM}\lambda_t=1.45(5)\times 10^{-4}\,,}
\ee
\be\label{CKMoutput3}
\boxed{\bar\varrho=0.165(12),\qquad \bar\eta=0.348(11)\,,}
\ee
where $\lambda_t=V_{ts}^*V_{td}$.
  
\begin{figure}[t]
\centering%
\includegraphics[width=0.6\textwidth]{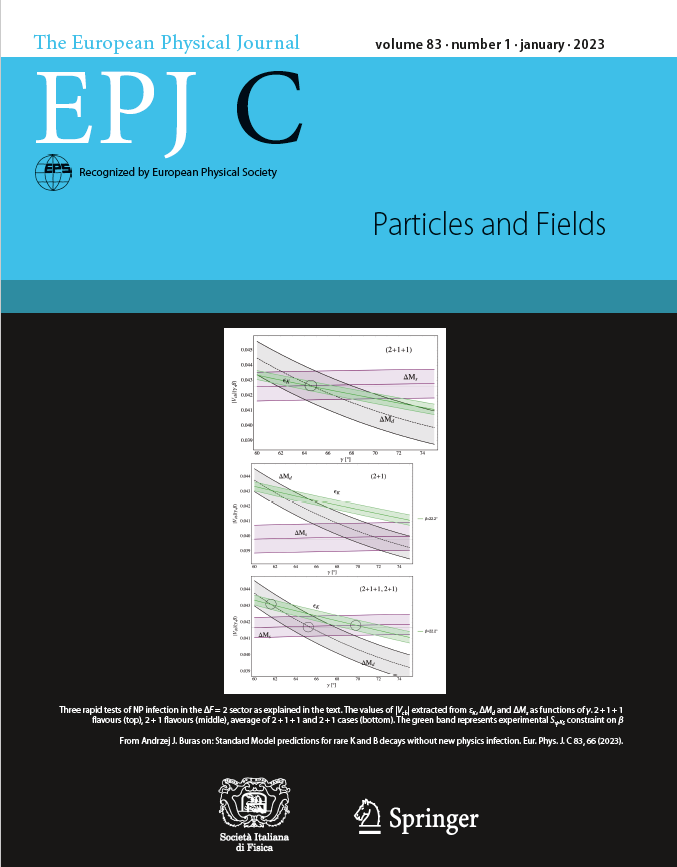}%
\caption{\it {The values of $\vcb$ extracted from $\varepsilon_K$, $\Delta M_d$ and  $\Delta M_s$ as functions of $\gamma$. $2+1+1$ flavours (top), $2+1$ flavours (middle), average of $2+1+1$ and $2+1$
    cases (bottom). The green band represents experimental $S_{\psi K_S}$ constraint on $\beta$.} From \cite{Buras:2022wpw}.\label{fig:10052}}
\end{figure}

 As shown in Fig.~\ref{fig:10052} this finding made it on the front page of the January 2023 Volume of EPJ C.

\boldmath
\section{Standard Model Predictions for Rare  K and B Decays  without New Physics Infection (2022)}
\unboldmath
Our finding that with the new hadronic matrix elements for $B^0_{s,d}-\bar B^0_{s,d}$ mixings from HPQCD no NP was required to fit the data
for $\Delta M_s$ and $\Delta M_d$ motivated me to write next a paper which
appeared in the fall of 2022 \cite{Buras:2022qip}. The reasoning in this paper was 
in my view  novel although not everybody liked it, in particular
those who do global fits. Here it comes.

The SM does not contain by definition any NP
  contributions to any observable but contains four CKM parameters which are not predicted   by this model. I pointed out that if these four parameters are determined in   a global fit which includes processes that are infected by NP and therefore by   sources outside the SM, the resulting so-called SM contributions to rare decay branching
  ratios cannot be considered as {{\em genuine} SM contributions to the latter.

    On the other hand as demonstrated using BV-strategy \cite{Buras:2021nns,Buras:2022wpw} {\em genuine}} SM predictions, that are {\em free} from the CKM dependence, can be obtained for suitable ratios of the $K$ and $B$ rare decay branching ratios  to $\Delta M_s$, $\Delta M_d$ and $|\varepsilon_K|$,  all calculated within the SM.   These three observables contain by now only small hadronic uncertainties and are already  well measured so that rather precise    SM predictions for the ratios in question can be obtained. I listed a few  in
  Section~\ref{BVfirst}.

    In this context 
  the {\em rapid test} of NP infection in the $\Delta F=2$ sector is provided
  by a $\vcb-\gamma$ plot, proposed in \cite{Buras:2022wpw}, that involves   
  $\Delta M_s$, $\Delta M_d$, $|\varepsilon_K|$, and the mixing induced   CP-asymmetry $S_{\psi K_S}$. As with the present hadronic matrix elements from HPQCD, as seen in Fig.~\ref{fig:10052}, this test turns out to be {\em negative},
  assuming negligible   NP infection in the $\Delta F=2$ sector and setting  the values of these four observables to
  the experimental  ones, allows to obtain SM predictions for all
  $K$ and $B$ rare decay branching ratios that are most accurate to date and
  as a byproduct, as listed above,  to obtain the full CKM matrix on the basis of $\Delta F=2$
  transitions {\em alone.}
  Using this strategy I obtained SM predictions for 26 branching
  ratios for rare semileptonic and leptonic $K$ and $B$ decays with the $\mu^+\mu^-$ pair   or the $\nu\bar\nu$ pair in the final state that are most precise to date. I hope many of them will disagree with future precise data
  indicating some pattern in deviations that would allow to find the
  viable NP scenarios.

A useful rapid test of $\Delta F=2$ is provided by the following, practically CKM free,
SM relation between the four $\Delta F=2$ observables \cite{Buras:2022qip} 
 \be\label{RNEW}
\boxed{\frac{ |\varepsilon_K|^{1.18}}{\Delta M_d\,\Delta M_s}=(8.37\pm 0.18)\times 10^{-5}\, \left(\frac{\sin \beta}{\sin 22.62^\circ}\right)^{1.027} K\,{\rm ps^2},}
\ee
where
\be
K=\left(\frac{\hat B_K}{0.7625}\right)^{1.18} 
\left[\frac{210.6\mev}{\sqrt{\hat B_{B_d}}F_{B_d}}\right]^2\,
\left[\frac{256.1\mev}{\sqrt{\hat B_{B_s}}F_{B_s}}\right]^2\,=\, 1.00\pm0.07\,.
\ee
The dependence on $\vcb$
drops out and the one on $\gamma$ being  negligible is included in the
uncertainty varying $\gamma$ in the range $60^\circ\le\gamma\le 70^\circ$.
Inserting the experimental values of the three $\Delta F=2$ observables on the l.h.s one finds for this ratio $(8.26\pm0.06)\times 10^{-5}$. Consequently,  with the presently known values of 
$\sqrt{\hat B_{B_d}}F_{B_d}$ and $\sqrt{\hat B_{B_s}}F_{B_s}$ from HPQCD, 
$\hat B_K$ from RBC-UKQCD 
and the present value of $\beta$ from $S_{\psi K_S}$, the SM is performing in the $\Delta F=2$ sector indeed very well. However with the $2+1$ flavours the central
value on the r.h.s of (\ref{RNEW}) decreases to $(6.29\pm 0.18)\times 10^{-5}$
so that the fact that this ratio  agrees with the data for present values of hadronic parameters with $2+1+1$ flavours and the experimental value of $\beta$ is remarkable.

The title having {\em Without New Physics Infection}  led to several invitations to medical conferences which I was tempted to accept but eventually rejected.
The BV-strategy has been reviewed
first in \cite{Venturini:2022sdf,Buras:2022irq} and in particular in my talk in Trani
at the workshop QCD@Work 2024 \cite{Buras:2024per}. But the talk of Elena at
the CKM2021 Workshop contains the best collection of the $\vcb$-independent
ratios \cite{Venturini:2022sdf}.
In this context also the review in \cite{Buras:2023qaf} should be useful.

\begin{table}
\centering
\renewcommand{\arraystretch}{1.4}
\resizebox{\columnwidth}{!}{
\begin{tabular}{|ll|l|}
\hline
Decay 
& SM Branching Ratio
& Data
\\
\hline \hline
 $B_s\to\mu^+\mu^-$ &  $(3.78^{+ 0.15}_{-0.10})\cdot 10^{-9}$      &  $(3.45\pm0.29)\cdot 10^{-9}$ \cite{HFLAV:2022pwe} 
\\
 $B_d\to\mu^+\mu^-$ &  ${(1.02^{+ 0.05}_{-0.03})}\ \cdot 10^{-10}$      & $\le 2.05\cdot 10^{-10}$ \cite{LHCb:2021awg}
\\
$B^+\to K^+\nu\bar\nu$ & $(4.99\pm 0.30)\cdot 10^{-6}$
&    $ (13\pm 4)\cdot 10^{-6}$ \cite{Belle-II:2023esi}
\\
$B^0\to K^{0*}\nu\bar\nu$ & ${(10.25\pm 0.92)}\cdot 10^{-6}$ &
 $\le 1.5\cdot 10^{-5}$ \cite{Grygier:2017tzo}
\\
\hline
$\kpn$ & $(8.65\pm 0.42)\cdot 10^{-11}$ &  $(13.0^{+3.3}_{- 3.0})\cdot 10^{-11}$ \cite{NA62:2024pjp} 
\\
 $\klpn$ & $(3.05\pm 0.17)\cdot 10^{-11}$ &   $\le 2.0\cdot 10^{-9}$ \cite{Ahn:2018mvc} 
\\
$(\ksm)_{\rm SD}$& {$(1.85\pm 0.12)\cdot 10^{-13}$} &   $\le 2.1\cdot 10^{-10}$
\cite{LHCb:2020ycd}
\\
\hline
\end{tabular}
}
\renewcommand{\arraystretch}{1.0}
\caption{\label{tab:SMBRs}
  \small
  Present most accurate  SM estimates  of the branching ratios  obtained using the BV-strategy. These are {\it the Magnificant Seven} decays, that being
  theoretically cleanest are optimal for the search for NP. Few numbers differ from the similar table in  \cite{Buras:2024per} due to the modification of $\beta$ and the NNLO QCD corrections to $X(x_t)$. 
    }
\end{table}

\part{Summary, Shopping List and Outlook}\label{outlook}\label{sec:7}
\section{Summary}
Despite the presence of various anomalies in the existing data, it is
clearly not evident which animalcula could be responsible for them. On the basis of the studies I have been involved in, possibly
the main candidates are $Z^\prime$ vector bosons, leptoquarks, vector-like
quarks and vector-like leptons but to find out without any doubts what they are  we need more data,
in particular for theoretically clean decays as the ones listed in Table~\ref{tab:SMBRs}. Precise measurements of branching ratios for these decays and of
various kinematical distributions should allow in this decade and next decade to find
out which animalcula are responsible for them. The strategies presented in
Parts~\ref{SBSM} and \ref{BV} of this Flavour Autobiography 
should be helpful in this respect. While, the ones in Part~\ref{SBSM} are more involved,
 the ones in Part~\ref{BV} are very simple. It should be stressed that  the ratios
listed there test the SM
independenty of whether NP effects are strongly suppressed in  $\Delta F=2$ processes or not. Having several measurements of these ratios and
seeing some pattern in deviations from their SM predictions would
already be very exciting. One could then return to the strategies presented
in Part~\ref{SBSM}  and to explicit models studied in my group and discussed in
Part~\ref{NPModels}.

\section{Shopping List}
Let me next present my shopping list for the coming years.
I list here only the entries related to indirect searches for NP, in particular
flavour physics. There is no question about that discovering a new particle
at the LHC would be most exciting as this would reduce the number of possible
extensions of the SM by much. Here comes my shopping list.
\begin{enumerate}
\item
  Precise measurements of the branching ratios for the 7 magnificant decays listed in Table~\ref{tab:SMBRs}  and as presented in Section~\ref{Harz}
  of missing energy distributions for four decays among them.  
Presently on a forefront are  the $B\to K^{(*)}\nu\bar\nu$ decays studied intensively by the Belle~II experiment~\cite{Belle-II:2023esi}, giving some
hints for NP contributions. There are very many recent analyses  of these data
but the review of them is beyond the scope of this story. A collection
of references to these analyses can be found in \cite{Aebischer:2025qhh} but surely some have been performed since then.
The same applies to $\kpn$ for which a very interesting result has 
been provided by the NA62 experiment listed in Table~\ref{tab:SMBRs}.
A new one should be available this year.

Taking the present experimental result for $B_s\to \mu^+\mu^-$ allows us
to determine the present $\vcb$-independent experimental values for the
the ratios in (\ref{R1}) and (\ref{R7})
that read 
\be
\boxed{\frac{\mathcal{B}(\kpn)_{\rm{exp}}}{\left[{\overline{\mathcal{B}}}(B_s\to\mu^+\mu^-)\right]^{1.4}_{\rm{exp}}}=91.4\pm22.8,\qquad    \text{SM}: 53.46\pm2.75}
\ee
\be
\boxed{\frac{\mathcal{B}(B^+\to K^+\nu\bar\nu)_{\rm{exp}}}{{\overline{\mathcal{B}}}(B_s\to\mu^+\mu^-)_{\rm{exp}}}=(3.8\pm1.2)\cdot 10^{3}, \qquad \text{SM}:(1.32\pm0.07)\cdot 10^{3}}
\ee
They differ significantly from very precise SM values. Let us hope that the experimental errors on these ratios will decrease in the coming years and
the data will depart even more from the SM predictions. I wish Belle II, NA62, LHCb, CMS and ATLAS experimentalists luck in measuring precisely the
involved branching ratios.
\item
  Precise measurements of the branching ratios and of other observables in $B\to K\ell^+\ell^-$ and $B\to K^*\ell^+\ell^-$ decays and the clarification of the
  anomalies in them.
\item
  Precise measurements of $\vcb$ and $\gamma$ in tree-level decays that
  would provide additional tests of the BV-strategy.
 \item
   Clarification of anomalies in $B\to\pi K$ decays \cite{Buras:2003dj,Buras:2004ub,Fleischer:2018bld,Berthiaume:2023kmp,Datta:2024zrl,Szabelski:2024cem} and
   continuation of the intensive studies of several other
   $B\to PP$ decays with $B=B_{d,s}^0,B^+$ and $P=\pi,K,K^*$ as described briefly
   in Section~\ref{NonL}.
 \item
   Completion of the calculations of $\epe$ and of the $\Delta I=1/2$ rule
   by the RBC-UKQCD lattice QCD collaboration and of corresponding calculations by another
   Lattice QCD group with $2+1+1$ flavours. In particular the isospin breaking contributions to QCD panguins are still missing in present lattice results for $\epe$.  A stressed in \cite{Buras:2023qaf},
   Jean-Marc G{\'e}rard and myself expect on the basis of DQCD developed
   with Bill Bardeen significant NP contributions to $\epe$ that as seen
   in the left plot in  Fig.~\ref{Fig:3} are consistent with the
   recent result on $\kpn$  of the NA62 collaboration. Despite my critical comments about their present results I do hope very much that they improve them
   and at some stage I have written a letter supporting their projects.
 \item
   Calculation of hadronic matrix elements relevant for $\Delta M_s$ and
   $\Delta M_d$ with  $2+1+1$ flavours by other lattice QCD collaborations in order to check HPQCD results
   for these matrix elements.
 \item
   Further searches for lepton flavour violation and EDMs.
\end{enumerate}

\begin{figure}[t]
\centering%
\includegraphics[width=0.8\textwidth]{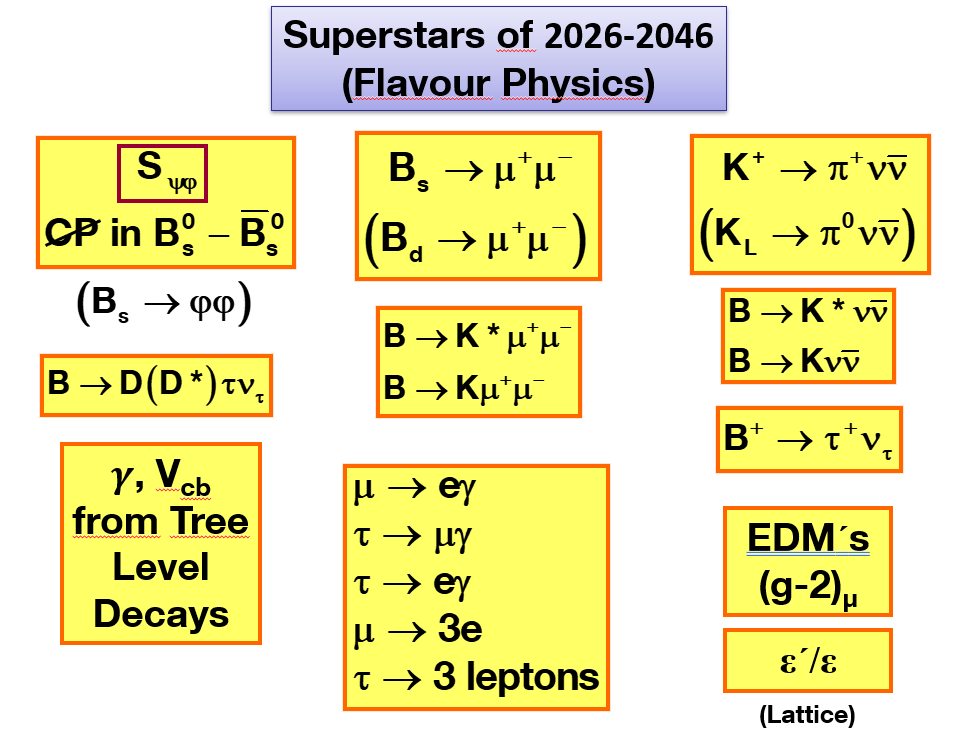}%
\caption{\it Coming Years.\label{Great1}}
\end{figure}
\begin{figure}[t]
\centering%
\includegraphics[width=0.8\textwidth]{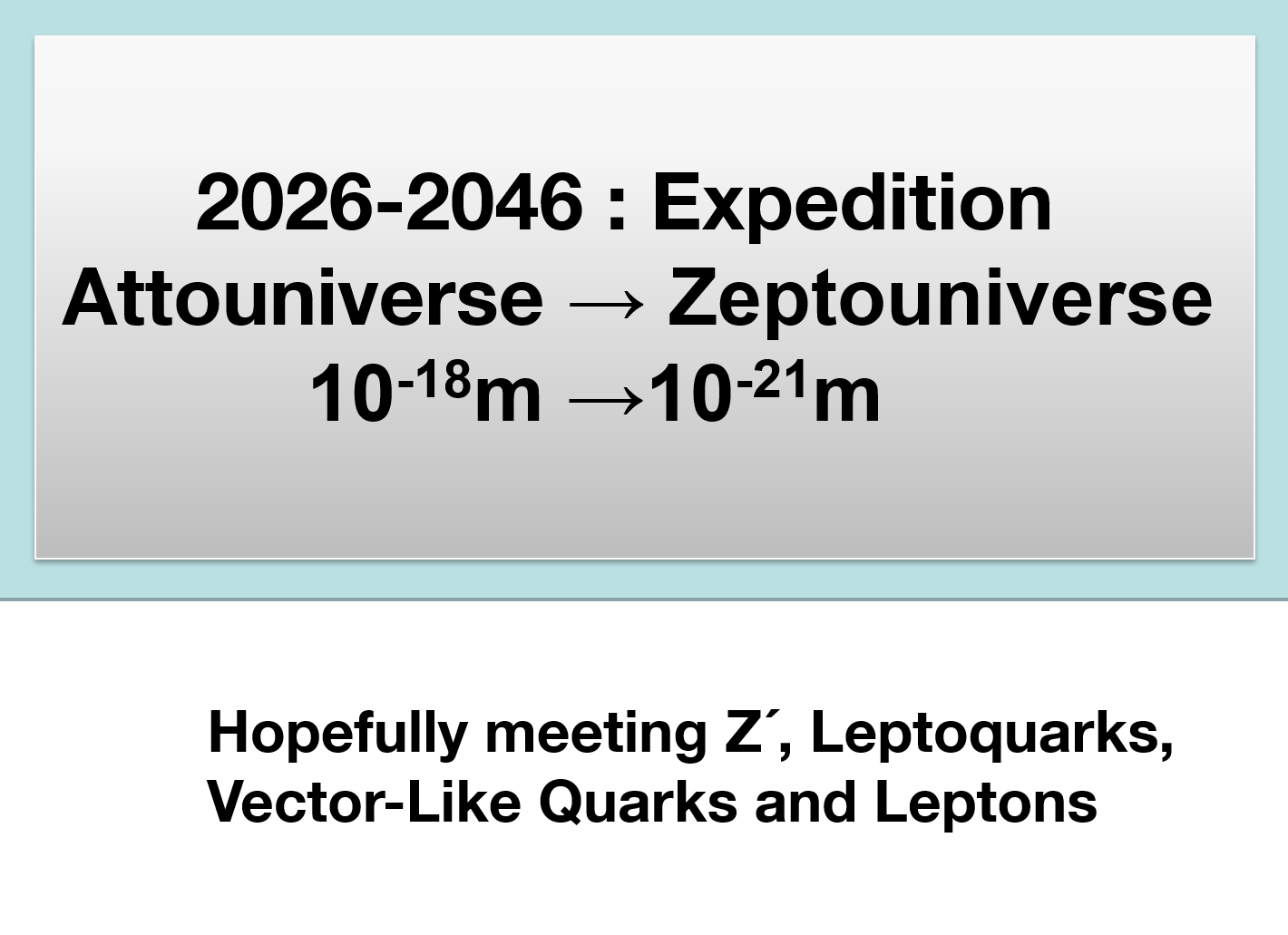}%
\caption{\it 2026-2046 Expedition.\label{Great2}}
\end{figure}

\section{Outlook}
Assuming that a least few of these entries will be realized in the coming 20 years, in particular the one on $\epe$, we should have then great time until 2046
 and this is expressed
in Figs.~\ref{Great1} and \ref{Great2}.

Unfortunately, it could take longer time. After all Enrico Fermi, already in the
1930s could sense something going on which was hidden in his $G_F$. But it
took first 30 years to suggest that it could be a heavy gauge boson $W^\pm$
behind these effects and another 20 years to discover it.
Let us hope that it will not take such a long time now and that we
will be able to narrow the number of possible extensions of the SM 
at least through indirect searches until 2046. The methods presented
in this Flavour Autobiography will hopefully play an important role in these
searches.

Clearly, I try to be optimistic here although the recently presented Physics Briefing Book as the input for the 2026 update of the European Strategy for
Particle Physics \cite{deBlas:2025gyz} does not allow for too much optimism for flavour physicists
of my generation, simply because of times involved and also because flavour
physics is not the top priority of the physics community for the coming decades.
Yet, as summarized in Chapter 5 of \cite{deBlas:2025gyz} and also in
{\em Projections for Key Measurements in Heavy Flavour Physics arXiv:2503.24346}
flavour physics will still be exciting and also other directions like
Higgs physics, EDMs among others. But from present perspective it is difficult
to predict how our field will develop in the coming decades. Indeed, let me
cite Niels Bohr: ``Prediction is very difficult, especially if it's about the future''.

\vspace{+0.5cm}

\section{Acknowlegments}

This flavour  story is dedicated to my collaborators with whom I had great time
in all our joint publication listed above.
 I hope this will continue  with some of them in the coming years. Many thanks
to all of them. Many thanks go also to various Institutions that supported
me during several decades, like DFG, BMBF, Humboldt Society, ERC Council TUM-IAS  and
the Excellence Clusters that started in 2005 and the present one that will
last until 2032. Many thanks to my colleagues in the Physics Department,
in particular to Martin Beneke,  Nora Brambilla, Laura Fabbietti, Stephan Paul, Elisa Resconi, Stefan Schönert,  Andi Weiler, Wolfram Weise and Giulia Zanderighi with whom I had most contacts after my retirement in April 2012.
Many thanks to the IAS-Team for creating such fantastic atmosphere at this
Institute. This will be clear from my last parts of this write-up.

I would also like to thank important personalities
at the TUM. In particular two presidents Wolfgang Herrmann and Thomas Hofmann,
the vicepresident Gerhard Kramer and the chancellor Albert Berger for great support over many years. In connection with the TUM Senior Excellence Faculty particular thanks go to Birgit Herbst-Gaebel and of course to MichaeL Molls and
J{\"o}rg Ebersp{\"a}cher, the spokesmen of this faculty.

Next, I want to thank my secretaries.  This is Marianne Glandien, my first secretary 1988-1998 with whom I still have some contact. However, particular thanks go to Elke Hutsteiner for a great 27 year collaboration (1998-2025)
 in constructing my power-point presentations in  all talks I have given in this period. I was the composer of
 these presentations and Elke was constructing them using power-point.
 Several figures presented in this story appeared in my talks and this is
 evident from their appearance. After Elke's retirement I was lucky
 that she was replaced by Alexandra Castrillon Perez who was very helpful
 with extracting these figures from my talks and modifying them if necessary.
 Many thanks to Alexandra as well.

{\bf But the main thanks go to my family for such a great support over many decades.}

However, this is not the end of my story as of 2026. I collected in the appendices
a number of memories that I hope are both interesting and amusing. Most are
related to physics, but not all.

\part{ Appendices}

\vspace{+0.5cm}
\appendix 
\vspace{+0.25cm}

\section{Basic Framework}\label{BFramework}

\begin{figure}[t]
\centering%
\includegraphics[width=0.8\textwidth]{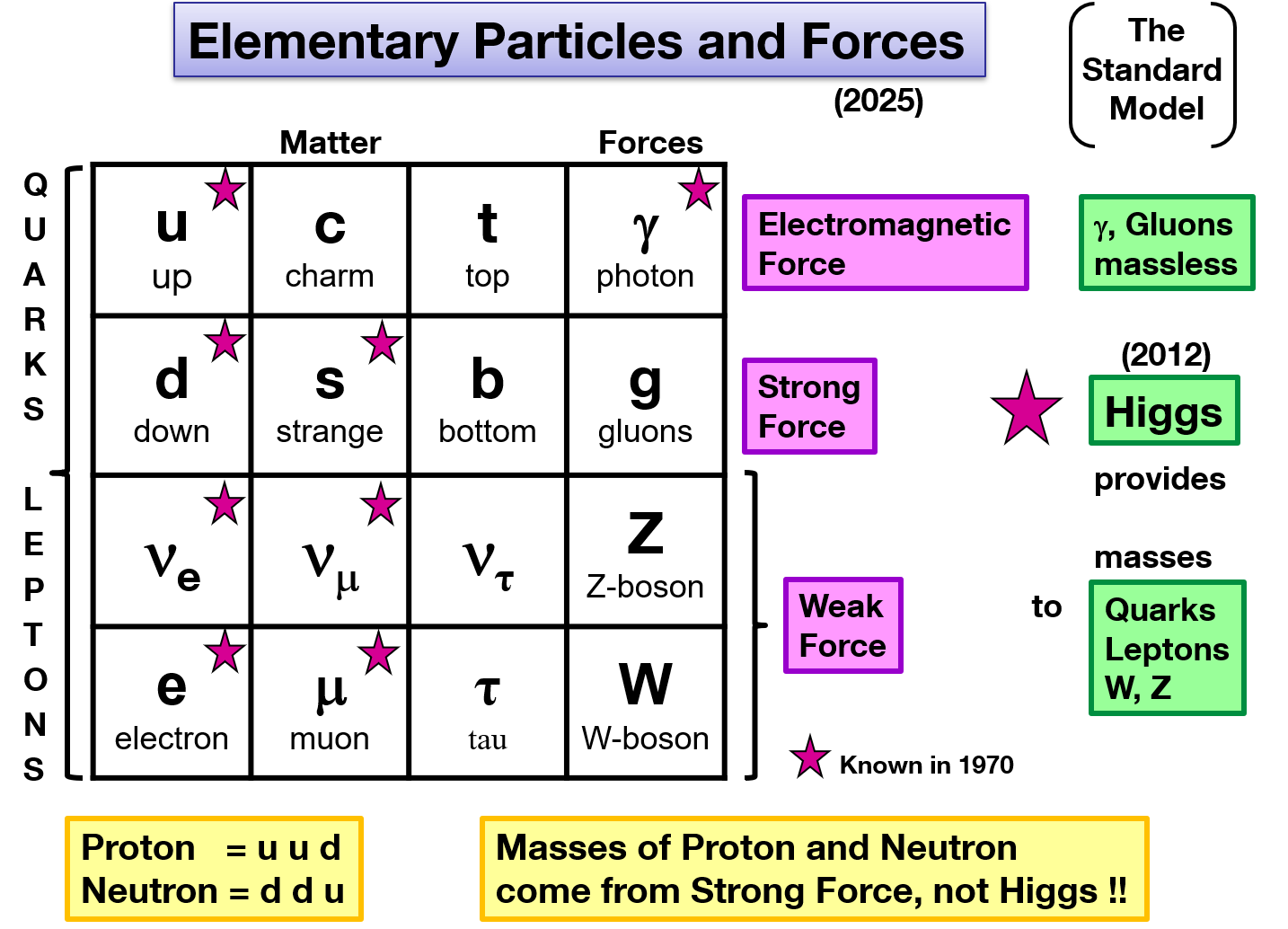}%
\caption{\it The Particles of the Standard Model.\label{SM}}
\end{figure}

 Let us recall the particles of the Standard Model that all have
 been discovered. The small red stars in Fig.~\ref{SM} indicate which particles were known
 in 1970 when I was doing my master thesis. The big red star is reserved for the
 Higgs.

\begin{figure}[t]
\centering%
\includegraphics[width=0.8\textwidth]{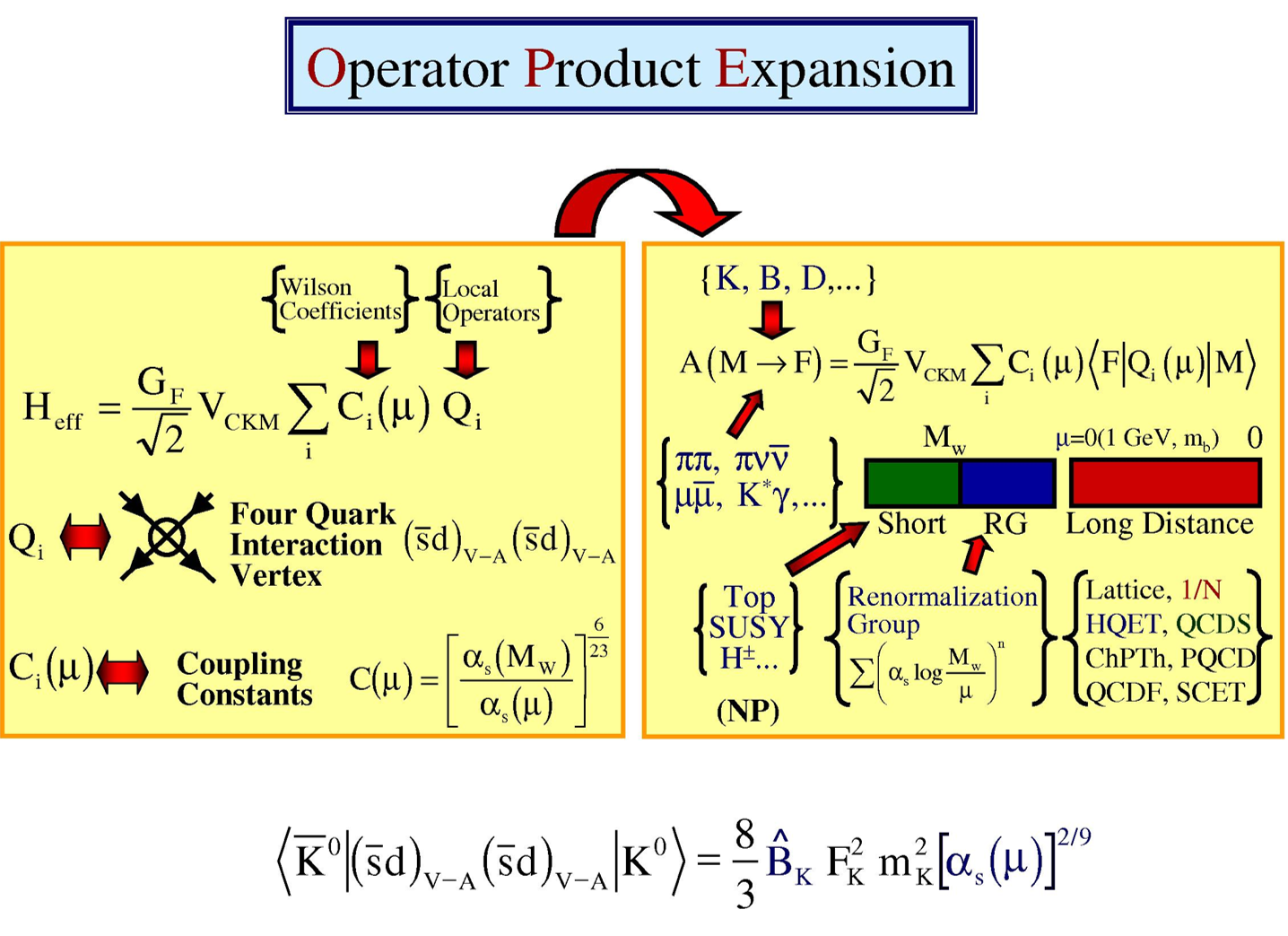}%
\caption{\it Operator Product Expansion. \label{OPE}}
\end{figure}

\begin{figure}[t]
\centering%
\includegraphics[width=0.8\textwidth]{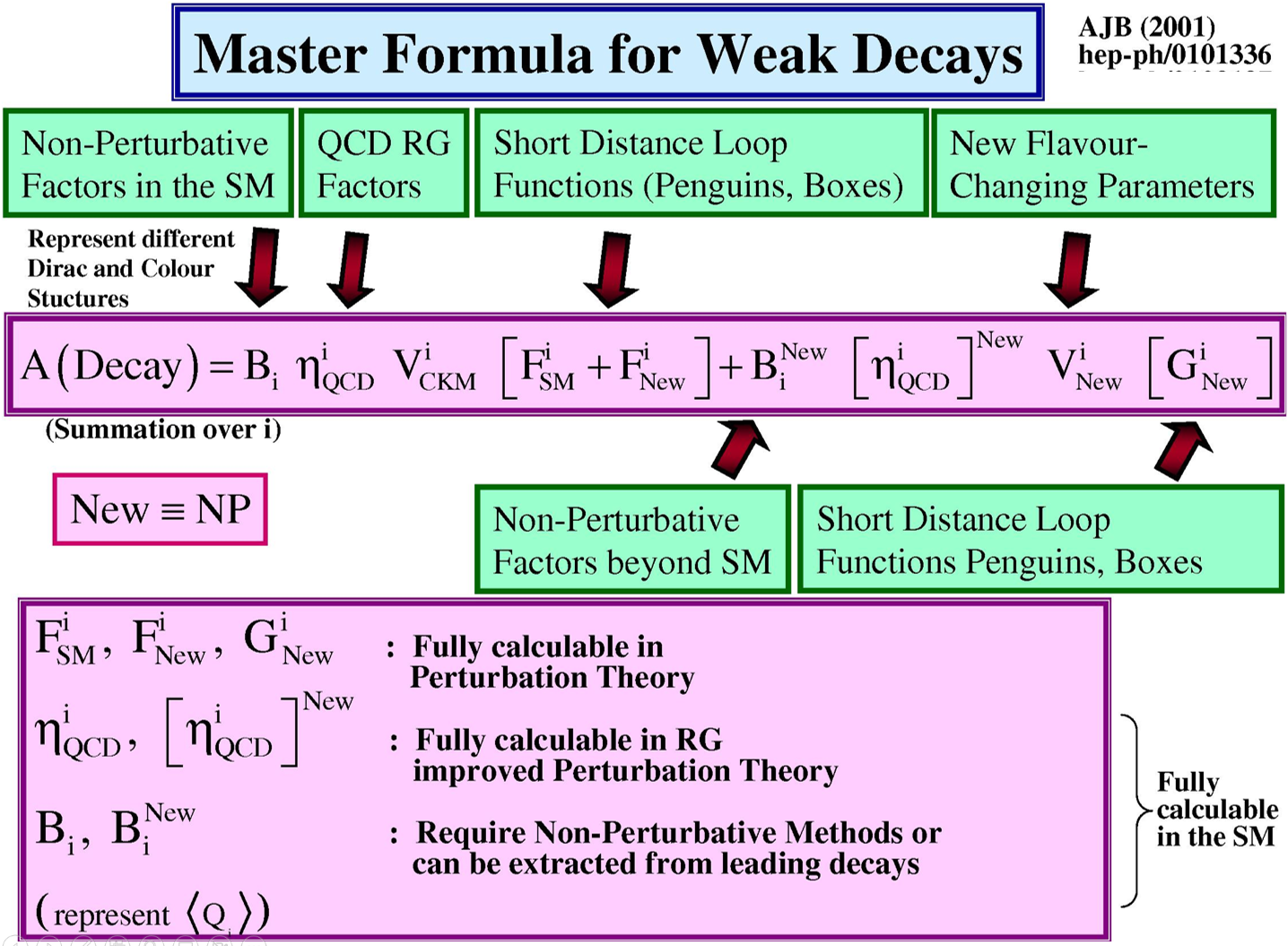}%
\caption{\it Master Formula for Weak Decay Amplitudes.\label{MASTERAP}}
\end{figure}

 In Fig.~\ref{OPE} I show the basis of effective theories which is the operator product expansion, 
 the sum of operators multiplied by the Wilson Coefficients
 \cite{Wilson:1969zs,Zimmermann:1972tv,Wilson:1972ee}.   In Fig.~\ref{MASTERAP} the master formula for weak decay amplitudes is shown and
 different contributions are briefly explained. Detailed  explanations
 can be found in my book \cite{Buras:2020xsm} and in other reviews already cited, in
 particular in my Les Houches lectures \cite{Buras:1998raa}.

\section{Operators of Effective Hamiltonians: SM}\label{HeffSM}
In this and the next appendix I have collected the operators that were mentioned in the text and a few more to stress that many operators could enter the game, in
particular in the SMEFT when RG effect are taken into account.
Full sets can be found in \cite{Buchalla:1995vs,Buras:1998raa,Buras:2020xsm} and
in \cite{Aebischer:2025qhh}.

\subsection{Nonleptonic Operators}
Of particular interest are the 
operators involving quarks only. In the case of the $\Delta B=1$
transitions the relevant set of operators is given as follows:

{\bf Current--Current:}
\begin{equation}\label{O1} 
Q_1 = (\bar c_{\alpha} b_{\beta})_{V-A}\;(\bar s_{\beta} c_{\alpha})_{V-A}
~~~~~~Q_2 = (\bar c b)_{V-A}\;(\bar s c)_{V-A} 
\end{equation}

{\bf QCD--Penguins:}
\begin{equation}\label{O2}
Q_3 = (\bar s b)_{V-A}\sum_{q=u,d,s,c,b}(\bar qq)_{V-A}~~~~~~   
 Q_4 = (\bar s_{\alpha} b_{\beta})_{V-A}\sum_{q=u,d,s,c,b}(\bar q_{\beta} 
       q_{\alpha})_{V-A} 
\end{equation}
\begin{equation}\label{O3}
 Q_5 = (\bar s b)_{V-A} \sum_{q=u,d,s,c,b}(\bar qq)_{V+A}~~~~~  
 Q_6 = (\bar s_{\alpha} b_{\beta})_{V-A}\sum_{q=u,d,s,c,b}
       (\bar q_{\beta} q_{\alpha})_{V+A} 
\end{equation}

{\bf Electroweak Penguins:}
\begin{equation} 
Q_7 = \frac{3}{2}\;(\bar s b)_{V-A}\sum_{q=u,d,s,c,b}e_q\;(\bar qq)_{V+A} 
\nnb
\ee
\be\label{O4}
Q_8 =  \frac{3}{2}\;(\bar s_{\alpha} b_{\beta})_{V-A}\sum_{q=u,d,s,c,b}e_q
        (\bar q_{\beta} q_{\alpha})_{V+A}
\end{equation}
\begin{equation}
 Q_9 =  \frac{3}{2}\;(\bar s b)_{V-A}\sum_{q=u,d,s,c,b}e_q(\bar q q)_{V-A}
\nnb
\ee
\be\label{O5}
Q_{10} = \frac{3}{2}\;
(\bar s_{\alpha} b_{\beta})_{V-A}\sum_{q=u,d,s,c,b}e_q\;
       (\bar q_{\beta}q_{\alpha})_{V-A} 
\end{equation}
Here, $\alpha,\beta$ denote colours and $e_q$ denotes the electric quark charges reflecting the
electroweak origin of $Q_7,\ldots,Q_{10}$. Finally,
$(\bar c b)_{V-A}\equiv \bar c_\alpha\gamma_\mu(1-\gamma_5) b_\alpha$.

These operators play a crucial role in non-leptonic decays of $B_s$ and $B_d$ 
mesons and have through mixing under renormalization also an impact on other 
processes as is evident from 
\cite{Buras:1998raa,Buchalla:1995vs}.
For non-leptonic $K$ decays the quark flavours have 
to be changed appropriately. 
Explicit expressions can be found in \cite{Buras:1998raa,Buchalla:1995vs}.
In particular the analogues of 
$Q_1$ and $Q_2$ govern the $\Delta I=1/2$ rule in $K_L\to\pi\pi$ decays, while 
the corresponding QCD penguins and electroweak penguins enter directly 
the ratio $\epe$.

\subsection{Magnetic Penguins}
In the case of $B\to X_s\gamma$  and  $B\to X_sl^+l^-$ decays 
and corresponding exclusive decays the crucial role is played by {\it 
magnetic} penguin operators:
\begin{equation}\label{O6}
Q_{7\gamma}  =  \frac{e}{8\pi^2} m_b \bar{s}_\alpha \sigma^{\mu\nu}
          (1+\gamma_5) b_\alpha F_{\mu\nu}\qquad            
Q_{8G}     =  \frac{g}{8\pi^2} m_b \bar{s}_\alpha \sigma^{\mu\nu}
   (1+\gamma_5)T^a_{\alpha\beta} b_\beta G^a_{\mu\nu}\,.
\end{equation}
It should be stressed that also the QCD mixing  of $Q_2$ operator into
$Q_{7\gamma}$ plays an important role.
The operator $Q_{8G}$ can 
also be relevant in nonleptonic decays. The magnetic operators are often 
called {\it dipole} operators. 

\boldmath
\subsection{$\Delta S=2$ and $\Delta B=2$ Operators}
\unboldmath
In the case of  $K^0-\bar K^0$ mixing and  $B_d^0-\bar B^0_d$ mixing
the relevant operators within the SM are
\begin{equation}\label{O7}
Q(\Delta S = 2)  = (\bar s d)_{V-A} (\bar s d)_{V-A}~~~~~
 Q(\Delta B = 2)  = (\bar b d)_{V-A} (\bar b d)_{V-A}~. 
\end{equation}
For $B_s^0-\bar B^0_s$ mixing one has to replace $d$ by $s$ in 
the last operator.

\boldmath
\subsection{Semileptonic Operators}
\unboldmath
In the case of $B\to X_sl^+l^-$, $B\to K l^+l^-$ and $B\to K^* l^+l^-$  also the following operators on top of 
magnetic penguins contribute
\begin{equation}\label{9V}
Q_{9V}  = (\bar s b  )_{V-A} (\bar \mu\mu)_{V}~~~~~
Q_{10A}  = (\bar s b )_{V-A} (\bar \mu\mu)_{A}.
\end{equation}
Changing approprately flavours one obtains the corresponding 
operators relevant for 
$B\to X_dl^+l^-$ and $K_L\to\pi^0l^+l^-$.

The rare decays $B\to X_s\nu\bar\nu$, $B\to K\nu\bar\nu$, 
$B\to K^*\nu\bar\nu$ and $B_s\to\bar\mu\mu$ are governed by
\begin{equation}\label{10V}
Q_{\nu\bar\nu}(B)  = (\bar s b  )_{V-A} (\bar \nu\nu)_{V-A}~~~~~
Q_{\mu\bar\mu}(B)  = (\bar s b )_{V-A} (\bar \mu\mu)_{V-A}~.
\end{equation}

The rare decays $K\to\pi\nu\bar\nu$ and $K_L\to\bar\mu\mu$ are governed on the 
other hand by
\begin{equation}\label{11V}
Q_{\nu\bar\nu}(K)  = (\bar s d  )_{V-A} (\bar \nu\nu)_{V-A}~~~~~
Q_{\mu\bar\mu}(K)  = (\bar s d )_{V-A} (\bar \mu\mu)_{V-A}~.
\end{equation}

\begin{figure}[t]
\centering%
\includegraphics[width=0.8\textwidth]{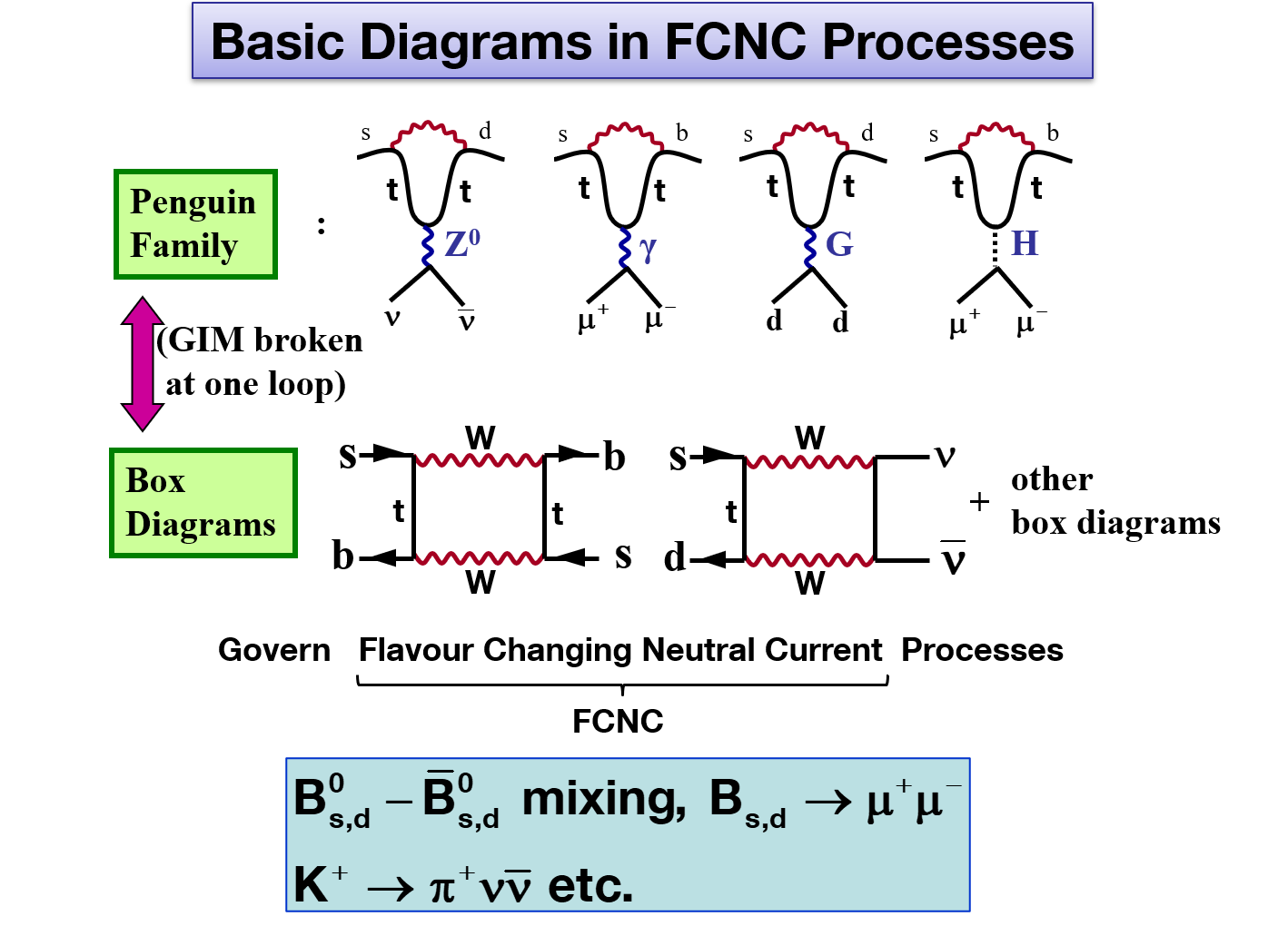}%
\caption{\it Basic Diagrams in FCNC Processes in the Standard Model.\label{Penguin}}
\end{figure}

As I did not show  any Feynman diagrams in my presentation let me just
indicate in Fig.~\ref{Penguin}  what are penguin and box diagrams.

\section{Selected Operators of Effective Hamiltonians: BSM}\label{HeffBSM}

NP can generate new operators. Typically new operators 
are generated through the presence of RH currents and 
{\it scalar} currents with the latter strongly suppressed within the SM.
New gauge bosons and scalar exchanges are at the origin of these operators 
that can have important impact on phenomenology. The two-loop anomalous 
dimensions of these operators have been calculated in 
\cite{Ciuchini:1997bw,Buras:2000if}

\boldmath
\subsection{$\Delta F=2$ Non-leptonic Operators}
\unboldmath

For definiteness, we shall consider here operators responsible for the
$K^0$--$\bar{K}^0$ mixing and consequently relevant also for $\varepsilon_K$. 
There are 8 such operators of dimension 6.
They can be split into 5 separate sectors, according to the chirality
of the quark fields they contain. The operators belonging to the first
three sectors (VLL, LR and SLL) read \cite{Buras:2000if}

\bea 
Q_1^{\rm VLL} &=& (\bar{s}^{\alpha} \gamma_{\mu}    P_L d^{\alpha})
              (\bar{s}^{ \beta} \gamma^{\mu}    P_L d^{ \beta}),
\nnb\\[4mm] 
Q_1^{\rm LR} &=&  (\bar{s}^{\alpha} \gamma_{\mu}    P_L d^{\alpha})
              (\bar{s}^{ \beta} \gamma^{\mu}    P_R d^{ \beta}),
\nnb\\
Q_2^{\rm LR} &=&  (\bar{s}^{\alpha}                 P_L d^{\alpha})
              (\bar{s}^{ \beta}                 P_R d^{ \beta}),
\nnb\\[4mm]
Q_1^{\rm SLL} &=& (\bar{s}^{\alpha}                 P_L d^{\alpha})
              (\bar{s}^{ \beta}                 P_L d^{ \beta}),
\nnb\\
Q_2^{\rm SLL} &=& (\bar{s}^{\alpha} \sigma_{\mu\nu} P_L d^{\alpha})
              (\bar{s}^{ \beta} \sigma^{\mu\nu} P_L d^{ \beta}),
\label{normal}
\eea
where $\sigma_{\mu\nu} = \frac{1}{2} [\gamma_{\mu}, \gamma_{\nu}]$ and
$P_{L,R} =\frac{1}{2} (1\mp \gamma_5)$. The operators belonging to the
two remaining sectors (VRR and SRR) are obtained from $Q_1^{\rm VLL}$ and
$Q_i^{\rm SLL}$ by interchanging $P_L$ and $P_R$. For $\Delta B=2$ the 
flavours have to be changed appropriately.

\boldmath
\subsection{$\Delta F=1$ Operators}
\unboldmath
The list of $\Delta F=1$ operators in the extensions of the SM is much 
longer and will not be given here. All the dimension six four-quark operators 
are discussed in \cite{Buras:2000if} where also their two-loop anomalous 
dimensions have been calculated. See also \cite{Ciuchini:1997bw}, where a 
different operator basis is used.

Concerning the semileptonic operators in the extensions of the SM the 
typical examples  of operators related to the presence of RH {\it currents} are

\begin{equation}\label{9VR}
\tilde Q_{9V}  = (\bar s b  )_{V+A} (\bar \mu\mu)_{V}~~~~~
\tilde Q_{10A}  = (\bar s b )_{V+A} (\bar \mu\mu)_{A}.
\end{equation}
\begin{equation}\label{10VR}
\tilde Q_{\nu\bar\nu}(B)  = (\bar s b  )_{V+A} (\bar \nu\nu)_{V-A}~~~~~
\tilde Q_{\mu\bar\mu}(B)  = (\bar s b )_{V+A} (\bar \mu\mu)_{V-A}~.
\end{equation}
\begin{equation}\label{11VR}
Q_{\nu\bar\nu}(K)  = (\bar s d  )_{V+A} (\bar \nu\nu)_{V-A}~~~~~
Q_{\mu\bar\mu}(K)  = (\bar s d )_{V+A} (\bar \mu\mu)_{V-A}~.
\end{equation}

If {\it scalar currents} resulting from scalar exchanges like the heavy 
Higgs in the 2HDM models or sparticles in the MSSM are present, scalar operators enter 
the game. The most prominent are the ones that govern the 
$B_s\to \mu^+\mu^-$ decay in 2HDMs and the MSSM at large $\tan\beta$:
\begin{equation}\label{scalarL}
Q_S  = (\bar s P_L b  ) (\bar \mu\mu)~~~~~
Q_P  = (\bar s P_L b ) (\bar \mu\gamma_5 \mu)
\end{equation}

\begin{equation}\label{scalarR}
\tilde Q_S  = (\bar s P_R b  ) (\bar \mu\mu)~~~~~
\tilde Q_P  = (\bar s P_R b ) (\bar \mu\gamma_5 \mu)
\end{equation}

These are clearly only selected sets of operators. The full sets of operators
in the SMEFT and WET can be found in our review \cite{Aebischer:2025qhh} and
in the original papers cited there.

\section{Useful Expressions}\label{UsefulEX}

In this appendix I list a number of useful expressions for the most interesting
branching ratios within the SM that being theoretically clean should play an important role in the indirect search for NP in the rest of this decade and in the
following decade.

\boldmath
\subsection{$\kpn$ and $\klpn$}
\unboldmath
\begin{align}
   { \mathcal{B}(\kpn) = {(8.65 \pm 0.24)} \times 10^{-11} \,
    \bigg[\frac{\left|V_{cb}\right|}{42.5\times 10^{-3}}\bigg]^{2.8}
    \bigg[\frac{\sin\gamma}{\sin 64.6^\circ}\bigg]^{1.39}},\label{kplusApprox}
\end{align}
\begin{align}
    \mathcal{B}(\klpn) ={(3.05 \pm 0.05)} \times 10^{-11} \,
    &{\bigg[\frac{\left|V_{cb}\right|}{42.5\times 10^{-3}}\bigg]^4
\bigg[\frac{\sin\gamma}{\sin(64.6^\circ)}\bigg]^{2}
    \bigg[\frac{\sin\beta}{\sin({22.6^\circ})}\bigg]^2},\label{k0Approx}
\end{align}
where the preliminary NNLO QCD corrections to $X(x_t)$ presented by Emmanuel
Stamou at Kaon25 have been included.

Inserting the values in (\ref{CKMBV}) one finds
\be\label{AJBBV}
\mathcal{B}(\kpn)_\text{SM}= (8.65\pm0.41)\times 10^{-11}, \quad 
\mathcal{B}(\klpn)_\text{SM}=(3.05\pm 0.21)\times 10^{-11}
\ee
that are consistent with (\ref{BV26}). The increased error in
$\klpn$ branching ratio is related to the fact that when using $\varepsilon_K$
also the $\beta$ and $\gamma$ dependences are weakened in addition to the
elimination of $\vcb$.

\boldmath
\subsection{$B_{s,d}\to\mu^+\mu^-$}
\unboldmath
Using the exact formulae {in \cite{Buras:2020xsm}} that are based
on the calculations over three decades by several groups
\cite{Buchalla:1993bv,Buchalla:1998ba, Buras:2012ru, Bobeth:2013uxa, Bobeth:2013tba, Hermann:2013kca,Beneke:2017vpq, Beneke:2019slt}
the dependence of the branching ratio for $B_s\to\mu^+\mu^-$
on the input parameters involved can be transparently summarized as follows 
 \cite{Bobeth:2013uxa} 
\begin{equation}
  \overline{\mathcal{B}}(B_{s}\to\mu^+\mu^-)_{\rm SM} = ({3.78}\pm0.06)\times 10^{-9}
\left(\frac{F_{B_s}}{{230.3}\mev}\right)^2 \left|\frac{V_{tb}^*V_{ts}}{{0.0418}}\right|^2\bar R_s
\label{BRtheoRpar}
\ee
\noindent
where
\be
\label{Rs}
\bar R_s=
\left(\frac{\tau_{B_s}}{{1.515} {\rm ps}}\right)\left(\frac{{0.935}}{r(y_s)}\right)
 \left(\frac{m_t(m_t)}{162.83 \gev}\right)^{3.02}\left(\frac{\alpha_s(M_Z)}{0.1184}\right)^{0.032} \,.
 \ee
 Here $r(y_s)$ summarizes $\Delta\Gamma_s$ effects with $r(y_s)={0.935}\pm0.007$ within the SM  \cite{DescotesGenon:2011pb,deBruyn:2012wj,deBruyn:2012wk}.

Similarly one finds \cite{Bobeth:2013uxa}
\begin{equation}
  {\mathcal{B}}(B_{d}\to\mu^+\mu^-)_{\rm SM} = ({1.02}\pm0.02)\times 10^{-10}
\left(\frac{F_{B_d}}{{190.0}\mev}\right)^2 \left|\frac{V_{tb}^*V_{td}}{0.00864}\right|^2\bar R_d \,.
\label{BRtheoRpard}
\ee
\noindent
As  to an excellent {accuracy}  $r(y_d)=1$, one has this time
\be
\label{Rd}
\bar R_d=
\left(\frac{\tau_{B_d}}{1.519 {\rm ps}}\right)
 \left(\frac{m_t(m_t)}{162.83 \gev}\right)^{3.02}\left(\frac{\alpha_s(M_Z)}{0.1184}\right)^{0.032} \,.
\ee

\boldmath
\subsection{$\kpn$ Correlations with $B_{s,d}\to\mu^+\mu^-$}
\unboldmath

Using (\ref{kplusApprox}) and (\ref{BRtheoRpar}) one can eliminate $\vcb$ to
find \cite{Buras:2015qea}
{\begin{align}
\mathcal{B}(\kpn) &= {(8.65\pm 0.34)}\times 10^{-11}   \left[\frac{\sin\gamma}{\sin 64.6^\circ}\right]^{1.39} \left[\frac{G(22.62^\circ,64.6^\circ)}{G(\beta,\gamma)}\right]^{2.8}\notag\\
&\qquad\qquad\qquad\quad\times\left[\frac{\overline{\mathcal{B}}(B_s\to\mu^+\mu^-)}{{3.78}\times 10^{-9}\bar R_s}\right]^{1.4}\left[\frac{{230.3}\mev}{F_{B_s}}\right]^{2.8}\label{master1},
\end{align}}
where we used
\be\label{vts}
\vts=G(\beta,\gamma)\vcb, 
\qquad G(\beta,\gamma)=
1 +\frac{\lambda^2}{2}(1-2 \sin\gamma\cos\beta)\,.
\ee
Relative to \cite{Buras:2015qea} we just adjusted the central value of $\gamma$ to previous formulae and included preliminary NNLO QCD corrections to $X(x_t)$
as in othe formulae.

Simlary, using (\ref{kplusApprox}) and (\ref{BRtheoRpard}) one can eliminate $\vcb$ to find
{\begin{align}
\mathcal{B}(\kpn) &= {(8.65\pm 0.36)}\times 10^{-11}     \left[\frac{\sin (64.6^\circ)}{\sin\gamma}\right]^{1.41}\notag\\
&\qquad\qquad\qquad\quad\times\left[\frac{{\mathcal{B}}(B_d\to\mu^+\mu^-)}{{1.02}\times 10^{-10}\bar R_d}\right]^{1.4}\left[\frac{{190.0}\mev}{F_{B_d}}\right]^{2.8}\label{master2}.
\end{align}}

\section{Reviews and Lectures at Summer Schools}\label{Lectures}
  I have given many plenary talks at various conferences in which I reviewed
  the field of Flavour Physics. But a real pleasure for me was writing reviews
  and lectures. One of the reasons was that it allowed me to remember later the relevant literature. Another one was the desire to summarize topics in
  a compact form that often is not possible in regular papers in which some
  derivations of formulae were necessary. Still many of my reviews and also
  my papers were quite long, 100 pages or more. While these writings were
  generally well accepted by the community, I got few years ago a comment from
  a female particle physicist that my papers and reviews are too long and she is not   reading them. It is a puzzle for me why she made this statement at all,
  but never mind because I found it rather amusing.
  
  On the other hand Vera Lüth, a well known SLAC experimentalist,
  praised my reviews  because of clarity so that one can understand everything.
  Vera told me this at a breakfast in a hotel  during the CKM
  Workshop   in 2014 in Vienna. Definitely this made me very happy because
  in the last 30 years I made some efforts so that also experimentalists can
  follow my often very technical work. This was not always the case before.

  Before listing the reviews and lectures on flavour physics let me mention
  my first review that dealt with 
deep-inelastic processes in the leading order and beyond
  \cite{Buras:1979yt}. I wrote it in 1979 partly at SLAC but dominantly   at Fermilab. This review
  was motivated by the papers on the NLO QCD corrections to deep-inelastic
  scattering \cite{Bardeen:1978yd} and photon structure functions \cite{Bardeen:1978hg} as well as my papers on parton distributions \cite{Buras:1977ra,Buras:1977yj}.

  This was my first review and was typed by a secretary at Fermilab.
  I did not know Latex at that time and wrote the review first by hand and
  gave it to this secretary who was very efficient in typing it. But this
  was my first review and I struggled with it. Bardeen was motivating me
  in completing it. It was completed in the Summer of 1979. I dedicated it
  to my mother who died suddenly on August 16th that year.
  This review was one of the first on this topic and was well accepted by the
  community. It collected   almost 800 citations by now.

Let me then move to several reviews on flavour physics.
   All these reviews were 
  of great help in writing my book \cite{Buras:2020xsm}, that took
  seven exciting years 2013-2020. Let me then list my most important reviews and lectures
  in chronological order.
  \begin{enumerate}
  \item
    ``The $\Delta I=1/2$ Rule and CP Violation in the $1/N$ Approach to
    the SM'',
    Contribution to International Europhysics Conference on High-Enery Physics
    1987 which took place in Uppsala, Sweden \cite{Buras:1987wp}.
   \item
     ``Strangeness and the Large N Expansion'', Contribution to International Symposium in Hadronic Matter in 1987 in Bad Honnef, Germany \cite{Buras:1987vm}.
      \item
 "The 1/N Approach to Non-leptonic Weak Interactions", published in 1989 in
        Adv. Ser. Direct. High Energy Phys. \cite{Buras:1988kp}.
      \item
        "Phenomenological Applications of the $1/N$ Expansion'', Lectures
        at the School in 1988 in Jaca, Spain \cite{Buras:1988ky}.
    \item
      ``A Top Quark Story: Quark Mixing, CP Violation and Rare Decays in the Standard Model" with Michaela Harlander, published in 1992 in
      Adv. Ser. Direct. High Energy Phys \cite{Buras:1992uf}.
    \item
      "Weak Decays beyond Leading Logarithms" with Gerhard Buchalla and
      Markus Lautenbacher, published in 1996 in  Rev. Mod. Phys. \cite{Buchalla:1995vs}. See more details below.
         \item
      "Quark Mixing, CP Violation and Rare Decays after the Top Quark Discovery" with Robert Fleischer, published in 1998 in
      Adv. Ser. Direct. High Energy Phys \cite{Buras:1997fb}.
\item
  "Weak Hamiltonian, CP violation and Rare Decays", published in 1998 in the proceedings of Les Houches Summer School in Theoretical Physics 1987 \cite{Buras:1998raa}. See some comments below and more details on the events at this school in Section~\ref{LesHouches97}.
\item
  ``OPE, Renormalization Group and Weak Decays'': Contribution to Ringberg Symposium on QFT (on the Occasion of Wolhart Zimmermann's 70th Birthday)
   \cite{Buras:1999rb}.
      \item
        ``Flavor dynamics: CP violation and rare decays'', published in 2002 in the proceedings of   the Erice School 2000, \cite{Buras:2001pn}.
      \item
"CP violation in B and K Decays: 2003", published in 2004 in the proceedings of Schladming  School in Theoretical Physics \cite{Buras:2003wd}.      
       \item 
         "Minimal flavor violation", published in  2003 in the proceedings of Zakopane School 2003 \cite{Buras:2003jf}. See Section~\ref{Zakopane2003} for the events at this school.
       \item
         "Flavor physics and CP violation", published in 2005 in the proceedings of the 2004 European School of High-Energy Physics in Saint Feliu de Guixols, \cite{Buras:2005xt}. See Section~\ref{Feliu} for the events at this school.
       \item
       "Waiting for precise measurements of $K^{+} \to \pi^{+} \nu \bar{\nu}$ and $K_{L} \to \pi^0 \nu \bar{\nu}$" with Felix  Schwab and Selma Uhlig,
         published in 2008 in Rev. Mod. Phys. \cite{Buras:2004uu}.
       \item  ``Flavour Physics in the ``Quark Sector'', Phys. Rep. 2009,
         \cite{Antonelli:2009ws}.
\item
  ``Minimal flavour violation and beyond: Towards a flavour code for short distance dynamics'', published in  the proceedings of Zakopane School 2010
  \cite{Buras:2010wr}.
\item
  ``Top and Flavour Physics in the LHC Era'' with G. Perez, T.A. Schwarz and T. M. P. Tait \cite{Buras:2012ub}.
\item
  ``BSM models facing the recent LHCb data: A first look'' with Jennifer Girrbach-Noe. Contribution to 18th Cracow Epiphany Conference on Present and Future of B Physics'' \cite{Buras:2012ts}. 
\item
  ``Towards the Identification of New Physics through Quark Flavour Violating Processes'' with Jennifer Girrbach-Noe, published in 2014 in Rept. Prog. Phys.
  \cite{Buras:2013ooa}.
\item
  ``Flavour Expedition to the Zeptouniverse''. Contribution to ``Flavorful Ways to New Physics'', 28-31 October 2014, Freudenstadt-Lauterbad, Germany
  \cite{Buras:2015nta}.
\item
``On the Importance of Rare Kaon Decays: A Snowmass 2021 White Paper,''
with Jason Aebischer and Jacky Kumar, published in Snowmass 2021 
  \cite{Aebischer:2022vky}. Proceedings can be found in \cite{Goudzovski:2022scl}.
\item
  "Climbing NLO and NNLO Summits of Weak Decays: 1988-2023", published in 2023
  in Phys.Rep. \cite{Buras:2011we}.
  \item
  Contribution to ``50 Years of Quantum Chromodynamics'', 2022,
  \cite{Gross:2022hyw}.
\item
  "Kaon Theory: `50 Years Later", published in 2023 in the proceedings of the KM50   symposium \cite{Buras:2023qaf}.
\item
  ``SMEFT ATLAS: The Landscape Beyond the Standard Model'', with  Jason\\ Aebischer and Jacky Kumar, to be published in 2026   in Phys.Rep. \cite{Aebischer:2025qhh}.
   \end{enumerate}

Let me describe five of them which in my view stand out.
\begin{itemize}
\item
  The review on weak decays beyond leading logarithms was written with two of my PhD students Gerhard Buchalla and Markus Lautenbacher \cite{Buchalla:1995vs}. Initially there was a plan
  to write a book for Springer Verlag but then I suggested to write first a review and it was a very good idea. We completed it in December 1995 so that it was published in Reviews of Modern Physics still before by 50th birthday. It turned out to be a bestseller, the standard reference to weak decays, cited more than 100
  times a year and reaching over 3000 citations in 2024. We celebrated it in
  the restaurant Nudelbrett in Munich, close to Viktualien Markt.
\item
  In 1997 I gave a series of lectures at the Les Houches, close to Chamonix
\cite{Buras:1998raa}. 
  The students were very impressed by my lectures which I
  did dominantly on the blackboard. This is the only time I got standing
  ovation at the end of my lectures. I completed writing them up in
  1998. It turned out to be another bestseller with 1000 citations reached in
  2025. While our review  \cite{Buchalla:1995vs} was rather technical, my
  Les Houches lectures were even more technical with explicit calculations.
  Several young particle theorists told me that they learned a lot from these
  lectures. I dedicated these lectures to my father who died on 22nd November
  1994. See more details on this school in Section~\ref{LesHouches97}.
\item
"Climbing NLO and NNLO Summits of Weak Decays: 1988-2023", published in 2023
  in Phys.Rep. \cite{Buras:2011we}. I have described there the story of NLO
  and NNLO QCD calculations performed in my group in more details than
  in Part~\ref{NLOQCD}. In particular this review contains many more equations than Part~\ref{NLOQCD}. 
\item
  The review celebrating the Kobayashi-Maskawa paper is I think a good
  summary of various aspects of Kaon Physics \cite{Buras:2023qaf} and of
  projections for the future.
\item
  The SMEFT ATLAS, to be  published hopefully in Phys. Rep. this year, could also turn out to be a bestseller.
\end{itemize}

In addition there were my contributions to collections of papers related
to a chosen topic \cite{Antonelli:2009ws,Appel:2008aa,Raidal:2008jk,Buchalla:2008jp} and three proceedings of the Ringberg Workshops 
\cite{Buras:1989er,Buras:1992qm,Buras:1998vd}.

As evident from this list I like writing reviews and eventually this lead to writing a book but before I describe this story let me stress that my much younger
collaborators, among them my PhD students, wrote recently a number of very good
reviews. These are
\begin{itemize}
\item
  ``Recent Progress in Model Building'' \cite{Altmannshofer:2024hmr},
\item
  ``Charm Physics'' \cite{Friday:2025gpj},
\item
  ``Flavour Physics Beyond the Standard Model'' \cite{Altmannshofer:2025rxc},
\item
  ``Meson-antimeson Mixing'' \cite{Nierste:2025jxe},
\item
  ``Electroweak precision physics'' \cite{Reina:2025suh}.
  \end{itemize}

\section{My  Book: Gauge Theory of Weak Decays}\label{Book}
Around the year 1993 I was asked by Cambridge University Press (CUP) whether I would be interested in writing a book on flavour physics. I have sent them an outline of the
book but told them that this could take still some time. Still I signed a contract with them.
I thought it would be better to write first a review on higher order QCD effects in weak decays. This resulted in the Reviews of Modern Physics article with
Gerhard Buchalla and Markus Lautenbacher \cite{Buchalla:1995vs} and subsequently
my Les Houches lectures \cite{Buras:1997fb} both described already in the
previous section. Both turned out to be some of my best cited papers.

In the meantime CUP was asking me on   the status of my book. I told them that I am making progress but it will still take some time because I am very busy.
In 1999 I informed them that the outline of my book has to be modified because
several things changed since 1993. Within a few minutes I received an email
that my contract with them has been cancelled.

In 2009 at the EPS High-Energy Conference in Cracow I was approached again
by a representativ of  CPU, with whom I had contact ten years earlier. I have given
a rapporteur talk on flavour physics \cite{Buras:2009if} and he thought
that maybe I could write a book after all. I waited still until 2011 and
signed a new contract with the deadline of January 2015. However, I really
started to write something in 2012 after I realized that Jennifer Girrbach would
be a great coauthor of this book. This was evident after I wrote two papers
with her \cite{Buras:2012gm,Buras:2012fs}. Jennifer agreed and constructed in
no time efficient files with the CUP format. The progress was slow because
we were busy with writing papers, in particular with Fulvia de Fazio, but
early 2014 we have written roughly 150 pages. While I have written more than
her, the Part II of the book, the description of the Standard Model, has been
almost entirely written by her. I extended the deadline to 2017 but unfortunately in
January 2015 Jennifer told me that she is leaving research and will not
continue writing our book. I waited still six months but failed to convince
her to be coauthor.

Still it was exciting to spend the next five years writing this book alone.
I completed it in September 2019 and it appeared in the bookstores in July 2020.
The book has 718 pages and 1347 references \cite{Buras:2020xsm}. Its photo is shown in Fig.~\ref{MBook}.


\begin{figure}
\centering
\includegraphics[width = 0.90\textwidth]{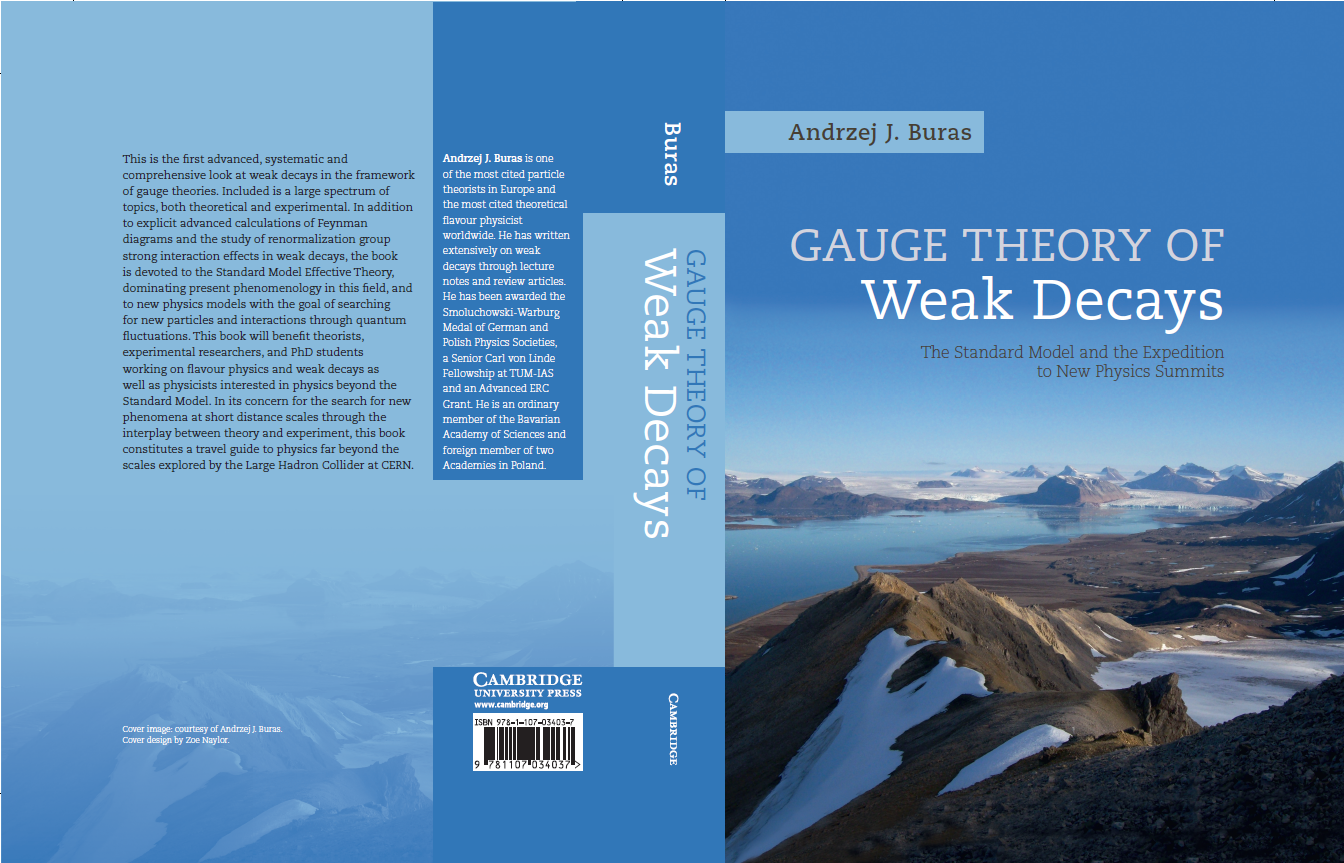}
 \caption{\it My Book. }\label{MBook}
\end{figure}

Many of my colleagues found this book very good. Jennifer was surprised
I succeeded to complete it. All together it took 7 years to write it.
Still parallel to it I succeeded to write $7\times 7$ papers. After all, I retired on April 1st 2012 and had plenty of time, not only to write this book but also to do research. I am very happy I completed this book.
It is for me my own bible. All topics discussed in this write-up are presented in
more details in this book.
I dedicated it to my family introduced at the beginning of my flavour story.

\section{Review Talks at Various Conference}\label{Conferences}

\subsection{Talks in Various Proceedings}
Over last 40 years I have given many review talks at various conferences and
workshops. I list only most important ones and 
first the ones which appeared in proceedings of these conferences and workshops. These are as follows.
   
\begin{enumerate}
\item
 `` B Meson Decays and $7\times7$ Features of Mixing and CP Violation in K and B
  Systems'', Moriond Meeting 1984 \cite{Buras:1984ur}.
\item
  "CP Violation in the Standard Model and Beyond'', Contribution to the Workshop on the Future of Intermediate Energy Physics in Europe, 1984 \cite{Buras:1984ch}.  
\item
  ``Weak Decays, Quark Mixing and CP Violation'' Contribution to International Europhysics Conference on High Energy Physics  1985, Bari, Italy \cite{Buras:1985wt}.
\item
  ``Strangeness and the Large N Expansion", in the Workshop on Strangeness,
  1987, Bad Honef, Germany, \cite{Buras:1987vm}. 
\item
``The $\Delta I=1/2$ rule and CP Violation in the $1/N$ Approach to the Standard Mode", at International Europhysics Conference on High-energy Physics 1987, Uppsala, Sweden, \cite{Buras:1987wp}.
\item
  "{Weak decays and QCD: eighteen years later}", at the {Workshop on QCD: 20 Years Later}, 1992, Aachen, Germany  \cite{Buras:1992aj}.
\item
  "{Theoretical progress in $K$ and $B$ decays}", at the International Symposium on Nuclear and Particle Physics with Meson Beams in the 1-GeV/c Region, 
Tsukuba, Japan, March 1994), \cite{Buras:1995ve}.
\item
  ``CP Violation: Present and Future'', Contribution to
  the First International Conference on Phenomenology of Unification from Present to Future, Rome, Italy, March 1994 \cite{Buras:1994sd}.
\item
  ``Towards precise determinations of the CKM matrix without hadronic uncertainties'', Contribution to ICHEP 1994, Glasgow, UK \cite{Buras:1994hm}.
\item
  ``Rare decays, CP violation and QCD", Contribution to
 Cracow Epiphany Conference on Heavy Quarks (In Honor of the 60th Birthday of Kacper Zalewski)  \cite{Buras:1995pp}.
\item
  ``Theoretical review of B physics'', Beauty 1995, Oxford, UK,
  \cite{Buras:1995iy}.
\item
  ``Theoretical review of K physics''
        Theoretical Summary Talk at the  Kaon 1996 conference that took place from May 30 to June 4, 1996 in Orsay, France 
      \cite{Buras:1996cw}.
    \item
      ``{Flavor changing neutral current processes}",
      Review of the status of FCNC processes at ICHEP 96, Warsaw, Poland \cite{Buras:1996wn}.
      \item
       "{CKM matrix: Present and future}", Review of the status of the CKM matrix'', the 7th International Symposium on Heavy Flavour Physics, 1997, Santa Barbara, USA \cite{Buras:1997th}. 
      \item
        ``CP violation and rare decays of K and B mesons'' at the 14th Lake Louise Winter Institute  (1998) \cite{Buras:1999tb}.
      \item
        ``Theoretical Status of $\epe$'',  Kaon 1999 , Chicago, USA \cite{Buras:1999if}.
      \item
        ``CP Violation 2000: Status and Perspectives'', 3rd International Symposium on Symmetries in Subatomib Physics, Adelaide, Australia, 2000 \cite{Buras:2000pj}.
      \item
        "{Flavor physics and CP violation in the SM}'',  Kaon 2001, Pisa, Italy  \cite{Buras:2001tc}.
        \item
      ``CP violation, rare decays and the CKM matrix'',  Lecture at 8th Mexican Workshop on Particles and Fields in Zacatecas, Mexico, 2001 \cite{Buras:2002ix}.
      \item
        "Unitarity triangle: 2002 and beyond", 14th Recontres de Blois on Matter-Antimatter, Blois, France \cite{Buras:2002sd}.
      \item
        ``Waiting for clear signals of new physics in B and K decays'' Adriatic Meeting, 2004, Dubrovnic, Croatia,  \cite{Buras:2004sc}.
      \item
        ``A  Guide to Flavour Changing Neutral Currents in the LHT Model'',
        with Monika Blanke, Flavianet 2006, Kazimierz, Poland,  \cite{Blanke:2007ww}.
      \item
        ``Quark and Lepton Flavour Physics in the LHT Model'', with Cecilia Tarantino, CKM 2006, Nagoya Japan   \cite{Buras:2007zt}.
      \item
        ``Searching for new physics with rare decays and CP Violation''
        Discrete 2008, Valencia, Spain, \cite{Buras:2009zz}. 
      \item
        ``Testing the CKM Picture of Flavour and CP Violation in Rare K and B Decays and Particle-Antparticle Mixing''
        Contribution on the occasion of the Nobel Prize awarded to Kobayashi and Maskawa   \cite{Buras:2009us}.
      \item
``Patterns of Flavour Violation in the RSc Model, the LHT Model and
Supersymmetric Flavour Models'', Kaon 2009, Tsukuba, Japan \cite{Buras:2009hr}.
              \item
                ``FCNC Processes Waiting for the Next Decade",
        the Flavianet topical workshop: Low energy constraints on extensions of the Standard Model \cite{Buras:2009ye}.
      \item
       ``Flavour Theory: 2009''  Contribution to EPS-HEP 2009, Cracow, Poland
  \cite{Buras:2009if}.
\item
 ``Flavour Theory and the LHC Era", Contribution to PLHC2010, \cite{Buras:2010xj}
\item
  ``Flavour Visions'', Contribution to Beauty 2011, Amsterdam,
\cite{Buras:2011fz},
\item
  ``Towards the Identification of NP through Correlations between Flavour Observables'' , Contribution to EPS-HEP 2013, Stockholm, Sweden \cite{Buras:2013dda}.
\item
``$\Delta I=1/2$ and $\hat B_K$: 2014'', 
  Contribution to QCD@Work 2014, Giovinazzo, Bari, \cite{Buras:2014apa}.
          \item
       `` Kaon Theory News'', talk presented at EPS-HEP2015, Vienna, \cite{Buras:2015hna}.
      \item
    ``The Renaissance of Kaon Flavour Physics'',   Contribution to Beauty 2016, Marseille   \cite{Buras:2016qia}.
          \item
    ``The Revival of Kaon Flavour Physics'', Contribution to QCD@Work 2016, Martina Franca, Italy,    \cite{Buras:2016bch}.
          \item
        ``Kaon Flavour Physics Strikes Back'',
            Contribution to Kaon 2016, Birmingham, England \cite{Buras:2016egb}.
            \item
              ``The Return of Kaon Flavour Physics'',      Contribution to 24th Cracow Epiphany Conference on Advances in Heavy Flavour Physics, Cracow, Poland \cite{Buras:2018wmb}. It contains an informative table of papers studying the
              impact of the possible $\epe$-anomaly on several NP models.
            \item
              ``The Dual QCD at Work: 2018'',  Contribution to QCD@Work 2018, Matera, Italy \cite{Buras:2018hze}.
            \item
              ``The $\epe$-Story: 1976-2021'' \cite{Buras:2020wyv}.
                          \item
``$\epe$ in the Standard Model and Beyond: 2021'', in 11th International
Workshop on the CKM Unitarity Triangle, Melbourne, Australia               
              \cite{Buras:2022cyc}.
            \item
            ``Hunting New Animalcula with Rare K and B Decays'',  Contribution to QCD@Work 2024, Trani, Italy \cite{Buras:2024per}.
\end{enumerate}

It is rather amusing that my contribution to the CKM 2021 Workshop in
Melbourne appears twice in the arXiv. One with me as a single author
listed above \cite{Buras:2022cyc} and the second one with a second
author Kind Regards \cite{Regards:2023fpf} who is even the first author.
I leave to the possible reader to find out what happened in the process
of publishing proceedings. One can check in INSPIRE that my coauthor has
one citation, the one from the present autobiography.

I will describe some of these conferences soon but let me first describe three
conferences 
at which I did not give any talk but they were interesting for me anyway.
\subsection{ICHEP 1974 London}\label{London74}
This was the first big conference I attended. As I will describe later
in Section~\ref{Deus}, this was also the first conference at which my
work has been cited. It was a paper on geometrical scaling in proton-proton collisions written with Jorge Dias de Deus \cite{DiasDeDeus:1973lde,Buras:1973km}.

The conference took place in London,  a few months before  {\em the November
  Revolution}, the discovery of charm. I do not remember much from this conference after so many years except that I listened for the first time to a talk of John Ellis and
that I met Cecilia Jarskog at the Conference Dinner. This was eleven years
before in 1985 Cecilia wrote two very important papers. One on the commutator of the quark mass matrices in SM and a measure of maximal CP nonconservation \cite{Jarlskog:1985ht} and the second one on
a basis independent formulation of the connection between quark mass matrices and a measure of maximal CP nonconservation \cite{Jarlskog:1985cw}.
Almost forty years later she received the EPS HEP Award 2023 for this work.
Well deserved. Her case shows that what really counts in our field are
simple ideas.

\subsection{ICHEP 1984 in Leipzig}\label{ICHEP84}
This was my first visit to East Germany. I traveled from Munich to Leipzig by train, where we had to leave the train at the border for a control. They collected a newspaper and a journal I carried. I stayed in a large hotel in the city center, close to the conference site. The corridors were dimly lit, which was quite different from Western hotels.

On the first day, I met Dimitri Nanopoulos after many years. Excitedly, he said, “Andrzej, how nice to see you again! Higgs with $9\gev$ mass has been discovered at DESY, supersymmetric one!” I was amused by the last remark. A few months later, it turned out to be a statistical fluctuation.

At this conference, I met Andrei Linde for the first time. He gave an excellent plenary talk on various aspects of cosmology. I enjoyed it so much that, in 1988, I proposed him as a plenary speaker at ICHEP in Munich, which I co-organized.

I also tried to talk with young students helping at the conference, but it was clear they were afraid of speaking with Western participants. This became even more evident when a more senior theorist from Leipzig, who had visited the Niels Bohr Institute ten years earlier, greeted me but admitted he was afraid to talk to me.

During the weekend, conference-organized bus excursions allowed visits to three cities: Dresden, Meissen, and Weimar. In 1984, they looked very different from today.

Dresden was largely destroyed during World War II and only partially rebuilt in 1984. The pride of Dresden, the Frauenkirche, was still a ruin. Interestingly, I was the only visitor there apart from Julius Wess, one of the fathers of supersymmetry. I introduced myself, and after a brief conversation about the ruin, I mentioned that we in Munich hoped he would accept an offer of a full professorship at TUM. Eventually, he rejected it in favour of a professorship at LMU and became a director at MPI. At the time, I had no idea I would one day obtain the very TUM position we were discussing. Beginning in 1998, I would visit Dresden frequently.

Meissen was the next city. The main goal was the famous porcelain museum, but I soon left for a walk. The streets were empty, with no people or cars. Buildings, while undamaged by the war, were old and dirty. For half an hour, I felt transported fifty years into the past. Thirty years later, Meissen looked completely different.

Weimar was in better shape. I visited the Goethe and Schiller museums, but missed Franz Liszt’s house. Thirty years later, visiting it, I found the museum profoundly moving—even more than Mozart’s house in Salzburg—despite listening to Mozart more often than Liszt.

These were very interesting excursions but I cannot resist to describe
briefly my second visit to East Germany which took place in the spring of
1985. I have been invited to a small workshop organized by Leipzig University
close to  a small town Cordobang placed roughly 350 km north of Munich. I went
their by train again, had to change the train once and eventually arrived
in Cordobang. Surprisingly, there was nobody at the station and as I did not
have any smartphone or even handy in 1985 it was unclear to me what I should do.
I decided to ask somebody to tell me how I could reach this meeting and
to call organizers to pick me up by car if necessary. Unfortunately,
the streets were empty. I decided to knock at a door of one house but the
woman who opened the door did not want to talk to me.

I continued walking and found finally a garage with a few workers. I asked them whether I could use their phone to call the person who invited me. This did not turn out to be obvious. They told me to wait and then I saw them discussing whether
they would allow me to use their phone. Finally, I gave them the phone number
of the meeting, they phoned and I was picked up. Rest was not sufficiently
interesting to describe it here.
\subsection{Photon-Lepton 1985 in Kyoto}\label{Japan1985}
In August 1985, I traveled to Japan for the first time, attending the Photon-Lepton Conference in Kyoto. I was also invited by my Fermilab collaborator Taizo Muta to spend a week in Hiroshima. My companion in Japan was Reinhold Rückl, making this one of my most memorable physics trips.

The journey from Munich to Kyoto took 22 hours: Munich → Paris → over the North Pole → Alaska → Osaka, with a stop in Tokyo. This was the only time I flew over the North Pole and visited Alaska, with an unforgettable view of
Denali (6160 m)\footnote{From 1917 till 2015 called Mount McKinley.} .

One day before departure, a terrible plane crash occurred in Japan, killing over 600 people. Only one small girl survived.  Fortunately, an American postdoc from Fermilab, who was meant to be on that plane, missed it, saving his life. While tragic, it reassured me statistically about my own flight.

Arriving in Kyoto exhausted, I found Reinhold in the hotel room wearing a kimono and watching TV. 
On the next day the conference began and being jetlagged I could not really follow
all talks. Fortunately, I was not giving any talk at this conference and was more relaxed than in Bari as described below. On the other hand as the conference took place from
19th to 24th August it was rather hot and one suffered from large humidity.

During the weekend we went on a excursion to Nara, which
old days was Japan's capitol. The main highlight was the visit to several
shrines.  During this excursion I met Monica Pepe for the first time, the future wife of Guido Altarelli. A very nice person despite the fact that in the following thirty years she addressed me with {\em Buras} and not
Andrzej. But it changed for some reasons later.

The highlights of this meeting were measurements of $\vcb$ and other aspects
like improved measurements of $W$ and $Z$ masses that were discovered two years
earlier. The plenary talk on weak decays was given by Ling Li Chau, the
author of the Standard Parametrization of the CKM matrix \cite{Chau:1984fp}. She ignored totally
our work on large $N$  \cite{Buras:1985yx,Buras:1985xv}. Fortunately I did not
have to inform her about our work. It was David Hitlin, an important experimentalist from SLAC, who during the discussion after her talk told her and the participants that it was an important work. Also Bill Bardeen was at this conference but only one year later when visiting MPI he got interested in this new direction which lead to the Dual QCD reported previously.

During the conference we stayed at a Western Hotel but afterwards staying still
a few days in Kyoto we decided to move to a Japanese hotel for a few days which
allowed us to see more shrines in Kyoto. The hotel was very different than
Western hotels. The shrines were truly impressive.

A special event was meeting a {\em maiko girl}, the sister of one of japanese
participants who visited Munich few weeks before. Maiko girls are trained to become geishas one day. We met her
together with an old Japanese lady who was taking care that we do not come
to close to her. Both ladies practically did not know english but our colleague
was also present so that we could have a ``small talk'' with them and drink
a bit of sake served by maiko girl.

Few days later we took a train to visit Muta in Hiroshima. From this vist
I mostly remember the museum related to the atomic bomb. It was very well
arranged with some music which brought a very special atmosphere. After visiting
few places close to Hiroshima and giving talks at the University we took
a fast train to Tokyo. While Tokyo itself did not impress me so much, especially
after being in Kyoto, the visit of the Toshogu-Shrine in Nikko  was very interesting. It is in my view one of the most impressive  shrines in Japan. It
consists of more than a dozen buildings set in a beautiful forest.

The time spent with Reinhold in Japan finished few days later and when Reinhold
left the hotel to fly to Hong-Kong and then to Munich I was really sad. One day later I took a plane directly to Paris and then to Munich. On the whole
this trip to Japan was very interesting and I returned to Japan several
times later. In 1995 with our son Robert and several times between 2005 and 2010
being a member of the advisory committee at KEK. This gave me the opportunity
to meet my friend Ikaros Bigi who was also a member of this committee.
In 2010 Gerhard Buchalla replaced me.

\subsection{Kaon, 1988, 1996, 1999, 2001, 2009, 2016, 2019, 2025}
\subsubsection{Kaon 1988 (Vancouver)}
This was the first Kaon conference. As I described already in Section~\ref{Anatomy1} I got there, after lunch with Fred Gilman, the idea to analyse the
ratio $\epe$ for arbitrary top quark mass. This lead to a series of papers
already described. It was a nice conference and my first trip to Canada.
Giuliano Preparata was also taking part in this conference and I remember
that he interrupted Gilman's talk by telling Fred that there is mounting
evidence against QCD and that he does not understand why Fred is still
talking about QCD.  I gave also a talk that involved some aspects of QCD
but somehow Preparata did not interrupt me.

At this conference we made a trip to Triumph laboratory where a Kaon factory
was planned and the director of the lab told me that he hopes I will spend
some time there when the factory will be built. This unfortunately did not
happened but I visited Vanouver in 1998 at ICHEP conference and in 2004 with
my family as I will describe later.

\subsubsection{Kaon 1996 (Orsay)}
  The next conference in the series of {\em Kaon} conferences I attended took plays in
  Orsay. It was an interesting conference. Let me just mention few highlights.
  There was a talk of Yossi Nir who talked about B physics stating that Kaon physics is going to the end and B physics will dominate from now on. He was
  strongly critized by the organizers after his talk. I made also a statement
  in the discussion after his talk, stressing the importance of rare Kaon decays. I repeated this in my talk that was the theory summary of the conference.
  Also Bruce Winstein, the experimental summary speaker, stressed the importance
  of Kaons. While definitely Nir was wrong concerning Kaon physics, which played
  an important role until today, he was definitely right that B physics plays these
  days even a bigger role than Kaon physics. This is mainly related to many
  more   observables one can measure.

  Yet, Yossi  learned something. Few months later he derived together with Yuval Grossmann an upper bound on the branching ratio for the $\klpn$ decay
  in terms of the $\kpn$ branching ratio \cite{Grossman:1997sk}. Gerhard Buchalla and me could
  have easily derived it few years earlier but we missed it. This bound
  is called these days Grossmann-Nir bound instead of Buchalla-Buras bound.
    \subsubsection{Kaon 1999 (Chicago)}
  The meeting took place in the southern part of  Chicago, quite close to a dangerous district of this city. Fortunately my hotel was close to the conference site. I gave a talk summarizing the status of $\epe$ \cite{Buras:1999if}. This was the time when only
  the first results from the second round of experiments at CERN (NA48) and Fermilab (KTEV)
  were known with the average value of $\epe=(21.2\pm4.6)\cdot10^{-4}$, significantly
  above the SM values presented that year by us \cite{Bosch:1999wr}. This
  has been even strengthened by the calculation of NNLO corrections to electroweak penguin contributions to electroweak penguins performed with Gambino and Haisch in November 1999 \cite{Buras:1999st}. There we found values in the ballpark
  of $\epe=6\cdot10^{-4}$ and after 26 years my estimate practically did not change.  
  \subsubsection{Kaon 2001 (Pisa)}
  I gave a talk on flavour physics and CP violation in the SM \cite{Buras:2001tc}. It was very well accepted. From this conference I remember only a walk with Konrad Kleinknecht one evening. There were some fireworks which we wanted to
  watch. Suddenly Konrad asked me whether I am already thinking about my death.
  I was approaching 55 in 2001 and definitely I did not think about my death.
  He was 61 and I was surprised about his question. I started to think a bit about it twenty years
  later when some of my colleagues in my age or only a bit older than me were
  passing away. This was also one of the motivations for writing this
  autobiography now.
  \subsubsection{Kaon 2009 (Tsukuba)}
 I summarized the results on patterns of flavour violation in RSc model and compared them with those identified in the LHT model and in a number of SUSY Flavour Models. While K decays play in this presentation a prominent role, the inclusion of B physics and lepton flavour violation is crucial in
the distinction between these three popular extensions of the SM by means of
flavour physics \cite{Buras:2009hr}.

\subsubsection{Kaon 2016 (Birmingham)}
  In my talk I stressed that Kaon physics is coming back \cite{Buras:2016egb}. This was
  the meeting at which I met Maria Cerd{\'a}-Sevilla for the first time. She gave a very
  nice talk on her work on NNLO QCD corrections to QCD penguin contributions
  to $\epe$ within the SM. Unfortunately, until today the results of this  calculation,
  in which Martin Gorbahn and Sebastian J\"ager were involved, has not apeared
  in print.  It can be found in various proceedings, in particular
  in \cite{Cerda-Sevilla:2016yzo,Cerda-Sevilla:2018hjk}. It is a pity it
  has not been published because they found an additional suppression of
  QCD penguin contributions to $\epe$ through NNLO QCD corrections and
  supporting my claim that the SM value of this ratio is below its experimental
  value. As Maria left physics research by now, I do hope that Martin and Sebastian
  make sure that their paper with Maria will be published one day after the
  input on hadronic matrix elements and CKM parameters has been updated.

\subsubsection{Kaon 2019 (Perugia)}
I gave a talk presenting the best strategy for SM prediction of $\epe$ 
\cite{Buras:2019vik}. RBC-UKQCD collaboration did not present their results
and as an example I predicted their value without isospin breaking corrections
to be $(9.4\pm3.5)\cdot 10^{-4}$ so that after the inclusion of these corrections they would find $(5.6\pm2.4)\cdot 10^{-4}$ in agreement with DQCD.

My prediction turned out
to be wrong as in 2020  RBC-UKQCD presented $(21.7 \pm 8.4) \cdot 10^{-4}$ still
without isospin breaking corrections that would bring it down to
 $(13.9 \pm 5.2) \cdot 10^{-4}$ \cite{Buras:2020pjp}. Yet, after the dramatic change of their result relative to their 2015 value, still everything is possible.
Hopefully, we will know it in this decade.

The worksop was very well organized and 
Perugia itself was very interesting. One of the highlights for me was the dinner
with  Veronique Bernard and  Marc Knecht.
  
 \subsubsection{Kaon 2025 (Mainz)}
 Being a member of the International Committee for this conference I did
 not want to give a talk but I was asked to take part in the round table
 discussion together with Augusto Ceccucci, Gino Isidori and Andreas Juttner.
 It was an interesting conference without spectacular results. Hopefully
 there will be some at Kaon 2028. I followed this conference per zoom.

\subsection{CKM Workshop 2002, 2003, 2005, 2006,  2014}
\subsubsection{CKM 2002 (CERN)}
This was the first CKM Workshop and I was asked to coordinate it with a few
experimental physicists. One of them was Achille Stocchi. There were various summaries of this workshop. I refer only to those which have material in the arXiv. These are: \cite{Flynn:2002fz,Proceedings:2003xqt}. One of the results of this workshop was a paper with Achille and Fabrizio Parodi, another experimentalist \cite{Buras:2002yj}.
In this paper we have investigated which is the most efficient strategy to determine the apex of the UT, that is   $(\bar\varrho,\bar\eta)$, by measuring 
any pair of two variables representing the sides and the angles of the UT.
It turned out that the measurement of $(\beta,\gamma)$ results in the most accurate values of $(\bar\varrho,\bar\eta)$. I described this strategy in Section~\ref{BPS}.

Personally, the most important event at the CKM 2002 was meeting Fulvia De Fazio
for the first time. She turned out to be one of the most frequent collaborators after I retired. Our papers  are described in Sections~\ref{ZprimeModels} and \ref{331Models} of this writing.

\subsubsection{CKM 2003 (Durham)}
The next CKM workshop took place in Durham. It was organized by Patricia Ball.
I gave a talk on my work with Achille and Fabrizio. It can be found in the
proceedings of this workshop \cite{Buras:2003bz}.
\subsubsection{CKM 2005 (San Diego)}
This workshop took place in March 2005 in San Diego. I traveled there with
Selma Uhlig and Felix Schwab. We had to take a bus from the hotel to the conference site. There are several photos from this meeting in which Patricia Ball,
Robert Fleischer and few other theorists, with whom I had contact at that time,
participated. I have shown these photos in my Christmas Talk in December that
year. During the weekend I rented a car and as a driver I have shown
to Selma, Felix and few others some parts of South California.

I gave the talk ``Probing New Physics through Rare Decays and CP Violation'' but
I did not find any write-up of my talk and any  proceedings of this workshop.
There was an interesting instruction from the Local Organizing Committee: ``Be provocative as much as possible''!  I followed it by making citations of papers
incomplete!

\subsubsection{CKM 2006 (Nagoya)}
I went to this workshop with Cecilia Tarantino. We reported in a duet our
work on the LHT model \cite{Buras:2007zt}. This was an interesting conference.
In particular, the CKM masters participated in this meeting. One evening, before
the conference dinner they all talked for few minutes with the following difference. Cabibbo 5 min, Kobayashi 2 min and Maskawa, probably 
30 seconds. I met Cabibbo few times before. The first time I think at MPI in Munich, where he was in the board of trustees in 1980s. This time I had more
contact with him. Nicola, Cecilia, Guido Martinelli and some other Italians
went to a Sushi place. I made some photos there. In addition a photo of Cecilia
at another Sushi place. I have shown these photos in my Christmas Talk in December that year.

Unfortunately, Cabibbo was not included in the list of Nobel Prize winners in physics of 2008. Half of the prize went to Yoichiro Nambu and the second half jointly to Kobayashi and Maskawa. This was clearly a big mistake of the Nobel Prize Committee
for two reasons. Nambu deserved the full prize and KM should have obtained
this prize together with Cabibbo one year later. I have written to Nicola
expressing my disappointment but he wrote back that it is not important
for him. I doubt this was true. Nicola received Dirac Medal together
with Sudarshan in 2010. The announcement came in August and I congratulated him.
Surprisingly he did not respond to my email which was unusual. I learned only few weeks later from Gino Isidori that he died eight days after the announcement of the prize
which is made usually on August 8th, Dirac's birthday. There is a video in YouTube of the ceremony of the 2010 Dirac Medal. It was in December that year.
His wife and his son were accepting it. At least he was alive when the prize
was announced. On the other hand if he received the Nobel Prize in 2008, he
would not be eligible for the Dirac Medal. 

The 16th of August on which Cabibbo died is very special for me.
In 1979 my mother died on that day.

\subsubsection{CKM 2014 (Vienna)}
I attended this workshop without giving a talk. Jennifer represented my group
presenting our work on $B\to K(K^*)\nu\bar\nu$ \cite{Buras:2014fpa} that became
the standard reference to be cited, in particular after in 2023 Belle II announced their results on $B^+\to K^+\nu\bar\nu$ that disagreed with the SM expectations analyzed
in detail in that paper. But possibly the discussion of these decays beyond
the SM were the most important results of our paper. Her talk can be found in
\cite{Girrbach-Noe:2014kea}.

Jennifer left unfortunately the research at the beginning of 2015, still being involved in a detailed analysis of
$\kpn$ and $\klpn$ decays in the SM \cite{Buras:2015qea}. The latter
paper is another classic from my group, regularly cited in connection with
NA62 and KOTO experiments on these decays. These last two Jennifer's papers turned out to be the  most cited papers by Jennifer in her career as a
theoretical physicist. 

The conference itself was very interesting, with many good talks. Visiting Vienna again was a real pleasure. We stayed in a hotel relative close to the conference site, which was at Vienna's Technical University, not far from Karlsplatz.
Being there reminded me of my father who in 1962 was staying in the
Karlshotel when visiting Atomic Agency and meeting my brother Jacek and me.
We stayed that year in Vienna without our parents but this is another story.

Another highlight of the conference was meeting Aida Elhadra, whom I knew
from my Fermilab visit in 2005. Still another highlight for me was a breakfast with Vera
Lüth, one of the leading experimentalist at that time. Vera told me that my
papers and reviews are very clear and she is reading them with pleasure.
This was a nice start of the day.

\subsection{EPS 1985, 2009, 2011, 2013, 2015}
\subsubsection{EPS 1985 (Bari)}
In 1985 I have been invited to give a rapporteur talk on weak decays and CP violation at the EPS conference in Bari, Italy. It was from 18th to 24th  July and rather hot
but still well below 40 degrees as can happen these days. The big boss of this
conference was Giuliano Preparata who was professor of the University there.

I was a bit worried giving a talk on weak decays in the presence of Preparata because I worked already on QCD effects in these decays, in particular I just started working on the first application of large N QCD in weak decays with Jean-Marc
G{\'e}rard and Reinhold R{\"u}ckl \cite{Buras:1985yx,Buras:1985xv}. Preparata was strongly against QCD. This was very clear already on the first day of the conference during a parallel session in which he very strongly attacked Guido
Altarelli for the same reason. My worries increased even more when 30 participants of the conference and a number of accompanying wives were robbed on the street. I remember still one wife shouting on Preparata telling him that he should
somehow protect the participants. This worked. During plenary seesions, which
took place in a theater, several policemen were standing in front of the
theater and street barricades  around it were built.

I am not claiming that Preparata would take revenge in this manner if I did not refer to him but his attack on Altarelli convinced me that I have to refer somehow to his work. I prepared therefore one transparency on his geometrical view
on strong interactions, just stating that other approaches to strong interactions
exist. When Preparata approached me after my talk, Roberto Peccei, my MPI
colleague, came immediately to us expecting that Preparata will complain
about only one transparency for Guliano's work. To Roberto's and my surprise
Preparata was very positive about my talk, stating that I referred properly
to everybody. Maybe this was the reason he did not attack me at
the Kaon conference in Vancouver in 1988.

However, not everybody was happy with my talk. Mary Kay Gaillard, who participated in this conference, told me that my talk was too technical for experimentalists. This criticism was very useful for my future talks. Another person who
did not like my talk was Soergel, DESY director at that time, who probably still remembered that I rejected DESY offer in 1981 as described in Section~\ref{1981P}. He was my Chairman. Those days such talks
were 60 min long but I stopped after 55 min in order to have some discussion.
There were only few question and he told me that I did not use properly the
time allocated to me. My contacts with him improved by much later. In 1990 he
asked me whether I would be interested in the highest position in the DESY Theory Group but I just moved to TUM and I did not want to do it.
Moreover after my father died in 1994 he told me several times how much he appreciated
his involvment at DESY. The three DESY Theory workshops which I organized in
1990, 1996 and 2000 (see Section~\ref{DESYW}) improved my relations with DESY as well as my participation in the
Physics Research Committee at DESY from 1998 to 2002.

At this conference I met also Svjetlana Fajfer, who spent two years as a Humboldt fellow at MPI 1987-1989 working with Jean-Marc G{\'e}rard and visited Munich several times later. I never worked
with her but we kept contacts regularly until today and I was honored to
give a talk at her 60th birthday in Ljubljana in  2014. She was very successful
in creating a very strong particle theory group at the University of Ljubljana 
and organizing very good workshops in Portoroz every second year. I gave
several talks at these workshops.

\subsubsection{EPS 2009 (Cracow)}
After long break in attending EPS conferences I went in 2009 to Cracow to give
a plenary talk on the status of flavour physics \cite{Buras:2009if}. Monika
Blanke who participated in this conference told me that it was one of the
best talks I gave until then. Indeed its write-up collected almost 100 citation
which is unusual as most proceedings are very poorly cited.

\subsubsection{EPS 2011 (Grenoble)}
I went to this conference flying first to Lyon and then taking train to Grenoble.
It looks like I was not giving any talk and it surprises me that I went at all
because I had problems with my back and could not walk well. After returning
to Munich I went to a doctor who gave me very strong drugs against pain.
This saved our vacation in Norway. 

At this conference the EPS-High Energy Prize was awarded to Glashow, Iliopoulos
and Maiani (GIM) which clearly they deserved. But for me the highlight
of the conference was meeting again Uta and Max Klein.

\subsubsection{EPS 2013 (Stockholm)}
In Stockholm I reported on my work with Jennifer that appeared in Reports on Progress in Physics \cite{Buras:2013ooa}. The talk itself, that can be found in
\cite{Buras:2013dda}, included also other papers dominantly with Fulvia and Jennifer \cite{Buras:2012jb,Buras:2012dp,Buras:2013uqa,Buras:2013rqa,Buras:2013raa}.
Fulvia was participating in this conference, Jennifer unfortunately not.

At this conference the first measurement of $B_s\to\mu^+\mu^-$ decay has been
presented. For me it was a bit disappointing because it turned out
to be in the ballpark of the SM value obtained by  Gerhard Buchalla and myself
twenty years earlier that included for the first time  NLO QCD corrections
\cite{Buchalla:1992zm,Buchalla:1993bv}. I hoped that NP would
show up in this decay. During the break
after the session at which this result has been presented, a number of
people congratulated me on this result so that for a moment I thought I
will be coming to Stockholm one day again, this time together with Gerhard. Yet, in view of substantial experimental
error there was still a significant room for NP contributions to this decay  which was the topic of two subsequent papers with Fulvia and Jennifer 
\cite{Buras:2013qja,Buras:2013dea}. One on general $Z^\prime$ models and models
with flavour violating $Z$ quark couplings and the other one on 331 models.
I described them in Sections~\ref{ZprimeModels} and \ref{331Models}.

As already stated at various places in this write-up precise measurements
of $B_{s,d}\to \mu^+\mu^-$ decays are very important and I hope we will
know  in this decade whether NP affects them.

\subsubsection{EPS 2015 (Vienna)}
It was great to be in Vienna again. The conference took place at the Campus of the Unversity which had more space than the Technical University. For flavour physics
the conference was very exciting because not only more data for the violation
of lepton universality in $b\to sl^+l^-$  and $b\to c l\nu$ were available but
in addition  new results from RBC-UKQCD collaborations indicated that the SM value of $\epe$ could be much lower than its experimental value as expected by me and
my collaborators already since 1999. I was reporting on this in a parallel session and arranged
that on this day three papers with my participation would appear among the
first in hep-arXiv. These were \cite{Buras:2015yba,Buras:2015xba,Blanke:2015wba}. Soon after the conference, still in July, the fourth paper appeared \cite{Buras:2015yca}. Two dealt with the SM, the third was the update of our
earlier analyses in the LHT model and fourth one, possibly the most interesting
among the four dealt with $\epe$, $\kpn$ and $\klpn$ in simplified models.
My talk can be found in \cite{Buras:2015hna}. The flavour session was very
lively with various views on the present status of $\epe$. The four papers
just mentioned have been already described at various places above.

One day when walking to my hotel I met Guido Altarelli and his wife Monica Pepe
whom I met for the first time at the Kyoto conference in 1985. Guido
congratulated me for my recent papers but he looked very tired. In fact
he died two months later at the age of 74.  Very sad.
He was first class physicist. In fact he received several awards like
Julius Wess medal, Sakurai Award and at this conference he was one of the
recipients of the High Energy HEP Prize. When I received the Max Planck Medal in 2020
and the Sakurai Award in 2024 Monica told me that he had a very high opinion
about my work. Great pleasure for me to hear this.

\subsection{QCD@Work, 2014, 2016, 2018, 2024}
These conferences are organized by the Bari University, in particular by my friend 
and great  collaborator Fulvia De Fazio. They are taking place near Bari, mostly at very nice places. The organization is perfect. I have always been picked up from
the Bari Airport and brought to the hotel. In 2018 I did not have to pay conference fee and in 2024 even my hotel room was paid by the organizers.

The atmosphere at the workshop is always very nice with dinners, beyond the conference dinner, at
various restaurants together with Fulvia, her collaborator Pietro
Collangelo and few other participants.

The main topic of this series of conferences is dominantly QCD and the name of
the conference states it explicitly: ``QCD at Work''. However, the topics
deal first of all with non-perturbative aspects of QCD. Therefore in 2014 and 2018 I talked dominantly about Dual QCD approach applied to Kaon physics,
in particular $\Delta I=1/2$ rule, non-perturbative parameters relevant for
$\varepsilon_K$ both within the SM and beyond it and of course $\epe$.
In these years we made significant progress in calculating these parameters
\cite{Buras:2014maa,Buras:2015xba,Buras:2016fys,Buras:2018evv,Buras:2018lgu,Aebischer:2018rrz} and I was reporting on these results and their implications.
In 2016 I concentrated on our analyses beyond the SM \cite{Buras:2015yca,Buras:2015kwd,Buras:2015jaq,Buras:2016dxz,Bobeth:2016llm} like 331 models and vector-like quark models. In 2024 my main topic were the strategies developed with Elena Venturini to avoid $\vcb$ and $\vub$ uncertainties in predicting rare K and B
branching ratios within the SM \cite{Buras:2021nns,Buras:2022wpw,Buras:2022qip}.

These reports can be found in the proceedings to these four conferences:
\begin{itemize}
\item
  Giovinazzo (2014) with the title `` $\Delta I=1/2$ Rule and $\hat B_K$: 2014''
  \cite{Buras:2014dxa}.
\item
  Martina Franca (2016) with the title ``The Rivival of Kaon Flavour Physics''
  \cite{Buras:2016bch}.
\item
  Matera (2018) with the title ``Dual QCD at Work: 2018'' \cite{Buras:2018hze}.
\item
  Trani (2024) with the title ``Hunting New Animalcula with Rare K and B Decays'' \cite{Buras:2024per}.
  \end{itemize}

In addition there are series of conferences which do not provide proceedings but
the talks can be found on the Webcites of these conferences. These are in particular Planck Conferences and the ones in Portoroz. I have given talks at several of them. Let me briefly describe my participation in these conferences.

\subsection{Planck 2002, 2006, 2007, 2008,  2011, 2012, 2017, 2023}
I participated in a number of Planck conferences in the years listed above.
These were not all Planck conferences because in some I did not want to partcipate.  The reason was that
I found them in some cases rather formal and I did not want to go to them.
However, I have been invited to some of them and in these cases it was difficult
to say no. Moreover, in some cases they took place in interesting cities and it was fun to participate.

Slight problem in describing these conferences is the lack of proceedings so
that after so many years it is hard to remember even what I was talking about.
I made some efforts to recall them but my description is in each case rather
brief.
\subsubsection{Planck 2002}
This conference took place in Kazimierz, a small town 100 km from Warsaw.
I have been there before. It was in 1968, I think, at a workshop organized
by my father. As in 1968, the Planck Conference 2002 took place in the Architect House, where I stayed with my father in 1968.

I do not recall whether I gave a talk there but this was the time I worked
with Janusz Rosiek, Slawianowska and Chankowski on some aspects of the MSSM
as described in Section~\ref{MSSM}.

Typically at each of Planck conferences 60th birthdays of important participants
were celebrated. This time it was Stefan Pokorski's birthday. This celebration  took place one evening and I was asked to be the moderator of this event.
I was told that the dean of the Physics Department would come from Warsaw to
say a few words about Stefan. I did not know who it was. It turned out to be
a lady in my age. When she saw me she ran to me screaming: ``Andrzejku (a nice
version of my name) how nice to see you again!'' and  started kissing me.
It was in fact my colleague from my physics studies: Katarzyna Chalasinska. We did not see each other for more than 30 years and I did not recognize her at first.

Another event of this conference, that I remember, was my chairmanship of
one afternoon session in which Marcela Carena gave her talk. It was planned
for 30 min. Unfortunately, Marcela did not want to keep the time allocated
to her despite of telling her to stop. After 35 min I found the solution.
I turned  the overhead projector off. But we still stayed friends until today.
At this conference I met Geraldine Servant for the first time. I recall
nice conversations with her.

\subsubsection{Planck 2006}

This workshop took place in Paris. Monika Blanke and Andi Weiler took part
in this meeting. Andi had problems with his leg after the accident in
Rome that I will describe in Section~\ref{Roma06}. Monika and me stayed at the same hotel not far
from Sorbonne where the workshop took place.

On the first evening I made a stupid mistake. Monika and me went to a shop to buy drink  water. Monika bought possibly 10 liters for herself and I as a gentleman carried not
only my two liters but also most of hers. As a result I got back problems
which lasted the full meeting. Indeed, I had some problems sitting in
the conference room and not only there. There is a photo with Monika in a
coffee
shop which definitely proves that I suffered. There is another photo in
the Paris metro with Monika binding my shoes. Monika has shown these photos
in a talk at my 70th birthday ten years ago.

During this meeting I met Ann Nelson for the first time. She was an excellent
theorist and a very nice person receiving Sakurai Award 2018 with Mike Dine.
Unfortunately she died in 2019 in a mountain accident.

\subsubsection{Planck 2007}
The conference took place this time in Warsaw. Several of my PhD students
took part in it. In particular Monika Blanke, Michaela Albrecht and Anton Poschenrieder. We all stayed in Harenda Hotel which was close to the venue of the conference, on the campus of Warsaw University. I stayed in this hotel previously,
when visiting Warsaw. It was a nice hotel a hundred meters from Krakowskie Przedmiescie and in a walking distance from the old town.
But I learned last year that it was closed.

This time I opened the conference with a talk which I think was well accepted.
This time 60th birthday of Stewart Raby  was celebrated. I have shown
my students a bit of Warsaw.

\subsubsection{Planck 2008}
This time the Planck meeting took place in Barcelona and I opened again the conference.
Among my PhD students was also Stefania Gori who participated in the conference
and gave a talk. This was the third time I was in Barcelona. First time with
Peter Hasenfratz that I described in Section~\ref{MPI}.
Second time  at FlavioNet Conference in 2006 that I cannot resist to describe
here.

This was the Inaugural Workshop of the European Flavour Physics Network Flavianet. It took place in Barcelona 2-4 November 2006. My great friends Monika Blanke and Fulvia de Fazio took part in this conference and we have  a joined photo proving this. But the highlight was probably a duet talk given by Monika and
me which I started, Monika followed and in the summary we listed the main points
together, interchangeably and finishing in a choir. The talk was very well accepted. It summarized our studies of FCNC processes in the Littlest Higgs Model with
T-parity and appeared in Acta Physica Polonica \cite{Blanke:2007ww}.

\subsubsection{Planck 2011}
The conference took place in Lisbon, this was the second time I was there.
The first one in 1997 at CPMASS97, which I will describe briefly in Section~\ref{Deus}. On June 3rd 2011 I gave again the summary talk with the title
``Hunting New Animalcula with Flavour'' describing our papers from
2008-2011 that I described in the main text. At this conference the 60th birthday of Hans Peter Nilles
was celebrated and I have shown some of his photos from the years 1988-1997
during which he was professor in our Physics Department at the TUM.
He moved to Bonn University subsequently. In my talk Supersymmetry
played an important role. This was still before Higgs has been discovered
in 2012. In 2011 several of my experimental colleagues were sure that SUSY
will be discovered before the Higgs. After 2012 the excitement about SUSY
decreased and in 2026 this direction does not belong to the most popular
ones. But this could change if LHC would discover some supersymmetric
particles in the coming decade. Then our detailed analyses could again be
useful.

\subsubsection{Planck 2012}
This time the conference took place in Warsaw again.  I stayed in the Metropol
hotel at the crossing of Marszalkowska street and Aleje Jerozolimskie. The
conference took place in  the Porczynsky Art Gallery and there was a direct connection by metro from
the hotel to the conference venue. The Gallery is located within a walking distance from the Old Town, which was useful.

My talk was one of the
last. I think Guido Altarelli was closing the meeting. I have sent my talk to the organizers one day before but when I was asked to give my talk they did not have it. Fortunately I had it on a USB-Stick and everything went well except
my final comment: ``I am looking forward to Planck 2022 when we will celebrate
80th birthday of Stefan Pokorski''. Nobody knew that Stefan just turned 70 and
surely he did not want to celebrate it at this meeting. Stefan was really angry
with me for a few hours after my talk because several participants congratulated
him for his birthday. Probably this was the reason why Planck 2022 did not
take place in Warsaw but first in 2023.

\subsubsection{Planck 2017}
The meeting took place again in Warsaw but this time in the new Warsaw University campus.
I stayed in the Reduta Hotel. It was a sad meeting because of a sudden death
of Maria Krawczyk with whom I studied physics in the 1960s. It was a heart attack.

In fact I met her one month earlier at the Portoroz meeting in Slovenia.
We stayed at the same hotel but probably already then she had some problems.
She was coming very late to the breakfasts. Eventually Planck 2017  conference was dedicated to her. She was a good physicist and a very nice person. 

I gave a plenary  talk on $Z$-mediated New Physics and Vector-Like Models.
I have also included some discussion of $Z^\prime$ models. This
talk was dominantly based on three papers: two papers with Bobeth, Celis and Jung on vector-like quark models \cite{Bobeth:2016llm,Bobeth:2017xry} and
an earlier solo paper by me on FCNC mediated by $Z^\prime$ and $Z$ gauge bosons
\cite{Buras:2015jaq}. I discussed rare Kaon decays, the ratio $\epe$  and of course $\varepsilon_K$ and the mass difference 
$\Delta M_K$. While my paper was very simple, the other two involved detailed
numerical analysis in the context of the SMEFT. I expect these papers to play
some role when the data improves and lattice made better calculations of $\Delta M_K$ and in particular of the ratio $\epe$. There is no write-up of this talk
but one can find my transparencies in the internet. In any case I described
these papers in the main part of this write-up.

\subsubsection{Planck 2023}
This time the conference took place again on the new University Campus of the Warsaw University but not
 in the new building of the Physics Department at Pasteur street, where  I was giving there talks before. First time in 2019 in the context of a collaboration of particle physicists at TUM and the Warsaw group.
Second occasion was the 100th Anniversary of the Physics Department in 2021 when I was asked to praise the contributions of this Department to particle physics.

In 2023 the Planck conference took part in a different building with a bigger lecture room. I talked dominantly about my work with Elena Venturini with some addition on papers with Jason Aebischer and Jacky Kumar.

At this meeting I met also a very nice Spanish postdoc from Ibarra's group: Mar
Ciscar. After returning to Munich we had two very interesting discussions. It was
a real pleasure to talk to her. She was always smiling and this was good in these
difficult times. Unfortunately similar to Jennifer and Elena she left research.
Really a pity.

\subsection{Portoroz, 2017, 2019, 2023}
The conferences in Portoroz are arranged by Ljubljana University, in particular by my friend Svjetlana Fajfer. They usually take place in the first two weeks of April. I participated in these meetings three times. In
2017 and 2019 the trip was very simple. There was a 45 min flight from Munich to Trieste
followed by a bus trip organized by the conference. In 2023 this flight was cancelled and flying to Trieste one would have to change the plane twice resulting in a 3 hour trip. In 2023 I flew therefore to Ljubljana from which I was brought
to Portoroz by car with the driver being Svjetlana herself.

These meetings were generally interesting with several flavour experts taking
part but there are no proceedings so that let me just summarize briefly what
I talked about.

{\bf 2017}

The title of my talk was ``Z-Mediated New Physics and Vector-Like Quark Models''. It was based on \cite{Bobeth:2016llm,Bobeth:2017xry}.

{\bf 2019}

The title of my talk was ``$\epe$: 2019''. It was based on five papers
\cite{Buras:2018evv,Buras:2018lgu,Aebischer:2018rrz,Aebischer:2018quc,Aebischer:2018csl}. It looks like we were very productive in 2018. In fact I have 12
papers in INSPIRE in this year. Only in 2009 I had more, 14. These days I am
glad when I write only two papers.

{\bf 2023}

The title of my talk was ``Standard Model Predictions for Rare K and B Decays without New Physics Infection and $Z^\prime$ at Work'' It was dominantly based
on  the strategies developed with Elena Venturini to avoid $\vcb$ and $\vub$ uncertainties in predicting rare K and B
branching ratios within the SM \cite{Buras:2021nns,Buras:2022wpw,Buras:2022qip}.
It included also the paper with Jason and Jacky on $Z^\prime$ contributions
to rare K decays and $K^0-\bar K^0$ mixing with purely imaginary $Z^\prime$ coupling to $\bar s d$ which implied no contributions to $\varepsilon_K$
but very interesting contributions to $\epe$, rare Kaon decays and $\Delta M_K$
that are correlated with each other \cite{Aebischer:2023mbz}. See Fig.~\ref{Fig:3}. I also
mentioned my solo paper on the $Z^\prime$-Tandem \cite{Buras:2023xby} which
until today did not find any supporters. Nierste and his student looked at
it but somehow did not write up any paper.

Svjetlana invited me to the 2025 edition but I decided not to go because
in my view I did not have much new to talk about. I could in principle talk
about \cite{Buras:2024ewl} and \cite{Buras:2024mnq} but being fully occupied with the SMEFT review \cite{Aebischer:2025qhh}
I simply did not have time and energy to prepare still
another talk and to travel to Portoroz. As I already described in the course of this write-up
both  papers took at least two years each but in retrospect they turned out
to be more valuable than if they had been finished within six months or even a year.

I expect that \cite{Buras:2024ewl} and \cite{Buras:2024mnq} not presented
at Portoroz 2025 and \cite{Buras:2021nns,Buras:2022wpw,Buras:2022qip,Aebischer:2023mbz} presented at Portoroz 2023 could play an important role when the data on the involved branching ratios like $\kpn$, $\klpn$, $B\to K(K^*)\mu^+\mu^-$, $B\to K(K^*)\nu\bar\nu$,
$B_s\to\mu^+\mu^-$ 
and the missing energy distributions improves.
I stressed it in my contribution to Bari 2024 proceedings \cite{Buras:2024per}.

\subsection{Talks in YouTube}\label{YouTube}
Some of my talks that were not published in proceedings appeared in YouTube.
In fact I discovered some of them only recently. Let me list them for completeness.
\begin{itemize}
\item
  ``Flavour Expedition to the Zeptouniverse'': talk given in Breslau at the 44th Annual Meeting of the Polish Physical Society  in September 2017.
\item
  ``Flavour Expedition to the Zeptouniverse'': Colloquium given in 2021 in the Department of Physics IIT Guwahati, India. I gave this talk sitting in my office at TUM-IAS. It was my first Zoom-Talk.
\item
 ``News from the Flavour Expedition to the Zeptouniverse'': the  talk related to the Max Planck Medal 2020
  given in Dresden at the 86th Annual Meeting of the German Physical Society  in March 2023.
\end{itemize}

I will next describe certain events that happened dominantly during various schools and
other meetings, but also few that had an impact on my life. I hope that some of them are not only interesting but also amusing. In fact several of them are not
known even to my best friends in physics.

\section{More Memories}\label{Memo}
\subsection{First Steps}
I was born on 26th  of October  1946 in Warsaw as a son of Hanna Kochanowicz Buras and Bronislaw Buras. My parents told me much later that in the morning of the 26th of October my mother after eating several cakes had to go rapidly to the hospital and I arrived at 14.00 that day. This morning event of that day had a significant impact on my approach to sweets in particular in the last sixty years during which I could decide by myself on the amount of sweets I was eating.

In 1946 Warsaw was, one year after the end of World War II, a big ruin. My parents lived in an appartment in the first floor at Zajecza street 7, in the walking
distance from the river Vistula. It was a five storeys building which before
the war belonged to my great grandfather Sroczynski.
It was partly destroyed and this was also the case of several buildings
around it. In  the early 1948 when I started to sense what is happening around me, I could see all these ruins but I guess I did not particularly wonder
why these buildings were in such a bad shape. In fact it took many years
until this part of Warsaw was rebuilt and in the first years of my life
I played in such surroundings.
My main companions in this period of my life was my brother Jacek Stanislaw Buras born one year before me and my sister Malgorzata Maria Buras
born in 1949. In 2026, this part of Warsaw looks quite different from how it did in those years. It is now one of the most modern areas of the city.

I was raised as a catholic although my father was an ateist and my mother,
although catholic, never went to church. She told me that the god is everywhere
and for her there is no need to go to church. One can equally well pray
at home or even better in the woods. She also was very negative about the
saints although I think she tolerated the pope. Yet, I think mainly because
of her father, who tested it every Sunday, Jacek, Malgosia and me were supposed
to go to church. But  my brother and in particular my sister were not particularly keen in
getting up early on Sundays and going to church and  I was going there alone. Looking back the main reason was in fact not a strong belief in the existence of god but rather the music which was played during the church services. I realized it many years later and this will be clear at the end of this write-up.

But whatever the reason was, I was going regularly to the nearby church  on Sundays and sometimes even during the week. The priests of that church must have noticed a boy coming regularly
every Sunday to church services and one of them invited me one day
to his home telling me how great it is to be a priest. I do not know whether
he had some bad plans in connection with me but looking back I find
it ridiculos that he talked to me about it. I was seven at that time.
But he must have been very convincing. When I came home I told my mother that
I wanted to be a priest. My mother answered: ``in that case your father will
jump from the bridge''. I loved my father and I did not talk to this priest
again and did not become a priest. When I once told  this story to Harald Fritzsch, my esteemed colleague from the Ludwig-Maximilian University in Munich,
he told me that I became a priest afterall, a priest of the Standard Model
of particle physics. This is indeed true.

Until the fall of 1953 I went to a nursery which was some distance away from my home, at the complex of the Warsaw University at Krakowskie Przedmiescie. My mother would take me there and later bring me home. One day, it must have been March 5th, 1953 we saw on the Tamka street a big procession with
people carrying pictures of Josef Stalin and various transparencies. Some of them were crying and also singing. I asked my mother what it was about and who was the person on the transparencies. My mother did not answer my question but took my hand and we went quickly to the side street. Then to my surprise after what I just saw she said, it is a great day, Josef Stalin died, he was a very bad man.

In the fall of 1953 I started going to school, first in the Nowy Swiat street
which again  was some distance away from my home and my mother would take me there and later bring me home. This turned out to be too complicated
for her and soon I was going to the nearby school on Dobra street, five minutes
by foot from my home. The school was apparently not as good as the one in
Nowy Swiat street but eventually I doubt this difference  had much impact on my life. What is however amusing that in this short period in the first school I
was in a class with Maria Krawczyk and Igor Szczyrba whom I met eleven years
later during my physics studies at the University.

An important impact on my life was the choice of vacations by my parents.
During my years in Poland from October 1946 til May 1971, when I emigrated to Denmark, with a break from
February 1959 til June 1962, where I was in Vienna, I spent $90\%$ of my vacations, both Summer and Winter, in High Tatra mountains.
Consequently my great love to the mountains rather than to the sea began already in these early years. We spent many summer
vacations in Zakopane and went there several times for Christmas. My father
was physics professor at the Warsaw University but for some reasons we
stayed always in a home for artists. This continued for many years,
allowing me to become acquainted with a completely different
community. A well-known painter
Wlodzimierz Jan Zakrzewski (1946-2025) was one of my closest  friends at that time.
My brother and I spent many days and nights playing bridge with him and his
girlfriend.

\subsection{July 14th (1967)}\label{July14}
As my love to the mountains and hiking  gradually increased, I decided in the spring of 1967 to become a member of Warsaw's Mountaineeiring Club. It was rather easy to
become a member. I had to answer a few questions about High Tatras, which I mastered sufficiently well, and contribute to the club with a rather small amount
of money. This entitled me to a place in the base camp near Lake Morskie Oko, in the vicinity of the highest peaks of the Polish High Tatras.

At that time, I was not very experienced in rope climbing, but I assumed that I would learn it quickly there. In fact, I first went for a weekend to the rocks near Cracow to practice a little. However, this turned out to be insufficient preparation for climbs significantly more demanding than ordinary hiking, even on the most difficult tourist routes, such as Orla Perć near Hala Gasienicowa. I had already completed all these routes in previous years and was eager to advance further.

I arrived at the base camp at Morskie Oko with a tent at the end of June, planning to stay for six weeks. It was an interesting experience to meet several climbers who were already well known at the time, some of whom had climbed in the Himalayas and the Alps.

After completing several easy routes with a few physicists who were also staying there, I decided one day—July 14th—to go on a solo hike in a region where I did not expect to meet anyone. My goal was to climb Cubryna (2376 m), but from the Slovak side, which meant crossing the Polish border. This was clearly illegal in those days unless one had made prior arrangements and possessed a special permit (przepustka). I did not have such a permit, but I assumed that I would not encounter anyone during the entire day and that no one would find out. This assumption turned out to be correct—I did not meet a single person.

I had chosen July 14th not because of the French Revolution, but because I did not want to climb on July 13th, which might have brought bad luck. As it turned out, this precaution was of no help.

I crossed the Polish border through a narrow pass called Wrota Chałubińskiego (2022 m) around 10 a.m., took a short rest at a lake, and then began a climb that initially did not seem particularly difficult. Eventually, however, I had to traverse snow. By then it was already approaching noon, the sun was shining, and the snow was beginning to melt. I did not have a rope with me, as studying the route beforehand had convinced me it would be unnecessary—and in any case, it would not have helped much.

At one point I slipped, but I managed to grab a rock immediately, preventing a fall of roughly 200 meters, which would almost certainly have resulted in death or a very serious injury. I was physically strong in those days, and eventually I succeeded in moving out of the snow and onto a small rocky ledge, no more than about $3 m^2$  in size, where I could finally sit down.

While everything I had done up to that point was totally stupid and irresponsible, my next decision was the right one. I gave up the attempt to climb Cubryna, even though there was still enough time to do so and return to the base camp before darkness. The real problem now was how to get down.

The wall on which I was sitting was too steep to attempt any descent, at least with my skills in climbing. Moreover, there was no one around, and since I had crossed the border illegally, I did not want to shout for help—assuming, in any case, that nobody would hear me. After some time, I noticed that if I could somehow reach a small platform three or four meters below the ledge on which I was sitting, I might be able to descend further, partly on snow and partly on rock.

The only way to reach this lower platform, however, was to jump, hoping to land there without breaking my legs. At that time, I was well trained in sports and confident that I could make the jump without injury—provided I landed on the platform itself, which was no larger than the one on which I was already sitting.

I spent an hour preparing for that jump, promising myself that in the future I would give up real climbing and confine my mountain activities to hiking. Eventually, I jumped—and this writing is proof that I survived.

The next steps were by no means easy. I injured my left hand, but not seriously enough to require a doctor. I returned to the base camp only to learn that one of my female physics colleagues had died a few hours earlier while climbing with two less experienced companions. Poor girl. I was truly shocked. The following day, someone else broke a leg while climbing Mnich, another mountain in the region.

Despite these events, I stayed at the camp for another two weeks, hiking around Morskie Oko. At the end of July, I returned to Warsaw. While my love of the mountains remained undiminished, from that point on—and for the next fifty-five years—I limited myself to hiking on tourist routes, not only in the High Tatras but later also in the Alps, Norway, and the Rocky Mountains in Colorado and British Columbia.

Several years later, I began climbing other summits, described in \cite{Buras:2011we}, which were safer than the ordinary ones. I found these endeavors far more successful—and far less dangerous—than my attempted climb of Cubryna.

\subsection{Meeting Leopold Infeld (1967)}

Albert Einstein became my hero when I began my physics studies in Warsaw in 1964. Einstein had died in 1955 in Princeton, so I never met him, although I was born in 1946. Nevertheless, I remember that on April 18, 1955, my mother showed me a photograph of an old man with long white hair and told me that a famous physicist had died. I was eight and a half years old at the time and did not fully appreciate the significance of the event.

By 1967, Albert Einstein had already been dead for twelve years, but his collaborator Leopold Infeld was still alive, although in poor health. He was a professor in the Physics Department of the University of Warsaw at Hoża 69. I saw him several times at seminars and colloquia, but I did not speak to him until one day in December 1967.

On that day my father, Bronisław Buras, a professor of experimental nuclear physics in the Physics Department, gave me some papers and asked me to deliver them to Infeld, who lived not far from our home. When I entered his room, he was lying in bed and looked unwell. He appeared very old to me, although in fact he was only 69 at the time. We spoke briefly. He asked me what I was doing, and I told him that I had decided to become a theoretical physicist. He replied that I should not do so, because theoretical physics was too difficult for me. How could he know?

My father later explained to me that Infeld had a very poor opinion of experimental physicists and, consequently, also of my father. From this he concluded that, as my father’s son, I must be incapable of doing theoretical physics. Of course, I ignored his remark. Unfortunately, I never had the opportunity to demonstrate to him that I was able to do theoretical physics. He died one month later. His funeral was a major event, and I took part in it.

\subsection{A Commited Autodidact}
At this point, I should mention that during my physics studies I had serious difficulties taking notes during lectures, even though many of them were excellent—particularly those given by my father in experimental physics, Jerzy Pniewski in nuclear physics, Iwo Białynicki-Birula in statistical physics, and Andrzej Trautman in both non-relativistic and relativistic quantum mechanics. Because of this difficulty, I attended lectures only occasionally, mainly to check the material that the professor had covered in a given session. The obvious exception was my father’s lectures—not only because they were exceptionally clear, but also because he would have been upset if I did not attend.

As a result, a large portion of my physics studies was spent at home, learning from textbooks and attending the obligatory exercises. In retrospect, this turned out to be an excellent strategy, at least in my case. By the end, I often knew more than my colleagues, simply because I usually read the full textbooks, whereas lectures could not cover, for example, all of the material in Landau and Lifshitz’s books. This habit of independent study became invaluable, providing training that greatly benefited my research for decades.
\subsection{Master Thesis in Warsaw (1970)}

In the summer of 1969, I decided to do my Master’s thesis in theoretical particle physics. My supervisor was one of the young professors, Józef Namysłowski, who had previously been in Kraków and had just returned from the USA. Let me briefly describe my Master’s thesis, which I completed in October 1970.

The topic, suggested by Józef, was single-pion production in proton-proton scattering, studied within the context of the generalized Veneziano model combined with some aspects of the Regge model. At that time, the Veneziano model, which provided an analytic formula for the amplitude of elastic proton-proton scattering, was one of the most popular directions in elementary particle physics. The Veneziano formula expressed the amplitude in terms of the Euler beta function, with its arguments being Regge trajectories. This successfully described the scattering process $pp\to pp$.
To describe the process $pp\to pp\pi$,
the model had to be generalized, and hypergeometric functions also came into play.

I had no prior knowledge of this model, and unfortunately, my supervisor’s understanding was quite limited. In fact, he did not explain the model to me and simply gave me a number of papers to read. Fortunately, my training in studying physics books during my undergraduate years allowed me to understand the papers without difficulty. Two weeks later, when I had my first discussion with Józef, it became clear that I already knew more about the model than he did. He was rather impressed and suggested that I present the Veneziano model at a regular weekly seminar of particle and nuclear physicists in the Physics Department just four weeks later.

I was shocked. I even spoke to both my father and my “uncle,” Jerzy Pniewski, a well-known experimental nuclear physicist, hoping they would convince Namysłowski that it was far too early for me to give this talk. But they refused to intervene. I still vividly remember Pniewski entering our apartment on Wilcza Street 62. The first thing he said was that he was looking forward to my talk in four weeks and that Namysłowski must hold a very high opinion of me if he had already suggested that I give the seminar.

It was 1969, and at that time there were no overhead projectors in the Physics Department—something that is hard to imagine today. Therefore, I was expected to give this talk entirely on the blackboard. For the next four weeks, I stayed at home reading numerous papers. There were no textbooks on the Veneziano model, which was only about two years old at the time. I eventually prepared several pages of formulas, each corresponding to a full blackboard. I then memorized all of them\footnote{After all, I was already well trained in memorization from preparing for the written Matura exam (the final high school exam) in Polish, for which I had memorized three essays, including every comma, written by a relative who was a high school teacher. Fortunately, one of these essays turned out to be very similar to the one on the exam.}.

The talk turned out to be a big success—certainly even more successful than the talk I gave a year later when presenting the results of my Master’s thesis.

After this seminar, I began working on my thesis. There is no need to describe the details here, but it eventually led to my first paper, which, amusingly, nobody cited until today and will be cited here for the first time \cite{Buras:1971mj}. At last!

I should emphasize that I am very grateful to Józef for all the support he gave me during my Master’s thesis. Working alone taught me how to work independently, a skill that proved invaluable in my later career.

I should also stress the importance of my discussions with Stefan Pokorski, who was an assistant professor in the Physics Department at that time. These discussions were extremely helpful. While I had very little contact with Józef after I emigrated from Poland in 1971, I met Stefan on many occasions in later years—most notably at the Planck conferences, as described above, and at TUM-IAS, as described below. We also wrote two papers together, one of which, written in 1993, was very well received by the community \cite{Buras:1993xp}. This work is discussed in Section~\ref{bsgamma}.

\subsection{Meeting Roy Glauber (1970)}

Let me now describe an event that was more interesting than my master’s thesis.

In the summer of 1970, the largest conference on high-energy physics (ICHEP) took place in Kiev, and a number of American physicists passed through Warsaw on their way there. One of them was my father’s friend Harry Palevsky from Brookhaven. He visited us in our apartment and was accompanied by a theorist, Roy Glauber, from Harvard.

Somehow my father, being an experimentalist, did not realize that Glauber was already at that time an important figure in theoretical physics. He even considered it somewhat inconvenient that Glauber had joined Palevsky, as he wanted to discuss certain aspects of the political situation in Poland with Palevsky alone. Consequently, he asked me to show Warsaw to Glauber.

I was, of course, delighted, because Glauber was a hero of Namysłowski, and a number of researchers in the Physics Department worked within the framework of the Glauber model. I showed him Łazienki Park and Chopin’s monument there. We then walked along Krakowskie Przedmieście and finally visited the Old Town. Glauber was very pleased with this tour. I met him again later in Copenhagen.

In the spring of 1979, when I visited SLAC and was already known for my work on QCD, he called me and invited me to spend a year at Harvard. He wanted me to teach him QCD. However, I was supposed to spend the following three years at Fermilab near Chicago, and moving for just one year to Boston with two small children (Robert and Karin) would have been too complicated. I therefore declined the invitation. This was clearly the correct decision.

On the other hand, I missed the opportunity to spend a year at Harvard, to meet several famous physicists there, and in particular to teach a future Nobel Prize winner. Glauber received the Nobel Prize in 2005, shortly after his 80th birthday. He died in 2018 at the age of 93.

\begin{figure}[!bt]
 \centering
\includegraphics[width = 0.90\textwidth]{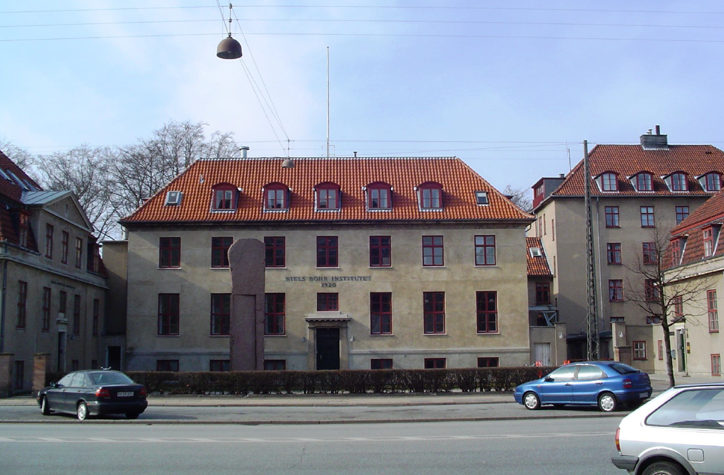}
\caption{Niels Bohr Institute. The second window from the right in the
  attick was where I performed my PhD. Nordita building on the left. During my
  three year fellowship I had an office in the attick of the toller building on the right.
}\label{fig:NBI}~\\[-2mm]\hrule
\end{figure}

\subsection{PhD Thesis in Copenhagen (1972)}
In May 1971 my parents, my sister, and I emigrated to Denmark, where we lived in Roskilde, about 30 km west of Copenhagen.

In July, my father’s colleague from Warsaw, Mieczysław Taube—who had already left Poland in 1968 and held a position at the Paul Scherrer Institute near Zurich—visited us in Roskilde. He arrived by car with his wife and, probably noticing that I was bored, suggested that I could join them on their return trip to Switzerland. He also told me that I could spend two weeks in Bern, staying free of charge in his son’s studio. I found this very exciting, partly because I had never been to Switzerland before.

Although I had visited Austria ten years earlier and seen the Alps—most notably during a drive to the Grossglockner, the highest mountain in Austria—I knew that the mountains in Switzerland were even higher, and the prospect of seeing them was particularly appealing.

Taube’s son not only offered me accommodation in his studio in Bern but also took his father’s car and showed me Interlaken, where I saw the Eiger for the first time. It was also in Interlaken that I encountered pizza for the first time in my life. I stayed in Bern for ten days, exploring the city and also joining organized bus trips to high mountain passes.

One day I even took the train to Geneva and visited CERN. This was possible thanks to Jacek Prentki, my father’s colleague who had left Poland many years earlier and was a member of the CERN theory group. I was deeply impressed by CERN and began to wonder whether I might one day work there myself. This was clearly just a dream at the time, and in July 1971 I certainly did not expect that it would become reality only four years later.

After ten days of being alone in Bern, I decided to return to Denmark. I began preparing for my PhD studies and tried to broaden my knowledge by learning various new topics.

Eventually, I went to the Niels Bohr Institute, shown in Fig.~\ref{fig:NBI}, to speak with Poul Olesen, who was roughly seven years older than me and was supposed to be my supervisor. I was curious about the topic of my PhD thesis, but this meeting brought a major surprise. He told me that he would supervise me, but that I should find the topic of my thesis myself.

This was clearly a challenge, as my knowledge of particle physics was very limited at that time. In particular, I was not aware of what the fashionable directions were beyond the Veneziano model, or more generally the Dual Resonance Model.

During the following two weeks, I spent most of my time in the library. Eventually, I identified a topic that was a natural generalization of my master’s thesis. I presented this idea to Olesen, and he approved it.

In 1971, the particle theory group at the Niels Bohr Institute consisted only of Ziro Koba, Poul Olesen, Holger Bech Nielsen, and a few postdocs. I had a desk in a room some distance away from them and, as a result, had very little contact with the group. Moreover, there were no lectures for PhD students, as I was in fact the first PhD student in particle physics at the Niels Bohr Institute.

At that time, the doctoral degree in Denmark corresponded to the habilitation in Germany, and all scientists who obtained it were over 30 years old.

On the one hand, this was a very challenging time for me, but on the other hand, I had already gained some training in independence during my physics studies in Warsaw, and I further perfected it at the Niels Bohr Institute. Yet another surprise awaited me. In June 1972, Poul Olesen entered my office and asked how I was doing. I presented my results and told him that I would like to publish them. He simply said that if I succeeded in publishing them, my PhD studies would be considered finished. Moreover, he added that it would be good to pass the exam in October of that year, because there was a good chance that I could be offered a three-year fellowship with a salary twice as high as my PhD scholarship.

I was thrilled by this development. I wrote a second paper, which was published, but like my first paper, it has not been cited by anyone to this day \cite{Buras:1972qq}. Two papers, yet 
$h=0$. One can only improve on this.

On October 9th, 1972—roughly one year after I had started my thesis—I passed the PhD exam, about two weeks before my 26th birthday. Strictly speaking, my PhD studies lasted two years, since I had completed my master’s thesis in Warsaw in the fall of 1970, but officially, they were considered to have lasted one year.

I should stress, like Josef Namysłowski, that I am very grateful to Poul for all the support he gave me during my PhD. I learned how to work independently, a skill that proved invaluable later in my career. Moreover, during my three-year fellowship at NBI, he recommended excellent literature that helped me deepen my knowledge of field theory. I still remember how excited he was when asymptotic freedom in QCD was discovered in 1973.

Before continuing my story, let me return to the fall of 1971 and describe in more detail a few events that occurred during the time of my thesis—not my thesis itself, but some memorable experiences around it.

On December 1st, 1971, I moved from Roskilde to a dormitory, Egmont Kollegium, just a ten-minute walk from NBI. As a PhD student, I was allowed to live there. During the first six months, I stayed in a room measuring $9 m^2$, sharing a bathroom and toilet with another student. Later, I moved to a larger room of $12m^2$. My mother was very sorry that I moved to Copenhagen, but commuting daily from Roskilde to NBI would have been too time-consuming. Living in Copenhagen also gave me the opportunity to learn Danish, although most of the students spoke to me in English, as it was a chance for them to practice their English. There were very few foreign students living in the dormitory.

Until then, I had never cooked, and now I had to. I solved this in the following way: for lunch, I ate sandwiches in the NBI cafeteria. For dinner, I had a simple routine—two potatoes, two meatballs, and either two tomatoes or two cucumbers. This became my standard dinner throughout my entire stay in the dormitory, until March 1st, 1973. The other students laughed at me, but I didn’t care. When I later told this story to my diploma students at the Technical University, one of them was very impressed and said he would try the same diet, believing it might bring success in physics. Unfortunately, it did not help him—he left physics after completing his diploma. Of course, on weekends, I visited my parents, and the dinners there were much better.

But the most important event during my time at Egmont Kollegium was meeting my wife, Gurli, which eventually led to three children and, as I mentioned at the beginning of this memoir, six wonderful grandchildren.

\subsection{Schladming Winterschool 1972}

In January 1972, Poul Olesen suggested that I attend the Schladming Winter School to learn about topics beyond my thesis, which focused on the Veneziano model. These were the early years when the future Standard Model of particle physics was beginning to gain attention, although QCD—the current theory of strong interactions—had not yet been formulated, and there were other candidate gauge groups for the electroweak interactions besides $\text{SU(2)}_L \times \text{U(1)}_Y$, which was only firmly established in 1977. The first experimental indications came in 1973 with the discovery of neutral currents at CERN, signaling the existence of the $Z$ boson. At that time, only the three lightest quarks $(u,d,s)$ and four leptons $(e, \mu, \nu_e, \nu_\mu)$ were known.

I was familiar with all of this only at a superficial level, so attending Schladming was very beneficial. In fact, in 1972 it was not really a school but more like a conference, featuring several well-known particle physicists who were, at that time, unfamiliar to me. This brings me to an amusing story that happened on the very first day of the school.

I entered the lecture room and, being nobody, chose a seat in one of the last rows. Next to my seat, someone had already reserved a spot with his conference map. I glanced at the map and realized it belonged to Heinrich Leutwyler. I had no idea who he was, though I had heard the name before, and concluded that he must be an important physicist. I was still puzzled, however, as to why he was sitting in one of the last rows, next to someone like me.

Near the podium, I noticed a group of physicists—mostly students—gathered around someone. Curious, I went over to see who it was. It was Murray Gell-Mann, whom I had seen in photos. But he didn’t impress me as much as a tall, strong man with a beard standing beside him, pushing away any students who came too close. I looked at his badge—it was Leutwyler. Suddenly, everything made sense: he wasn’t just another physicist; he had come from Caltech as Gell-Mann’s bodyguard.

When I shared my discovery with some of my Polish colleagues, who were also there, they couldn’t stop laughing and told me that the “bodyguard” had already authored several important papers. Clearly, I could not know at the time that I would meet him again many times at various conferences in the years to come.

\subsection{Erice 1972}

I learned a great deal during the Schladming School, but the Erice School in July of the same year was even more fascinating. The number of famous physicists lecturing there was larger than at Schladming. Gell-Mann was again present, discussing his attempts to find the theory of strong interactions. The stars of prewar theoretical physics, Wigner and Rabi, were there, along with slightly younger figures like Nambu and Fubini. Several students who would later have illustrious careers in physics also attended, including Bob Jaffe, Hank Thacker, Peter Hasenfratz, and Henry Tye. I shared a room with Sergio Ferrara, another student who made a illustrious career in physics.

This school made me realize that I was far behind the best students, like Jaffe, Thacker, and Tye. But I wasn’t discouraged—after all, they had the opportunity to work at places like Harvard, MIT, and Cornell without the difficulties I had faced in Poland, and my early steps in Denmark had also been challenging. But that is another story.

There were a few amusing events I want to recall. One evening, Rabi’s 80th birthday was celebrated with a fancy dinner in the school building itself. The tables were arranged to form a large quadrangle, so everyone could see everyone else in the room. I had a rather good seat next to Bob Jaffe and not far from the guest of honor, with the stars among the lecturers sitting nearby.

Before dessert, Rabi stood up and gave a rather long speech, offering advice for the future and pointing his finger quite frequently. Suddenly, I noticed that a seat on the other side of the table was empty, and I could hear some noise coming from that direction. It was one of my Danish colleagues, someone who was easily amused. He had gone under the table to hide his laughter at Rabi’s finger, as he later explained to me.

Seeing him already under the table, I struggled not to laugh myself, which would have been disastrous given my exposed position. Fortunately, Rabi finished his talk soon after, and I was immensely relieved.

On another occasion, I happened to sit at lunch with Gell-Mann and his wife, Froissart, Telegdi, and a few other well-known physicists. We were eating pizza when, suddenly, Gell-Mann announced that he couldn’t finish his whole pizza and wanted to share it with someone. I was quick enough, so today I can proudly say that I shared a pizza with the great man.

Gell-Mann then asked me my name. When I told him, he informed me that I was pronouncing it incorrectly. According to him, the correct pronunciation was “Burash.” I explained that this might be true in Hungary, but not in Poland, where I was from, and where “Buras” was actually the nickname of a gray cat. I doubt, however, that I convinced him.

Interestingly, I experienced a similar Gell-Mann-style pronunciation of my name ten years later, when I moved permanently to Munich and the Bavarians called me “Burash” or “Borash.”

The last time I met Gell-Mann was in 2002 at a meeting in Munich celebrating the 60th birthday of my eminent LMU colleague Harald Fritzsch, who had worked with Gell-Mann in the early 1970s on the development of the theory of strong interactions. Many well-known physicists gave talks at this conference, including Heinrich Leutwyler, Bill Bardeen, and Peter Minkowski. Yet the final talk by Gell-Mann was the most impressive, particularly because he did not seem to be in the best of health.

In his talk, he candidly admitted that he and his collaborators, including Harald, had missed the discovery of asymptotic freedom in QCD, and that the credit for this breakthrough rightly belonged to Gross, Politzer, and Wilczek. It is worth noting that in February 2002, these three gentlemen had not yet received the Nobel Prize for asymptotic freedom—it was only in 2004 that they were awarded it.

However, I believe that Fritzsch, Gell-Mann, and Leutwyler also deserved significant recognition—even a Nobel Prize—because they were the first to suggest the non-abelian structure of QCD, which ultimately implied asymptotic freedom \cite{Fritzsch:1973pi}. Fortunately, Gell-Mann had already received the Nobel Prize in 1969 for predicting the $\Omega$ baryon, along with many other honors, and Leutwyler was awarded the J.J. Sakurai Prize in 2023 and the EPS High-Energy Prize in 2025.

In summary, the Erice School of 1972 was a remarkable experience. It left me highly motivated to one day catch up with the Harvard, Princeton, and MIT students who were far ahead of me in publications, citations, and overall experience. I returned to Copenhagen, and on October 9th, I successfully passed my PhD examination. Shortly afterward, I began a three-year fellowship at the Niels Bohr Institute.

\subsection{My First Collaborator: Ziro Koba (1973)}

In those years, the study of multiparticle distributions in proton–proton collisions was rather fashionable. These distributions generally depend on the number 
n of particles produced in a collision. In the fall of 1972, shortly after I had completed my PhD at the Niels Bohr Institute in Copenhagen, Koba, Nielsen, and Olesen (KNO) \cite{Koba:1972ng} proposed that properly normalized multiparticle distributions should, at sufficiently high collision energies, exhibit scaling behavior. Specifically, they suggested that such distributions should depend only on a single variable
$x=n/\langle n\rangle$ where $\langle n\rangle$
denotes the average number of particles produced. While 
 $\langle n\rangle$
clearly increases with energy, the product of the distribution and  $\langle n\rangle$ would depend only on $x$.

Similar scaling behavior had already been observed at SLAC in deep-inelastic electron–proton scattering, and a number of quite different scaling phenomena were known earlier in condensed-matter physics, particularly in the context of phase transitions.

KNO scaling was a very general and appealing idea, and Koba suggested to me that we construct a simple model for the corresponding scaling function. We wrote a short paper together, and in this way he became my first scientific collaborator \cite{Buras:1973jj}. The paper was noticed by the community, our collaboration worked very well, and we planned to continue it.

Unfortunately, Koba died suddenly in September 1973, just two weeks after a multiparticle conference in Pavia that was attended by him, Olesen, Nielsen, and myself. It later turned out that he was suffering from tuberculosis and, fearing cholera—which at that time was considered dangerous mainly in southern Italy, while Pavia is close to Milan—he vaccinated himself against it. The combination of tuberculosis and the cholera vaccination proved fatal.
 
\begin{figure}
\centering
\includegraphics[width = 0.90\textwidth]{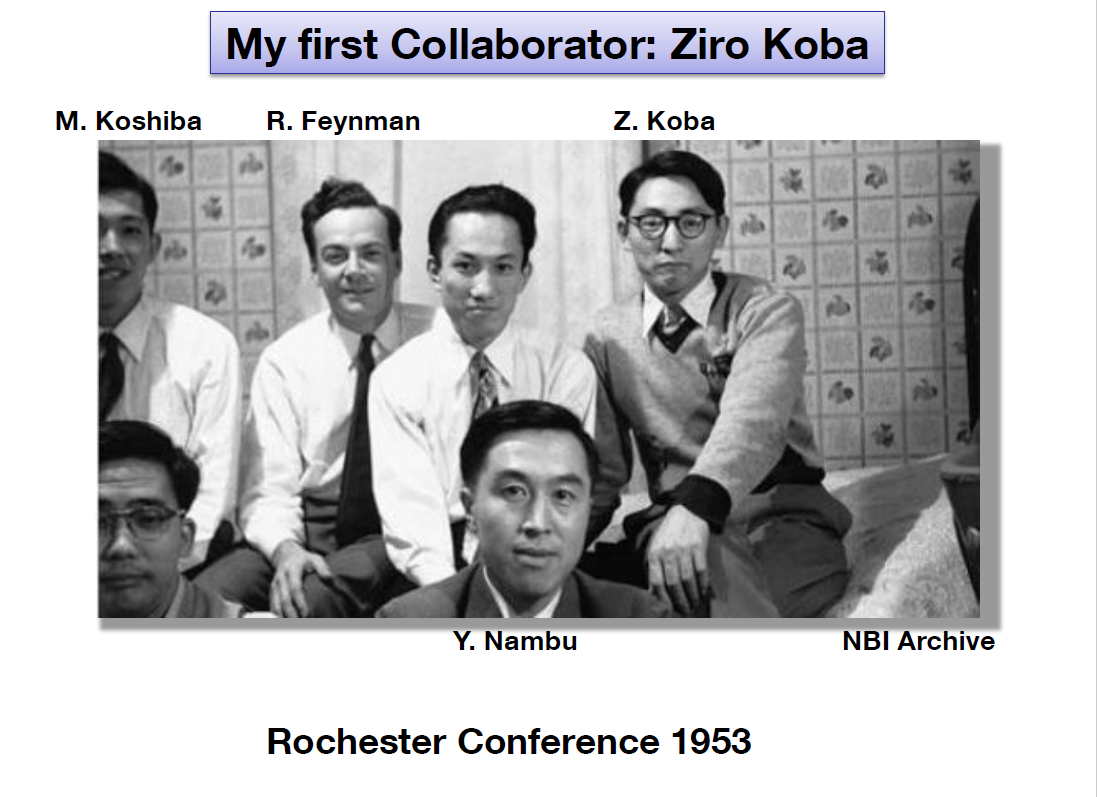}
 \caption{\it Ziro Koba together with a number of fameous physicists.}\label{Koba}
\end{figure}

I have to admit that when I was working with Koba in 1973, I did not know that in Japan he was a member of Tomonaga’s group. As is well known, Tomonaga received the Nobel Prize in Physics in 1957 together with Feynman and Schwinger for the development of quantum electrodynamics. He was the leading figure in quantum field theory in Japan and one of the leading figures worldwide. Yoichiro Nambu, who was awarded the Nobel Prize in Physics in 2008, held Koba in very high regard. In one of his writings, he states:

 ``Among those people around me, I was particularly influenced by Ziro Koba.
From him I absorbed the Tomonaga theory''

In Fig.~\ref{Koba}, one can see Koba together with a number of famous physicists, including Feynman, Koshiba, and Nambu—all three Nobel Prize laureates. Masatoshi Koshiba received the prize in 2002, Nambu in 2008, and Feynman already in 1957, four years after this photograph was taken.

Koba, however, was a committed communist. He spent some time in prison in Japan and subsequently suffered from tuberculosis. He later left Japan and, after spending one year in Warsaw, moved to Denmark a few years before I arrived there. In fact, I briefly met him in Warsaw in 1969.

As for Nambu, I met him for the first time in Chicago in 1978, when I gave a colloquium on QCD effects in deep-inelastic scattering. I had a short discussion with him afterward. In 1985, he received the Max Planck Medal of the German Physical Society, and I met him again at the award ceremony during the Society’s annual meeting. Our discussions, however, were always very brief—although still somewhat longer than the one I had with Paul Dirac, which I describe in Section~\ref{Dirac}.

\subsection{My Second Collaborator: Jorge Dias de Deus (1973)}\label{Deus}
Koba’s death was a shock for me. Fortunately, in the fall of 1972 a new postdoctoral fellow arrived at the Niels Bohr Institute, Jorge Dias de Deus, who became my second scientific collaborator in the fall of 1973. Our most important joint work concerned what later became known as geometrical scaling in elastic proton–proton scattering. This idea was very much in the spirit of KNO scaling and related scaling phenomena.

The differential branching ratio $d\sigma/dt$ for proton-proton elastic scattering depends usually on two variables,
 $t$ and $s$ but again properly normalized was predicted in this framework to depend on a single
 dimensionless variable \cite{DiasDeDeus:1973lde,Buras:1973km}.
The main idea was due to Jorge, and our joint paper explored its consequences. I considered it a nice piece of work, but in 1973 I did not expect the physics community to pay much attention to it. As a result, when I attended an ICHEP conference for the first time, in London in 1974, I did not submit an abstract to present it in a parallel session.

To my great surprise and joy, our paper was mentioned by someone in one of the parallel sessions, and—more importantly—Vernon Barger devoted nearly one third of his plenary talk to discussing the implications of our proposal.\footnote{In fact, even 50 years later our paper still plays a role in studies of elastic proton–proton scattering \cite{Praszalowicz:2025twy}.}

Unfortunately, I had essentially no contact with Jorge after he left NBI in the fall of 1974. I met him only once more before he passed away in 2021 at the 16th Autumn School and Workshop on Fermion Masses, Mixing, and CP Violation (South European Schools on Elementary Particle Physics, CPMASS 97) in Lisbon, held from 6–15 October 1997. There exists a photograph of me with him and Gustavo Branco taken at the reception of this meeting.

At this same meeting, I also met Gabriela Barenboim, whose uncle, Daniel Barenboim, is a famous pianist and conductor. We recently had a very pleasant email exchange, and I still recall being impressed by Lisbon. I also remember a long walk with Luca Silvestrini, who attended the meeting as well. This was just the beginning of our collaboration and friendship.

Going back to 1974, after returning to Copenhagen from London and motivated by the success of geometrical scaling, I decided to derive it from Reggeon Calculus—a field theory in two dimensions, formulated in impact-parameter space. I worked on this project until the summer of 1975, but ultimately I did not succeed. However, the effort was far from wasted. In the process, I learned a great deal about the physics of phase transitions, critical exponents, and, most importantly, the renormalization group.

This turned out to be a significant step in my research. On one hand, it prepared me to give courses in statistical physics at TUM, which were typically taught by condensed-matter physicists. On the other hand—more importantly—it provided me with a solid grasp of renormalization-group techniques, which I applied to QCD a year later and continued to use for the next fifty years, as described in this autobiography, first within the Standard Model and later in its various extensions, particularly in the context of the SMEFT. My long-standing commitment as an autodidact finally paid off.

But 1974 was important for other reasons as well.

\subsection{1974: An Important Year}\label{1974}
1974 was a very important year for both particle physics and for me personally. In November, the charm quark was discovered—a major milestone in the field—and I received my first postdoctoral offer, from none other than CERN.

\subsubsection{Charm Discovery}
The particle-physics community was electrified by the discovery of the charm quark, and a flood of papers—of vastly differing quality—was published. Being immersed in Reggeon Calculus, I did not write a single paper on charm until 1976. It was never my style to jump on the “bandwagon” and follow the crowd. Nevertheless, I followed the developments closely, and many important papers appeared during this period. The same cannot necessarily be said for other occasions, such as the so-called “high-y anomaly” I discuss later, or the purported observation of a $750\gev$ resonance at CERN some forty years later.

Unfortunately, similar caution would have been warranted for the violation of lepton-flavor universality reported by the LHCb experiment in 2013, which disappeared in December 2022, and for the 
$(g-2)_\mu$ anomaly, which vanished last year. I did not work on any of these topics, as they were too crowded with competition.

My strategy was often to explore directions that others considered premature. This approach had the advantage of avoiding the need to rush, as there was little competition, but it also required strong motivation and confidence in one’s ideas—often in the face of intense criticism from the community. The development of Dual QCD (DQCD) is one example. Similarly, the BV strategy, described in Part~\ref{BV}, has not always been fully appreciated, especially by global fitters.

\subsubsection{Getting the CERN Fellowship}
My fellowship at the Niels Bohr Institute continued until September 1975. Unlike many young postdocs today, who apply to twenty or more institutions, I applied in September 1974 only for a fellowship in the CERN Theory Group—and I got it! I was very happy, though a bit surprised, because I did not have many publications at the time, although our paper on geometrical scaling had gained attention following ICHEP 1974. Probably, it was a combination of several factors. Jacek Prentki, a staff member in the CERN Theory Group, knew me from my visit to CERN in 1971, and I met John Ellis for the first time when he visited the Niels Bohr Institute in October 1974. Moreover, there were only two applicants from Denmark. Although being Danish was not a requirement, I was chosen despite not having Danish citizenship at the time—I was stateless. I am very grateful to my colleagues at the Niels Bohr Institute, particularly to Poul Olesen, my PhD advisor, for their strong support during this stateless period.

On the whole, my stay at NBI was extremely stimulating, also because of the many distinguished physicists who visited. In 1973, Richard Feynman received the Niels Bohr International Gold Medal from the Danish Queen. I also briefly spoke with Hans Bethe (1906–2005)\footnote{In 2009, when visiting Cornell, I got his office.} and met Bill Bardeen, who was accompanying his father, John Bardeen (1908–1991), to Stockholm, where John received the 1972 Nobel Prize in Physics. I do not recall the exact topics Bill discussed, but probably not the Triangle Anomalies with Steve Adler \cite{Adler:1969er}, which were already three years old by then. I also met Christian M\"oller (1904–1980), known for M\"oller scattering, Jerry Brown, and several other notable nuclear physicists.

\boldmath
\subsection{$e^+e^-\to W^+W^-$ (1976)}\label{Alles}
\unboldmath
In the early summer of 1976 the  LEP 200 workshop has been organized at CERN
with the goal of investigating the physics of an $e^+ e^-$ collider with the
energy of $200 \gev$. One of the important key questions was whether LEP 100
would be able to discover $W^\pm$ gauge boson in the process $e^+ e^-\to W^+W^-$
with one $W$ being off-shell, that is, only entering virtually  so that
$100\gev$ would be sufficient for producing the second $W$.

W. Alles, C. Boyer
and I were assigned to perform this calculation. Boyer, a PhD student of Mary Kay
Gaillard and I  first calculated the process
$e^+ e^-\to W^+W^-$ with both $W$ bosons on-shell, which could only be studied
at LEP 200. It took us only a few days and we initially thought we were the first to
do it, since previous papers had been quite rough, focusing
primarily on demonstrating the cancellation of $s$ and $s^2$
terms present in the three separate diagrams contributing to this process, as required by unitarity.

Soon after we completed our calculation, someone informed us that Kriplovitch and his collaborators had already published the result for this cross-section a year earlier \cite{Flambaum:1974wp}. This came as a bit of a shock. I vividly remember sitting in the living room of our apartment in Meyrin, nervously comparing our result with the Russian one. To my relief, our results agreed. Moreover, we calculated differential distributions, which Kriplovitch et al. must have done as well, but they did not present them in their paper.

This was just a warming up for the more difficult calculation of one $W$
being virtual so that one $W$ and two leptons were in the final state.
This calculation can certainly be done by hand but Alles suggested
that we use SCHOONSHIP, a computer program developed by Veltman for
analytic calculations. He has given us an example of a calculation from
which we were supposed to learn SCHOONSHIP, explained to us a few things
and left CERN for Bologna where he had permanent position. Being older
than Boyer I was leading this project, working even a few nights at CERN.
Boyer was doing very well and we finally completed the calculation, I think
in a month. Unfortunately, I worked probably too much and he got so depressed
that he left CERN and I never met him again. It is a pity because he was
a very nice person.

Our result implied that the cross-section for this process being around
$10^{-26} cm^2$ was simply too small to be measured so that LEP 200
was necessary to discover this process and in fact it was observed at LEP 200
in 1993. John Ellis used to say that because of this calculation
LEP 200 was built. He urged me to make a regular paper out of it.
But Boyer disappeared and Alles was in Bologna and could not
contribute to the write-up because he was not doing these calculations.
Therefore, I had to do it alone. I have to admit that I did it with joy.

After completing the draft, which John Ellis found to be good, the question arose as to whether Alles deserved to be a coauthor of the paper. John was not entirely sure and suggested that I call Alles to clarify the matter. I did so, and Alles told me that he really did not deserve to be on the paper. I was relieved. However, five minutes later he called me again and said that he wanted to sign after all. In the end, I became the last author on the paper, despite having done much more work than my collaborators \cite{Alles:1976qv}.

But this is not the end of the story.

The paper was typed by one of the secretaries, and I had to insert the equations by hand—Latex did not exist at that time. I submitted the manuscript to Nuclear Physics B and waited for acceptance. In the meantime, Dimitri Nanopoulos visited CERN and, to my surprise, asked me whether everything in the paper was correct. I told him that, unfortunately, while writing many equations by hand, I had introduced three misprints, which I planned to correct as soon as I received a positive response from Nuclear Physics B. He then asked me what these misprints were.

To my great surprise, two weeks later a response from Nuclear Physics B arrived, signed by the chief editor, Hector Rubinstein. Our paper was rejected, with the argument that, when this process would eventually be measured, people would construct much simpler models than the Standard Model, and the result would therefore be much simpler. Moreover, the referee pointed out three errors in our result—precisely the three misprints I had mentioned earlier.

I never asked Dimitri whether he was the referee who identified these three errors, partly because I found the situation rather amusing. I was less amused, however, by the rejection itself and by the ridiculous statement about the Standard Model made by the chief editor of Nuclear Physics B. This statement certainly could not have come from Dimitri.

I showed the rejection letter to Mary Kay, who then contacted Tiny Veltman, a member of the editorial board of Nuclear Physics B. One week later, I received another letter from the journal. I was informed that, following the intervention of my colleagues, our paper would be accepted, provided that the three errors identified by the referee were corrected.

\subsection{Les Houches (1976)}

In July and August 1976 I took part in the well-known Les Houches Summer School. The school lasted six weeks—perhaps a bit too long—but I felt it was a great opportunity to learn a lot. Gurli did not really protest, partly because she had frequent contact with other wives of theorists working at CERN. Still, for the first two weekends I drove from Les Houches to Meyrin to spend time with Gurli and Robert.

For the remaining four weeks of the school, Gurli went with Robert to Denmark to stay with her parents. Robert was still very small, only one year old, and when I picked them up at Geneva airport four weeks later, I had the impression that he did not recognize me. I was quite sorry about this, and only twice in the future—in 1998 (CERN) and 1999 (Fermilab)—did I leave my family for a whole month. By then our children were much older, and this was no longer a problem.

The Les Houches school itself was very good, with several well-known physicists teaching modern particle physics. One of them was Richard Feynman. However, he had some health problems during the school, and I must admit that I was not particularly impressed by his lectures—except for his style, which was very different from that of the other lecturers.

I should also mention Leonard Susskind, who impressed me not only as a physicist but also as a runner. One day he, Willy Fischler (a CERN fellow like me), Chris Llewellyn Smith, and I ran together from the school down to the town of Les Houches and back. As the school is situated at a significantly higher altitude than the town, this was clearly a challenge.

At that time I was a good runner, and I knew that both Willy and Chris were very good runners as well. But only Willy could keep up with Susskind. On the way back, Chris and I were far behind them and completely exhausted. In fact, we walked part of the way, and only when we were approaching the school did we start running again—to show the students that we were still alive.

Chris Llewellyn Smith, a well-known theorist and, years later, CERN director, was the director of the school and was accompanied by his son. At that time, I had even wilder hair than I do now. One day, when I met them, Chris pointed at me and said, “He has even more hair than Uncle John.” John, of course, was John Ellis, who also gave a set of lectures.

I liked John’s lectures very much and studied their write-up in detail after returning to CERN, even making suggestions to him. In fact, while reading these lectures, I began to dream about the reviews I might one day write myself. Fortunately, those dreams came true as early as 1979 \cite{Buras:1979yt}.

At Les Houches, I also met Keith Ellis, with whom I shared an apartment, as well as Nathan Isgur and Andrzej Czarnecki, who would later become an excellent collaborator on two of the papers mentioned above.

After returning from Les Houches, I decided to study neutrino-proton and antineutrino-proton deep-inelastic scattering. This was motivated by the so-called “high-y anomaly,” an unexpected increase in the cross-section at high values of the variable $y$.
To tackle this, I had to refresh my knowledge of renormalization group techniques, which I had gained while studying Reggeon calculus in Copenhagen. Some modifications were needed to apply these methods to deep-inelastic scattering, but after a few weeks, I was ready to investigate whether the high-y anomaly could be related to the violation of Bjorken scaling by strong interactions—described by Quantum Chromodynamics, which had been gaining importance after the discovery of asymptotic freedom in 1973 by Gross, Politzer, and Wilczek.

This brings me to the next story.

\subsection{More on Parton Distributions in QCD (1977)}\label{PD77}

As already mentioned at the beginning of my story, to my surprise there was nobody at CERN studying scaling violations in deep-inelastic scattering. The main activities in this field were taking place at Harvard and in Rome, with Altarelli and Parisi developing their equations. In addition, Gl{\"u}ck and Reya in Dortmund were generating numerically the $Q^2$
dependence of parton distributions by performing Mellin transformation of the moments of these distributions.

At that time, essentially all the literature on scaling violations in deep-inelastic scattering—beginning with the 1973 discovery of asymptotic freedom—was formulated in terms of moments of deep-inelastic structure functions. This was because the operator product expansion could be applied directly to these moments. In the absence of the Altarelli–Parisi equations—which had in fact been independently derived by Russian physicists, a fact not known at that time—the only method available for a numerical study of scaling violations was the numerical inversion of moments using Mellin transformation, as employed in particular by Gl{\"u}ck and Reya \cite{Gluck:1976iz,Gluck:1976nm}.

At this point an important lesson emerges: certain deficiencies can turn out to be a stroke of luck. Had I been strong in computer calculations, I would have simply followed the approach of Gl{\"u}ck and Reya and inverted the moments numerically by means of Mellin transformation to obtain the parton distributions. However, this was beyond my computer skills at the time. Fortunately, around this period Feynman and Field \cite{Field:1976ve} published a very long paper presenting analytical formulae for parton distributions without any $Q^2$
 dependence—simple functions of the Bjorken scaling variable $x$.
The simplest of these formulae was the one for the valence quarks, which according to Feynman and Field was 
\be
V(x)= x^{1/2}(1-x)^3.
\ee

Seeing this I got the idea to generalize this expression to
\be
V(x,Q^2)= x^{a(Q^2)}(1-x)^{b(Q^2)},
\ee
fit $a$ and $b$ to data at a given value of $Q^2$ finding 
$a\approx 1/2$ and $b\approx 3$ in agreement with Feynman and Field.
Subsequently knowing that the violation of Bjorken scaling by strong interactions is logarithmic in $Q^2$ I parametrized appropriately  $a$ and $b$ as functions of  $Q^2$
introducing two free parameters. Integrating over $x$ I could then find
the moments of $V(x,Q^2)$:
\be
M(n,Q^2)=\int^1_0 x^n V(x,Q^2)
\ee
and comparing them with the $Q^2$ dependence of  moments predicted by QCD allowed me to
determine the functions $a(Q^2)$ and $b(Q^2)$ and thereby the analytic formula $V(x,Q^2)$.  Inserting it 
into known formulae for deep-inelastic structure functions given in terms of parton distributions allowed me right away to find the $Q^2$ dependence
of the relevant cross-section for deep-inelastic neutrino-proton scattering.

Having these analytic formulae allowed me to demonstrate that the high-y anomaly could not be explained by scaling violations as predicted by QCD. Interestingly, before my analysis appeared as a CERN preprint, Altarelli and collaborators had claimed that this anomaly could be explained by charm-mass effects. On the other hand, a Harvard group led by Howard Georgi and Alvaro de Rújula, making various approximations, found the possible QCD scaling violations to be very small.

My results lay between these two estimates, and I was naturally curious whether there would be any reaction from these two groups, consisting of theorists far more experienced than I was, possibly challenging my analysis. I was therefore truly delighted when Mary Kay Gaillard, returning from a conference in Spain, informed me about a talk by Guido Altarelli in which he stated that my analysis was the best among the three \cite{Buras:1977ra}.

Motivated by this success, I generalized the method to gluon and sea-quark distributions. As this required more extensive numerical work, I carried it out in collaboration with a CERN fellow from Leyden, Karel Gaemers, who was much stronger than I was in numerical calculations \cite{Buras:1977yj,Buras:1977hj}.

Our first paper \cite{Buras:1977yj} soon became very well known. In fact, Chris Sachrajda, returning from the EPS Conference in Stockholm in 1977, told me about it by saying: ``Andrzej, you made it!''. Indeed, for several years our parametrization was referred to as ``the Buras–Gaemers parametrization'' and the paper became one of the most cited in particle physics in the last years of the 1970s. In the 1980s, however, more precise parametrizations were developed, and over the last 25 years it has been cited only about 40 times.

I did not have much contact with Karel for many years, until April 2024, when he called me to congratulate me on receiving the Sakurai Award 2024. We then decided to compensate for this long break in our contact by speaking on WhatsApp every Monday for 15 minutes at 11:00 a.m. We named this time slot Standard Buras–Gaemers Time (SBGT), as the name Buras–Gaemers parametrization has practically disappeared from the literature.

\subsection{Meeting T.D. Lee (1977)}

I already mentioned that during my stay at CERN there were some reservations about QCD and asymptotic freedom. Here is another story illustrating this atmosphere.

In August 1977 I attended the 8th Photon–Lepton Conference in Hamburg. A plenary talk on QCD and the implications of asymptotic freedom was given by Otto Nachtmann. It was a very nice talk, not least because Otto referred to my CERN papers, in particular to those written with Karel Gaemers. Yet not everyone was pleased. The famous T. D. Lee—who, together with C. N. Yang, had received the Nobel Prize for predicting parity violation—was among the participants and was sitting in the first row of the large auditorium. Roughly five minutes after Nachtmann had begun his presentation, T. D. Lee stood up and left the room. He returned only after the talk had finished.

I noticed this at the time, but initially it did not worry me. This changed during the coffee break, when I met Aage Bohr, the son of Niels Bohr and himself a Nobel laureate (together with Ben Mottelson in 1975). Aage Bohr told me that he was looking forward to my two lectures on asymptotic freedom scheduled for September at the Niels Bohr Institute, especially because T. D. Lee would be visiting him at the same time. He expected interesting discussions, since T. D. Lee did not really believe in asymptotic freedom.

This was definitely not good news. We were hoping to return to Copenhagen at some stage, and I had been counting on these two lectures to help me move closer to this goal in the coming years. Being taken apart by T. D. Lee in front of the NBI researchers would have been disastrous. Unsurprisingly, I became rather nervous,  as T. D. Lee was known to be a very tough person. However, while travelling by train from Hamburg to Copenhagen, I unexpectedly saw T. D. Lee and his wife on the ferry from Puttgarden to R{\"o}dby. I had never met T. D. Lee before, and I doubt that he knew who I was, but I decided spontaneously to introduce myself. I told him that I had heard from Aage Bohr that he would be visiting the Niels Bohr Institute and that I would also be there. I do not recall what we talked about afterward, as I did not want to disturb him.

Two days later I began my lecture with T. D. Lee again sitting in the first row. I feared that he would stand up and leave the room after five minutes, but this did not happen. Instead, after about five minutes he interrupted me, arguing—quite forcefully—that it was nonsense to apply asymptotic freedom at values of
 $Q^2$ of a few $\gev^2$  because asymptotic freedom means $Q^2=\infty$.
I replied that these were currently the only experimentally accessible values, and that it was important to investigate what happens at such scales. After all, Bjorken scaling at very low $Q^2$
 had been puzzling for some time, and its logarithmic $Q^2$
violation in the presence of asymptotic freedom could provide the correct explanation of this puzzle. He was not satisfied with this answer, but fortunately the chairman intervened, saying: ``Let him talk.''

It worked—there was no further interruption from T. D. Lee until I finished. He did not even ask a question at the end. However, once everyone had left the auditorium, he came up to me and proceeded to give a 30-minute lecture at the blackboard about his views on the subject. I do not recall exactly what he said; I probably understood little of what he was trying to teach me. After my second lecture, he gave me another 30-minute lecture on his perspective on strong interactions.

Yet, there was a happy ending. At the conclusion of his second lecture, he told me that my lectures, despite his criticism, were much better than Nachtmann’s talk in Hamburg—a somewhat odd comment, considering he had listened to only five minutes of it.

As far as I remember, this was the last time I saw T. D. Lee in person. He was almost exactly twenty years older than me and passed away in August 2024 at the age of 97. His collaborator from the 1950s,
C. N. Yang, was four years older than him and died in October 2025 at the age of 103. It seems that academicians tend to live very long lives—I have observed this also among members of the Bavarian Academy of Sciences. I met C. N. Yang only once, in 1988 at the MPI for Physics, but our conversation was rather brief.

\boldmath
\subsection{The $\overline{\text{MS}}$ Story (1978)}\label{MSbar}
\unboldmath
While only a few CERN fellows were studying the implications of asymptotic freedom for deep-inelastic scattering (DIS), three of them—Manolis Floratos, Douglas Ross, and Chris Sachrajda—decided to push the calculations beyond the leading order (LO) in QCD perturbation theory. Their goal was to generalize the existing renormalization-group-improved LO results to next-to-leading order (NLO).

The most challenging part of this effort was the computation of two-loop anomalous dimensions of the operators relevant for DIS. By early summer 1977, they had successfully completed this calculation \cite{Floratos:1977au}.
Importantly, they emphasized that two-loop anomalous dimensions depend on the choice of renormalization scheme. Consequently, to make physical predictions for structure functions, these anomalous dimensions must be combined consistently with the $\ord(\alpha_s)$ corrections to the Wilson coefficients computed in the same renormalization scheme. This combination ensures that the resulting DIS structure functions are scheme-independent, as required by physics.

After completing their calculation and releasing it as a CERN preprint, their strategy was to quickly apply it to phenomenology before others could. At that time, a calculation of the $\ord(\alpha_s)$ 
corrections to Wilson coefficients had been performed by Alvaro de Rujula, Howard Georgi, and David Politzer (RGP) at Harvard \cite{DeRujula:1976baf}.
There was a risk that RGP could combine their results with those of Floratos, Ross, and Sachrajda, potentially being the first to present a full NLO QCD calculation for DIS.

Since I had already gained experience through my solo paper and two papers with Karel, I was invited to perform a next-to-leading order (NLO) QCD analysis together with Floratos, Ross, and Sachrajda, which we completed by July 1977
\cite{Buras:1977qg}.
At the time, we did not check the Harvard calculation because all three Harvard authors had already produced several significant papers, so we trusted their results. Unfortunately, this turned out to be a mistake.

Only in March 1978, while reproducing the Harvard calculation in the Feynman gauge, Taizo Muta and I found a result that differed from the RGP result, which had been obtained in the Landau gauge. Their paper also had other problems that we did not notice when using it in our earlier work \cite{Buras:1977qg}.
As a result, my only paper with Floratos, Ross, and Sachrajda turned out to contain incorrect results. Let me explain what went wrong in the RGP paper.

By Christmas 1977, I was already at Fermilab and decided to check the Harvard calculation of the $\ord(\alpha_s)$ 
corrections to deep-inelastic structure functions. Taizo Muta, roughly ten years my senior, became interested in joining me. We performed the one-loop calculation using dimensional regularization in the Feynman gauge for the three relevant diagrams and found results that differed from the Harvard calculation, which had been done in the Landau gauge.

We informed Bill Bardeen about our findings. Until then, neither of us had much contact with him, but he became interested in the problem. A few days later, I noticed that he was also calculating these diagrams. After discussing it with Taizo, we suggested that Bill join us, and he agreed.

A week later, Bill  informed us that he had calculated only the gauge-dependent part of the diagrams for an arbitrary gauge. By adding his result to ours and then choosing the Landau gauge, he was able to reproduce the Harvard calculation. The implication was clear: the result was gauge dependent and therefore could not be considered the $\ord(\alpha_s)$
correction to the Wilson coefficient of a gauge-invariant operator. Moreover, the result was sensitive to the external momentum $p$ through $\ln \mu^2/-p^2$  signaling an infrared divergence.

Today, we know that this calculation was performed in the full theory, and the coefficient obtained in our first calculation was not the Wilson coefficient of the relevant operator. It took us at least six weeks to realize that, to obtain the Wilson coefficient, we had to calculate the $\ord(\alpha_s)$
corrections to the matrix elements of the involved operators. These matrix elements were themselves gauge dependent and sensitive to the infrared 
regulator $-p^2$.
 Crucially, the gauge and infrared dependences of the matrix elements exactly canceled those in the original calculation, leaving the Wilson coefficients gauge independent and  $-p^2$
 independent. In other words, these unphysical dependences belonged to the matrix elements, not to the Wilson coefficients.

This procedure, which we called the ``proper calculation of the Wilson coefficients'', was the first such calculation in QCD for any high-energy process. Today, it is recognized as matching the full theory to an effective theory, and it has become the standard method for computing Wilson coefficients in the Standard Model and in new physics models beyond it. Yet, I must admit that at the time, I did not fully appreciate the importance of this result, because new puzzles arose that we did not yet understand.

While the coefficients of the logarithm $\ln (Q^2/\mu^2)$
matched the relevant elements of the one-loop anomalous dimension matrix, the non-logarithmic contributions contained a rather strange combination, 
$\ln 4\pi-\gamma_E$ with $\gamma_E=0.57721$ being Euler–Mascheroni constant. We were convinced that our result was still non-physical.

I decided to call Chris Sachrajda at CERN, reasoning that since he, Douglas, and Manolis had performed the two-loop calculations of the anomalous dimensions of the operators we were studying, he could likely help us. To my surprise, Chris told me that they did not find any $\ln 4\pi-\gamma_E$
terms in their two-loop calculations and was confident that we must have made an error in our one-loop calculation.

By April, I was growing increasingly frustrated with the situation. Despite my successes at CERN, I feared that I might have to leave research altogether. Although my position at Fermilab had been extended for a second year, I still had to apply for new positions in the fall of 1978, and I had not produced any papers at Fermilab up to that point. My father visited in April on his way to a conference in California and encouraged me not to give up, which helped immensely.

My spirits improved significantly in early May, when Gurli informed me that she was expecting a second child. On October 24, Karin was born, a moment far more important to me than any achievement in physics.

In addition, I came across a comment by David Gross in his 1975 Les Houches lectures, which I thought might provide a clue to our difficulties. I no longer remember the precise wording, but on one Friday I informed Bill, Taizo, and Dennis Duke, who had meanwhile joined our team, about what I had found. During a party the next day, Bill told us that my observation was indeed useful and that he understood what was going on.

On the following Monday, we realized that the terms $\ln 4\pi-\gamma_E$
were artifacts of dimensional regularization. They naturally appear in ’t Hooft’s minimal subtraction (MS) scheme, in which only the
the divergence $1/\epsilon$
is subtracted during renormalization. This insight led us to the idea of subtracting these additional terms as well, which effectively amounts to a redefinition of the QCD coupling constant. This was the birth 
of the $\overline{\text{MS}}$ scheme, which, 
incidentally, is as old as our daughter.

Bill was initially reluctant to move away from ’t Hooft’s MS scheme to the 
 $\overline{\text{MS}}$ scheme. However, Taizo and I eventually convinced him, helped decisively by our computer expert Dennis Duke, who found that the NLO QCD corrections were significantly smaller in the $\overline{\text{MS}}$
than in MS. I believe it was also Dennis who, in the very first plots, added a ``bar'' over MS to distinguish our scheme from ’t Hooft’s original one.

The main results of our paper were then:
\begin{itemize}
\item
  The first correct matching of the full theory to the effective theory which
  in the following years has been generalized to other processes and these days
  is central to applications of quantum field theory.
\item
  Realization that the QCD coupling constant is renormalization scheme
  dependent which was a shock for experimentalists who by then tried to
  extract its value from scaling violation in deep-inelastic scattering.
  While I am sure this fact must have been known in the case of QED
  coupling constant, in the case of QCD it was pointed out to my knowledge for the first time
  by us.
\end{itemize}

With all this in hand, I decided to write up the paper, which I managed to complete within a week. I showed the first draft to my collaborators. Dennis and Taizo liked it very much, but Bill did not like it at all. I therefore suggested that the two of us go through the manuscript page by page, so that I could better understand what he objected to.

One afternoon, we met in the theory seminar room and spent several hours discussing the text. Bill had a very different view of how the paper should be written. In those sections where we disagreed on the presentation, I suggested that he rewrites the text himself, but Bill declined. I therefore decided not to change anything in those parts.

There was, however, one important point related to infrared divergences that was missing from my draft, and Bill strongly felt that the paper should include a separate subsection on this issue. I told him that I did not fully understand his argument and would therefore be unable to write it up properly. Bill agreed to write this subsection himself before leaving Fermilab a week later for a two-week trip.

A week later, Bill left Fermilab without giving us his text. I began to worry that we would not finish the paper in time for the conference in Iowa in early July, where I was scheduled to give a talk and present our new results. Bill’s office, however, was not locked, and when I entered it, I found on his desk the text he had promised to write. It was only half a page long and ended with an unfinished equation, terminating with a minus sign. Had it ended with a plus sign, one might have hoped that the remaining terms were negligible—but a minus sign suggested otherwise.

Over the following days, Taizo and I somehow managed to complete this subsection. It eventually became Section 3.4 of our paper, with the equation in question labeled (3.4). Surprisingly, after Bill returned, he introduced only minor corrections to our construction, and the paper was finally completed. Nevertheless, several colleagues who were otherwise very enthusiastic about the paper asked us what we were actually trying to say in that section. I usually replied that they should talk to Bill if they wanted a deeper understanding.

Our paper appeared in June 1978 and eventually became a classic in higher-order QCD calculations, applicable to essentially all processes. If everyone who relied on it had cited it, we would have far more than the 1870 citations it currently has. But after all, the Dirac equation is not cited by anyone, and one should not complain about such things.

Attentive readers may have noticed that so far I have shown only one photograph of physicists (Fig.~\ref{Koba}) and only few photographs will be shown as we proceed. 
In Fig.~\ref{F2005} I show the photo of the $\overline{\text{MS}}$ team
taken in 2005 during the celebration of Bill's 64th birthday at Fermilab.

\begin{figure}
\centering
\includegraphics[width = 0.80\textwidth]{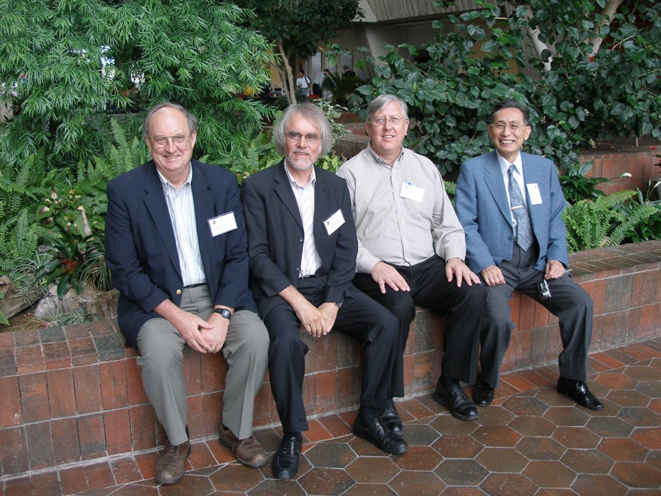}
 \caption{\it The $\overline{\text{MS}}$ team in 2005 at Fermilab. Bill's 64th birthday. }\label{F2005}
\end{figure}

Already in June, I reported on our paper at a workshop in Iowa. My plenary talk was  well received. At this meeting, I met Frank Wilczek and his wife Betsy for the first time. Frank was at Princeton at the time and asked whether I would be interested in applying for a three-year assistant professor position there. I told him that I was certainly interested, but since my wife was expecting our second child, I preferred to make a decision only toward the end of the year. In the fall, however, Fermilab offered me a five-year position, and we decided to stay at Fermilab.

The conference in Iowa was quite interesting, but instead of describing it in detail, let me mention one amusing episode. In a parallel session attended by about ten theorists, among them Herbert Pietschmann from Vienna, a speaker was giving a rather technical talk. Suddenly, Pietschmann interrupted him and asked the audience whether anyone understood what the speaker was talking about, adding that he himself understood nothing at all. The audience was so surprised that no one reacted, which Pietschmann evidently interpreted as confirmation that no one understood it either.

``Then you should interrupt him and ask questions. You should really do it,'' he said—and left the room.

In 2010, I met Pietschmann again in Vienna. As a Schr{\"o}dinger Visiting Professor, I was giving a series of lectures, but this time nothing of that sort happened, because Pietschmann did not attend any of my talks.

\subsection{More about Bill Bardeen}

The other direction I explored with Bill was DQCD, which I 
described in Section~\ref{DQCD} and Part~\ref{RDQCD}. Unfortunately, our 2014
paper, was the last one I have written
with Bill. In the following years Bill was traveling to many Institutions around the world 
giving talks not about DQCD but about his father, John Bardeen, who remains
the only physicist who received  the Physics Nobel Prize twice.
It is  amusing that he carried both Nobel Prize Medals with
him. There is even a photo of Bill giving a seminar at KIT in Karlsruhe in 2016 with
the two medals displayed on a table.

As this is the last time Bill appears prominently  in  my flavour autobiography let
me next share  two anecdotes  I recently got from Jean-Marc:
\begin{itemize}
  \item
    Bill, knife in hand, standing in front of him, waiting for an answer regarding the non-linear sigma-model;
    \item
      Bill asking him the purpose of each of his published articles before writing a recommendation letter for him. These days this would not be necessary as this   could be solved by AI including writing this letter. Still I did not use this method until now.
\end{itemize}
Let me next share two anecdotes from my own experience:

\begin{itemize}
\item
As described above, in the spring of 1978, when we encountered some problems with our Fermilab paper, I became worried about my future in physics, having not written a single paper in six months. One day, while collecting vegetables from Bill’s garden, I shared my concerns with him, hoping for some encouragement. His reaction was brief: ``That’s too bad.'' Yet, a year later, when I was struggling with my first review article, it was Bill’s encouragement that helped me push through and complete it.
\item
Shortly after Bill’s arrival at MPI in 1986, I was finishing a paper that I had first written by hand. It was a Friday, and I mentioned to Bill that I would type it over the weekend.
Bill came to MPI from Fermilab        with a small
computer which looked rather different than present laptops.
Bill told me that I should instead spend the weekend with my family and suggested that he would type the paper for me. I refused, of course—but in retrospect, that was a mistake. Otherwise, I could have thanked the great Bill Bardeen for typing my paper.
\end{itemize}

Bill was a great physicist-truly remarkable. He was rather special but I liked him very much
and we collaborated very well. Moreover, I learned a lot from him.
Unfortunately, on November 18th 2025 Bill passed away, far too early.
I had still an opportunity to highlight his contributions to $\overline{\text{MS}}$ and DQCD at the celebration
of his 50th anniversary at Fermilab, delivering a short comment via Zoom
after two in-person talks by Chris Quigg and
Chris Hill on September 4th 2025. I am grateful to Marcela Carena for informing me about this event and sending me the Zoom link. Bill was apparently very happy about
my comments and I was fortunate to speak with him for 30 minutes via Zoom on September 16th, 2025. He was already very weak, but I had hoped to talk with him again.

Unfortunately, when I tried to organize another Zoom meeting in the first half of November, he was already too weak to talk. I did, however, manage to write a message to him, which was read by Marge Bardeen just two and a half hours before he passed away.

In this note I have stressed that among my 126 collaborators he was the best. I was told by his family that Bill regarded me as his best collaborator, which
he also confirmed in our last conversation. Considering that he worked with several very well known scientists like Murray Gell-Mann and Steve 
Adler this was of course a wonderful complement. Of
course his gauge anomaly paper with Adler \cite{Adler:1969er} had a far greater impact on particle physics than his papers with me.

On the other hand my statement that Bill was my best collaborator can be underlined by the following facts:
\begin{itemize}
\item
  As just described Bill played the crucial role in the $\overline{\text{MS}}$ Story.
\item
Even though Jean-Marc and I carried out the bulk of the calculations in DQCD, Bill was the true inventor of DQCD.
\item
  If one removed my review with Gerhard Buchalla and Markus Lautenbacher \cite{Buchalla:1995vs} from Table~\ref{tab:ranking}, Bill would be number one in this table as far as the number
  of citations is concerned.
\end{itemize}

It is appropriate to mention here other Bill's contributions beyond those made in collaboration with me. In this context
the obituary for Bill written by Chris Hill for Physics Today and the information in the Wikipedia turned out to be useful. Here we go.

\begin{itemize}
\item
  He was renowned for his  work on the chiral anomaly \cite{Adler:1969er} 
  which is of foundational importance in modern theoretical physics. In this context he developed with Stephen L. Adler the "non-renormalization theorem", known as {\em the Adler–Bardeen theorem.}  As is well known, gauge anomalies are quantum mechanical violations of classical currents, typically “axial currents” as in the theory of mesons. This Adler-Bardeen theorem establishes that the anomaly coefficient is not subject to renormalization to all orders in perturbation theory, ultimately implying a deeper connection to topology.
  
Subsequently, Bill  computed the full, nontrivial structure of the chiral anomaly in non-abelian Yang-Mills gauge theories \cite{Bardeen:1969md} which was cited 1296 times until now. This  involved tedious computation of triangle, square and pentagon Feynman diagrams and  lead to the distinction between the "consistent anomaly" and the "covariant anomaly." The covariant anomaly is gauge invariant and arises in physical applications, such as pion decay, while the consistent anomaly is symmetric between left- and right-handed chiral fermions and was subsequently related to topology.   Bill’s paper was initially controversial, and rejected by a referee, but it was followed by the  "Wess-Zumino consistency conditions"  that confirmed Bill’s results.  This  opened the door to the connection of field theory to topology, e.g., the structure and coefficient obtained by Bill's calculation of the consistent anomaly turns out to be equivalent to that generated by a topological Chern-Simons term built of Yang-Mills fields in 5 dimensions. These issues are crucial in describing the anomalous physics of pseudoscalar and vector mesons, and topological effects in Yang-Mills gauge theories, such as the instanton
physics in QCD.
\item
  With Bruno Zumino, Bill coformulated the theory of the gravitational anomaly which is of fundamental importance to string theory \cite{Bardeen:1984pm}.
\item
  On sabbatical, working at CERN in 1971, Bill collaborated with Murray Gell-Mann and  Harald Fritzsch. They considered a gauge theory of quark interactions, in which each quark comes in one of three  colours.  They wrote down the theory now known these days as QCD and, through the chiral anomaly, established the existence of the three quark colours from the rate of $\pi^0$ decay \cite{Bardeen:1972xk}. Bill was also involved in discussion with Gell-Mann, Fritzsch and
  Leutwyler in which these three great theoriests gave the arguments
  for non-abelian structure of strong interactions \cite{Fritzsch:1973pi} leading to QCD. But apparently Bill didn't like the IR divergences in QCD and
  didn't put his name on this paper.  Chris Hill told me that Bill  realized later that was
  a mistake.  But according to Chris, who was Gell-Mann's student, Gell-Mann had a very high opinion about Bill. One day he said to Chris: ``Bardeen is one of the best in the business''.
    \item
  He also developed a theory of dynamical breaking of electroweak symmetry via top quark condensates, leading to one of the first composite Higgs boson models.
  In the early 1990s it was becoming clear that the top quark was heavy and this
  could imply new strong interactions at very short distances. Hence,  pairs of top quarks could in principle form a composite Higgs boson.  This led Bill to formulate, in collaboration with Chris Hill and Manfred Lindner \cite{Bardeen:1989ds}, the theory of  top quark condensates,  a dynamical approach to electroweak symmetry breaking, in analogy to (his father’s) Bardeen-Cooper-Schrieffer theory of superconductivity.  This predicted a heavy top quark, but it also predicted too heavy a Higgs boson, almost twice the observed mass of 125 GeV. Nonetheless, this was the first composite Higgs boson model and the general idea, with modern reformulation, the theory comes into concordance with experiemant and makes predictions potentially accessible to the LHC.
\item
  His work on the chiral dynamics of heavy-light quark bound states correctly predicted abnormally long-lived resonances.
Bill and Chris Hill \cite{Bardeen:1993ae}, in 1994, recognized that heavy-light mesons, which contain a heavy quark and a light anti-quark, provide a unique window on the chiral dynamics of a single light quark. They showed that the (spin)parity ground states are split from the parity partners by a universal mass gap of about $350$ MeV due to the light quark chiral symmetry breaking. This correctly predicted an abnormally long-lived resonance ten years before it was discovered by the BABAR collaboration. The theory was further developed by Bill, Chris and Estia Eichten \cite{Bardeen:2003kt}, and various decay modes were predicted that have been confirmed by experiment. Similar phenomena should be seen in the mesons and heavy-heavy-light baryons.
\end{itemize}

As one can see in INSPIRE, Bill still has several well-cited papers with other authors, but summarizing them here would be beyond the scope of my autobiography.

\subsection{Meeting Bjorken (1978)}
In September 1978 Jonathan Rosner visited Fermilab and wanted to understand the work of Edward Witten on QCD corrections to photon structure functions \cite{Witten:1977ju}. The paper was very well written, and I think I could explain to Jonathan what Witten had done. It was a leading-order analysis in renormalization-group–improved perturbation theory, and after completing a careful reading of this paper I got the idea to generalize it to the NLO level.

At that time Bill was travelling and Muta had already returned to Japan. Therefore I carried out the calculation by myself. In fact, it took me only three weeks not only to complete the calculation but also to write up the paper. Nevertheless, to be sure, I asked Bill upon his return whether he would be interested in checking my calculation. He was, and in October we were able to send our paper to the publisher \cite{Bardeen:1978hg}.

I was very satisfied with this paper and told Bjorken, who was at Fermilab at that time, about it. To my surprise and disappointment Bj simply said: ``The process has to be measured first before one starts calculating higher-order QCD corrections.'' I knew that he was opposed to QCD as the theory of strong interactions, but I was still surprised by his attitude.

This fits well with a few remarks about my stay at SLAC in April and May 1979. I must admit that it was somewhat disappointing from the physics point of view, because I was quite ahead of almost all SLAC theorists as far as QCD was concerned. I can give two examples.

When I entered the office of Stan Brodsky for the first time and asked him what he was working on, he told me that he was learning QCD and was presently doing so by studying my papers with Karel Geamers. This was, of course, pleasing, but I had no real scientific discussions with Stan during my stay.

Much worse was the reaction of the SLAC theorists to my seminar on NLO QCD corrections to photon structure functions. After my talk there were basically no questions. I then went to Bj’s office, which was next to mine, in order to get his view on my talk. Bj told me only that he did not understand why I was doing all this, which was fully consistent with his earlier remarks at Fermilab concerning my work with Bill on photon structure functions. I was still so surprised that, without saying anything, I left his office and went to Helen Quinn, hoping for a more constructive reaction.

Helen told me that my talk had probably been too technical for the SLAC audience, as there were no real QCD experts there at that time. Moreover, she explained that Bj was generally more discouraging than encouraging toward people, as he was extremely critical of himself.

I returned home that day rather discouraged, but my spirits improved significantly the following day. Roy Glauber, whom I had first met during his visit to Warsaw in 1970 and later again in Copenhagen, called me from Harvard and invited me to spend a year there. He told me that he would like me to teach him QCD during that year. I was very pleased by this invitation, but unfortunately I had to decline. I had just started a five-year position at Fermilab, and with Karin still a baby, a move from Fermilab to Harvard would have been rather complicated.

Returning to Bjorken, it should be emphasized that his views on QCD evolved when he began spending extended periods at Fermilab starting in 1980. In fact, I later coauthored a paper with him and P. Q. Hung on Petite Unification \cite{Hung:1981pd}, which I will discuss later on.

Despite my not entirely positive recollections of some of our earlier interactions, there is no question that Bj was an outstanding physicist. I fully agree with Chris Hill, who once told me that Bj deserved a Nobel Prize. In this respect his case is similar to that of Nicola Cabibbo. Fortunately, Bj was recognized with the High Energy and Particle Physics Prize of the European Physical Society in 2015, among other distinctions.

\subsection{Boulder (1979)}
In July 1979 I finally completed the review on asymptotic freedom in DIS \cite{Buras:1979yt} and submitted it for publication to Reviews of Modern Physics. While waiting for the referee report, I went to the Boulder Summer School to give two lectures on higher-order QCD corrections. The lectures were well received, although some students—and even some teachers—told me that they were too technical. I was slowly learning that there is little point in giving talks or lectures that focus on the detailed aspects of higher-order QCD corrections.

I do not remember much about the other lectures, but I would like to mention three events that clearly remained in my memory. Boulder is located at an altitude of about 2000 m, whereas Fermilab is essentially at sea level. Therefore, before engaging in any sports activities in Boulder, I should have waited at least a few days to acclimatize. Unfortunately, after arriving on a Saturday, I went the very next day with some students and teachers on a hike that reached an altitude of 3000 m. On the way back I felt very unwell, to the point that some students feared I might not even make it back to the school site, and they worried that I would not be able to give my lectures two days later. Fortunately, I recovered the following day.

However, one day later I made a second mistake by playing tennis against Steve Ellis, a very nice physicist from Seattle. At this altitude one needs special tennis balls, heavier than the usual ones, otherwise they fly too fast. But even these special balls were still flying faster than at Fermilab. I adapted rather quickly and, despite the fact that Steve was a solid player, I was leading 5:1 after half an hour. Then I became exhausted again and lost 5:7. This was unfortunate, as it turned out to be my only tennis victory in Boulder that slipped away. Rick Field, Feynman’s collaborator, and Howard Haber, a well-known theorist, were simply playing much better than I did.

One day I decided to call my parents in Denmark. I used to call them regularly, a few times a month, from Fermilab, but now I wanted to call from Colorado. I spoke only with my mother; my father was not at home. It was a very pleasant conversation, but I did not know that this would be the last time I talked to her.

Two weeks later, on August 16th at 5 a.m., my father called us to tell us that my mother had died early that morning in Denmark, a few hours earlier. My father found her lying on the kitchen floor. She was clearly feeling unwell and had tried to boil water to make tea, but then collapsed. It was a heart attack. She died on the way to the hospital. She was only 58 years old.

After my review was accepted in Reviews of Modern Physics, I dedicated it to my mother.

\subsection{The East Coast Trip: Syracuse, Cornell, Princeton (1979)}

In October 1979 I made a tour of the East Coast. I was taken to O'Hare Airport by a limousine, and during the drive the Nobel Prize in Physics awarded to Glashow, Salam, and Weinberg was announced. I explained to the driver for which achievement they had received the prize. While handing me my bag at the airport, he told me that one day I would receive a Nobel Prize myself. I have been waiting for it now for 46 years.

First I flew to Syracuse, where I gave a seminar. At the Syracuse airport I was supposed to rent a car in order to drive to Cornell the next day. It turned out that no mid-size car was available, so I could only rent a Cadillac. I then drove to Ithaca in a Cadillac. When Peter Lepage, who had invited me, saw the car I had rented, he immediately began to worry about who would pay for it. The seminar at Cornell went very well; I spoke about QCD, and the discussion was very good—much better than in Syracuse, where only a small group had attended. Peter Lepage showed me the campus, which I found much more interesting than Fermilab. Two days later I drove back to the Syracuse airport and flew first to Newark.

Before continuing, let me mention that while driving a Cadillac in 1979 I could not help thinking about an event from my childhood in Warsaw in the early 1950s. My father had some contacts with members of the Polish government. One whom I remember in particular was Stanisław Skrzeszewski, who served as Minister of Foreign Affairs from 1951 to 1956. He and his wife visited us at our home several times, and one summer day he arrived with his chauffeur in an open Cadillac and took Jacek, Małgosia, and me for a ride along Aleje Ujazdowskie, an important street in Warsaw. I still remember how proud I was to be sitting in that car. This time, many years later, I was driving it myself.

Back to the 1979 story.

I will never forget the next flight, from Newark to Princeton. At Newark Airport I went to the gate for the flight to Princeton, but to my surprise there was nobody there except me. After a while a pilot arrived and told me to follow him. We walked to a small propeller plane. We boarded it, and I turned out to be the only passenger. He turned the key like in a car, and soon we were flying roughly one hundred meters above the ground.

I am usually not afraid of flying, but this flight was rather scary. I kept thinking about the possibility that the pilot might get a heart attack. Moreover, the engine was very noisy. The only good thing was that one could see quite precisely what was happening on the ground, which is of course impossible during normal flights at an altitude of about 10,000 meters. The flight lasted about half an hour.

At the Princeton airport David Gross, whom I knew from Aspen, picked me up. It was actually Frank Wilczek who had invited me during the meeting in Iowa one year earlier.

My stay in Princeton was clearly the most interesting among the three places I visited on this tour. In addition to David Gross, I spoke with Frank Wilczek, Curtis Callan, and Roger Dashen at the university. I also visited the famous Institute for Advanced Study, where I met again Eugene Wigner and a number of other theorists working there. My seminar went very well. Roger Dashen, whenever he met me in the corridor, kept telling me how good my talk was. This was perhaps a bit overdone, but certainly pleasant.

I was also very pleased by a small party at Frank and Betsy Wilczek’s home, attended by several well-known physicists, including David Gross, Curtis Callan, Roger Dashen, and others, together with their wives. Altogether it was a very nice and interesting stay. It was not the last time I met David, Frank, and Betsy.

{\bf It is interesting that I am not the only member of my family who visited Princeton and enjoyed the visit very much. My father, Bronislaw Buras, was invited by Robert Oppenheimer in 1960 to spend a few days at the Institute for Advanced Study. This story, this time told by my father, can be found in Appendix ~\ref{Oppenheimer}.}

\subsection{A Strange Workshop in San Francisco (1980)}\label{EST}

In the spring of 1980 I was invited to a conference in San Francisco sponsored by the millionaire—or perhaps even billionaire—Werner Erhard. The meeting was organized by Howard Georgi and focused on QCD. It took place in Erhard’s villa. One of the highlights was the participation of Richard Feynman. This was the third time I met him, and his health seemed to be much better than during our first meeting at the Les Houches Summer School in 1976. There were about thirty participants, all experts in QCD. One of them was Steven Wolfram, the future author of MATHEMATICA, who at that time was a postdoc at Caltech.

We stayed at a very good hotel, to which I was driven from the airport by one of Erhard’s drivers. Although the conference was supposed to start the next day, tea and coffee were already served in the hotel a few hours after my arrival, and I met all the participants there. When I entered the coffee room, I immediately saw Feynman surrounded by several participants of my generation, while Wolfram was lying on the floor studying large computer printouts—possibly already developing MATHEMATICA. He was about ten years younger than me and by far the youngest participant. This was impressive, but at the same time I did not feel that he was smarter than the rest of us as far as physics was concerned. Computer skills are another story, and in particular his idea of developing a program like MATHEMATICA was clearly excellent.

The next day we were supposed to have breakfast not at the hotel but at Erhard’s villa, which was within walking distance, about 20–30 minutes away. Most participants were driven there by car, but I decided to walk. I found the villa and rang the bell. What happened next was a bit of a shock. The door opened and about ten very elegant ladies, all dressed in long black dresses, greeted me. Apparently I was the first participant to arrive that morning.

The meeting itself was not particularly interesting from the physics point of view. Perhaps I was already becoming somewhat tired of QCD after having written a long review on the subject. Still, there were two events that I clearly remember and would like to describe.

One was the conference dinner. There was a very long table. On one side, in the middle, sat Feynman and Erhard. On the opposite side, facing them, sat Wolfram. For reasons I did not fully understand, I was seated next to Feynman. Perhaps because of my beard and long hair the organizers thought I was the third-best physicist at the meeting, after Feynman and Howard Georgi. Howard, as far as I remember, was sitting next to Erhard on his right. Feynman was on my right, and Bob Jaffe was on my left. In 1972, at the dinner in Erice celebrating Rabi’s 80th birthday, Jaffe had been sitting on my right, one place closer to Rabi and Antonio Zichichi. I concluded that I had made progress over the eight years in between.

There were seven courses, each introduced by Erhard’s personal chef. One fish was flown in from Alaska on the same day, another from somewhere else. I did not talk very much with Feynman. Partly this was because he was busy talking to Erhard and, in particular, telling Wolfram to stop criticizing the millionaire for his personal development program, EST (Erhard Sensitivity Training). However, when we did talk about parton distributions, I got the impression that Feynman had a high opinion of my work with Karel Gaemers \cite{Buras:1977yj}, in which we generalized the Feynman–Field parton distributions in the quark model \cite{Field:1976ve} by including their $Q^2$ dependence, as described earlier in Appendix~\ref{PD77}. In fact, this is precisely what Rick Field had already told me at Feynman’s 60th birthday in September 1978 at Caltech.

The next day several participants, including Feynman, went to a first-class Japanese restaurant for dinner, and I was of course among them. The evening was completely dominated by Feynman, who told us many stories, some of them very interesting. The only other time I witnessed such dominance was several years later at a dinner in Munich with Rudolf Mössbauer, another Nobel Prize winner. However, one should not conclude from this that all Nobel laureates are so talkative. If Paul Dirac had been with us at this dinner, he would most probably have said nothing—although apparently from time to time he could also talk for hours without a pause.

Returning to the dinner at the fancy Japanese restaurant, I began to worry that the bill might be very expensive. Then came a surprise. When we asked for the bill, the owner told us that our friend (the millionaire) had already paid for everything. Encouraged by this, I went the following day with another participant—if I remember correctly, it was Ikaros Bigi—for dinner. To make sure that our friend knew where we were dining, we asked one of his drivers to take us to another restaurant, not as fancy as the Japanese one but still very good. Unfortunately, to our disappointment, our friend did not pay for our dinner this time. Quite simply, Feynman was not with us.

\subsection{Meeting Paul Dirac (1980)}\label{Dirac}

In May 1980 there was an international symposium on the history of particle physics at Fermilab, focusing on the beginnings of particle physics before the era of large accelerators. Among the speakers were such luminaries as P. A. M. Dirac, Carl Anderson, Victor Weisskopf, Willis Lamb, Julian Schwinger, Richard Feynman, and Murray Gell-Mann. I was particularly impressed by Dirac’s talk. He did not use any transparencies or notes. He simply spoke for 45 minutes in a very impressive manner, even though he was already 78 at the time. The audience responded with a long standing ovation.

The next day there was a round-table discussion in which Dirac, Schwinger, and several of the other physicists mentioned above participated. As a member of the Fermilab theory group, I took part in this discussion. As is well known, Dirac was strongly against renormalization and expressed this during the discussion. Although some participants tried to convince him of the importance and successes of renormalization, they were unsuccessful.

Dirac, together with Max Planck, Albert Einstein, Niels Bohr, and Werner Heisenberg, belonged to the greatest physicists of the first half of the 20th century. Nevertheless, he was certainly wrong regarding the renormalization of quantum field theories. Yet, even Einstein and Heisenberg, in the last twenty years of their lives, pursued directions that ultimately proved unfruitful: Einstein attempted to unify electromagnetism with gravity, and Heisenberg worked on constructing a spinor theory.

In December 1980, we were invited by Dennis Duke and his wife to spend Christmas with them in Tallahassee. Dennis, coauthor of the paper on NLO QCD corrections to deep-inelastic scattering, had spent a year at the Rutherford Laboratory in England before moving to Tallahassee, where he eventually obtained a permanent position at Florida State University.

We drove to Florida by car, visiting other parts of the United States that were very different from the places we had visited before. We met Duke’s family and thoroughly enjoyed our stay. On the way back, we visited Louisiana, which was very different from other American cities and reminded us of some French cities. Visiting locations where slavery had once flourished was both shocking and deeply interesting. Yet the most memorable event of this trip was a party organized by one of Dennis’s colleagues, where I was introduced to Paul Dirac, who had moved to Florida in 1973 from Cambridge. Our conversation was extremely brief. He said, “How are you?” I replied, ``How are you?'' and that was the end. This was entirely consistent with his biography {\em The Strangest Man} by Graham Farmelo, which I read only twenty years later. The book is very interesting. In fact, in July 2025, Farmelo gave a long talk on Dirac at the Max Planck Institute for Physics, now in Garching, very close to the TUM-IAS building where I have my office. I attended the talk and could not resist telling this story afterward.

Recently, Karel Gemmers shared another story about Dirac. At an important dinner, a lady sitting next to him remarked that she would manage to get three words out of him that evening. His answer was simply: ``You lose.''

The last time I saw Dirac was a few months later at the 1981 QCD conference in Tallahassee, where he gave another talk criticizing renormalization. It was almost amusing, as his talk was placed between two presentations on higher-order QCD calculations for high-energy processes.

\subsection{Petite Unification of Quarks and Leptons (1981,2003,2004)}\label{Petite}
In 1981 I explored, together with Bjorken and P.Q. Hung, an early unification of quarks and leptons based on the Pati-Salam idea, in which the lepton number is treated as a fourth color and the QCD gauge group
$\text{SU(3)}_c$ is extended to $\text{SU(4)}_\text{PS}$. But while Pati-Salam did
not specify in their paper the electroweak part of the gauge group, in
subsequent papers on left-right symmetric models the group $\text{SU(2)}_L\times\text{SU(2)}_R$ replaced the SM gauge group
$\text{SU(2)}_L\times\text{U(1)}_Y$. Now without a $\text{U(1)}$ factor the value
of $\sin^2\theta_W$, in contrast to the SM case, can be predicted at the unification scale. In the case at hand it turns out to be $1/2$ and the unification scale
must be of $\ord(10^{10})\gev$ in order to obtain through renormalization group
effects  $\sin^2\theta_W\approx 0.23$ at the electroweak scale as measured in experiment. This is clearly
not an early unification.

Bjorken and P.Q. Hung noticed that by enlarging the weak gauge group, one could lower the value of $\sin^2\theta_W$ at the unification scale and thereby reduce the unification scale. P. Q. told me about this idea, and I asked whether I could join them, as any application of the renormalization group was like a magnet to me. They agreed, which proved fortunate: a few weeks later Bjorken became busy with other matters and, I believe, lost interest in the project. P.Q., trained in model building, naturally led the project, and I learned a great deal from him.

There are many useful formulae in the paper we wrote, particularly for $\sin^2\theta_W$ for an arbitrary weak gauge group $G_W$ in this 
{\em Petite Unification}
\be
\text{SU(4)}_\text{PS}\times G_W,
\ee
as we named this construction. The study of rare processes mediated by leptoquarks, which became popular in the 2010s, can also be found in this paper.

P.Q. and I wrote the paper and gave it to Bjorken for comments, but it took several months before we could discuss it with him. We had considered extending the analysis further, but Bjorken showed no interest and said, ``There is no reason to beat up a dead horse.'' After the work we had done, this was not a pleasant response, but I was already accustomed to it, after his reaction to my paper with Bill on photon structure functions and his comment on my SLAC talk two years earlier. P.Q. was unhappy, as he had led the project but, alphabetically, would appear last on the author list. I suggested to Bjorken that P.Q. should be first and I last; he agreed but recommended reversing the order, and the paper was published accordingly \cite{Hung:1981pd}.

The weak gauge group we studied in detail in that paper was $\text{SU(2)}^4$, which gave $\sin^2\theta_W = 1/4$ at the unification scale and about $0.21$ at the electroweak scale—consistent with experiment at that time. P.Q. and I returned to this idea 22 years later \cite{Buras:2003du,Buras:2004ij}. By then, the $\text{SU(2)}^4$ group had been ruled out, so we considered two other possibilities, $\text{SU(2)}^3$ and $\text{SU(3)}^2$, which give $0.333$ and $0.375$ for $\sin^2\theta_W$ at the unification scale and can be made consistent with its measured value for unification scales of $1~\tev$ and $3-10~\tev$, respectively.

After that period, I met P.Q. only once in person, in 2010, at a meeting I organized at Starnberg Lake south of Munich. About one hundred participants, mostly my former collaborators, attended. Janusz Rosiek was there, and at 1.99 m he replaced Bjorken in a photo with P.Q. (1.55 m) and me (1.72 m).

I think the idea of Petite Unification of quarks and leptons is interesting, but judging from the number of citations of the three papers we wrote since 1981—83 in total—it was clearly not a major success. Nevertheless, I do not consider it a dead horse, and it may yet become more popular. In fact, P.Q. and I attempted to revive it in 2015, but being busy with other projects, no paper materialized. Unfortunately, in 2024, both of my collaborators on Petite Unification passed away, leaving the future of this idea entirely in my hands.

\subsection{Technicolour and Composite Models (1982)}
My last two papers at Fermilab and the first two at MPI focused on model building in the context of technicolor \cite{Buras:1982ff} and composite models \cite{Buras:1982bn}, in collaboration with Sally Dawson and Bert Schellekens at Fermilab, and on supersymmetric technicolor at MPI—one with Yanagida \cite{Buras:1982nb} and one with Wojtek Słomiński \cite{Buras:1983yg}. I learned a great deal from these works, but they were clearly far from successful. In my view, the Petite Unification idea remains far superior to the concepts explored in these four papers, so there is no reason to discuss them further.

\subsection{Seeking a Permanent Position: 1981}\label{1981P}
Over the years several colleagues have asked why, after spending four years in Denmark, two years in Switzerland at CERN, five years at Fermilab, and being frequently at DESY in Hamburg before 1981, I finally ended up at the Max Planck Institute for Physics in Munich, which I first visited only in October 1981.

Let me briefly describe the reasons for this, which will also allow me to thank those who took part in these events that unfolded in 1981. They took place 45 years ago, and perhaps I have waited too long to describe them, as several of the actors in this drama are no longer among us.

In February 1981, Gurli told me that she missed her family in Denmark and was tired of the long winters at Fermilab, indicating that it was time to return to Europe. I wrote a letter to Aage Bohr asking about the possibility of a permanent position at the Niels Bohr Institute. He replied and invited me to spend several months there, promising to investigate the prospects for a position in the meantime. We stayed at the Institute in September, October, and November, which turned out to be very useful and in fact had a major impact on my life, as we shall see later.

In February I also wrote a letter to John Ellis asking about the possibility of visiting CERN for a year, with the goal of finding a permanent position in Europe. John did not respond to this letter, but a few weeks later I received a letter from DESY practically offering me a position in the DESY theory group. In order to discuss the details, Professor Erich Lohrmann invited me to spend ten days at DESY.

I went to DESY in April, gave a talk on my work with Bjorken and Hung that I have just described, and met for the first time the DESY director, Soergel, who asked me to sign several forms indicating my interest in a position at DESY. During these ten days at DESY I realized that, with the exception of Tom Walsh (who was absent during my visit), the theory group was not interested in hiring me. The main advocates for my appointment were in fact experimentalists, in particular Björn Wick, who were dissatisfied with the performance of the theory group in phenomenological aspects of particle physics.

This impression was further strengthened by several personal observations. I was not invited to a dinner at a restaurant, neither after my talk nor at any time during my stay. Moreover, there was another seminar speaker during my visit for whom a party was organized, to which I was not invited. I tried to arrange a dinner with a member of the theory group, but he said he was too busy. Finally, in conversations with Joos, the most senior theorist in the DESY theory group, I realized that his main goal was to secure a permanent position for Ahmed Ali rather than for me.

This visit was a major disappointment, but there was one particular event that convinced me even more strongly that DESY was not the place where I wanted to work. This was my meeting with Kurt Symanzik, the famous field theorist. I went to his office, introduced myself, and told him that there were plans to hire me at DESY. I expected that he would ask me what I was working on, but he did not. Therefore, I asked him what he was working on. He responded with a 15-minute lecture on his research. When he finished, I again expected that he would now ask me about my work, but this did not happen. Already somewhat annoyed, I asked him whether he was working on anything beyond what he had just described. The answer was another 15-minute lecture. When this second lecture ended, it was clear to me that the audience with the famous field theorist was over. I thanked him for the two lectures and left his office. I never spoke with him again. He died in October 1983 at the age of 59.

Three weeks later, already back at Fermilab, I received a formal offer of a position at DESY. Gurli was very happy about it, but I told her that it was not certain that I would accept this offer, given my experience during the visit. At the same time, having such an offer would increase my chances of obtaining a position at the Niels Bohr Institute. However, my visit there was scheduled only four months later, and it was not clear whether DESY would wait that long for my decision. At this point my father helped me greatly. He had contacts at DESY in connection with synchrotron radiation and visited DESY frequently from Denmark.

DESY offered me a BAT Ib position, which at that time corresponded to a C2 position (assistant professorship) at the university level. Knowing my standing in particle physics, and also being familiar with young German scientists at DESY who held BAT Ib positions, my father told me that DESY should definitely offer me a BAT Ia position, corresponding to a C3 position (associate professorship). I wrote back to DESY, explicitly referring to my father. I received a reply stating that my request would be investigated, but that this could take several months. This was, in fact, precisely what I wanted.

Fermilab did not react to the DESY offer, and Bill and Chris Quigg did not expect that I would leave Fermilab for DESY, since the Fermilab Theory Group was much stronger than the DESY Theory Group at that time. However, the situation changed in July at the EPS High Energy Conference in Bonn, where both Bill and I gave plenary talks—Bill on photon structure functions in QCD and I a general talk on perturbative QCD. At this conference I was informed by Erich Lohrmann that DESY would offer me a BAT Ia position. At breakfast the next morning I told Bill that if Fermilab did not offer me a permanent position, I would accept the DESY offer.

A few weeks later, after my return to Fermilab, Bill—who was head of the Theory Group that year—made my position at Fermilab permanent. I became a Scientist I at Fermilab. However, this did not resolve the situation, because Gurli insisted that we return to Europe and clearly favoured DESY over Fermilab. I told her that I preferred to stay at Fermilab and that, moreover, having two offers might help convince the Niels Bohr Institute to offer me a permanent position.

In September we went to Denmark, but during the first weeks I was unable to speak with Aage Bohr for reasons I no longer remember. In the second half of September there was another QCD workshop, this time at DESY, where I again gave a plenary talk. At this workshop I met Carl and Cindy Albright, who had spent the first half of 1981 at the Max Planck Institute (MPI) in Munich and the second half at DESY. I knew them from Fermilab and told them about my situation. They told me that life in Munich was, for them, much more exciting than in Hamburg and suggested that I investigate the possibility of an offer from MPI, as there was apparently a vacant position there.

In the DESY canteen I then met Leo Stodolsky, one of the directors of the Max Planck Institute for Physics in Munich, whom I knew from conferences. I told him that I was deciding between Fermilab and DESY and, to his surprise, asked whether there might be possibilities at MPI. He did not answer directly, but instead asked whether I would be interested in giving a seminar at MPI in October. I said yes. The following week, already at the Niels Bohr Institute, I received a phone call from Roberto Peccei inviting me to give a seminar at MPI and to spend ten days there. At the DESY meeting I also met Reinhold Rückl for the first time. He was at LMU Munich at that time, and I was impressed by his friendly manner—very different from what I had experienced at DESY half a year earlier.

Even before going to Munich I began receiving phone calls from Fermilab, trying to convince me to stay. In particular, Bill and Bjorken called me. I told them, however, that this would be difficult because of Gurli, although at that point I had not yet made a final decision.

I went to Munich by train and arrived on the morning of October 15. I stayed in the MPI guesthouse on the Institute grounds. It was a Thursday, and not much happened during the first two days. On Saturday I walked through the English Garden to Marienplatz, which took, I think, about 40 minutes. I then walked around the city center and was very impressed by Munich, especially in comparison with the surroundings of Fermilab.

I visited a bookstore and bought the book Eichtheorien (Gauge Theories) by Becher, Böhm, and Joos. It was written in German, and I thought it would be a good opportunity to refresh my German after so many years. Since leaving Vienna in 1962, I had read German books from time to time, but if I were to move to Germany it would clearly be useful to improve my command of the language. In fact, I began reading the book immediately after returning to the guesthouse, again walking through the English Garden.

On Sunday I met Reinhold Rückl and Dorthe, his girlfriend at the time and later his wife. We first went to the Michaelskirche near Marienplatz and listened to a Bruckner Mass. Afterwards we had lunch with Nürnberger Würstchen, and then drove by car to Spitzingsee. We took the cable car, climbed the Rotwand, hiked a bit in the surrounding area, and on the way back to Munich stopped for dinner. It was a very exciting day.

On Monday I gave my talk on the work with Bjorken and Hung on ``Petite Unification.'' The talk was well received, although Peccei did not think that this was a particularly promising direction, and he was partly right. More importantly, he asked me whether I would be interested in a permanent BAT Ia position at MPI. I said yes, and he promised that I would receive an offer before Christmas, that is, within two months.

I told Reinhold about this possibility. He was very excited, and in the following days we went to several jazz clubs and other music venues, which convinced me completely that this was the place where I wanted to spend the rest of my life. On Saturday, the day I was returning overnight to Denmark on a very late train, I went hiking with Harald Fritzsch and Hans Kühn in the area near Kufstein. Afterwards Hans invited me for dinner at his home. These ten days were truly amazing.

After returning to Denmark, I did not tell Gurli about the possible offer from Munich. However, with offers now from DESY and Fermilab, and effectively from MPI, I finally managed to explore the possibility of a permanent position at the Niels Bohr Institute by talking to Ben Mottelson and Aage Bohr. Ben told me that they could offer me a six-year professorship at Nordita, the Institute for the Nordic countries—Denmark, Finland, Norway, and Sweden. The building can be seen in Fig.~\ref{fig:NBI}, to the left of the main NBI building. I was 35 years old at the time, and I did not think it was a good idea to accept such an offer, particularly in view of the three other possibilities for permanent positions that I had.

The conversation with Aage Bohr was different. He congratulated me on all the offers I had received and told me that it would be very nice if I accepted the DESY offer, as I would then be closer to Denmark and able to visit NBI more often. On the other hand, he said that by no means should I accept the MPI offer, because his father, Niels, had had a difficult relationship with Heisenberg during the Second World War. Heisenberg had died in 1976, and I could not accept that someone who had not offered me a position at his own Institute would tell me which offer I should reject—regardless of how great physicists the two Bohrs were. This conversation with Aage Bohr convinced me that I should accept the Munich offer if I received it.

Before returning to Fermilab, I spent a week at Helsinki University. Masud Chaichian, whom I had known from my CERN days, invited me to give a talk. It was November, and the days were rather short, but it was an interesting stay. I also met Kajantie, a professor at the University whom I had encountered several times at both NBI and CERN. Meeting him there was quite amusing. One peculiar feature of the Institute was the traffic lights on each office door, indicating whether you should not enter (red), wait (yellow), or enter (green). I did not notice this at first and walked into Kajantie’s office on red, finding him talking on the phone. He was clearly surprised, but fortunately he did not get angry.

Back at Fermilab, I informed Bill that I might receive an offer from MPI in Munich. He simply said, ``Shit,''and did not speak to me for the rest of the day. Bill knew MPI and, in particular, Munich. He had visited the city several times, especially during Oktoberfest, and he knew that, despite the stronger research environment at Fermilab, it would be hard to compete with this offer—especially since Gurli wanted to return to Europe.

On December 20, I received a telegram from Roberto Peccei, informing me that he was delighted that the MPI theory group had decided to offer me a permanent position starting in September 1982. I was delighted as well. The formal offer from the MPI director, Hans Peter Dürr, arrived in early January 1982. It was an excellent offer, with a salary higher than my salary at Fermilab, although Munich was certainly a much more expensive place than Fermilab’s surroundings.

Shortly after receiving Roberto’s telegram, I informed Gurli about the MPI offer. She was a bit surprised that I had kept this possibility a secret, but she considered it an acceptable solution, since she wanted to spend the rest of her life in Europe, closer to her family. I liked Fermilab, but today I am very thankful to her that we left the USA. This decision proved particularly valuable because it allowed easier contact with our families in Poland and Denmark, and also because of the political climate in the United States. Moreover, theoretical particle physics in Europe—especially research closely connected with experiments—became stronger than in the USA already by the 1990s.

I accepted the Munich offer soon after receiving the letter from Hans Peter Dürr in January 1982. I informed Bill that I would be leaving Fermilab, and he did not speak to me for the next three days—but we remained friends.

The case of DESY was quite different. I called the director, Soergel, who was somewhat annoyed. In fact, during the 1980s, even after I had moved to Munich, I was invited to DESY only once, by Roberto Peccei, who had moved there in 1984. On one occasion, an experimentalist wanted to invite me for a major colloquium, but this was blocked by someone—apparently not Soergel, though I do not know who it was.

The situation changed in the 1990s, when DESY asked me to organize their Theory Workshops in 1990, 1996, and 2000, which I will describe in the next section. In addition, I was invited to a number of seminars. This was partly due to the fact that the DESY Theory Group had improved considerably, hiring many excellent theorists over the past thirty years, among them three with whom I continue to have contact from time to time: Wilfried Buchmüller, Géraldine Servant, and Christophe Grojean.

Definitely, Roberto Peccei and Reinhold Rückl played a crucial role in my landing in Munich. But without the offer from DESY, the disappointing visit to the DESY Theory Group, the offer from Fermilab, and the conversation with Aage Bohr, I would not have even considered a permanent position in Munich. Therefore, in fact, I should thank them all for what happened.

Looking back, this was indeed the best decision—not only because of MPI and the Bavarian Alps, but in particular because of the Technical University of Munich, which in 2025 was classified as the best university in the European Union. Most of my PhD students, whom we met on my “expedition” above, were excellent. But the main reason this was the best decision was personal: if we had moved to Hamburg, our two sons, Robert and Allan, would not have met Franzi and Ute, respectively, and we would not have these six wonderful grandchildren we now have.

\subsection{DESY Theory Workshops}\label{DESYW}
Since 1987, DESY has been organizing annual Theory Workshops. The organizing committee consisted of non-DESY theorists, with possibly one exception. I was selected three times to serve as chairman of the committee—in 1990, 1996, and 2000.

The organization was relatively straightforward. My main responsibility was to select the members of the committee, who would then set up the program for the three-day workshop. Today, such committee meetings would take place online via Zoom, but at that time, I do not remember exactly how we communicated—I believe most arrangements were made by email. The rest of the logistics, such as housing and lecture rooms for parallel sessions, were handled by the DESY Theory Secretariat. Plenary talks were given in the main DESY auditorium.

There are a few events from these workshops that I particularly recall.

\subsubsection{1990}
In 1990, before the Workshop began on October 3, DESY Director Volker Soergel gave a short talk related to German unification.

During this workshop, there was a parallel session devoted to QCD calculations. The work by Jamin, Weisz, and myself on NLO QCD corrections to quark mixing \cite{Buras:1990fn} was presented by Matthias, with my participation in the audience. Also present were Emanuel Paschos and York-Peng Yao. Paschos and Yao had independently performed similar calculations prior to ours, but their results were incomplete—they did not include two-loop anomalous dimensions, and in the case of Paschos, the necessary matching to the effective theory had not been carried out. I describe this in detail in \cite{Buras:2011we}.

Paschos did not agree with our criticisms of their paper. After Jamin’s talk at DESY, he continued to dispute the results and even asked Yao whether he agreed with him that our paper was wrong. Yao’s answer was quite surprising: ``No, their paper is correct and our papers are wrong.'' That effectively closed the chapter on the discussion.

Needless to say, both Paschos and Yao made important contributions to our field, but I leave it to them to describe these contributions in detail.

Among the plenary speakers at the workshop were Altarelli, Bardeen, Denner, Haidt, Marciano, Peccei, Peskin, Rosner, Wiik, and Zerwas.

\subsubsection{1996}
The chairpersons of the plenary sessions were the participants of the workshops. One memorable moment during the 1996 workshop involved Graham Ross, who was giving a talk while Harald Fritzsch, serving as his chairman, sat in the front row. When Graham finished his talk, he humorously informed the audience that Harald had fallen asleep, allowing him to continue his presentation. Of course, everyone laughed, Harald woke up, and Graham had to stop.

This was the first story of an important physicist falling asleep during a seminar. Two more such stories will be recounted in Section~\ref{Stech}.

\subsubsection{2000}
Among the plenary speakers of the 2000 Workshop were, unsurprisingly, several of my former PhD students and collaborators. After all, the title of the workshop was Standard Model Analyses and Effects from Physics beyond the Standard Model. These included Buchalla, Fleischer, Isidori, Jamin, Lindner, Martinelli, and Silvestrini. Other prominent physicists such as Barbieri, Buchmüller, Greub, Mannel, Nir, Peccei, Quigg, Graham Ross, and a few others also contributed plenary talks.

One of the parallel sessions of the 2000 workshop was led by Matthias Jamin, my great collaborator. I could not attend this session, which included only five talks and an audience consisting solely of the session speakers. The first talk was given by Jörg Urban, another close collaborator of mine. Apparently, during his talk the remaining four speakers were not listening, being preoccupied with reviewing their own upcoming presentations. Consequently, after Jörg’s talk there was no discussion, except for a single question from Matthias. Jörg’s reaction was immediate: he announced to the audience that he would not listen to the remaining talks, as nobody had paid attention to his, and then left the room.

The workshop was a great success, but it could have been even bigger if it had been organized in 2001 or 2002, after the discovery of CP violation in B physics.

\subsection{Ringberg Workshops}
Not far from Tegernsee, about a 40-minute drive from Munich, stands Rinberg Castle, a charming venue where small workshops for up to 35 participants can be held. The castle belongs to the Max Planck Society, and over the years I had the pleasure of organizing several workshops there, starting in 1988 and continuing until 2009.

The early workshops, in 1988 and 1996, led to published proceedings \cite{Buras:1989er,Buras:1998vd}. The very first one also inspired a collection of articles on Heavy Quark Physics \cite{Buras:1992qm}, in which my review article with Michaela Harlander appeared \cite{Buras:1992uf}. I have always considered it a strong review, though, unfortunately, it never appeared on the hep-arXiv and garnered only 29 citations.

Later workshops, in 2003, 2006, and 2009, which I organized together with Andre Hoang, did not result in proceedings. Nevertheless, they provided an excellent setting for discussion and exchange among a small, focused group of participants—a hallmark of Rinberg Castle workshops.

\subsection{Workshop in Santa Fe and  New Mexico in 1990}

In the spring of 1990, I was invited to a three-week workshop in Santa Fe, New Mexico. The workshop was organized by Bob Jaffe, whom I had already met at the Erice School in 1972 and at the “billionaire meeting” in San Francisco in 1980. Most of the participants came from Harvard, MIT, Princeton, and Caltech, leaving space for only a few Europeans. Julius Wess, who was somehow involved in organizing the workshop, asked me to invite two European physicists. I chose Chris Sachrajda and Manfred Lindner, with whom I had close contacts at the time.

The workshop was financed by Los Alamos. What made it truly extraordinary was that not only were participants’ trips to Santa Fe covered, but also the hotel stays—including accommodations for accompanying family members. Thanks to this generous arrangement, my family joined me. In addition to the three weeks in Santa Fe, we were able to visit several other places in New Mexico, Arizona, and Nevada. Our five-week vacation ended up being far more affordable than it would have been otherwise.

The workshop itself was quite engaging, with several talks scheduled each day. My own talk, on NLO QCD corrections to $K^0-\bar K^0$ and $B^0_{s,d}-\bar B^0_{s,d}$  mixing, calculated in collaboration with Matthias Jamin and Peter Weisz \cite{Buras:1990fn}, was very well received—particularly by Bob Jaffe, who recognized its importance. Fortunately, he was right: to date, this calculation, completed in April 1990, has collected almost 800 citations.

During the workshop, I also met Ben Grinstein and Lisa Randall for the first time; both were at Harvard then. I discussed with Ben my earlier paper with Peter Weisz \cite{Buras:1989xd}, where we emphasized the importance of evanescent operators. Subsequently, he coauthored a paper with Mike Dugan \cite{Dugan:1990df}, defining these operators slightly differently. As my PhD students Stefan Herrlich and Uli Nierste demonstrated four years later \cite{Herrlich:1994kh}, there are indeed multiple ways to define evanescent operators, and the choice of definition is part of specifying the renormalization scheme. However, they also showed that the definition used by Peter and me remains the most practical for explicit calculations (see Sections~\ref{NLO1} and \ref{NLO2}).

I also had the opportunity to discuss with Lisa her work with Jonathan Flynn on heavy top-quark effects in $\epe$
from Spring 1989. This work I corrected in collaboration with Gerhard Buchalla and Michaela Harlander in the fall of 1989, as discussed in some detail in my review \cite{Buras:2011we} and referenced here in Section~\ref{Anatomy1}. I met Lisa several times afterwards and followed her remarkable career, particularly her work on the RSc model, which was studied intensively in my group (see Section~\ref{RSM}).

I was honored to have a desk in the same office as Bob Jaffe, who was the organizer of the workshop. One day, Steven Weinberg arrived. He was originally expected to spend at least a week at the workshop, but in the end he came for only a single day. He gave a talk, the content of which I no longer recall. What made the moment unforgettable for me was that he arrived just one day after my talk. To my surprise, when he entered Bob’s office and saw me, he apologized for not being able to attend my presentation. Weinberg had been one of my heroes, and I was clearly delighted by this small but unforgettable gesture.

\subsection{Bertold Stech's Retirement (1992) and Sheldon Glashow}\label{Stech}
I met Berthold Stech for the first time at a meeting at MPI in November 1982. As I mentioned in Section~\ref{LargeN}, his papers with Bauer and Wirbel \cite{Wirbel:1985ji,Bauer:1986bm} had motivated Jean-Marc, Reinhold, and me to study QCD in the large-N limit, eventually leading to the development of DQCD.

Although I never collaborated directly with Berthold, we enjoyed very good relations. I would like to recount one event that some of my colleagues found rather amusing. It took place in 1992 in Heidelberg, at the celebration of Berthold’s retirement. I had been invited to give a 45-minute talk on the occasion. It was an honor, particularly because there were only three talks in total—one by the well-known Swiss astrophysicist Norbert Straumann, and the other by the Nobel Prize winner Sheldon Glashow, whom I had seen at one of the Grand Unification Conferences but had never spoken to until that day.

Otto Nachtmann, who organized the meeting, had asked me to give my talk in German, as there would be students and other officials present. I was a bit surprised, but I prepared the full talk in German—including all my transparencies and even several jokes.

A second surprise came just five minutes before my talk, which was the first of the meeting. Otto informed me that I should give the talk in English because Glashow did not understand German. I warned him that this would be tough, especially since my jokes were all in German, but he insisted.

I began my talk in English, but about five minutes later, I noticed Glashow dozing off—probably jet-lagged, or perhaps he found my talk less than thrilling. I switched to German. Glashow remained asleep until I delivered my first joke, which made the audience laugh and finally woke him up. I switched back to English, but five minutes later, Glashow was asleep again, so I went back to German until the next joke.

Overall, it was a lot of fun to give this talk—I believe not only for me but for the audience as well.

{\bf Recently, I came across a remarkably similar story in Weinberg’s book ``A Life in Physics''. On page 85, he recounts an incident in which Sidney Coleman was giving a physics seminar at Harvard and Glashow fell asleep. Coleman paused mid-lecture, pointed at Glashow, and made a witty comment that caused the room to erupt in laughter, finally waking him up.

Curious about the exact comment? You’ll have to read the book—but I can say it is well worth it, both for this anecdote and for the many fascinating insights it contains.}

As far as I recall, at Stech’s retirement, Glashow was originally asked to speak about Grand Unification, in particular the 
$\text{SU(5)}$ model that he had invented with Howard Georgi \cite{Georgi:1974sy} back in 1974. However, during his flight to Europe, he became intrigued by the dynamics of billiard balls and ended up presenting his findings on that topic instead. Some participants were disappointed that he did not discuss his famous paper, but since I had already studied 
Grand Unification in 1977 \cite{Buras:1977yy} and was aware that the model had encountered some issues by 1992, I was pleasantly surprised by his talk.

At this meeting, I also met Matthias Neubert, Berthold’s PhD student, and his wife Tina for the first time. Matthias has since gone on to have a remarkable career.

While I continued to meet Berthold at various conferences after his retirement, there are two particular events that left a strong impression on me.

The first memorable event was the 2003 Schladming Winter School \cite{Meissner:2004hx}, where I gave two lectures documented in \cite{Buras:2003wd}. Truth be told, I do not remember much from the scientific content of this school, except for skiing with Berthold\footnote{This does not mean the school was uninteresting, but it did not have the same impact on me as the Schladming School in 1972.}. He was 78 years old at the time, and I was truly impressed that he could ski so well at that age. I was not a bad skier myself, but I had stopped skiing at 61.

The second event was his 90th birthday in 2013, at which he gave a very impressive talk—remarkable at that age. Over the years, I had occasional email contact with him, and only a few years before his death in 2022, I asked him a physics question and received a long, detailed reply. I once told an older colleague, who knew Berthold, that I hoped to be as fit as Berthold at his age. His response was perfectly honest: ``Andrzej, first you have to reach this age.'' I can only hope.

Berthold passed away on June 29, 2022, and did not reach his 100th birthday as I had hoped. However, on December 8, 2020, he celebrated a symbolic 100th birthday in collaboration with my granddaughter Inga, who was born on December 8, 2016 as well. I congratulated him on this occasion. My last contact with him was in January 2022.

\subsection{Second Japan Expedition (1995)}
In the spring of 1995 I was invited to a conference in Tokyo. Ten years earlier I had visited Japan for the first time. I described it Section~\ref{Japan1985}. This time I did not want to travel to Japan alone, and our son Robert joined me on this expedition.

We first flew to Tokyo, this time over Siberia, which reduced the flight time to only ten hours. We spent five days there attending the conference, as Robert was studying physics at TUM at that time. We also visited a few shrines. The remainder of our trip was similar to the one with Reinhold Rückl in 1985: we visited Kyoto and Hiroshima and subsequently returned to Tokyo as in 1985. But this time I was invited to give a talk, this time at the annual meeting of the Japanese Physical Society. This was a rather special experience.

Indeed, Robert and I were the only non-Japanese physicists participating in the meeting. I was the second speaker on the morning of the first day. All talks except mine were in Japanese. We listened only to the talk before me and left after my own talk. My talk went well—at least I think so—but after it nobody asked any questions until the chairman said: ``Please ask some questions, he is good.'' A few questions were then asked, one of them by Muta, who I suspect had suggested to the organizers that they invite me.

During the subsequent days we visited Kamakura to see the huge Buddha monument, a very impressive place. In the evening we returned to Tokyo, having been invited by Muta and a few Japanese physicists to a fish restaurant. There were eight fish dishes—far too many to be able to eat. It reminded me of the dinner at the strange meeting in San Francisco in 1980.

During this visit I learned that one does not give tips in restaurants in Japan, at least not in those we visited. On one occasion we left some money on the table, and one of the waiters ran out of the restaurant to tell us that we had forgotten our money and returned it to us. This was quite the opposite of an experience in a restaurant in Chinatown in San Francisco in 1989, where I was together with Jean-Marc Gérard and Reinhold Rückl. We did not leave a tip, and after Reinhold and I left the restaurant first, poor Jean-Marc was confronted by an elderly lady who demanded a tip. I have never been to China, and it is possible that tipping is not customary there, but in Chinatowns in the US, as in other restaurants, tipping is expected.

As already mentioned, I visited Japan several times until 2010, both for conferences and as a member of the KEK Advisory Committee. Now, because of the political situation, the flight from Munich to Tokyo takes about fourteen hours, and I doubt that I will visit Japan again.

\subsection{Beauty Conference  1995}
In the summer of 1995, from July 10 to 14, the Beauty Conference took place in Oxford. The venue was the Oxford University campus. As this was during the summer break, the participants were accommodated in university student apartments. The conference was dominated by experimentalists.

I was invited to summarize the theoretical status of  B-physics
at that time \cite{Buras:1995iy}. There were extensive discussions of the prospects for measuring rare 
B-meson decays, in particular $B_s\to\mu^+\mu^-$.
As I have already mentioned, in 1993 Gerhard Buchalla and I calculated the NLO QCD corrections to this decay \cite{Buchalla:1993bv}. While our original goal was the calculation of these corrections, the main achievement turned out to be Gerhard’s identification of an error in previous calculations, which had missed a factor of two in the branching ratio.

The authors of these earlier papers obtained their results by taking the formula for $K_L\to\mu^+\mu^-$
and replacing the kaon parameters with those of the B meson, 
in particular the masses and weak decay constants. Unfortunately, they overlooked the fact that the definition of $K_L$
contains an additional factor of $1/\sqrt{2}$ which $B_s$ meson does not have. Consequently, their Standard Model prediction for the branching ratio was wrong by a factor of two. At the time, however, the experimentalists were not aware of this error.

When I pointed this out during a talk on the prospects for measuring this decay, my remark turned out to be the highlight of the day.

Thirty years later it is evident that our 1993 result was very important. The present data from ATLAS, CDF, and LHCb agree with the Standard Model prediction calculated by us at the level of about $10\%$.
While a small new-physics contribution to this branching ratio is not yet excluded, had nobody identified this error over the last three decades, a fake anomaly in this decay would surely have been claimed—one much larger than any anomaly currently discussed. Reading the proceedings of this workshop, one finds that impressive progress in 
B physics has been made since 1995 \cite{Harnew:1995pa}.

This is surely the appropriate place to mention Peter Schlein, who, as far as I know, was the father of the Beauty meetings. I met him several times, and it was always a pleasure to discuss physics with him. Unfortunately, he passed away in 2009, much too early.

However, the B-physicist with whom I had even closer contacts than with Peter was Sheldon Stone, who also participated in this edition of the Beauty conference. I met him repeatedly in the early 1990s at Fermilab and in Aspen in the context of the Fermilab Program Advisory Committee. The last time we met was at the EPS–High Energy Physics Conference in 2015 in Vienna. There is a photograph of us together with Marina Artuso taken at the conference dinner, which I showed in my keynote talk at the annual LHCb workshop ``Implications of LHCb Measurements and Future Prospects (2021)'' at CERN, noting that Sheldon was ``a great flavour physicist and a wonderful person with a special sense of humour.''

Unfortunately, Sheldon passed away on October 6 of that year, much too early. My talk took place on October 22.

\subsection{East Coast USA Trip with Robert (1996)}
In the spring of 1996, I was invited to give seminars at several institutions on the U.S. East Coast, namely the University of North Carolina at Chapel Hill, the College of William \& Mary in Williamsburg, the Thomas Jefferson National Accelerator Facility, and the University of Virginia. I was pleased to accept these invitations, as I was not familiar with this part of the United States. The same was true for our son Robert, who joined me on this trip.

At all of these institutions I presented the same seminar, devoted to our work on rare K an B 
 decays and on CP violation. I will not discuss the scientific aspects here; instead, I will recall a few events unrelated to my seminars.

 \subsubsection{Chapel Hill}
We first traveled to Chapel Hill, where we were picked up at the airport by Paul Frampton, whom I have already mentioned in connection with the 331 model \cite{Pisano:1992bxx,Frampton:1992wt}. He drove a rather expensive car that he had received from Anne-Marie, his French wife, as a present for his 52nd birthday.

Paul took care of us almost every day, showing us the university and the city, and nearly every evening we were invited to his home. Anne-Marie prepared very tasty dinners for us. Although she was not a physicist, she was very interested in the number of citations of physicists. In 1996 it was far from trivial to obtain such information, but she had access to some Stanford account, and one evening she informed me of the total number of citations of my papers—a number I did not know at that time. This was both truly surprising and rather amusing.

The following day, while we were walking around the university campus, Paul asked Robert to take a photograph of the two of us using his camera. That same evening we were again invited for dinner at his home. Already upon our arrival, Paul told me that his wife had found additional citations to my papers. He then showed us two rooms whose walls were covered with photographs of him standing next to well-known physicists. In one room these were Nobel Prize winners, such as Weinberg, ’t Hooft, and a few others. In the second room, according to Paul, these were candidates for the Nobel Prize. I recall that one of them was David Gross, who indeed received the Nobel Prize together with David Politzer and Frank Wilczek in 2004.

What was truly amazing, however, was to find among these photographs the picture of Paul and me that Robert had taken earlier that very day. This was a very nice gesture and reminded me of a comment made by a taxi driver in 1979 at O’Hare Airport. But until today, these particular predictions have not yet been confirmed.

Recently I was again in contact with Paul, who still hopes that the bilepton predicted by his 331 model will be discovered one day. The LHC searched for it in 2021 but observed no excess in same-sign leptons. Paul nevertheless believes that it could still show up at the HL-LHC as the first crack in the 
$\text{SU(2)}\times\text{U(1)}$
Standard Model. Even Glashow, with whom Paul wrote twelve papers, appears to share this view, as suggested by a video interview (search for “Keating Glashow” on Google), which I found very interesting.

On my side, since I studied only the $Z^\prime$
sector of the 331 model (see Section~\ref{331Models}), my hope is that a  $Z^\prime$
will be discovered one day. In fact, from my perspective it would be equally interesting if a  $Z^\prime$
with right-handed couplings to quarks—absent in the 331 models (see Section~\ref{ZprimeModels})—were to be found.

  \subsubsection{Virginia and Washington}

At the end of our stay in Chapel Hill, we rented a car and drove to Virginia, where I was scheduled to give three talks: one at the Thomas Jefferson National Accelerator Facility, the second at the University in Williamsburg, and the third at the University of Virginia in Charlottesville. The drive from North Carolina to Virginia was very interesting. In particular, we visited the house of Thomas Jefferson, the third President of the United States.

Our stay at the University in Williamsburg was short, but the physicists there were very friendly. A particular highlight, however, was meeting Nathan Isgur again, this time at Jefferson Lab. I had first met Nathan at the Les Houches Summer School in 1976, where I lost to him in the quarterfinal of a table-tennis championship. Nathan made important contributions to heavy-quark physics and, twenty-five years later, received the prestigious J. J. Sakurai Prize in 2001 jointly with Mark Wise and Mikhail Voloshin. Unfortunately, he passed away in July 2001, only a few months after receiving this award.

In 1996, however, he was in very good shape and showed us the laboratory, where he was head of the theory division. I met him again two and a half years later at CERN, where he told me that he had cancer and that it would kill him. I was deeply impressed by the calm and direct way in which he conveyed this to me. It is truly a pity that he died at the age of 54. He was not only a first-class physicist but also a very kind person.

Our stay in Charlottesville was also very pleasant, as I met there, after many years, P. Q. Hung, with whom I had worked at Fermilab in 1981, as described in Section~\ref{Petite}. We also met his very nice Italian wife, Simonetta Liuti, a nuclear theorist. Together with their children and Simonetta’s parents, they visited us in Ottobrunn in June 1998. After that visit, I met P. Q. in person only once more, in 2010, at a meeting organized by me at Lake Starnberg, south of Munich.

Our final drive was to Washington, D.C. During our three-day stay there, we visited many places, in particular the United States Capitol, the National Mall, the Washington Monument, and, of course, the White House—although only from a distance. We also went to Chinatown. Altogether, it was a very interesting expedition.

  \subsection{ICHEP 1996 in Warsaw}
In July 1996, the ICHEP conference took place in Warsaw, with parallel sessions held at the main campus of Warsaw University and the plenary sessions in the Congress Hall of the Palace of Culture. This hall can accommodate 3,000 participants. At that time, the number of ICHEP attendees was around 1,500, so only half of the hall was filled; in the 2020s, this would correspond to about a quarter.

I was invited to give a plenary talk on flavour physics. As can be found in \cite{Buras:1996wn}, my talk covered topics such as the Unitarity Triangle, the ratio $\epe$, CP asymmetries in B decays, the rare decays 
$\kpn$ and 
$\klpn$, and a glimpse of physics beyond the Standard Model. Remarkably, 30 years later, these topics remain highly relevant.

My talk was well received, but something unusual happened—an incident that was repeated ten years later at a conference in the Galápagos. The chairman, Lincoln Wolfenstein, who had been watching my talk attentively, suddenly looked at his watch and mistakenly informed me that I had only five minutes left, although in reality I had fifteen. As a result, I had to skip several transparencies.

Today, such a situation would be disastrous, as talk durations have been reduced over the years, typically to 30 minutes. In 1996, however, I had at least 50 minutes for my talk, followed by discussion, so cutting ten minutes was unpleasant but manageable—it was not a disaster.

 \subsection{Les Houches 1997}\label{LesHouches97}
Twenty-one years after attending the Les Houches Summer School in 1976 as a student, I was invited back to Les Houches to give a series of lectures on flavour physics, effective Hamiltonians, renormalization group techniques, and QCD corrections. I was well prepared for these lectures, having previously given a series of lectures on renormalization group methods at the Technical University in the first half of the 1990s and conducted related research. Initially, I was asked to give twelve 45-minute lectures, but the students were so enthusiastic that I eventually delivered sixteen lectures.

Les Houches is close to Chamonix, and I drove there in our Toyota Camry. This time I had a large room to myself, in contrast to 1976, when I shared a small room with Keith Ellis. My interactions with the students were very rewarding: we went hiking, danced in the evenings, and played table tennis. This time, I reached the semi-finals, defeating Andrzej Czarnecki in the quarterfinal. He was already far ahead of the other students and an expert in higher-order QCD calculations. Realizing this, I suggested a collaboration on NLO QCD calculations for the 
 $B\to X_s\gamma$ decay\footnote{There is even a photo in the proceedings of the school showing the two Andrzejs—Czarnecki and myself—deep in discussion about this project.}. This collaboration came to fruition a few years later in two papers completed in 2001 and 2002, together with Mikołaj Misiak and Jörg Urban, who were postdocs in my group at the time. I have described these papers in Section~\ref{bsgamma}.

Anesh Manohar was also lecturing at this school. I had met him previously at Fermilab, but at Les Houches I also met his wife, Elizabeth Jenkins. I would meet her again a few years later at a summer school in Mexico. Both Anesh and Elizabeth made important contributions to particle physics, particularly to the Weak Effective Theory (WET) and the Standard Model Effective Field Theory (SMEFT) in the 2010s, although they had already authored other significant papers earlier in their careers. For all these contributions, they received the Sakurai Prize in 2025.

Overall, these two weeks were extremely enjoyable. But what eventually proved most important for me were the lecture notes, which had no page limit. By June 1998, I had written a 250-page opus, which became the most cited paper I have written alone, surpassing 1,000 citations in 2025 \cite{Buras:1998raa}. I had simply decided to write up my lectures at the University, which contained far more detail than the material presented at Les Houches. These notes also turned out to be very useful later when I wrote my book \cite{Buras:2020xsm}.

 \subsection{Australia with Robert (2000)}
At the turn of the millennium, in March 2000, I was invited to the 3rd International Symposium on Symmetries in Subatomic Physics in Adelaide, Australia \cite{Buras:2000pj}. This conference, straddling the border of particle and nuclear physics, was organized by Anthony Thomas, a colleague of Wolfram Weise. I will highlight a few events from this trip, which I hope will be amusing.

It turned out to be one of the most interesting trips I have taken in connection with physics conferences. Our son Robert joined me on this journey. We decided to arrive ten days before the conference to explore the country, since attending only the four days of the conference made little sense given the length of the trip.

The flight to Adelaide was even longer than my journey to Kyoto in 1985. Our route first took us to Kuala Lumpur, with a stopover in London. The flight from London to Kuala Lumpur lasted 12 hours. We stayed there for seven hours to rest in a hotel, which was not very close to the airport—about 40 minutes by taxi. Apparently, the travel company had made a mistake in the arrangements. On the bright side, staying at that hotel allowed us to see a bit of the country, which was very different from Europe, the USA, and Japan.

The subsequent flight to Adelaide took around 10 hours. By the time we arrived, we were quite exhausted. In Adelaide, we stayed in a dormitory for several days, visiting the university and even having lunch with Wolfram Weise, who happened to be visiting Adelaide at the same time.

While in Adelaide, we rented a car and drove to various places, including Kangaroo Island south of the city. However, the most memorable encounter with kangaroos happened in a park north of Adelaide. On Kangaroo Island, there were many tourists, but in this park, we were practically alone—except for the kangaroos. During the first half-hour of our walk, we did not see any. Suddenly, about 50 meters ahead, a group of more than ten kangaroos appeared. One of them, quite tall, was standing directly on our path, watching us very seriously. Fearing it might attack, we carefully stepped aside and continued our walk through the trees, constantly under the watchful eyes of the tall kangaroo.

Once we had put a safe distance between ourselves and the kangaroos, we returned to the main path and did not encounter them again. During the walk, we discovered the entrance to an underground tunnel. Robert decided to explore it, and I could not stop him, so I followed. The tunnel was at most one meter high, forcing us to crawl on all fours. Fortunately, it was only about 30 meters long, and not completely dark. I was genuinely relieved once we emerged on the other side.

The next day, we flew to Sydney. We had booked a room in a hotel, but were dissatisfied because the air conditioning was very loud and could not be switched off. After complaining, we were offered no ordinary rooms, but instead an admiral suite at the same price. It was very spacious, with two bedrooms on two floors. I slept downstairs, while Robert took the upstairs room.

The next day we visited the well-known sights of Sydney, in particular taking a guided tour of the Sydney Opera House. In the evening, we decided to attend a concert there. While preparing for the trip, I had made sure that I had appropriate clothing for the opera, and the same was true for Robert. Unfortunately, he changed his mind at the last moment, although this was not the real problem.

The real problem was that Gurli had packed our clothes and, instead of putting aside Robert’s trousers, she put aside mine. As a result, I discovered in Sydney that I had no trousers suitable for the opera. Since Robert is slimmer and taller than I am, borrowing his was not an option.

The concert itself was not particularly interesting. Moreover, it was rather annoying that some Japanese tourists took photographs during the performance, despite repeated requests by the opera authorities not to do so. Still, we can at least say that we attended a concert at the Sydney Opera House.

The following day, we rented a car and drove to the Blue Mountains, west of Sydney, where we stayed for a few days. This was the only time I have driven a car sitting on the right side and driving on the left lane. I found it more difficult than driving in England and Scotland a few years earlier. With the help of Robert—who screamed whenever I drove too far to the left—we nevertheless survived the trip.

In the Blue Mountains we went on two hikes. They were quite different from hiking in the Alps. Instead of climbing uphill, we were mostly going down, almost like stepping into the sea, but with trees instead of water. These hikes were very interesting and memorable.

After returning to Sydney and spending one more night in the same hotel—this time in an ordinary room—we flew back to Adelaide. Over the following four days, I attended the conference, while Robert continued exploring the city, except for attending my talk. After the conference, we flew back to Munich, stopping for a few hours in Kuala Lumpur and in London.

This was the only time I visited this part of the world. A few years later, there was an opportunity to attend a conference in New Zealand, but for various reasons I did not go. I regret this decision, as it is a beautiful country. On the other hand, Robert not only visited other parts of Australia several years later, but also New Zealand.

 \subsection{Meeting Gerardus 't Hooft (2000)}
A few days after returning from Australia, I traveled in the last days of March to Dresden to give a talk at the Annual Meeting of the German Physical Society. This was in fact the second time I had been in Dresden. The first visit was in 1984, when, as described in Section~\ref{ICHEP84}, I met Julius Wess at the ruins of the Frauenkirche. It turned out that in 2000 I met another great physicist in Dresden.

One evening, a concert was organized in the Semper Opera for the participants of the meeting. For some reason, however, I was not interested in attending and decided instead to stay in the hotel and have dinner at the restaurant on the top floor. When I arrived, I was surprised to find that there was only one other person dining there. It was Gerardus ’t Hooft, whom I had met many years earlier in Rome at breakfast in the hotel where we were both staying. He knew me, possibly also from our work on DQCD based on his large-N ideas, as well as from our 
$\overline{\text{MS}}$ paper.
 I asked him whether I could join him, and he agreed.

We spent about two hours talking about many things, but let me mention just two of them. One was his discovery in 1972 of the negative sign of the $\beta$ function
in non-abelian gauge theories and the reaction of his PhD supervisor, Tiny Veltman, who simply said, “So what?” I already knew this story from Tiny during my time at Fermilab. I had also seen a footnote by Kurt Symanzik in the proceedings of the EPS High-Energy Conference in Marseille in 1972, which suggested that Symanzik, too, did not fully appreciate the importance of this result at the time.

Since ’t Hooft had received the Nobel Prize together with Tiny just four months earlier, he told me a great deal about his days in Stockholm and about how he and the other Nobel laureates were treated there. This was the last topic we discussed, and he concluded it with a rather amusing remark: “I can really recommend getting the Nobel Prize to you.”

 \subsection{Erice School 2000}
In January 1999 I was contacted by the secretary of Antonio Zichichi, inviting me to give a lecture on $\epe$
 physics at the Erice School, which I had attended back in 1972. I should add that in 1972 I had some difficulty following the lectures in detail, owing to my rather poor knowledge of particle physics at that time. Consequently, at the end of the school I voiced some mild criticism of the lectures. Apparently, this had not been forgotten even 27 years later. When I told the secretary that I had participated in the 1972 school, she reacted very strongly: “Yes, we know!”

However, I already had other commitments in 1999, and we agreed that I would give the lecture in 2000 instead. I flew to Palermo, with a stopover in Rome. After landing in Palermo, I could not find my suitcase. There were two baggage belts on which it was supposed to appear, but both eventually turned out to be empty, and my suitcase was missing.

I was to be picked up by a driver who would take me to Erice. When he arrived, I told him about the problem and explained that I had to report to the airport personnel that my suitcase was probably still in Rome. To my surprise, the driver immediately went to the personnel standing near the relevant baggage belts. I only heard the name “Zichichi,” and suddenly the driver asked me whether the single suitcase that had just appeared on one of the belts—five minutes earlier still empty—was mine. It was. Rather amazing.

My lecture was among the first of the school. I began with a historical introduction, but Zichichi interrupted me almost immediately and started correcting my statements about the history. I no longer remember exactly  what it was. However, during a break he brought his book SUBNUCLEAR PHYSICS: The First Fifty Years, Highlights from Erice to ELN \cite{Zichichi:1998nz} and explained to me what, in his view, was wrong with my historical account. At the end of our discussion he gave me a copy of the book.

After returning to Munich I did not read it, being busy with other matters, and eventually I forgot about it altogether. Many years later, while going through my books, I opened it again and, to my surprise, discovered a dedication that I had missed in 2000: “To my friend A. J. Buras, who knows how to compute $\epe$.” Possibly, had I read this dedication already in the year 2000, I would have been even more motivated to make further progress on $\epe$.
The proceedings of this school can be found in \cite{Zichichi:2001yz} and my lecture in \cite{Buras:2001pn}.

 \subsection{Mexico 2001}\label{Mexico2001}
In 2001 I was invited to give two lectures at the VIIIth Mexican Workshop, which took place from 14 to 20 November in Zacatecas, a town in the northern part of Mexico. I flew from Munich to Mexico City, where Piotr Kielanowski picked me up at the airport and brought me to his apartment, where I stayed during my visit to Mexico City.

I knew Piotr from my physics studies in Poland; he was a few years older than me. Since I had left Poland in 1971, we had practically no contact with each other, and I was pleasantly surprised to be staying in his home. His daughter Janka was living there as well; she was about 20 years old at the time.

During the first days Piotr showed me the center of the city and the Mayan pyramids outside the city. It was my first visit to this part of the world, and I was very impressed by the differences compared to European cities. One of the highlights was a visit to Frida Kahlo’s home, a very interesting museum. In addition to her paintings, one could also visit the room of Leo Trotsky, who spent some time there. At the end of this visit I bought a poster of one of her self-portraits. It hung in my office in the physics department until I retired in April 2012, and it is now hanging in my office in the IAS building.

After spending several days in Mexico City, we drove to Mérida in Yucatán, where I gave a colloquium in the physics department of one of the universities. Somehow nobody had told me that there were only three particle physicists in the audience. My colloquium was clearly too difficult for this audience, but apparently I was regarded there as a famous physicist, and simply listening to me was already considered a major experience. This impression was confirmed by my hotel room, which was the most impressive one I have ever stayed in during my life.

After this visit we drove to Cancún and Playa del Carmen, holiday towns on the east coast of Yucatán. On the other side of the water was Cuba. We then drove south to visit Chichén Itzá, where we once again climbed Mayan pyramids. Next we continued to the border with Guatemala to take a boat trip on a river there. The boat was relatively small, and seeing crocodiles on the shore was an interesting, but also somewhat scary, experience.

After this very interesting trip we returned to Mexico City for two days, and then drove on to Zacatecas, where the workshop was held. On the way we stopped at a hotel. Our room, including the bathroom, was rather dirty, and for a while we considered complaining and changing the hotel, but in the end we stayed.

The next day we arrived in Zacatecas. My room there was at least 
$40 m^2$  large and could almost compete in quality with the one in Mérida.

The workshop itself was very enjoyable. I met Elisabeth Jenkins again and was able to talk to her much more than in Les Houches. I also became acquainted with two very nice CDF experimentalists, Robin Erbacher and her husband John Conway. I spent most of the free time with Elisabeth, Robin, and John. This was particularly the case during an excursion and during a dinner at the restaurant Mama Inés in Zacatecas.

In this restaurant it was possible to draw paintings directly on the wall. I drew a QCD penguin diagram, and John added the corresponding penguin. This drawing, together with a photo of us, is documented at the beginning of the proceedings of the workshop. Apparently, the next day people were talking about two Australians who had visited the restaurant and drawn these penguins.

My lectures, published in \cite{Buras:2002ix}, went well. Looking at them recently, it is clear that impressive progress has been made on the experimental side since then. In particular, the measurements of the branching ratios for the decays  $B_{s,d}\to\mu^+\mu^-$
have improved by about three orders of magnitude. The CKM matrix is also much better known today, thanks not only to experimentalists but also to theorists.

After the workshop we drove back to Mexico City, where I spent two more days before taking my flight back to Munich. It was one of the most interesting physics trips of my life. Possibly only my trips to Japan, Australia, the Galápagos, and British Columbia could compete with it. This was certainly thanks to Piotr, who showed me the country, and to the great hospitality of the workshop organizers.

\subsection{Zakopane School 2003}\label{Zakopane2003}
In 2003 I was invited for the first time to the Zakopane School, which had been organized for 43 years by particle theorists from the Jagiellonian University in Kraków. I first flew to Kraków to join the school bus that would take us to Zakopane.

Zakopane, a town in the High Tatra Mountains, was where I had spent most of my summer and winter vacations while living in Poland, the last time being in 1970, 33 years earlier. While it was moving to see Zakopane again after so many years, I was somewhat disappointed that the town no longer resembled the one I remembered from 1970. There were far more tourists than back then, but what bothered me most were the many new modern shops and cafés, which had altered the appearance of the town considerably.

Fortunately, the mountains remained as I remembered them in 1970. In my free time, I took several short hikes by myself, and one full-day hike with several other teachers and students. We hiked to Hala Gasienicowa and then through the mountain pass between Kościelec and Mały Kościelec. In fact, this was the first time I had crossed this pass, despite having visited Hala Gasienicowa surely more than twenty times.

One of the lecturers at this school was Bob Jaffe from MIT, whom I had already met in Erice in 1972 and at the peculiar EST meeting in San Francisco in 1980, which I described in Appendix~\ref{EST}. During the hike to Hala Gasienicowa, Bob realized that I had never been to Boston, and after the school he invited me, together with Frank Wilczek, to spend some time at MIT.

Although I had actually been planning to visit MIT in one of the coming years, for various reasons this trip never materialized, and it now seems unlikely that I will ever visit MIT and Harvard. As is clear from this autobiography, the years following the Zakopane School were extremely busy for me. I attended numerous conferences and participated in several workshops. Moreover, the number of my diploma and PhD students increased dramatically: from 2004 to 2010 I supervised more than ten PhD students, and a similar number of diploma students. I therefore had practically no time to go to Boston.

While it would certainly have been interesting to visit Boston as a city—not just Harvard and MIT, but also Boston University—the research topics explored by physicists there at that time were quite different from mine.

Another lecturer at this school was Andrzej Czarnecki, whom I had met at the Les Houches Summer School in 1997. I had also collaborated with him, Mikołaj Misiak, and Jörg Urban on NLO QCD corrections to the $B\to X_S\gamma$ decay
in 2001 and 2002, as mentioned above. While sitting with him in a coffee shop on the main street in Zakopane, Krupówki, and enjoying ice cream, Andrzej suggested that I visit him as a “Distinguished Lecturer of the University of Alberta.”

This is the subject of the next story.

My lectures from this school can be found in \cite{Buras:2003jf}, and the proceedings of the school in \cite{Praszalowicz:2003na}.

\subsection{British Columbia and Edmonton  2004}\label{BritishC}
After I received the official invitation from Andrzej in the spring of 2004 to spend three weeks in Edmonton, where the University of Alberta is located, and give a few lectures, my family decided to join me. However, we soon realized that spending three weeks in Edmonton alone would not be optimal, since the most beautiful part of Canada is British Columbia. We therefore decided to go first in August for three weeks to Canada, spending most of the time in British Columbia, and I would return alone in late September to Edmonton for two weeks to give a series of lectures.

Actually, only Gurli, Karin, and Allan were able to make the trip to British Columbia, as Robert had to attend a workshop in Seattle at the same time. A detailed report of our expedition to British Columbia was written by Gurli, but as it is in Danish, I will not present it here. Suffice it to say, it was a fantastic trip.

At the end of September I flew to Edmonton, where Andrzej picked me up at the airport and brought me to a university apartment of about $60 m^2$—far too large for one person. The next day I visited the university, and my first lecture was scheduled for the following day.

In July, Andrzej had informed me that I would replace him in giving a series of lectures on neutrino physics, a topic rather poorly known to me. Still, I knew a little. Indeed, when applying for the professorship at TUM in 1987, I had been asked by Rudolf Mössbauer to give a lecture on neutrino physics, which had been the focus of his research in the final years of his life. I therefore had some notes on the topic, but they were already 17 years old and far from sufficient for three lectures in 2004.

Although I was a bit surprised by this request, I thought it was a great opportunity to finally learn more about this important field. I began preparing the lectures in Munich, but I still had to refine them in Edmonton. Jet-lagged after the long trip, I found myself waking up as early as 1 a.m. to work on them. In the end, the lectures went well, although they were apparently a bit too technical for some of the fifteen students who attended. But, as in Mérida, according to Andrzej, it was important for the students simply to hear a known physicist speak. Nice!

One afternoon there was a casual meeting of all the professors, and I was invited to attend. One of the professors, whom I did not know, approached me and introduced himself as the dean of the physics department at the University of Alberta. He asked who I was. My answer was very simple: “Distinguished Lecturer of the University of Alberta.” He seemed a bit surprised, but Andrzej found it perfect. He was amused and kept repeating this short story to several people afterward.

During one weekend, Andrzej invited me to his summer house, where I met his wife and three children—a very nice family. One of the highlights of the visit was seeing the devastation of their vegetable garden by bears, who had apparently visited during their absence. During the second weekend, I was invited to a party at the home of Andrzej’s parents. It was a very pleasant evening, except for one discovery after dinner: apparently, the meat we had eaten was bear, shot either by Andrzej’s father or by Andrzej himself.

At first I was a bit surprised, but I quickly realized that I am used to eating pork and beef simply because that is what is commonly served in Europe. Therefore, unless one is vegetarian, one does not really have the right to criticize people for eating bears, kangaroos, or even dogs. All animals are equal, after all.

\subsection{100 Years of the Photon: 2005}\label{Photon05}

In 2005, an international conference on the structure of the photon, celebrating its 100th birthday, was organized by Zygmunt Ajduk, Marysia Krawczyk, and Andrzej K. Wróblewski. The first part of the conference took place in Warsaw, and the second in Kazimierz, a very charming town about 100 km from Warsaw.

I was invited to give a plenary talk on QCD corrections to photon structure functions. In 1978, I had calculated NLO QCD corrections to these functions in collaboration with Bill Bardeen \cite{Bardeen:1978hg}. The LO contributions had been calculated by Witten \cite{Witten:1977ju}, and I had hoped he would attend the meeting. I first met him in 1977 at a conference at Argonne National Laboratory, just one month after my arrival at Fermilab. Unfortunately, Witten did not come to this conference, so I had to report on his work as well. I asked him to send me his photo, which he kindly did; it can be found in my talk, though not in the written version \cite{Buras:2005nj}. The full proceedings of this conference are published in \cite{Ajduk:2006gb}.

At this meeting I also met David Gross again, whom I had seen several times before\footnote{The first time was in Aspen in 1978, where his daughter was babysitting Robert, who appears several times in my story. Later meetings included 1979 in Princeton, 1980 in Copenhagen, 1988 at ICHEP in Munich, and much later in Santa Barbara. In the last five years, I met him twice a year via Zoom for the Committee of the Max Planck Medal of the German Physical Society. All very pleasant meetings.}. This time he came, as far as I remember, with his wife. Marysia Krawczyk took care of them, and one evening we all went together to a charming café in a side street off Nowy Świat, an important street in the center of Warsaw. It was a very enjoyable evening. Another highlight for me was the concert at the Warsaw Philharmonic, as it was the first time I had visited it since leaving Poland 34 years earlier.

At this meeting I also met Witek Augustyniak, with whom I had studied physics. In 2006, all of us, along with our physics colleagues, would turn 60, and I had the idea that we should meet in Warsaw the following year to celebrate our 60th birthdays. Witek thought this was an excellent idea, and although it was my suggestion, it was he who brought it to fruition. We held two gatherings: one in 2006, and a second in 2014, celebrating the 50th anniversary of the beginning of our studies in 1964. It was wonderful to reconnect with so many colleagues.

During my physics studies, I had very good contacts with Zygmunt Ajduk, who was an excellent assistant, and even closer contacts with Marysia Krawczyk, with whom I studied physics and whom I met at several conferences afterward. The last time was in April 2017, at a conference in Portorož.

Unfortunately, Zygmunt died in a car accident in 2012, and Marysia passed away just before the Planck Conference 2017, which took place in Warsaw only one month after I had seen her in Portorož. It was a heart attack. She must already have been having problems in Portorož, because despite staying at the same hotel, we barely spoke. She would arrive late to breakfast and often sat alone.

Marysia was a very kind person and very active in organizing conferences, particularly the Scalars meetings in 2011, 2013, and 2015. It is truly a pity that she passed away so early.

\subsection{CERN School at Saint Feliu de Guixols (2005)}\label{Feliu}
This school was organized by Robert Fleischer together with a few experimentalists. I gave three lectures, which were well received, with one exception that I will report below. There were a few episodes that I would like to mention here.

First, while all other lecturers were teaching using laptops, I was the only one relying on hard copies of transparencies prepared in PowerPoint by my secretary. This turned out to be a complication, as the overhead projector was not of particularly high quality. In fact, this was the last time I gave a talk using transparencies. In retrospect, this proved to be a big advantage. Until then, I had to carry my transparencies—sometimes collected in two folders—with me as hand luggage. I had to take them onto the plane, and occasionally their weight was close to the maximum allowed. I do not really know why I kept doing this. On the other hand, it did prompt me to buy a special, very nice leather bag that could accommodate two folders. After this school, with my talks stored either on a USB stick or on my laptop, I could use this bag on subsequent trips to schools and conferences to carry many more things as hand luggage.

In the evenings, the students played table tennis. At the time I was approaching 59, and they were a bit surprised when I suggested that I could join them. After I won the first match, several students wanted to play against me, so that I ended up playing a series of matches and winning all of them. At some point I was so exhausted that I stopped playing and told them that if I continued, I would not be able to give my lecture the next day.

Finally, not everybody liked my lectures. After my last lecture there were a few reasonable questions. Then a female student, who throughout the school appeared very confident, asked me what I was actually trying to tell the students in my lectures. Fortunately, I found an immediate answer. I told her that I wanted to convince her to work with me. This left her speechless. The write-up of my lectures can be found in \cite{Buras:2005xt}.

\subsection{Rome 2006 and 2009}\label{Roma06}
In 2006 I was invited by Luca Silvestrini to spend three weeks at the Physics Department of Roma 1. Gurli accompanied me for the first ten days. We rented a nice apartment in Trastevere. Although I had visited Rome several times before, I did not know this part of the city very well.

We arrived on Friday, April 21, a few days before the Italian holiday on April 25, and I did not go to the University during the first days of our stay. Instead, we spent this time together visiting various places in Rome and recalling our previous visits in 1974 and 2001.

During the first week of my stay at the University, I worked in the office of Guido Martinelli, who was attending a Ringberg workshop near Munich that week. It was a spacious office. However, there was a drawback. Guido was called many times during the day and answered the calls in Italian, often with the help of the answering machine. After a few days, I knew these sentences by heart. From time to time students knocked on the door and were either surprised by my presence in Guido Martinelli’s office or, if they did not know him personally, started talking to me in Italian.

Nevertheless, this office was on the whole significantly better than the one I had during the rest of my stay. After Guido’s return, I was assigned a new office with just two tables and two chairs. I shared it with Andi Weiler, who had come to Rome to make progress on our flavour analysis in the LHT model, which involved several other PhD students \cite{Blanke:2006sb}. I describe this work in Section~\ref{LHT1}.

Not much happened that week apart from working on our paper and one accident. One day Andi decided to stay longer at the University to carry out some numerical calculations. At around 23:00 he decided to go home, but to his surprise the building was locked. He could not leave through any door and could not find anyone who could help him.

Being used to climbing in the mountains, he decided to leave the building via a balcony above the main entrance of the Roma 1 building. This required climbing down a wall that was surely about five meters high. The attempt ended with Andi breaking his leg. This was bad enough in itself, but in addition we were both supposed to go at the end of May to the Planck 2006 conference in Paris and later to the Galápagos Islands. It is impressive that Andi nevertheless managed to go to both places despite his broken leg.

In 2009 I went to Rome again. Cecilia Tarantino invited Monika and me to spend a week at Roma 3 with the goal of writing a review on the LHT model, which we had studied so extensively a few years earlier when Cecilia was working at TUM. This project was not realized, but it was nevertheless a very nice week.

We met Guido Altarelli, Marco Ciuchini, and Vittorio Lubich. This time Monika and I gave two separate talks. I still remember discussing $\epe$
 with Guido. In his view, similarly to $(g-2)_\mu$,
it would take a very long time to calculate $\epe$
using LQCD. He was right. In this decade, LQCD has made very significant progress in calculating $(g-2)_\mu$.
 Let us hope that similar progress will be achieved for $\epe$
still within this decade.

\subsection{Galapagos World Summit on Physics Beyond the Standard Model: 2006}

In June 2006, a conference took place on San Cristóbal Island. It started on Thursday, June 22, and ended on Sunday, June 25. The meeting was organized by Ecuadorian experimentalists working at Fermilab, with the purpose stated on the conference webpage:

“We hope the Galápagos meeting will be to Particle Physics what the Copenhagen meeting was to Quantum Mechanics. A new paradigm lies beyond the Standard Model. The clues hide in experiment (high $p_t$
searches, precision measurements, rare decays, neutrinos, cosmology, astrophysics, cosmic rays) and in a multitude of theoretical ideas. Let us advance the solution of the puzzle!”

Because of the location, many excellent scientists participated, including two Nobel Prize winners, Leon Lederman and Frank Wilczek. I was invited to give a 90-minute talk, which opened the meeting. Among the speakers, I knew several from previous conferences. Beyond the two Nobel laureates, there were Guido Altarelli, Gabriela Barenboim, Riccardo Barbieri, Álvaro de Rújula, John Ellis, Gordon Kane, Ulrich Nierste, Chris Quigg, Roberto Peccei, Stuart Raby, Lisa Randall, and Stan Wojcicki.

As I was approaching 60, I thought it would be safer to travel to this meeting with someone. Since no family members could or wanted to join me, I chose Andi Weiler as my companion.

We first traveled to Quito, the capital of Ecuador, where we spent a few days visiting the University. We were picked up at the Quito airport. Already after leaving the plane, two men were waiting at the stairs holding a piece of paper with our names. They told us they would handle our baggage and asked us to wait in a luxurious lounge with TV and drinks. We were rather impressed with this service—until we realized that Ecuador was playing against Colombia in the 2006 World Cup, and there was no way to get our baggage in the usual way because everyone was watching the game. So we had to wait as well.

After the match, I got my baggage, but Andi’s luggage did not arrive. This was particularly unfortunate—not only because of his clothes, but also because his camera cable was in the missing luggage. Consequently, for several days, until his baggage arrived, he had to limit his photography.

We were taken to the Holiday Inn, and the first thing we did was go to a shop to buy some clothes for Andi. The next day, we were invited to lunch with the president of Quito University. The meal was served by students from the cooking faculty and consisted of French food, very nicely prepared. After lunch, the president showed us around the campus, and in particular the bookstore, where we could see his book on cooking recipes. He was a physicist, and apparently, when teaching statistical physics, he also presented his cooking recipes to students—much to the criticism of his colleagues. Subsequently, we went for a walk in downtown Quito.

The next day we went to the airport and took a flight to the Galápagos Islands, landing on Baltra Island. We spent one night there exploring the surroundings, and the following day a car brought us to a ship, which became our home for the next four days. We shared a cabin, which was sufficient for us, as most of the time during those four days we spent on deck or participating in excursions on various islands. In this way, we were able to see animals that are rarely encountered outside the Galápagos, in particular the Galápagos tortoise, marine and land iguanas, and many others—the full list can be found on Google.

Those four days on the ship were a new experience for me. I found the tours interesting, but otherwise, the routine—breakfasts, lunches, and dinners in the same restaurant—was rather monotonous. Eventually, we were both glad that the tour did not last longer.

On June 21, we arrived at San Cristóbal Island. The hotel where we and the other participants were accommodated was, by European standards, a two-star hotel. Only Leon Lederman and Frank Wilczek were placed in a better hotel, probably at a three-star level. But this was perfectly fine.

The conference began the next day, with me as the first speaker. The talk went well, except for a small misunderstanding with the chair, Giulia Ricciardi. In the final part of my presentation, Giulia signaled the remaining time using one of
her hands and one of the fingers of the second hand.
She meant 15 minutes, but I interpreted it as six minutes, which prompted me to skip several transparencies. However, since my talk was one hour long, followed by a 30-minute discussion, this was not really a problem.

The entire conference was very interesting, with several different perspectives expressed by the speakers. Moreover, I had the opportunity to meet several physicists whom I had not seen for years. It also allowed for some photos, and in Fig.~\ref{AJBWilczek} I show one with Frank Wilczek, whom I had already met in Iowa (1978) and Princeton (1979), as well as at the celebration of Heisenberg’s 100th birthday in December 2001 in Munich, where he gave a talk and I served as his chairman.

\begin{figure}[t]
\centering%
\includegraphics[width=0.8\textwidth]{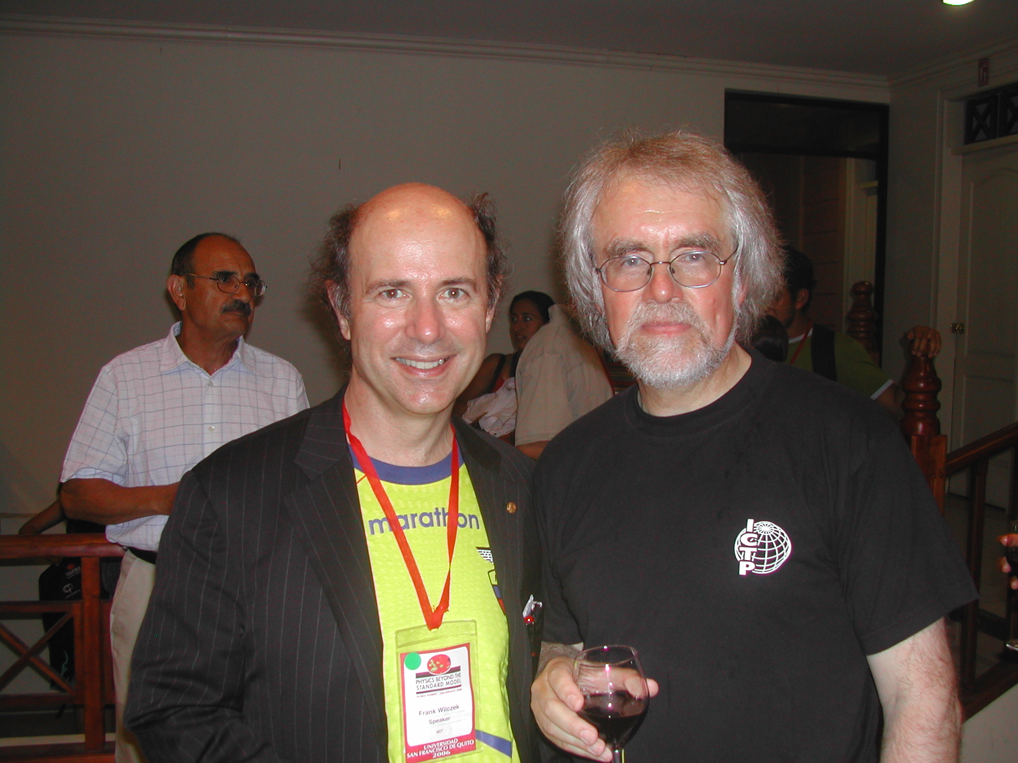}%
\caption{\it Frank Wilczek and AJB at Galapagos Islands (2006).\label{AJBWilczek}}
\end{figure}

But let me now report on two events that had nothing to do with physics.

One of the days of the meeting was devoted to excursions. We chose the one with a small boat that took us to other islands. Frank and Betsy Wilczek were on the boat as well. At one point, the boat stopped, and people were invited to dive into the water. Several participants did so, including Frank and Betsy. Being a poor swimmer, I did not join them. One should add that one highlight of this dive was the possibility of encountering sharks. In most parts of the world, this would be extremely dangerous, but the sharks around the Galápagos Islands are very different—they do not attack humans. Still, the main organizer of the meeting told me that one day he had seen a group of eight sharks swimming toward him and had gotten scared, momentarily thinking it was the end. Fortunately, the sharks simply passed by.

One of the evenings featured dancing, dominated by Mexican music. In my younger years, I was a good disco dancer. While those skills do not necessarily translate to Mexican dancing, in my case they helped. I danced with an eighteen-year-old daughter of one of the Fermilab experimentalists—unfortunately, I do not remember her or her father’s name. We ended up winning third place, after two French couples, who surely had better training than we did. Perhaps I could have done even better if I had danced with the wife of the Japanese physicist Hisakazu Minakata. She and her husband had been visiting the Galápagos before the conference on the same ship that Andi and I were on, and one evening I had danced with her. But the young girl I danced with that night was also a very good dancer. I wonder whether, twenty years later, she still remembers our achievement in this little competition.

On the whole, it was a great experience being there. On Monday, 26th June, all participants who did not stay longer on the Galápagos Islands were brought back to Quito by plane organized by the conference. During the flight, a few prizes were announced to participants. One of them was for the most energetic dancer — and that was me.

We spent one night in the same Holiday Inn in Quito. On the way back, the plane stopped on an island north of Venezuela. During that stop, somebody took a photo of us together with John Ellis.

\subsection{Durham, Edinburgh, 2017}\label{Durham}

In 2017, Alex Lenz invited me to spend ten days at the Institute for Particle Physics Phenomenology in Durham. I first flew to Edinburgh and then took a train to Durham, finally reaching the hotel by taxi. This was already the third time I visited Durham: first with my family in 1991, and then for the CKM Workshop in 2003. I found the city very pleasant.

I walked to the Institute every day, a journey of roughly 30 minutes along the river and partly through a park. I met Keith Ellis, who was the head of the Department. I knew him from the Les Houches School in 1976 and from Fermilab. Otherwise, I practically did not meet anyone except Alex, of course. While this was somewhat disappointing socially, it allowed me to make significant progress on my book, particularly through discussions with Alex. Chapter 8 was largely written during this stay.

This is also an appropriate place to highlight the very important contributions of Alex to flavour physics. I regret that we never collaborated, despite the fact that he was my PhD student (collaborating with Ulrich Nierste) and, in 2011, served as my substitute professor during my IAS fellowship. His contributions can be explored on INSPIRE and are also mentioned in Section~\ref{NLOfamily}. Here, I will just highlight two important review articles \cite{Artuso:2015swg,Friday:2025gpj}, the first of which was particularly helpful in writing Chapter 8 of my book.

After my colloquium in Durham, which was well received, I left for Edinburgh to give a talk at the Higgs seminar. The discussions there, particularly with Roman Zwicky, were very engaging. This was my second visit to Edinburgh, the first being with my family in 1991. I spent four days there, including a weekend to explore the city. I found it interesting once again, but unfortunately I was not feeling well, possibly due to the humidity. I was genuinely worried that I might have to cancel my upcoming trip to Hong Kong, scheduled for just one week after my return to Munich.

Fortunately, I recovered in Munich and did not have to cancel the Hong Kong trip, which greatly improved my spirits. Perhaps the climate in Bavaria is simply better than in Durham or Edinburgh, as I immediately began to feel better.

\subsection{Hong Kong 2017}
I was asked to organize a flavour session in the context of the so-called Gordon Conferences. In 2017, it took place in Hong Kong, and since I had never been there, I gladly accepted the invitation. I selected Christoph Bobeth, Andreas Crivellin, and Vincenzo Cirigliano as speakers. Christoph and I flew together from Munich to Hong Kong. From the airport, we took a taxi to the University, which was practically on top of a mountain. The journey was quite long, but along the way we could already see the city, with its many tall buildings. Our hotel was very close to the conference venue. We arrived a few days before the meeting began, and since I was not giving a talk during the conference itself, the organizers asked me to give a one-hour colloquium two days beforehand.

While I generally felt much better in Hong Kong than I had in Durham and Edinburgh, I developed some neck pain on the third day of the conference. The culprit was the extremely strong air conditioning in both the hotel and the lecture room, which kept the temperature at $18^\circ$ C while the outside temperature was above $25^\circ$ C. Andreas Crivellin experienced the same problem, but using Voltaren Forte he was able to reduce his pain significantly. He kindly gave me some, and after a few days of use my neck pain almost disappeared. This is not an advertisement for Voltaren Forte, but I have used it occasionally since then, and it has always helped.

During the weekend, Christoph and I took a bus to downtown Hong Kong. We walked around and found it a very interesting experience. Overall, it was a highly stimulating stay. The organizers told me that it would be great if I could visit them again for a longer period.  However, after the political situation in Hong Kong changed a few years later, I did not hear from them again, and under the new circumstances, I would not go there anyway.

\subsection{Physics Trips with Allan (2007,2022,2024)}
I have already described three physics trips with Robert—to Japan, the U.S. East Coast, and Australia. But Allan, our younger son, was also active. This time, his involvement was unrelated to special conferences; it was due to three awards I received for my contributions to particle physics, primarily in flavour physics. Allan represented our family by joining me on these three trips, all of which will remain in my memory forever.

Some readers, if they are still following my story, may criticize me for mentioning these awards. Yet this is also an opportunity to thank those who nominated me. In particular, I am grateful to Konrad Kleinknecht, Gerhard Buchalla, and Ulrich Nierste in connection with the Max Planck Medal 2020, and to Ikaros Bigi, Andreas Kronfeld, Gino Isidori, Monika Blanke, Aida El-Khadra, Chris Quigg, and Tom Browder in connection with the Sakurai Award 2024. Special thanks go to Allan, who accompanied me on all these trips.

I do not recall who nominated me for the Smoluchowski-Warburg Medal of the German and Polish Physical Societies—possibly it was Konrad Kleinknecht again.

\subsubsection{Smoluchowski-Warburg Medal 2007}
I had to wait until 2007 for the first prize. It was the Smoluchowski-Warburg Medal of the German and Polish Physical Societies. To receive it, I had to travel in September 2007 to the annual meeting of the Polish Physical Society, held in Szczecin, a city in northern Poland near the German border. As Allan was staying in Rügen at the time, we met in a city close to the border and took the train to Szczecin. We arrived on Saturday and spent Sunday exploring the city, including a cemetery where we found many graves of Russian soldiers who had died in World War II. The award ceremony took place on Monday. In addition to the medal, I received 1000 Euro.

After the ceremony, I gave a talk on flavour physics. I had prepared it expecting an audience of particle physicists, but there were practically none; a large portion of the audience were high school teachers. My talk was relatively simple, yet still too difficult for much of the audience. This was a bit disappointing, but Allan thought it was a good presentation. He returned to Rügen already on Monday evening, leaving a short and very kind message, which I discovered after returning to the hotel from dinner.

\subsubsection{Max Planck Medal of the German Physical Society 2020}

This medal, first awarded to Einstein and Max Planck in 1929, is the highest honor of the German Physical Society (GPS) in theoretical physics. It is given to physicists in all fields, with only one condition: they must have worked, even briefly, in Germany.

It is very difficult to receive this medal because theoretical physics encompasses many different fields. The selection committee consists of former Max Planck medalists. After the Second World War, most medals were awarded to formal quantum field theorists, so the chances for an applied quantum field theorist like me were not high. A well-known German experimentalist, Konrad Kleinknecht, nominated me for this medal in 2010, and a few years later my two PhD students, Gerhard Buchalla and Ulrich Nierste, also nominated me.

Only in November 2019 did I receive an email informing me that I was the winner of the Max Planck Medal 2020. This was a real surprise, as after so many years of nominations, I did not really expect to receive it. The prestige of this medal is indeed very high. In fact, I am the second Dane, after Niels Bohr (1930), and the first Pole to receive this honor. Being listed among medalists such as Einstein, Planck, Bohr, Heisenberg, and Dirac is a truly special feeling.

I was supposed to receive the medal at the annual GPS meeting, scheduled for March 2020 in Bonn. Allan was, as usual, supposed to accompany me, and  our trip  and hotel were all arranged by February. Unfortunately, the meeting was canceled due to the COVID-19 pandemic.

Similarly, in 2021 and 2022, no meetings of this type took place. The GPS eventually decided to hold a combined award ceremony for all medals from 2020, 2021, and 2022 over three days at the Bad Honnef Physics Center near Bonn. The audience consisted of medalists and a few officials—about fifty people in total—making the ceremony very different from the usual one, which typically attracts several hundred physicists. There was a Zoom connection, allowing my family to watch, although only a few other participants joined online.

While this was somewhat disappointing, I met Feli Arias and Robert Lebedka for the first time, two truly nice people involved not only in organizing this ceremony but also in many GPS events. I had already been in email contact with them in 2019 and 2020. They arranged a boat trip, which was a real highlight.

This time, we traveled by car—our Audi Q3—so that Allan could attend the entire meeting. I gave a 20-minute talk prior to receiving the medal, making sure that physicists outside my field could follow it. With only first-class scientists and no high school teachers in the audience, it was much easier to present my work.

Finally, in March 2023, I gave the talk I had originally planned for Bonn in 2020. This was at the annual GPS meeting in Dresden, where, as expected, I met Feli Arias again.

\subsubsection{J.J. Sakurai Award  of the American Physical Society 2024}

The J.J. Sakurai Award of the APS is the highest honor in theoretical particle physics awarded by the APS. Sakurai was a Japanese physicist who died suddenly in 1984 in Geneva, shortly after a visit to MPI in Munich, where I met him for the last time. I knew him from my CERN years and later worked with his student P.Q. Hung and Bj on the Petite Unification of Quarks and Leptons at Fermilab, and much later with P.Q. when he was in Williamsburg, as I have described earlier.

After several nominations beginning in 2010, I received in October 2023 the message from the APS that I would be the recipient of the 2024 Sakurai Award. This was the first time in the forty-year history of the award that it was given to a theorist working in Germany, to a Dane and a Pole, and only the fourth time that it was awarded individually to a European theorist. The previous three were Cabibbo (1989), Gribov (1991), and Leutwyler (2023). At first glance, receiving this award seemed an even bigger achievement than the Max Planck Medal. On the other hand, looking at the list of Max Planck Medal recipients—Einstein, Planck, Bohr, Heisenberg, Dirac—it is clear which prize carries more historical prestige.

What is also amusing is that, as of 2026, I am the only living scientist who has received both the Max Planck Medal and the J.J. Sakurai Award. The only person who achieved this before me was Nambu, who died in 2016. Of course, he received many other prizes, including the Nobel Prize. Still, it is a strange feeling to be the only living person on our planet who has climbed these two summits.

Now comes an amusing story. Stephan Paul, one of the leaders of the Excellence Cluster Origin, informed the members of the cluster about this award. He did so using his smartphone, which autocorrected the name of the award, so that apparently I had received the “Samurai Award.” This nickname stuck among the members of the cluster, and several even thought I would have to travel to Japan to receive it. Moreover, at the annual cluster meeting in December 2023 at Irsee Kloster, Max Gicklhorn, the son of Alice Smith-Gicklhorn, used AI to construct a photo of me as a samurai. I show it in Fig.~\ref{Samurai}.

\begin{figure}
\centering
\includegraphics[width = 0.90\textwidth]{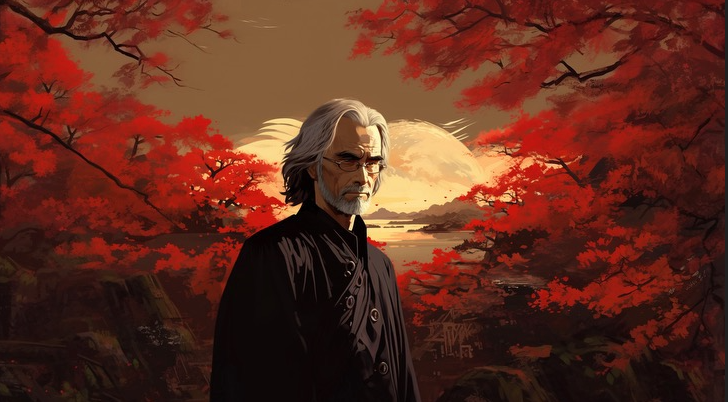}
 \caption{\it Andrzej Buras as a Samurai. Constructed by Max Gicklhorn }\label{Samurai}
\end{figure}

I received the award on April 4th, 2024, at the APS Annual Meeting in Sacramento, just north of San Francisco. Once again, Allan accompanied me, and he came up with the idea that after the award ceremony we should fly to Toronto to watch the solar eclipse on April 8th. At first, I considered this idea a bit crazy. But since Allan was accompanying me to Sacramento—which turned out to be a great help, as we will see—and the award included a prize of 10,000 US dollars, I accepted his proposal with genuine joy! As I have described previously, I had already made three long trips with Robert, and this time it was Allan’s turn to join me. I have recounted the trip in detail elsewhere; here, I will mention only a few possibly amusing aspects.

The preparation for this expedition, which took place from Friday, March 29th, to April 10th, required some work beforehand, including arranging flights, hotels, and rental cars. Everything went smoothly except for the online check-in for our United Airlines flight from Munich to San Francisco. When entering our names, we added our titles, which the computer interpreted as part of our first names. As a result, our tickets listed us as “Allandr” and “AndrzejProfdr,” which did not match the names on our passports. In my case, I also forgot to include my middle name, which was on the passport but missing from the ticket.

Fortunately, the next day everything went smoothly at Munich airport, and after a 12-hour flight we landed at San Francisco International Airport around 4pm on Friday, March 29th. The first demonstration of bureaucratic absurdity came at passport control, where we were asked whether we were lovers.

The next event, however, was much better. While renting a car—which was supposed to be a small Toyota Camry—we discovered that the Sixt agent had spent the first sixteen years of his life in Poland and spoke perfect Polish. When he asked me why I had come to San Francisco, and I told him it was to receive the APS prize, he stood up, congratulated me, and offered us a better car at the same price: a very modern BMW. It was more comfortable but also much larger. We had a bit of trouble starting it at first, as we could not find the ignition button. However, after consulting the instructions, Allan figured it out, and after a 35-minute drive, guided by the car’s perfect GPS, we arrived at the Oceano Hotel and Spa at Half Moon Bay Harbor on the Pacific coast, south of San Francisco.

We were given a very nice room with a view of the harbor. I must admit that if I had been alone, I probably would not have accepted the BMW, as I would have struggled with its size. But Allan, who has four children and drives a much larger car at home, had no difficulty at all and served as the sole driver throughout our expedition.

Our stay on the Pacific coast was very enjoyable and brought back memories of a trip in 1979 when we visited Stanford. We drove about 45 minutes south to explore Big Basin, particularly to see the towering Redwood trees, whose impressive height left a lasting impression. The long walk that followed was also remarkable—one of the most pleasant walks I have ever experienced.

On April 2nd, on our way to Sacramento, we stopped several times to take photos of the beautiful beaches, notably Grey-Whale Cove, and of the Golden Gate Bridge as we drove through San Francisco. One of these photos now serves as the background on my laptop; it is shown in Fig.~\ref{Cove}.

\begin{figure}
\centering
\includegraphics[width = 0.90\textwidth]{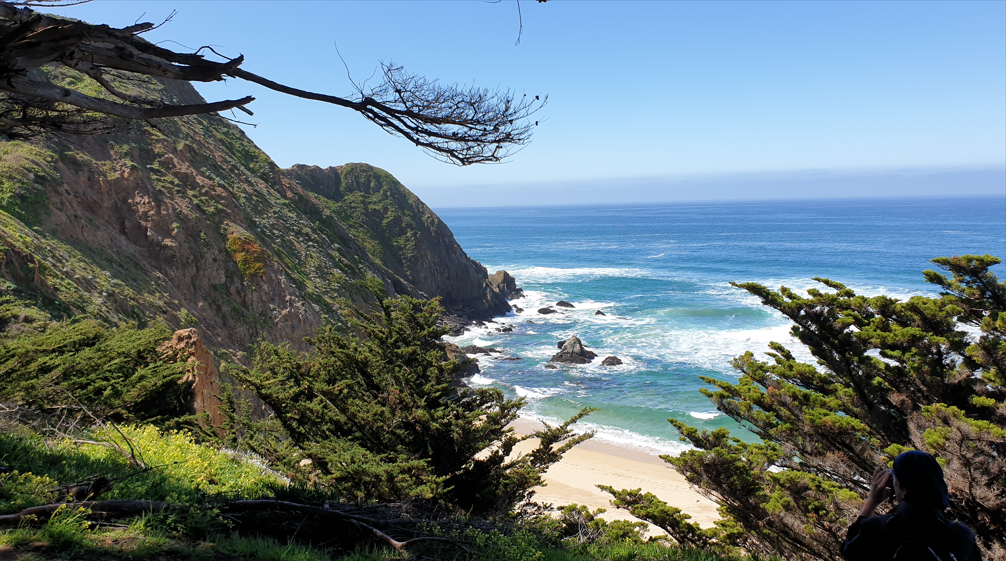}
 \caption{\it Grey-Whale Cove. Allan in the right down corner. }\label{Cove}
\end{figure}

At the APS meeting, I met several physicists whom I had not seen for years. In particular, Robin and John, whom I knew from the school in Zacatecas in 2001 (described in Section~\ref{Mexico2001}), it was very nice to talk to them again. I also met Ritchie Patterson, Mike Barnett and Steve Gottlieb. Jacky Kumar, my great collaborator who was then working at Los Alamos, also attended the meeting.

I gave my award-winning talk on the morning of April 4th. It went well, but I had expected around 100 participants, whereas only roughly 40 attended. Although the audiences for the two prize talks before mine were even smaller, it was still a bit disappointing. Flavour physics in the USA is simply not as popular as it once was and is much less prominent than in Europe.

In the evening, there was finally the celebration of the APS 2024 prizes, completing the story of the J.J. Sakurai Prize. It was a very nice event. There were 23 prizes, and their prestige seemed to increase progressively. My prize was the last among the research scientist awards, highlighting the prestige of the J.J. Sakurai Prize. Overall, it was presented next to the last among all prizes; the final award was the American Institute of Physics Tate (AIPT) Medal for International Leadership in Physics, given to Mahouton Norbert Hounkonnou from the Republic of Benin. This medal recognizes non-U.S. citizens for international leadership in physics, emphasizing leadership, statesmanship, and service to the physics community rather than research achievement.

The next day, we drove to San Francisco Airport and flew to Toronto. Over the following days, we explored the surrounding area, in particular Niagara Falls. Finally, on April 8th at 3.15 pm, we witnessed the impressive solar eclipse. Altogether, it was a fantastic expedition, and I am very grateful to Allan for making this collaboration possible.

\subsection{Meeting Artificial Inteligence}

I think this writing would be incomplete if I did not say anything about Artificial Intelligence (AI). I did not use it frequently, but one day, partly out of curiosity, I decided to ask AI for some advice about my future prospects and posed the following question.

 {\bf AJB:}

 I am an expert in Particle Physics, in particular Flavour Physics and the search for New Physics beyond the Standard Model. But I have no ideas for new projects\footnote{Not really true.}. Could you suggest some projects and collaborators?

 {\bf AI:}
 
Here are {\bf project directions} and {\bf collaborator profiles} that are genuinely promising in 2026 for someone with your background in {\bf flavor physics}  and {\bf new physics searches} but without tying to specific people or positions (to avoid giving you outdated affiliations). If you want, I can later generate a list of current researchers once you tell me which subareas you want collaborators for, and I can look them up live.

{\bf High-Impact Project Directions in Flavour Physics (2026)}

  {\bf 1. Precision global fits incorporating upcoming HL-LHC + Belle II data}
  New tensions may appear or disappear as HL-LHC and Belle II datasets grow.
  
{\bf Idea:} Build a modular global-fit framework that can:
combine flavor anomalies (rare B decays, LFU ratios, CPV, charm anomalies, strange sector, top FCNCs), incorporate SMEFT matching consistently through 1-loop,
propagate correlations correctly when combining experiments.

{\bf Why now:} Belle II datasets are finally large enough to test LHCb anomalies; many fits are not fully correlated or not consistently matched to SMEFT.

{\bf Collaborator profiles:} People working on SMEFT global fits, flavor EFTs, lattice inputs.

{\bf 2. Connecting Flavour Physics with Future Muon Collider Signals}

Muon colliders will be uniquely sensitive to flavour-violating EFT operators.
Just map the flavor-violating operator space ($\mu\to e, \tau\to\mu$, $b\to s$, etc.) into measurable collider signals:
\begin{itemize}
\item
  $\mu^+\mu^-\to\tau\mu, e\mu$,
\item
  four-muon contact interactions,
\item
  heavy flavor tagging at multi-TeV energies,
\item
deviations in Higgs-muon couplings tied to flavor models.
\end{itemize}

{\bf Why now:} The theory community is trying to define benchmarks. There is an open niche here.

{\bf Collaborator profiles:} People working on muon collider phenomenology + flavor EFT.

{\bf 3. Flavour structure of Dark Sector portals}

{\bf Idea:} Explore how dark sector mediators can generate:
\begin{itemize}
  \item
LFU violations in $b\to s \ell\ell$ 
\item
  anomalies in $K\to\pi\nu\bar\nu$
\item
  rare $\tau$ decays
\item
  electron/muon $g-2$ differences
\item
  dark photon flavor textures
  \end{itemize}
Could produce sharp predictions for upcoming NA62, KOTO-II, Belle II, PADME, VEPP-2000.

{\bf Collaborator profiles:}
Groups doing dark photon physics, ALPs, hidden sectors with flavor charges.

{\bf 4. Charm and Hyperon Flavour Physics as New Physics Probes}

{\bf Idea:} Explore poorly constrained CP-violating and rare processes:
\begin{itemize}
  \item
CP asymmetry in D decays (theory predictions tightening)
\item
$\Delta C = 1, 2$ operators (especially with lattice advances)
\item
hyperon CPV at BESIII / future STCF
\end{itemize}
These areas have far fewer theorists but promising experimental programs.

{\bf 5. High-precision lattice inputs + EFT matching for rare decays}

{\bf Idea:} Work with lattice and EFT groups to reduce theory uncertainties for:
\begin{itemize}
  \item
    $B\to K^*\ell\ell$
  \item
    $K\to\pi\nu\bar\nu$
    \item
D mixing and $\Delta C=2$
\item
$B_s$ mixing beyond the $V-A$ structure
\end{itemize} 
Even a single improved matrix element can unlock large progress.

{\bf Collaborator profiles:}

Lattice QCD groups developing multi-hadron methodologies or heavy-quark improvements.

{\bf 6. Flavour-symmetric or modular flavor models with predictive patterns}

{\bf Idea:} Use modular flavor symmetries, gauged U(1)’s, or minimal flavor violation extensions to generate:
\begin{itemize}
  \item
controlled LFU violation,
\item
neutrino-mixing-driven quark flavor anomalies,
\item
predictive correlations between $g-2$, $b\to s\mu^+\mu^-$ and rare kaon modes.
\end{itemize}
The mathematical side (modular forms, texture zeros, anomaly cancellation) is wide open.

{\bf 7. Machine Learning for Flavour Amplitude Analysis}

{\bf Idea:} ML tools to:
\begin{itemize}
\item
classify long-distance vs short-distance contributions,
\item
  emulate QCD uncertainties in rare decays,
  \item
    optimize experimental binned observables for new physics sensitivity.
    \end{itemize}

{\bf Why now}: ML is underused in flavor physics relative to collider physics.

{\bf 8. Reinterpreting existing collider data for flavor-violating mediators}

{\bf Idea:} Build reinterpretation pipelines for:
\begin{itemize}
  \item
$Z^\prime$ with family non-universal couplings,
\item
leptoquarks with hierarchical textures,
\item
  heavy vector-like quarks contributing to FCNCs
  \end{itemize}

You can use open LHC analysis toolkits to produce public reinterpretations.

{\large \bf Types of Collaborators You Might Seek}

{\bf A. EFT/SMEFT experts}

For global fits, matching, renormalization, and Wilson-coefficient evolution.

{\bf B. Belle II or LHCb experimentalists}

To shape theory proposals into real measurable observables.

{\bf C. Lattice QCD theorists}

For matrix-element improvements, especially in B, K, and D sectors.

{\bf D. Dark-sector model builders}

To explore flavored dark matter, ALP flavor couplings, $Z^\prime$ textures.

{\bf E. Muon physics community}

Muon collider advocates, muon $g-2$, $\mu\to e\gamma$, $\mu\to 3e$ groups.

{\bf F. Neutrino/flavor model builders}

Cross-talk between quark and lepton flavor sectors can produce highly predictive ideas.

{\bf Want concrete names and institutions?}

I can generate a curated, current list of active researchers and groups based on your project choice. Just tell me:
which subareas you’re most interested in (e.g., B anomalies, charm, kaons, EFT, dark sector, lattice collaboration), your region preference (Europe, US, Asia, global), whether you prefer theorists, experimentalists, or mixed groups.
Then I can fetch up-to-date people and groups for you.

{\bf AJB:} subareas: rare B decays, K decays, Lepton Flavour violation as search for new physics, region can be  global but preferably theorists.

{\bf AI} : Wolfgang Altmannshofer, Christoph Bobeth, Joahim Brod,  Andrzej J. Buras, Andreas Crivellin, Martin Gorbahn, Yuval Grossmann, Diego Guadagnoli,
Joaquim Matias, Gino Isidori, David Straub.

Except for Joahim, Yuval and Joaquim, all my collaborators. The answer was rather long
with statements what these people are good at but this you can find yourself.

As AI wanted to contact this list of physicists which would require
giving her my name, I thanked and stopped the conversation.

I think it was fun. I was impressed by the speed these answers were generated.
Most of the points listed were of course known to me but could be of interest
to younger readers. I wished AI existed when I was looking for the topic
of my PhD thesis in 1971 in Copenhagen.

One of the most popular directions in particle physics these days is the point
{\bf 7.} on this list. Collection of references on this topic can be found in 
\cite{Giarnetti:2025mit}.

\subsection{TUM Senior Excellence Faculty}\label{SEF}
Since 2007, the Technical University of Munich has awarded the honorary title TUM Emerita of Excellence or TUM Emeritus of Excellence to outstanding retired female and male scientists and has actively involved them in key tasks of the university.

As members of the TUM Senior Excellence Faculty (SEF), the TUM Emeriti of Excellence (EoE) contribute their scientific expertise, international networks, and high academic reputation to the strategic development of TUM—independently, on a voluntary basis, and with a strong personal commitment to the university.

I became a member of this distinguished faculty after my retirement in 2012. I have been very impressed by the engagement of several of my colleagues—many of them older than myself—in a wide range of faculty activities. These efforts have undoubtedly contributed in an important way to the international standing of TUM as one of the leading universities worldwide. For my part, in addition to mentoring outstanding young foreign students and serving on committees responsible for hiring new professors, my main contribution has been my research activity, which has resulted in roughly 90 publications since 2012.

As can be seen on the faculty’s website, it is highly active and regularly publishes impressive documents of various kinds. In addition to the spokespersons Georg Fäber, Michael Molls, Jörg Eberspächer, and Winfried Petry, one should particularly emphasize the remarkable contributions of Birgit Herbst-Gaebel. I very much enjoy our meetings, which take place twice  a year in the Faculty Club on the fourth floor of the TUM-IAS building, where I have my office.

\subsection{Excellence Cluster and MIAPP}
A very important role in particle physics, astrophysics and biosciences in the Garching's
research center north of Munich plays the Excellence Cluster which combines
the efforts in these fields of TUM and LMU Universities and several  Max Planck Institutes. These are MPI for Physics (MPP), MPI for Astrophysics (MPA), MPI for Extraterrestrial physics (MPE), MPI for Biochemistry in Martinsried (MPIB) and MPI for Plasmaphysics (IPP). Moreover ESO-European Southern Observarory and the
Leibniz-Rechenzentrum (LRZ) are also involved.

The Excellence Cluster started on 15th October 2006 as {\em the Universe Cluster} and in 2019 continued as
 the {\em the Origin Cluster} involving MPIB, IPP and LRZ in
addition to other institutions listed above 
and will continue at least until 2032.
It is divided in various research groups and I was the leader in one of these groups
from 2005 until my retirement in 2012. Since then I participate in the
activities of the cluster with the only duty being the member of the
award comittee for the PhD theses performed at the cluster. The annual workshops
of the cluster, arranged usually in November or early December in
Seeon or Irsee Monastery close to Munich are very interesting events summarizing
the progress made by different groups.

MIAPP, related to the Excellence Cluster (EC) but placed in a separate building
next to the EC's one, is the institution organizing workshops devoted to different subjects explored within the EC. It was born in 2011 and I was supposed to
be one of the directors of this Institute but after receiving the Advanced
ERC grant in 2011 I thought it would be too much time consuming for me and
Martin Beneke replaced me. The second director from the beginning is the
Astrophysicist Rolf Kudritzki. In 2019 Martin was replaced by Andi Weiler,
my great PhD student and collaborator.

All the details about our Excellence Cluster and MIAPP can be found in the Internet where also the members of the administration are listed.
Over the years I had in particular interactions with the directors
Andi Burkert (LMU) and Stephan Paul (TUM) and the general menagers Birgit Schaffhauser,
Katharina Langosch, Ina Haneburger. Moreover with Alice M. Smith-Gicklhorn (scientific coordinator) and the managers for science communication  Odele Straub and Stefan Waldenmaier. Michael Nies, system administrator was also very helpful in arranging laptops and monitors for me.

I could mention still other names but the ones listed above were the ones
I had most contact with until now and I wanted to thank them for everything.
I should stress that there is an excellent
atmosphere at this cluster and I enjoy very much visiting it from time to
time and taking part in the annual workshops. I am sure that in the coming
years I will also have close contacts with new directors Barbara Ercolano and Stefan
Schönert as well as with Sandra Schmid-Willers who repaced Ina Haneburger, Odele Straub who replaced Alice, Stefan W, Petra Regenscheit, Christiane Reichart, Susanne Rittel,  Sofie Kronseder
and Bernadette Rothmeier. I am looking forward to the coming 7 years.

\subsection{My New Home: TUM Institute for Advanced Study (IAS)}
It is appropriate to complete my physics memories with my new home, where
my office is now. It gives me also a great pleasure to thank the
members of the IAS-Team for fantastic atmosphare at this institute.
Let me just mention those members of IAS with whom I had most contacts
over last 18 years. These are
\begin{itemize}
\item
  Directors of TUM-IAS:   Michael Molls, Ernst Rank, Gerhard Abstreiter, Patrick DeWilde.
\item
  Managing Directors: Ulrich Marsch, Ana Santos.
\item
Senior Program Managers: Susanne Wagenbauer, Anna Kohut, Tatjana Steinberger.
\item
  Program Managers: Annette Gr{\"o}tler, Daniela H{\"a}gele, Morwenna Joubin, Clara Pappenberger.
\item
  Senior Event Manager (IAS Events) and Web Coordinator: Sigrid Wagner.
\item
  Assistants: Simone Schmidt, Priscilla Pellegrini, Nadya Neumann.
  \end{itemize}

Also the Wednesday
Coffee talks (per zoom since Corona pandemic) is a nice occassion to learn
about topics beyond physics.

Among the scientist with whom I have very much contact at IAS is Fridtjof N{\"u}sslin, TUM distinguished professor of Biomedical Physics.

There is still one person I would like to mention, with whom I have had the most contact at the IAS since May 2022. This is Julia Yamnenko, one of the ten Ukrainian scientists who received TUM fellowships in 2022 after Russia started the war. Julia was the Head of the Department of Electronic Devices and Systems at the Faculty of Electronics of the National Technical University of Ukraine in Kyiv.

In addition to electronics and electrical engineering, she is an expert in AI and machine learning, which allowed me to broaden and improve my own knowledge in these areas. Despite the fact that our research fields are very different, we found many common topics to talk about. I am particularly impressed by how quickly she learned German. While at the beginning we communicated in English, during the last two years we have used German exclusively.

Fortunately, Julia’s three children and her father live in the West. Recently, she obtained a two-year position at the Max Planck Institute for Quantum Optics, within walking distance of the IAS building, and I very much hope that she will one day secure a permanent position in Garching or Munich.

My story at TUM-IAS begins in 2006, twenty years ago—another anniversary to celebrate this year. In that year, I was contacted by an official from TUM who asked whether I would be interested in obtaining a Carl von Linde Senior Fellowship at the TUM Institute for Advanced Study. The Institute had been founded only a year earlier, in 2005, and, truth be told, I was not aware of its existence at that time.

The phone call reached me during a discussion with my collaborators in my office at the Physics Department. As I did not know what he was talking about, I asked him to send me some details about this fellowship before I could answer his question.

After receiving his e-mail in the evening, I was so excited about this possibility that I drank two glasses of red wine instead of the usual one. This fellowship had a very important impact on my research, not only until my retirement in 2012, but also well beyond it. It had the following benefits.

\begin{itemize}
\item
  I was free from teaching for three years.
\item
  I was free from the administration at my chair for three years still having my secretary for letters etc.
\item
A substitute professor was responsible for both teaching and administrative duties, except for the supervision of my diploma and PhD students, which I preferred to handle myself. The fellowship allowed me more time to dedicate to them.
  \item
  I could nominate somebody for the Junior Carl von Linde fellowship.
\item
  50.000 Euro for arranging small conferences.
\end{itemize}

I could have started this fellowship already in October 2007 but postponed
it by a year so that it lasted from October 2008 till September 2011.
In this manner only one Semester was left for teaching until my retirement.
For this semester I asked my colleagues to fullfy my ``last wish''. Namely
teaching my renormalization group course that did not require any preparations.

The substitute professors where Thorsten Feldmann (2008-2010) and Alex Lenz (2010-2011). This was a full professor position (C4 at that time). As Thorsten and Alex did
not have permanent positions at that time,  this serving as a substitute professor
was helpful in their careers. Thorsten received an associate professorship
at Siegen University in 2010 and Alex permanent position at Durham University in
2011. Eventually ten years later Alex received a full professorship (W3) at
the Siegen University as well. For the Junior Carl von Linde fellowship
I have chosen Martin Gorbahn, which I expect helped him in obtaining permanent position at the Liverpool University.

I served later at the committee which was deciding about this fellowship
and other IAS fellowships. The procedure was rather involved. But
in my case I practically did not have to do anything except for writing
two pages about my research plans during the fellowship. I was told later
that this simple procedure in my case resulted from a beer conversation of the
TUM president Wolfgang Herrmann with  my colleague Gerhard Abstreiter. Herrmann thought
that it would be good to have a theoretical physicist at IAS and wanted to know
from Gerhard who has the most citations among theorists in the Physics Department. It was me and in this manner I became one of the first IAS fellows and in
fact until today the only Carl von Linde Senior Fellow in our Theory Group.

But there was still another very important benefit which I will describe next.
\begin{figure}
\centering
\includegraphics[width = 0.90\textwidth]{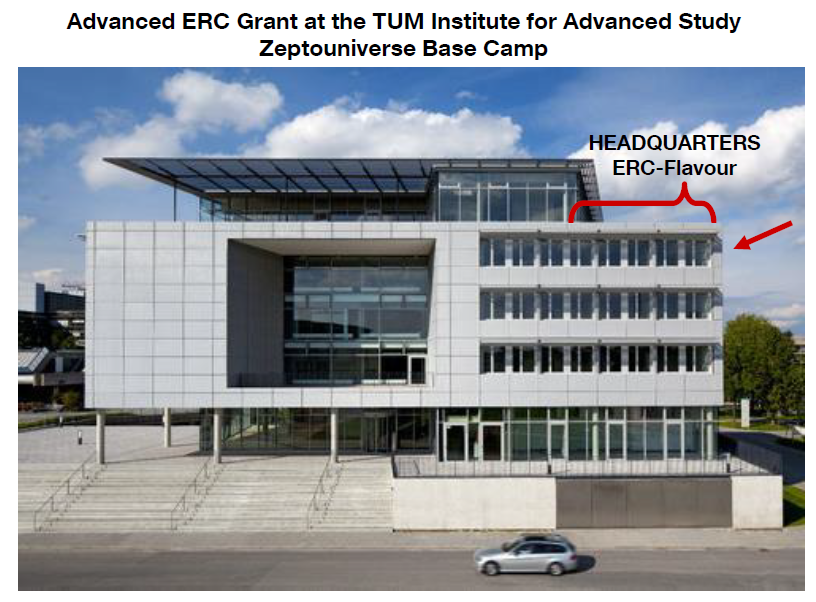}
 \caption{\it TUM-IAS. My office on the third floor has the first two windows on the right. }\label{NEWHOME}
\end{figure}

In December 2010, the IAS building was completed, constructed by BMW. A celebratory dinner was held that month in the Faculty Club on the building’s 4th floor. The offices were scheduled to be occupied a few weeks later, with each fellow assigned a space. During the dinner, I stood up and told those at my table that I would go around to find an office for myself. Everyone laughed, but it turned out to be a good idea.

Soon afterward, I found an office in the southeast corner of the 3rd floor, with a beautiful view of the Bavarian Alps. On clear days, one could see both Wendelstein and Zugspitze. It worked out perfectly: I got the office and still have it today. It is shown in Fig.~\ref{NEWHOME}, with the two windows on the right being mine.

From December 2010 until my retirement on April 1, 2012, I occupied both this office at IAS and my office in the Physics Department, gradually moving my papers and books from the Physics Department to IAS. I knew that securing a single office in the department as an Emeritus would be non-trivial, and in this way I could have a permanent office not far from the Physics Department—with the added bonus of a view of the Bavarian Alps. From my Physics Department office, I could see only the reactor. Moreover, my IAS office had air conditioning, whereas the Physics Department none.

Maintaining contacts with IAS proved crucial when I won the Advanced ERC Grant, which I began in the spring of 2011. The grant amounted to 1.6 million Euro, and I planned to host several PostDocs. I asked the Dean of the Physics Department how many offices I could have there for my ERC team. The answer was: one desk for me.

At first, this seemed absurd, given that it was the first Advanced ERC Grant in our department. However, it eventually worked to my advantage. I then approached the Excellence Cluster, which agreed to provide several rooms, though at a cost covered by my ERC overhead. Since my office was already at IAS, I asked the IAS administration whether additional offices could be made available there at no cost. They agreed, and in fact, until 2016, my ERC team occupied several offices on the third floor.

In addition to my own office ($18 m^2$), my group—called at IAS Fundamental Physics\footnote{Most of the groups at IAS these days are not related to physics at all.}—occupied five offices with a total area of $124 m^2$.

After 2016, with still some ERC overhead, I continued to host a few postdocs, though fewer rooms were occupied. The last postdoc, Jacky Kumar, left in the fall of 2022, leaving only my own office. Nevertheless, I can still occasionally arrange offices for visitors.

In addition to the Carl von Linde Fellowship for TUM professors, another notable fellowship at IAS is the Hans Fischer Fellowship for distinguished foreign scientists. I was delighted when two of my collaborators, Stefan Pokorski and Gino Isidori, received this fellowship and each spent approximately a year at IAS. Other Hans Fischer fellows included Andreas Kronfeld (Fermilab), while Humboldt Prize winners Estia Eichten (Fermilab) and Chris Quigg (Fermilab) also spent time at IAS, with Quigg visiting multiple times to give highly impressive pedagogical talks both at the Physics Department and in our Excellence Cluster.

Among my collaborators, Jennifer spent the longest period at IAS—almost four years—during which we wrote 19 papers together. All 11 papers with Fulvia were also written at IAS, with Fulvia visiting me twice.

The situation changed in 2021, after Christoph Bobeth left IAS. Today, at both the Physics Department and the MPI Institute for Physics—now located very close to the IAS building—practically no one works on BSM flavor physics. I now collaborate exclusively with scientists outside Munich, relying heavily on Zoom, Teams, and email. While this may seem obvious to younger generations, it is worth remembering that in the 1980s, when we developed DQCD, email was the only available tool for remote discussions.

\section{Meeting Robert Oppenheimer}\label{Oppenheimer}
During the process of revising Version 1 of this write-up, I reread the unpublished memoirs of my father, Bronisław Buras (1915–1994), an experimental nuclear physicist. These memoirs were originally intended only for our family. However, I came across one story that may be of interest to a broader audience, and I present it here. It was written in 1991.

The story was originally written in Polish and has been translated by me into English using ChatGPT. What follows are my father’s words, not mine.

From 1959 to 1961 I was Head of the Physics Section at the International Atomic Energy Agency in Vienna. During that time Professor Robert Marshak was organizing a major international Conference on High-Energy Physics in Rochester, in the United States. On the sidelines of this conference, discussions were planned concerning the possible construction—jointly involving the United States and the Soviet Union—of an international particle accelerator of very high energy. The Director General of the Agency, the American Sterling Cole, wished, in some as yet unspecified way, to involve the Agency in this project. He discussed the matter with me and proposed that I take part in the Rochester conference as the Agency’s official representative. The proposal was tempting, but I explained that I could not accept it, since I had little to do with that field of physics. It would be more appropriate to delegate a specialist. Cole, however, insisted that it had to be an Agency staff member and proposed that an expert accompany me. To my great astonishment, he mentioned the name… Robert Oppenheimer.

Robert Oppenheimer (1904-1967) was an outstanding physicist and at that time the director of one of the world’s leading scientific institutions, the Institute for Advanced Study in Princeton. He was widely known above all as the “father” of the atomic bomb. He had directed its construction in the years 1942–1945 at the secret nuclear research center in Los Alamos. He opposed the construction of the hydrogen bomb and was known for his left-wing sympathies. In 1953, a Congressional Commission, on the basis of an investigation conducted by the Federal Bureau of Investigation (FBI), declared Oppenheimer a “security risk” and deprived him of access to classified documents. For this reason, the choice of Oppenheimer surprised me greatly, especially since Cole himself had been a member of that Congressional Commission. The decision in the Commission had been taken by a majority vote of four to one. Perhaps Cole had voted against it?

I could not refuse the proposal that I travel together with Oppenheimer; I felt honored by it. That very day Cole wrote a letter to Oppenheimer inviting him to Vienna. It began with the words, “Dear Oppi.”

Some time later I picked Oppenheimer up at the airport and brought him to the Agency. The first part of Oppenheimer’s conversation with Cole took place in private; later I was invited to join them. Oppenheimer agreed to co-represent the Agency at the Rochester conference, and the three of us discussed the possible role of the Agency should the project be realized. Later, without Cole, we discussed practical issues connected with participation in the conference. In the evening we were guests of Cole for dinner. The next day I spoke with Oppenheimer a little about physics, a little about the current international situation, and a
little about this and that, after which I drove him to the airport.

Oppenheimer made a strong impression on me. Slim, tall, with a slender face, a penetrating gaze, and sharply sculpted features testifying to a strong character. I quickly sensed that I was speaking with an outstanding individual of extraordinary intelligence. This impression deepened further during my brief stay at his home in Princeton and during our conversations in Rochester. Above all, I was impressed by his vast knowledge in fields far removed from physics, by the precision of his thinking, and by his original and often surprising way of viewing things. One difficulty in conversation for me was his extraordinarily refined and rich vocabulary. I tried to remember some of the unfamiliar words and look them up later in a dictionary. Unfortunately, I could not find many of them. The English–Polish dictionary I owned—otherwise a good one—proved too limited. Not to mention the one encoded in my own brain! This troubled me in nearly every intellectual conversation.

On a certain day I flew to New York and from there went on to Princeton, where Oppenheimer was waiting for me at the Institute, as arranged. The host’s first act was to show me around the Institute. It was an unusual tour. We must have visited nearly every office, and the visit to each followed the same procedure. Oppenheimer would knock, open the door, introduce me to the occupant, who would mutter something under his breath that was presumably his name. If I happened to know the name from elsewhere, I could guess whom I was facing; otherwise it was difficult. As it soon turned out, the purpose of the tour was not to introduce me to people working in fields of physics unfamiliar to me. The purpose revealed itself in Oppenheimer’s comments: “In this room Albert Einstein worked; in this one, Hans Bethe; and in that one, for a short time, Max Born, and so on.” Unfortunately, after more than thirty years I do not remember all the names of the great physicists who worked in Princeton for shorter or longer periods. In any case, the most outstanding figures passed through there.

After discussing some details related to the conference, Oppenheimer took me to dinner at his home, where I was also to spend the night. After dinner there was another “tour,” this time of his house. What struck me most were the numerous objects originating mainly from the Near and Far East. These regions of the world, especially India, greatly interested Oppenheimer. This was his—perhaps not his only—hobby, and if I understood correctly, Sanskrit, the root of all European languages, was not foreign to him. I met his wife, and the three of us spent a pleasant and interesting evening, at least for me. Knowing that I was Polish, my hosts asked me about life in Poland—about the political, social, and economic situation, and about food supplies. The last topic particularly interested Mrs. Oppenheimer. To illustrate her method of provisioning the household, she took me to the basement, where two large top-opening freezers stood. She lifted the lid of one of them, and before my eyes appeared large cuts of meat in partially transparent plastic bags.

The next day came a surprise—on the front page of the New York Times, in large letters: Oppenheimer to Participate in the Rochester Conference with a Communist. I do not remember the exact title of the article, and the newspaper clipping I kept was lost somewhere during several moves. Oppenheimer gave no indication that it bothered him in any way, but for me the article somewhat clouded my stay in the United States. The conference concerned a field of physics with which I was not very familiar, so I listened with interest to several plenary lectures presenting the directions of research and the main achievements of high-energy and elementary-particle physics. The closed-session discussion concerning the large international accelerator came to nothing, as might have been expected during the period of the “Cold War.” For me, the most interesting aspect of the conference lay on its margins—the meeting with Robert Oppenheimer.
\vspace{0.6cm}

This brings my father’s story to a close, and with it the physics-related part of my recollections. Yet I would like to conclude this write-up by briefly describing two additional passions that have accompanied me throughout my life: mountains and classical music.

\section{Climbing Highest World Summits at Home}\label{Climb}
\subsection{Preface}
After the accident on July 14th 1967, described in Appendix~\ref{July14}, I reduced my climbing to hiking. First in High Tatra mountains and then in the Alps,
Rocky Mountains and Norway. But my interest in climbing did not
disappear.

During my summer vacation 1988 I read several books about the Himalaya 
expeditions. Among them the one by colonel Hunt, in which he described in 
detail the well known 1953 Mount Everest expedition that he organized. From 
these books I learned also about the competition between Reinhold Messner 
and polish climber Janusz Kukuczka to conquer the fourteen highest
Himalayan summits, the ones over 8000 m \footnote{Messner won this competition 
but Kukuczka was the second to reach all these highest summits by October 
1988. Unfortunately he died on October 24th, 1989 in an avalanche on the South 
Face of Lhotse.}. 
These were truly fantastic achievements but I wondered whether the difficulty 
of climbing a 8000 m high mountain by an experienced mountainer could be 
comparable in 1988 to the difficulty of a NLO calculation of weak decays 
performed by 
an experienced physicist like me. This comparison is not fully idiotic. 
After all the  difficult and often pioneering NLO and NNLO calculations in the last 
37 years required not only high technical skills
but also certain planning in advance and first of all psychological and 
physical strength. I mean here the ability to be involved in a calculation 
that results sometimes in a single number but lasts at least six months and 
often a year or longer. The air during  NLO and NNLO 
calculations can be very thin indeed.

These thoughts prompted me to perform the  QCD NLO analyses  of all relevant $K$ and $B$ decays including rare, 
radiative and in particular CP-violating decays. In 1988 this field was, with 
respect to NLO QCD corrections, essentially unexplored and all NLO summits were 
still waiting to be conquered. Being the first to complete all these 
calculations would certainly be an achievement with a lasting impact on the 
phenomenology of weak decays.

These were my dreams of 1988. Feeling like colonel Hunt before the Mount 
Everest expedition I made a list of 
most interesting decays and the corresponding operators. This list is given 
in Section~2.2 of my 2023 report on the results of this grand project 
\cite{Buras:2011we} and also in Table~\ref{TAB11} together with the list of 
climbers of various expeditions. Some details of various expeditions
have been given in Part~\ref{NLOQCD} of this write-up. I think this project
was successful.

As my last difficult QCD climb was performed 20 years ago 
\cite{Buras:2005gr,Buras:2006gb}, in the last 20 years I was involved dominantly
in searching for new animalcula which could also be considered as summits of a different type. These expeditions have been described in Parts~\ref{SBSM} and \ref{NPModels}.

Moreover, in these years I read numerous books about climbing
in Polish, German and English.  I will not list these
books but they dealt not only with climbing Himalayas but also with high
mountains at other places on earth. In particular with climbing the seven
summits, the highest summits on each continent. These are:
\begin{enumerate}
\item
  Mount Everest (8848 m) in Himalaya (China, Nepal).
\item
  Aconcagua (6961 m) in Andes moutains (Argentina, South America).
\item
  Denali (Mount McKinley) (6190 m) (Alaska, USA).
\item
  Kibo (5895 m) (Kilimandscharo, Tanzania).
\item
  Elbrus (5642 m) (Kaukasus, Russia).
\item
  Mount Vinson (4892 m) (Sentinel Range, Antarctis).
\item
  Puncak Jaya (4884 m) (Sudirman Mountains, Oceania/Australia).
\end{enumerate}

The description of all these summits can be found in Google so that I will not
describe them. 
Instead I will decribe  two events which one cannot find in
Google, at least until now.

As described at various places in this autobiography I met over the years several famous physicists like Dirac, Feynman, Weinberg, Glashow, Dirac, Gross, Wilczek, Nambu among others who receiving Nobel Prize climbed the highest summit
of particle physics. Indeed the Nobel Prize can be considered as Mount Everest
with two significant differences:
\begin{itemize}
\item
  In a given year at most three climbers are allowed to stand on this summit
  instead of many climbers on Mount Everest,
\item
  Instead of paying thousands of dollars for getting sharpas to achieve this
  goal one receives the prize of several hundred thousands of dollars.
  \end{itemize}
Yet, the climbing of high mountains is a big achievement and
I thought it would be interesting to have some contact with real climbers, in particular with those who climbed Mount Everest and 
Seven Summits. In fact I succeeded to have contact with 
two female climbers
who separately climbed the seven summites including therefore Mount Everest.
Let me describe the meetings with these two ladies.

\subsection{Helga Hengge}

In 2020 I read two books by Helga:
\begin{itemize}
\item
  ``Abenteuer SEVEN SUMMITS'' (2015). 
\item
  ``Nur der Himmel ist Höher'' (2017). English version: ``My journey to the top of the world''
\end{itemize}

The first  book describes Helga's climbing of the seven summits from
1997 til 2011. The second concentrates on Mount Everest. They are very well written, better than many books
on climbing I read, possibly because Helga is a journalist and a very inspiring
keynote speaker. After I read both books and listened to her talks online
and several interviews I could not resist to contact her per email. We
exchanged a number of emails in 2020. Recently I had contact with her
again. She congratulated me for reaching my Sakurai Summit and stated
that this climb  must be at least as difficult as climbing the east wall of Mount Everest. I doubt this is true but definitely I cannot judge it and definitely it was safer to climb the Sakurai Summit than this wall. Helga is writing a new book which she considers as her ``Mount Everest 2026'' although as far as I understan has nothing to do Himalayas. I am looking forward to reading it.

\subsection{Ania Lichota}

Many years ago, I do not remember when, I met Eva Klaput, the mother of one
of PhD students of Manfred Lindner. She was also climbing at various places
in the world but mainly mountains below 6000 m. She was a friend of Wanda
Rutkiewicz, {\bf the first European woman and the third woman in the world
who climed Mount Everest (1978).} She was  the one who
climed several Himalaya Mountains over 8000 m. She could have been the first
female who climed all 14 mountains over 8000 m but in 1992 she died
while climbing the third highest mountain on earth, Kanchenjunga, after
Mount Everest and K2. I never met Wanda although possibly she was at
Morskie Oko in 1967 when I was there. She was 24 at that time.

But Eva told me about a much younger climber, Ania Lichota, and somehow
I got an Email contact with her and even got an Email from her one month
before she climed Mount Everest in 2010. She wrote one book that is not only in
polish but also in english:
\begin{itemize}
\item
  ``Why the hell Bother?: How climbing the Seven Summits changed my life''
\end{itemize}
A very interesting book. It describes her climbing of seven summits from 2006
til 2010.

Finally on the 8th of December 2025 I talked to Ania using
WhatsApp. We talked only 30 min but it was a very nice meeting.
After the 7 summits she still climbed Lhotse (8516 m) and Makalu (8485 m),
the fourth and the fifth highest mountain on the earth. Lhotse is considered
as one of the most difficult mountains to climb, more difficult than Mount Everest
and possibly only K2 is more difficult but this is cleary far beyond
my knowledge.

\section{Addicted to Classical Music: 1982-2046}
My memories would be incomplete if I did not mention at the end my second
passion beyond  physics. It is classical
music. I do not play any instrument but I listen to classical music all the time
when I am alone, in particular when I work. In addition in the last 40 years I
read many more books about great composers, violinists, pianists and conductors  than those about famous physicists and famous climbers together.

Still in Poland, Denmark, Switzerland and USA I got interested in it, 
listened to classical music from time to time and going occassionaly to
concerts, but my interest in classical music
increased by much once we moved to Munich. This is not  surprising because
Munich has much more to offer in this case than other places I have been before.
But going just to concerts is definitely insufficient for me. Fortunately,
since we moved to Munich, I always had a single office and could also listen
to classical music at MPI, Physics Department and also IAS after my retirement without using
head phones which as far as I know are not optimal for your hearing.

Most of my colleagues do not really understand how I can work and listen to
classical music at the same time. On the other hand there are some claims that listening to Mozarts music increases your productivty, which if true would have significant impact on
my research, Mozart is one of my favourite composers. But there is no scientific proof of it. I learned it from a Hans Fischer fellow of IAS, an expert in the History
of music in Princeton one day.

That day being in my office at IAS and not inspecting anybody in the adjacent offices
to mine I played Beethoven symphonies louder than usually.
Suddenly somebody knocked at my door, entered my office and said: ``it is a beautiful
music but could you please put it a bit down.'' Of course I did it but
then it turned out that this person was a real expert in classical music and we had
a nice conversation so that it payed off to play loudly afterall.

{\bf In any case, in my talk at my 70th birthday I thanked Beethoven, Chopin, and Mozart for all the great support they provided during my research, and advised the audience: if you have a difficult problem to solve, listen to Beethoven’s Triple Concerto and you will solve it.}

Until 2015 we rarely went to the opera. This changed when our daughter Karin began working at the Bavarian State Opera as a stage dancer and performer in 2015 and in 2023 became member of a semiprofessional choir. For this reason,  beyond usual concerts with classical music, we now attend the opera or other performances several times a year to watch her.

Until October 2020 I did not meet any physicist who was as passionated  about classical
music as me. But in October 2020 I got contact with a postdoc in Andreas Weiler group,
Elena Venturini, who turned out to be as crazy about the classical music as
I am. In fact since our first meeting we exchange emails with music every
day, although unfortunately since April 2022 she is not in Munich but  in Trieste teaching physics and mathematics in a high school. Still we succeeded
to compose
two papers together on the flavour phenomenology without
$V_{cb}$ uncertainties \cite{Buras:2021nns,Buras:2022wpw} which I described
in Part~\ref{BV}.

Now, similarly to my interest in mountain climbing, I wanted to make some contact with musicians. In 2020, I attended a concert where Patricia Kopatchinskaja played Schumann's Violin Concerto. I was not familiar with this concerto, and there are reasons for that, despite my view that it is one of the finest violin concertos. Apparently, Schumann composed it in 1853, just a few years before his death in 1856, and its premiere was not realized during his lifetime. Moreover, after his death, his wife Clara Schumann and the well-known violinist Joseph Joachim (1831–1907) judged the concerto to be of insufficient quality, so it remained unknown for 84 years—until the Nazis, having banned Mendelssohn’s Violin Concerto, replaced it with Schumann’s. The premiere eventually took place on 26 November 1937 in Berlin, with the renowned violinist Georg Kulenkampff and the Berlin Philharmonic, conducted by Karl Böhm.

I was so impressed by this concerto that I could not resist sending an email to Patricia. A few days later, I received a reply—not from her, but from her husband, Lukas Fierz, the son of the famous physicist Markus Fierz (1912–2006), whom we know from the Fierz transformation mentioned several times in this write-up. Markus Fierz had won the Max Planck Medal in 1979. Lukas, on the other hand, is a well-known writer living in Bern. His email was very kind; he even invited me to spend some time at his home should I ever be in Bern, as his house is quite close to the Albert Einstein Center for Fundamental Physics at Bern University. I did not realize it at the time, but perhaps it may work out one day. He also sent me one of his books, along with a photograph of his father playing the violin at the age of 92.

Most importantly, Lukas checked online who I was, and the next day Patricia herself contacted me, thanking me for praising her concert. This seemed to help her, as she had been dissatisfied with her performance that day.

We exchanged few Emails in which she, among other things, criticized the performance of this
concerto by some known violinists whose names I will not mention here.
She wrote this email during the night from Moscow where she performed
together with the conductor Teodor Currentzis.
Certainly her way of playing is very innovative. One can check it by
listening to her performance of the Beethoven's violin concerto.

But what impressed me in addition by her, was  her personality, in particular
the way she formulated
her emails to me. After all she did not know me. But already her
first Email started with ``Lieber Andrzej'', rather untippical in
Germany for the first Email to a person one does not know. I did not
have contact with her already for several years but maybe it will work out
one day to meet her in person.

        \renewcommand{\refname}{R\lowercase{eferences}}

\addcontentsline{toc}{section}{References}
 
\bibliographystyle{JHEP}
\bibliography{flavour}
\end{document}